\documentclass{aastex63}
\received{}
\revised{}
\accepted{2023 October 2}
\submitjournal{ApJ}

\shorttitle{Blue wing asymmetries in chromospheric lines during mid M dwarf flares}
\shortauthors{Notsu et al.}
\graphicspath{{figures/}}

\begin{document}

\title{ 
APO \& SMARTS flare star campaign observations I.
Blue wing asymmetries in chromospheric lines during mid M dwarf flares from simultaneous spectroscopic and photometric observation data
}

\correspondingauthor{Yuta Notsu}
\email{Yuta.Notsu@colorado.edu}

\author[0000-0002-0412-0849]{Yuta Notsu}
\affil{Laboratory for Atmospheric and Space Physics, University of Colorado Boulder, 3665 Discovery Drive, Boulder, CO 80303, USA}
\affil{National Solar Observatory, 3665 Discovery Drive, Boulder, CO 80303, USA}

\author[0000-0001-7458-1176]{Adam F. Kowalski}
\affil{Laboratory for Atmospheric and Space Physics, University of Colorado Boulder, 3665 Discovery Drive, Boulder, CO 80303, USA}
\affil{National Solar Observatory, 3665 Discovery Drive, Boulder, CO 80303, USA}
\affil{Department of Astrophysical and Planetary Sciences, University of Colorado Boulder, 2000 Colorado Ave, Boulder, CO 80305, USA}

\author[0000-0003-0332-0811]{Hiroyuki Maehara}
\affil{Okayama Branch Office, Subaru Telescope, National Astronomical Observatory of Japan, NINS, Kamogata, Asakuchi, Okayama 719-0232, Japan}

\author[0000-0002-1297-9485]{Kosuke Namekata}
\affil{ALMA Project, National Astronomical Observatory of Japan, NINS, Mitaka, Tokyo, 181-8588, Japan}

\author[0000-0001-7515-2779]{Kenji Hamaguchi}
\affiliation{CRESST II and X-ray Astrophysics Laboratory, NASA/GSFC, Greenbelt, MD 20771, USA}
\affiliation{Department of Physics, University of Maryland, Baltimore County, 1000 Hilltop Circle, Baltimore, MD 21250, USA}

\author[0000-0003-1244-3100]{Teruaki Enoto}
\affil{Department of Physics, Kyoto University, Sakyo, Kyoto 606-8502, Japan}
\affiliation{RIKEN Cluster for Pioneering Research, 2-1 Hirosawa, Wako, Saitama 351-0198, Japan}

\author[0000-0001-5974-4758]{Isaiah I.Tristan}
\affil{Laboratory for Atmospheric and Space Physics, University of Colorado Boulder, 3665 Discovery Drive, Boulder, CO 80303, USA}
\affil{National Solar Observatory, 3665 Discovery Drive, Boulder, CO 80303, USA}
\affil{Department of Astrophysical and Planetary Sciences, University of Colorado Boulder, 2000 Colorado Ave, Boulder, CO 80305, USA}

\author[0000-0002-6629-4182]{Suzanne L. Hawley}
\affil{Department of Astronomy, University of Washington, Seattle, WA 98195, USA}

\author[0000-0002-0637-835X]{James R. A. Davenport}
\affil{Department of Astronomy, University of Washington, Seattle, WA 98195, USA}

\author[0000-0001-6653-8741]{Satoshi Honda}
\affil{Nishi-Harima Astronomical Observatory, Center for Astronomy, University of Hyogo, Sayo, Hyogo 679-5313, Japan}

\author[0000-0002-5978-057X]{Kai Ikuta}
\affil{Department of Multidisciplinary Sciences, The University of Tokyo, 3-8-1 Komaba, Meguro, Tokyo 153-8902, Japan}

\author{Shun Inoue}
\affil{Department of Physics, Kyoto University, Sakyo, Kyoto 606-8502, Japan}

\author{Keiichi Namizaki}
\affil{Department of Astronomy, Kyoto University, Sakyo, Kyoto 606-8502, Japan}

\author{Daisaku Nogami}
\affil{Department of Astronomy, Kyoto University, Sakyo, Kyoto 606-8502, Japan}

\author{Kazunari Shibata}
\affil{Kwasan Observatory, Kyoto University, Yamashina, Kyoto 607-8471, Japan}
\affil{School of Science and Engineering, Doshisha University, Kyotanabe, Kyoto 610-0321, Japan}



\begin{abstract}

We conducted the time-resolved simultaneous optical spectroscopic and photometric observations of mid M dwarf flare stars YZ CMi, EV Lac, and AD Leo.
Spectroscopic observations were obtained using Apache Point Observatory 3.5m and Small \& Moderate Aperture Research Telescope System 1.5m telescopes during 31 nights.
Among the \color{black}\textrm{41 }\color{black} detected flares, 
seven flares showed clear blue wing asymmetries in the H$\alpha$ line, 
with various correspondences in flare properties.
The duration of the blue wing asymmetries range from 20 min to 2.5 hours, 
including a flare showing the shift from blue to red wing asymmetry.
\color{black}\textrm{
Blue wing asymmetries can be observed during both white-light and candidate non white-light flares. }\color{black}
All of the seven flares showed blue wing asymmetries also in the H$\beta$ line, 
but there are large varieties on which other chromospheric lines 
showed blue wing asymmetries. 
One among the 7 flares was also observed with soft X-ray spectroscopy, 
which enabled us to estimate 
the flare magnetic field and length of the flare loop.  
The line-of-sight velocities 
of the blue-shifted components range from -73 to -122 km s$^{-1}$. 
Assuming that the blue-shifts were caused by prominence eruptions,
the mass of upward moving plasma was estimated to be 
10$^{15}$ -- 10$^{19}$ g, which are roughly on the relation 
between flare energy and erupting mass expected from solar coronal mass ejections (CMEs). 
Although further investigations are necessary for understanding the observed various properties, 
these possible prominence eruptions on M-dwarfs could evolve into CMEs, 
\color{black}\textrm{assuming the similar acceleration mechanism with solar eruptions.}\color{black}

\end{abstract}

\keywords{Stellar flares (1603); Stellar coronal mass ejections (1881); Optical flares (1166); Stellar x-ray flares (1637); M dwarf stars (982); Flare stars (540); Red dwarf flare stars (1367); Stellar chromospheres (230)}


\section{Introduction} \label{sec:intro}
Solar flares are sudden brightness increases resulting from impulsive magnetic energy release through magnetic reconnection in the solar atmosphere (e.g., \citealt{Shibata+2011} and references therein).
They generate strong emissions at various wavelengths ranging from radio to high-energy X-rays/gamma-rays.
In the flaring solar atmosphere, part of the magnetic energy released by the magnetic reconnection in the corona is transported into the lower atmosphere (chromosphere and upper photosphere) through thermal conduction, radiative backwarming, and high-energy particles (e.g., high energy electrons). This process causes chromospheric evaporations and chromospheric condensations, producing bright coronal (e.g., X-ray), chromospheric (e.g., H$\alpha$), and photospheric radiation (e.g., \citealt{Fisher+1985}; \citealt{Allred+2005}).

It is also known that various types of stars produce flares (``stellar flares"). They are observed mainly as rapid increases and slow decays of the intensity in various wavelength bands (radio, optical, ultraviolet (UV), X-ray).
In particular, young rapidly-rotating stars, close binary stars, and magnetically active M-type main sequence stars (dMe stars) show frequent and energetic flares (e.g., \citealt{Lacy+1976}; \citealt{Hawley+1991}; \citealt{Shibata_Yokoyama+2002}; \citealt{Gershberg+2005}; \citealt{Reid+2005}; \citealt{Benz+2010}). 
Recent photometry from \textit{Kepler}/\textit{TESS} space telescopes (e.g., \citealt{Maehara+2012} \& \citeyear{Maehara+2021}; \citealt{Shibayama+2013}; \citealt{Notsu+2013} \&  \citeyear{Notsu+2019}; \citealt{Okamoto+2021}; \citealt{Hawley+2014}; \citealt{Davenport+2016}; \citealt{Davenport+2020}; \citealt{Paudel+2019}; \citealt{Feinstein+2020};  \citealt{Medina+2020}) and ground-based surveys (e.g., \citealt{Howard+2019}; \citealt{Jackman+2021}) have helped to refine statistical properties of stellar flares. 
Because of the similarities in observational properties between solar and stellar flares such as Neupert effect (\color{black} 
\textrm{\citealt{Neupert+1968};
}
\color{black}\citealt{Hawley+1995}; \citealt{Guedel+1996} \& \citeyear{Guedel+2004}; \citealt{Mitra-Kraev+2005a}; 
\color{black}\textrm{\citealt{Tristan+2023_ApJ}}\color{black}), they are considered to be caused by the same physical processes (i.e., plasma heating by accelerated particles and subsequent chromospheric evaporation, e.g., \citealt{Hawley+1992}; \citealt{Allred+2006}; \citealt{Kowalski2016}; \citealt{Namekata+2020_PASJ}).
Recently, stellar flares have been also getting more and more attention in terms of the effects on the exoplanet atmosphere/habitability (e.g., \citealt{Lammer+2007}; \citealt{Segura+2010}; \citealt{Airapetian+2016} \& \citeyear{Airapetian+2020}; \citealt{Linsky2019}) and possible extreme events on our Sun (\citealt{Aulanier+2013}; \citealt{Shibata+2013}; \citealt{Battersby+2019};  \citealt{Miyake+2019}; \citealt{Notsu+2019}; \citealt{Okamoto+2021}; \citealt{Usoskin+2021}; \citealt{Cliver+2022}; \citealt{Buzulukova-Tsurutani_2022}).

Spectroscopic studies of solar and stellar flares have been carried out in order to understand the dynamics of plasma and the radiation mechanisms during flares.
Past spectroscopic observations of solar flares have shown that chromospheric lines (e.g., H$\alpha$, H$\beta$, Ca II, Mg II) often exhibit asymmetric line profiles during flares. In particular, red wing asymmetries (enhancement of the red wing) have often been observed during solar flares (e.g., \citealt{Svestka+1962}; \citealt{Ichimoto+1984}; \citealt{Canfield+1990}; \citealt{Shoji+1995}; \citealt{Berlicki+2007}; \citealt{Graham+2015}; \citealt{Kuridze+2015}; \citealt{Kowalski+2017}; \citealt{Graham+2020}; \citealt{Namekata+2022_ApJ}; \citealt{Otsu+2022}).
This is thought to be caused by the process known as the chromospheric condensation, which is the downward flow of cool plasma in the chromosphere, while \citet{Kuridze+2015} also reported \color{black} 
\textrm{
the enhancement of red side of the line profile around the line center (line center red asymmetry) 
}
\color{black}
caused by upflows.
Blue wing asymmetries (enhancement of the blue wing) have also been observed mainly in the early phase of solar flares (e.g.,  \citealt{Svestka+1962}; \citealt{Canfield+1990}; \citealt{Heinzel+1994_A&A}; \citealt{Kuridze+2016}; \citealt{Tei+2018}; \citealt{Panos+2018}; \citealt{Huang+2019}; \citealt{Li+2019}; \citealt{Libbrecht+2019}). 
\citet{Svestka+1962} found that among the 92 studied solar flares, only 23\% showed a blue asymmetry, while 80\% exhibited a red asymmetry. \citet{Tang+1983} reported the similar result that an even lower fraction of 5\% of flares with blue asymmetries.
As one possibility it is suggested that blue asymmetry is caused by an upward-moving cool plasma blob, which is lifted up by expanding hot plasma caused by the deep penetration of non-thermal electrons into the chromosphere during a flare (\citealt{Tei+2018}; \citealt{Huang+2019}; \citealt{Zhu+2019}; \citealt{Hong+2020}). 
However, the detailed origins and properties of these blue asymmetries are still controversial.
\color{black} 
\textrm{For example, these asymmetries could be also caused by either excess or lack of flux on one side of the line profile without upward-moving plasma.
}
\color{black}

Similar line asymmetries in chromospheric lines (especially H$\alpha$ line) have been observed during stellar flares on late-type stars (e.g., M-dwarf flare stars). 
For example, \citet{Houdebine+1993}, \citet{Crespo-Chacon+2006}, \citet{Wollmann+2023_A&A}, and \citet{Namizaki+2023_ApJ} reported red asymmetries in Balmer lines during flares on M-dwarf flare stars. 
Moreover, in the case of M dwarf flare stars,
various blue asymmetries have been widely observed (e.g., \citealt{Houdebine+1990}; \citealt{Eason+1992}; \citealt{Gunn+1994};
\citealt{Crespo-Chacon+2006}; \citealt{Hawley+2007}; 
\citealt{Fuhrmeister+2008} \& \citeyear{Fuhrmeister+2011}; \citealt{Vida+2016} \& \textrm{\citeyear{Vida+2019}}; \citealt{Honda+2018}; \citealt{Muheki+2020} \& \citeyear{Muheki+2020_EVLac} \citealt{Maehara+2021}; \citealt{Johnson+2021}).
Not only on M-dwarfs, but blue asymmetries or blue-shifted absorption during flares have been also observed during flares on the young active K dwarf LQ Hydrae \citep{FloresSoriano+2017} and young active G dwarf (solar-type star) EK Dra \citep{Namekata+2022_NatAst}.

These previous studies have clarified large varieties of blue asymmetries. \citeauthor{Fuhrmeister+2008} (\citeyear{Fuhrmeister+2008} \& \citeyear{Fuhrmeister+2011}) investigated flares on M5.5 dwarfs Proxima Centauri and CN Leo, respectively, and they both
reported blue asymmetry of H$\alpha$ line during flare onset and red asymmetry during decay, along with temporal evolution in the asymmetry pattern on the scale of minutes.
\citet{Vida+2016} reported several H$\alpha$ flares on
the M4 dwarf V374 Peg showing blue wing asymmetries of H$\alpha$, H$\beta$, and H$\gamma$
lines with a maximum line-of-sight velocity of -675 km s$^{-1}$.
\citet{Vida+2016} also found that red-wing enhancements in the H$\alpha$ line were observed after blue wing asymmetries, which can suggest that some of the erupted cool plasma fell back on the stellar surface.
\citet{Honda+2018} reported a long-duration H$\alpha$ flare on the M4.5 dwarf EV Lac. During this flare, a blue wing asymmetry in the H$\alpha$ line with the velocity of $\sim$ -100 km s$^{-1}$ has been observed for $\gtrsim$2 h (almost from flare start to end). 
\citet{Maehara+2021} reported a H$\alpha$ flare without clear brightening in continuum,
which exhibited blue wing asymmetry lasting for $\sim$1 hour.
The line-of-sight motions of cool plasma such as prominence eruptions 
can cause blue/red wing asymmetries of stellar spectral lines.
Solar prominences and filaments are cool plasma blobs ($\sim$10,000K)
in the hot corona ($\gtrsim10^{6}$K), 
and they are observed above the limb (prominences) and on the disk (filaments), respectively (\citealt{Parenti_2014_LRSP}; \citealt{Vial+2015_ASSL}). In the case of the Sun, prominences are commonly observed in emission of Balmer lines (especially H$\alpha$ line), while filaments show absorption lines (\citealt{Parenti_2014_LRSP}; \citealt{Otsu+2022}).
The eruptions of such solar prominences and filaments are often observed, 
and they are often associated with solar flares (e.g., \citealt{Shibata+2011}; \citealt{Sinha+2019_ApJ}).
Such filament/prominence eruptions can evolve into CMEs (coronal mass ejections) if the erupted plasma is accelerated enough 
\color{black}\textrm{until }\color{black} the velocity exceeds the escape velocity 
(e.g., \citealt{Gopalswamy+2003}; \citealt{Seki+2021}).
In analogy with solar prominence eruptions, 
if the cool stellar plasma is launched upward and seen above the limb \footnote{
\color{black}\textrm{
\citet{Leitzinger+2022} discussed that even filaments can cause enhancements in Balmer lines of M-dwarfs. It may be speculated that the plasma, which causes the enhancements, is not necessary to be seen above the limb, but is possible to be seen also on the disk.
}
}, 
the emission can cause blue-shifted or red-shifted enhancements 
of the Balmer lines (e.g., \citealt{Odert+2020}), and this can be eventually related with CMEs.

Several recent studies have discussed the blue wing asymmetries 
assuming they can be related with stellar mass ejections.
\citet{Vida+2019} reported a statistical study of 478 events 
with asymmetries in Balmer lines of M-dwarfs, which were found from
more than \color{black} \textrm{5500 spectra (similar events were also reported in \citealt{Fuhrmeister+2018})}\color{black}.  
The velocity and mass of the possible ejected materials estimated from the blue-shifted or red-shifted excess in Balmer lines range 
from 100--300 km s$^{-1}$ and $10^{15} - 10^{18}$ g, respectively.
The correlations between the mass/kinetic energy of CMEs 
and the flare energy of associated flares on various types of stars 
have been
discussed (e.g., \citealt{Moschou+2019}; \citealt{Maehara+2021}; \citealt{Namekata+2022_NatAst}).
They found that estimated stellar flare CME masses are consistent 
with the trends extrapolated from solar events 
but \color{black}\textrm{kinetic }\color{black} energies are roughly two orders of magnitude smaller than expected. 
\citet{Maehara+2021} and \citet{Namekata+2022_NatAst} suggested that this could be understood by the difference in the velocity between CMEs and prominence eruptions.

It is important to understand the properties of stellar CMEs in order to evaluate their effects not only on the mass and angular momentum loss of the star (e.g., \citealt{Osten+2015}; \citealt{Odert+2017}; \citealt{Cranmer2017}; \citealt{Wood+2021}), but also on the habitability of exoplanets (e.g., loss of atmosphere, atmospheric chemistry, radiation dose as described in \citealt{Lammer+2007}; \citealt{Segura+2010}; \citealt{Airapetian+2016} \& \citeyear{Airapetian+2020}; \citealt{Scheucher+2018}; \citealt{Tilley+2019}; \citealt{Yamashiki+2019}; \citealt{Chen+2021}). 
Attempts to detect stellar CMEs have not yet found the expected Type-II radio signatures (e.g., \citealt{Crosley+2018}; \citealt{Villadsen+2019}), but several candidates have been reported: line blue asymmetries (described above), post-flare UV / X-ray dimmings (\citealt{Veronig+2021}; \citealt{Loyd+2022}), and a possible Type-IV radio burst (\citealt{Zic+2020}).
However, our understanding of blue/red asymmetries in chromospheric lines and their connections with stellar flares/CMEs is still limited by the low number of samples observed in time-resolved spectroscopy simultaneously with high-precision photometry 
(see also \citealt{Namekata+2022_IAUS} and \citealt{Leitzinger_Odert_2022_Review} for brief reviews of current stellar CME observations).

In order to investigate the connection between the blue/red asymmetries in Balmer lines and the properties of flares more in detail,
we conducted the time-resolved simultaneous optical spectroscopic and photometric observations of mid M dwarf flare stars, during the 31 nights over two years (2019 January -- 2021 February). 
Spectroscopic observations were conducted using the high-dispersion spectrographs of the Apache Point Observatory (APO) 3.5m telescope and the Cerro Tololo Inter-American Observatory (CTIO) Small and Moderate Aperture Research Telescope System (SMARTS) 1.5m telescope. Photometric observations were obtained from the 0.5m telescope of APO (ARCSAT: Astrophysical Research Consortium Small Aperture Telescope) and 1m \&~0.4m telescopes of the Las Cumbres Observatory Global Telescope (LCOGT) network, while 5 nights are also covered with the observation window of the Transiting Exoplanet Survey Satellite (\textit{TESS}).
For the 3 nights among the total 31 nights, 
we also conducted the soft X-ray spectroscopic observations with Neutron Star Interior Composition ExploreR (\textit{NICER}).
In Section \ref{sec:data-methods}, the details of our campaign monitoring observations and data analyses are described.
In Section \ref{sec:results}, we report 
the light curves and H$\alpha$ \& H$\beta$ spectra of the flares detected in our campaign observations. We investigate whether the flares show blue wing asymmetries in H$\alpha$ \& H$\beta$ lines. If the blue wing asymmetries are observed in H$\alpha$ \& H$\beta$ lines, we also investigate whether other chromospheric lines (e.g., H$\gamma$, H$\delta$, Ca II K) also show blue 
\color{black} \textrm{wing }\color{black} 
asymmetries. 
In Section \ref{sec:discussions}, we discuss 
the various properties of flares with blue wing asymmetries, and 
the implications from the observed blue \color{black} \textrm{wing }\color{black} asymmetries on the possible stellar mass ejections.

\section{Data and Methods} \label{sec:data-methods}
\subsection{Target Stars} \label{subsec:data-methods}
During the 31 nights over two years (2019 January -- 2021 February), we conducted time-resolved simultaneous optical spectroscopic and photometric observations of the three nearby mid M dwarf flare stars YZ CMi, EV Lac, and AD Leo.
The basic parameters of these three target flare stars are listed in Table \ref{table:targets_basic_para}. The log of the observations is summarized in Table \ref{table:obs_log}. 
These three stars have been known to flare frequently (e.g., \citealt{Lacy+1976}; \citealt{Hawley+1991}; \citealt{Kowalski+2013}; \citealt{Namekata+2020_PASJ}; \citealt{Muheki+2020_EVLac}; \citealt{Maehara+2021}; \color{black} \textrm{\citealt{Paudel+2021_ApJ}; \citealt{Ikuta+2023_ApJ}}\color{black}).
Zeeman-broadening and Zeeman-Doppler Imaging observations of these stars have shown the existence of strong magnetic fields on the stellar surface (e.g., \citealt{Saar+1985}; \citealt{Johns-Krull+2000}; \citealt{Reiners+2007}; \citealt{Morin+2008}; \citealt{Kochukhov+2021}).

\begin{deluxetable*}{lcccccccccc}[htbp]
   \tablecaption{Target star basic parameters}
   \tablewidth{0pt}
   \tablehead{
     \colhead{Starname} & \colhead{Spectral Type\color{black}\tablenotemark{\rm \dag}} & \colhead{$U$\tablenotemark{\rm \dag}}  & 
     \colhead{$g$\tablenotemark{\rm \dag}}  & \colhead{$V$\tablenotemark{\rm \dag}} &
     \colhead{$d_{\rm{Gaia}}$\tablenotemark{\rm \ddag}} &
     \colhead{$R_{\rm star}$\tablenotemark{\rm \dag}} &
     \colhead{$P_{\rm{rot}}$\tablenotemark{\rm \#}} &
     \colhead{$v\sin i$\tablenotemark{\rm *}} &
     \colhead{$i$\tablenotemark{\rm *}} \\
     \colhead{} & \colhead{} &  \colhead{(mag)}   & \colhead{(mag)}  & \colhead{(mag)} & \colhead{(pc)} & \colhead{($R_{\odot}$)} & \colhead{(day)}  & \colhead{(km s$^{-1}$)} &
     (deg)}
   \startdata
   YZ CMi (Gl 285) & dM4.5e & 13.77 & 11.76 & 11.19 & 5.99 & 0.30 & 2.77 & 5.0 & 60 \\
   EV Lac (Gl 873) & dM3.5e & 12.96 & 10.99 & 10.28 & 5.05 & 0.36 & 4.30 & 4.0 & 60 \\
   AD Leo (Gl 388) & dM3e & 11.91 & 10.12 & 9.32 & 4.97 & 0.43 & 2.24 & 3.0 & 20 \\
   \enddata
 \tablenotetext{\dag}{
The \textrm{\color{black}Spectral type,\color{black}} $U$ and $V$-band magnitudes, and stellar radius ($R_{\rm star}$) values are from Table 1 of \citet{Kowalski+2013}.
The $g$-band magnitudes are from \citet{Zacharias+2013}. 
   }
\tablenotetext{\ddag}{
Stellar distance from Gaia DR2 catalog \citep{GaiaCollaboration+2018}.
   }
\tablenotetext{\#}{
Rotation Period values ($P_{\rm rot}$). The values of YZ CMi and EV Lac are those estimated from \textit{TESS} data in \citet{Maehara+2021} and \citet{Muheki+2020_EVLac}, respectively. The $P_{\rm rot}$ value of AD Leo is from \citet{Morin+2008}.}
\tablenotetext{*}{
Projected rotational velocities ($v\sin i$) and 
stellar inclination angle values ($i$), reported in \citet{Morin+2008}.
}
   \label{table:targets_basic_para}
 \end{deluxetable*}

\startlongtable 
\begin{deluxetable*}{lccc}
   \tablecaption{Observation log}
   \tablewidth{0pt}
   \tablehead{
     \colhead{Telescope/Instrument} & \colhead{UT date (MJD)} & \colhead{Time \tablenotemark{\rm \dag}} & \colhead{Exp. Time}  \\
 \colhead{(Data Type)} & \colhead{} & \colhead{(h)} & \colhead{[sec]}  
     }
   \startdata
 \multicolumn{4}{c}{YZ CMi}  \\
    \hline 
   ARC 3.5m/ARCES & 2019 Jan 26 (58509) & 7.2 & 600, 900  \\
   (3800--10000\AA; $\lambda/\Delta\lambda \sim 32000$) & 2019 Jan 27 (58510) & 7.5 & 300   \\
    & 2019 Jan 28 (58511) & 7.2 & 300, 600    \\
    & 2019 Dec 02 (58819) & 1.8 & 300  \\
    & 2019 Dec 08 (58825) & 2.4 & 300  \\
    & 2019 Dec 12 (58829) & 5.8 & 300   \\
    & 2019 Dec 15 (58832) & 1.9 & 300  \\
    & 2020 Jan 14 (58862) & 3.9 & 300   \\
    & 2020 Jan 18 (58866) & 4.9 & 300, 600    \\
    & 2020 Jan 20 (58868) & 4.9 & 300, 450, 600  \\
    & 2020 Dec 03 (59186) & 4.1 & 300, 600  \\
    & 2020 Dec 06 (59189) & 5.3  & 300, 450  \\
    & 2020 Dec 07 (59190) & 5.7 & 300, 360  \\
    & 2021 Jan 31 (59245) & 9.9 & 450, 600, 900  \\
    & 2021 Feb 04 (59249) & 1.8 & 900  \\
    \hline 
  SMARTS 1.5m/CHIRON & 2020 Jan 16 (58864) & 5.2 & 600  \\
   (4500--8900\AA; $\lambda/\Delta\lambda \sim 25000$) & 2020 Jan 17 (58865) & 4.2 & 600 \\
   & 2020 Jan 18 (58866) & 6.0 & 600  \\
   & 2020 Jan 19 (58867) & 5.0 & 600  \\
   & 2020 Jan 20 (58868) & 6.0 & 600  \\
   & 2020 Jan 21 (58869) & 6.0 & 600  \\
   & 2020 Jan 22 (58870) & 6.0 & 600  \\
   & 2020 Jan 23 (58871) & 6.0 & 600  \\
    \hline 
  ARCSAT 0.5m/flarecam & 2019 Jan 26 (58509) & 1.0 ($u$), 7.0 ($g$) & 30 ($u$), 4, 15, 30 ($g$)  \\
   ($u, g$-band photometry)\tablenotemark{\rm \ddag} & 2019 Jan 27 (58510) &  7.3 ($u$), 7.3 ($g$) & 30 ($u$), 4 ($g$)   \\
   & 2019 Jan 28 (58511) & 7.0 ($u$), 7.3 ($g$) & 30 ($u$), 4, 6, 12, 20 ($g$)   \\
   & 2019 Dec 02 (58819) & 1.2 ($u$), 1.2 ($g$) & 30 ($u$), 6 ($g$)  \\
   & 2019 Dec 12 (58829) & 7.2 ($u$), 7.2 ($g$) & 30 ($u$), 6 ($g$)  \\
   & 2019 Dec 15 (58829) & 2.6 ($u$), 2.6 ($g$) & 30 ($u$), 6 ($g$)  \\
   & 2020 Jan 14 (58862) & 5.6 ($u$), 5.6 ($g$) & 30 ($u$), 6 ($g$)  \\
   & 2020 Jan 18 (58866) & 6.2 ($u$), 6.2 ($g$) & 30 ($u$), 6 ($g$)   \\
   & 2020 Jan 19 (58867) & 7.5 ($u$), 7.5 ($g$) & 30 ($u$), 6 ($g$) \\
   & 2020 Jan 20 (58868) & 7.4 ($u$), 7.4 ($g$) & 30 ($u$), 6 ($g$)  \\
   & 2020 Dec 03 (59186) & 4.1 ($g$) & 6 ($g$)  \\
   & 2020 Dec 06 (59189) & 6.4 ($u$), 6.4 ($g$) & 30 ($u$), 6 ($g$)  \\
   & 2020 Dec 07 (59190) & 6.1 ($u$), 6.1 ($g$) & 30 ($u$), 6 ($g$)  \\
   & 2021 Jan 31 (59245) & 7.4 ($u$), 7.4 ($g$) & 30 ($u$), 6 ($g$)  \\
   & 2021 Feb 04 (59249) & 0.6 ($u$), 0.6 ($g$) & 30 ($u$), 6 ($g$)  \\
     \hline 
 LCO 1m/Sinistro & 2020 Jan 16 (58864) &  5.1 & 25  \\
   ($U$-band photometry)\tablenotemark{\rm \ddag} & 2020 Jan 17 (58865) & 0.8 & 25  \\
 & 2020 Jan 18 (58866) & 8.1 & 10, 25  \\
 & 2020 Jan 19 (58867) & 4.2 & 10  \\
 & 2020 Jan 20 (58868) & 4.5 & 10  \\
 & 2020 Jan 21 (58869) & 4.1 & 10  \\
 & 2020 Jan 22 (58870) & 4.0 & 10  \\
 & 2020 Jan 23 (58871) & 6.2 & 10  \\
 & 2020 Jan 24 (58872) & 3.2 & 10  \\
 & 2020 Jan 25 (58873) & 5.3 & 10  \\
 & 2020 Jan 26 (58874) & 5.4 & 10  \\
 & 2020 Jan 27 (58875) & 0.9 & 10  \\
     \hline 
 LCO 0.4m/SBIG & 2020 Jan 16 (58864) & 5.4 & 6  \\
   ($V$-band photometry)\tablenotemark{\rm \ddag} &  2020 Jan 17 (58865) & 2.3 & 6 \\
 & 2020 Jan 18 (58866) & 9.6 & 6  \\
 & 2020 Jan 19 (58867) & 4.9 & 6  \\
 & 2020 Jan 20 (58868) & 10.1 & 6  \\
 & 2020 Jan 21 (58869) & 6.7 & 6  \\
 & 2020 Jan 22 (58870) & 5.7 & 6  \\
 & 2020 Jan 23 (58871) & 5.9 & 6  \\
 & 2020 Jan 24 (58872) & 3.3 & 6  \\
     \hline 
   \textit{TESS} &   \multicolumn{2}{c}{Covering our observations} & 120 \\
   (\textit{TESS}-band photometry)\tablenotemark{\rm \ddag} &   \multicolumn{2}{c}{on 2019 Jan 26 -- 28}  &   \\
     &  
   \multicolumn{2}{c}{and 2021 Jan 31 -- Feb 04}  &   \\
   \hline 
   \textit{NICER} &  2019 Jan 26 (58509)  &  $\sim$0.5$\times$3  & --  \\
   (0.2--12 keV X-ray) &  2019 Jan 27 (58510) &  $\sim$0.5$\times$4 & --  \\
    & 2019 Jan 28 (58511) & $\sim$0.5$\times$3 & -- \\
     \hline 
   \hline 
  \multicolumn{4}{c}{EV Lac} \\
    \hline
   ARC 3.5m/ARCES & 2019 Dec 15 (58832) & 5.1 & 240, 250, 300 \\
   (3800--10000\AA; $\lambda/\Delta\lambda \sim 32000$) & 2020 Aug 26 (59087) & 4.4 & 240, 300, 340, 360, 400 \\
   & 2020 Aug 27 (59088) & 4.2 & 300 \\
   & 2020 Aug 29 (59090) & 4.3 & 300, 360, 600 \\
   & 2020 Aug 30 (59091) & 0.7 & 300, 600 \\
   & 2020 Sep 01 (59093) & 3.0 & 300  \\
   & 2020 Sep 02 (59094) & 4.4 & 300  \\
    \hline 
   ARCSAT 0.5m/flarecam & 2019 Dec 15 (58832) & 2.8 ($u$), 2.9 ($g$) & 20 ($u$), 3 ($g$) \\
   ($u, g$-band photometry)\tablenotemark{\rm \ddag}  & 2020 Aug 26 (59087) &  8.0 ($u$), 8.0 ($g$) & 20 ($u$), 3 ($g$) \\
    & 2020 Aug 27 (59088) & 7.9 ($u$), 7.9 ($g$) & 20 ($u$), 3 ($g$)  \\
    & 2020 Aug 29 (59090) & 4.2 ($u$), 4.2 ($g$) & 20 ($u$), 3 ($g$)  \\
    & 2020 Aug 30 (59091) & 0.5 ($u$), 0.5 ($g$) & 20 ($u$), 3 ($g$)  \\
    & 2020 Sep 01 (59093) & 2.1 ($u$), 2.1 ($g$) & 20 ($u$), 3 ($g$)  \\
    & 2020 Sep 02 (59094) & 8.5 ($u$), 8.5 ($g$) & 20 ($u$), 3 ($g$)  \\
   \hline
    \hline 
   \multicolumn{4}{c}{AD Leo} \\
    \hline
   ARC 3.5m/ARCES & 2019 May 17 (58620) & 3.6 & 180, 200, 300 \\
   (3800--10000\AA; $\lambda/\Delta\lambda \sim 32000$) & 2019 May 18 (58621) & 3.6 & 200, 240, 300  \\
   & 2019 May 19 (58622) & 3.6 & 200, 240, 300  \\
    \hline 
   ARCSAT 0.5m/flarecam & 2019 May 17 (58620) & 2.8 ($u$), 2.8 ($g$) & 20 ($u$), 1 ($g$)  \\
   ($u, g$-band photometry)\tablenotemark{\rm \ddag} & 2019 May 18 (58621) &   2.8 ($u$), 3.0 ($g$) & 20 ($u$), 1 ($g$)  \\
   & 2019 May 19 (58622) & 2.9 ($u$), 2.9 ($g$) & 20 ($u$), 1 ($g$) \\
   \enddata
    \tablenotetext{\dag}{
   Time is the total monitoring time for the night.
   }
    \tablenotetext{\ddag}{
Filter profiles of these bands are shown in Figure \ref{fig:filter_YZCMi}.
}
   \label{table:obs_log}
 \end{deluxetable*}
 
\subsection{Spectroscopic Data} \label{subsec:data-spec}
Time-resolved spectroscopic observations were obtained at two facilities.
For the 25 nights among the total 31 nights (Table \ref{table:obs_log}), we conducted spectroscopic observations of the three target stars,
using the ARC Echelle Spectrograph (ARCES; \citealt{Wang+2003}) on the ARC 3.5 m telescope at Apache Point Observatory (APO). The wavelength resolution ($R=\lambda/\Delta\lambda$) is $\sim$32000, and the spectral coverage is 3800 -- 10000~\AA. This wavelength range includes H$\alpha$, H$\beta$, H$\delta$, H$\gamma$, H$\epsilon$, Ca II H\&K, Ca II 8542\AA, He I D3 5876, and Na I D1\&D2 lines.
The exposure times are listed in Table \ref{table:obs_log}, which were determined to achieve signal-to-noise (S/N) ratio values $\sim$ 40 -- 50 at the continuum level around the H$\alpha$ 6563\AA~line. We note that the APO/ARCES spectroscopic data have relatively long overhead and read-out time: $\sim$180 sec in total.
The data reduction methods of APO3.5m/ARCES spectroscopic data are the same as in \citet{Notsu+2019}.
We conducted standard image reduction procedures such as bias subtraction, flat-fielding, and scattered light subtraction, using the ECHELLE package in IRAF\footnote{IRAF is distributed by the National Optical Astronomy Observatories, which are operated by the Association of Universities for Research in Astronomy, Inc., under cooperate agreement with the National Science Foundation.} and PyRAF\footnote{PyRAF is part of the stscipython package of astronomical data analysis tools and is a product of the Science Software Branch at the Space Telescope Science Institute.} software. We used a Th/Ar lamp for wavelength calibration. We also applied the heliocentric radial velocity correction using the ECHELLE package.

For the 8 nights among the total 31 nights (Table \ref{table:obs_log}), we conducted spectroscopic observations of one of the target stars YZ CMi, using the cross-dispersed, fiber-fed echelle CTIO HIgh ResolutiON (CHIRON) spectrogragh \citep{Tokovinin+2013} attached to the Small and Moderate Aperture Telescope Research System (SMARTS) 1.5m telescope at Cerro Tololo Interamerican Observatory (CTIO). For the 2 among these 8 nights (Table \ref{table:obs_log}), we also observed YZ CMi, using APO3.5m/ARCES. The wavelength range and wavelength resolution of our CHIRON data are 4500--8900 \AA~and $R\sim$25000, respectively. 
This wavelength range includes H$\alpha$, H$\beta$, Ca II 8542\AA, He I D3 5876, and Na I D1\&D2 lines.
The exposure time was 600 sec (Table \ref{table:obs_log}), which was determined to achieve signal-to-noise (S/N) ratio values $\sim$40 at the continuum level around the H$\alpha$ 6563\AA~line.  The spectra were reduced using the CHIRON pipeline described in \citet{Tokovinin+2013}.

\subsection{Photometric Data} \label{subsec:data-phot}

Time-resolved photometric observations were done by using two ground-based facilities (ARCSAT \&~LCOGT) and \textit{TESS} satellite. 
We conducted ground-based photometric observations using 0.5m Astrophysical Research Consortium Small Aperture Telescope (ARCSAT) for the 24 nights (Table \ref{table:obs_log}), simultaneously with the spectroscopic observations using APO3.5m/ARCES. We note that among the 25 nights when we conducted APO3.5m/ARCES spectroscopic observations, we have no ARCSAT0.5m photometric data on 2019 Dec 08 because of the bad weather condition.
We carried out $u\&g$-band photometric observations using the Flarecam instrument of ARCSAT0.5m (\citealt{Hilton2011}; \citealt{Kowalski+2013}), which has enhanced UV sensitivity and rapid filter wheel rotation. 
The exposure times are listed in Table \ref{table:obs_log}.
Considering the filter wheel rotation time, the typical time cadence for each band is 50--60 sec.
Dark subtraction and flat-fielding were performed using PyRAF software in the standard manner before the photometry.
Aperture photometry was performed using AstroimageJ \citep{Collins+2017}.
We used nearby stars as the magnitude references.

We also conducted ground-based photometric observations of one of the target stars YZ CMi, using the Las Cumbres Observatory Global Telescope (LCOGT) network (\citealt{Brown+2013}). These LCO (Las Cumbres Observatory) observations \color{black}\textrm{were} \color{black} conducted for 12 nights (Table \ref{table:obs_log}) to support SMARTS1.5m/CHIRON spectroscopic observations. Using LCO 1m telescopes with the Sinistro cameras, we carried out U-band photometric observations with exposure times of 10\&25 seconds (Table \ref{table:obs_log}). V-band photometric observations were conducted using LCO 0.4m telescopes with the SBIG STL-6303 cameras and the exposure times are 6 seconds (Table \ref{table:obs_log}).
The data were reduced with the LCOGT automatic pipeline 
BANZAI\footnote{\url{https://github.com/LCOGT/banzai}},
which masks bad-pixels, applies an astrometric solution, and performs bias \& dark subtraction.
Aperture photometry was performed using AstroimageJ \citep{Collins+2017}, 
and we used nearby stars as the magnitude references.

Among the 31 nights we conducted the above ground-based spectroscopic and photomteric observations, the two terms observing one of the target star YZ CMi (2019 Jan 26 - 28 and 2021 Jan 31 - Feb 04) were also covered with the observation window of the Transiting Exoplanet Survey Satellite (\textit{TESS}; \citealt{Ricker+2015}) (Table \ref{table:obs_log}).
We used the 120-sec time cadence \textit{TESS} Sectors 7 \& 34 Pre-search Data Conditioned Simple Aperture Photometry (PDC-SAP) light curve data (\citealt{Vanderspek+2018}) of YZ CMi, retrieved from the Multimission Archive at the Space Telescope (MAST) Portal 
site\footnote{\url{https://mast.stsci.edu/portal/Mashup/Clients/Mast/Portal.html}}, as we have done in \citet{Maehara+2021}.
The data release (DR) numbers of Sectors 7 and 34 data we used are
DR9 (\citealt{Fausnaugh+2019}) and DR50 (\citealt{Fausnaugh+2021}), respectively.

\subsection{X-ray Data} \label{subsec:data-Xray}

The X-ray instrument {\textit NICER} (Neutron star Interior Composition ExploreR, \citealt{Gendreau+2016}) onboard the International Space Station (ISS) conducted monitoring observations of YZ CMi on 2021 Januray 26, 27, \& 28 (Observation ID: 1200510101 -- 1200510103). This was scheduled for simultaneously observations with the ARC 3.5m/ARCES spectroscopy, ARCSAT 0.5m photometry, and \textit{TESS} photometry of YZ CMi (Table \ref{table:obs_log}).
{\textit NICER} observed YZ CMi for about $\sim$2 ks for each ISS orbit (about 90 min) and 
 3-4 times every day (Table \ref{table:obs_log}).

{\textit NICER} X-ray Timing Instrument (XTI) is an array of aligned 56 X-ray modules, each of which consists of a set of an X-ray concentrator
\citep[XRC, ][]{Okajima2016a} and a silicon drift detector \citep[SDD,][]{Prigozhin2016a}.
Each XRC concentrates X-rays within a $\sim$3 arcmin radius field of view to the paired SDD,
which detects each photon at accuracy at $\sim$84~ns.
The XTI as a whole has one of the largest collecting areas among X-ray instruments between 0.2$-$12 keV ($\sim$1900~cm$^{-2}$ at 1.5 keV).
We use 50 XTI modules as the remaining six (ID: 11, 14, 20, 22, 34, 60) are inactive or noisy.

As also done in \citet{Hamaguchi+2023_ApJ}, 
we reprocess the datasets with the \ NIC\ calibration ver.\ CALDB XTI(20210707)
using {\tt nicerl2} in HEASoft ver.\ 6.29c and NICERDAS ver.\ V008c.
Since {\textit NICER} is not an imaging instrument,
we evaluate particle background level using nibackgen3C50 ver.\ v7b with the parameters 
dtmin=10.0, dtmax=60.0, hbgcut=0.1, s0cut=30.0 \citep{Remillard+2022}.

\subsection{Flare luminosities and energies} \label{subsec:quiescent-ene}
In the following sections, 
the flare luminosities and energies are calculated for continuum bands and the chromospheric emission lines. 
In this process, the distance of the target stars (Table \ref{table:targets_basic_para}), the quiescent luminosities of photometric bands ($L_{\rm{band},\rm{q}}$), and the quiescent flux densities at the continuum levels around the lines ($F_{\rm{line},\rm{q}}^{\rm{cont}}$) are used.

\begin{deluxetable*}{lcccccccc}[ht!]
   \tablecaption{Quiescent luminosities of continuum bands
   and quiescent flux densities around lines}
   \tablewidth{0pt}
   \tablehead{
     \colhead{Starname} & \colhead{$\log L_{U,\rm{q}}$\tablenotemark{\rm \dag}} & \colhead{$\log L_{u,\rm{q}}$\tablenotemark{\rm \dag}} & \colhead{$\log L_{g,\rm{q}}$\tablenotemark{\rm \dag}} & \colhead{$\log L_{V,\rm{q}}$\tablenotemark{\rm \dag}} & \colhead{$\log L_{TESS,\rm{q}}$\tablenotemark{\rm \dag}} & \colhead{$F_{\rm{H}\alpha,\rm{q}}^{\rm{cont}}$\tablenotemark{\rm \ddag}}
     & \colhead{$F_{\rm{H}\beta,\rm{q}}^{\rm{cont}}$\tablenotemark{\rm \ddag}}  
     }
   \startdata
   YZ CMi & 28.6 & 28.5 & 29.57 & 29.65 & 30.99 & 25.2 & 8.3 
   \\
   EV Lac & 28.8 & 28.7 & 29.80 & 29.87 & 31.11 & 57.0 & 20.3 
   \\ 
   AD Leo & 29.2 & 29.1 & 30.17 & 30.23 & 31.37 & 128.5 & 48.1 
   \\
   \enddata
    \tablenotetext{\dag}{
The quiescent luminosity values in $U$, $u$, $g$, $V$, and \textit{TESS} bands (cf. Figure \ref{fig:filter_YZCMi}). Units are erg s$^{-1}$.
}
 \tablenotetext{\ddag}{
Quiescent flux densities at the continuum levels around the lines (H$\alpha$, H$\beta$
lines). 
Units are 10$^{-14}$ erg s$^{-1}$ cm$^{-2}$ \AA$^{-1}$.
\color{black}
\textrm{
The continuum regions are determined by using the definitions in Table 3 of \citet{Kowalski+2013}.
}
\color{black}
   }
   \label{table:targets_quiescent_flux}
 \end{deluxetable*}

 \begin{figure}[ht!]
   \begin{center}
      \gridline{
    \fig{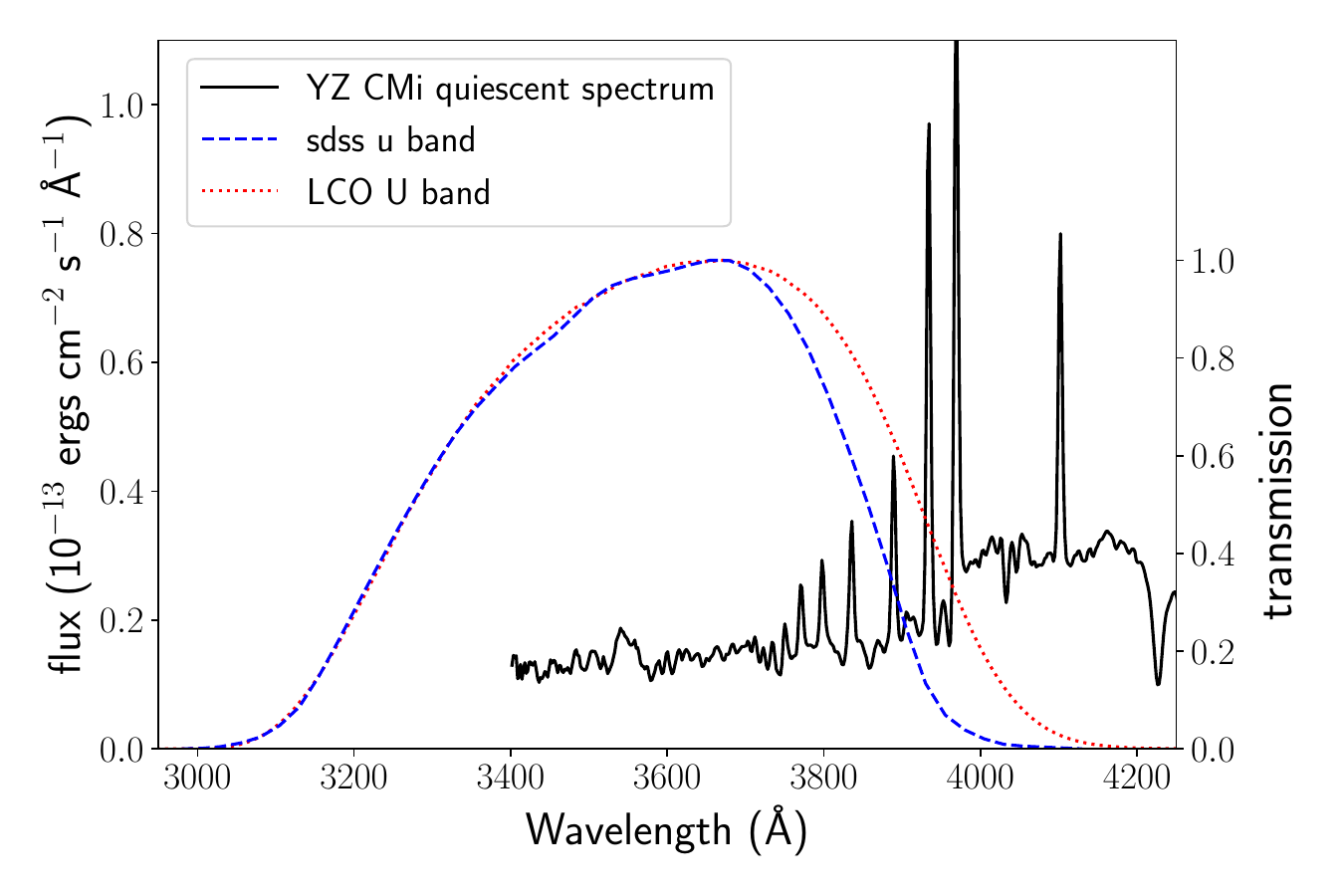}{0.45\textwidth}{\vspace{-5mm}(a)}
    \fig{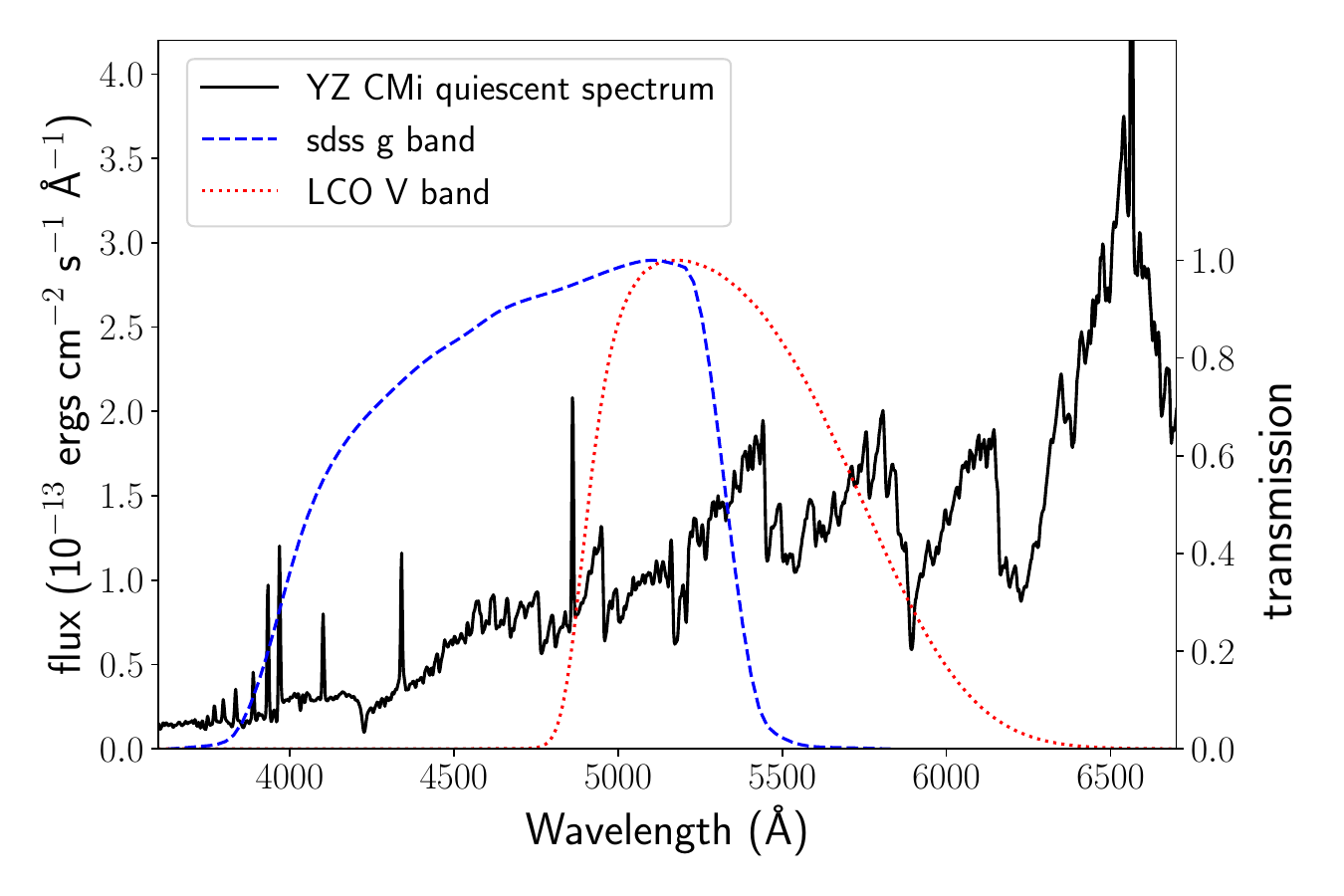}{0.45\textwidth}{\vspace{-5mm}(b)}
    }
    \gridline{
    \fig{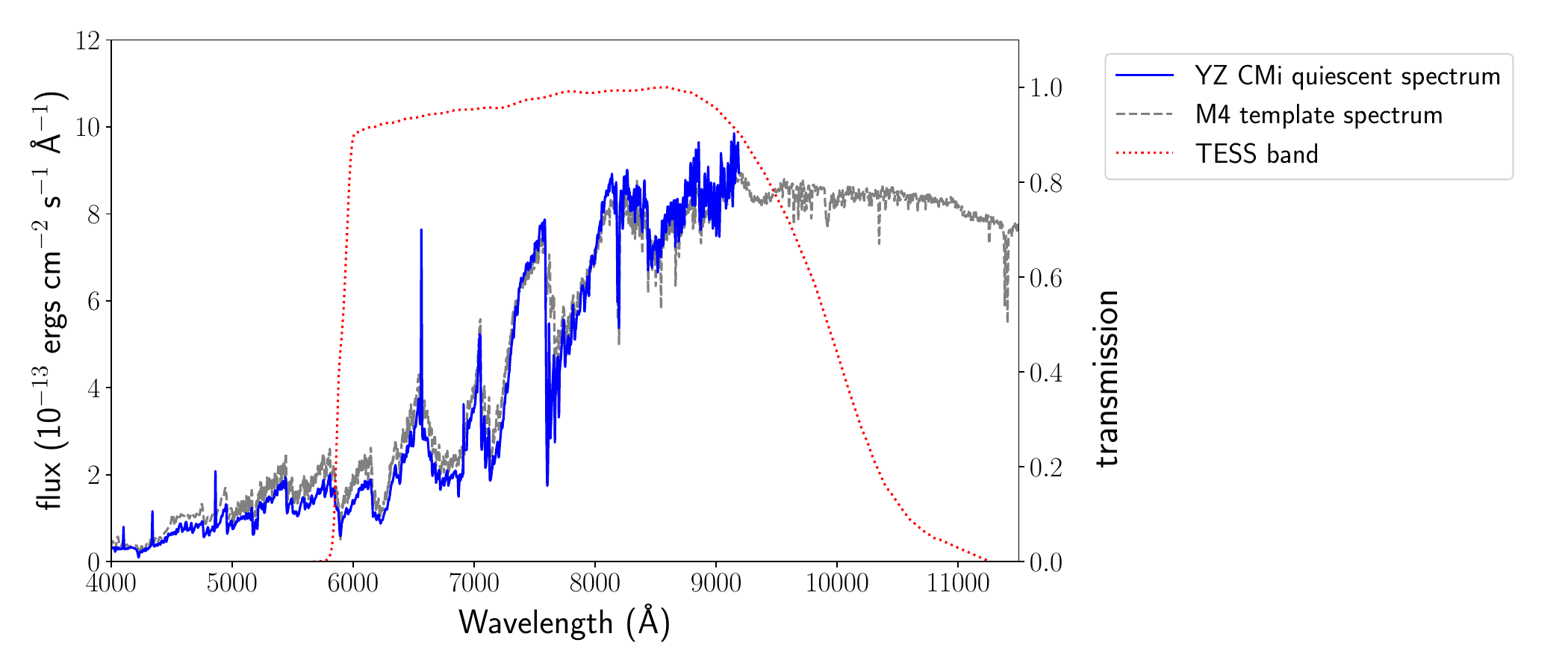}{0.7\textwidth}{\vspace{-5mm}(c)}
    }
     \vspace{0mm}
     \caption{
(a) Flux-calibrated quiescent spectrum for YZ CMi (black filled line) from \citet{Kowalski+2013}, with the photon transmission bandpass data of sdss $u$-band  and LCO $U$-band.
(b) The same as (a), but with the bandpass data of sdss $g$-band and LCO $V$-band.
(c) Flux-calibrated quiescent spectrum for YZ CMi (blue filled line) from \citet{Kowalski+2013}, with the scaled NUV-NIR M4 template (thin gray dashed line) from \citet{Davenport+2012},
which is used to calculate the quiescent luminosity for YZ CMi in the \textit{TESS} bandpass (red dotted line). The scaling normalization of the template spectrum is done by using the wavelength regions of 7000--7500\AA~and 8000--9000\AA.
     }
   \label{fig:filter_YZCMi}
   \end{center}
 \end{figure}

Quiescent luminosities of photometric bands ($L_{U,\rm{q}}$, $L_{u,\rm{q}}$, $L_{g,\rm{q}}$, $L_{V,\rm{q}}$, $L_{TESS,\rm{q}}$) are estimated as in the following and are listed in Table \ref{table:targets_quiescent_flux}. The $U$-band quiescent luminosities $L_{U,\rm{q}}$ are taken from Table 1 of \citet{Kowalski+2013}.
The $u$-band quiescent luminosities $L_{u,\rm{q}}$ are converted from $L_{U,\rm{q}}$, using the flux-calibrated quiescent spectroscopic data of the three target stars reported in \citet{Kowalski+2013}\footnote{
The spectroscopic data are available at \url{https://doi.org//10.26093/cds/vizier.22070015}.
} and the bandpass data of sdss $u$-band (used in ARCSAT) and LCO $U$-band\footnote{The bandpass data (LCO $U$, sdss $u$, sdss $g$, LCO $V$, and $TESS$) used in this study, which are also shown in Figure \ref{fig:filter_YZCMi}, are taken from the SVO Filter Profile Service \url{http://svo2.cab.inta-csic.es/theory/fps/} (\citealt{Rodrigo+2012}; \citealt{Rodrigo+2020}).}.
As an \color{black}\textrm{example}\color{black}, the quiescent spectrum of YZ CMi taken from \citet{Kowalski+2013} is shown with the $u$- \& $U$-bands filter data in Figure \ref{fig:filter_YZCMi} (a). 
The $g$- \& $V$-bands quiescent luminosities $L_{g,\rm{q}}$ and $L_{V,\rm{q}}$ are calculated from the same flux-calibrated quiescent spectroscopic data \citep{Kowalski+2013}, the band-pass data of sdss $g$-band (used in ARCSAT) and LCO $V$-band, and the stellar distance $d_{\rm Gaia}$ (Table \ref{table:targets_basic_para}).
In Figure \ref{fig:filter_YZCMi} (b), the $g$- \& $V$-bands filter data are shown with the quiescent spectrum of YZ CMi as an example.
The \textit{TESS}-band quiescent luminosities $L_{TESS,\rm{q}}$ are also calculated by using the flux-calibrated quiescent spectra, the filter data, and and the stellar distance $d_{\rm Gaia}$. 
In the case of YZ CMi, as shown in Figure \ref{fig:filter_YZCMi} (c), the filter curve is convolved with an M4 NUV-NIR spectral template from \citet{Davenport+2012}, which is normalized to the above flux-calibrated spectrum of YZ CMi from \citet{Kowalski+2013}.
The normalization is done by using the wavelength regions of 7000--7500\AA~and 8000--9000\AA.
\color{black}\textrm{
It is noted that when calculating $L_{TESS,\rm{q}}$, the actual data are used for 
calculation at the wavelength where the data exist ($<$9168\AA), while the templates are only used in the remaining redder part ($>$9168\AA).}
\color{black}
This method is basically the same as that done for a M4-dwarf flare star GJ1243 in \citet{Davenport+2020}. 
As for the other target stars EV Lac and AD Leo, we estimated $L_{TESS,\rm{q}}$ values with the basically same method 
using the flux-calibrated quiescent spectra, the filter data, and the stellar distances.
The data of flux-calibrated spectra of EV Lac and AD Leo were taken from \citet{Kowalski+2013}.
The M3 spectral template from \citet{Davenport+2012} is used for AD Leo instead of the M4 template for YZ CMi and EV Lac.

Flare luminosities in the photometric bands ($L_{\rm{band},\rm{flare}}(t)$) are 
calculated from the quiescent luminosities $L_{\rm{band},\rm{q}}$ (Table \ref{table:targets_quiescent_flux}) and the relative fluxes during the flares 
($\Delta f_{\rm{band, flare}}(t)$): 
  \begin{eqnarray}   \label{eq:L_band_flare}
  L_{\rm{band},\rm{flare}}(t) = L_{\rm{band},\rm{q}} \times \Delta f_{\rm{band, flare}}(t) \ .
  \end{eqnarray}
Relative flux (cf. Figures \ref{fig:rot1_YZCMi_2019Q1}, \ref{fig:Ha_rot_YZCMi_2019Q4to2020Q1}, \ref{fig:alllc1_2020Q4_YZCMi}, \ref{fig:alllc1_2021Q1_YZCMi}, \ref{fig:rot_all1_EVLac_2019Q4_2020Q3}, \& \ref{fig:rot_all1_ADLeo_2020Q2} in Section \ref{subsec:results:obs-summary}) is here defined as $\Delta f(t)= (f(t) - f_{\rm{ave}})/f_{\rm{ave}}$, where $f(t)$ is flux and $f_{\rm{ave}}$ is average flux of the non-flare phase.
Flare energies in the photometric bands ($E_{\rm{band}, \rm{flare}}$) are calculated by integrating $L_{\rm{band},\rm{flare}}$ over the flare duration:
  \begin{eqnarray}   \label{eq:E_band_flare}
  E_{\rm{band},\rm{flare}} &=& \int L_{\rm{band},\rm{flare}}(t) dt \\
   &=& L_{\rm{band},\rm{q}} \times \int \Delta f_{\rm{band, flare}}(t) dt \\
   &\equiv& L_{\rm{band},\rm{q}} \times ED_{\rm{band, flare}} \ ,
   \label{eq:E_band_flare_ED}
  \end{eqnarray}
where $ED_{\rm{band}}$ are equivalent durations (cf. \citealt{Hunt-Walker+2012}).

In this study, we identified flares in the photometric bands
when the relative flux $\Delta f_{\rm{band, flare}}(t)$ is 
larger than $3\times \sigma_{\rm{band}}$ at around the flare peak for multiple data points, and the light curve shape looks consistent with stellar optical flares (i.e. rapid increase and gradual decay as in \citealt{Davenport+2014}, \citealt{Okamoto+2021}).
$\sigma_{\rm{band}}$ is the standard deviation of the relative flux in each band on each night for the phases without flares.
\color{black}\textrm{
If no clear flares are identified }\color{black} in the photometric bands during the flares in H$\alpha$ and H$\beta$ lines, the upper limit of flare peak luminosity ($L_{\rm{band}}$ in Table \ref{table:list1_flares}) is estimated by
applying this detection threshold $\Delta f_{\rm{band, flare}}<3\times \sigma_{\rm{band}}$ to Equation (\ref{eq:L_band_flare}). The upper limit of flare energy ($E_{\rm{band}}$ in Table \ref{table:list1_flares}) is 
calculated by assuming the light curve 
shows the linear decay with the peak amplitude $3\times\sigma_{\rm{band}}$ and 
the decay time comparable to the H$\alpha$ flare duration ($\Delta t^{\rm{flare}}_{\rm{H}\alpha}$ in Table \ref{table:list1_flares}).
Then we apply $ED_{\rm{band, flare}}=0.5 \times (3\times\sigma_{\rm{band}}) \times \Delta t^{\rm{flare}}_{\rm{H}\alpha}$ in Equation (\ref{eq:E_band_flare_ED}) for 
estimating the upper limit of flare energy.

The quiescent flux densities at the continuum levels around the H$\alpha$ \&  H$\beta$
lines ($F_{\rm{H}\alpha,\rm{q}}^{\rm{cont}}$ and $F_{\rm{H}\beta,\rm{q}}^{\rm{cont}}$
) are calculated 
based on the flux-calibrated quiescent spectra of the target stars from \citet{Kowalski+2013} (cf. Figure \ref{fig:filter_YZCMi}), 
and the values are listed in Table \ref{table:targets_quiescent_flux}.
In this process, the continuum regions are determined by using the definitions in Table 3 of \citet{Kowalski+2013}.
Flare luminosities in H$\alpha$ \& H$\beta$ lines ($L_{\rm{line}, \rm{flare}}(t)$) can be calculated from the quiescent fluxes ($F_{\rm{line}, \rm{q}}^{\rm{cont}}$ in Table \ref{table:targets_quiescent_flux}\color{black}\textrm{, and $F_{\rm{line}}^{\rm{cont}}(t)$ is the flux at the continuum level}\color{black}), relative fluxes at the continuum level around the lines $\Delta f^{\rm{cont}}_{\rm{line}}(t) = (f^{\rm{cont}}_{\rm{line}}(t) - f^{\rm{cont}}_{\rm{ave, line}})/f^{\rm{cont}}_{\rm{ave, line}}$, the stellar distance $d_{\rm Gaia}$ (Table \ref{table:targets_basic_para}), and the equivalent width of the flare component 
($EW_{\rm{line}, \rm{flare}}(t)$ = $EW_{\rm{line}}(t)$ - $EW_{\rm{line, q}}$ where $EW_{\rm{line}, \rm{flare}}(t)$ is the equivalent width (EW) of the flare component, $EW_{\rm{line}}(t)$ is the total equivalent width, and $EW_{\rm{line, q}}$ is the equivalent width at the quiescent level. See also Figures \ref{fig:rot1_YZCMi_2019Q1}, \ref{fig:Ha_rot_YZCMi_2019Q4to2020Q1}, \ref{fig:alllc1_2020Q4_YZCMi}, \ref{fig:alllc1_2021Q1_YZCMi}, \ref{fig:rot_all1_EVLac_2019Q4_2020Q3}, \& \ref{fig:rot_all1_ADLeo_2020Q2} in Section \ref{subsec:results:obs-summary})\footnote{
We note that equivalent width of a spectral line is defined as an area of the line on a plot of the continuum-normalized intensity as a function of wavelength, and in this study we define that the positive value of the EW indicates line emission so that an increase (positive change) of equivalent width indicates an increase of emission line flux. 
This positive EW definition is the same as one of our previous papers \citet{Honda+2018} but is opposite to our other previous papers \citet{Namekata+2020_PASJ} \& \citet{Maehara+2021}.
}:
  \begin{eqnarray}  
  L_{\rm{line},\rm{flare}}(t) 
  &=&  4\pi d_{\rm Gaia}^{2} \times 
  \Bigl(F_{\rm{line}}^{\rm{cont}}(t)\times EW_{\rm{line}}(t) 
  - F_{\rm{line}, \rm{q}}^{\rm{cont}} \times EW_{\rm{line, q}}\Bigr) \\
  &=& 4\pi d_{\rm Gaia}^{2} \times F_{\rm{line}, \rm{q}}^{\rm{cont}}  \times 
    \Biggl[
    \Bigl(1.0 + \Delta f^{\rm{cont}}_{\rm{line}}(t)\Bigr) \times 
    \Bigl(EW_{\rm{line}, \rm{flare}}(t) + EW_{\rm{line, q}}\Bigr)
    - EW_{\rm{line, q}}\Biggr]\\
  &=& 4\pi d_{\rm Gaia}^{2} \times F_{\rm{line}, \rm{q}}^{\rm{cont}}  \times 
    \Biggl[
    \Bigl(1.0 + \Delta f^{\rm{cont}}_{\rm{line}}(t)\Bigr) \times 
    EW_{\rm{line}, \rm{flare}}(t) 
    +  \Delta f^{\rm{cont}}_{\rm{line}}(t) \times EW_{\rm{line, q}}\Biggr] \ .
   \label{eq:L_line_flare}
  \end{eqnarray}
\noindent
Flare energies in H$\alpha$ \& H$\beta$ lines ($E_{\rm{line}, \rm{flare}}$) can be calculated by integrating $L_{\rm{line},\rm{flare}}$ over the flare duration:
  \begin{eqnarray}   
  E_{\rm{line},\rm{flare}} &=& \int L_{\rm{line},\rm{flare}}(t) dt \\
   \label{eq:E_line_flare_2}
   &=&  4\pi d_{\rm Gaia}^{2} \times F_{\rm{line}, \rm{q}}^{\rm{cont}} 
   \times \int \Bigl[(1.0 + \Delta f^{\rm{cont}}_{\rm{line}}(t)) 
   \times EW_{\rm{line}, \rm{flare}}(t) + \Delta f^{\rm{cont}}_{\rm{line}}(t) \times EW_{\rm{line, q}} \Bigr]dt \ . 
   \label{eq:E_line_flare}
   \end{eqnarray}

\rm\color{black}
In this study, $g$-band flux observations are mainly used for estimating $\Delta f^{\rm{cont}}_{\rm{line}}(t)$ at the continuum level around the H$\alpha$ and H$\beta$ lines. 
Since the $g$-band flux changes can be larger than the changes of the real local continuum levels around H$\alpha$ \& H$\beta$ lines considering the typical M-dwarf flare 
spectra (cf. \citealt{Kowalski+2013}), the resultant values are shown with the ranges 
(e.g., $L_{\rm{H}\beta}= \color{black} 2.0-2.3 \color{black}\times 10^{27}$erg s$^{-1}$ for Flare Y1) : 
the lower values do not take into account 
any continuum flux changes ($\Delta f^{\rm{cont}}_{\rm{line}}(t)=0$ in Equations (\ref{eq:L_line_flare})\&(\ref{eq:E_line_flare_2})) 
and the upper values correspond to the values incorporating $g$-band flux changes ($\Delta f^{\rm{cont}}_{\rm{line}}(t)$ taken from $g$-band light curves). We use the same method when we estimate the flare 
peak luminosities ($L$) and flare energies ($E$) in the H$\alpha$ \& H$\beta$ lines of all 
the other flares listed in Table \ref{table:list1_flares}. 
The \textit{TESS}-band continuum fluxes are not used in this process even for flares with TESS data (e.g., Flare Y1), 
since most of the flares in this study have no \textit{TESS} data (Table \ref{table:list1_flares}).

In addition, as for the peak luminosities in photometric bands ($U$-, $u$-, $g$-, $V$-, and $TESS$-bands) in Table \ref{table:list1_flares}, 
we selected the peaks that are considered to be most physically associated with the flare 
peaks in the H$\alpha$ \& H$\beta$ lines.  This means that the largest flare peaks in photometric bands are not necessarily selected, but those closest in time with the flare main peaks in the H$\alpha$ \& H$\beta$ lines are basically selected. 
The detailed descriptions for the individual flares are in Sections \ref{subsec:results:2019-Jan-27} -- \ref{subsec:results:2019-May-19} and Appendix \ref{subsec:results:2019-Jan-26} -- \ref{subsec:results:2019-May-18}.
In contrast, all changes (peaks) of the \color{black}\textrm{photometric}\color{black}-band luminosity (not only the highest peaks) during the whole flare duration in H$\alpha$ line ($\Delta t^{\rm{flare}}_{\rm{H}\alpha}$) 
are taken into account for calculating the flare energies following Equation (\ref{eq:E_line_flare_2}).

It is noted that some flares reported in this study seem to be superimposed on potential decay tails of previous flares, and this could affect the values of flare luminosities and energies.
However, 
we do not correct for this issue,
since it is difficult to correctly subtract the component of previous flares. Potential errors caused from this point should be kept in mind when discussing flare luminosities (see also descriptions for each flare in the following sections of this paper).
Moreover, some flares 
show an extra blue/red-shifted component (see the discussions in the following sections), but the flare luminosity/energy values were not corrected for this. The emission contributions from the blue/red-shifted extra components 
are included in the resultant flare luminosity/energy values. The reason is that 
main discussions are not affected without any correction
since the purpose of this paper is not discussing the detailed energetic of flares 
and order of magnitude estimate of flare energies are sufficient for the purposes of this paper.

\section{Flare light curves and spectra} \label{sec:results}
\subsection{Observational Summary} \label{subsec:results:obs-summary}

\begin{figure}[ht!]
   \begin{center}
      \gridline{
    \fig{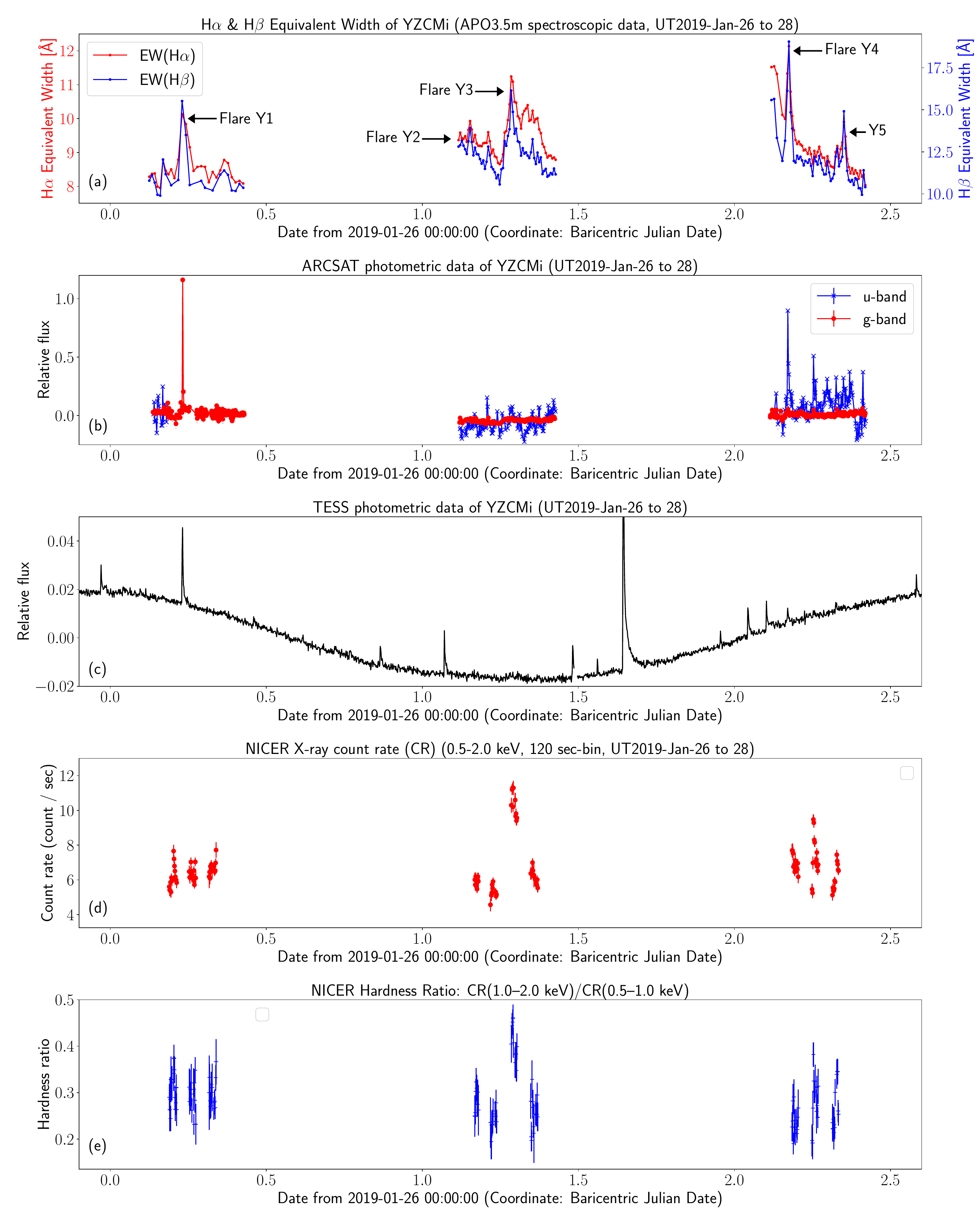}{0.85\textwidth}{\vspace{0mm}}
    }
     \vspace{-5mm}
     \caption{
Summary light curves of YZ CMi during the campaign season [i] 2019 January 26 -- 28. The horizontal axes represent the observation time in Barycentric Julian Date (BJD).
\color{black}\textrm{
The data points correspond to the middle time of each exposure, and this is the same for all the figures of the light curves in the following of this paper.
}\color{black}
(a) H$\alpha$ \& H$\beta$ equivalent width values from APO3.5m spectroscopic data. Red and Blue symbols correspond to H$\alpha$ \& H$\beta$ EWs, respectively. Black arrows indicate flares.
(b) $u$- \& $g$-band relative flux light curves from ARCSAT photometric data. 
Blue asterisks and red circles correspond to $u$- \& $g$-band data, respectively.
Relative flux is here defined as $(f - f_{\rm{ave}})/f_{\rm{ave}}$, where $f$ is flux and $f_{\rm{ave}}$ is average flux of the non-flare phase. This definition is the same for the following figures of this paper.
(c) \textit{TESS}-band relative flux light curve from \textit{TESS} photometric data.
(d) Background-subtracted X-ray count rates [count s$^{-1}$] from \textit{NICER} data in 0.5--2.0 keV. 
(e) X-ray Hardness ratio (count rate (1.0--2.0 keV) / count rate (0.5-1.0 keV)) from \textit{NICER} data.
     }
   \label{fig:rot1_YZCMi_2019Q1}
   \end{center}
 \end{figure}

\begin{figure}[ht!]
   \begin{center}
   \gridline{
    \fig{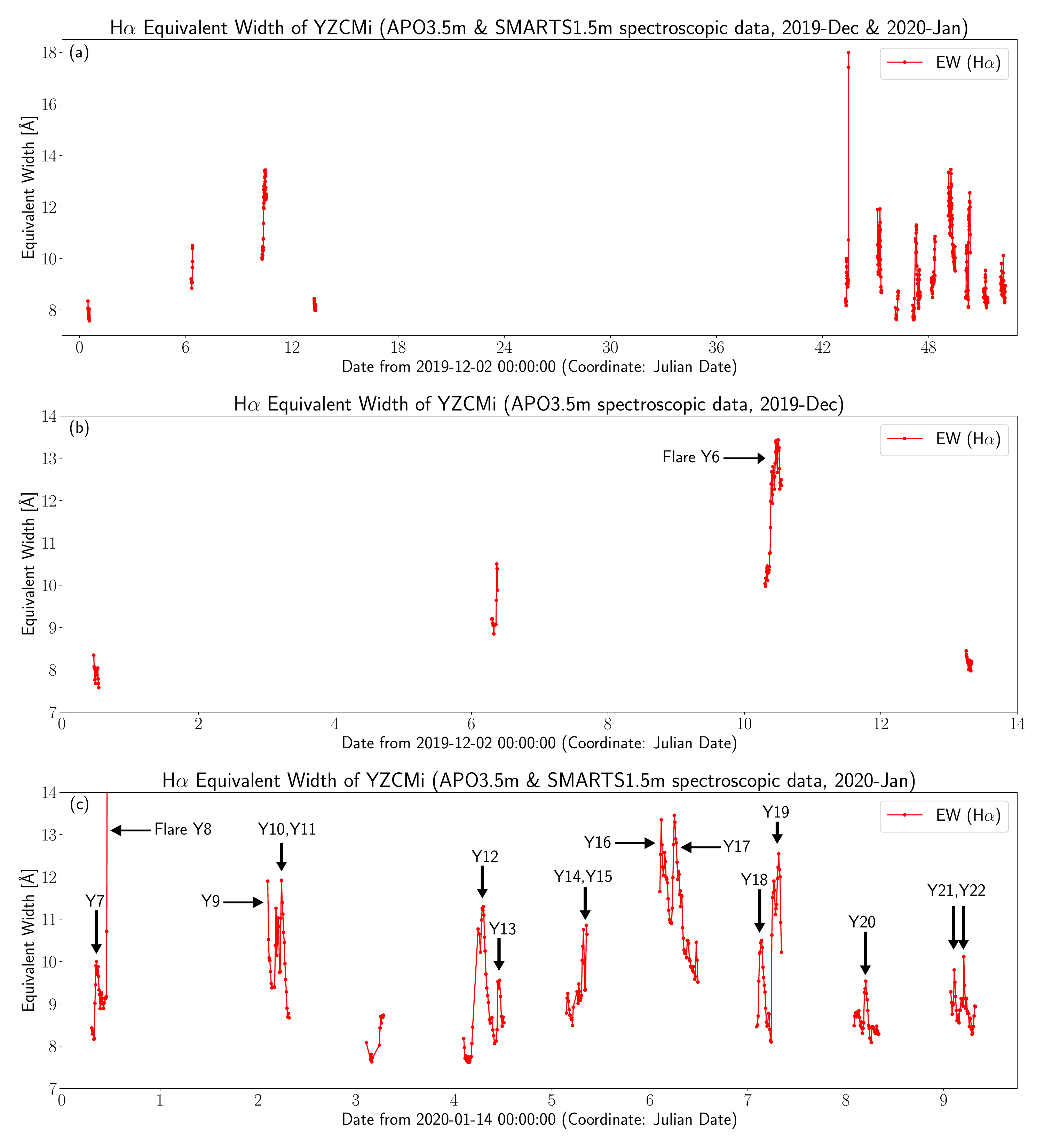}{0.5\textwidth}{\vspace{0mm}}
    \fig{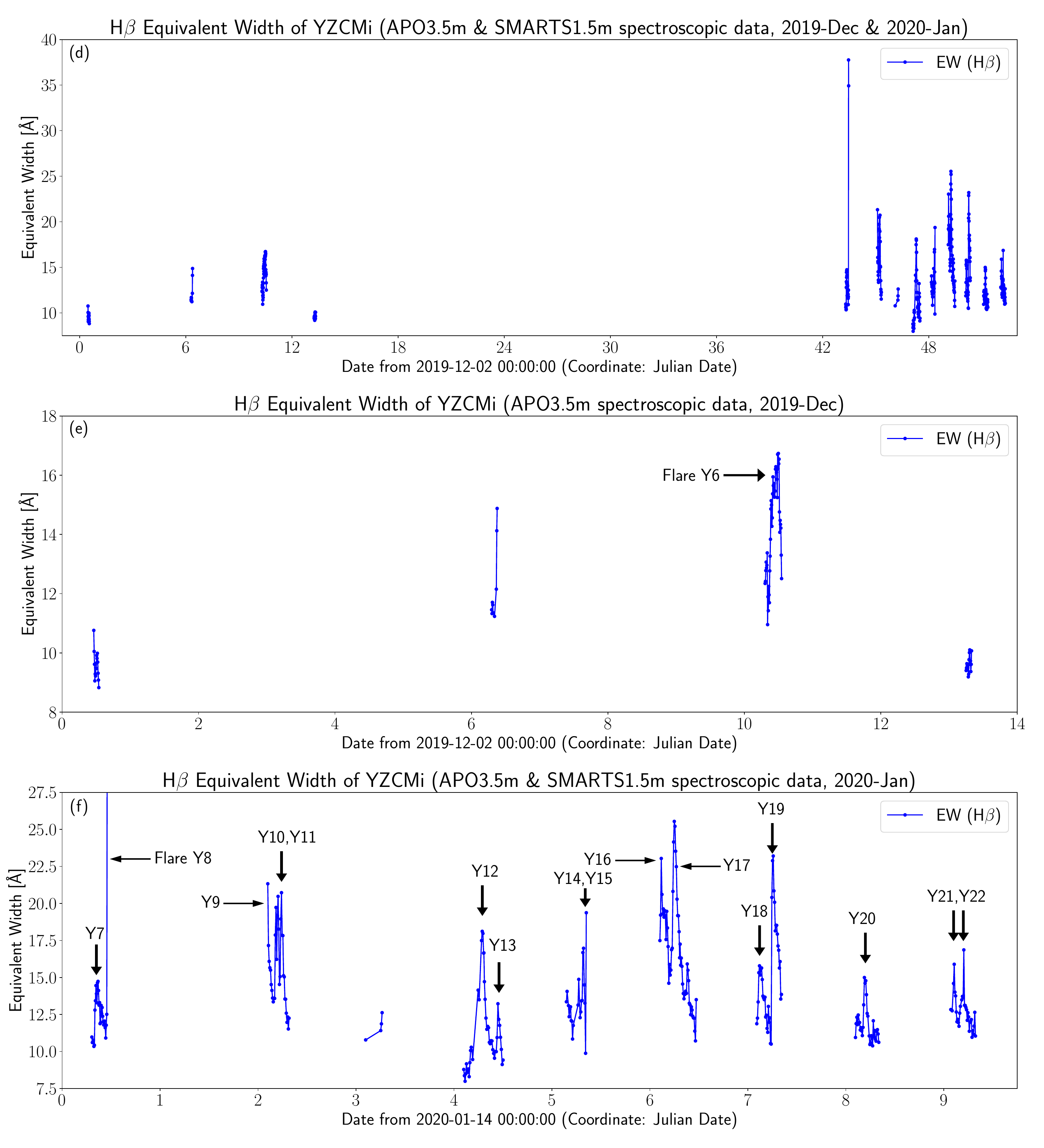}{0.5\textwidth}{\vspace{0mm}}
    }
   \gridline{
    \fig{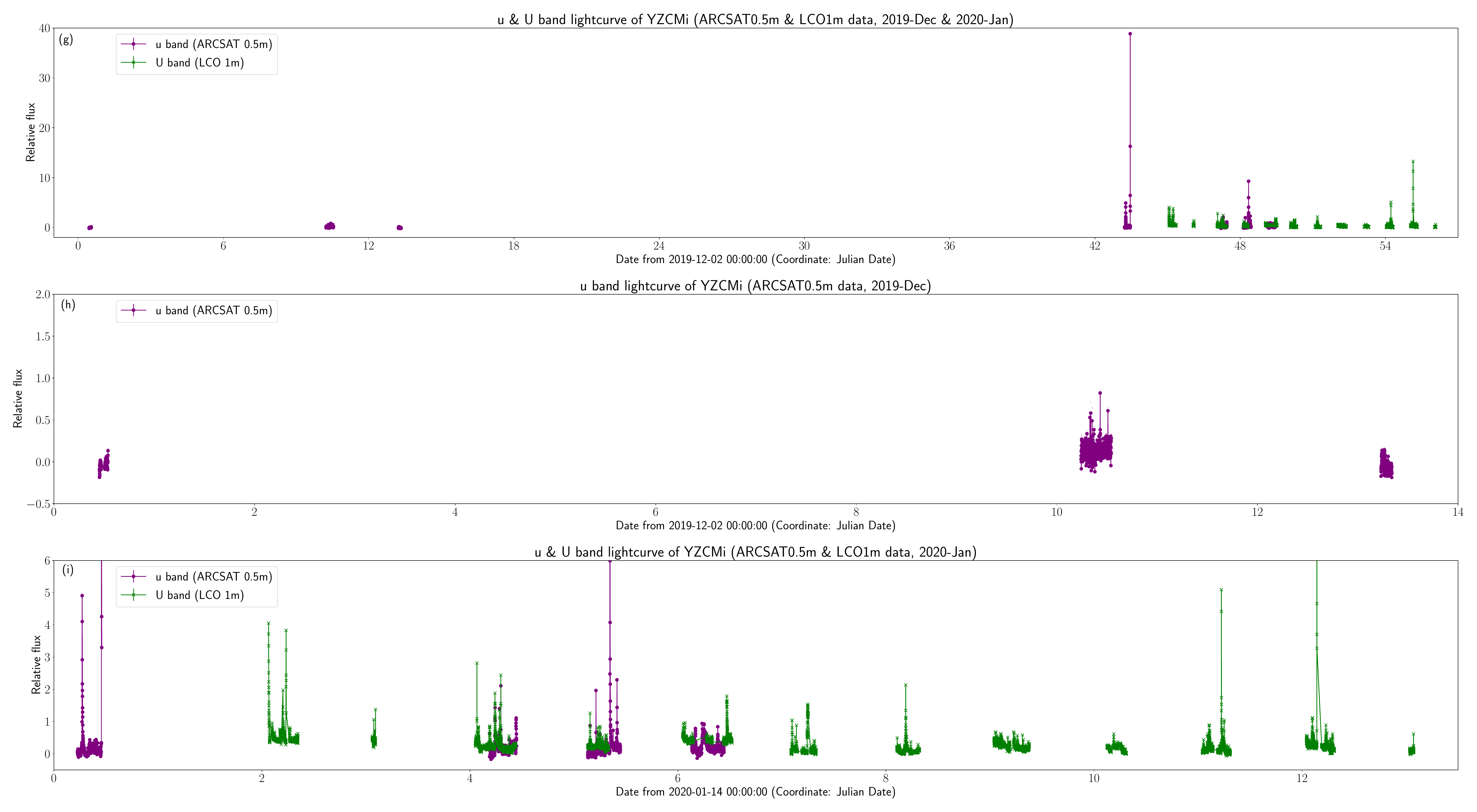}{1.0\textwidth}{\vspace{0mm}}
    }
     \vspace{0mm}
     \caption{
Summary light curves of YZ CMi during the campaign seasons [ii] 2019 December and [iii] 2020 January. The horizontal axes represent the observation time in Julian Date (JD).
(a)\&(d) H$\alpha$ \& H$\beta$ equivalent width values from APO3.5m and SMARTS1.5m spectroscopic data for the both campaign seasons [ii] \& [iii]. (b)\&(e) The same as (a)\&(d), but only for the campaign season [ii]. 
(c)\&(f) The same as (a)\&(d), but only for the campaign season [iii]. 
Black arrows in (b)\&(c) indicate flares.
(g) $u$-band \& $U$-band relative light curves from ARCSAT 0.5m and LCO 1m photometric data (purple and green symbols, respectively) for the both campaign seasons [ii] \& [iii].
(h) The same as (g) but only for the campaign season [ii]. 
(i) The same as (h) but only for the campaign season [iii]. 
     }
   \label{fig:Ha_rot_YZCMi_2019Q4to2020Q1}
   \end{center}
 \end{figure}

\begin{figure}[ht!]
   \begin{center}
   \gridline{
    \fig{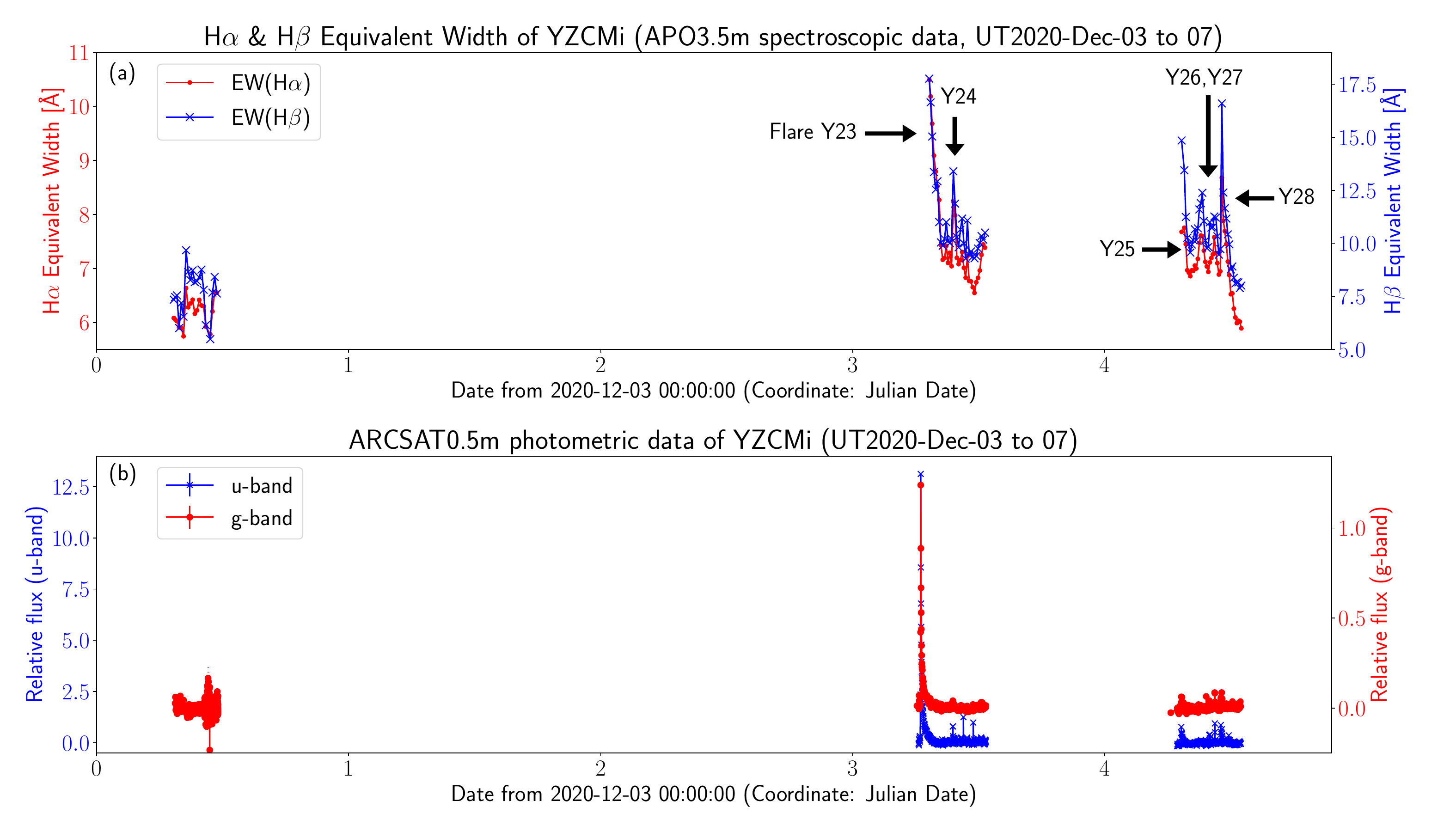}{0.9\textwidth}{\vspace{0mm}}
    }
     \vspace{-5mm}
     \caption{
Summary light curves of YZ CMi during the campaign season [iv] 2020 December 3 -- 7. The horizontal axes represent the observation time in the time coordinate of Julian Date (JD). 
(a) H$\alpha$ \& H$\beta$ equivalent width values from APO3.5m spectroscopic data. Red and Blue symbols correspond to H$\alpha$ \& H$\beta$ EWs, respectively. Black arrows indicate flares.
(b) $u$- \& $g$-band relative flux light curves from ARCSAT0.5m photometric data. Blue asterisks and red circles correspond to $u$- \& $g$-band data, respectively.
     }
   \label{fig:alllc1_2020Q4_YZCMi}
   \end{center}
 \end{figure}

\begin{figure}[ht!]
   \begin{center}
   \gridline{
    \fig{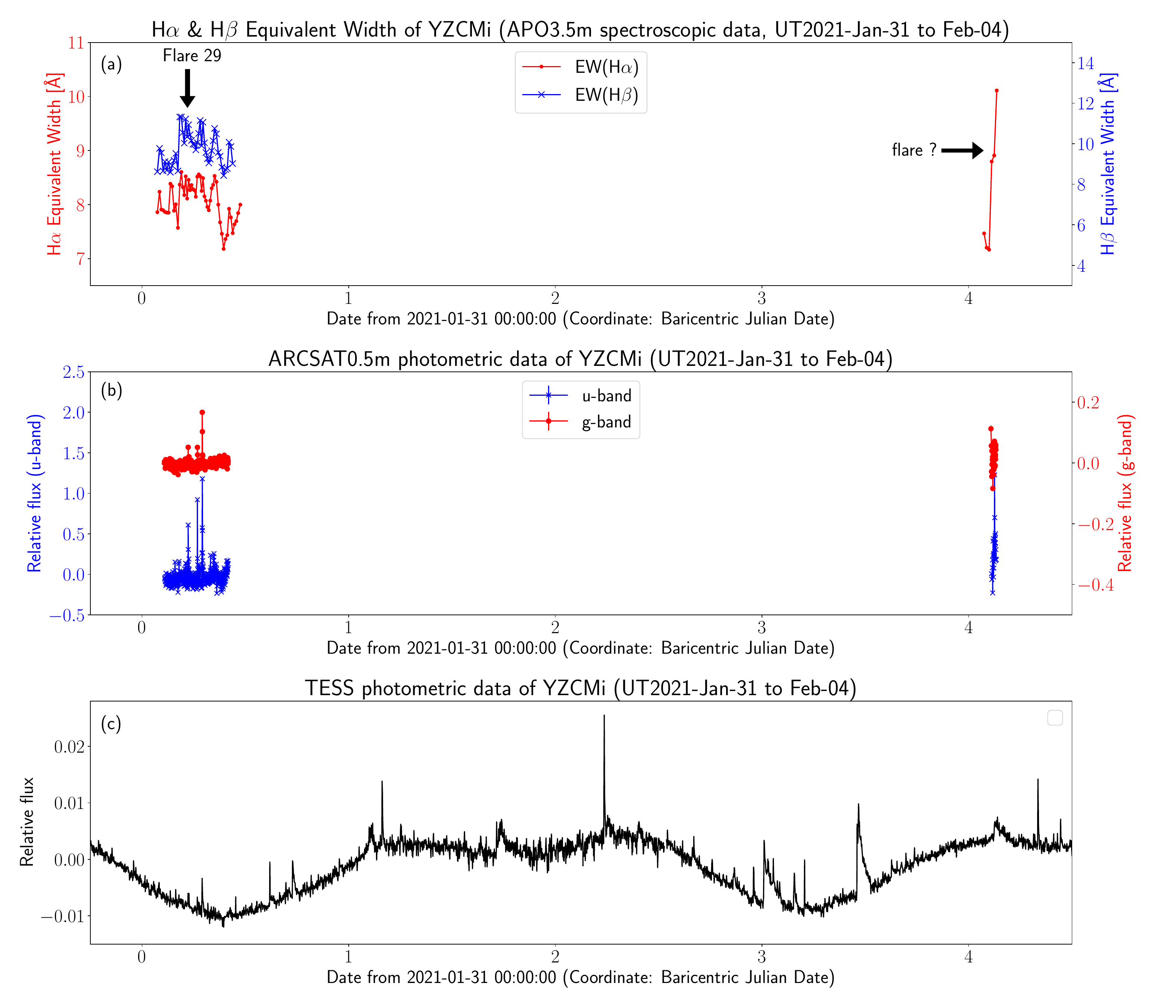}{0.9\textwidth}{\vspace{0mm}}
    }
     \vspace{-5mm}
     \caption{
Summary light curves of YZ CMi during the campaign season [v] 2021 January 31 -- February 4. 
The horizontal axes represent the observation time in Barycentric Julian Date (BJD).
\color{black}\textrm{
(a) \& (b) 
H$\alpha$ \& H$\beta$ equivalent width and $u$- \& $g$-band relative flux light curves 
are plotted with the same symbols as Figure \ref{fig:alllc1_2020Q4_YZCMi}.
 } \color{black}
(c) \textit{TESS}-band relative flux light curve from \textit{TESS} photometric data.
     }
   \label{fig:alllc1_2021Q1_YZCMi}
   \end{center}
 \end{figure}

\begin{figure}[ht!]
   \begin{center}
   \gridline{
    \fig{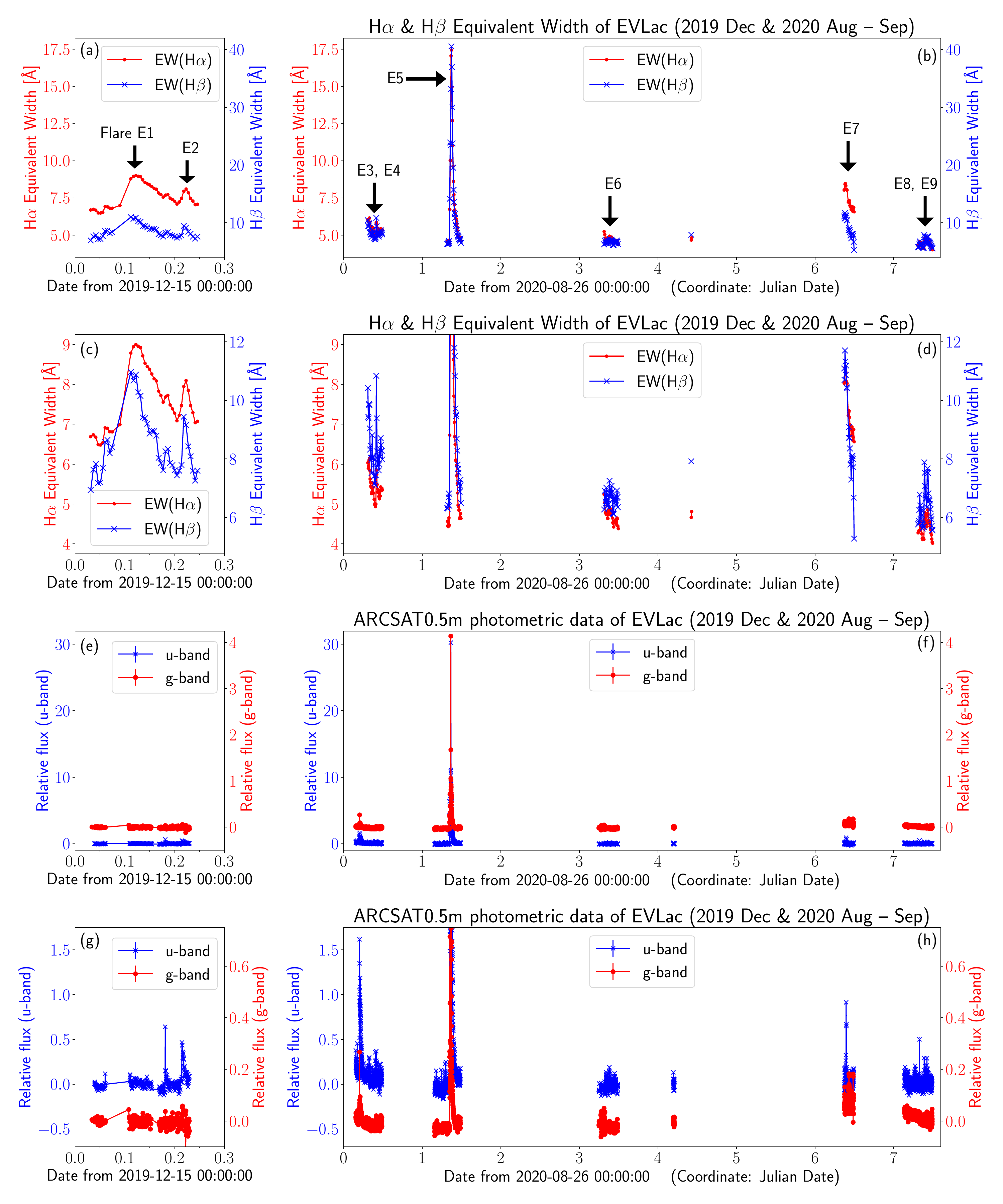}{0.9\textwidth}{\vspace{0mm}}
    }
     \vspace{-5mm}
     \caption{
Summary light curves of EV Lac during the two campaign seasons [vi] 2019 December 15 and [vii] 2020 August 26 -- September 2. 
\color{black}\textrm{
The horizontal axes represent the observation time in Julian Date (JD).
H$\alpha$ \& H$\beta$ equivalent width and $u$- \& $g$-band relative flux light curves 
are plotted with the same symbols as Figure \ref{fig:alllc1_2020Q4_YZCMi}. 
Panels (a), (c), (e), \& (g) are the data from the campaign season \color{black}\textrm{[vi]}\color{black}, while panels (b), (d), (f), \& (h) are  from the campaign season [vii].
Panels (c), (d), (g), \& (h) are vertical axis enlarged figures of (a), (b), (e), \& (f), respectively. 
     }
     }
   \label{fig:rot_all1_EVLac_2019Q4_2020Q3}
   \end{center}
 \end{figure}

\begin{figure}[ht!]
   \begin{center}
   \gridline{
    \fig{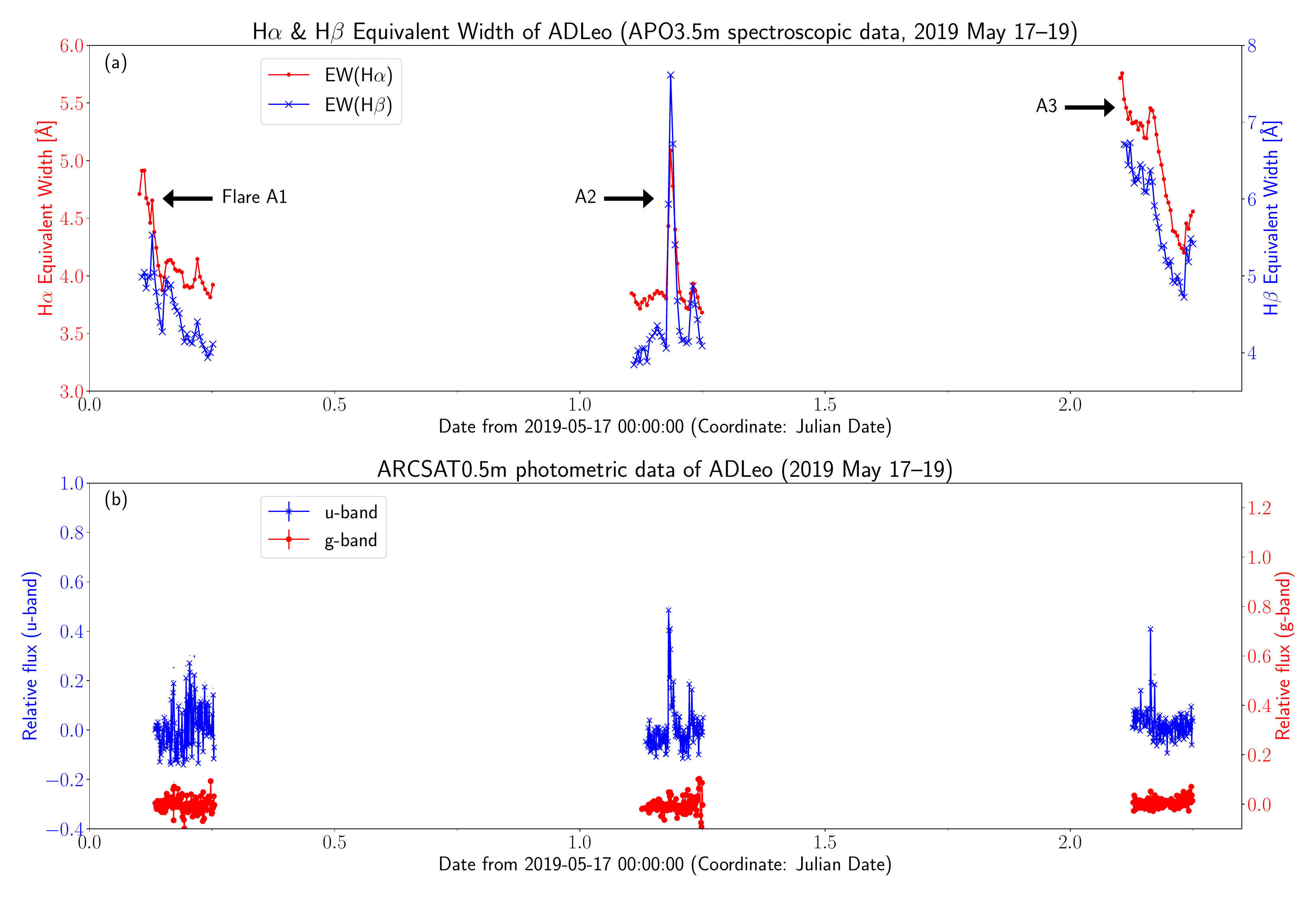}{0.9\textwidth}{\vspace{0mm}}
    }
     \vspace{-5mm}
     \caption{
\color{black}\textrm{
Summary light curves of AD Leo during the campaign season [viii] 2019 May 17 -- 19. The horizontal axes represent the observation date in the time coordinate of Julian Date (JD). 
H$\alpha$ \& H$\beta$ equivalent width and $u$- \& $g$-band relative flux light curves 
are plotted with the same symbols as Figure \ref{fig:alllc1_2020Q4_YZCMi}.
} \color{black}
     }
   \label{fig:rot_all1_ADLeo_2020Q2}
   \end{center}
 \end{figure}

As described in Section \ref{sec:data-methods}, we conducted the time-resolved simultaneous optical spectroscopic and photometric observations of the three target stars YZ CMi,  EV Lac, and AD Leo  during the 31 nights in total (Table \ref{table:obs_log}).
YZ CMi was observed during the five campaign seasons: [i] 2019 January (3 nights), [ii] 2019 December (4 nights), [iii] 2020 January (9 nights), [iv] 2020 December (3 nights), and [v] 2021 January -- February (2 nights). EV Lac was observed during the two campaign seasons: [vi] 2019 December (1 night) and [vii] 2020 August -- September (6 nights). AD Leo was observed during the one campaign season [viii] 2019 May (3 nights).
Figures \ref{fig:rot1_YZCMi_2019Q1}, \ref{fig:Ha_rot_YZCMi_2019Q4to2020Q1}, \ref{fig:alllc1_2020Q4_YZCMi}, \ref{fig:alllc1_2021Q1_YZCMi}, \ref{fig:rot_all1_EVLac_2019Q4_2020Q3}, \& \ref{fig:rot_all1_ADLeo_2020Q2} show the all light curves from the campaigns.

We note that flares are defined from the H$\alpha$ \& H$\beta$ data since the main purpose of this paper is to investigate the blue/red asymmetries of Balmer lines during flares, 
as described in the Introduction section.
It could be possible by definition that blue/red asymmetries occur with (i) flare emissions of Balmer lines below the detectable level, or (ii) 
in absence of flare-enhanced Balmer emission.
For example, if the prominence eruption causes the line asymmetries (see the references in the Introduction section), the \color{black}\textrm{detectability }\color{black} of prominence eruption could be unrelated with that of the flare emission itself in Balmer lines.
However, 
distinguishing among these alternative scenarios is beyond the scope of the current paper, 
since the main purpose of this paper is to report blue asymmetries of Balmer lines associated with clear flares, 
and to discuss the properties of blue asymmetries with flare properties. 
We note that there are no clear blue/red line asymmetries without clear flares in Balmer lines (see figures in the following part of this paper), and all line asymmetries in our observations are associated with flares in Balmer lines. Therefore 
our approach does not ultimately cause any major ambiguities.

In total, 41 
flares are detected as shown in Figures \ref{fig:rot1_YZCMi_2019Q1} -- \ref{fig:rot_all1_ADLeo_2020Q2} and listed in Table \ref{table:list1_flares}.
We label the 41  
flares by the first character of each the target star: Flares Y1-\color{black}\textrm{Y29 }\color{black} on YZ CMi, Flares E1-\color{black} \textrm{E9} \color{black} 
on EV Lac, and Flares A1-A3 on AD Leo.
As can be seen in Figures \ref{fig:rot1_YZCMi_2019Q1} -- \ref{fig:rot_all1_ADLeo_2020Q2}, 
the H$\alpha$ and H$\beta$ light curves are almost always variable (e.g., Figure \ref{fig:Ha_rot_YZCMi_2019Q4to2020Q1}(c))
compared with the photometric data, and this makes difficult 
to define ``non-flare" or ``quiescent" phases for many nights.
Since the duration of flares in Balmer lines can be relatively long (e.g., up to a few hours) in many cases, there can be a lot of flares overlapping with other flares, or in other words, the other flare starts before the preceding flare emission completely decays (e.g., Flare Y16\&Y17 in Figure \ref{fig:Ha_rot_YZCMi_2019Q4to2020Q1}(c)). 
Moreover, there are also many ``partial" flares (e.g., Flare A3 in Figure \ref{fig:rot_all1_ADLeo_2020Q2}(a)), and 
their observed flare properties (e.g., flare energies) could include various uncertainties since only the portions of flare phases were observed.
The main purpose of this paper is 
to understand the existence of various blue asymmetry events among various Balmer line flares, and some uncertainty of definitions of each flare could be left, as long as they are not expected to cause a serious problem for the main conclusion of this paper. 
Then we defined flares with rough definition as a phase having clear emission ``increase": 
the EW amplitude of H$\alpha$ $\gtrsim+$(0.5--1)\AA, 
compared with nearby ``quiescent" phase (or the phase having locally smaller emission compared with nearby data points).  The threshold values are
roughly determined for each observation period, considering the data S/N and quiescent level modulations 
($\gtrsim+$(0.5--1)\AA~for YZ CMi; 
$\gtrsim+$0.5\AA~for EV Lac and AD Leo), 
and the values are also described in the following paragraphs of this subsection.
There are flares with multiple peaks (e.g., Flare Y3 in Figure \ref{fig:rot1_YZCMi_2019Q1} (a))
but they are basically broadly classified into one long-timescale flare if the H$\alpha$ EW amplitude of these multiple peaks are smaller than the threshold 
$\lesssim+$(0.5--1)\AA~(e.g., Flare Y6).
We briefly describe how each flare is defined in the following paragraphs of this section.
All of the uncertainties of the flare definition described in this subsection should be kept in mind for the remainder analyses and discussions in this paper.

Figure \ref{fig:rot1_YZCMi_2019Q1} shows the light curves of YZ CMi during the campaign season [i] 2019 January 26 -- 28. 
During this campaign season [i], YZ CMi was observed with APO3.5m optical high-dispersion spectroscopy, ARCSAT ground-based photometry ($u$\&$g$-bands), \textit{TESS} space photometry Sector 7, and \textit{NICER} X-ray spectroscopy. 
Five flares (Y1 -- Y5) were detected in the H$\alpha$ \& H$\beta$ EW data
in Figure \ref{fig:rot1_YZCMi_2019Q1}(a).
\color{black}\textrm{These five flares were defined as phases showing the H$\alpha$ EW increase of 
$\gtrsim$ 1 \AA~compared with nearby local ``quiescent" phase on each night (EW of H$\alpha$ $\sim$8.0--8.6\AA~in the period of  2019 January 26 -- 28).
With this definition, a small amplitude increase after Flare Y1 at $\sim$0.4d in Figure \ref{fig:rot1_YZCMi_2019Q1} (a) was not counted as a flare. 
Flares Y2 and Y3 consist of multiple peaks but we only broadly classified into two flares since
the peaks during Flare Y2 and Y3 have H$\alpha$ EW amplitudes smaller than $\sim$1\AA,
and the lightcurve returned below the threshold only once at  $\sim$1.25d in Figure \ref{fig:rot1_YZCMi_2019Q1} (a) during the observation on 2019 January 27th (See also Section \ref{subsec:results:2019-Jan-27}). 
Flares Y4 and Y5 could be classified into one flare since there are continuous decreasing trend 
over the observation period of this night (2019 January 28th), 
but we classified them into two flares 
since both peaks have amplitudes larger than $\sim$1 \AA. 
It is then probable that independent flares can cause the time evolution of the EWs and there can be some meanings to separately classify them (as Flares Y4 and Y5) and investigate whether each peak has line asymmetries, considering the main purpose of this paper.
There is also a H$\alpha$ emission increase at around the beginning of the observation data on 2019 January 28 before Flare Y4 (at $\sim$2.13d in Figure \ref{fig:rot1_YZCMi_2019Q1} (a)),  
but we do not define this as a flare since only three data points with relatively low S/N exist and it is difficult to judge whether it showed line asymmetries for these data points (see also figures in Appendix \ref{subsec:results:2019-Jan-28}). In other words, this event cannot contribute to the main purpose of this paper even if it is counted as a flare, since it cannot be used for the line asymmetry classification.
}\color{black}
Among these five flares, only Flare Y3 was 
observed with \textit{NICER} X-ray data (Figure \ref{fig:rot1_YZCMi_2019Q1} (d) \& (e)).
The parameters of these five flares are listed in Table \ref{table:list1_flares} and these flares are described in detail in Section \ref{subsec:results:2019-Jan-27} and Appendix \ref{subsec:results:2019-Jan-26} -- \ref{subsec:results:2019-Jan-28}.

Figure \ref{fig:Ha_rot_YZCMi_2019Q4to2020Q1} shows the light curves of YZ CMi during the two campaign seasons [ii] 2019 December and [iii] 2020 January. During the campaign season [ii], 
YZ CMi was observed with APO3.5m optical high-dispersion spectroscopy and ARCSAT ground-based photometry ($u$\&$g$-bands)  (Table \ref{table:obs_log}). 
During the latter season [iii], YZ CMi was observed with APO3.5m spectroscopy on 2020 Jan 14, 18, and 20, and with SMARTS1.5m spectroscopy on every nights from 2020 Jan 16 to Jan 23 (Table \ref{table:obs_log}). ARCSAT photometry ($u$\&$g$-bands) and LCO photometry  ($U$\&$V$-bands) were conducted during the nights with APO3.5m spectroscopy and SMARTS 1.5m spectroscopy, respectively. We note that LCO observations continued until Jan 27 after the SMARTS 1.5m spectroscopic observations finished on 2020 Jan 27 (Table \ref{table:obs_log} and Figure \ref{fig:Ha_rot_YZCMi_2019Q4to2020Q1}(c)).

As a result, 17 flares (Flares Y6 -- Y22) were detected in H$\alpha$ \& H$\beta$ data during these extensive campaign seasons (Figure \ref{fig:Ha_rot_YZCMi_2019Q4to2020Q1} (a)\&(b)). 
\color{black}\textrm{
These 17 flares were defined as phases showing the H$\alpha$ EW increase of 
$\gtrsim$ 1 \AA~compared with nearby local ``quiescent" phase on each night.
During the campaign seasons [ii]\&[iii], there are flare-like increases at $\sim$6.2--6.3d in Figure  
\ref{fig:Ha_rot_YZCMi_2019Q4to2020Q1}(b) (2019 December 8) and 
at $\sim$3.2--3.3d in Figure  
\ref{fig:Ha_rot_YZCMi_2019Q4to2020Q1}(c) (2020 January 17), 
but we did not classify them into flares since 
a few data points with low S/N only exist and it is not possible to discuss whether they showed line asymmetries. In other words, these events cannot contribute to the main purpose of this paper even if they are counted as flares.
Flare Y6 has multiple peaks but these peaks are not separated into multiple independent flares,  
since the amplitude of each peak is only $\lesssim$1\AA~and 
the H$\alpha$ EW value was continuously much larger ($>$2.5-3.0\AA) than the local ``quiescent" level ($\sim$10.0--10.2\AA) (See also Section \ref{subsec:results:2019-Dec-12}).
Although Flare Y8 has only three data points after the flare start, this is counted as a flare since this has a very larger amplitude ($\sim$9 \AA) compared with other flares and it is possible to discuss the line profiles (See also Appendix \ref{subsec:results:2020-Jan-14}).
As for Flares Y10 and Y11, the H$\alpha$ EW did not come back to the quiescent level between these two flares and it was still at $\sim$9.7--9.8\AA, but we classified these two events into two flares (Y10 and Y11), since both of the two have duration of $\gtrsim$1 hour
and the emission has a clear local minimum between the two whose amplitude is $\gtrsim$1\AA~ (See also Appendix \ref{subsec:results:2020-Jan-16}).
It is then probable that independent flares can cause these time evolution of the EWs and there can be some meanings to separately classify them (as Flares Y10 and Y11) and investigate whether each peak has line asymmetries, considering the main purpose of this paper. 
Moreover, Flares Y14\&15 and Y16\&Y17 are classified into separated two flares, respectively, 
because of the same reason with the above Flares Y10\&Y11 
(See also Appendix \ref{subsec:results:2020-Jan-19} \& \ref{subsec:results:2020-Jan-20}).
In addition, Flares Y16\&Y17 are similar to Flares Y4 and Y5 mentioned above for the point that
they showed a continuous decreasing trend over the whole observation period of this night (2019 January 20), but they are treated as two flares because of the same reason with Flares Y4\&Y5 in the above.
}\color{black}

These \color{black}
\textrm{17 flares (Flares Y6 -- Y22) }\color{black}
are listed in Table \ref{table:list1_flares} and are described with more detailed figures in Section \ref{subsec:results:2019-Dec-12} 
 -- \ref{subsec:results:2020-Jan-21} and Appendix
\ref{subsec:results:2020-Jan-14} -- \ref{subsec:results:2020-Jan-23}. Figure \ref{fig:Ha_rot_YZCMi_2019Q4to2020Q1} (a)\&(b) also show that the H$\alpha$ \& H$\beta$ equivalent width EW values of the quiescent phase (non-flare phase) exhibit variability among the observation dates; this could be related with the rotational modulations, considering the rotation period of 2.77 days (Table \ref{table:targets_basic_para}).

Figure \ref{fig:alllc1_2020Q4_YZCMi} shows the light curves of YZ CMi during the campaign season [iv] 2020 December 3 -- 7. 
During this campaign season [iv], YZ CMi was observed with APO3.5m optical high-dispersion spectroscopy and ARCSAT ground-based photometry ($u$\&$g$-bands).
Six flares (Y23 -- Y28) were detected in the H$\alpha$ \& H$\beta$ equivalent width data
in Figure \ref{fig:alllc1_2020Q4_YZCMi} (a). 
\color{black}\textrm{
These six flares were defined as phases showing the H$\alpha$ EW increase of 
$\gtrsim$ 1 \AA~compared with nearby local ``quiescent" phase on each night (H$\alpha$ EW $\sim$6--7\AA).
As for the data of 2020 December 7,
three H$\alpha$ emission increase peaks with smaller amplitude (EW amplitude of H$\alpha$ $\gtrsim$0.5 \AA) were classified into separated flares (Flares Y25, Y26, and Y27). 
This is because there are peaks whose duration is $\gtrsim$1 hour and the emission of each peak clearly come back to local ``quiescent" phase (See also figures in Appendix \ref{subsec:results:2020-Dec-07}).
It is then probable that independent flares can cause the time evolution of the EWs and there can be some meanings to separately classify them (as Flares Y25, Y26, and Y27) and investigate whether each peak has line asymmetries, for the main purpose of this paper. 
}\color{black}
These flares are listed in Table \ref{table:list1_flares} and are described with more detailed figures in Section \ref{subsec:results:2020-Dec-06} and Appendix \ref{subsec:results:2020-Dec-07}. 

Figure \ref{fig:alllc1_2021Q1_YZCMi} shows the light curves of YZ CMi during the campaign season [v] 2021 January 31 -- February 4. 
During this campaign season [v], YZ CMi was observed with APO3.5m optical high-dispersion spectroscopy, ARCSAT ground-based photometry ($u$\&$g$-bands), and \textit{TESS} space photometry Sector 34. 
\color{black}\textrm{One flare (Y29)} \color{black} was detected in the H$\alpha$ \& H$\beta$ equivalent width data
in Figure \ref{fig:alllc1_2021Q1_YZCMi} (a). 
There could be a flare on February 4 as shown with the description ``flare ?" in 
Figure \ref{fig:alllc1_2021Q1_YZCMi} (a), 
but the S/N was low due to bad weather.
These flares are listed in Table \ref{table:list1_flares} and are described with more detailed figures in Appendix \ref{subsec:results:2021-Jan-31}. 

Figure \ref{fig:rot_all1_EVLac_2019Q4_2020Q3} shows the light curves of EV Lac during the two campaign seasons [vi] 2019 December 15 and [vii] 2020 August 26 -- September 2. During these two campaign seasons [vi] \& [vii], EV Lac was observed with APO3.5m optical high-dispersion spectroscopy and ARCSAT ground-based photometry ($u$\&$g$-bands).  
Nine flares (E1 -- E9) were detected in the H$\alpha$ \& H$\beta$ equivalent width data in Figures \ref{fig:rot_all1_EVLac_2019Q4_2020Q3} (a) \& (b). 
\color{black}\textrm{
These nine flares were defined as phases showing the H$\alpha$ EW increase of 
$\gtrsim$ 0.5 \AA~compared with nearby local ``quiescent" phase on each night (H$\alpha$ EW $\sim$5--7\AA). 
} \color{black}
These flares are listed in Table \ref{table:list1_flares} and are described with more detailed figures in Section \ref{subsec:results:2019-Dec-15} and Appendix \ref{subsec:results:2020-Aug-26} -- \ref{subsec:results:2020-Sep-02}. 

Figure \ref{fig:rot_all1_ADLeo_2020Q2} shows the light curves of AD Leo during the campaign season [viii] 2019 May 17 -- 19. 
During this campaign season [viii], AD Leo was observed with APO3.5m optical high-dispersion spectroscopy and ARCSAT ground-based photometry ($u$\&$g$-bands). 
Three flares (A1 -- A3) were detected in the H$\alpha$ \& H$\beta$ equivalent width data
in Figure \ref{fig:rot_all1_ADLeo_2020Q2} (a). 
\color{black}\textrm{
These three flares were defined as phases showing the H$\alpha$ EW increase of 
$\gtrsim$ 0.5 \AA~compared with nearby local ``quiescent" phase on each night (H$\alpha$ EW $\sim$5--7\AA).
} \color{black}
These flares are listed in Table \ref{table:list1_flares} and are described with more detailed figures in Section \ref{subsec:results:2019-May-19} and Appendix \ref{subsec:results:2019-May-17} -- \ref{subsec:results:2019-May-18}.

Next, 
we investigate whether blue wing asymmetries (enhancements of blue wings) are seen in the Hydrogen Balmer H$\alpha$ \& H$\beta$ lines.
If the blue wing asymmetries are observed during a flare in H$\alpha$ \& H$\beta$ lines, we also investigate whether other major chromospheric lines (H$\gamma$, H$\delta$, Ca II K, Ca II 8542, He I D3, Na I D1\&D2, H$\epsilon+$Ca II H) also show blue wing asymmetries.
As reported in the following subsections and listed in 
Tables \ref{table:list1_flares} \& \ref{table:list_blue_flares},
seven flares (Flares Y3, Y6, Y18, Y23, E1, E2, and A3) among all the 
\color{black}\textrm{41 }\color{black} flares 
showed clear blue wing asymmetries in H$\alpha$ \& H$\beta$ lines.  
These seven flares are also marked as ``(B)" in Table \ref{table:list1_flares}.
In Section \ref{subsec:results:2019-Jan-27} -- \ref{subsec:results:2019-May-19},
we discuss the detailed flare light curves and flare chromospheric line spectra from the observation dates when blue wing asymmetries in H$\alpha$ \& H$\beta$ lines 
were detected (YZCMi: 2019 January 27, 2019 December 12, 2020 January 18, \& 2020 December 6. 
EVLac: 2019 December 15. ADLeo: 2019 May 19). 
The data of the observation dates when blue wing asymmetries 
were not detected are shown in Appendix \ref{sec:app:non-blue-lcspec}. 
In this paper, we focus our analysis on the flares with blue wing asymmetries; 
the flares without blue asymmetries (e.g., flares only with red asymmetries and symmetric broadening) are briefly summarized in Section \ref{subsec:dis:flares-redsym}
and will be discussed in detail in our future papers.

\begin{longrotatetable}
\begin{deluxetable*}{lccccccccccccccccc}
   \tablecaption{List of all flares detected in H$\alpha$/H$\beta$ lines}
   \tablewidth{0pt}
   \tablehead{
     \colhead{ID \tablenotemark{\rm a}} &  \colhead{\color{black}WLF\tablenotemark{\rm b}} &  \colhead{UT date} & \colhead{$t^{\rm{flare}}_{\rm{start}}$ \tablenotemark{\rm d}} &   \multicolumn{6}{c}{Peak luminosity \tablenotemark{\color{black}\rm e,f}}  & &  \multicolumn{6}{c}{Energy \tablenotemark{\color{black}\rm g,h}} &
     \colhead{$\Delta t^{\rm{flare}}_{\rm{H}\alpha}$ } \\
     \colhead{} &   \colhead{\color{black}Assym.\tablenotemark{\rm c}} & \colhead{}  & \colhead{(JD/BJD} &   \multicolumn{6}{c}{($10^{27}$ erg s$^{-1}$)}  & &  \multicolumn{6}{c}{($10^{31}$ erg)} &
     \colhead{(hour)} \\
     \colhead{} &   \colhead{} & & \colhead{-2400000)} &   \multicolumn{6}{c}{}  & &  \multicolumn{6}{c}{} &
     \colhead{\tablenotemark{\rm d,h}} \\
           \cline{5-10}   \cline{12-17} 
      &  &  &  & \colhead{$L_{U}$} & \colhead{$L_{u}$} & \colhead{$L_{g}$} &
       \colhead{$L_{TESS}$} &  \colhead{$L_{\rm{H}\alpha}$} &  \colhead{$L_{\rm{H}\beta}$}  & &
      \colhead{$E_{U}$} & \colhead{$E_{u}$} & \colhead{$E_{g}$} &
      \colhead{$E_{TESS}$} &  \colhead{$E_{\rm{H}\alpha}$} &  \colhead{$E_{\rm{H}\beta}$} &  \colhead{} \\
      &  &  & & \colhead{} & \colhead{} & \colhead{} &
       \colhead{$L_{V}$ \tablenotemark{\color{black}\rm i}} &  \colhead{} &  \colhead{}  & &
      \colhead{} & \colhead{} & \colhead{} &
      \colhead{$E_{V}$ \tablenotemark{\color{black}\rm i}} &  \colhead{} &  \colhead{} &  \colhead{} 
     }
   \startdata
   Y1 & \color{black}WL\color{black} & 2019 Jan 26 & 58509.706 & -- & -- & 419 & 320 & \color {black}2.4--2.8 & \color {black}2.0--2.3 & & --  & -- & 11 & 15 & \color {black}0.73--0.84 & \color {black}0.37--0.42 & 1.5  \\
   Y2 & \color{black}NEP\color{black} & 2019 Jan 27 & before obs. & -- & $<$8 & $<$16 & $<$33 & 1.3 & \color {black}1.0 & & -- & $<$4.1 & $<$8.9 & $<$19 & \color {black}$>$\color{black}0.74 & \color {black}$>$0.49 & $>$3.2  \\
   Y3 & NWL, & 2019 Jan 27 &  58510.748 & -- &  $<$8 & \color{black} $<$16 & \color{black} $<$33 & 2.8 & 1.8 & & -- & \color{black} $<$5.6 & \color{black} $<$12 & \color{black} $<$26 & 1.7 &  \color {black}0.76 & 4.3  \\
 & (B)  &  &  &  & &  &  &  &  &  & & & &  & & & \\
   Y4 & \color{black}WL, \color{black}  & 2019 Jan 28 & 58511.654 & -- & 22 & 16 & 73 & \color {black}4.2--4.3 & \color {black}3.0--3.1 & & -- & 0.8 & 0.7 & 6.0 & \color {black}0.92--0.93 & \color {black}0.51--0.52 & 1.1  \\
 & (R)  &  &  &  & &  &  &  &  &  & & & &  & & & \\
   Y5 & \color{black}NWL\color{black}& 2019 Jan 28 & 58511.819 & -- & \color{black}$<$5 & \color{black}$<$9 & \color{black}$<$29 & \color {black}1.7--1.8 & 1.5--1.6 & & -- & \color{black}$<$1.0 & \color{black}$<$1.9 & \color{black}$<$6.5 & \color {black}0.38--0.40 &  \color {black}0.26--0.27 & 1.3  \\
   Y6 & NWL, & 2019 Dec 12 & 58829.841 & -- & \color{black} $<$8 & \color{black}$<$1.8 & -- & \color {black}3.6--3.8 & \color {black}1.8--1.9 &  & -- & $<$7.2 & $<$16 & -- &  \color {black}$>$3.9 &  \color {black}$>$1.8 & $>$4.9 \\
 & (B)(R)  &  &  &  & &  &  &  &  &  & & & &  & & & \\
   Y7 & \color{black}WL, \color{black} & 2020 Jan 14 & 58862.830 & -- & 15 & 19 & -- & \color {black}1.8--2.1 & \color {black}1.4--1.6 & & -- & 4.9 & 3.9 & -- & \color {black}1.1--1.2 & \color {black}0.76--0.82 & 2.9  \\
 & (B?)  &  &  &  & &  &  &  &  &  & & & &  & & & \\
   Y8  & \color{black}WL\color{black} & 2020 Jan 14 & 58862.951 & -- & 1249 & 1492 & -- & \color {black}10 & \color {black}9.6 & & -- & 12 & 13 & -- & \color {black}$>$0.88 & \color {black}$>$0.71 & $>$0.4 \\
   Y9 & \color{black}WL\color{black}  & 2020 Jan 16 & before obs. & 103 & -- & -- & 202 & 3.4 & 3.3 &   & 3.7 & --  & -- & 6.0 & \color {black}$>$\color{black}0.72 & \color {black}$>$\color{black}0.66 & $>$1.7  \\
   Y10  & \color{black}WL, \color{black} & 2020 Jan 16 & 58864.670 & 44 & -- & -- &  \color{black} -- & 2.8 & 3.0 & & 1.8 & -- & -- & \color{black} -- & 0.84 & 0.89 & 1.2 \\
 & (R)  &  &  &  & &  &  &  &  &  & & & &  & & & \\
   Y11 & \color{black}WL, \color{black} & 2020 Jan 16 & 58864.719 & 97 & -- & -- & 144 & \color {black}3.5--3.9 &  \color {black}3.1--3.3 & & 1.8 & -- & -- & 3.2 & \color {black}1.2--1.4 &  \color {black}0.89--0.97 & 2.0 \\
 & (R)  &  &  &  & &  &  &  &  &  & & & &  & & & \\
   Y12  & \color{black}WL, \color{black} & 2020 Jan 18 & 58866.677 & 84 & 71 & 63 & 47 & \color {black}3.9--4.5 & \color {black}3.4--3.6 & & 5.6 & 11 & 16 & 4.2 & \color {black}4.2--4.8 & \color {black}3.0--3.2 & 5.7 \\
 & (R)  &  &  &  & &  &  &  &  &  & & & &  & & & \\
   Y13  & \color{black}WL\color{black} & 2020 Jan 18 & 58866.914 & -- & 38 & 63 & 32 & 2.0 & 1.7 & & -- & 1.5 & 1.9 & 0.8 & 0.97 & 0.60 & 2.3 \\
   Y14 & \color{black}WL\color{black} & 2020 Jan 19 & 58867.788 & 21 & 14 & 21 & 22 & \color {black}2.3--2.7 & \color {black}1.9--2.1 & & 2.1 & 1.9 & 4.2 & 3.3 & \color {black}0.54--0.67 & 0.40-0.47 & 1.2 \\
   Y15 & \color{black}WL\color{black} & 2020 Jan 19 & 58867.844 & -- & 292 & 547  & -- & \color {black}2.4--3.7 & \color {black}2.7--3.5 & & -- & 8.0 & 13 & -- & \color {black}$>$0.24  & \color {black}$>$0.07 & $>$0.3 \\
   Y16  & \color{black}WL, \color{black} & 2020 Jan 20 & before obs. & -- & 24 & 21 & 30 & \color {black}4.0 & \color {black}4.3 & & -- & 2.4 & 1.7 & 0.7 & \color {black}$>$2.4 & \color {black}$>$2.4 & $>$2.5  \\
 & (R)  &  &  &  & &  &  &  &  &  & & & &  & & & \\
   Y17 & \color{black}WL, \color{black} & 2020 Jan 20 & 58868.707 & -- & 29 & 25 & 34 & \color {black}4.2--5.0 & \color {black}5.2--5.7 & & -- & 15 & 6.3 & 9.8 & \color {black}3.4--3.7 & \color {black}4.2--4.3 & 6.0  \\
 & (R)  &  &  &  & &  &  &  &  &  & & & &  & & & \\
   Y18  & \color{black}WL, \color{black} & 2020 Jan 21 & 58869.591 & 41 & -- & -- & 41 & \color {black}2.6--2.8 & \color {black}1.9 & & 4.6 & -- & -- & 2.4 & \color {black}1.5--1.6 & \color {black}1.3 & 3.4 \\
 & (B)  &  &  &  & &  &  &  &  &  & & & &  & & & \\
   Y19  & \color{black}WL, \color{black}  & 2020 Jan 21 & 58869.738 & 59 & -- & -- & 41 & \color {black}4.8--5.1 &  \color {black}4.5--4.9 & & 11 & -- & -- & 4.4 & \color {black}$>$3.3 & \color {black}$>$2.5 & $>$2.5  \\
 & (R)  &  &  &  & &  &  &  &  &  & & & &  & & & \\
   Y20  & \color{black}WL\color{black} & 2020 Jan 22 & 58870.669 & 87 & -- & -- & 46 & \color {black}1.5--1.9 & \color {black}1.3--1.7 & & 5.8 & -- & -- & 2.2 & \color {black}0.52--0.56 & \color {black}0.26--0.28 & 2.0 \\
   Y21  & \color{black}WL\color{black} & 2020 Jan 23 & 58871.586 & 21 & -- & -- & $<$17 & 1.6 & 1.6 & & 3.8 & -- & -- & \color{black} $<$5.0  & 0.44 & 0.40 & 1.7   \\
   Y22 & \color{black}NEP\color{black}  & 2020 Jan 23 & 58871.656 & -- & -- & -- & -- & 1.9 & 1.9 & & -- & -- & -- & -- & 0.74 & 0.59 & 3.2  \\
   Y23  & \color{black}WL, \color{black} & 2020 Dec 06 & before obs. & -- & 397 & 461 & -- & \color {black}4.1--4.5 & \color {black}3.0--3.2 & & -- & 16 & 22 & -- & \color {black}$>$1.1 & \color {black}$>$0.59 & $>$1.3 \\
 & (B)  &  &  &  & &  &  &  &  &  & & & &  & & & \\
   Y24 & \color{black}WL\color{black} & 2020 Dec 06 & 59189.892 & -- & 23 & 16 & -- & \color {black}1.7--1.8 & \color {black}1.4--1.5 & & -- & 0.5 & 0.6 & -- & \color {black}0.25--0.26 & \color {black}0.17--0.18 &  0.7 \\
   Y25  & \color{black}WL\color{black} & 2020 Dec 07 & before obs. & -- & 27 & 20 & -- & \color {black}1.9--2.0 &  \color {black}2.4--2.5 & & -- & 0.8 & 0.5 & -- & \color {black}$>$0.43 & \color {black}$>$0.39 & $>$0.8   \\
   Y26 & \color{black}NWL\color{black} & 2020 Dec 07 & 59190.839 & -- & \color{black}$<$8 & $<$18 & -- & \color {black}1.8 & \color {black}1.5 & & -- & $<$7.2 & $<$16 & -- & \color {black}0.82--0.86 & \color {black}0.59--0.62 & 1.9  \\
   Y27 & \color{black}WL\color{black} & 2020 Dec 07 & 59190.916 & -- & 32 & 28 & -- & \color {black}1.7 & \color {black}1.1 & & -- & 1.2 & 0.4 & -- & \color {black}0.41--0.42 & \color {black}0.29--0.30 & 0.8  \\
   Y28 & \color{black}WL, \color{black} & 2020 Dec 07 & 59190.952 & -- & 30 & 28 & -- & \color {black}2.9--3.3 & \color {black}3.0--3.3 & & -- & 1.7 & 1.0 & -- & \color {black}0.71 & \color {black}0.52 & 1.7  \\
 & (R)  &  &  &  & &  &  &  &  &  & & & &  & & & \\
   \color{black}Y29 & \color{black}WL\color{black} &  \color{black}2021 Jan 31 & \color{black}59245.674 & -- & \color{black}43 & \color{black}63 & \color{black}57 & \color {black}1.3--1.4 & \color {black}0.94--0.95 & & -- & \color{black}3.3 & \color{black}1.8 & \color{black}2.5 & \color {black}1.7 & \color {black}0.95 &  \color{black}5.3  \\
   E1 & \color{black}NEP, \color{black} & 2019 Dec 15 & 58832.556 & -- & -- & -- & -- & 4.3 & 2.2 & & -- & -- & -- & -- & 2.9 & 1.4 & 3.6  \\
 & (B)  &  &  &  & &  &  &  &  &  & & & &  & & & \\
   E2 & \color{black}WL, \color{black} & 2019 Dec 15 & 58832.705 & -- & 23 & 37 & -- & \color {black}2.8--2.9 & \color {black}1.3--1.4 & & -- & 1.1 & 0.7 & -- & 0.58 &  0.16--0.17 & 0.9  \\
 & (B)  &  &  &  & &  &  &  &  &  & & & &  & & & \\
   E3  & \color{black}WL\color{black} & 2020 Aug 26 & before obs. & -- & 13 & $<$8 & -- & \color {black}2.0--2.2 & \color {black}1.9--2.1 & & -- & 1.0 & $<$3.3 & -- & \color {black}$>$0.74 & \color {black}$>$0.55 & $>$2.3  \\
   E4  & \color{black}WL\color{black} & 2020 Aug 26 & 59087.906 & -- & 16 & 29 & -- & \color {black}1.5--1.9 & \color {black}2.2--2.5 & & -- & 1.4 & 2.0 & -- & \color {black}$>$0.47 & \color {black}$>$0.47 & $>$2.1  \\
   E5  & \color{black}WL, \color{black} & 2020 Aug 27 & 59088.846 & -- & 1618 & 2708 & -- & \color {black}22--47 & \color {black}21--35 & & -- & 84 & 106 & -- & \color {black}7.3--10.4 & \color {black}6.7--9.1 & 3.5  \\
 & (R)  &  &  &  & &  &  &  &  &  & & & &  & & & \\
   E6  & \color{black}WL\color{black} & 2020 Aug 29 & before obs. & -- & 10 & 14 & -- &  \color {black}1.3--1.9 & \color {black}0.70--0.74 & &  -- & 0.3 & 0.7 & -- & \color {black}$>$0.53 & \color {black}$>$0.35 & $>$2.7  \\
   E7 & \color{black}NEP\color{black} & 2020 Sep 01 & before obs. & -- & \color{black}-- & \color{black}-- & -- & \color {black}3.0--3.2 &  \color {black}2.3--2.4 & & -- & \color{black}-- & \color{black}-- & -- & \color {black}$>$1.2 & \color {black}$>$0.88 & $>$2.1  \\
   E8 & \color{black}WL\color{black} & 2020 Sep 02 & 59094.811 & -- & 24 & \color{black}$<$9 & -- & \color {black}0.78--0.81 & \color {black}0.68--0.70 & & -- & 0.1 & \color{black}$<$2.4  & -- & \color {black}0.17--0.19 & \color {black}0.12--0.13 & 1.4  \\
   E9 & \color{black}WL\color{black} & 2020 Sep 02 & 59094.884 & -- & 14 & \color{black}$<$9 & -- & \color {black}1.2--1.3 & \color {black}1.4 & & -- & 0.9 & \color{black}$<$4.5 & -- & \color {black}0.55--0.61 & \color {black}0.55--0.58 & 2.7  \\
   A1 & \color{black}NEP\color{black} & 2019 May 17 & before obs. & -- & -- & -- & -- & 3.7 & 1.9 & & -- & -- & -- & -- & \color {black}$>$\color{black}0.85 & \color {black}$>$\color{black}0.35 & $>$1.1  \\
   A2 & \color{black}WL, \color{black} & 2019 May 18 & 58621.676 & -- & 61 & 72 & -- & \color {black}4.5--4.9 &  \color {black}4.8--5.0 & & -- & 3.0 & 1.6 & -- & \color {black}0.52--0.58 &  \color {black}0.55--0.57 &  1.0 \\
 & (R)  &  &  &  & &  &  &  &  &  & & & &  & & & \\
   A3 &  \color{black}NEP, \color{black} & 2019 May 19 & before obs. & -- & \color{black}-- & \color{black}-- & -- & \color {black}7.4 & \color {black}3.7 & & -- & \color{black}$>$2.7 & \color{black}$>$1.4 & -- & $>$5.3 & $>$2.5 & $>$3.1  \\
 & (B)  &  &  &  & &  &  &  &  &  & & & &  & & & \\
   \enddata
   \tablenotetext{\rm a}{
  \color{black}\textrm{Flare ID. }\color{black}
 Flares Y1 -- \color{black}\textrm{Y29 }\color{black} are on YZ CMi,
 Flares E1 -- E9 are on EV Lac, and 
 Flares A1 -- A3 are on AD Leo.
   }
    \tablenotetext{\rm b}{
The 6 flares with the mark ``NEP" do Not have Enough Photometric data to 
judge whether the flares are white-light (WL) or non white-light (NWL) flares.
The 31 flares  with the mark ``WL" are identified as WL flares.
The 4 flares with the mark ``NWL" are identified as ``candidate" NWL flares. 
\color {black}\textrm{
See Section \ref{subsec:dis:flare-energy} for the classification criteria.
} }
\tablenotetext{\rm c}{
 Flares with ``(B)" showed clear blue wing asymmetries in H$\alpha$ line, while those with ``(B?)" showed possible blue wing asymmetries but not so clear.
 Flares with ``(R)" showed clear red wing asymmetries in H$\alpha$ line.
   }
       \tablenotetext{\rm d}{
  The flare start time ($t^{\rm{flare}}_{\rm{start}}$) and flare duration ($\Delta t^{\rm{flare}}_{\rm{H}\alpha}$) are measured from the H$\alpha$ light curve. Flares Y1 -- Y5 and \color{black}\textrm{Y29 }\color{black} are shown with BJD. Other flares are shown with JD.
  }
\tablenotetext{\rm e}{         
  \color{black}\textrm{
   As for the peak luminosities in photometric bands ($L_{U}$, $L_{u}$, $L_{g}$, $L_{V}$, and $L_{TESS}$), we selected the peaks that are considered to be most physically associated with the flare peaks in the H$\alpha$ \& H$\beta$ lines.  This means that the largest flare peaks in photometric bands are not necessarily selected, but those most closest in time with the flare peaks in the H$\alpha$ \& H$\beta$ lines are basically selected.
  } \color{black}
  }
\tablenotetext{\rm f}{         
  \color{black}\textrm{
The upper limit marks ``$<$" for the peak luminosities in photometric bands ($L_{U}$, $L_{u}$, $L_{g}$, $L_{V}$, and $L_{TESS}$) show that 
the clear flare emission is not identified in this band, 
and the detection threshold value ``$\Delta f_{\rm{band, flare}}<3\times \sigma_{\rm{band}}$" 
is shown (see Section \ref{subsec:quiescent-ene} for the details).
  } \color{black}
   }
\tablenotetext{\rm g}{         
The upper limit marks ``$<$" for the flare energies in photometric bands ($E_{U}$, $E_{u}$, $E_{g}$, $E_{V}$, and $E_{TESS}$) show that 
the clear flare emission is not identified in this band, 
and the upper limit flare energies are shown (see Section \ref{subsec:quiescent-ene} for the estimation method). 
As for Flare A3, the lower limit of flare energies are listed with the marks ``$>$" to show the energy values 
\color {black}\textrm{
from the available data, since only the late phase of this Flare A3 was observed in photometric bands and it is highly possible that there are additional white-light emission (see also Section \ref{subsec:results:2019-May-19})}\color{black}.
Most of the LCO $U$-\& $V$-band photometric data (except for Flare Y14) have some data gaps. Because of this, $E_{U}$ and $E_{V}$ values 
can be lower limit values except for Flare Y14.
   }   
\tablenotetext{\rm h}{  
 The lower limit marks (``$>$") for the $E_{\rm{H}\alpha}$, $E_{\rm{H}\beta}$ and 
 $\Delta t^{\rm{flare}}_{\rm{H}\alpha}$ show that the total flare phases 
 were not observed in the Balmer lines, 
 and only lower limit of these values are measured. 
 In these cases, the flare energy values in photometric bands ($E_{U}$, $E_{u}$, $E_{g}$,  $E_{V}$, $E_{TESS}$)
 can be larger than the listed values, but the lower limit marks (``$>$") are 
 not added for these photometric flare energy values, since it is not necessarily obvious whether there are additional (unobserved) white-light emissions even if the total flare phases 
 were not observed in the Balmer lines (except for Flare A3 described in the footnote g).
 As for Flares Y2, Y6, and E3, the upper limit values of photometric flare energies ($E_{u}$, $E_{g}$, and $E_{TESS}$) with the marks ``$<$" are shown (cf. footnote g), but the real upper limit values of $E_{u}$, $E_{g}$, and $E_{TESS}$ could be larger than the listed values, as the total flare phases of these flares 
 were not observed in the Balmer lines (cf. the lower limit marks are shown for the $E_{\rm{H}\alpha}$, $E_{\rm{H}\beta}$ and  $\Delta t^{\rm{flare}}_{\rm{H}\alpha}$ values of these flares). 
   }  
       \tablenotetext{\rm i}{
As for Flares Y1 -- Y5 and \color{black}\textrm{Y29}\color{black}, \textit{TESS}-band data are listed. 
As for Flares Y9 -- Y22,  $V$-band data are listed.
   }
   \label{table:list1_flares}
 \end{deluxetable*}
\end{longrotatetable}

 \clearpage

\subsection{Flares Y2 \& Y3 (Blue wing asymmetry) observed on 2019 January 27} \label{subsec:results:2019-Jan-27}

\begin{figure}[ht!]
   \begin{center}
   \gridline{
    \fig{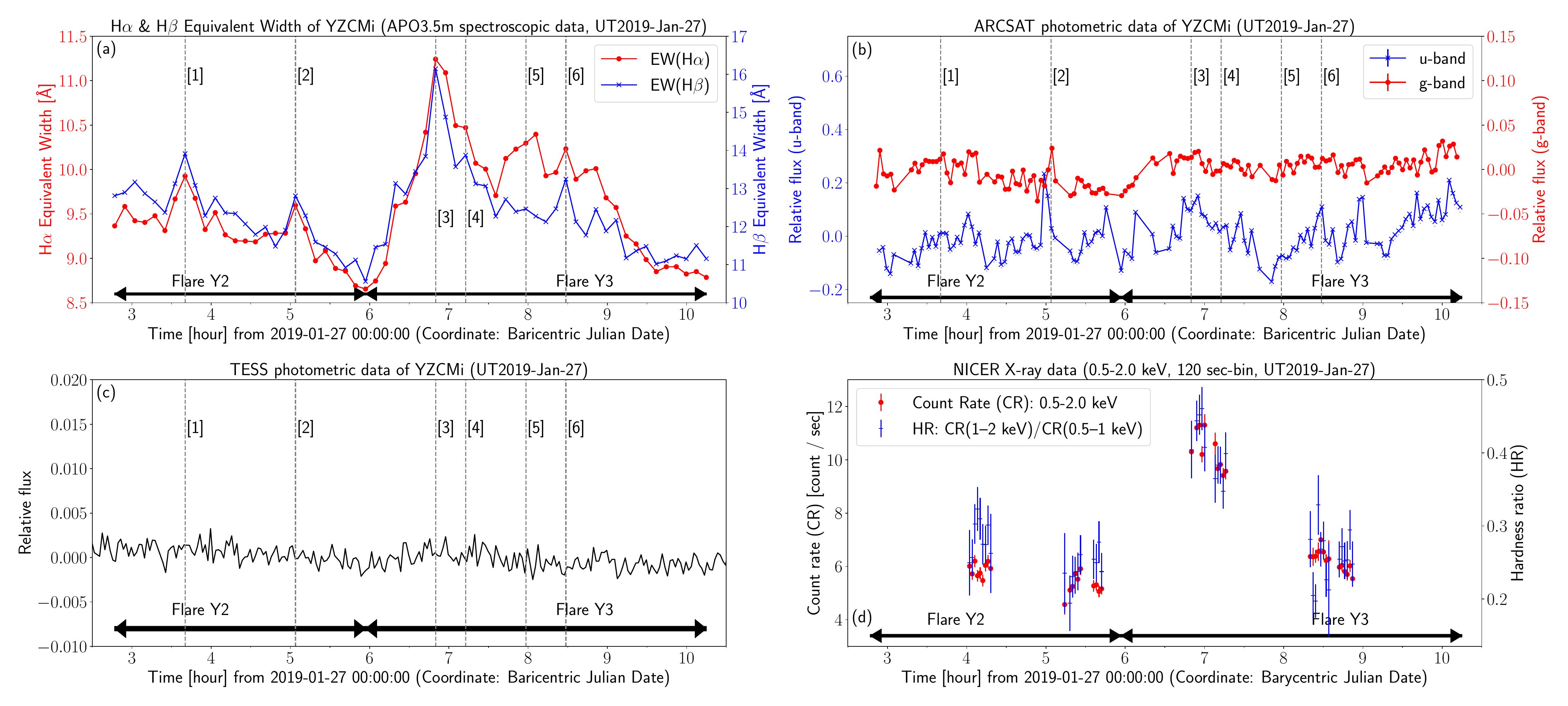}{1.0\textwidth}{\vspace{0mm}}
    }
     \vspace{-10mm}
   \gridline{
       \hspace{0.00\textwidth}
    \fig{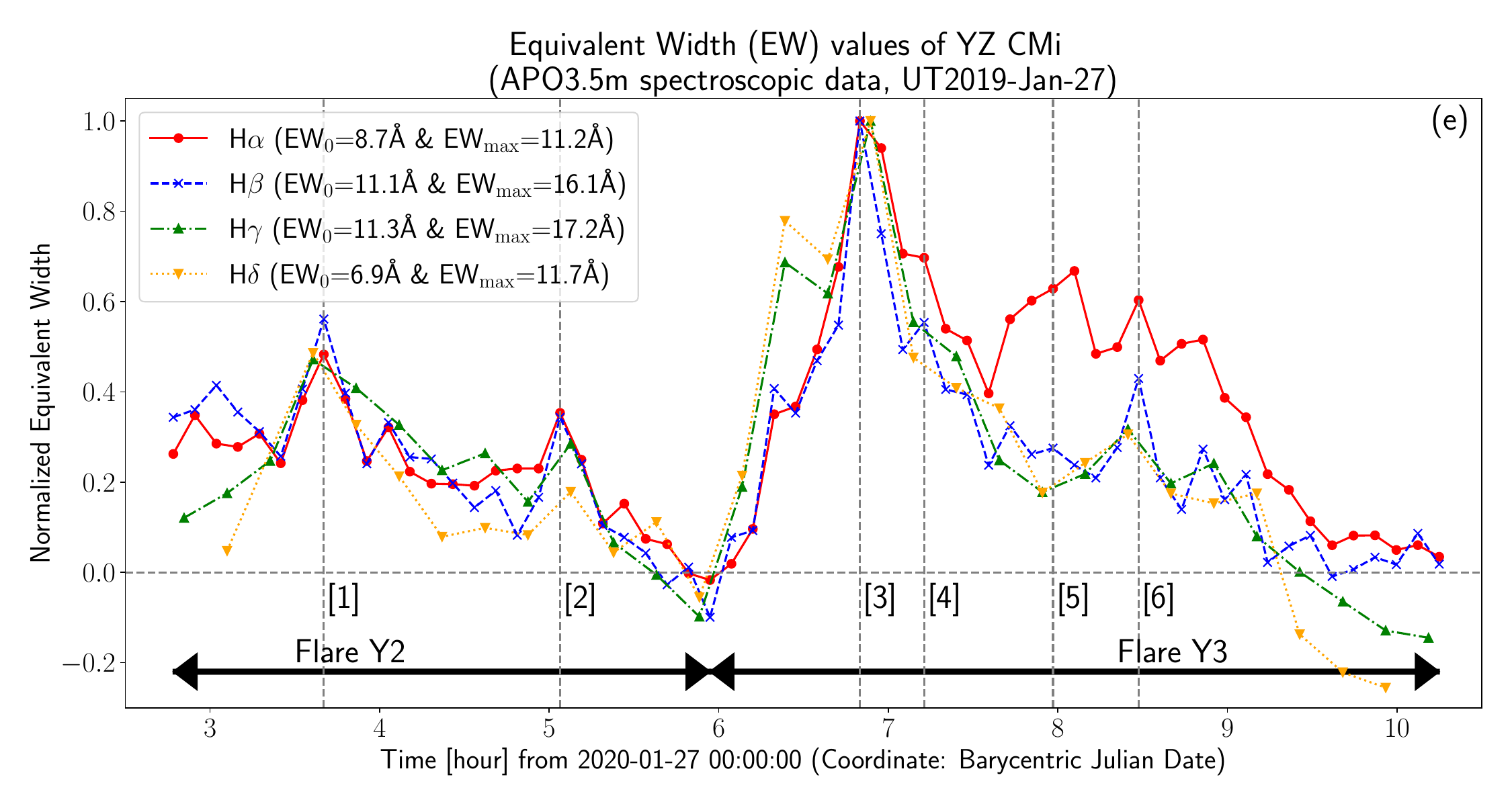}{0.5\textwidth}{\vspace{0mm}}\hspace{-0.00\textwidth}
 \fig{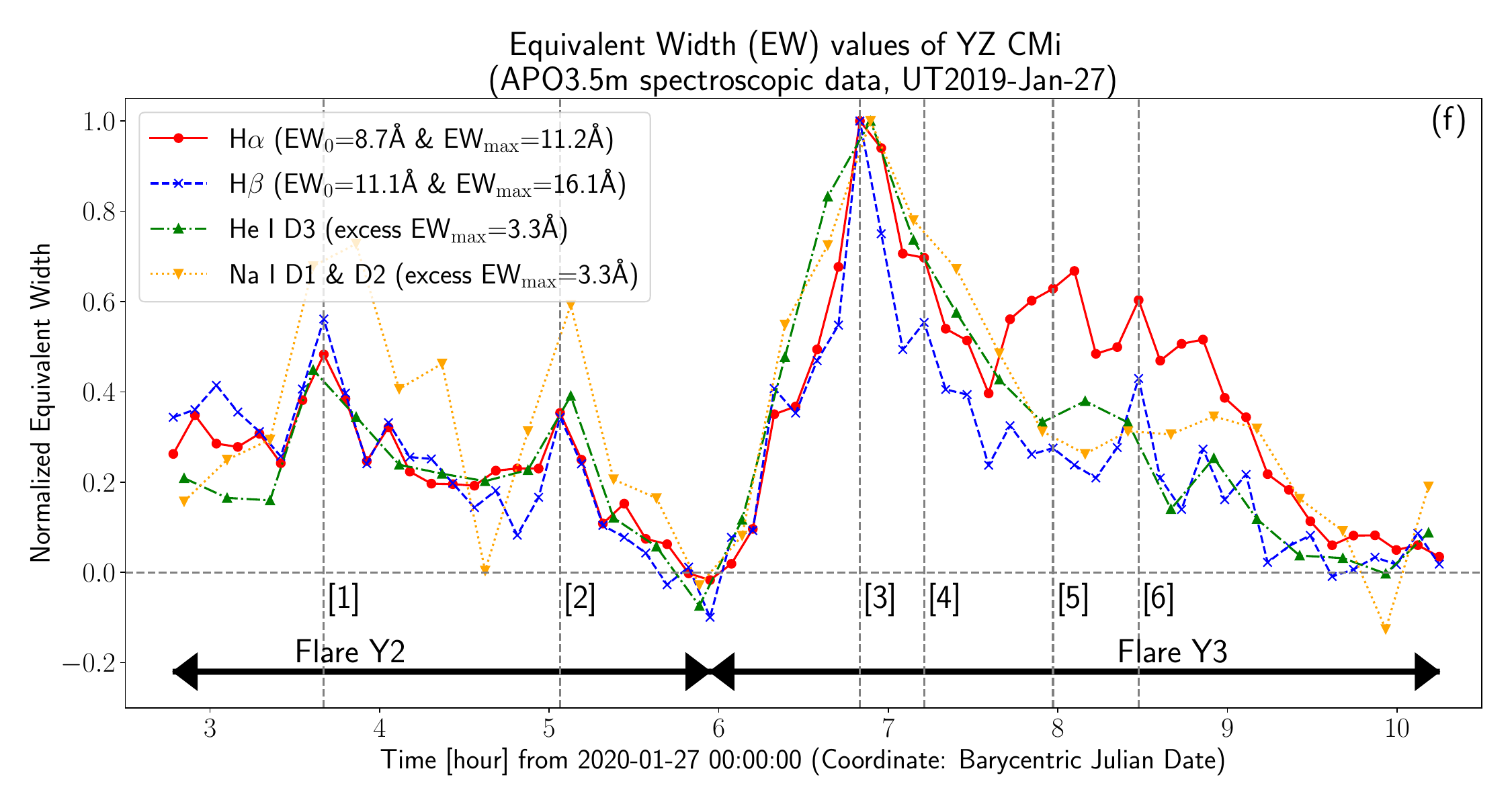}{0.5\textwidth}{\vspace{0mm}}
    }
         \vspace{-10mm}
       \gridline{
       \hspace{-0.02\textwidth}
       \hspace{0.00\textwidth}
    \fig{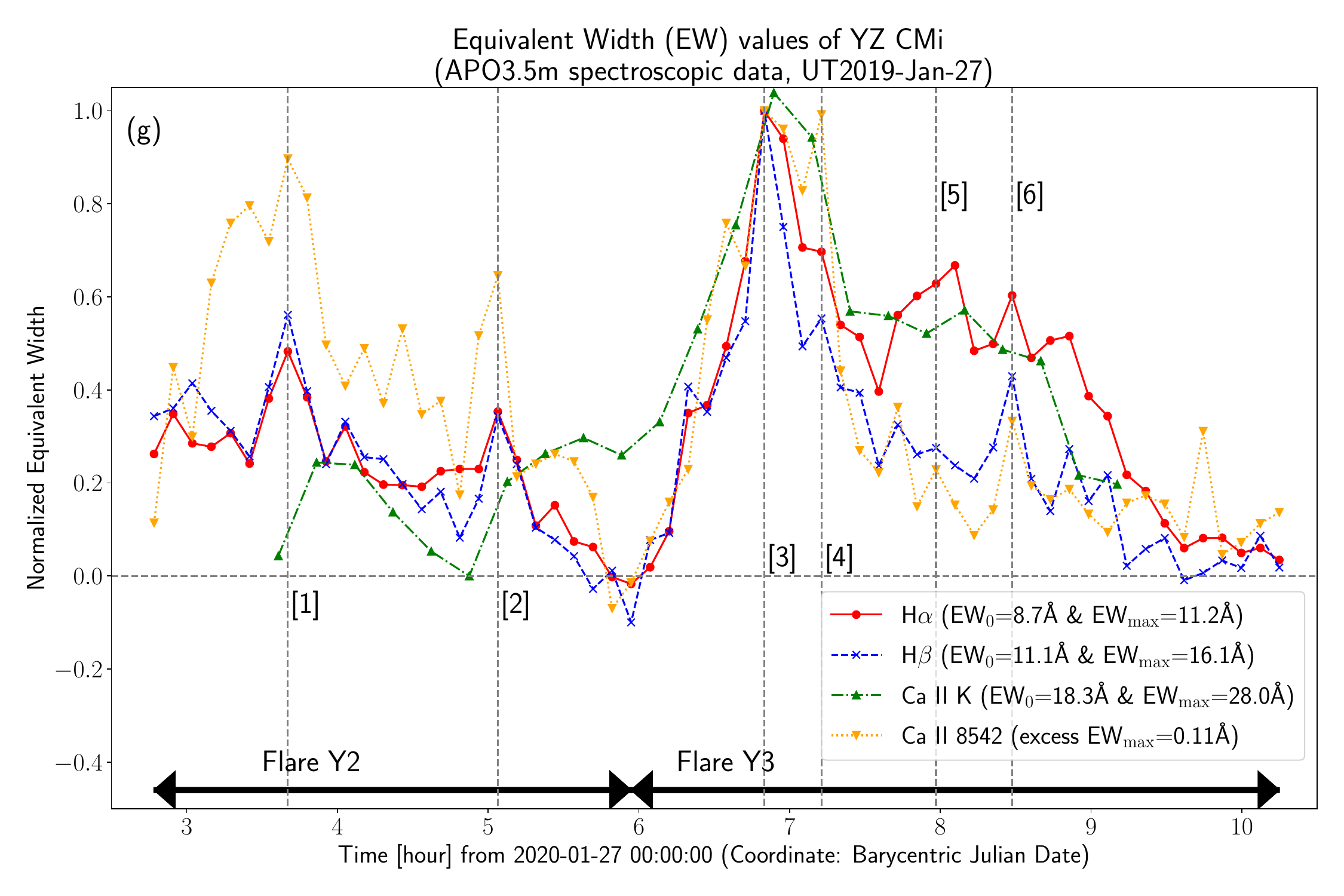}{0.5\textwidth}{\vspace{0mm}}
    }
     \vspace{-5mm}
     \caption{
Light curves of YZ CMi on 2019 January 27 showing Flares Y2 \& Y3. 
The horizontal axes represent the observation time in the time coordinate of Barycentric Julian Date (BJD). 
The grey dashed lines with numbers ([1]--[6]) correspond to the time shown with the same numbers in Figures \ref{fig:spec_HaHb_YZCMi_UT190127}, \ref{fig:map_HaHb_YZCMi_UT190127}, \& \ref{fig:spec_HcHd_YZCMi_UT190127}.
(a) H$\alpha$ \& H$\beta$ equivalent width light curves from APO 3.5m spectroscopic data. 
Red and Blue symbols correspond to H$\alpha$ \& H$\beta$ EWs, respectively. The black double-headed arrows indicate the start and end time of Flares Y2 \& Y3 (cf. $\Delta t^{\rm{flare}}_{\rm{H}\alpha}$ in Table \ref{table:list1_flares}).
(b) $u$- \& $g$-band relative flux light curves from ARCSAT0.5m photometric data.  Blue asterisks and red circles correspond to $u$- \& $g$-band data, respectively.
(c)
\textit{TESS}-band relative flux light curve from \textit{TESS} photometric data.
(d)
Red circles are background-subtracted X-ray count rates [count s$^{-1}$] from \textit{NICER} data in 0.5--2.0 keV. Blue plus marks are X-ray Hardness ratio (count rate (1.0--2.0 keV) / count rate (0.5--1.0 keV)) from \textit{NICER} data.
(e) EW light curves of H$\alpha$ (Red symbols), H$\beta$ (Blue), \color{black}\textrm{H$\gamma$ (Green), \& H$\delta$ (Orange)} \color{black} from APO 3.5m spectroscopic data. The EW values are normalized with their peak and quiescent values (EW$_{\rm max}$ and EW$_{0}$). 
(f) Same as (e), but for H$\alpha$ (Red symbols), H$\beta$ (Blue), He I D3 (Green), \& Na I D1\&D2 (Orange). As for He I D3 and Na I D1\&D2 lines, excess EWs (differences from the quiescent components) are plotted.
(g) Same as (e), but for H$\alpha$ (Red symbols), H$\beta$ (Blue), Ca II K (Green), \& Ca II 8542 (Orange). As for Ca II 8542 line, excess EWs (differences from the quiescent components) are plotted.
}
\label{fig:lcEW_HaHb_YZCMi_UT190127}
   \end{center}
 \end{figure}

On 2019 January 27, two flares (Flares Y2 \& Y3) were detected in H$\alpha$ \& H$\beta$ lines as shown in Figure \ref{fig:lcEW_HaHb_YZCMi_UT190127} (a).  During Flare Y2,
the H$\alpha$ \& H$\beta$ equivalent widths increased to 9.9\AA~and 13.9\AA, respectively, and the flare duration in H$\alpha$ ($\Delta t^{\rm{flare}}_{\rm{H}\alpha}$) is $>$3.2 hours (Table \ref{table:list1_flares}). We note that Flare Y2 was in progress when the observation was started.
Flare Y3 has the larger amplitude than Flare Y2. During Flare Y3, the H$\alpha$ \& H$\beta$ equivalent widths increased to 11.2\AA~and 16.1\AA, respectively, and the flare duration $\Delta t^{\rm{flare}}_{\rm{H}\alpha}$ is 4.3 hours (Table \ref{table:list1_flares}).
In addition to chromospheric lines, Flare Y3 is detected also in \textit{NICER} X-ray data (Figure \ref{fig:lcEW_HaHb_YZCMi_UT190127} (d)).
\color{black}\textrm{ 
The white-light flux observed by ARCSAT $u$- \& $g$-bands and \textit{TESS} did not show clear enhancements above the photometric errors of the data ($3\sigma_{u}$=22.9\% , $3\sigma_{g}$=4.2\%, and $3\sigma_{TESS}$=0.34\%)  during these two flares (Figures \ref{fig:lcEW_HaHb_YZCMi_UT190127} (b) \& (c)). 
As described in Section \ref{subsec:quiescent-ene}, 3 $\times$ the standard deviation ($3\sigma_{u}$, $3\sigma_{g}$, and $3\sigma_{TESS}$) of the relative flux in the quiescent phase for each night 
is used for the detection threshold of the white-light flare emission.
} \color{black}
There are very small ``suggestive" increases in $u$- \& $g$-bands and \textit{TESS} data around time 6--8h in Figures \ref{fig:lcEW_HaHb_YZCMi_UT190127} (b) \& (c),
\color{black}\textrm{
although this is still smaller than the threshold ($3\sigma_{TESS}$=0.34\%).
} \color{black}
We also note that these small increases could be caused 
by the emission lines (e.g., Balmer lines) included in $u$-, $g$-, and \textit{TESS}-bands.

\color{black}\textrm{  
We estimated the upper limits to the flare component peak luminosities and flare energies in $u$-, $g$-, and \textit{TESS}-bands, 
following the method described in Section \ref{subsec:quiescent-ene}, 
and the resultant values ($L_{u}$, $L_{g}$, $L_{TESS}$, $E_{u}$, $E_{g}$, and $E_{TESS}$) are in Table \ref{table:list1_flares}.
The flare component peak luminosities and flare energies of H$\alpha$ \& H$\beta$ lines
($L_{\rm{H}\alpha}$, $L_{\rm{H}\beta}$, $E_{\rm{H}\alpha}$, and $E_{\rm{H}\beta}$) are also estimated and listed in Table \ref{table:list1_flares}, following the method described in Section \ref{subsec:quiescent-ene}.
Since Flare Y2 already started when the observation was started, 
the real flare energy values could be larger than the values listed here.
} \color{black}

\begin{figure}[ht!]
   \begin{center}
           \gridline{  
     \hspace{-0.06\textwidth}
    \fig{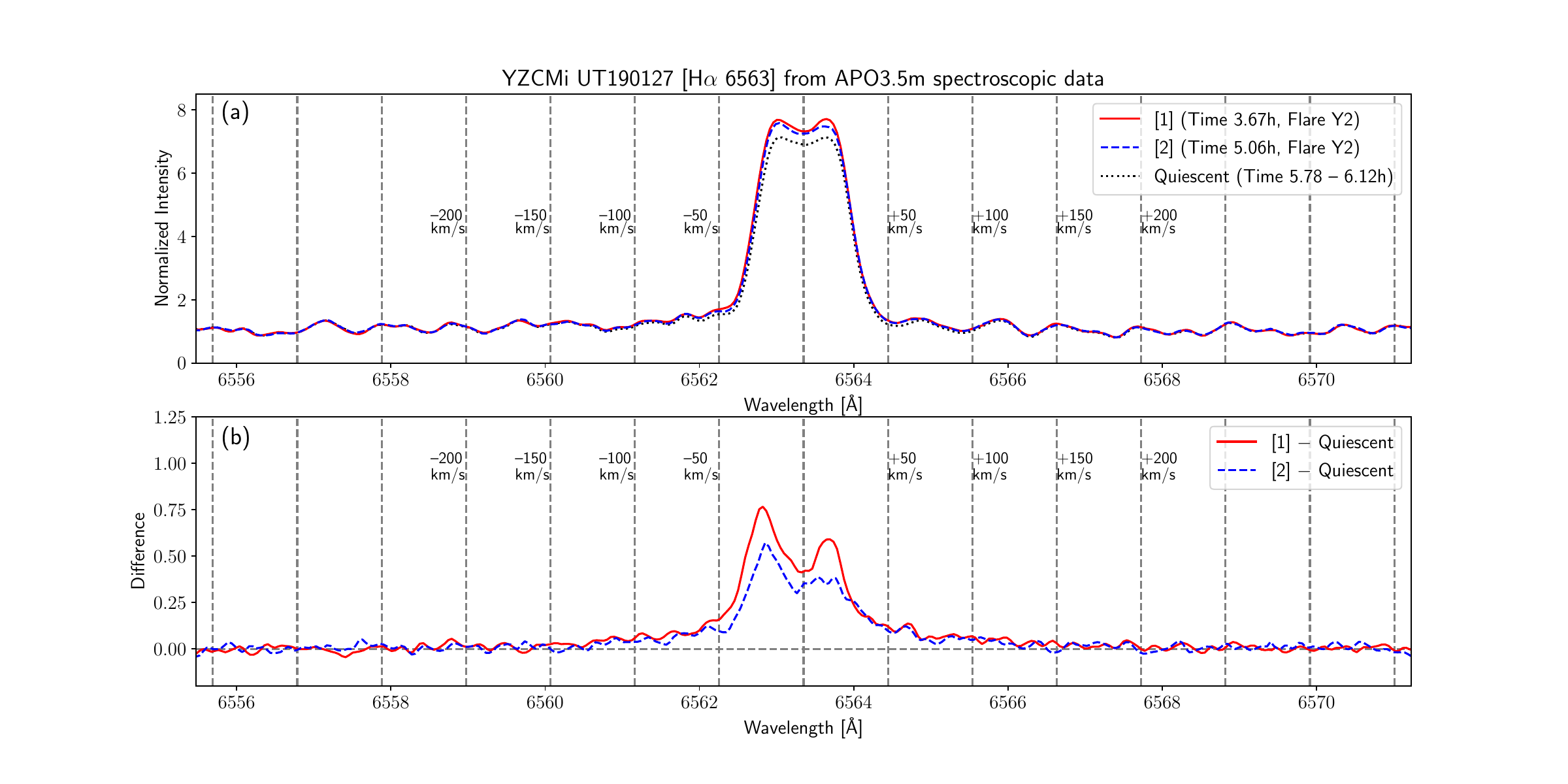}{0.58\textwidth}{\vspace{0mm}}
     \hspace{-0.06\textwidth}
       \fig{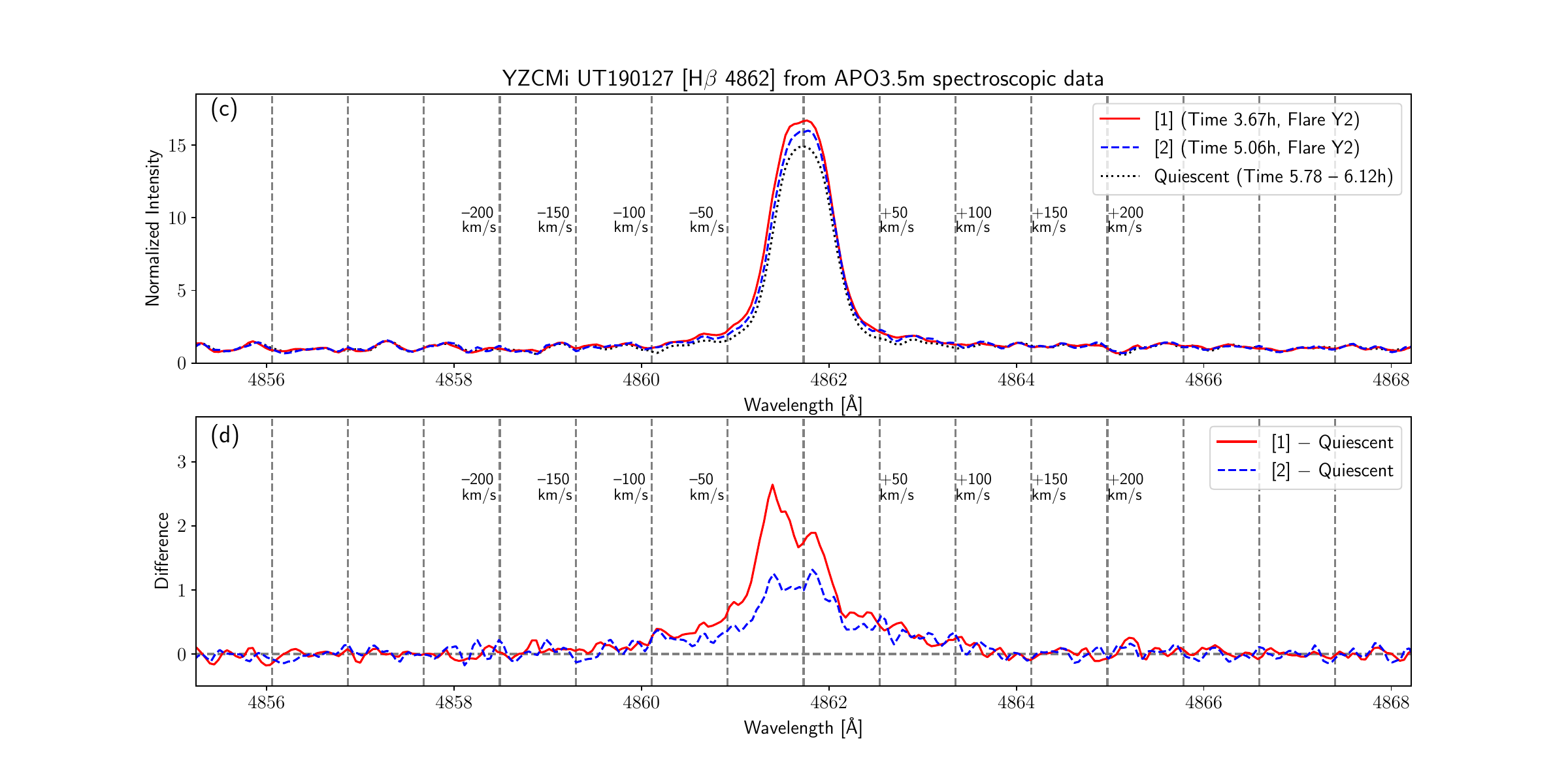}{0.58\textwidth}{\vspace{0mm}}
    }
    \vspace{-1.2cm}
        \gridline{  
     \hspace{-0.06\textwidth}
    \fig{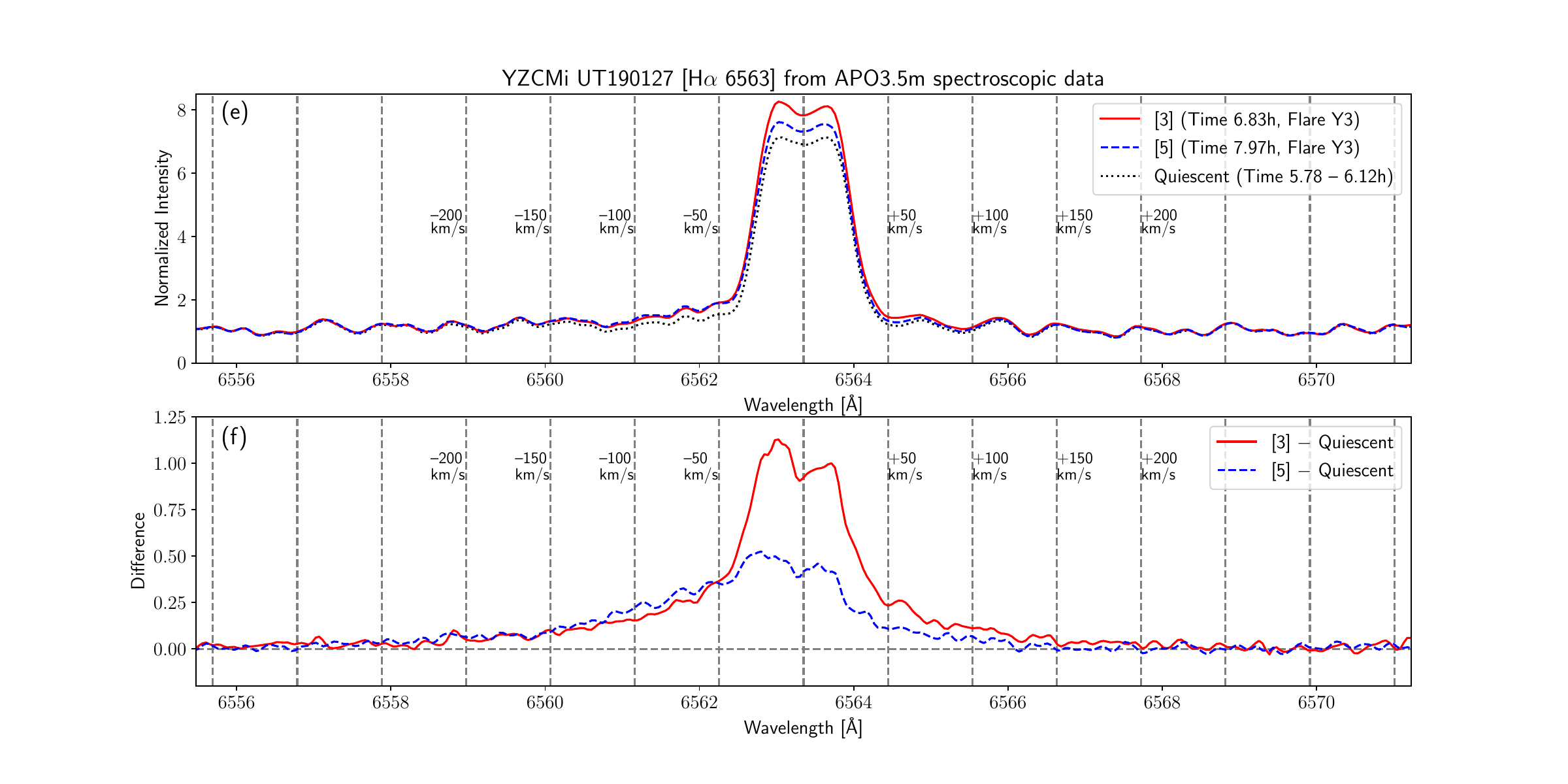}{0.58\textwidth}{\vspace{0mm}}
     \hspace{-0.06\textwidth}
       \fig{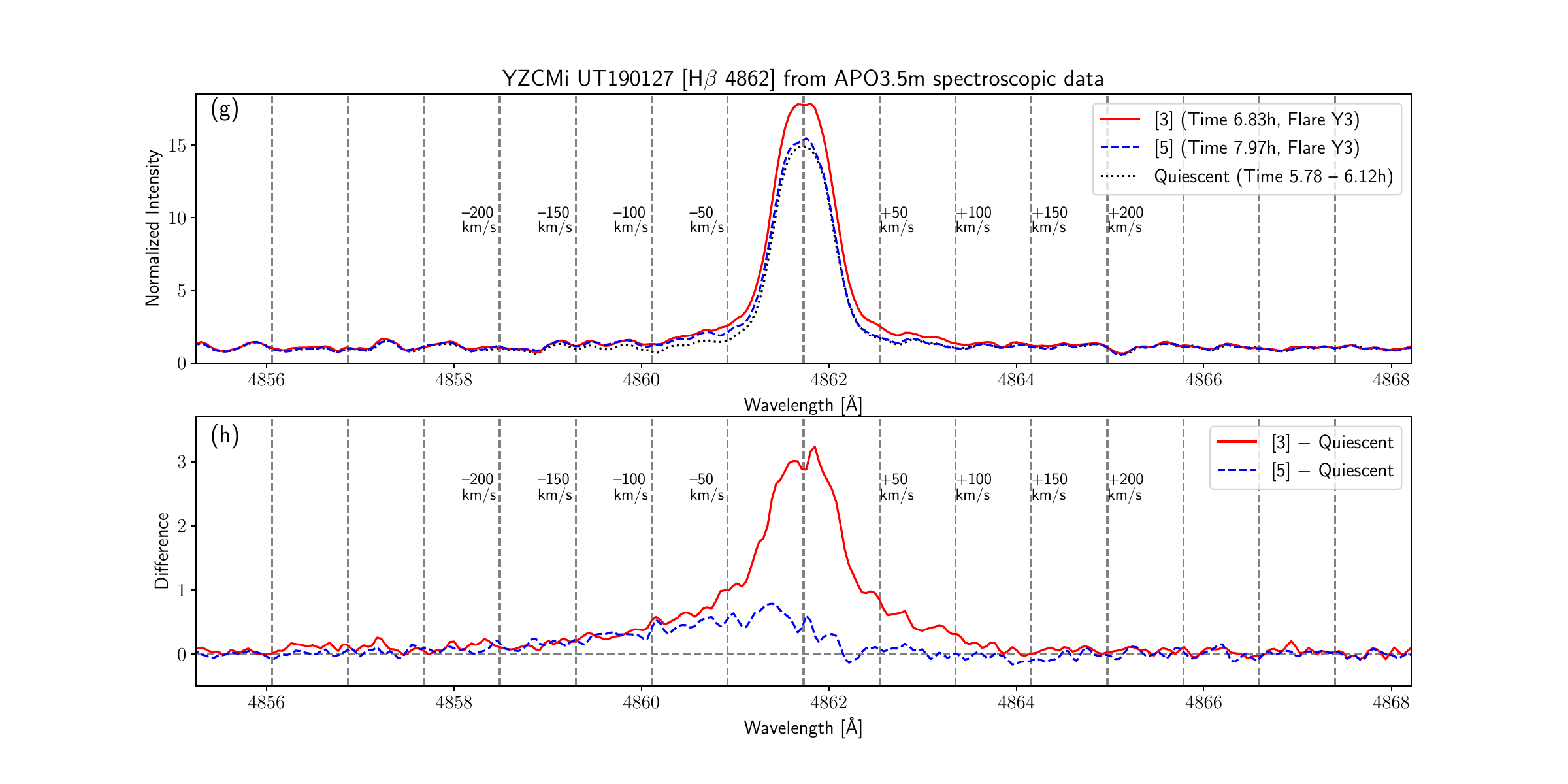}{0.58\textwidth}{\vspace{0mm}}
    }
    \vspace{-1.2cm}
        \gridline{  
     \hspace{-0.06\textwidth}
    \fig{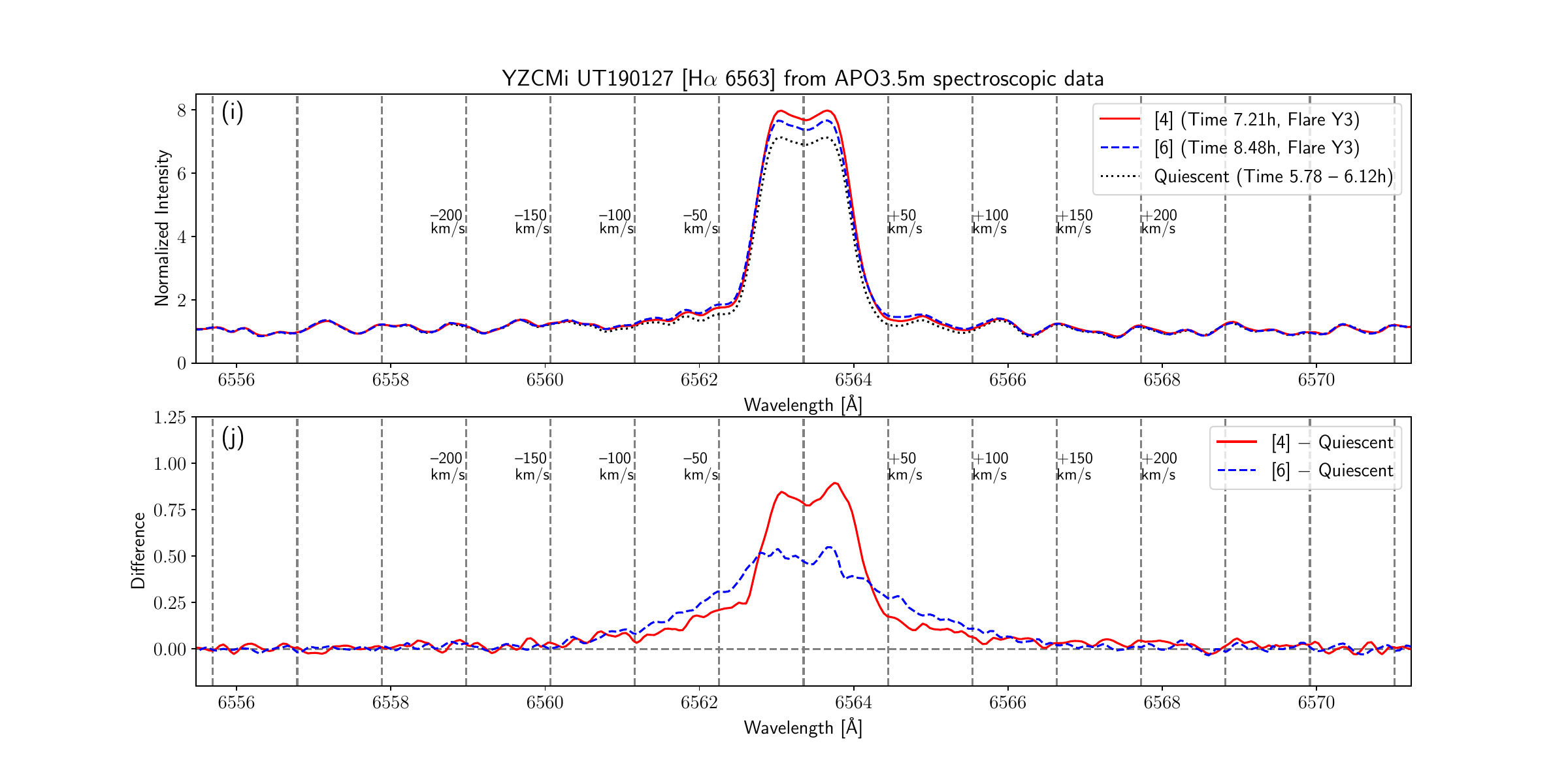}{0.58\textwidth}{\vspace{0mm}}
     \hspace{-0.06\textwidth}
       \fig{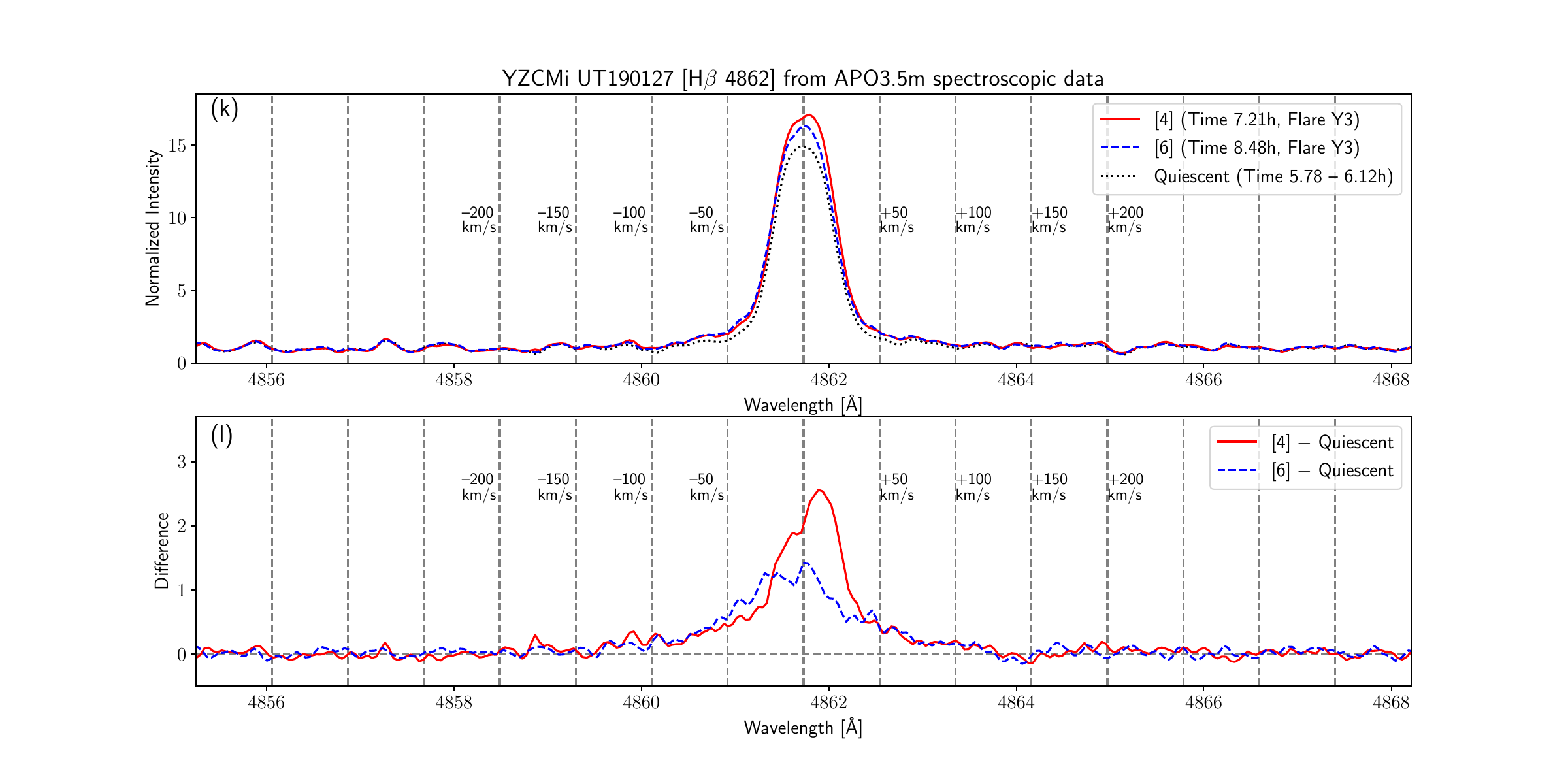}{0.58\textwidth}{\vspace{0mm}}
    }
    \vspace{-0.6cm}
     \caption{
(a) Line profiles of the H$\alpha$ emission line during Flare Y2 on 2019 January 27 from APO3.5m spectroscopic data. The horizontal and vertical axes represent the wavelength and flux normalized by the continuum. The grey vertical dashed lines with velocity values represent the Doppler velocities from the H$\alpha$ line center. The red solid and blue dashed lines indicate the line profiles at the time [1] and [2], respectively, which are indicated in Figure \ref{fig:lcEW_HaHb_YZCMi_UT190127} (light curves) and are during Flare Y2.
The black dotted line indicates the line profiles in quiescent phase, which are the average profile during 5.78h -- 6.12h on this date (cf. Figure \ref{fig:lcEW_HaHb_YZCMi_UT190127}(a)).
\color{black}\textrm{
(e) Same as panel (a), but the line profiles at the time [3] and [5] during Flare Y3. 
(i) Same as panel (a), but the line profiles at the time [4] and [6] during Flare Y3.
It is noted that the profiles at Time [3] -- [6] are not plotted in order (Those at [3]\&[5] are in (e)--(h), while those at [4]\&[6] are in (i)--(l)), so that the blue asymmetries at Time [3]\&[5] are plotted in the same panels.
(c), (g), \& (k) Same as panels (a), (e), \& (i), respectively, 
but the line profiles for the H$\beta$ emission line.
(b), (d), (f), (h), (j), \& (l) Same as panels (a), (c), (e), (g), (i), \& (k), respectively, 
but the line profile \color{black}\textrm{differences }\color{black} from the quiescent phase.
}
     }
   \label{fig:spec_HaHb_YZCMi_UT190127}
   \end{center}
 \end{figure}

\begin{figure}[ht!]
   \begin{center}
      \gridline{  
     \hspace{-0.07\textwidth}
    \fig{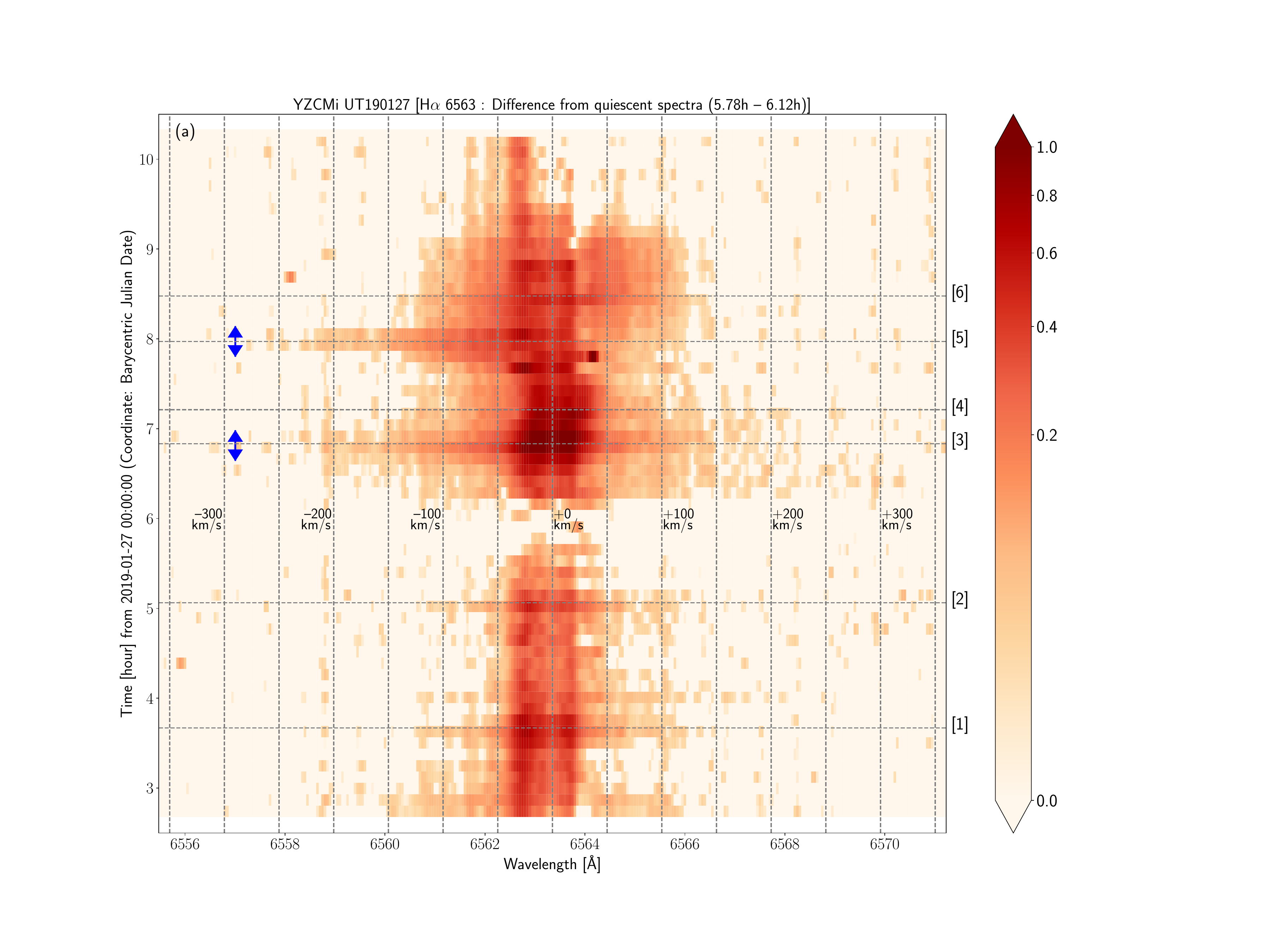}{0.63\textwidth}{\vspace{0mm}}
     \hspace{-0.11\textwidth}
    \fig{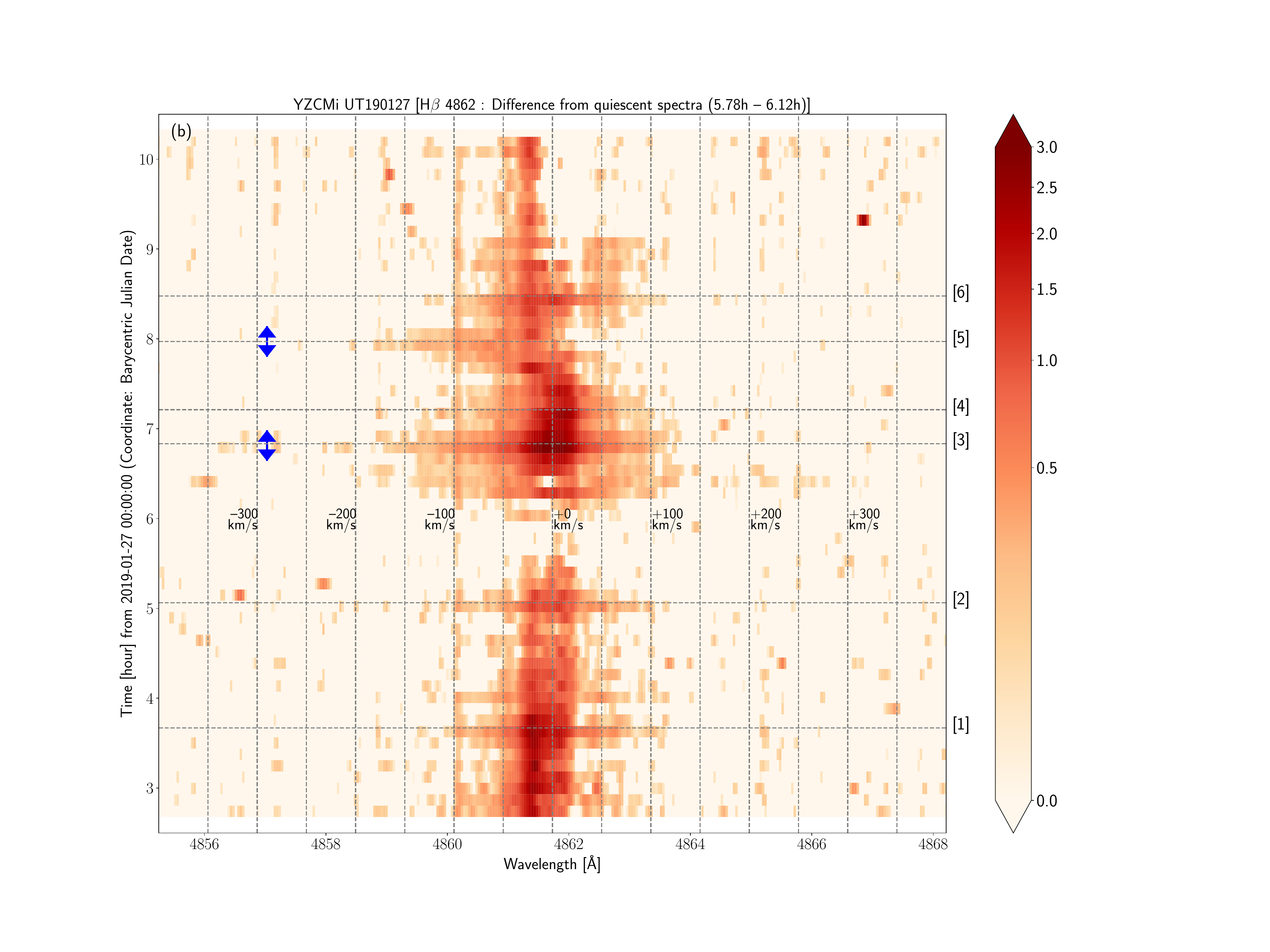}{0.63\textwidth}{\vspace{0mm}}
    }
     \vspace{-1cm}
     \caption{
(a)
Time evolution of the H$\alpha$ line profile covering Flares Y2 \& Y3 on 2019 January 27. The horizontal and vertical axes represent the wavelength and the observation time in the time coordinate of Barycentric Julian Date (BJD). The grey vertical dashed lines with velocity values represent the Doppler velocities from the H$\alpha$ line center.
The grey horizontal dashed lines indicate the time [1] -- [6], which are shown in Figure \color{black}\textrm{ \ref{fig:lcEW_HaHb_YZCMi_UT190127} (light curves)} \color{black} and Figure \ref{fig:spec_HaHb_YZCMi_UT190127} (line profiles).
The color map represents the line profile changes from the quiescent profile (cf. Figures \ref{fig:spec_HaHb_YZCMi_UT190127}(b), (f), \& (j)).
\color{black}\textrm{The blue double-headed arrows roughly represent the times when blue wing asymmetries were clearly seen and used for estimating the duration of blue wing asymmetries (cf. $\Delta t_{\rm{H}\alpha}^{\rm{blueasym}}$ in Table \ref{table:list_blue_flares}). }\color{black}
(b) 
Same as panel (a), but for the H$\beta$ line profile.
     }
   \label{fig:map_HaHb_YZCMi_UT190127}
   \end{center}
 \end{figure}

The H$\alpha$ \& H$\beta$ line profiles during Flares Y2 and Y3 are shown in
Figures \ref{fig:spec_HaHb_YZCMi_UT190127} \& \ref{fig:map_HaHb_YZCMi_UT190127}. The clear H$\alpha$ blue wing asymmetries with blue wing enhancements up to $\sim -$200 km s$^{-1}$ were seen twice at around the time [3] and [5] during Flare Y3 (Figures \ref{fig:lcEW_HaHb_YZCMi_UT190127},
\ref{fig:spec_HaHb_YZCMi_UT190127}, \&
\ref{fig:map_HaHb_YZCMi_UT190127}).
The durations of these two blue wing asymmetries were both only $\sim$20 min.
As for H$\beta$ line the blue wing asymmetry was not so clear at around the time [3], while the blue wing asymmetry with wing enhancements up to $\sim -$150 km s$^{-1}$ was clearly seen at around the time [5].
In addition to blue wing asymmetries, we note that red wing components of H$\alpha$ \& H$\beta$ lines show some enhancemnents up to $\sim +$150 km s$^{-1}$ (e.g., see the time [3], [4], and [6] in Figures \ref{fig:spec_HaHb_YZCMi_UT190127} \& \ref{fig:map_HaHb_YZCMi_UT190127}), 
and it can be interpreted that almost symmetric broadened wing components are seen at around these times.

\begin{figure}[ht!]
   \begin{center}
    \vspace{-0.5cm}
           \gridline{  
     \hspace{-0.06\textwidth}
    \fig{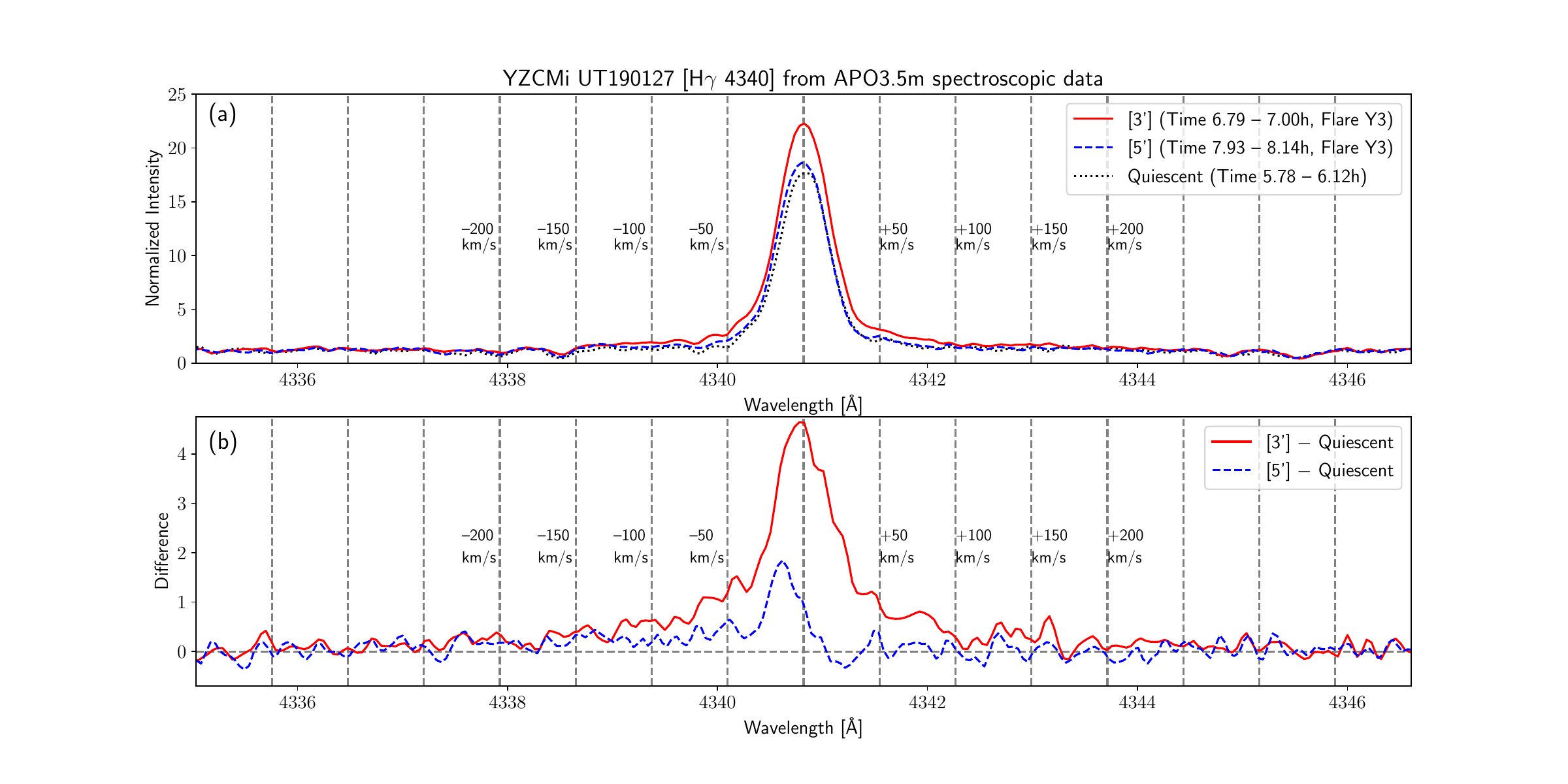}{0.58\textwidth}{\vspace{0mm}}
     \hspace{-0.06\textwidth}
       \fig{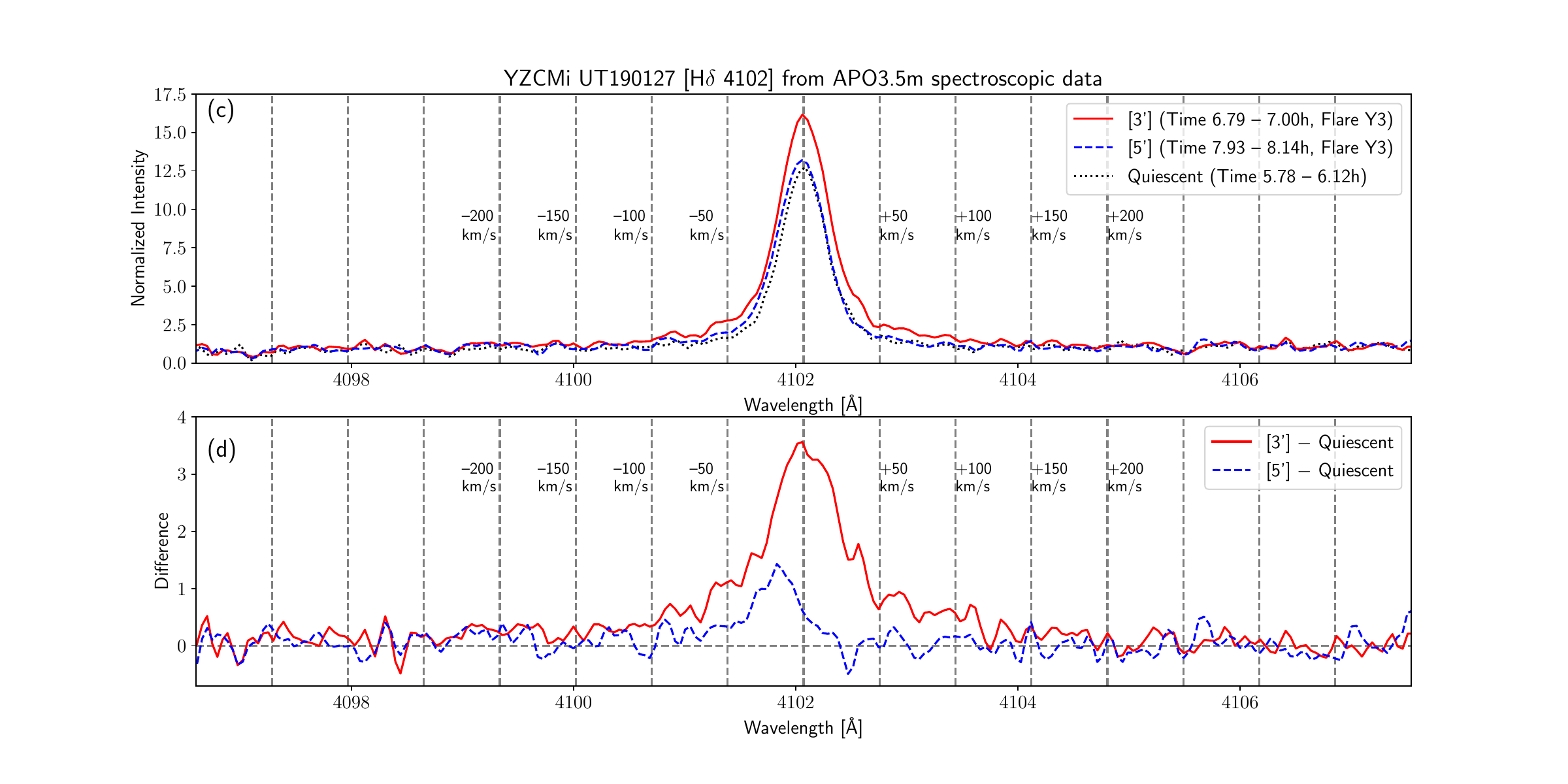}{0.58\textwidth}{\vspace{0mm}}
    }
    \vspace{-1.0cm}
        \gridline{  
     \hspace{-0.06\textwidth}
    \fig{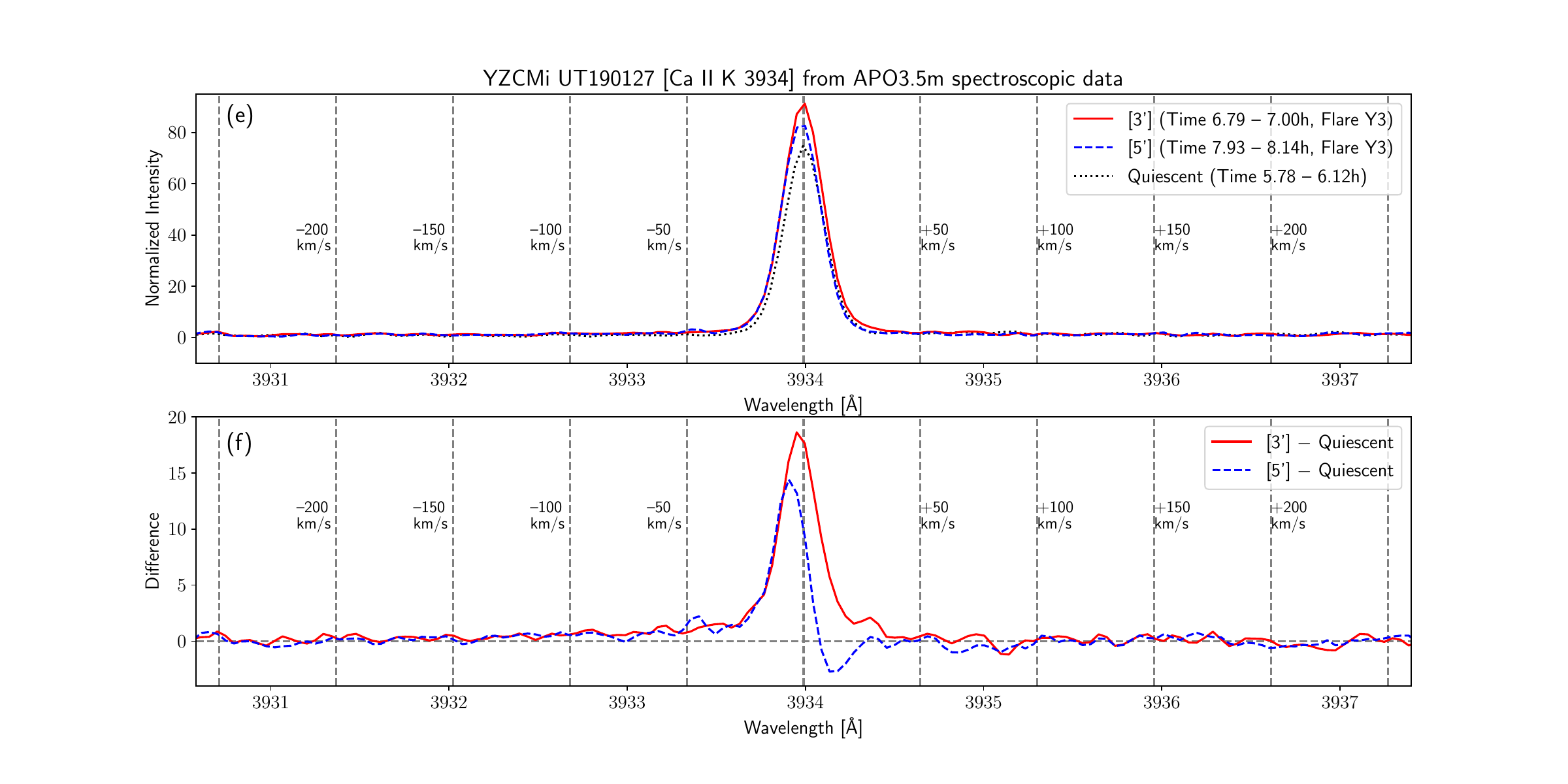}{0.58\textwidth}{\vspace{0mm}}
     \hspace{-0.06\textwidth}
       \fig{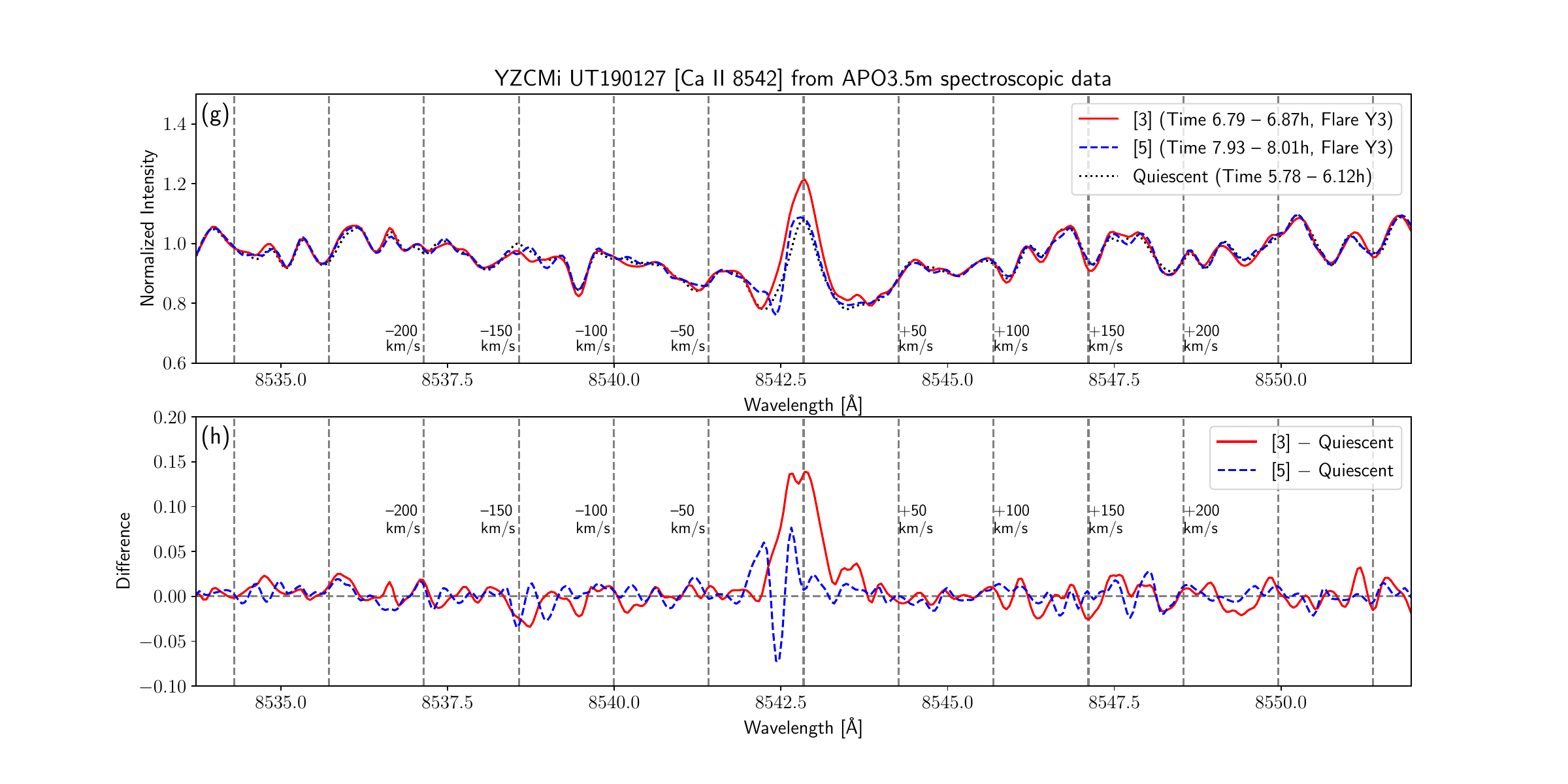}{0.58\textwidth}{\vspace{0mm}}
    }
    \vspace{-1.0cm}
        \gridline{  
     \hspace{-0.02\textwidth}
    \fig{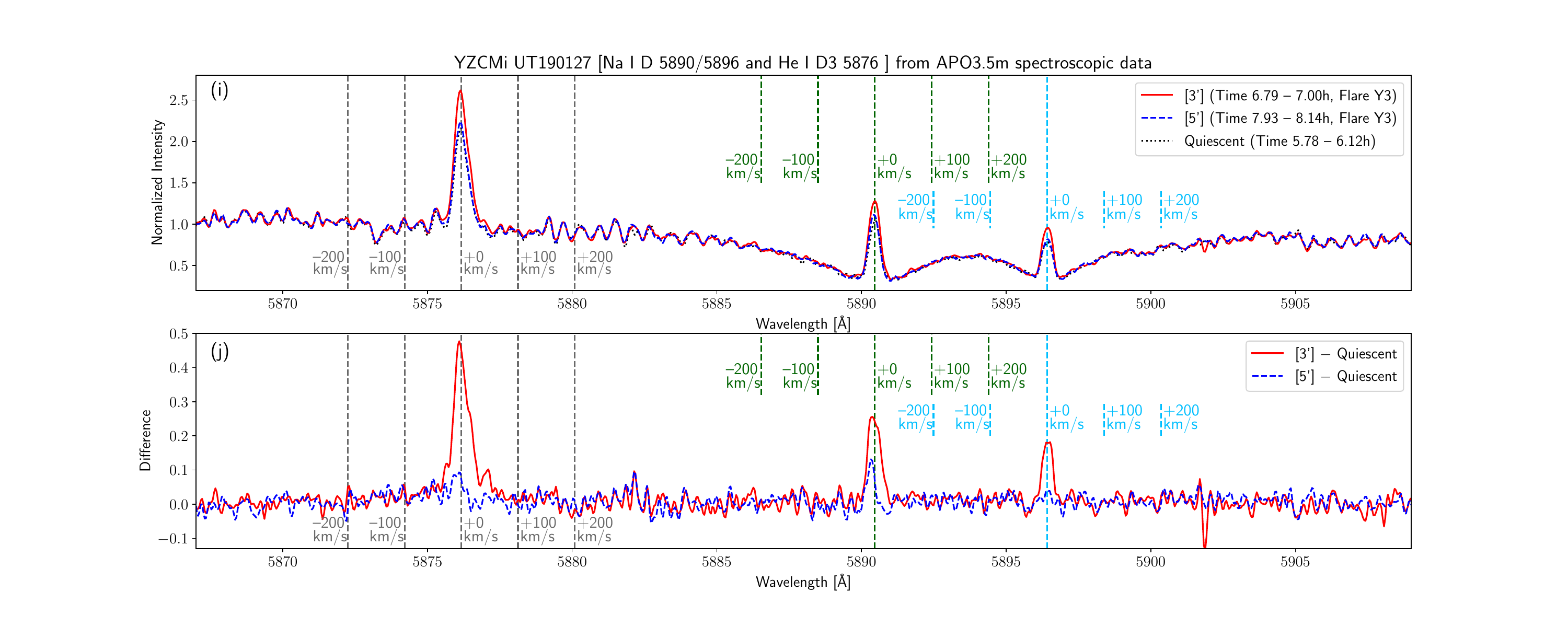}{0.75\textwidth}{\vspace{0mm}}
    }
    \vspace{-1.0cm}
        \gridline{  
     \hspace{-0.02\textwidth}
    \fig{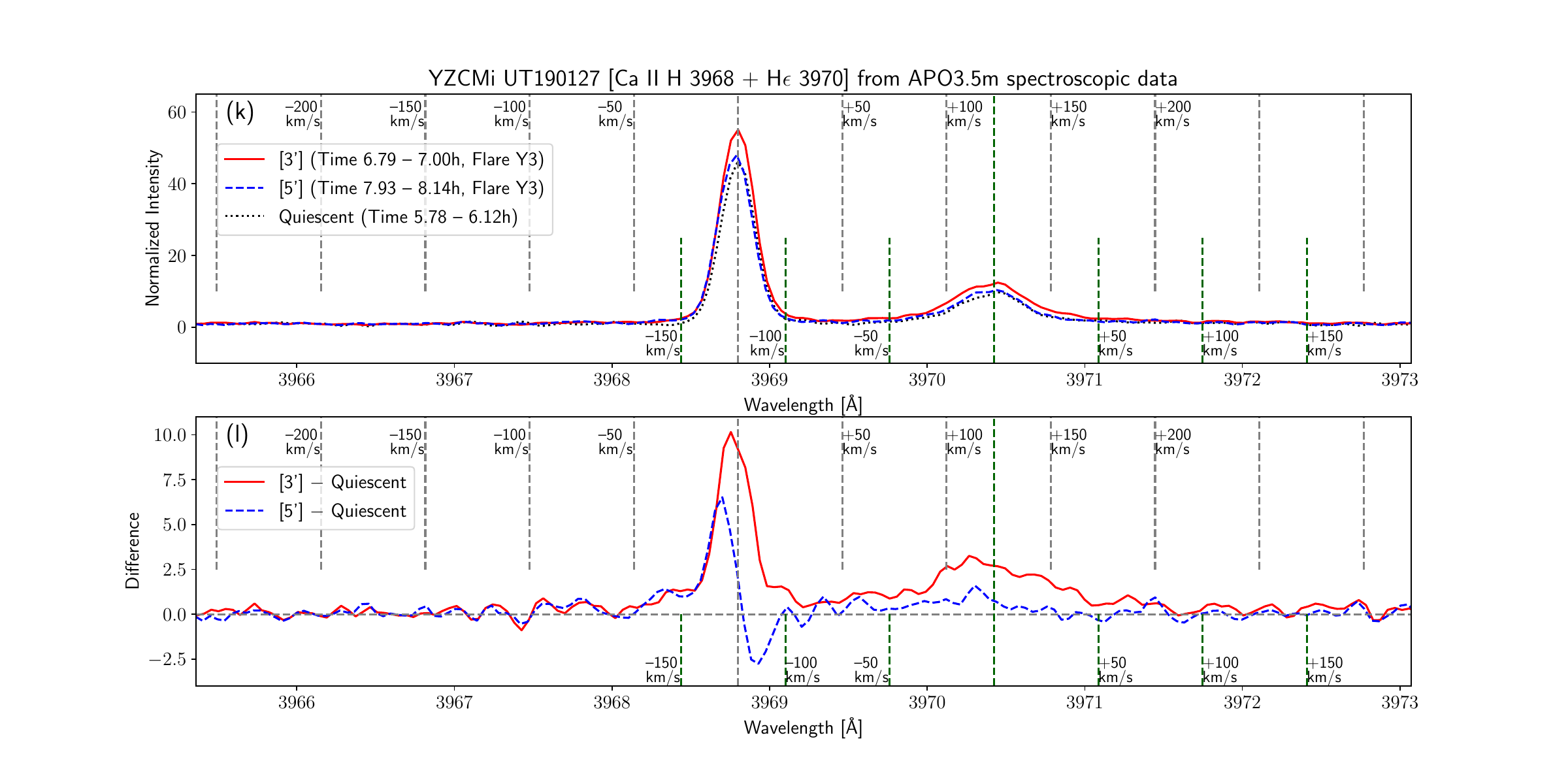}{0.58\textwidth}{\vspace{0mm}}
    }
     \vspace{-1.0cm}
     \caption{
     \color{black}\textrm{
(a) Line profiles of the H$\gamma$ line when the H$\alpha$ line show blue wing asymmetries (cf. Figure \ref{fig:spec_HaHb_YZCMi_UT190127} (e)\&(f)) 
during Flare Y3 on 2019 January 27 from APO3.5m spectroscopic data.
The horizontal and vertical axes represent the wavelength and flux normalized by the continuum. The grey vertical dashed lines with velocity values represent the Doppler velocities from the H$\gamma$ line center. The red solid and blue dashed lines indicate the integrated line profiles over the time [3$^{\prime}$] (Time 6.79 -- 7.00h) and [5$^{\prime}$] (Time 7.93 -- 8.14h) on this date, which include the time [3] and [5] in Figure \ref{fig:lcEW_HaHb_YZCMi_UT190127} (light curves), respectively.
(c), (e), (g), (i), \& (k)
Same as panel (a), but for H$\delta$, Ca II K, Ca II 8542, Na I D1 \& D2 (5890 \& 5896) \& He I D3 5876, and H$\epsilon+$Ca II H lines, respectively. 
As for the Ca II 8542 line, the data at the time [3] and [5] are plotted (not [3$^{\prime}$] and [5$^{\prime}$).
(b), (d), (f), (h), (j), \& (l) Same as panels (a), (c), (e), (g), (i), \& (k), respectively, 
but the line profile changes from the quiescent phase.
} \color{black}
}
   \label{fig:spec_HcHd_YZCMi_UT190127}
   \end{center}
 \end{figure}

The equivalent width light curves of H$\gamma$, H$\delta$, Ca II K, Ca II 8542, Na I D1 \& D2, and He I D3 5876 lines are also shown in Figures \ref{fig:lcEW_HaHb_YZCMi_UT190127} (e), (f), \& (g)\footnote{In this paper, the EW light curve of H$\epsilon+$Ca II H lines are not plotted and only snapshot spectra of H$\epsilon+$Ca II H lines are shown as in Figure Figure \ref{fig:spec_HcHd_YZCMi_UT190127}(k) \& (l), since these two lines can overlap with each other. }.
The profiles of these lines and H$\epsilon+$Ca II H lines during Flare Y3 are shown in Figure \ref{fig:spec_HcHd_YZCMi_UT190127}. 
Since S/N ratio of the spectroscopic data around \color{black}\textrm{these} \color{black} lines are smaller than those of H$\alpha$ \& H$\beta$ lines,
we integrate two or three temporally adjacent spectra into one spectra (see time resolution of the H$\gamma$, H$\delta$, Ca II K, Na I D1 \& D2, and He I D3 5876 lightcurve data in Figure 
\ref{fig:lcEW_HaHb_YZCMi_UT190127}).  
In Figure \ref{fig:spec_HcHd_YZCMi_UT190127}, 
such \color{black}\textrm{time-integrated data are }\color{black} shown with the prime mark, and 
for example, the time [3$^{\prime}$] 
in Figure \ref{fig:spec_HcHd_YZCMi_UT190127} shows the Time 6.79 - 7.00h, which include Time [3] (Figures \ref{fig:lcEW_HaHb_YZCMi_UT190127} \& \ref{fig:spec_HaHb_YZCMi_UT190127}). 
The same way of using the prime mark 
is applied to all figures in the following of this paper.

As for H$\gamma$, H$\delta$, H$\epsilon$, and Ca II H\&K lines, 
the blue wing asymmetries were not so clear at around the time [3$^{\prime}$], while the blue asymmetries with wing enhancements of up to -50 -- -100 km s$^{-1}$ were seen at around the time [5$^{\prime}$]. 
At around the time [3$^{\prime}$], blue wing asymmetries are not clearly seen for these lines,  
but almost symmetric broadened wing components ($\pm$150 km s$^{-1}$) can be seen and these symmetric wing components can be similar to those seen in H$\alpha$.
In addition, the red-shifted absorption components were seen especially in Ca II H\&K lines together with blue wing asymmetries at \color{black}\textrm{around }\color{black} the time [5$^{\prime}$], and this component looks larger than the noise level. Similar red-shifted components have been observed 
in the previous observation of H$\alpha$ blue wing asymmetry (\citealt{Honda+2018}), 
but currently the physical origin of them is still unclear.
\color{black}\textrm{
It is also important to discuss the origin of the red-shifted absorption in the future studies }\color{black}
(see also Section \ref{subsec:dis:flares-redsym} for future prospects of red-shifts of lines).

The clear blue wing asymmetries were not seen in  Ca II 8542, Na I D1 \& D2, and He I D3 5876 lines, while Na I D1 \& D2, and He I D3 5876 lines show slight (-10 -- -20 km s$^{-1}$) blue shifts at around the time [5$^{\prime}$] (Figure \ref{fig:spec_HcHd_YZCMi_UT190127}(j)).
The EW light curves in Figures \ref{fig:lcEW_HaHb_YZCMi_UT190127} (e), (f), \& (g) show that H$\alpha$ and Ca II K evolve similarly while other Balmer lines decrease faster during Flare Y3. We also note that Ca II 8542 and Na D \color{black}\textrm{lines} \color{black} show relatively large responses in Flare Y2, while other lines show smaller responses compared to Flare Y3.

    \begin{figure}[ht!]
   \begin{center}
   \gridline{
    \fig{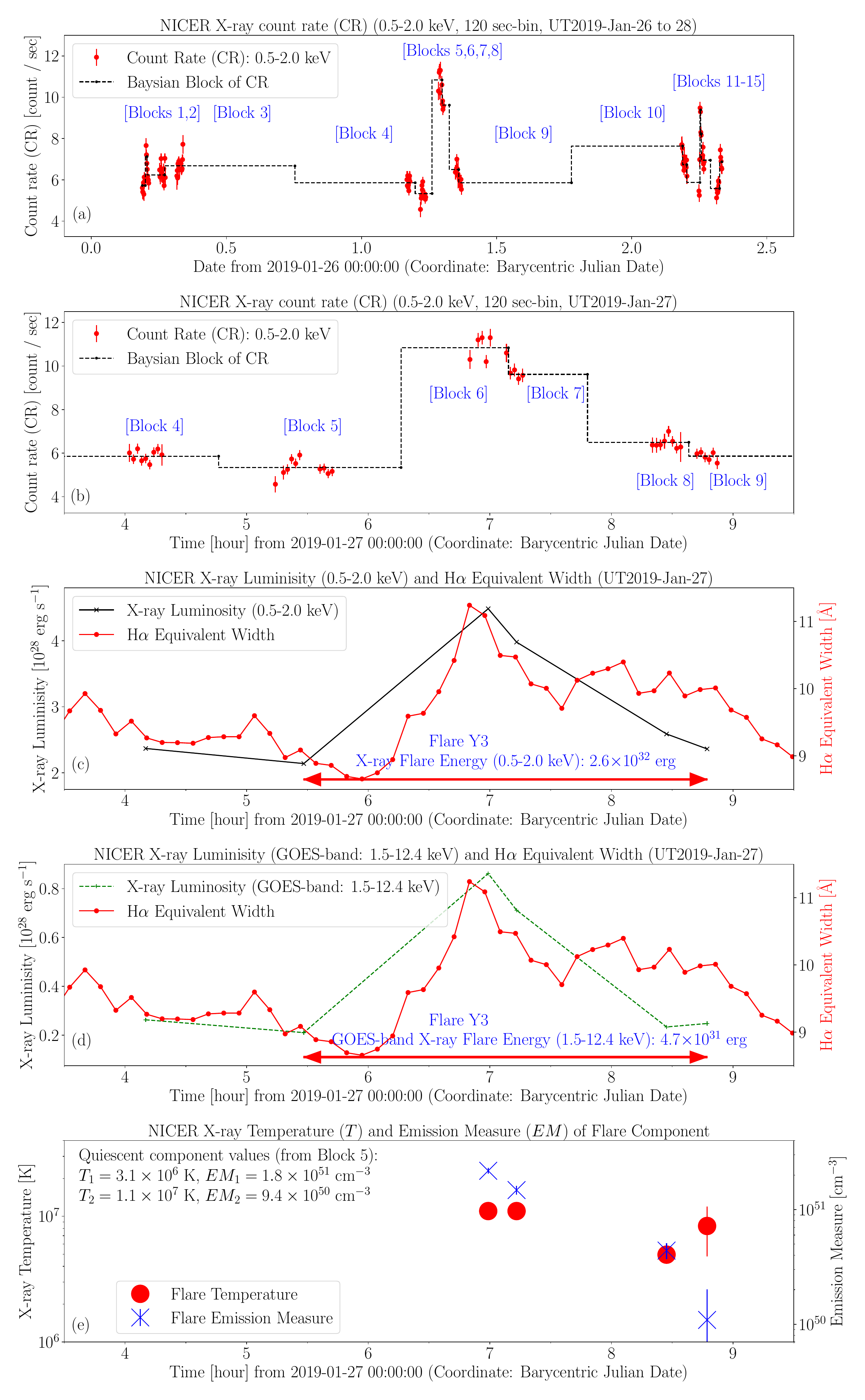}{0.68\textwidth}{\vspace{0mm}}
    }
     \vspace{-5mm}
     \caption{
(a) \textit{NICER} X-ray light curve of YZ CMi over the whole 3-day \textit{NICER} observation period from UT 2019 Jan 26 to 28.
The red circles are \textit{NICER} X-ray count rates [count s$^{-1}$] in 0.5--2.0 keV. 
The black dotted line shows Bayesian block light curve (cf. \citealt{Scargle+2013}) of the count rates, and each block is shown with the number. 
(b)
\textit{NICER} X-ray light curve of YZ CMi on 2019 Jan 27 showing Flare Y3. 
The symbols are plotted in the same way as (a).
(c)
Light curves of \textit{NICER} X-ray luminosity (0.5--2.0 keV, black x-marks) and H$\alpha$ Equivalent Width (red circles) on UT 2019 Jan 27, showing Flare Y3. 
(d)
Light curves of \textit{NICER} X-ray luminosity in \textit{GOES}-band (1.5--12.4 keV, green plus marks) and H$\alpha$ Equivalent Width (red circles) on UT 2019 Jan 27, showing Flare Y3. 
(e)
Light curves of \textit{NICER} X-ray temperature (red circles) and emission measure (blue asterisks) of the flare component during Flare Y3 on UT 2019 Jan 27.
The values are derived from the X-ray spectral fitting shown in Figure \ref{fig:NICER_190127_spec1}.
     }
   \label{fig:NICER_BBlc_190127}
   \end{center}
 \end{figure}

We generate a background-subtracted X-ray light curve between 0.5$-$2~keV for the {\textit NICER} X-ray data. The light curve shows a count rate increase by a factor of two on the second day, which coincides with Flare Y3 in the H$\alpha$ band (Figure \ref{fig:lcEW_HaHb_YZCMi_UT190127}). The photon count in the other intervals hovers around 6$-$7 cts s$^{-1}$ with a few small flare-like variations.

We applied an adaptive binning to the light curve with a Bayesian block algorithm \citep[][see Figure~\ref{fig:NICER_BBlc_190127} (a) \& (b)]{Scargle+2013} and used those blocks for spectral analysis of Flare Y3. We assume Block 5 just before the flare onset to represent the flare's non-flaring (quiescent) emission. The spectrum shows a hump between 0.5$-$1 keV, which originates from the Fe L and O K lines and the weak Mg and Si K lines. We reproduce the spectral shape by an optically-thin thermal plasma emission \citep[{\tt apec}][]{Smith+2001} model with two temperature components at 0.26 and 0.97~keV ($3.1\times10^{6}$ and $1.1\times10^{7}$K) and an elemental abundance at 0.52~solar (see also temperature and emission measure values of the quiescent component shown in Figure \ref{fig:NICER_BBlc_190127} (e)). However, the fit is not statistically acceptable at above 3$\sigma$ due to the line-like excesses at 0.51 and 1.22~keV.

\begin{figure}[ht!]
   \begin{center}
   \gridline{
        \hspace{-0.04\textwidth}
    \fig{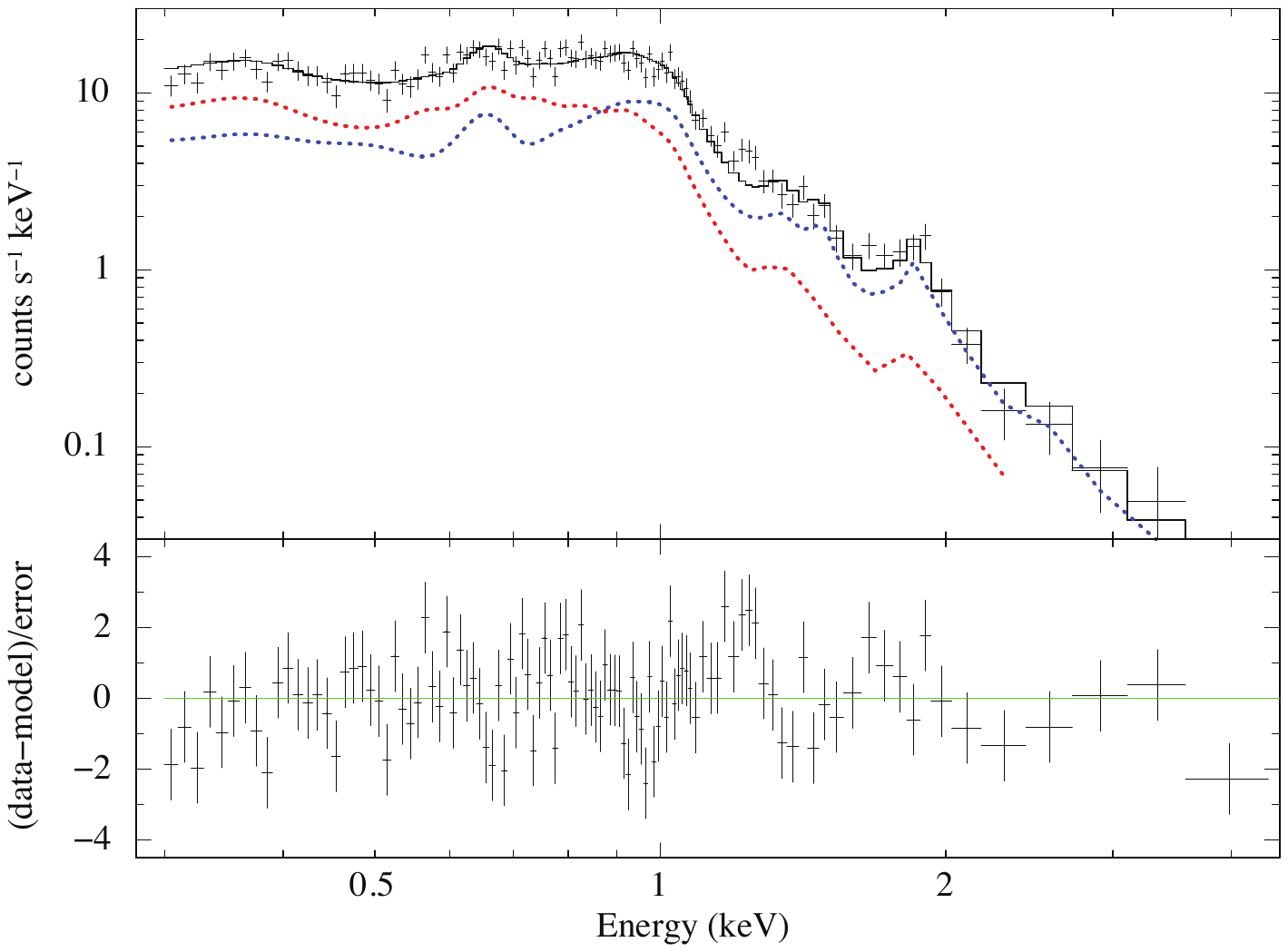}{0.57\textwidth}{\vspace{0mm}}
     \hspace{-0.06\textwidth}
    \fig{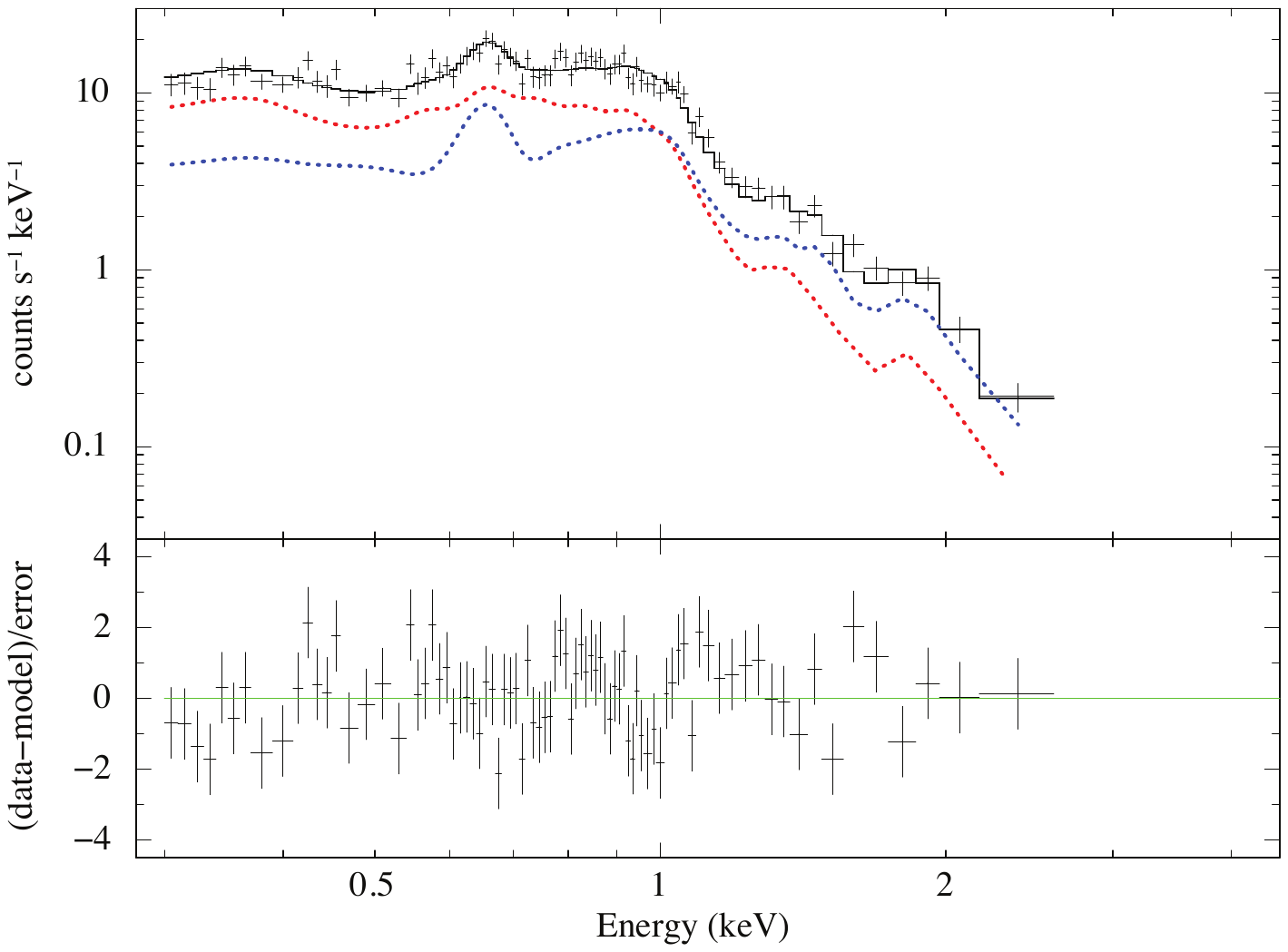}{0.57\textwidth}{\vspace{0mm}}
    }
            \vspace*{-15mm}
       \gridline{
            \hspace{-0.04\textwidth}
    \fig{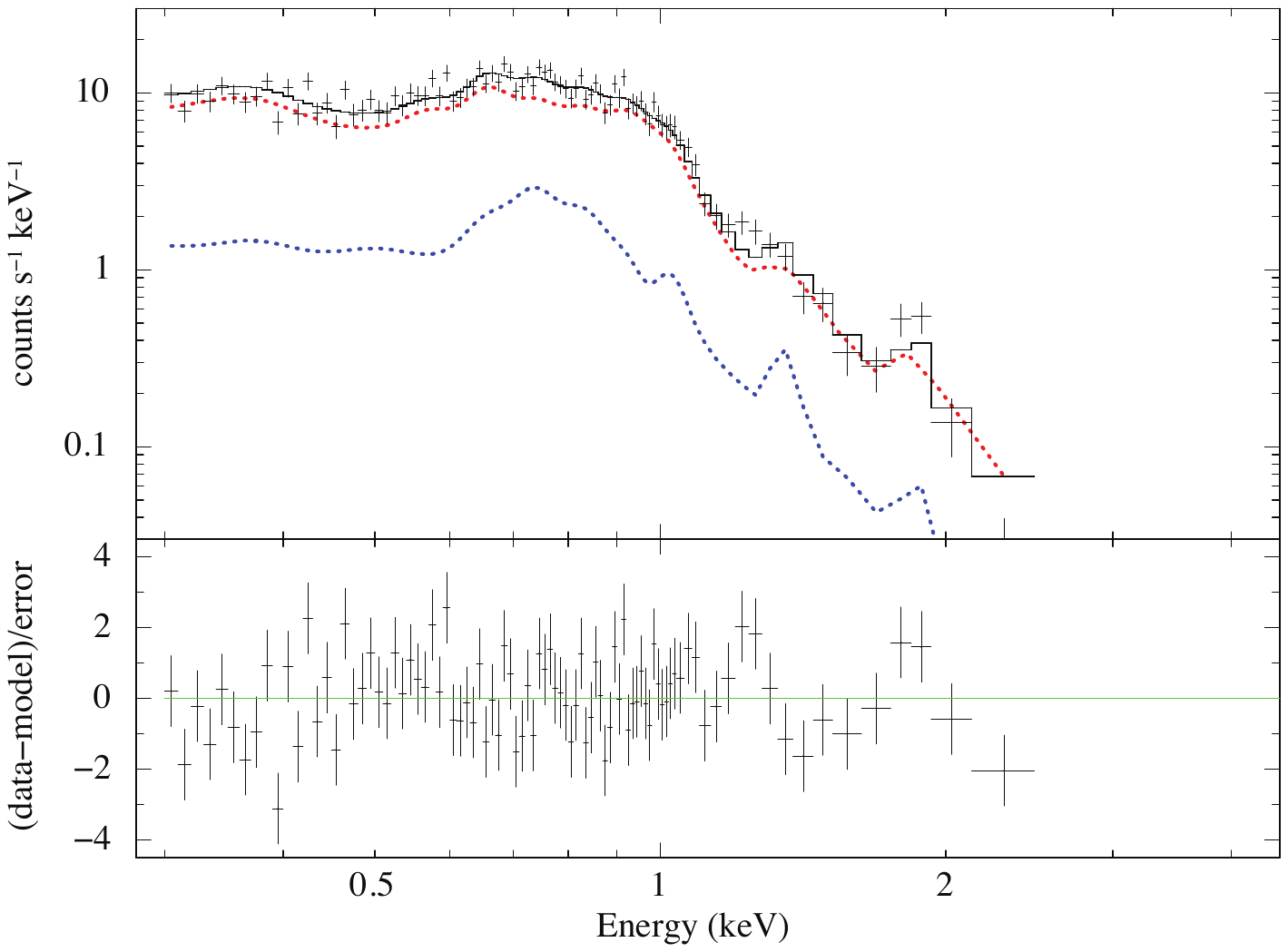}{0.57\textwidth}{\vspace{0mm}}
     \hspace{-0.06\textwidth}
    \fig{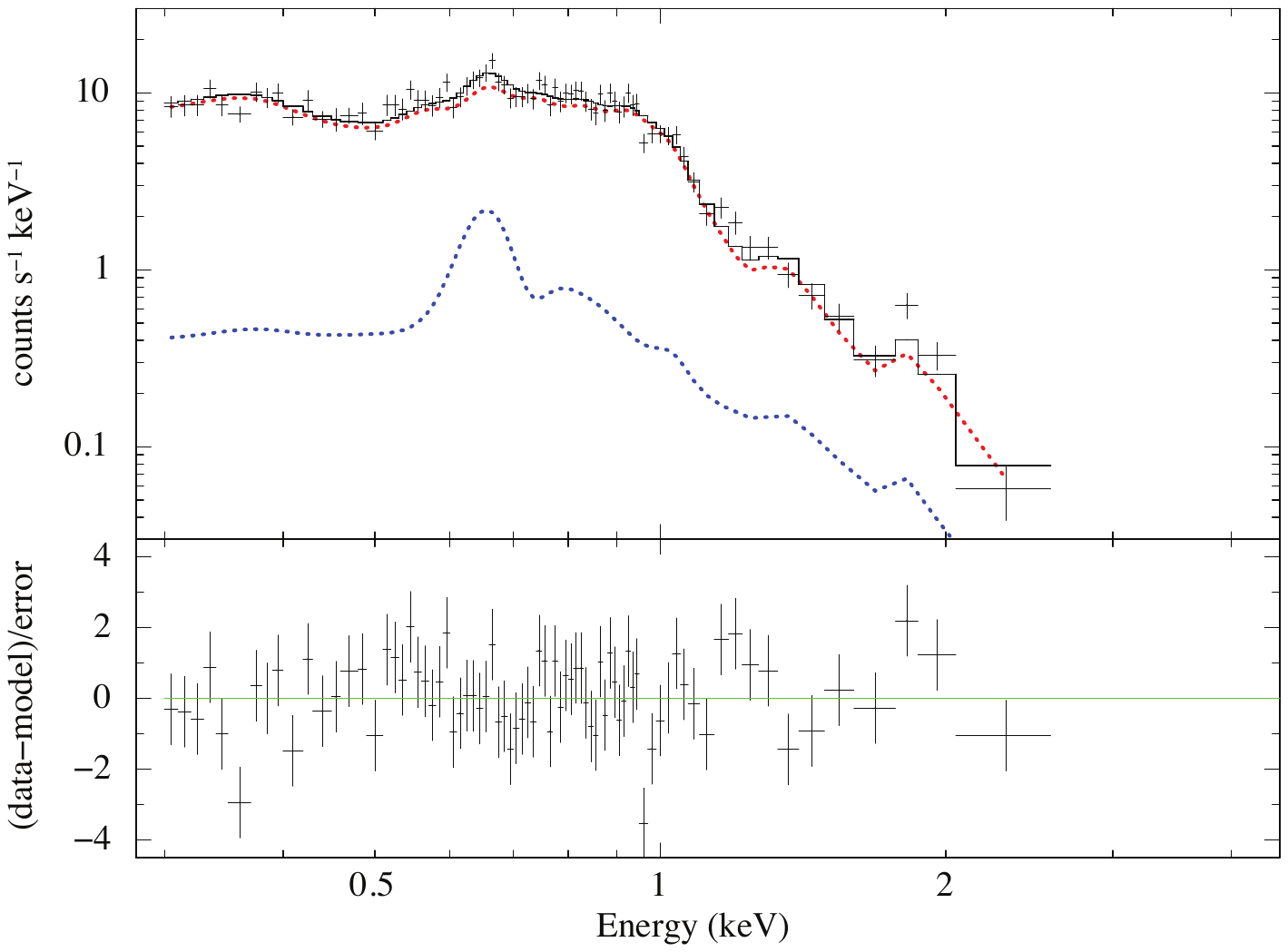}{0.57\textwidth}{\vspace{0mm}}
    }
     \caption{
{\textit NICER} X-ray spectra of Flare Y3: observation + model (Block 6: top left, Block 7: top right, Block 8: bottom left, Block 9: bottom right). The red and blue dotted lines show the preflare and flare components in the model. The black solid line shows the sum of these components.
     }
   \label{fig:NICER_190127_spec1}
   \end{center}
 \end{figure}

Figure~\ref{fig:NICER_190127_spec1} shows time-resolved X-ray spectra during Flare Y3. The flare spectrum near the peak is significantly harder than the quiescent spectrum, with a strong oxygen K line at 0.64 keV. Since we measure the X-ray flux variation during the flare in this study, we fit each spectrum by a one-temperature apec model with independent oxygen and iron elemental abundances on top of the best-fit quiescent spectrum model.

The resultant values of the temperatures ($T$) and emission measures ($EM=n^{2}V$) are shown in 
Figure \ref{fig:NICER_BBlc_190127} (e). 
Here \color{black}\textrm{$n$ }\color{black} is the electron density and $V$ is the volume.
Using the modeling results,
the X-ray luminosities in the 0.5--2.0 keV band ($L_{\rm{Xray, flare}}$(0.5--2.0 keV)) and the \textit{GOES}-band (1.5--12.4 keV \color{black}\textrm{= 1--8 \AA} \color{black}, $L_{\rm{Xray, flare}}$(\textit{GOES}-band))\footnote{
\color{black}\textrm{\textit{GOES}-band is the soft X-ray band used for the solar soft X-ray flux observation with the Geostationary Operational Environmental Satellite.} \color{black}
}
were calculated and shown in 
Figures \ref{fig:NICER_BBlc_190127} (c) \& (d) with H$\alpha$ light curve. 
From this figure, we can see that H$\alpha$ flare duration is longer than that of soft X-ray. 
The X-ray energy of Flare Y3 in the 0.5--2.0 keV band ($E_{\rm{Xray, flare}}$(0.5--2.0 keV)) and the \textit{GOES}-band \color{black}\textrm{($E_{\rm{Xray, flare}}$(\textit{GOES}-band)) }\color{black} are also calculated to be $2.6 \times 10^{32}$ erg and $4.7 \times 10^{31}$ erg.
$E_{\rm{Xray, flare}}$(0.5--2.0 keV) is $\sim$15 times larger than the H$\alpha$ flare energy ($E_{\rm{H}\alpha}=1.7\times 10^{31}$erg) and
\color{black}\textrm{  
at least slightly larger than the upper-limit of \textit{TESS} white-light flare energy ($E_{TESS}<2.6\times 10^{32}$erg). 
 } \color{black}

\clearpage

\subsection{Flare Y6 (Blue wing asymmetry) observed on 2019 December 12} \label{subsec:results:2019-Dec-12}

      \begin{figure}[ht!]
   \begin{center}
   \gridline{
    \fig{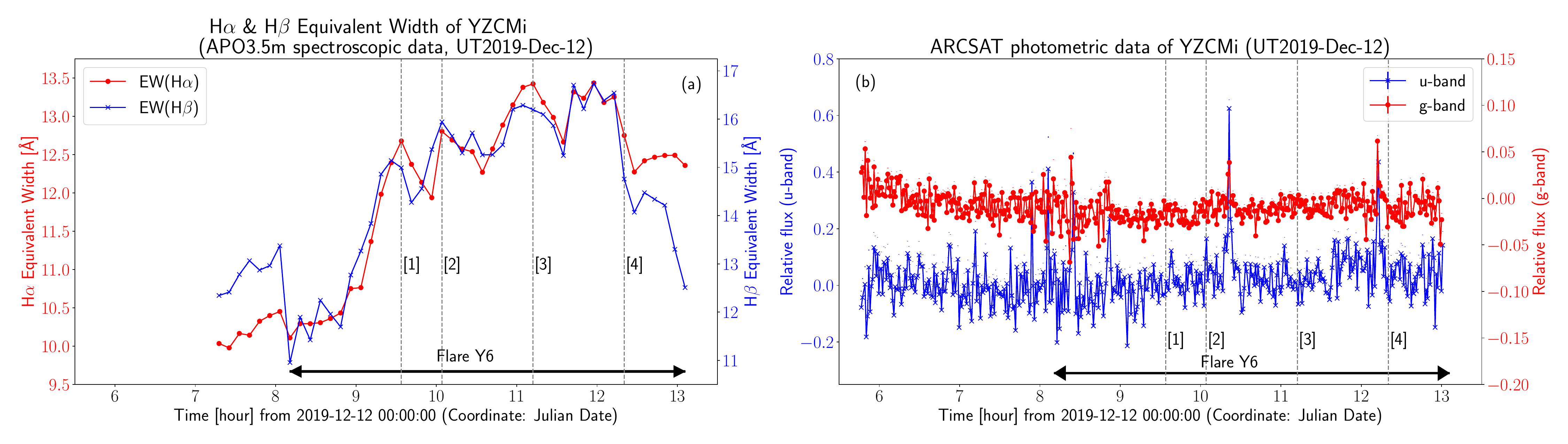}{1.0\textwidth}{\vspace{0mm}}
    }
     \vspace{-1cm}
              \gridline{  
     \hspace{-0.02\textwidth}
    \fig{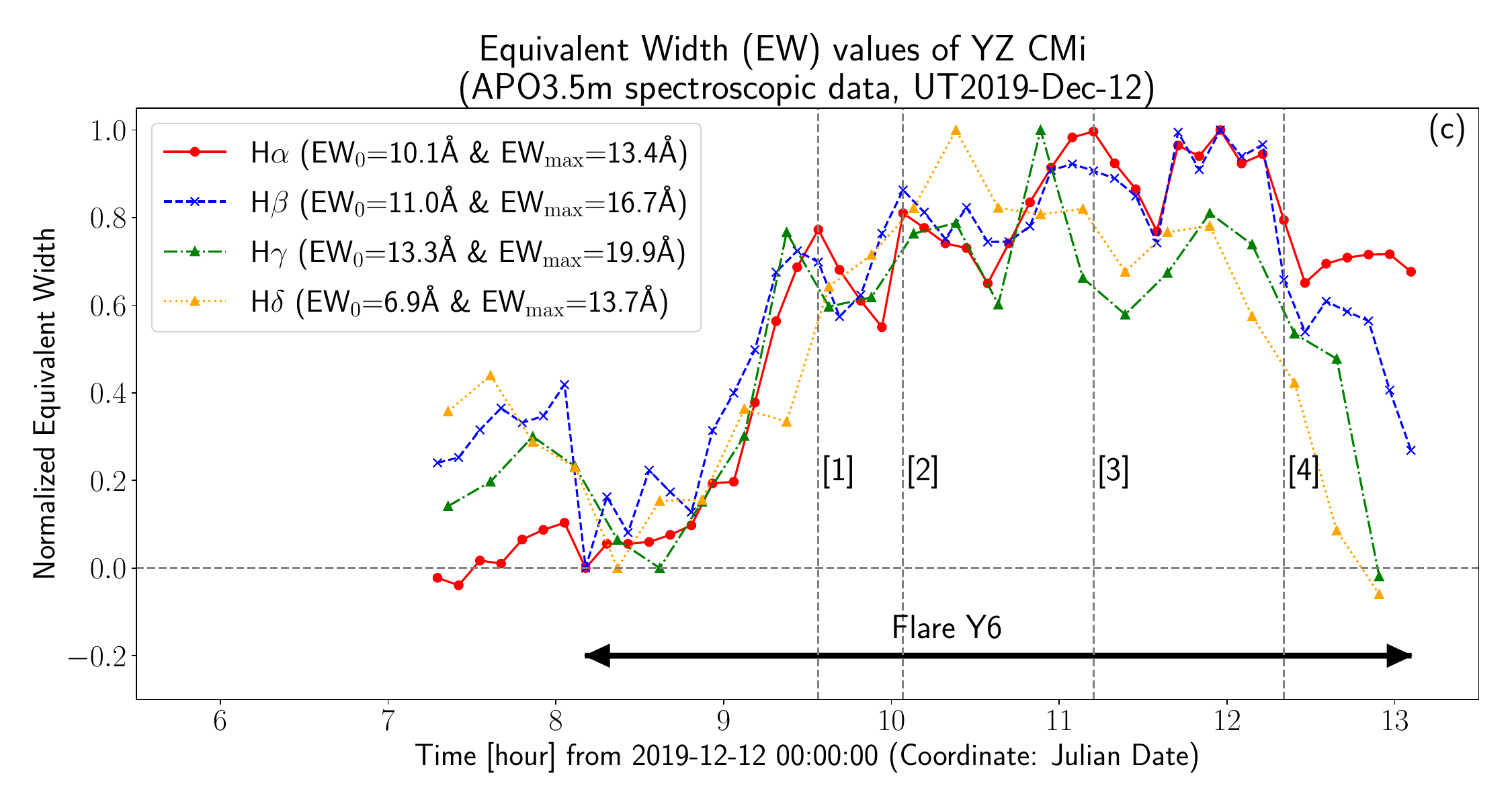}{0.5\textwidth}{\vspace{0mm}}
     \hspace{-0.02\textwidth}
    \fig{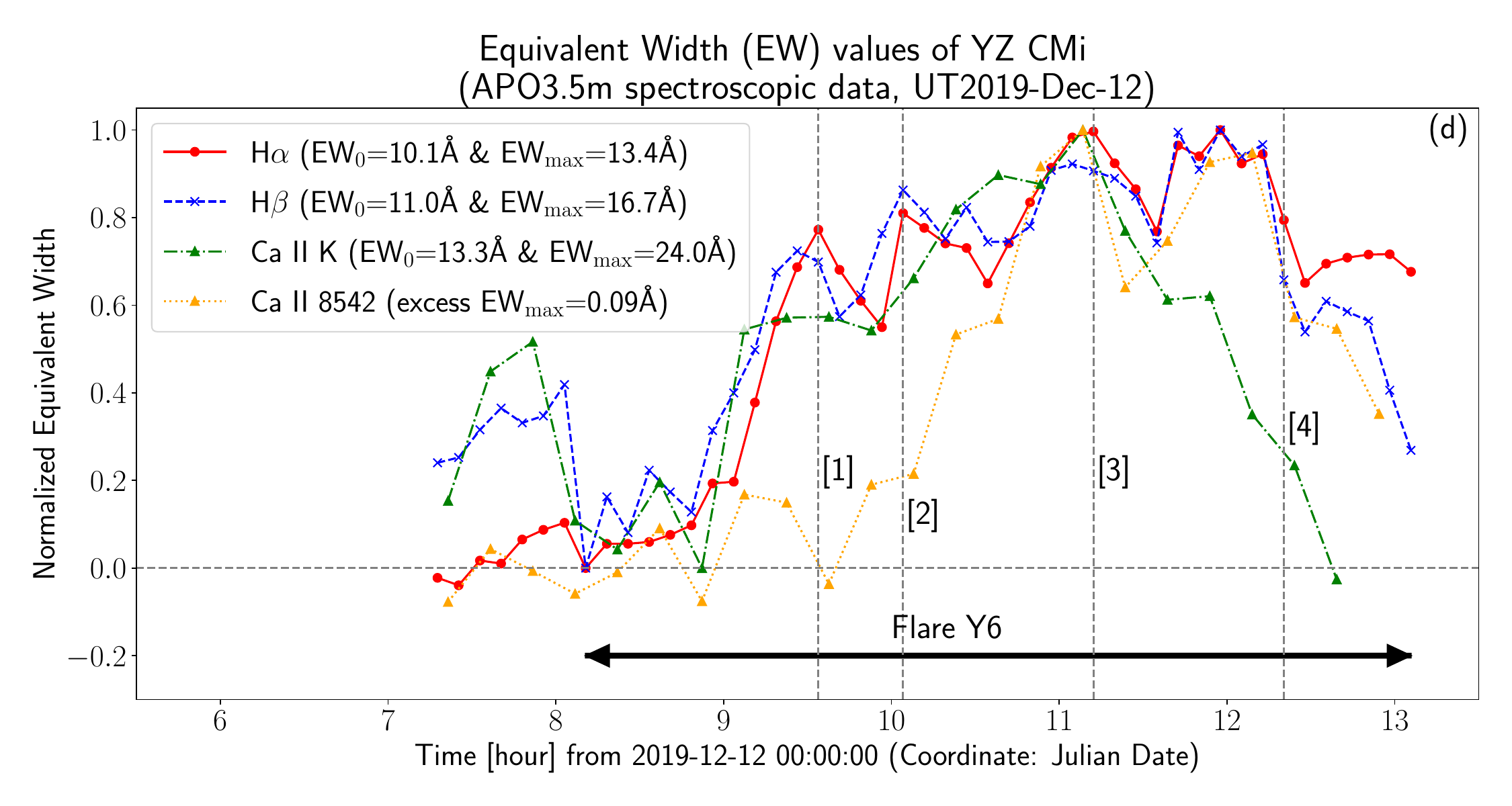}{0.5\textwidth}{\vspace{0mm}}
    }
     \vspace{-1cm}
              \gridline{  
     \hspace{-0.02\textwidth}
    \fig{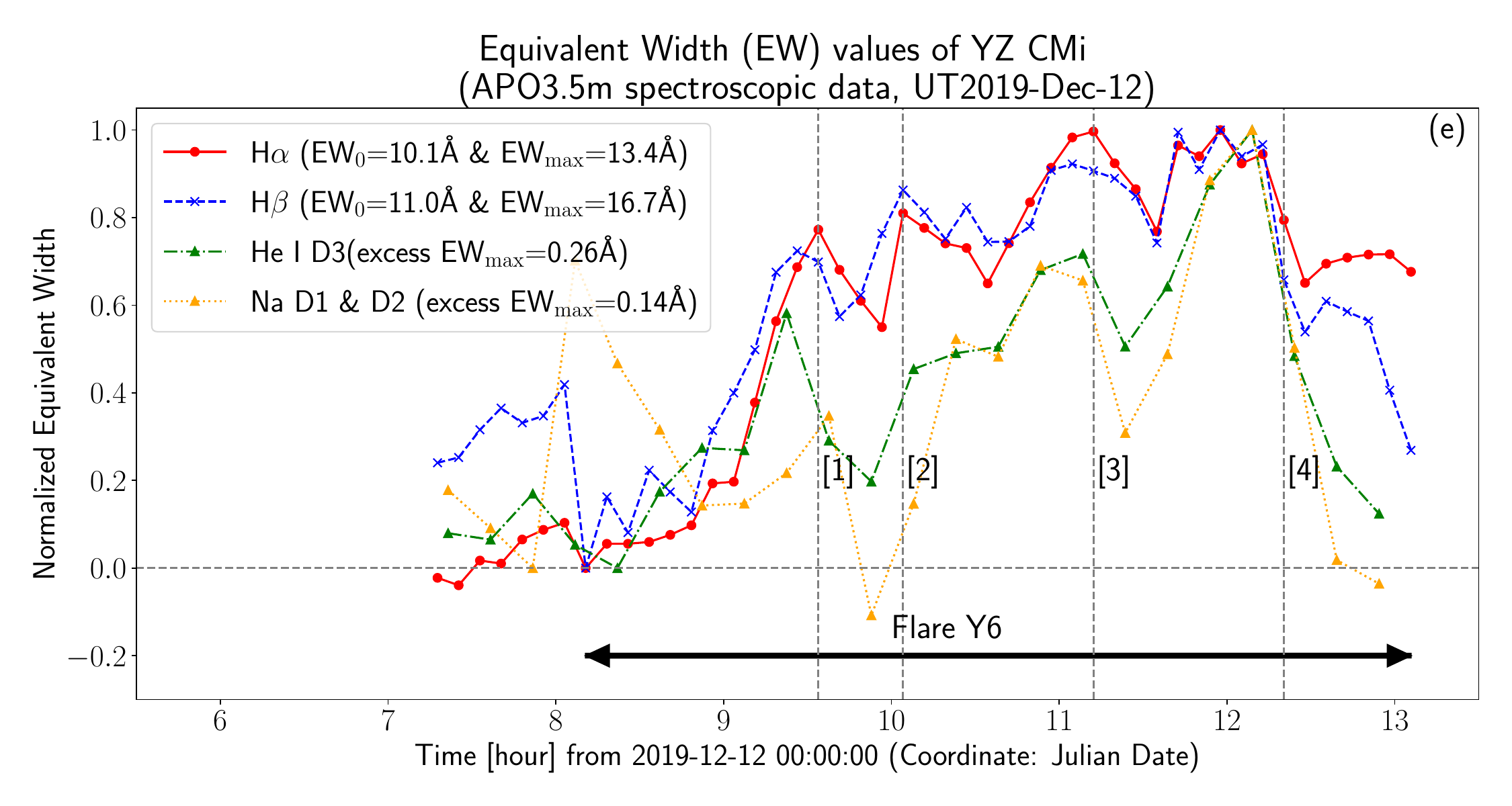}{0.52\textwidth}{\vspace{0mm}}
    }
        
     \vspace{-0.9cm}
     \caption{
     \color{black}\textrm{ 
Light curves of YZ CMi on 2019 December 12 summarizing the multi-line behaviour of Flare Y6. 
The data are plotted in (a), (b), (c), (d), \& (e) similarly with Figures \ref{fig:lcEW_HaHb_YZCMi_UT190127} (a), (b), (e), (g), \& (f), respectively, but
the horizontal axes are in the time coordinate of Julian Date (JD).
The grey dashed lines with numbers ([1]--[4]) correspond to the time shown with the same numbers in Figures \ref{fig:spec_HaHb_YZCMi_UT191212} \& \ref{fig:map_HaHb_YZCMi_UT191212}.
  } \color{black}
     }
   \label{fig:lcEW_HaHb_YZCMi_UT191212}
   \end{center}
 \end{figure}
 
     \begin{figure}[ht!]
   \begin{center}
            \gridline{  
     \hspace{-0.06\textwidth}
    \fig{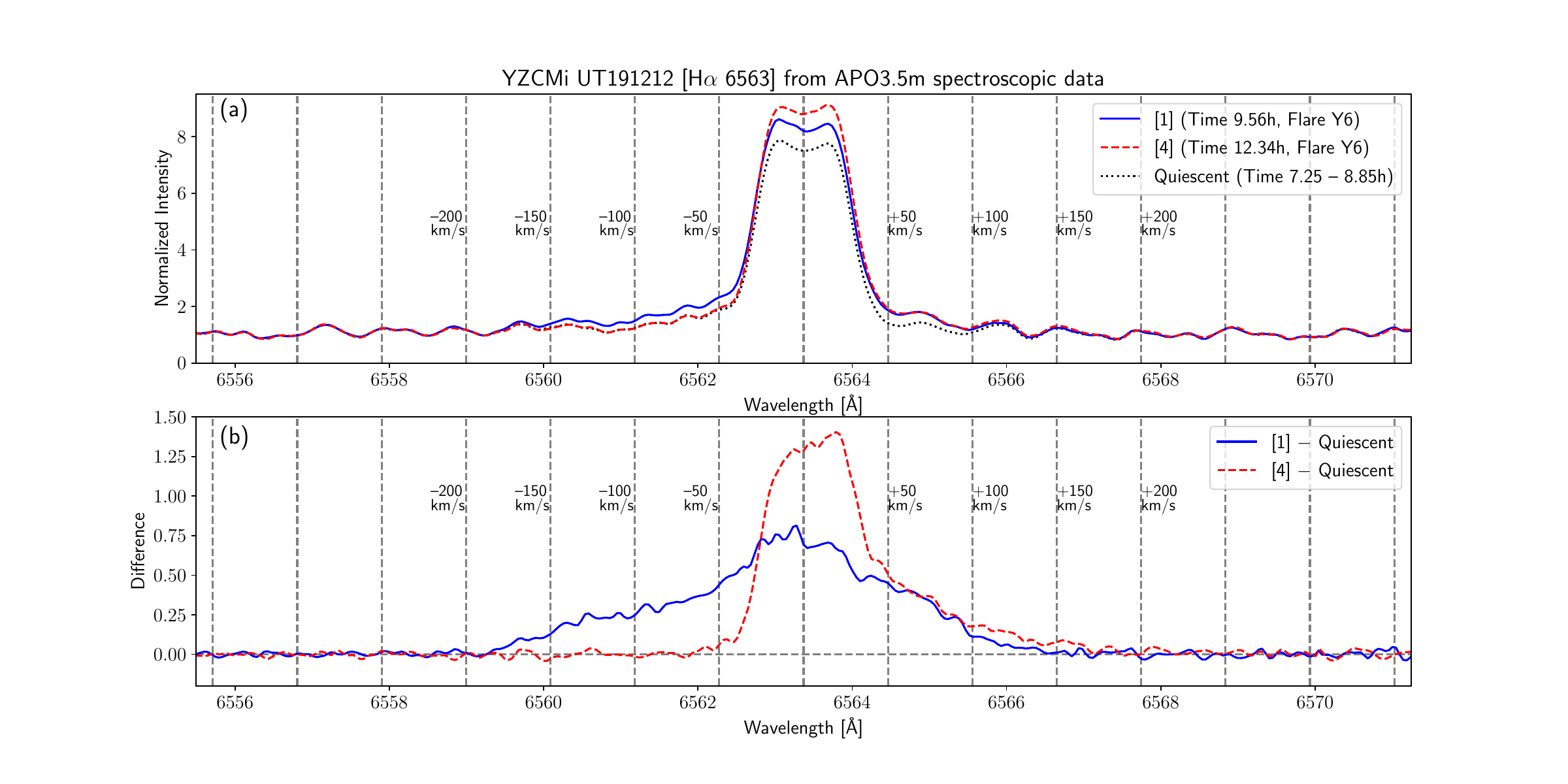}{0.58\textwidth}{\vspace{0mm}}
     \hspace{-0.06\textwidth}
       \fig{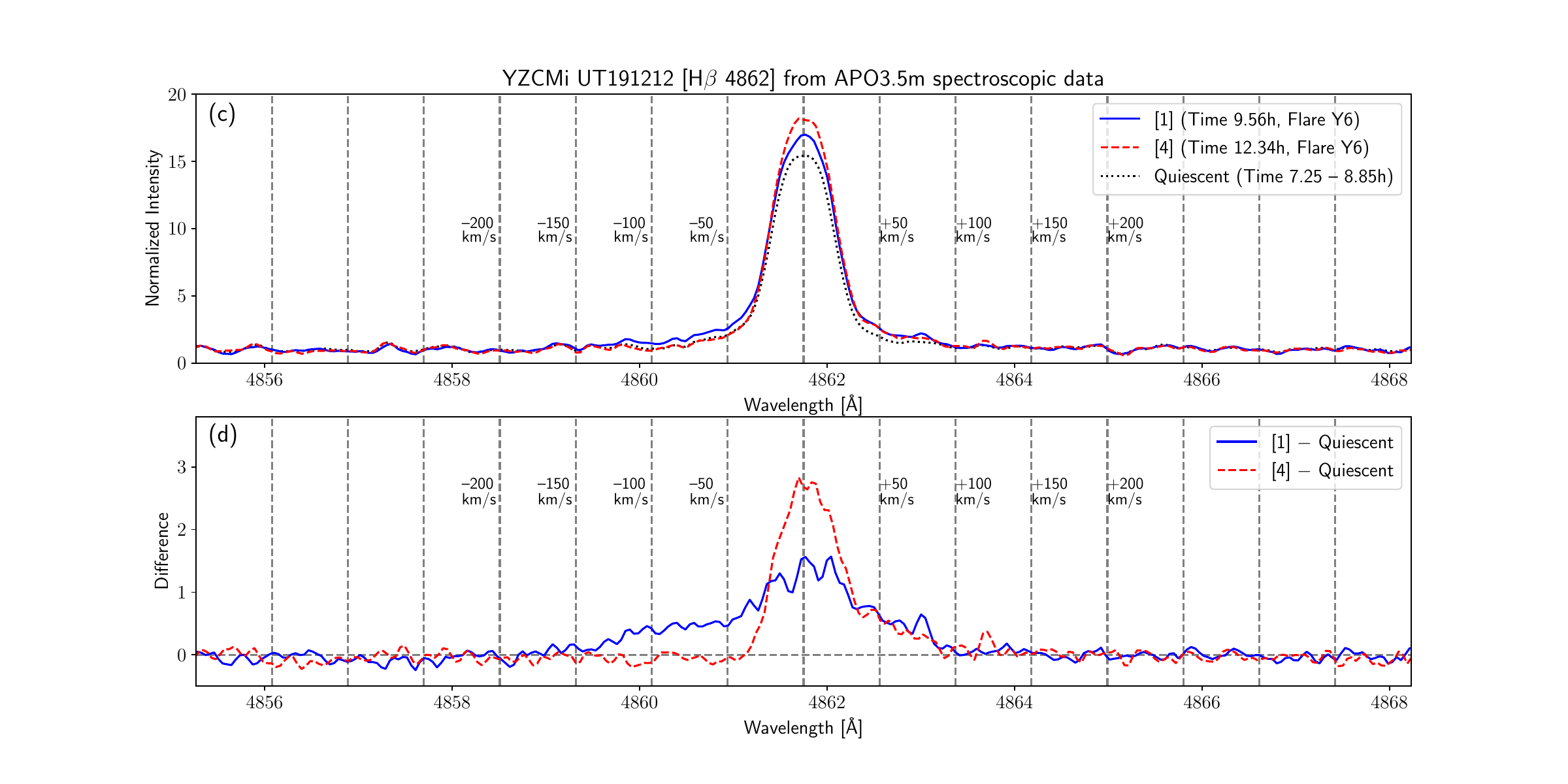}{0.58\textwidth}{\vspace{0mm}}
    }
    \vspace{-1.0cm}
     \gridline{  
     \hspace{-0.06\textwidth}
    \fig{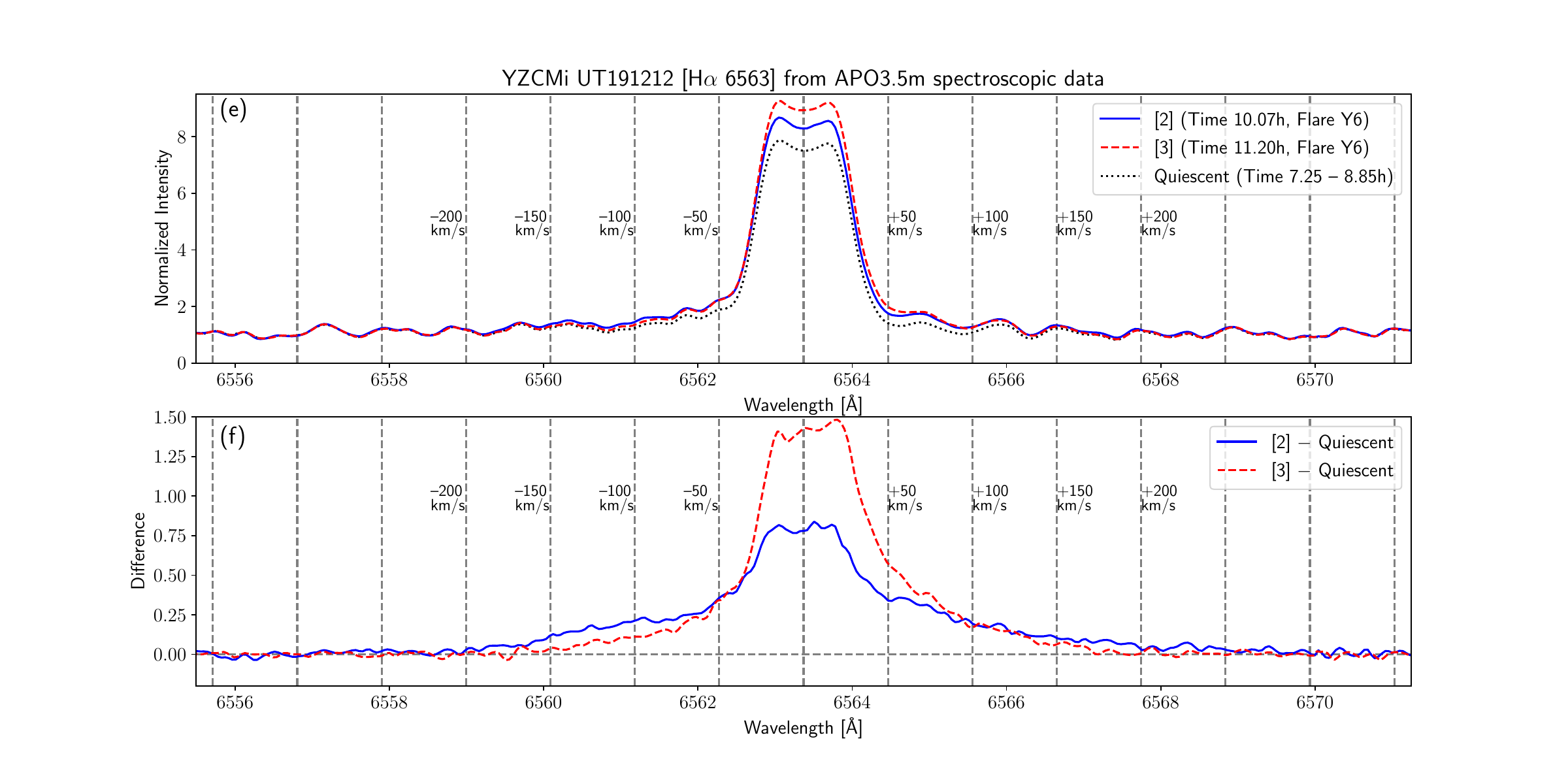}{0.58\textwidth}{\vspace{0mm}}
     \hspace{-0.06\textwidth}
       \fig{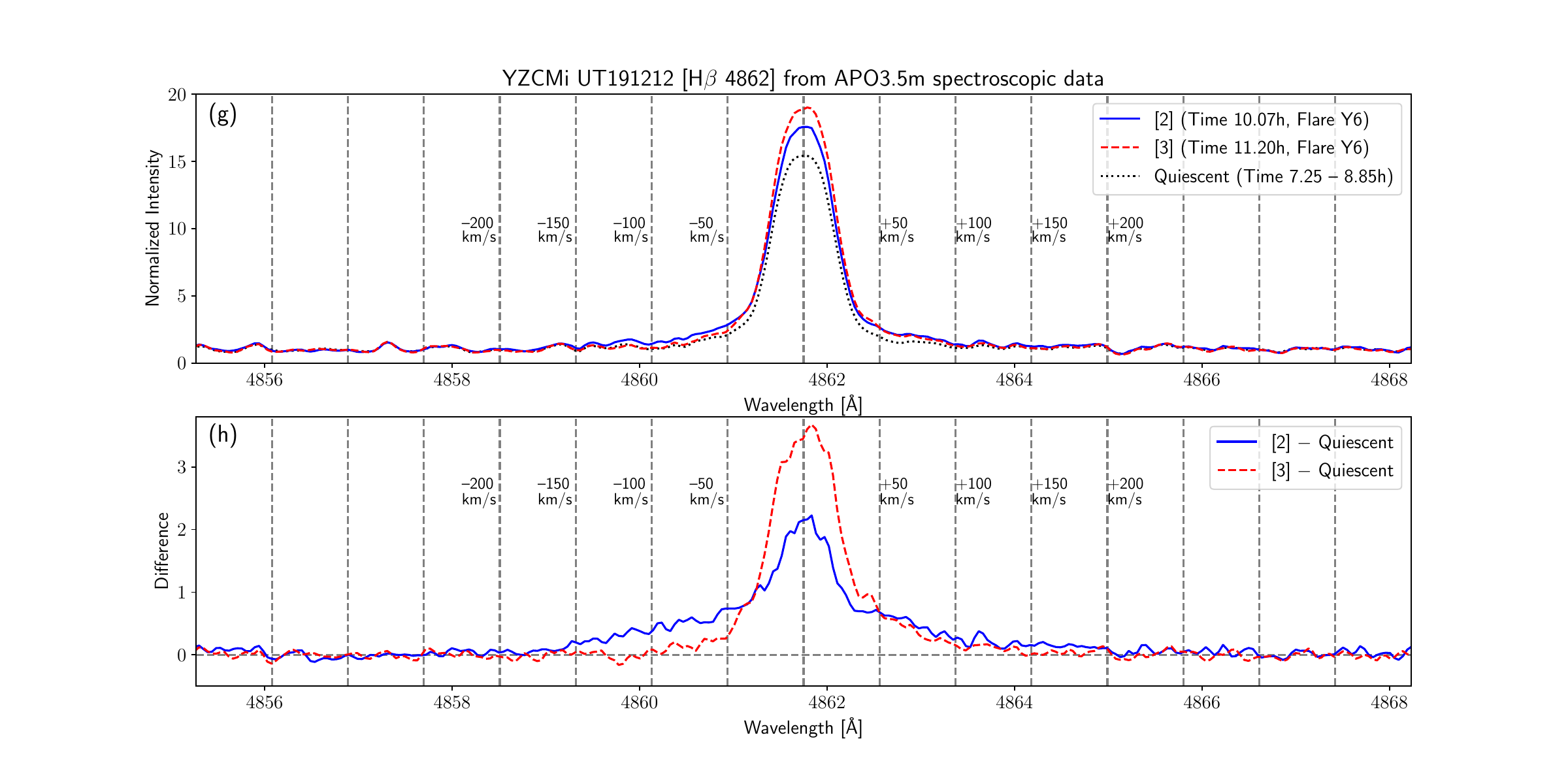}{0.58\textwidth}{\vspace{0mm}}
    }
      \vspace{-1cm}  
     \caption{
     \color{black}\textrm{
 Line profiles of the H$\alpha$ \& H$\beta$ emission lines during Flare Y6 on 2019 December 12 from APO3.5m spectroscopic data, which are plotted similarly with Figure \ref{fig:spec_HaHb_YZCMi_UT190127}.
 It is noted that the line profiles at the time [1] and [4] are in panels (a)--(d),
 while those at the time [2] and [3] are in (e)--(h).
} \color{black}
}
   \label{fig:spec_HaHb_YZCMi_UT191212}
   \end{center}
 \end{figure}

      \begin{figure}[ht!]
   \begin{center}
          \gridline{  
     \hspace{-0.07\textwidth}
    \fig{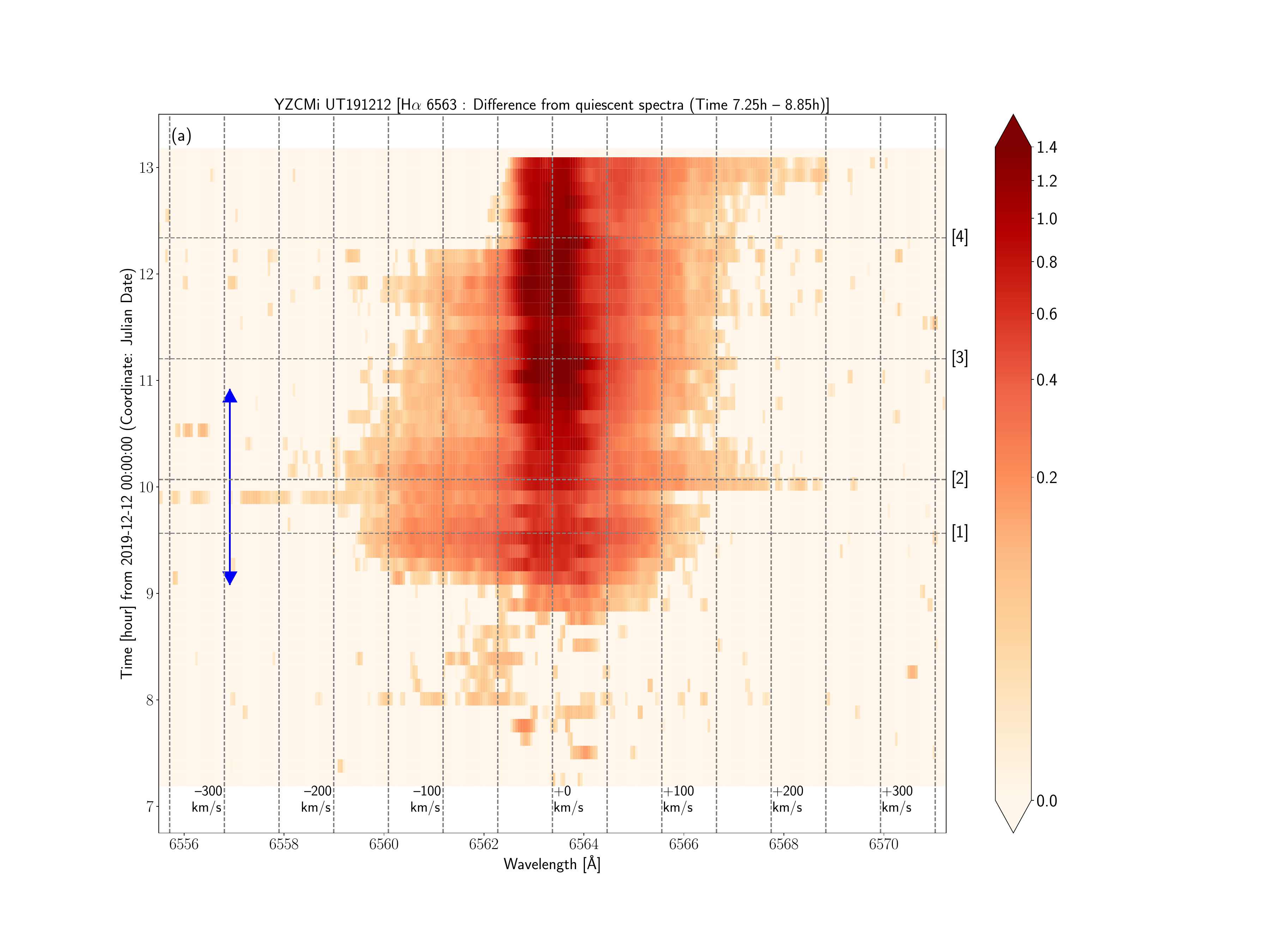}{0.63\textwidth}{\vspace{0mm}}
     \hspace{-0.11\textwidth}
    \fig{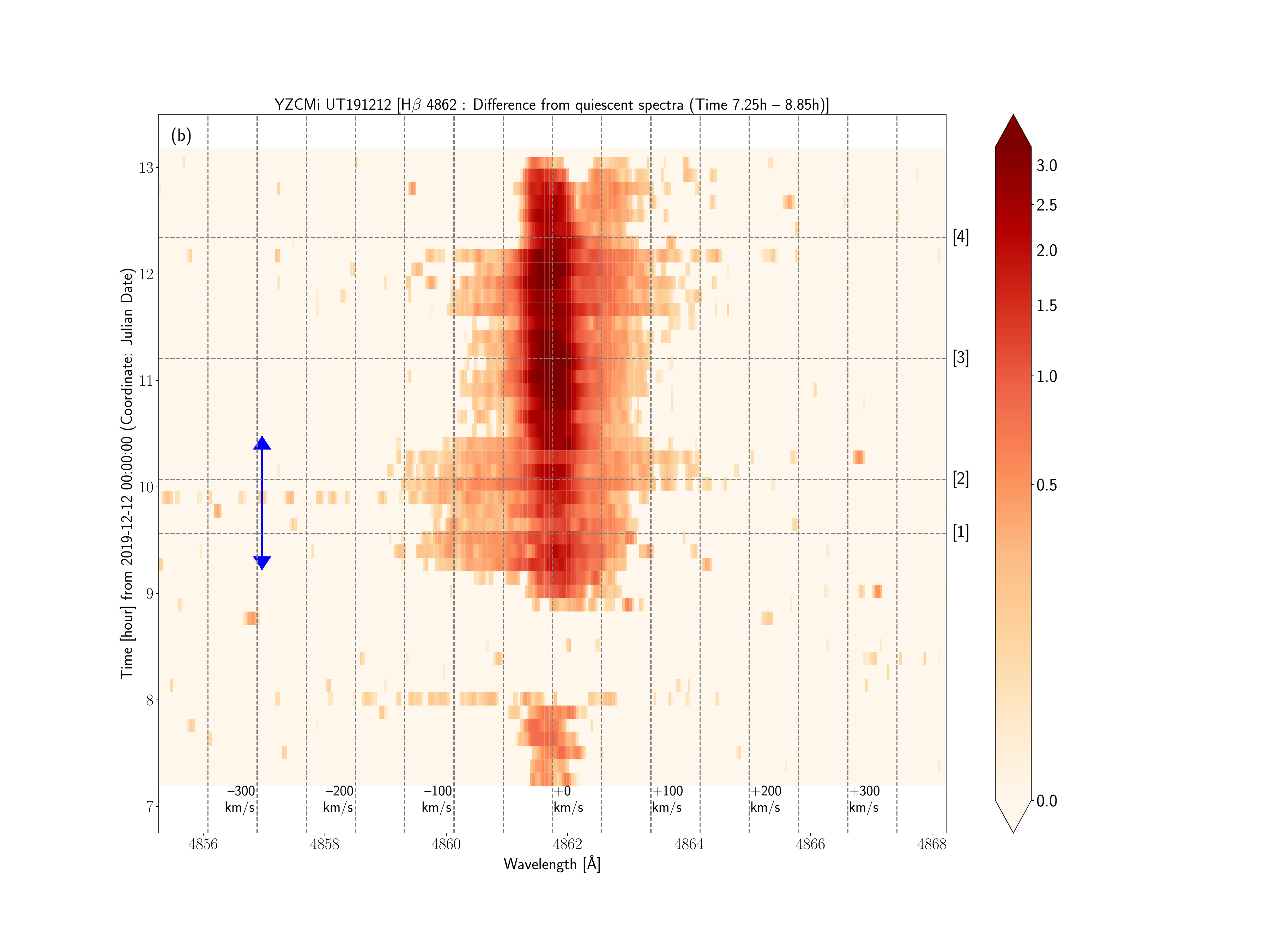}{0.63\textwidth}{\vspace{0mm}}
    }
    \vspace{-1cm}
     \caption{
     \color{black}\textrm{  
Time evolution of the H$\alpha$ \& H$\beta$ line profiles covering Flare Y6 on 2019 December 12, which are plotted similarly with Figure \ref{fig:map_HaHb_YZCMi_UT190127}, but
the vertical axes are in the time coordinate of Julian Date (JD).
The grey horizontal dashed lines indicate the time [1] -- [4], which are shown in Figure \ref{fig:lcEW_HaHb_YZCMi_UT191212} (light curves) and Figure \ref{fig:spec_HaHb_YZCMi_UT191212} (line profiles).
     }
     }
   \label{fig:map_HaHb_YZCMi_UT191212}
   \end{center}
 \end{figure}

On 2019 December 12, a flare (Flare Y6) was detected in H$\alpha$ \& H$\beta$ lines as shown in Figure \ref{fig:lcEW_HaHb_YZCMi_UT191212} (a).  During Flare Y6,
the H$\alpha$ \& H$\beta$ equivalent widths increased to 13.5\AA~and 16.8\AA, respectively, and $\Delta t^{\rm{flare}}_{\rm{H}\alpha}$ is $>$4.9 hours (Table \ref{table:list1_flares}). 
The observation ended in the decay of Flare Y6.  
The continuum flux observed by ARCSAT $u$- \& $g$-bands shows (at least two)  short ($\lesssim$10 min each) enhancements during Flare Y6 (Figures \ref{fig:lcEW_HaHb_YZCMi_UT191212} (b) \& (c)). The amplitudes of these short continuum enhancements in $u$-band are $\sim$60\% (around the time 10.3h -- 10.4h) and $\sim$40\% (around the time 12.2h -- 12.3h), and those in $g$-band are $\sim$ 4 -- 5\% (around the time 10.3h -- 10.4h) and $\sim$5 -- 6\% (around the time 12.2h -- 12.3h).

Although there are continuum enhancements around the times  10.3h -- 10.4h and 12.2h -- 12.3h, there are no clear white-light emissions that are considered to be physically associated with the early increasing phase of the 
H$\alpha$ and H$\beta$ flare emission (time before $\sim$10h).
Considering this, the flare peak luminosities of Flare Y6 listed in Table \ref{table:list1_flares}
are upper limit values considering the photometric error values 
(3$\sigma_{u}$=26\% and 3$\sigma_{g}$=4.9\%): 
$L_{u}<8\times 10^{27}$erg s$^{-1}$ and $L_{g}<1.8\times 10^{28}$erg s$^{-1}$ 
(cf. Section \ref{subsec:quiescent-ene}).
As for flare energies, only the upper limit values are calculated from these upper limit peak luminosities following the method in Section \ref{subsec:quiescent-ene}:
$E_{u}<7.2\times 10^{31}$erg and $E_{g}<1.6\times 10^{32}$erg. These upper limit values 
larger than the values that can be estimated by only integrating the clear peaks around the times 10.3h -- 10.4h and 12.2h -- 12.3h ($E_{u}=6\times 10^{30}$erg, and $E_{g}=3\times 10^{30}$erg). 
We also estimated $L_{\rm{H}\alpha}$, $L_{\rm{H}\beta}$, $E_{\rm{H}\alpha}$, and $E_{\rm{H}\beta}$ values,
which are listed in Table \ref{table:list1_flares}.
Since the observation ended before Flare Y6 ended, the real flare energy values can be larger than the values listed here.

      \begin{figure}[ht!]
   \begin{center}
            \gridline{  
     \hspace{-0.06\textwidth}
    \fig{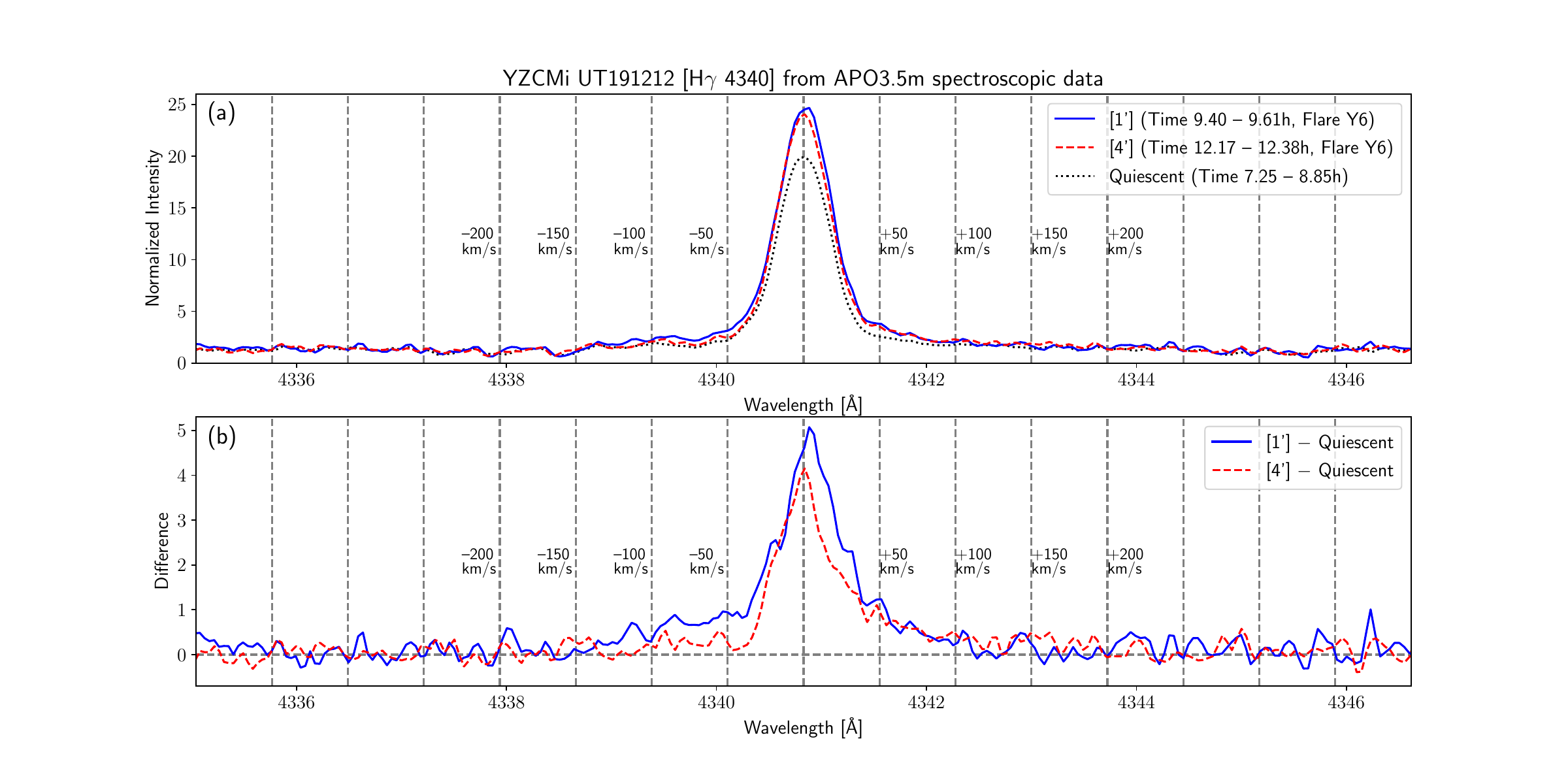}{0.58\textwidth}{\vspace{0mm}}
     \hspace{-0.06\textwidth}
       \fig{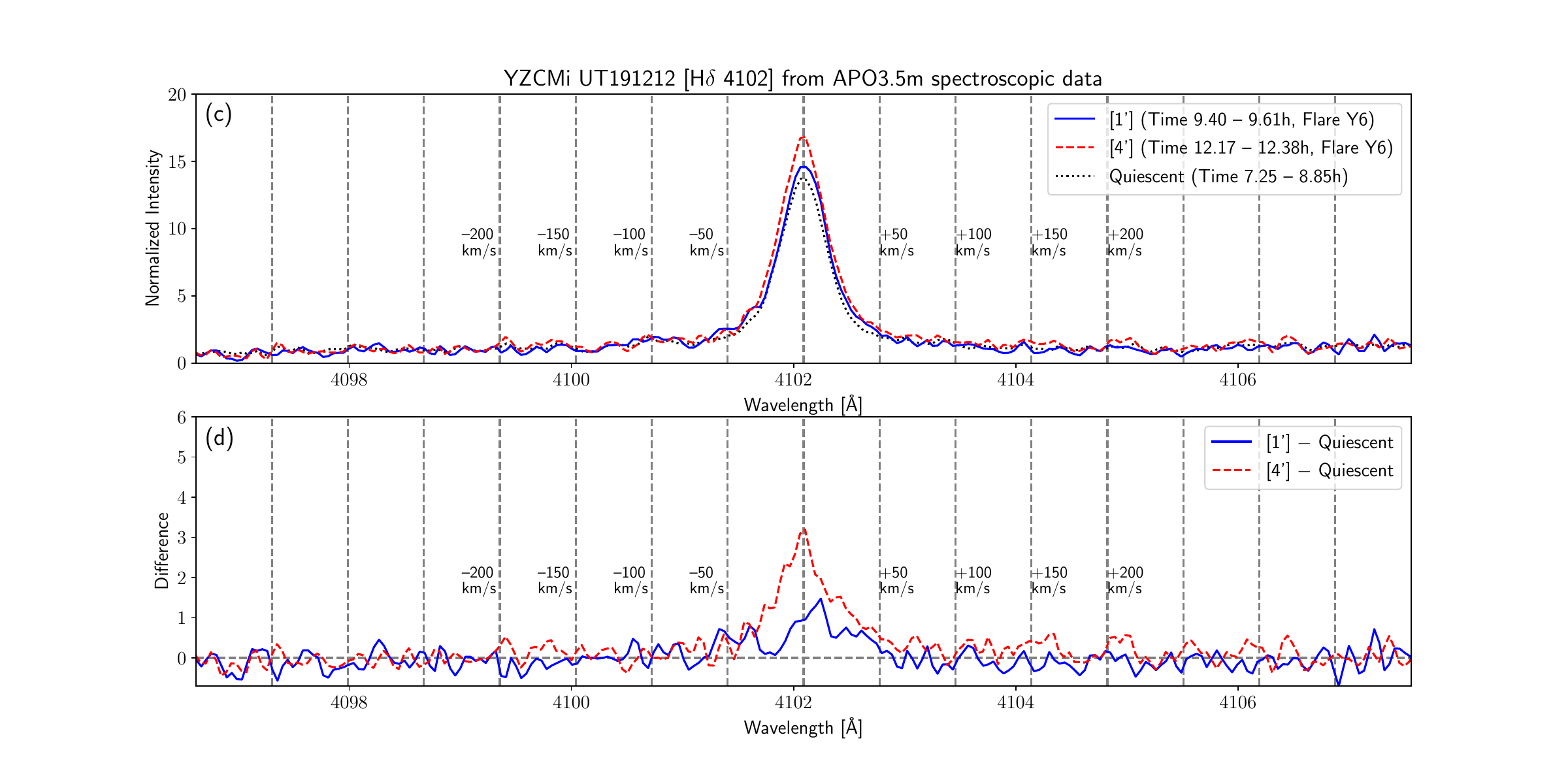}{0.58\textwidth}{\vspace{0mm}}
    }    
   \vspace{-1.0cm}
    \gridline{  
     \hspace{-0.06\textwidth}
    \fig{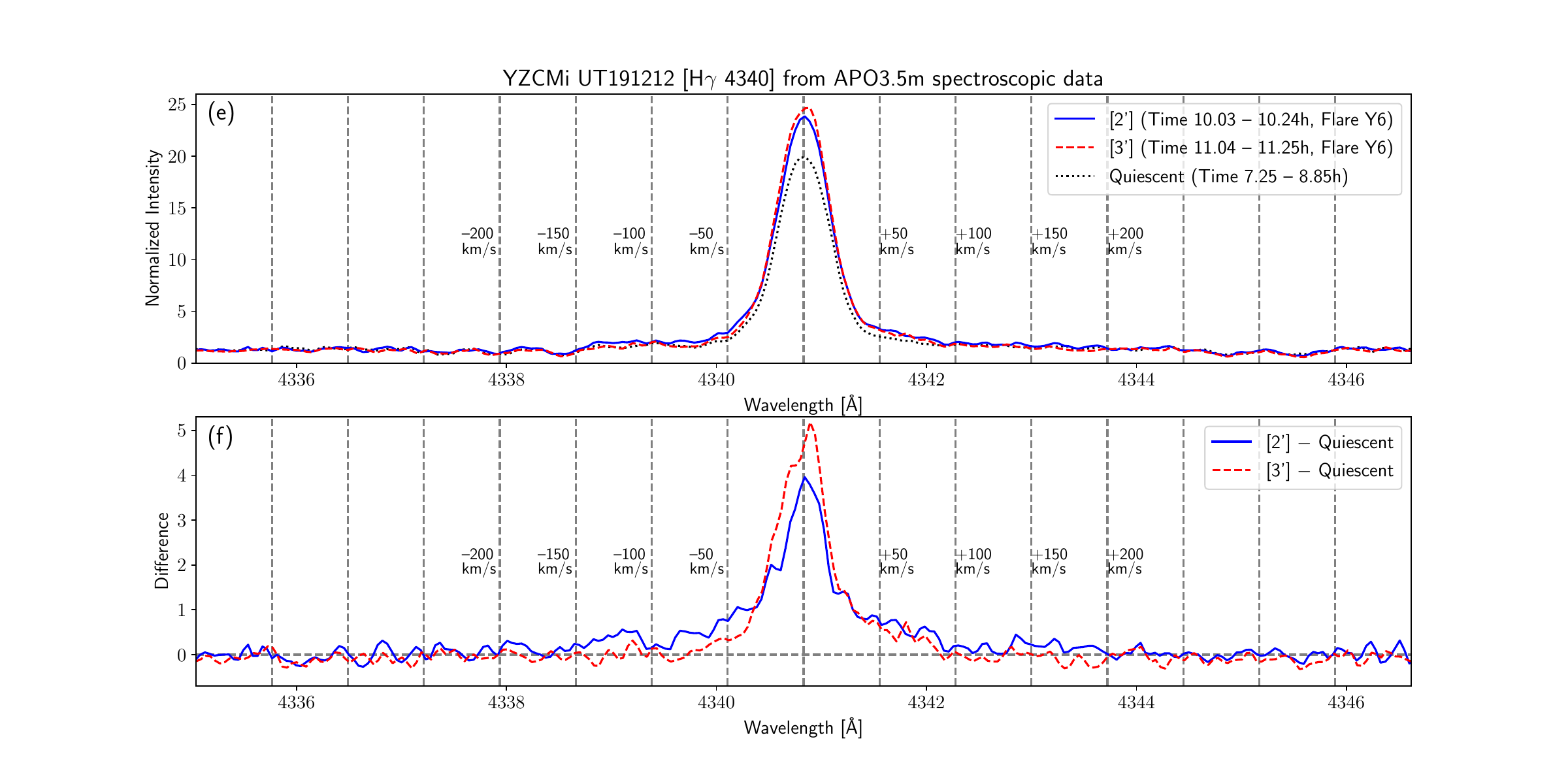}{0.58\textwidth}{\vspace{0mm}}
     \hspace{-0.06\textwidth}
       \fig{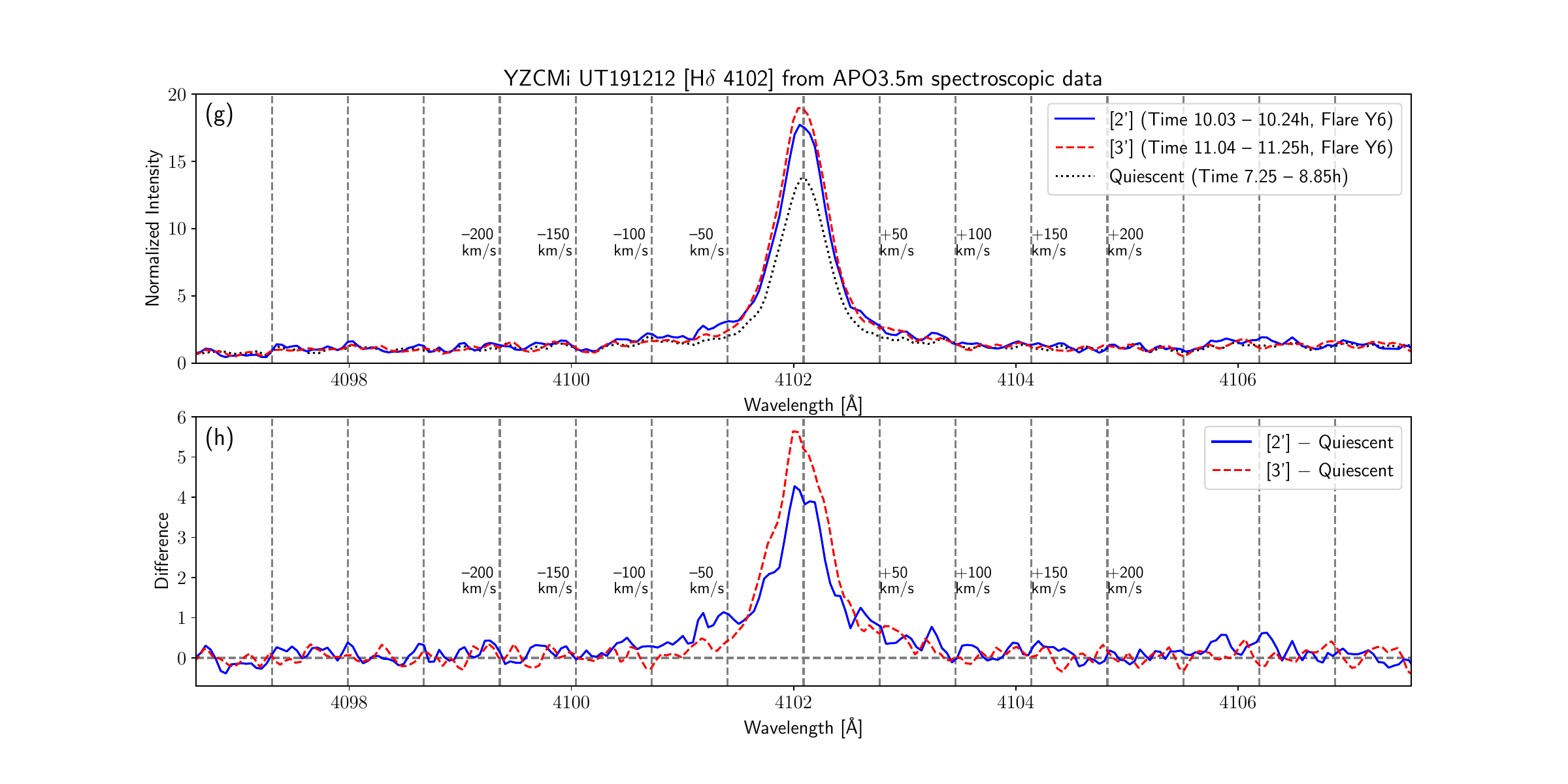}{0.58\textwidth}{\vspace{0mm}}
    }
    \vspace{-1cm}
     \caption{
     \color{black}\textrm{  
    (a) -- (d)
Line profiles of the H$\gamma$ \& H$\delta$ emission lines during Flare Y6 on 2019 December 12 from APO3.5m spectroscopic data, which are similarly plotted with Figure \ref{fig:spec_HaHb_YZCMi_UT191212}.
The blue solid and red dashed lines indicate 
the integrated line profiles over the time [1$^{\prime}$] (Time 9.40 -- 9.61h) and [4$^{\prime}$] (Time 12.17 -- 12.38h) on this date, which include the time [1] and [4] in Figure \ref{fig:lcEW_HaHb_YZCMi_UT191212} (light curves), respectively.
(e) -- (h) Same as (a) -- (d), but the line profiles integrated over the time [2$^{\prime}$] (Time 10.03 -- 10.24h) and [3$^{\prime}$] (Time 11.04 -- 11.25h), which include the time [2] and [3] in Figure \ref{fig:lcEW_HaHb_YZCMi_UT191212} (light curves), respectively.
} \color{black}
}
   \label{fig:spec_HcHd_YZCMi_UT191212}
   \end{center}
 \end{figure}
 
       \begin{figure}[ht!]
   \begin{center}
            \gridline{  
     \hspace{-0.06\textwidth}
    \fig{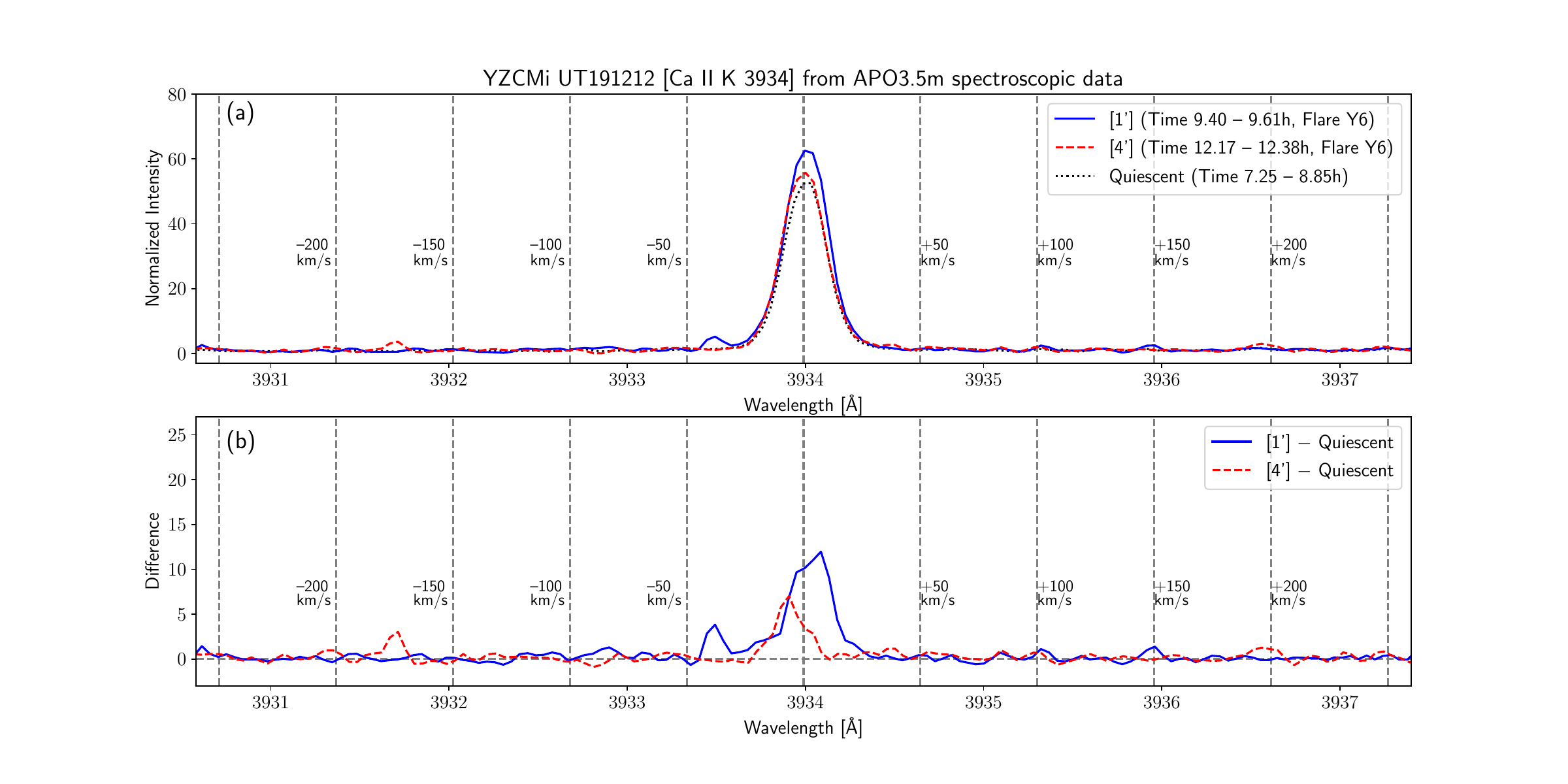}{0.58\textwidth}{\vspace{0mm}}
     \hspace{-0.06\textwidth}
         \fig{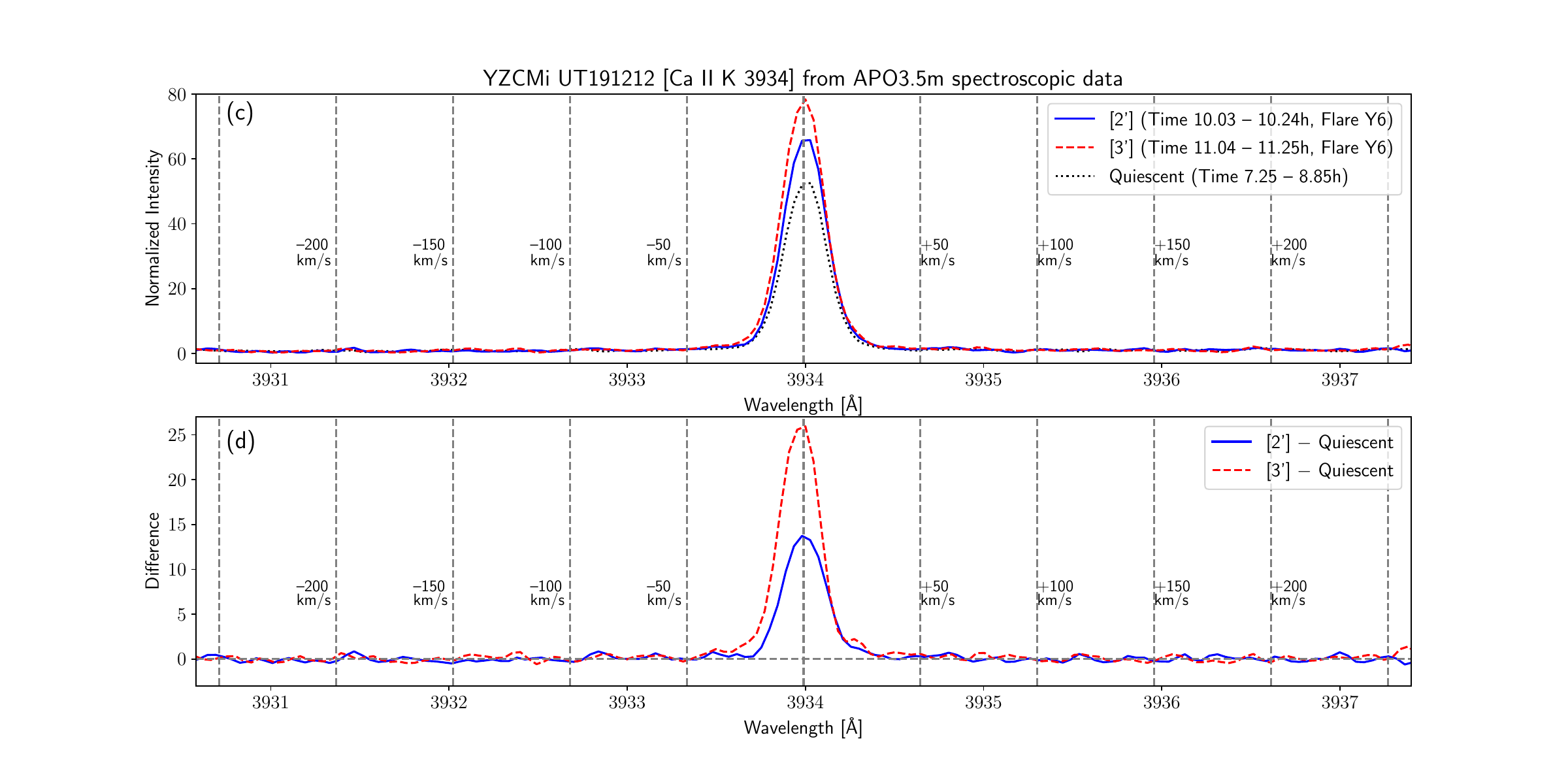}{0.58\textwidth}{\vspace{0mm}}
    }    
   \vspace{-1.0cm}
    \gridline{  
     \hspace{-0.06\textwidth}
        \fig{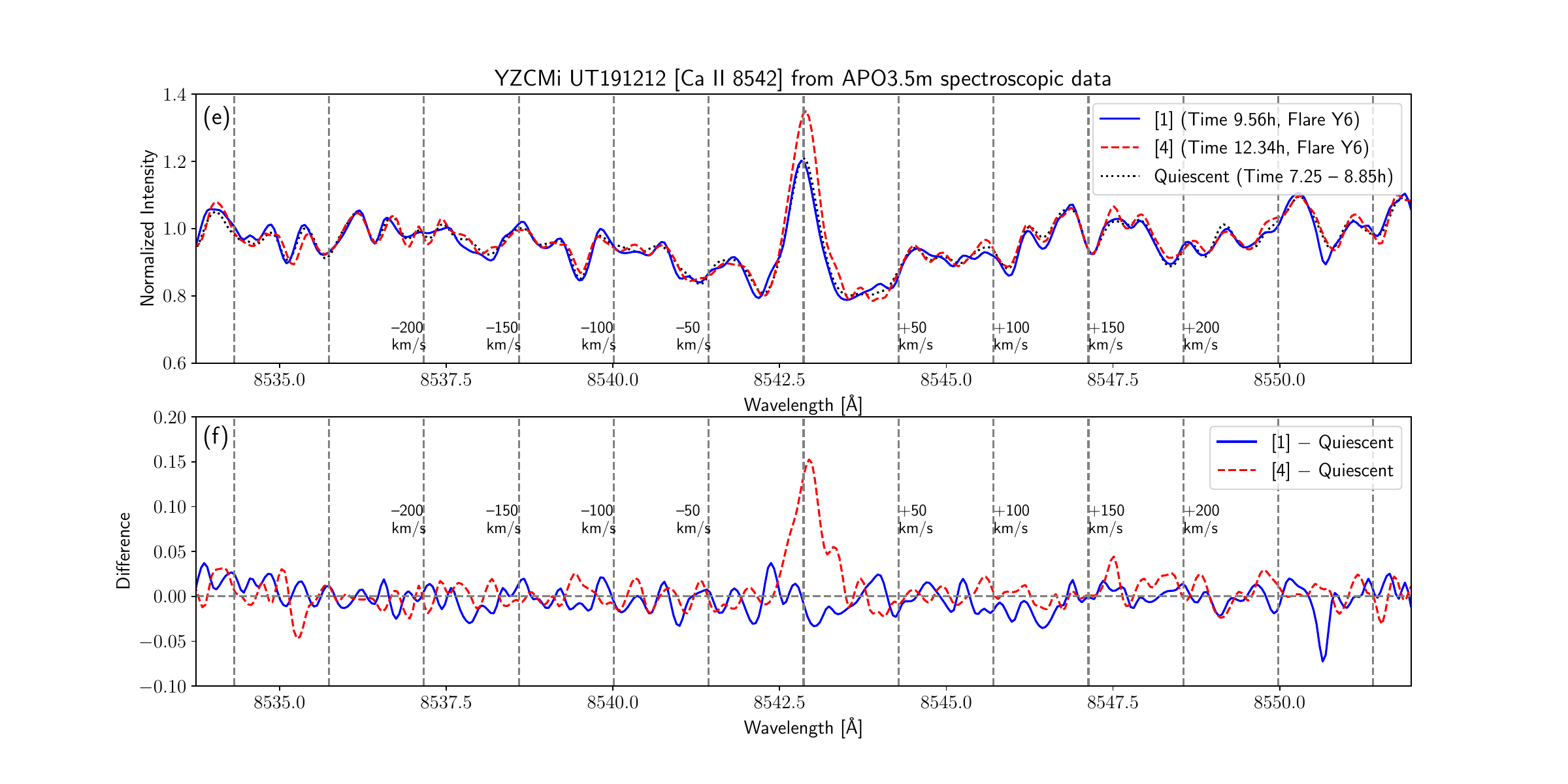}{0.58\textwidth}{\vspace{0mm}}
     \hspace{-0.06\textwidth}
         \fig{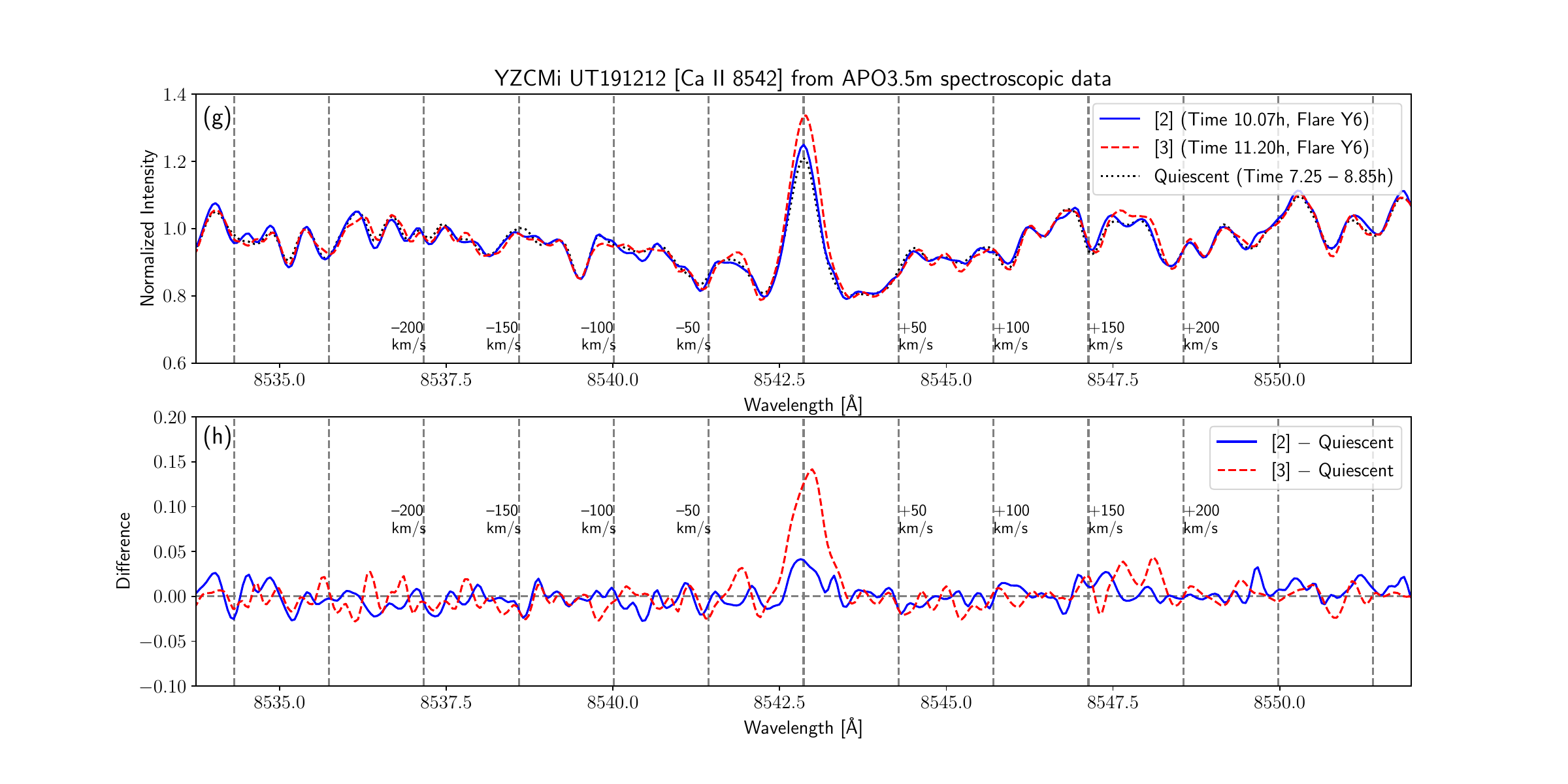}{0.58\textwidth}{\vspace{0mm}}
    }
   \vspace{-1.0cm}
    \gridline{  
     \hspace{-0.06\textwidth}
    \fig{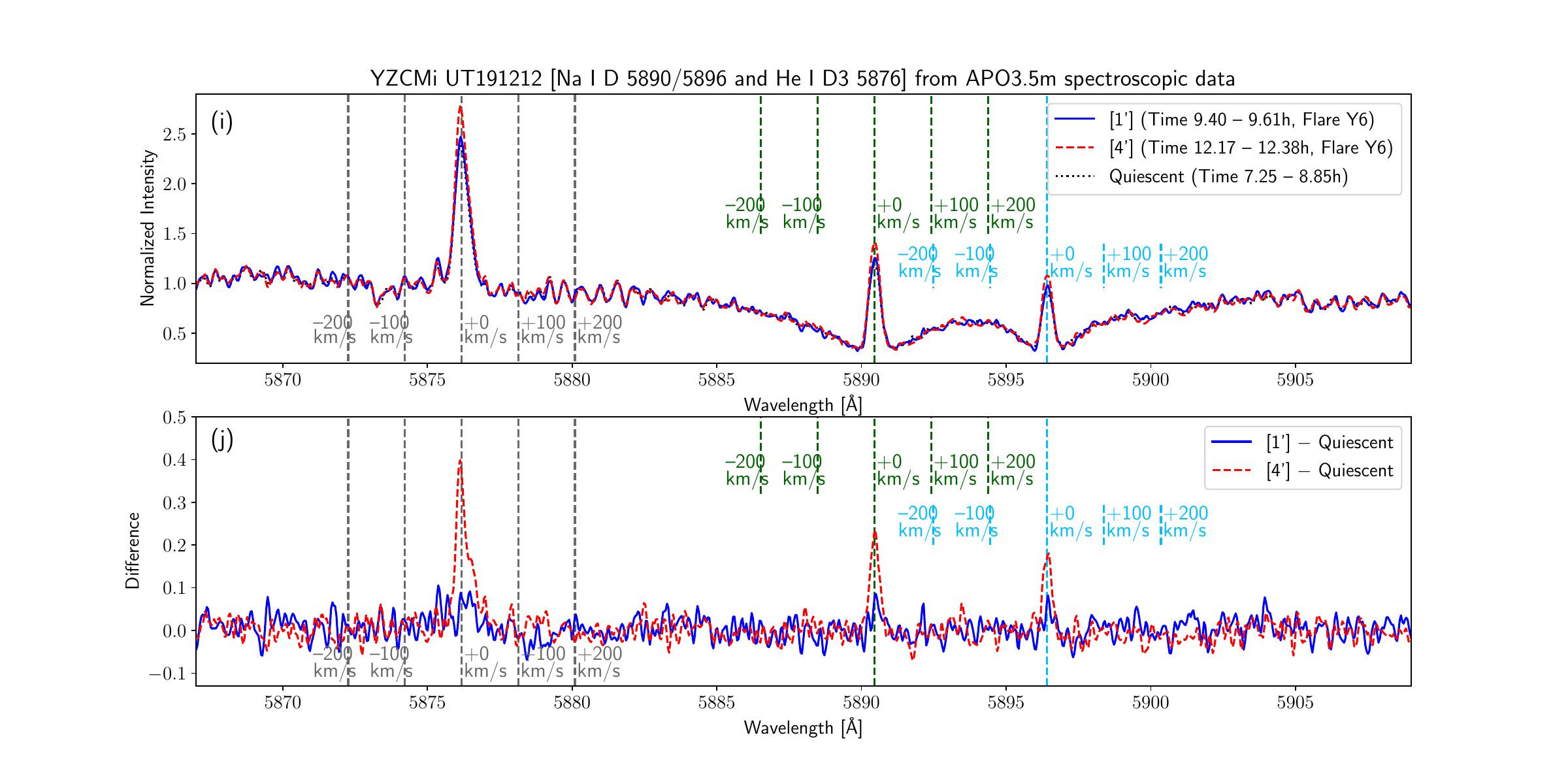}{0.58\textwidth}{\vspace{0mm}}
     \hspace{-0.06\textwidth}
     \fig{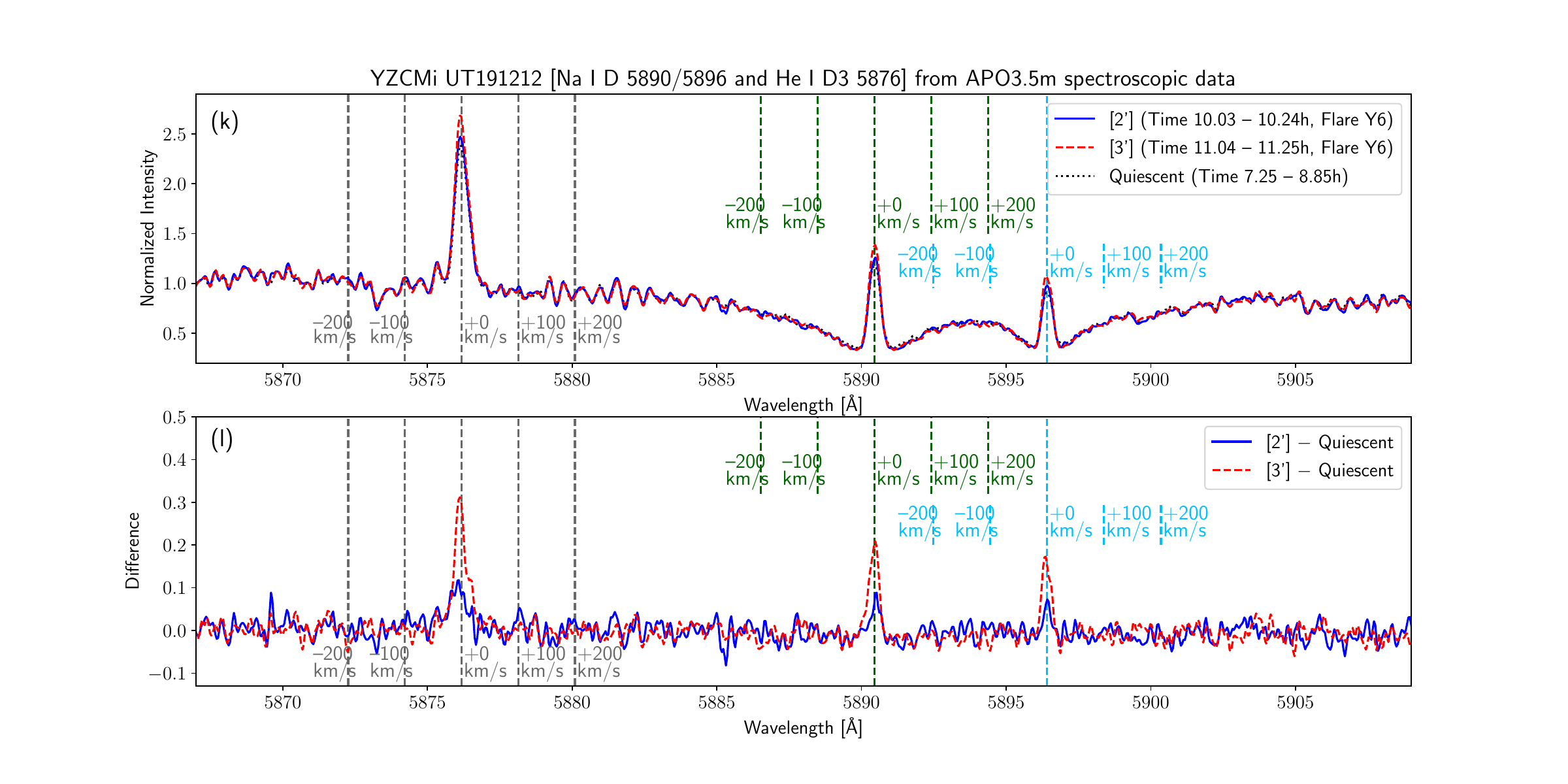}{0.58\textwidth}{\vspace{0mm}}
    }
   \vspace{-1.0cm}
    \gridline{  
     \hspace{-0.06\textwidth}
    \fig{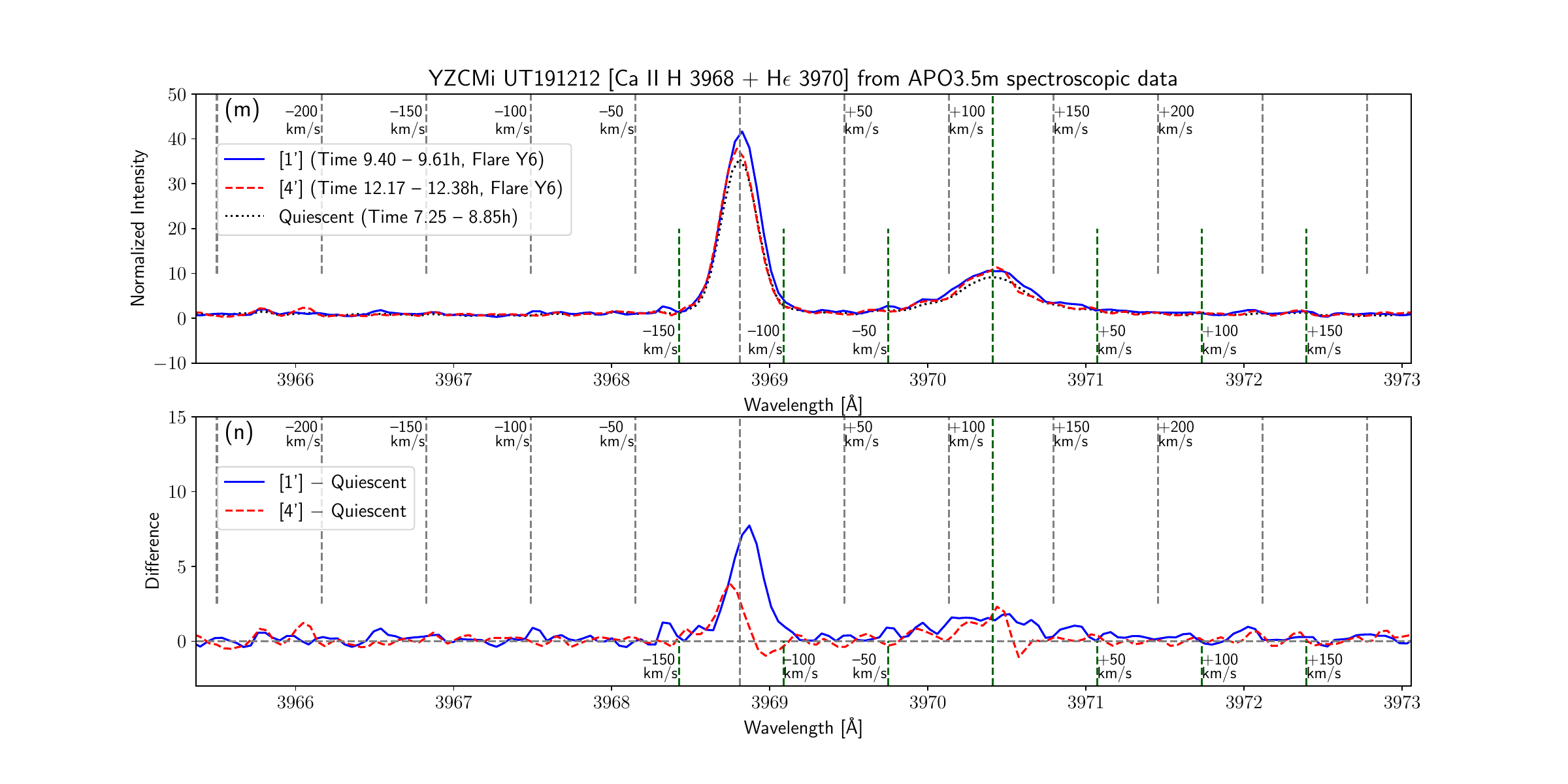}{0.58\textwidth}{\vspace{0mm}}
     \hspace{-0.06\textwidth}
    \fig{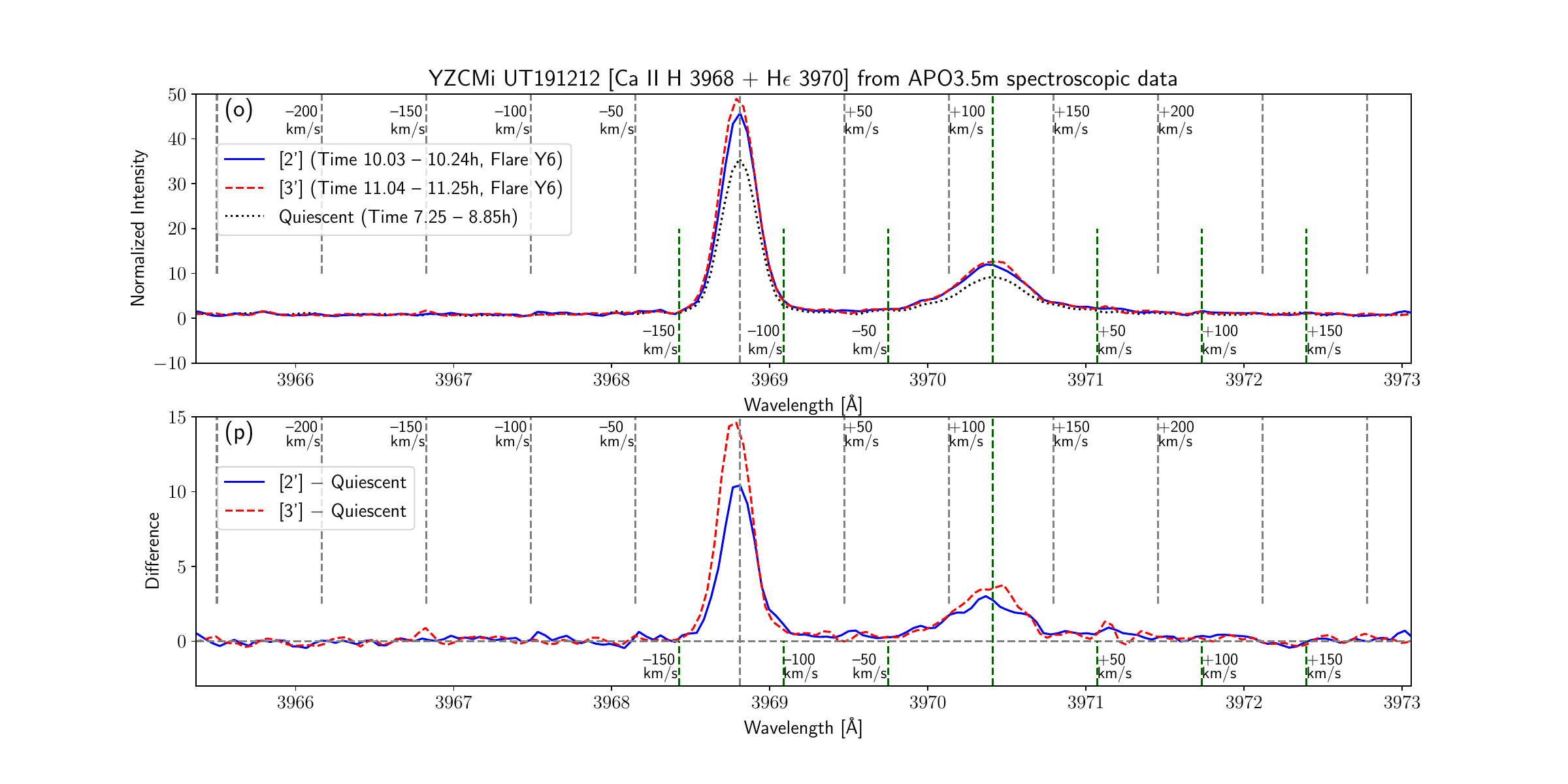}{0.58\textwidth}{\vspace{0mm}}
    }
   \vspace{-1cm}
     \caption{
\color{black}\textrm{  
Same as Figure \ref{fig:spec_HcHd_YZCMi_UT191212} but for 
Ca II K, Ca II 8542, Na I D1 \& D2 (5890 \& 5896)$+$He I D3 5876, and H$\epsilon+$Ca II H lines, respectively.
As for the Ca II 8542 line, the data at the time [1]--[4] are plotted (not [1$^{\prime}$] -- [4$^{\prime}$]).
 } \color{black}
     }
   \label{fig:spec_CaKCa8542HeD3NaD1D2_YZCMi_UT191212}
   \end{center}
 \end{figure}

The H$\alpha$ \& H$\beta$ line profiles during Flare Y6 are shown in
Figures \ref{fig:spec_HaHb_YZCMi_UT191212} \& \ref{fig:map_HaHb_YZCMi_UT191212}. The clear H$\alpha$ blue wing enhancement (blue wing asymmetry) up to $\sim -$200 km s$^{-1}$ was seen 
in early phase of the flare (e.g., time [1]), while the line profile gradually shifted to the red wing enhancement (red wing asymmetry) up to $\sim +$200 km s$^{-1}$ (e.g., time \color{black}\textrm{[4]} \color{black})
(Figures \ref{fig:lcEW_HaHb_YZCMi_UT191212},
\ref{fig:spec_HaHb_YZCMi_UT191212}, \&
\ref{fig:map_HaHb_YZCMi_UT191212}).
The evolution of the H$\alpha$ line from blue to red shifted line wing asymmetry is particularly evident.
H$\beta$ line also shows the time evolution from the blue wing asymmetry to the red wing asymmetry (Figures \ref{fig:spec_HaHb_YZCMi_UT191212} \&
\ref{fig:map_HaHb_YZCMi_UT191212}), 
which is very similar to that of H$\alpha$ line.
The wing enhancements of these blue and red wing asymmetries in H$\beta$ line are slightly smaller  than those in H$\alpha$ line: from $\sim -$200 km s$^{-1}$ to $\sim +$200 km s$^{-1}$ in H$\alpha$ line, and from $\sim -$150 km s$^{-1}$ to $\sim +$150 km s$^{-1}$ in H$\beta$ line.

The EW light curves of H$\gamma$, H$\delta$, Ca II K, Ca II 8542, Na I D1 \& D2, and He I D3 5876 lines are also shown in Figures \ref{fig:lcEW_HaHb_YZCMi_UT191212} (c), (d), \& (e).
The profiles of these lines and H$\epsilon+$Ca II H lines during Flare Y6 
are shown in Figures \ref{fig:spec_HcHd_YZCMi_UT191212} \& 
\ref{fig:spec_CaKCa8542HeD3NaD1D2_YZCMi_UT191212}.
As for H$\gamma$ and H$\delta$ lines, 
the blue wing asymmetries in the early phase of the flare are not so evident (time [1$^{\prime}$], [2$^{\prime}$], and [3$^{\prime}$] in Figure \ref{fig:spec_HcHd_YZCMi_UT191212}) while 
the red wing asymmetries in the later phase are seen (time [4$^{\prime}$] in 
Figure \ref{fig:spec_HcHd_YZCMi_UT191212}).
Similar time evolution of red wing asymmetry was seen also in He I D3 5876 line, 
but the possible 
red wing asymmetry component at time [4$^{\prime}$] was very small ($\lesssim$50 km s$^{-1}$)
(Figure \ref{fig:spec_CaKCa8542HeD3NaD1D2_YZCMi_UT191212}(j)).
As for Ca II H\&K, Ca II 8542, Na I D1 \& D2, and H$\epsilon$ lines, 
the line asymmetries are not readily detected (Figure \ref{fig:spec_CaKCa8542HeD3NaD1D2_YZCMi_UT191212}).

\clearpage
\subsection{Flares Y18 (Blue wing asymmetry) \& Y19 observed on 2020 January 21} 
\label{subsec:results:2020-Jan-21} 

       \begin{figure}[ht!]
   \begin{center}
   \gridline{
    \fig{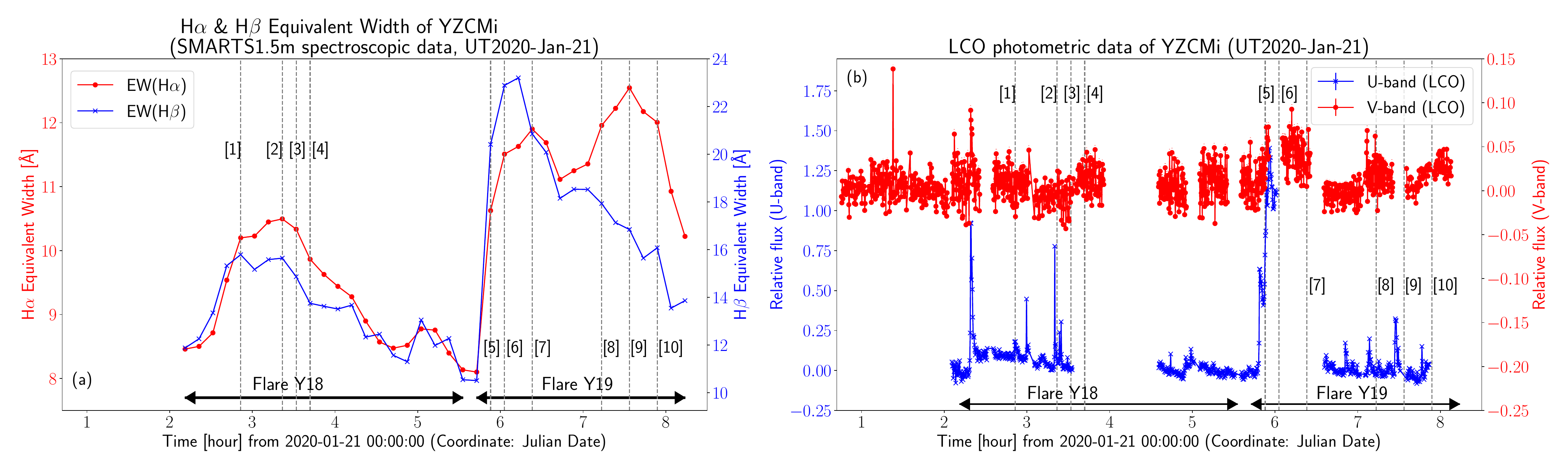}{1.0\textwidth}{\vspace{0mm}}}
     \vspace{-5mm}
   \gridline{
    \fig{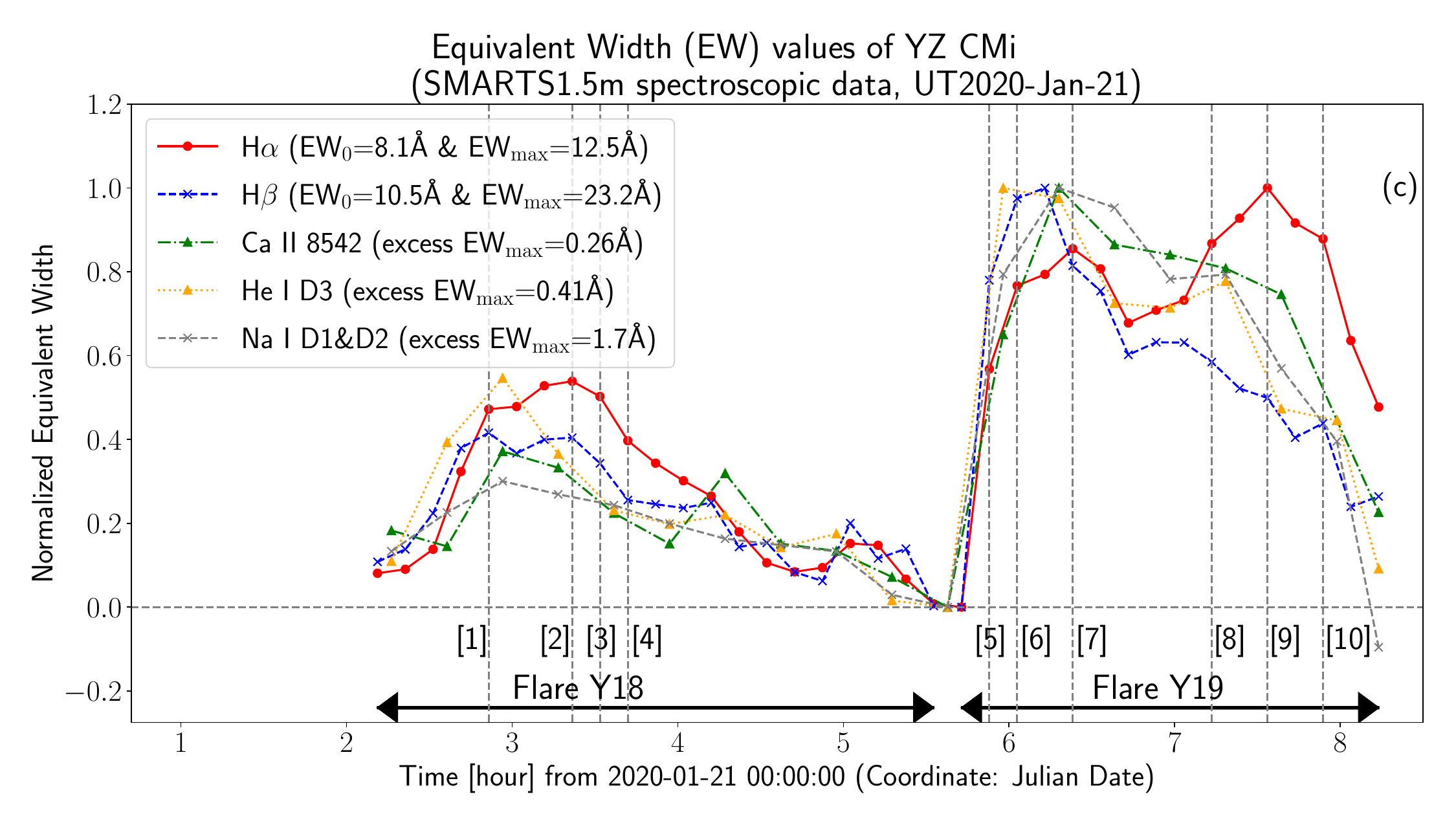}{0.5\textwidth}{\vspace{0mm}}}
     \vspace{-5mm}
     \caption{
     \color{black}\textrm{ 
Light curves of YZ CMi on 2020 January 21 showing Flares Y18 \& Y19. 
The data are plotted similarly with Figure \ref{fig:lcEW_HaHb_YZCMi_UT191212},
but the the chromospheric line emission data are from the SMARTS1.5m spectroscopic data,
which include only H$\alpha$, H$\beta$, Ca II 8542, He I D3, and Na I D1 \& D2 lines.
Different from Figure \ref{fig:lcEW_HaHb_YZCMi_UT191212}(b), 
the LCO $U$- \& $V$-band photometric data are plotted in (b).
The grey dashed lines with numbers ([1]--[10]) correspond to the time shown with the same numbers in Figures \ref{fig:spec_HaHb_YZCMi_UT200121_Y18}, \ref{fig:spec_HaHb_YZCMi_UT200121_Y19}, \& \ref{fig:map_HaHb_YZCMi_UT200121}.
}
     }
   \label{fig:lcEW_HaHb_YZCMi_UT200121}
   \end{center}
 \end{figure}

On 2020 January 21, two flares (Flares Y18 \& Y19) were detected in H$\alpha$ \& H$\beta$ lines as shown in Figure \ref{fig:lcEW_HaHb_YZCMi_UT200121} (a).  
As for Flare Y18, the H$\alpha$ \& H$\beta$ equivalent widths increased up to 10.5\AA~and 15.8\AA, respectively, and $\Delta t^{\rm{flare}}_{\rm{H}\alpha}$ is 3.4 hours (Table \ref{table:list1_flares}).
In addition to these enhancements in Balmer emission lines, the continuum flux from the LCO $U$- \& $V$-band increase is at least
$\sim$90\%, and $\sim$5\%, 
respectively, during Flare Y18 (Figure \ref{fig:lcEW_HaHb_YZCMi_UT200121} (b)). 
We note that the LCO photometric observation has gaps in the later phase of Flare Y18, and it could be possible that 
we missed the continuum flux increase during this time.
As for Flare Y19, the H$\alpha$ \& H$\beta$ equivalent widths increased to 12.5\AA~and 23.2\AA, respectively, and $\Delta t^{\rm{flare}}_{\rm{H}\alpha}$ is $>$2.5 hours (Table \ref{table:list1_flares}).
We note that the observation finished before Flare Y19 ended.
In addition to these enhancements in Balmer emission lines, the continuum flux observed with LCO $U$- \& $V$-band increased at least by
$\sim$ 130 -- 140\%, and $\sim$ 5 -- 10\%, 
respectively, during Flare Y19 (Figure \ref{fig:lcEW_HaHb_YZCMi_UT200121} (b)). 
The LCO photometric observation has gaps during Flare Y19, 
and it \color{black}\textrm{is} \color{black} possible that we missed the continuum brightness increase \color{black}\textrm{components} \color{black} during the gap time.
 
             \begin{figure}[ht!]
   \begin{center}
            \gridline{  
     \hspace{-0.06\textwidth}
    \fig{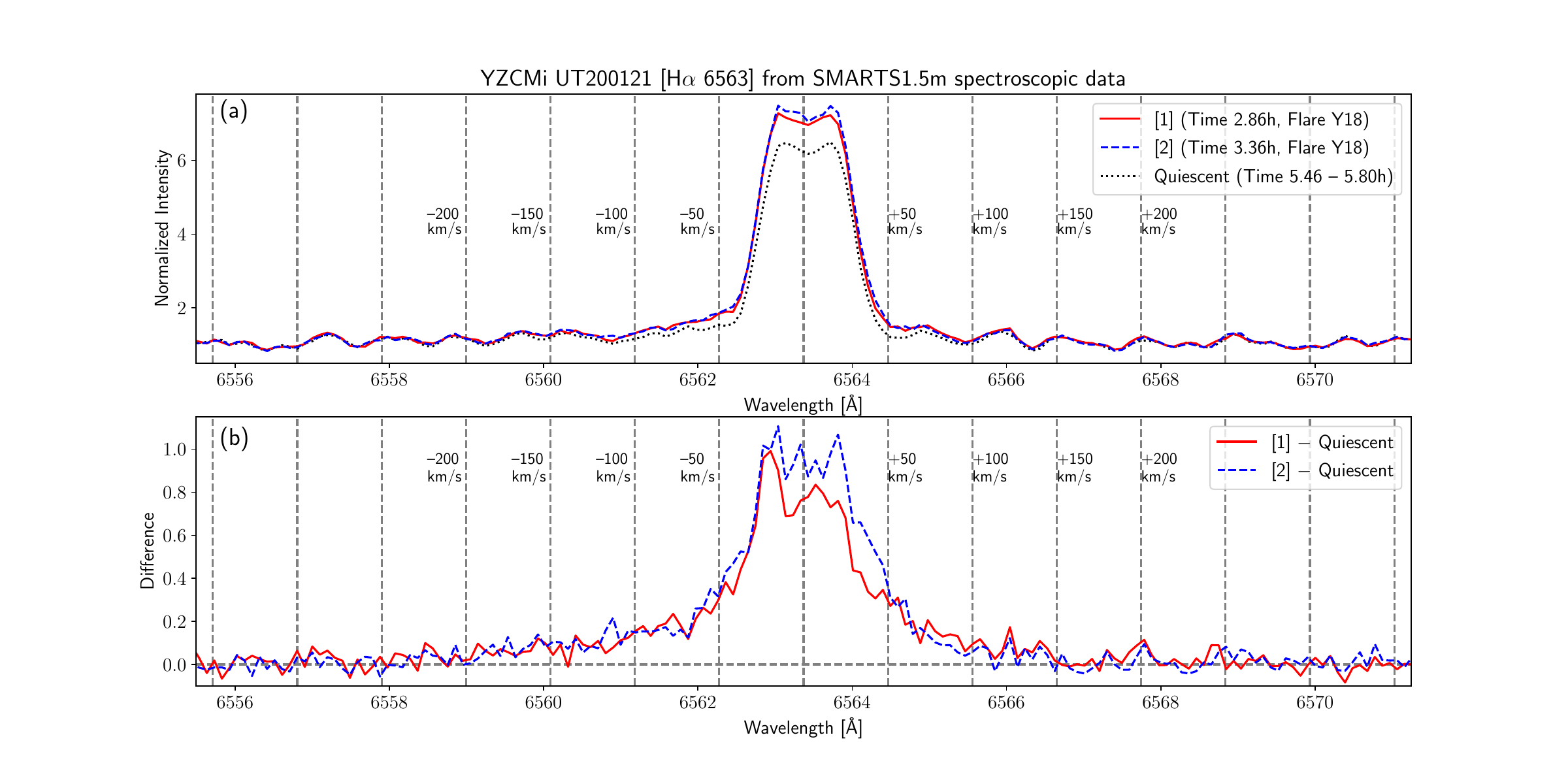}{0.58\textwidth}{\vspace{0mm}}
     \hspace{-0.06\textwidth}
       \fig{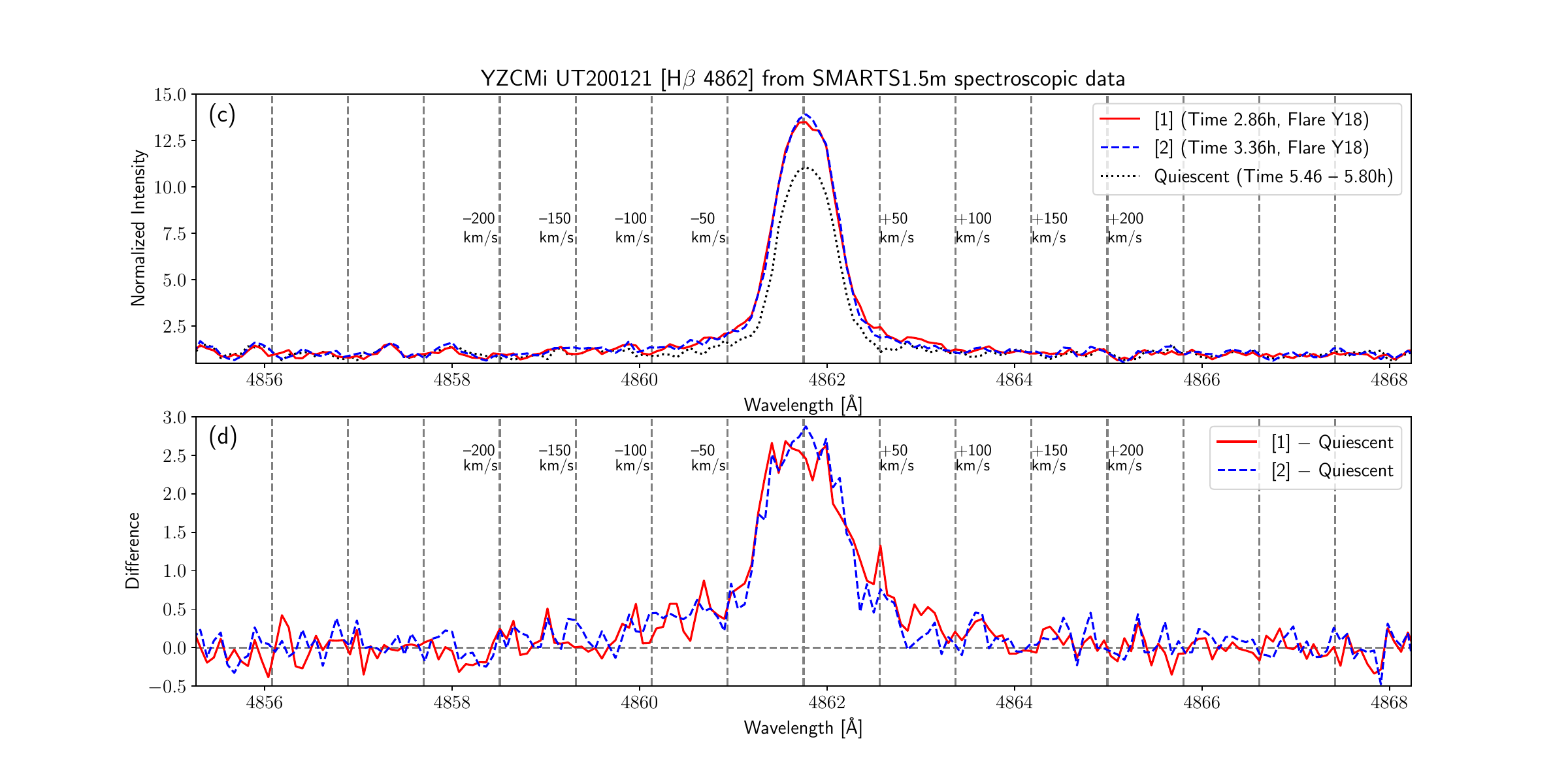}{0.58\textwidth}{\vspace{0mm}}
    }
     \vspace{-1.0cm}
            \gridline{  
     \hspace{-0.06\textwidth}
    \fig{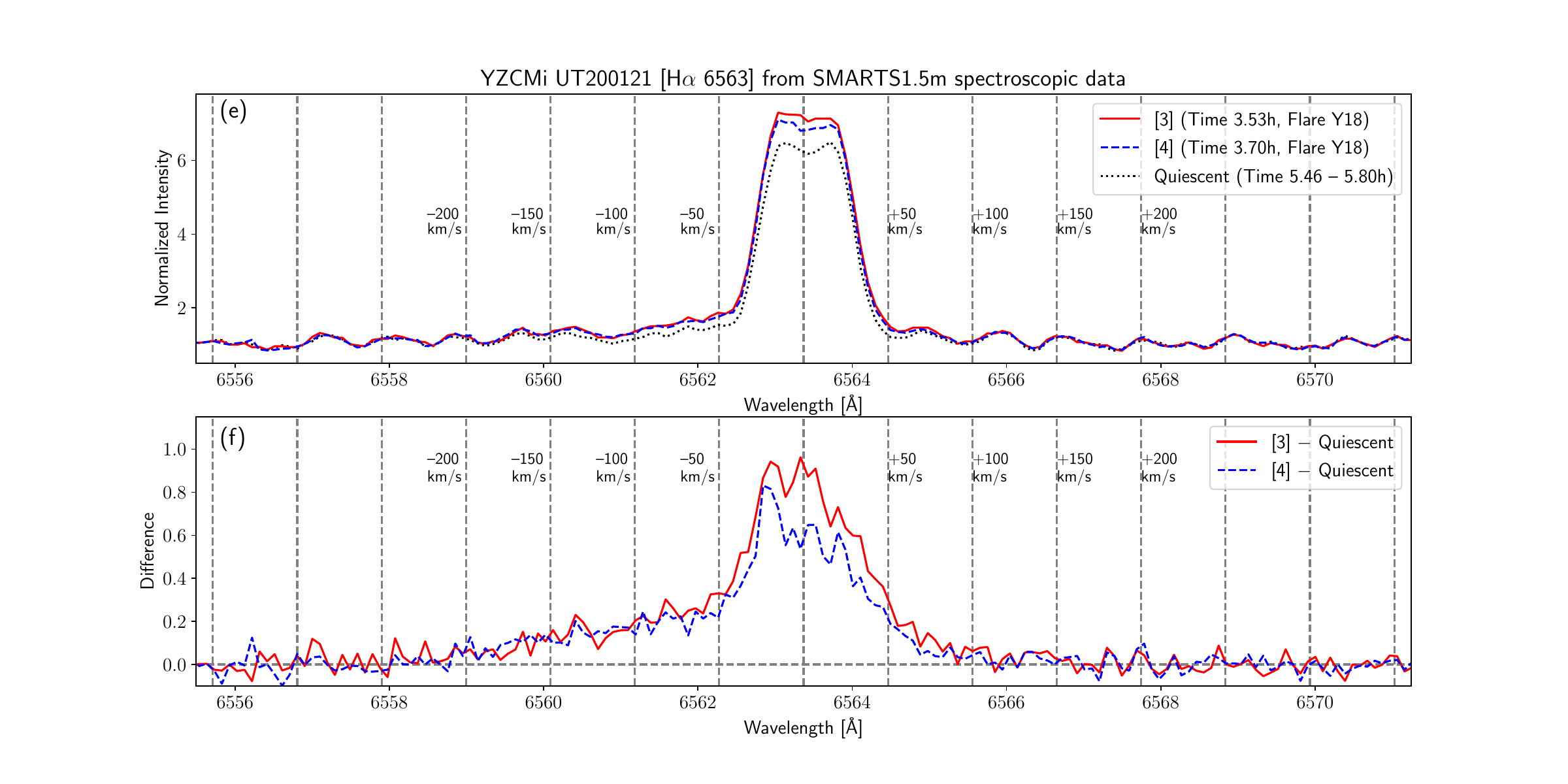}{0.58\textwidth}{\vspace{0mm}}
     \hspace{-0.06\textwidth}
       \fig{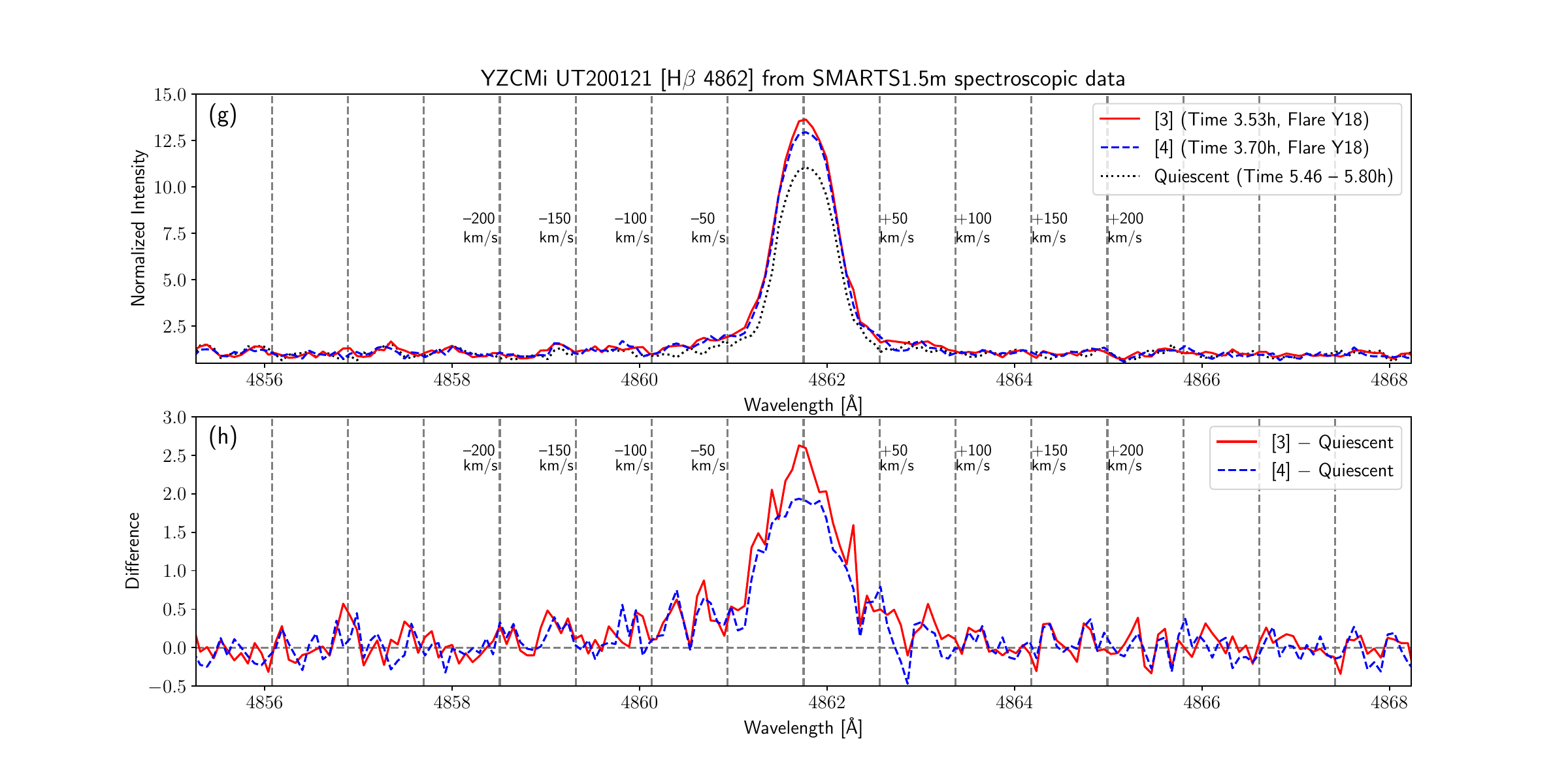}{0.58\textwidth}{\vspace{0mm}}
    }
     \vspace{-1cm}
     \caption{\color{black}\textrm{ 
Line profiles of the H$\alpha$ \& H$\beta$ emission lines during Flare Y18 (at the time [1]--[4]) on 2020 January 21 from SMARTS 1.5m spectroscopic data, which are plotted similarly with Figure \ref{fig:spec_HaHb_YZCMi_UT190127}.
}
     }
   \label{fig:spec_HaHb_YZCMi_UT200121_Y18}
   \end{center}
 \end{figure}

             \begin{figure}[ht!]
   \begin{center}
            \gridline{  
     \hspace{-0.06\textwidth}
    \fig{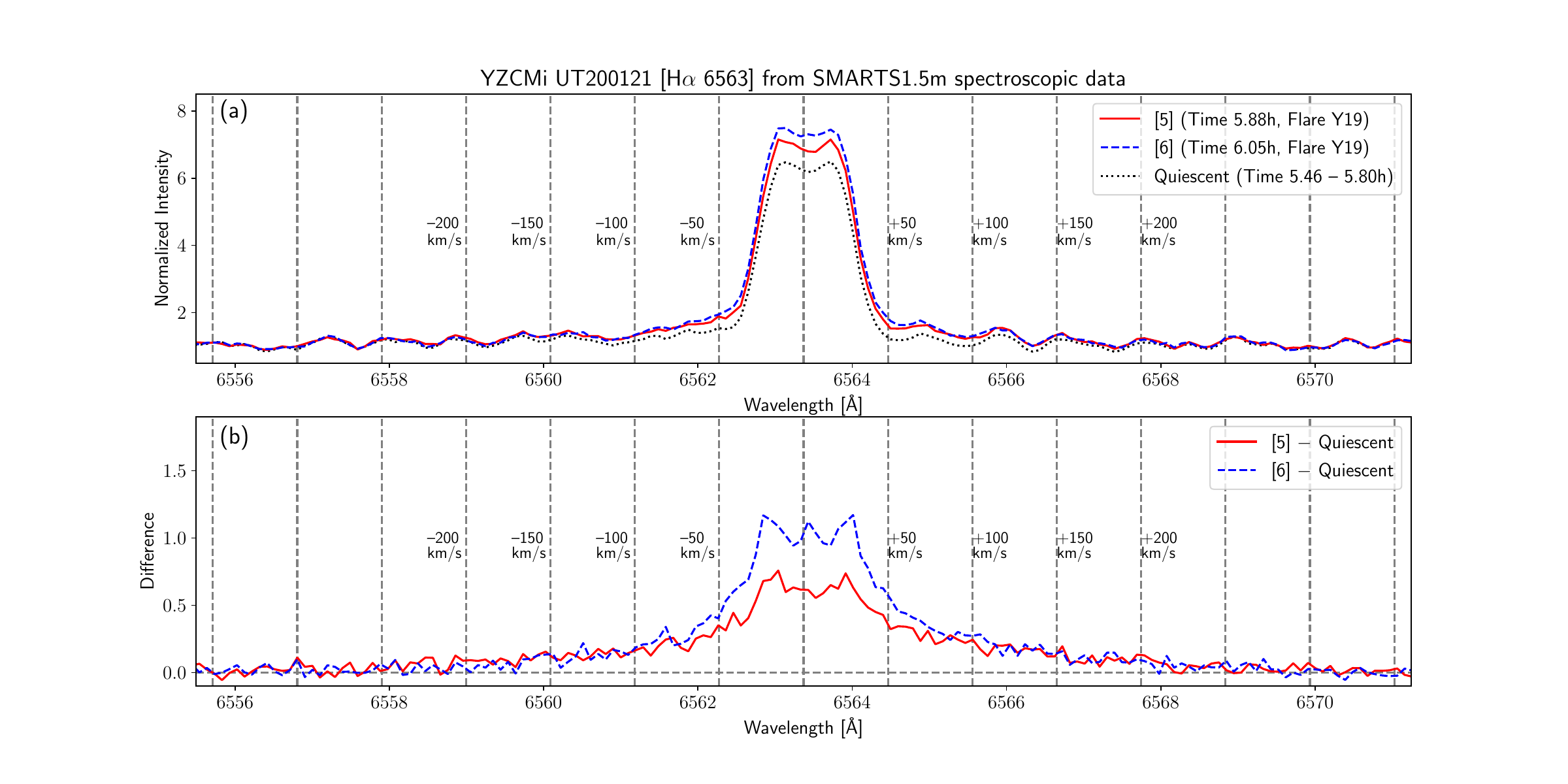}{0.58\textwidth}{\vspace{0mm}}
     \hspace{-0.06\textwidth}
       \fig{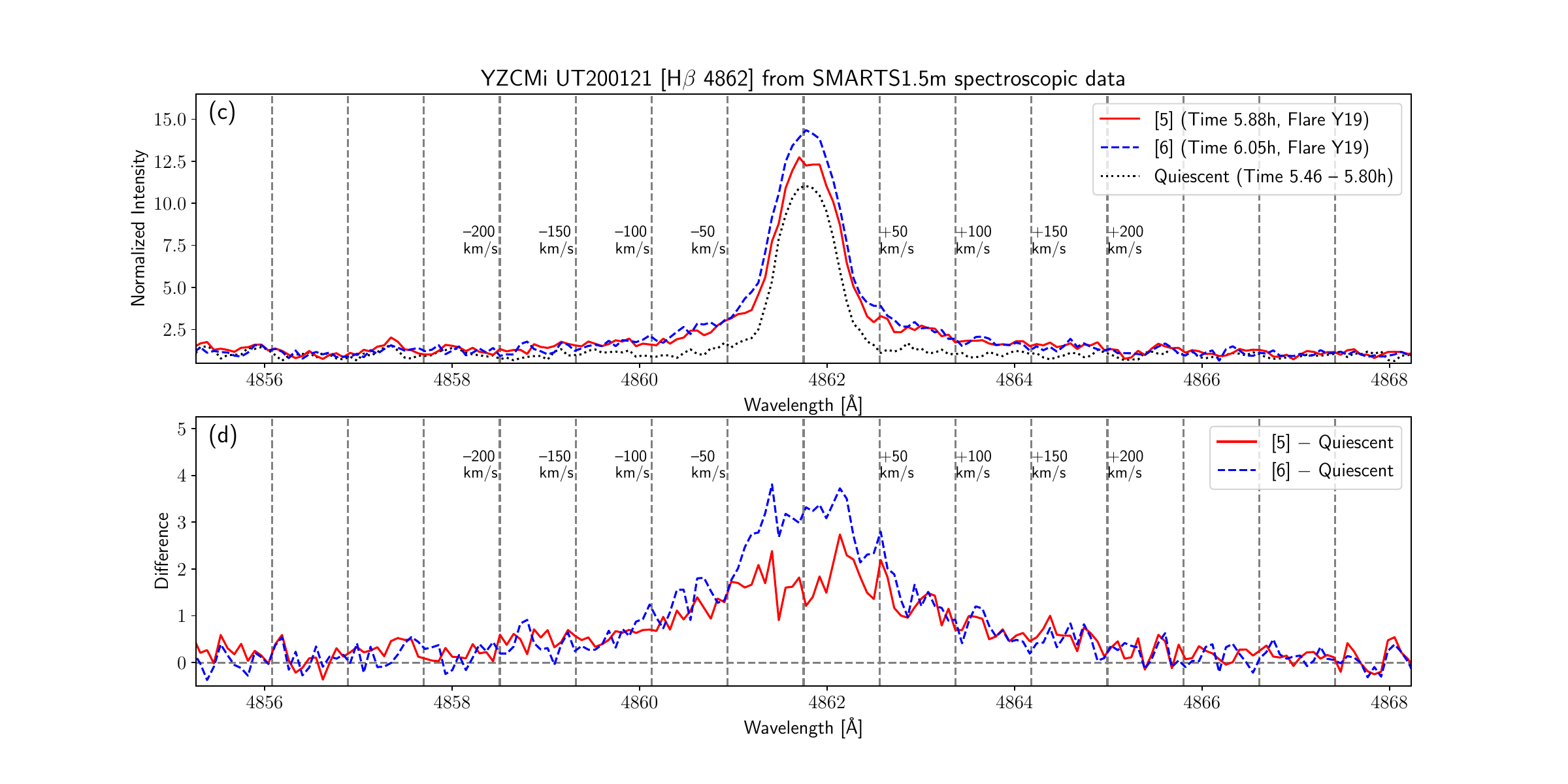}{0.58\textwidth}{\vspace{0mm}}
    }
         \vspace{-1.0cm}
            \gridline{  
     \hspace{-0.06\textwidth}
    \fig{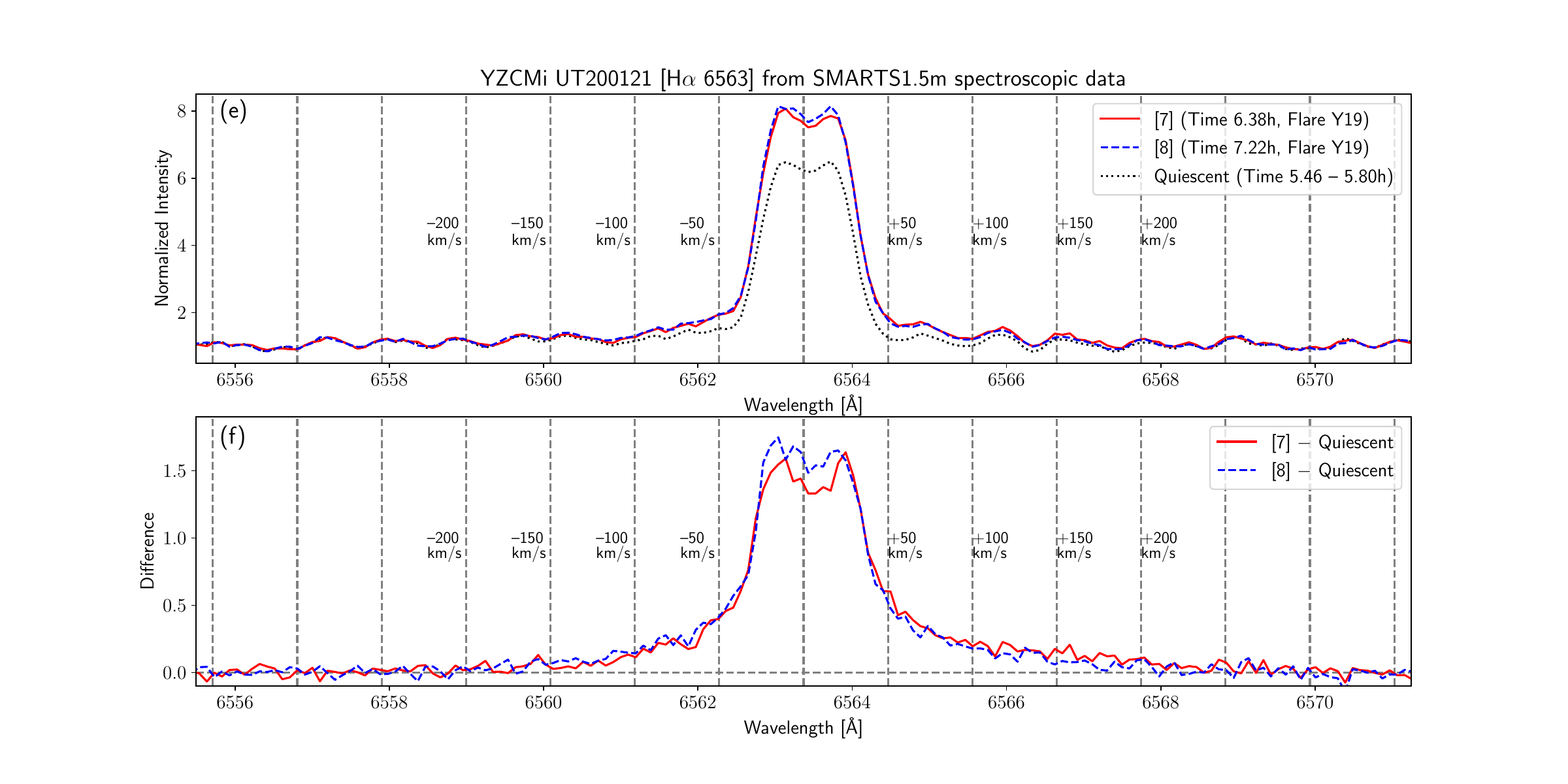}{0.58\textwidth}{\vspace{0mm}}
     \hspace{-0.06\textwidth}
       \fig{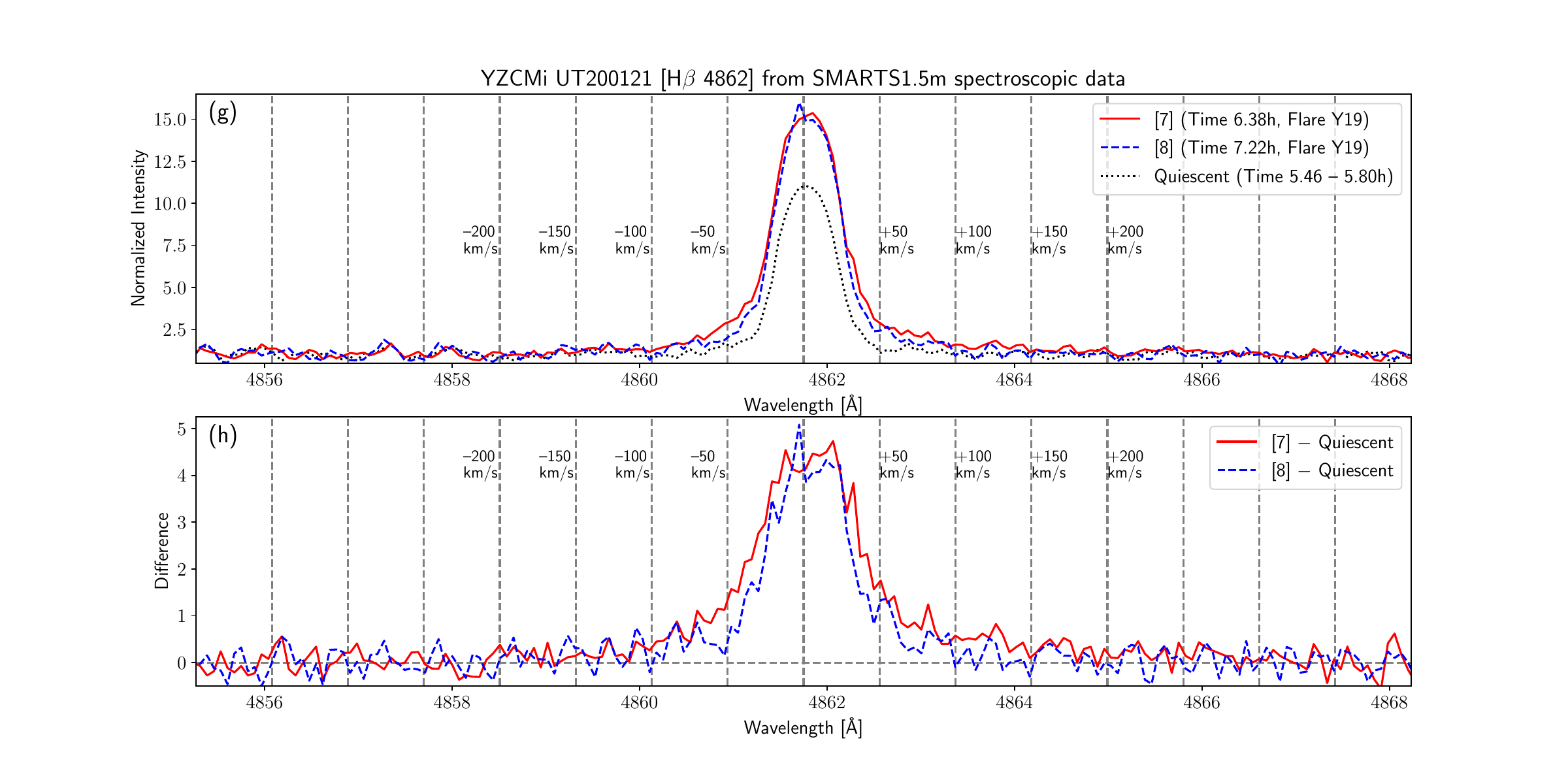}{0.58\textwidth}{\vspace{0mm}}
    }
      \vspace{-1.0cm}
            \gridline{  
     \hspace{-0.06\textwidth}
    \fig{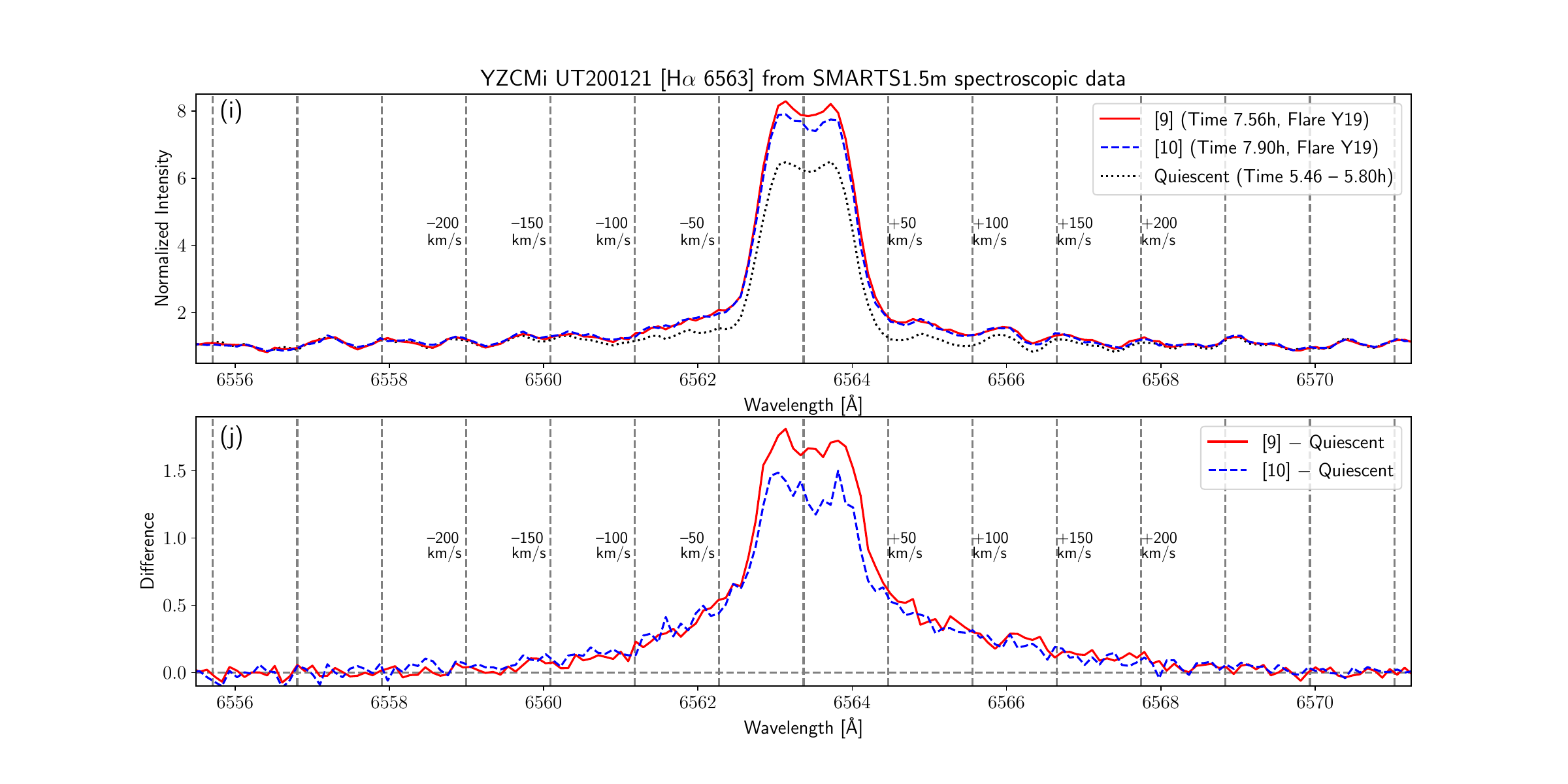}{0.58\textwidth}{\vspace{0mm}}
     \hspace{-0.06\textwidth}
       \fig{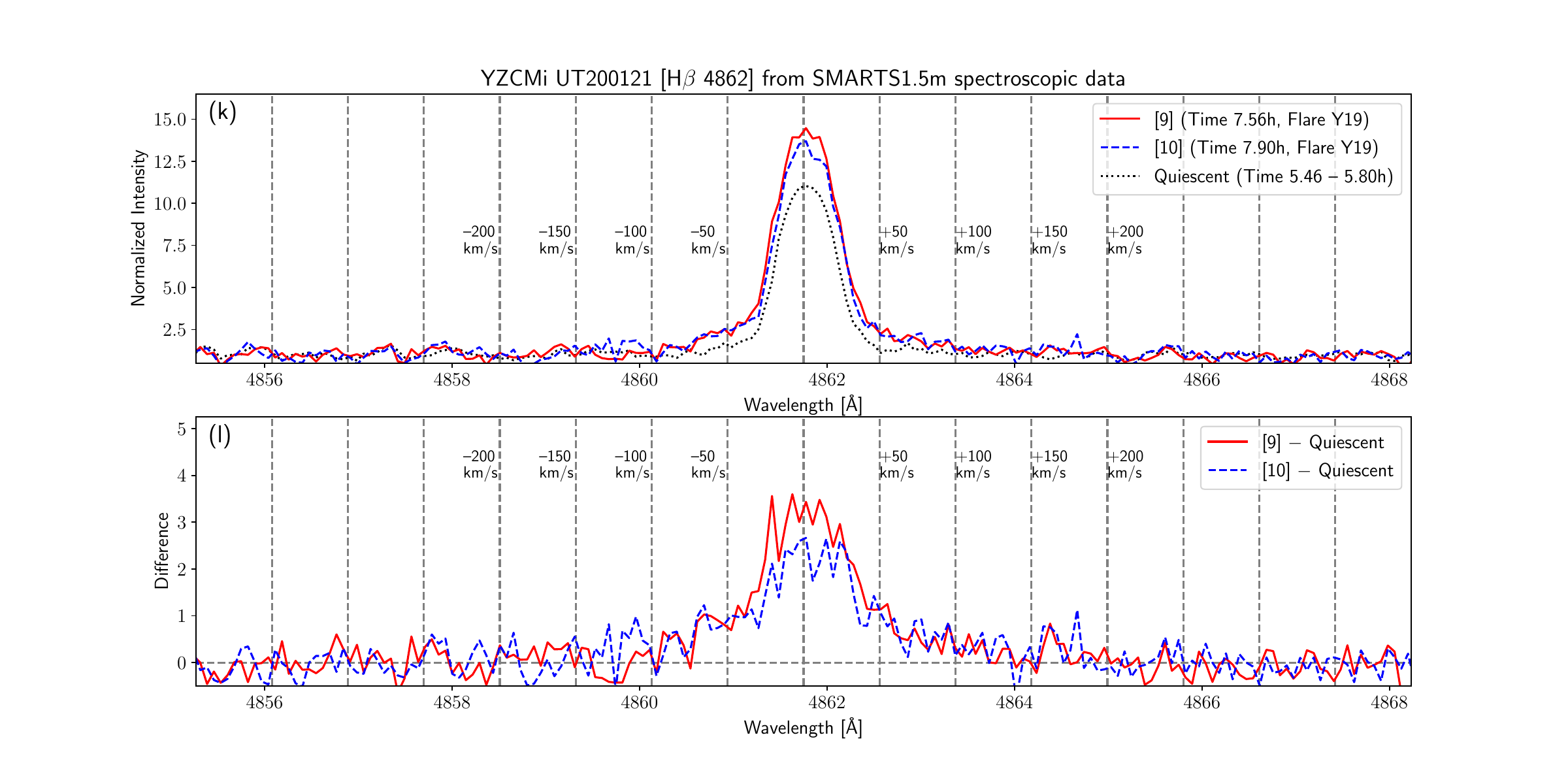}{0.58\textwidth}{\vspace{0mm}}
    }
     \vspace{-1cm}
     \caption{
\color{black}\textrm{ 
Line profiles of the H$\alpha$ \& H$\beta$ emission lines during Flare Y19 (at the time [5]--[10]) on 2020 January 21 from SMARTS 1.5m spectroscopic data, which are plotted similarly with Figure \ref{fig:spec_HaHb_YZCMi_UT190127}.
}
     }
   \label{fig:spec_HaHb_YZCMi_UT200121_Y19}
   \end{center}
 \end{figure}

         \begin{figure}[ht!]
   \begin{center}
      \gridline{
         \hspace{-0.07\textwidth}
      \fig{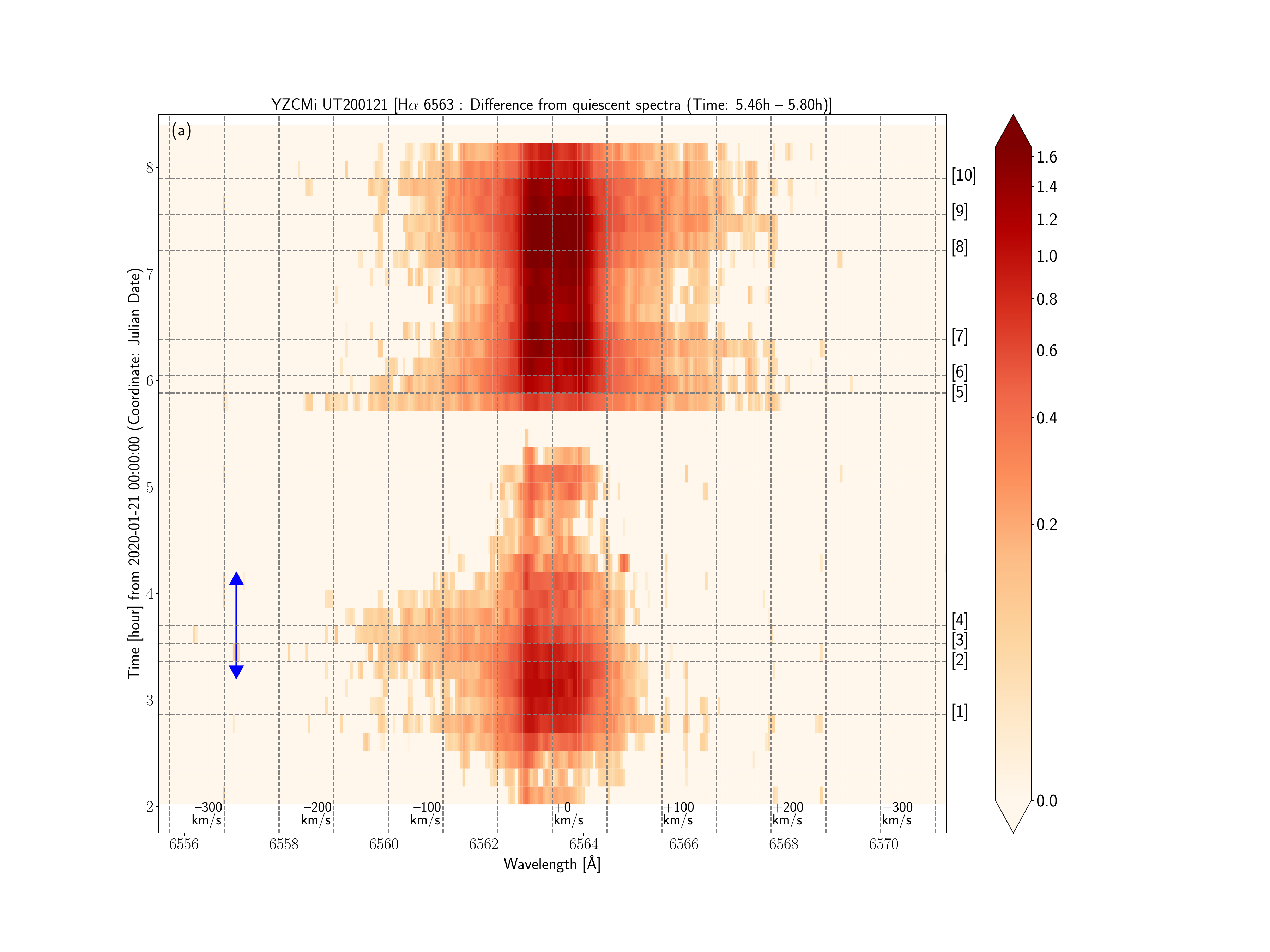}{0.63\textwidth}{\vspace{0mm}}
     \hspace{-0.11\textwidth}
    \fig{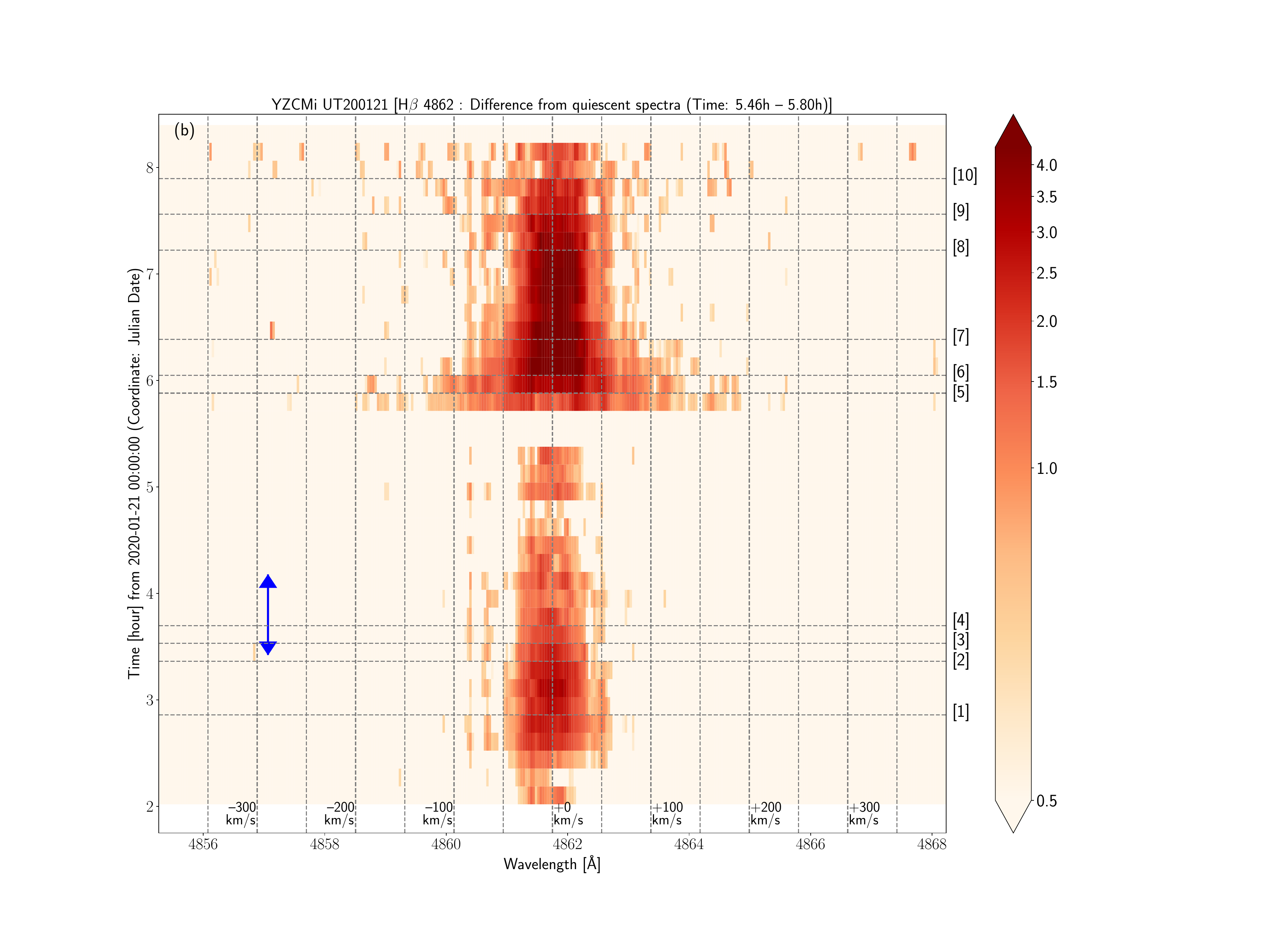}{0.63\textwidth}{\vspace{0mm}}
    }
     \vspace{-1cm}
     \caption{
Time evolution of the H$\alpha$ \& H$\beta$ line profiles 
covering Flares Y18 \& Y19 on 2020 January 21,
which are similarly plotted with Figure \ref{fig:map_HaHb_YZCMi_UT191212} (from the \color{black}\textrm{SMARTS1.5m } spectroscopic data).
The grey horizontal dashed lines indicate the time [1] -- [10], which are shown in Figure \ref{fig:lcEW_HaHb_YZCMi_UT200121} (light curves) and Figures \ref{fig:spec_HaHb_YZCMi_UT200121_Y18} \& \ref{fig:spec_HaHb_YZCMi_UT200121_Y19} 
(line profiles).
     }
   \label{fig:map_HaHb_YZCMi_UT200121}
   \end{center}
 \end{figure}

      \begin{figure}[ht!]
   \begin{center}
            \gridline{  
     \hspace{-0.06\textwidth}
    \fig{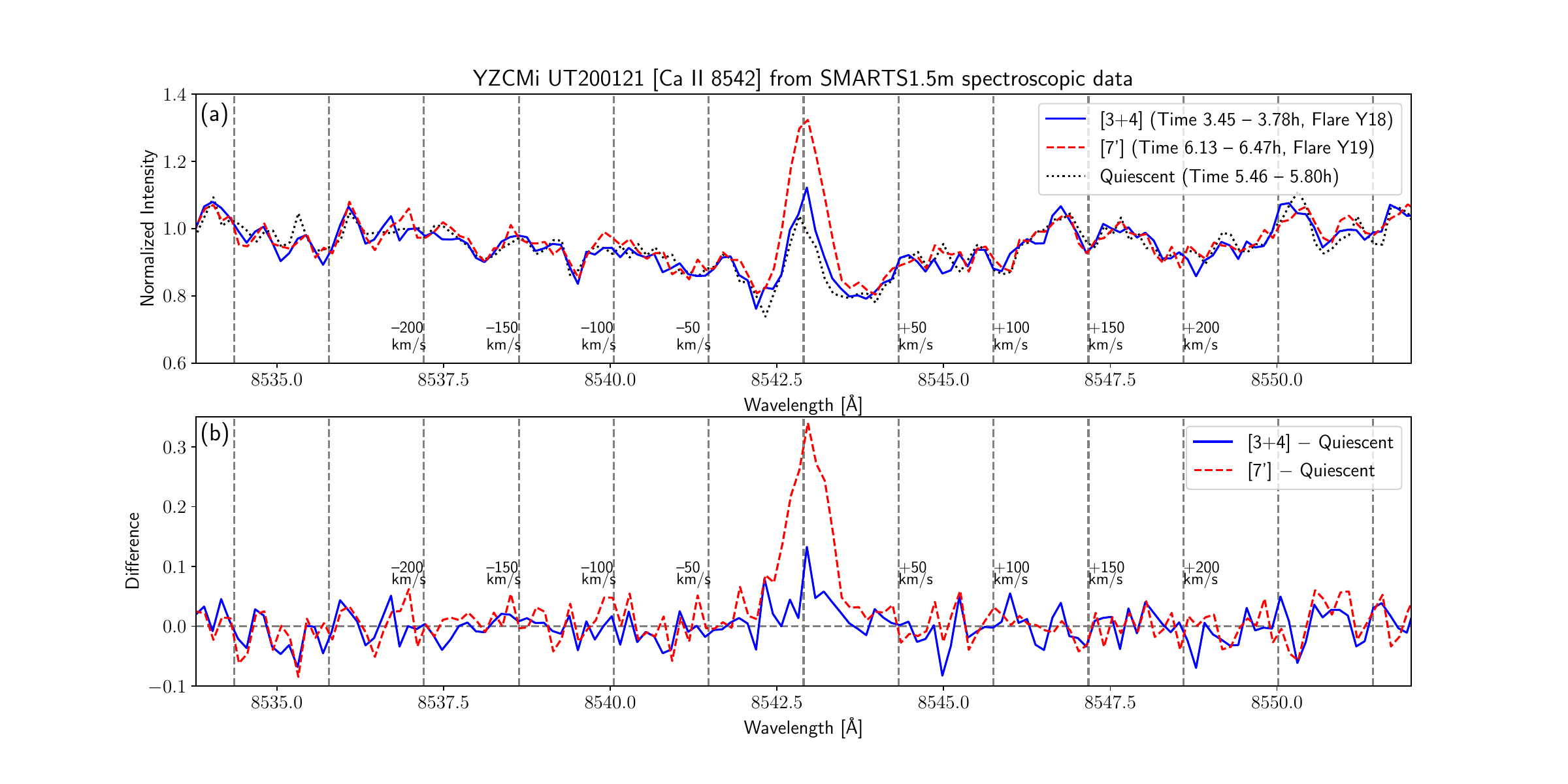}{0.58\textwidth}{\vspace{0mm}}
     \hspace{-0.06\textwidth}
    \fig{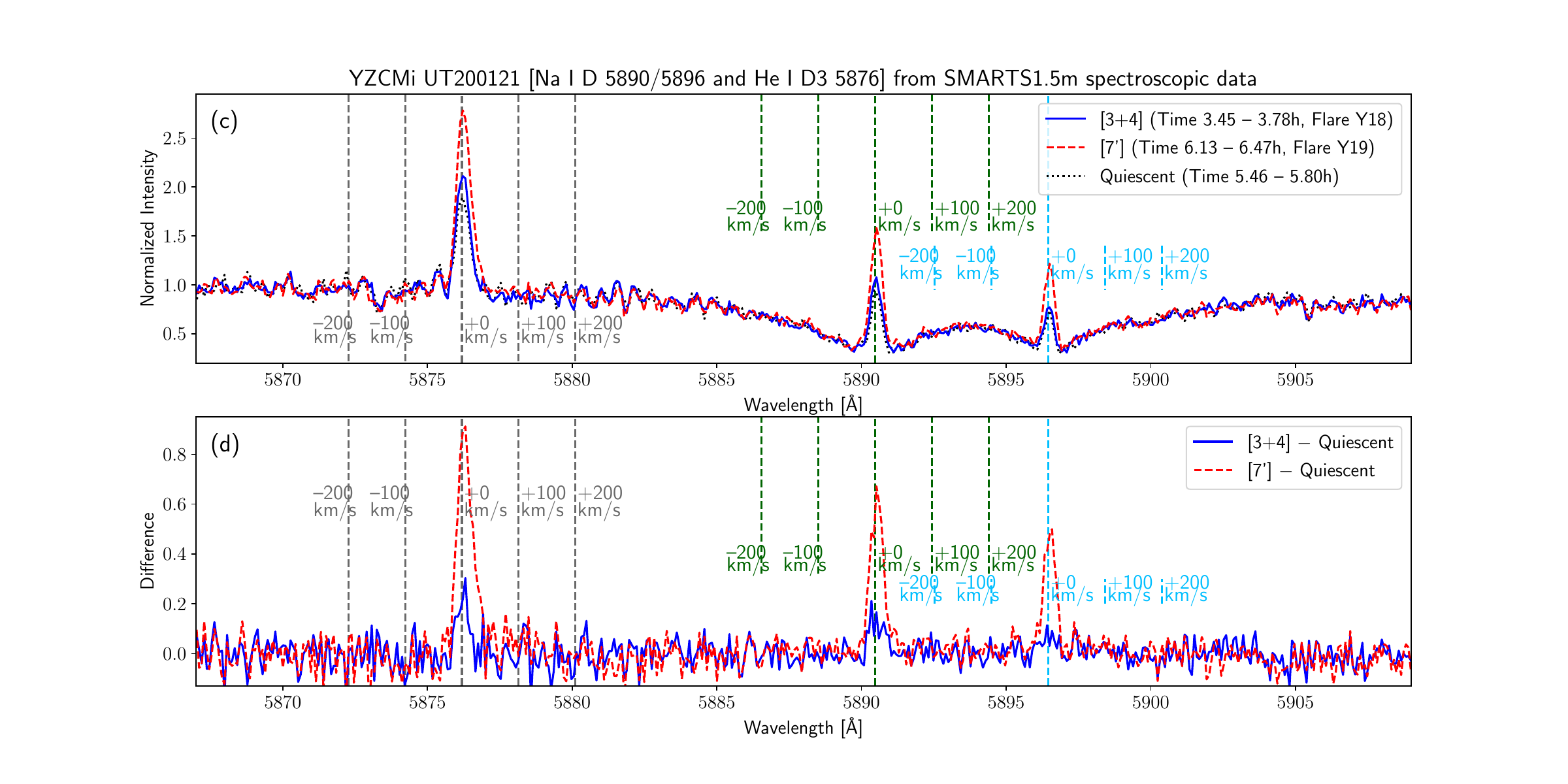}{0.58\textwidth}{\vspace{0mm}} 
    }
     \vspace{-1cm}
     \caption{
\color{black}\textrm{
(a) \& (b)   
Line profiles of the Ca II 8542 line during Flares Y18 \& Y19 on 2020 January 21 from SMARTS1.5m spectroscopic data, similarly plotted with Figures 
\ref{fig:spec_HaHb_YZCMi_UT200121_Y19} \& \ref{fig:spec_HaHb_YZCMi_UT200121_Y19}. 
The blue solid and red dashed lines indicate 
the integrated line profiles over the time [3$+$4] (Time 3.45 -- 3.78h) and [7$^{\prime}$] (Time 6.13 -- 6.47h) on this date, which include the time [3] \& [4] and [7] in Figure \ref{fig:lcEW_HaHb_YZCMi_UT200121} (light curves), respectively.
(c)\&(d) Same as (a)\&(b), but the line profiles for the Na I D1 \& D2 (5890 \& 5896) and He I D3 5876 line.
} \color{black}
}
   \label{fig:spec_Ca8542NaHe_YZCMi_UT200121}
   \end{center}
 \end{figure}

\color{black}\textrm{  
We estimated the flare component peak luminosities and flare energies in $U$- and $V$-bands, and the resultant values ($L_{U}$, $L_{V}$, $E_{U}$, and 
$E_{V}$) are in Table \ref{table:list1_flares}.
The values listed here could be only the lower limit values, 
since the LCO observation has gaps during the both Flares Y18 and Y19,  (Figure \ref{fig:lcEW_HaHb_YZCMi_UT200121} (b)), 
and in the case of Flare Y19, the observation also finished before the flare ended.
The $L_{\rm{H}\alpha}$, $L_{\rm{H}\beta}$, $E_{\rm{H}\alpha}$, and $E_{\rm{H}\beta}$ values are also estimated and listed in Table \ref{table:list1_flares}.
The H$\alpha$ \& H$\beta$ energy values of Flare Y19 listed here are
only the lower limit values, since the observation finished before Flare Y19 ended.
} \color{black}

The H$\alpha$ \& H$\beta$ line profiles during Flares Y18 and Y19 are shown in
Figures \ref{fig:spec_HaHb_YZCMi_UT200121_Y18}, \ref{fig:spec_HaHb_YZCMi_UT200121_Y19},  \& \ref{fig:map_HaHb_YZCMi_UT200121}. 
During Flare Y18, 
the blue wing asymmetry of H$\alpha$ line was detected over about 1 hour around the flare peak (Figure \ref{fig:map_HaHb_YZCMi_UT200121}).
The enhancements of the blue wing of H$\alpha$ line were the largest at around the beginning of the decay phase (time [3]\&[4] in Figures \ref{fig:lcEW_HaHb_YZCMi_UT200121} (a), \ref{fig:spec_HaHb_YZCMi_UT200121_Y18}, \& \ref{fig:map_HaHb_YZCMi_UT200121}) and 
we can see the enhancements up to $\sim$ -200 km s$^{-1}$ then. 
These blue wing enhancements up to $\sim$ -150 km s$^{-1}$ were also possibly seen in H$\beta$ line (time [3]\&[4] in Figure \ref{fig:spec_HaHb_YZCMi_UT200121_Y18}(h)), though the signal-to-noise ratio of the data is relatively low.
During Flare Y19, 
the H$\alpha$ line showed the line wing broadenings ($\sim \pm $150 -- 200 km s$^{-1}$) twice during flares: one at around the time [5]--[7] and the other at around the time [8]--[10] (Figures \ref{fig:spec_HaHb_YZCMi_UT200121_Y19} \& \ref{fig:map_HaHb_YZCMi_UT200121}).
During these broadenings, red wing of the H$\alpha$ line was slightly enhanced.
The H$\beta$ line showed the similar line wing broadening ($\sim \pm $150 -- 200 km s$^{-1}$) at around the time [5]--[7], but the wing broadening in H$\beta$ line was not seen in later phase at around the time [8]--[10] (Figures \ref{fig:spec_HaHb_YZCMi_UT200121_Y19} \& \ref{fig:map_HaHb_YZCMi_UT200121}).

The EW light curves of Ca II 8542, Na I D1 \& D2, and He I D3 5876 lines are also shown in Figure \ref{fig:lcEW_HaHb_YZCMi_UT200121} (c).
The profiles of these lines during Flare Y19 
are shown in Figure \ref{fig:spec_Ca8542NaHe_YZCMi_UT200121}.
Line asymmetries were not clearly seen in these lines.

\clearpage

\subsection{Flares Y23 (Blue wing asymmetry) \& Y24 observed on 2020 December 6} 
\label{subsec:results:2020-Dec-06}

        \begin{figure}[ht!]
   \begin{center}
   \gridline{
    \fig{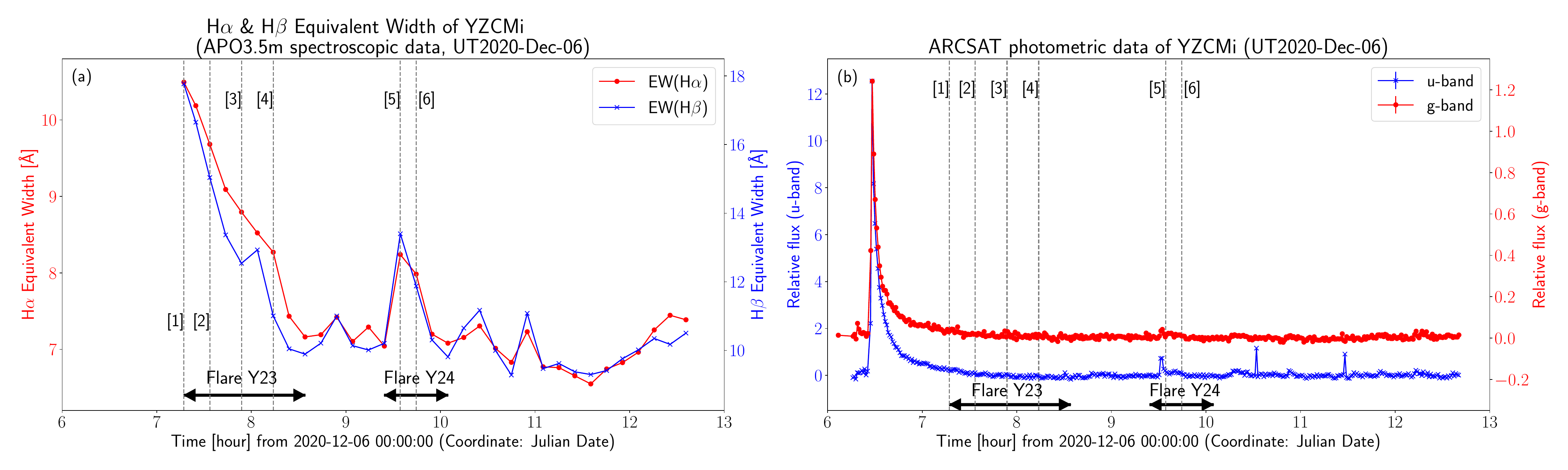}{1.0\textwidth}{\vspace{0mm}}}
     \vspace{-1cm}
              \gridline{  
     \hspace{-0.02\textwidth}
    \fig{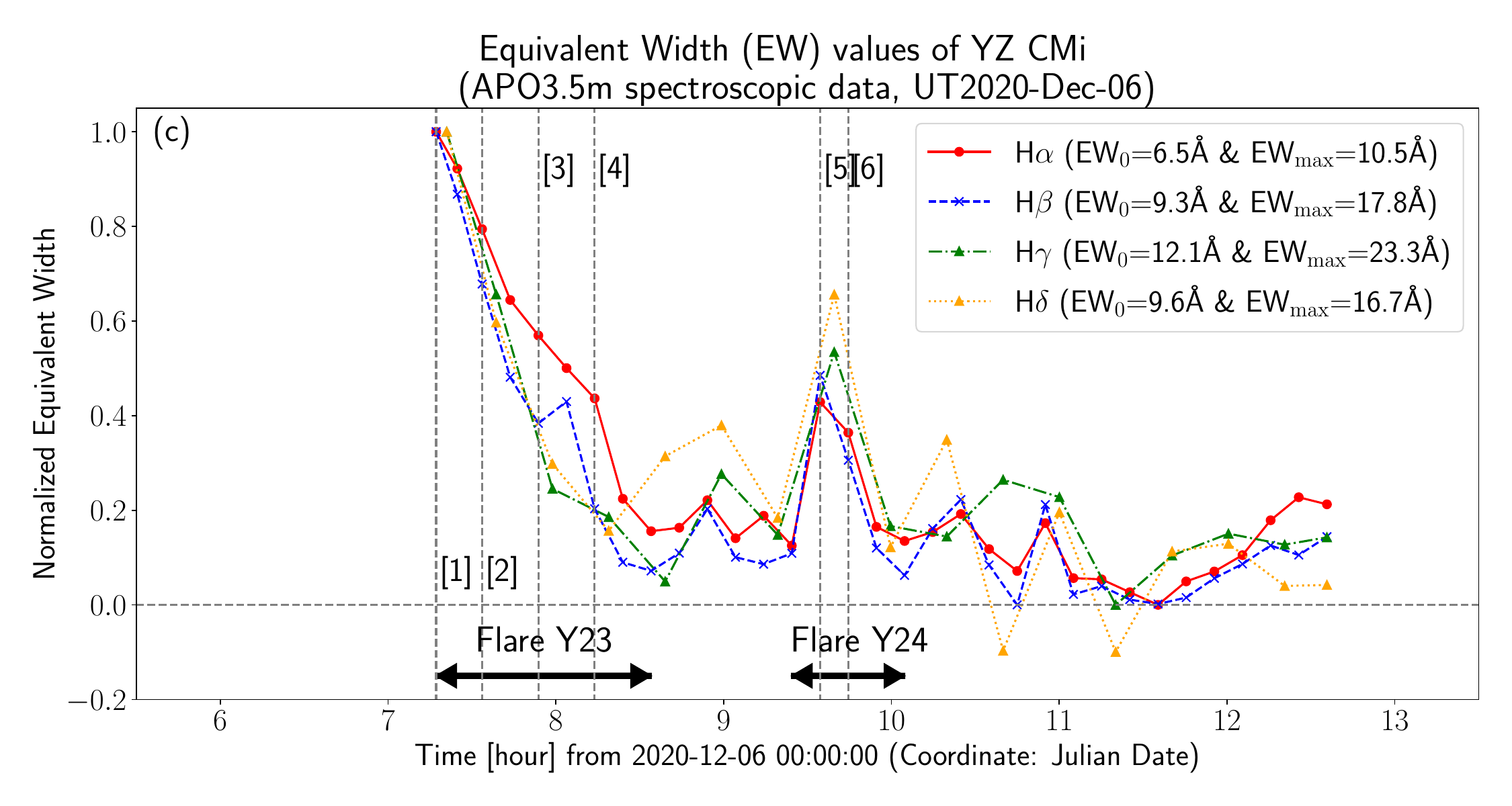}{0.5\textwidth}{\vspace{0mm}}
     \hspace{-0.02\textwidth}
    \fig{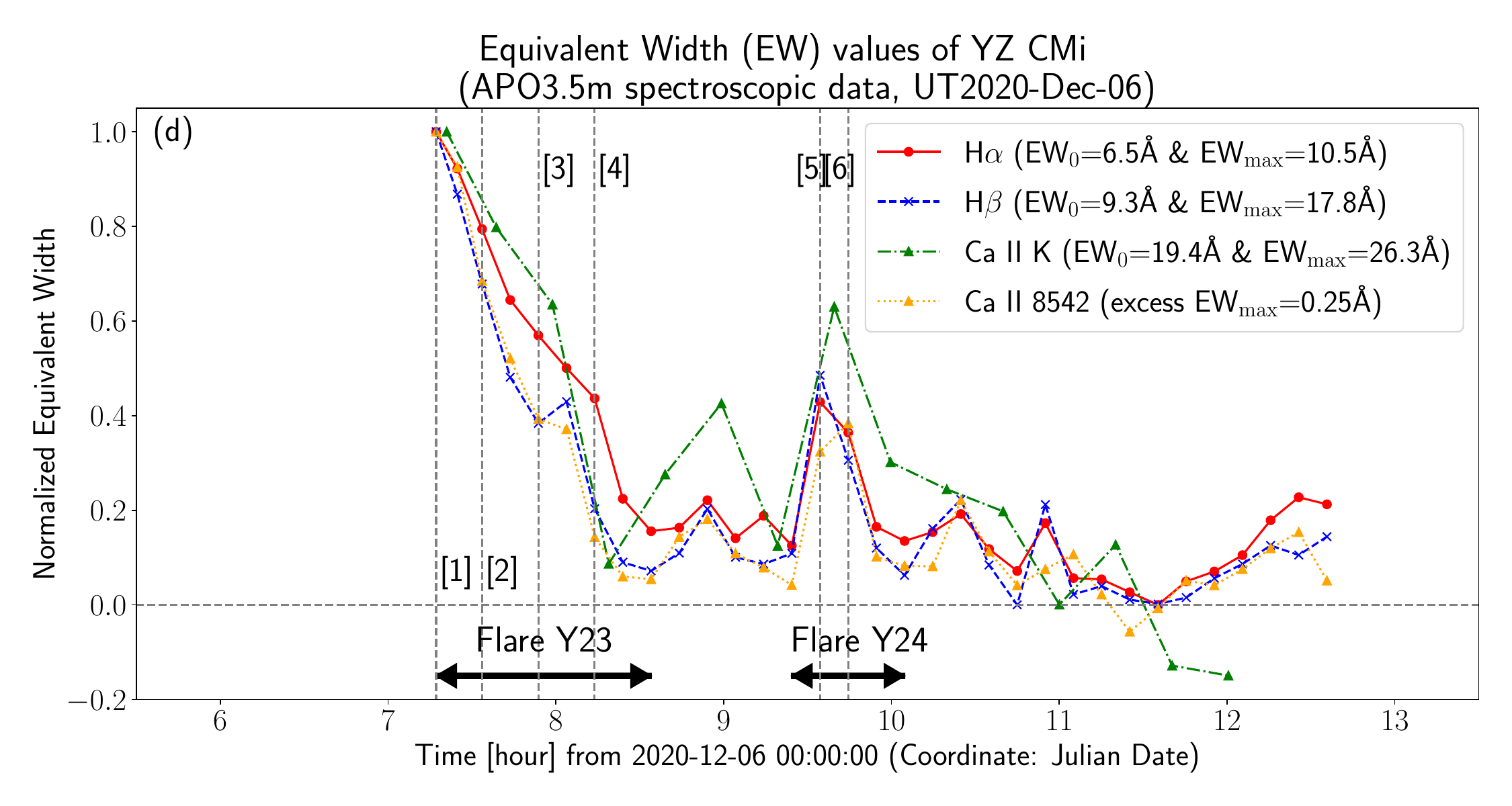}{0.5\textwidth}{\vspace{0mm}}
    }
     \vspace{-1cm}
              \gridline{  
     \hspace{-0.02\textwidth}
    \fig{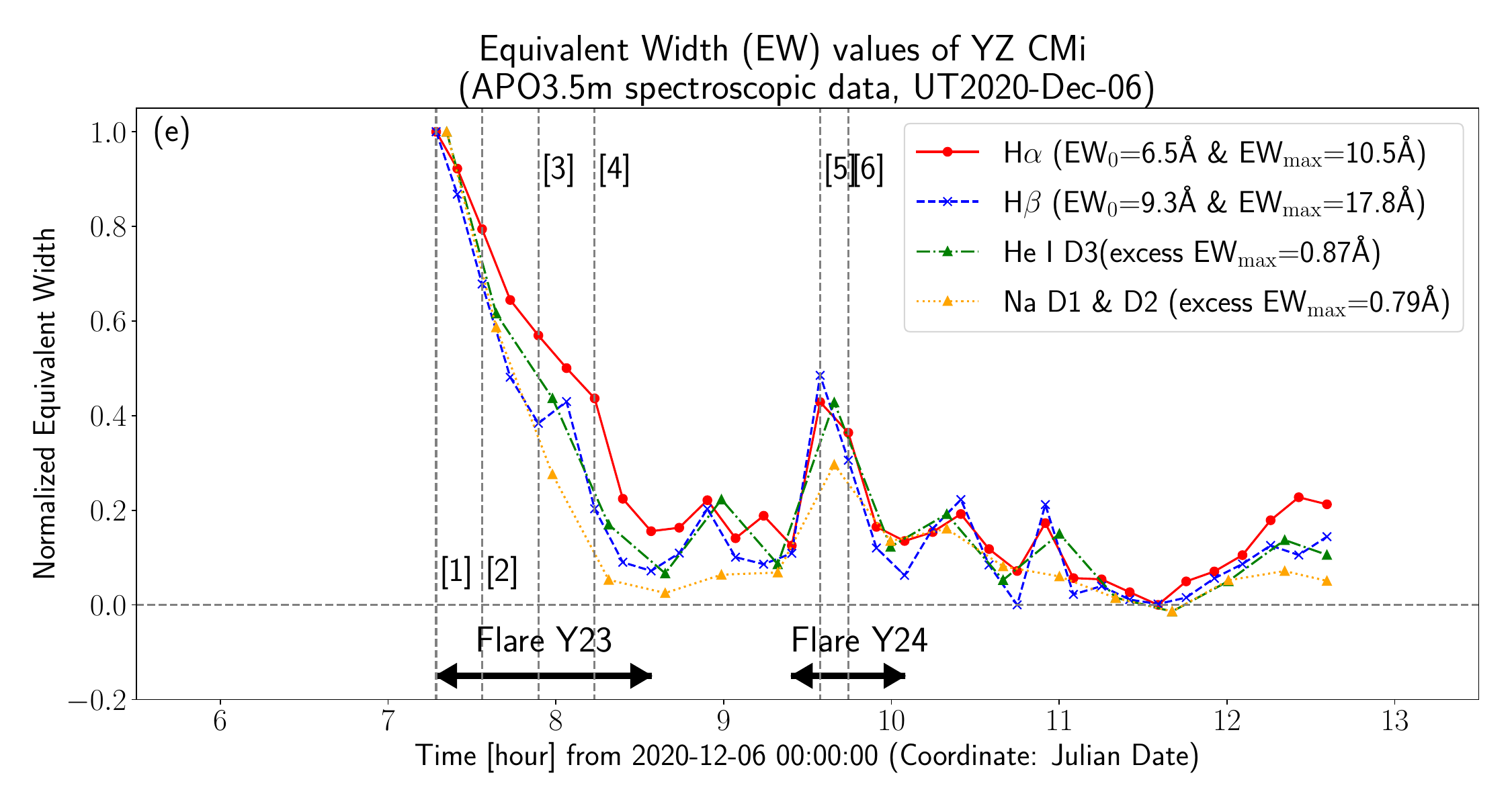}{0.5\textwidth}{\vspace{0mm}}
    }
     \vspace{-0.8cm}
     \caption{
     \color{black}\textrm{  
Light curves of YZ CMi on 2020 December 6 showing Flares Y23 \& Y24, which are plotted 
similarly with Figure \ref{fig:lcEW_HaHb_YZCMi_UT191212}.
The grey dashed lines with numbers ([1]--[6]) correspond to the time shown with the same numbers in Figures \ref{fig:spec_HaHb_YZCMi_UT201206} \& \ref{fig:map_HaHb_YZCMi_UT201206}.
 } \color{black}
    }
   \label{fig:lcEW_HaHb_YZCMi_UT201206}
   \end{center}
 \end{figure}

              \begin{figure}[ht!]
   \begin{center}
            \gridline{  
     \hspace{-0.06\textwidth}
    \fig{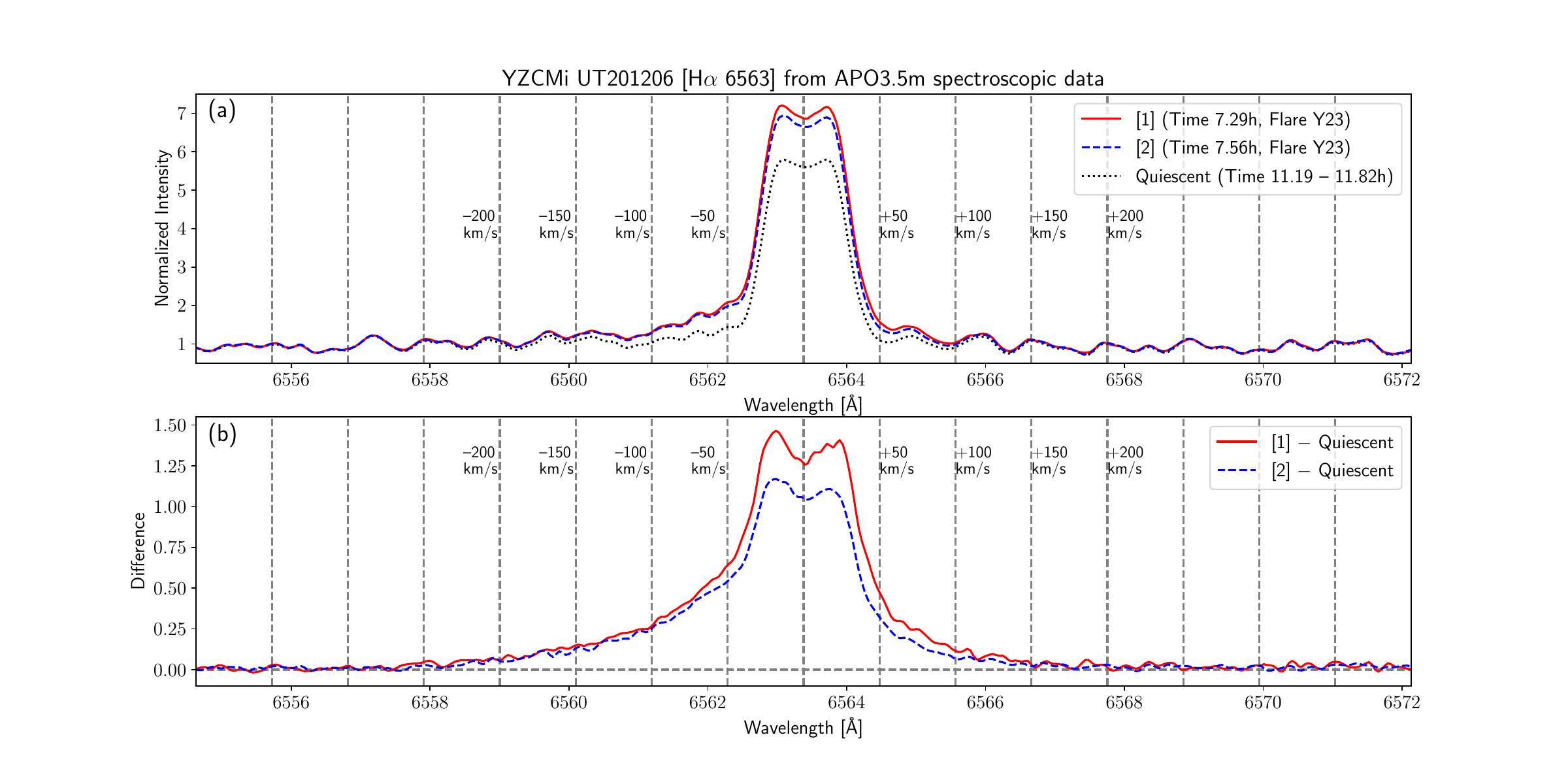}{0.58\textwidth}{\vspace{0mm}}
     \hspace{-0.06\textwidth}
       \fig{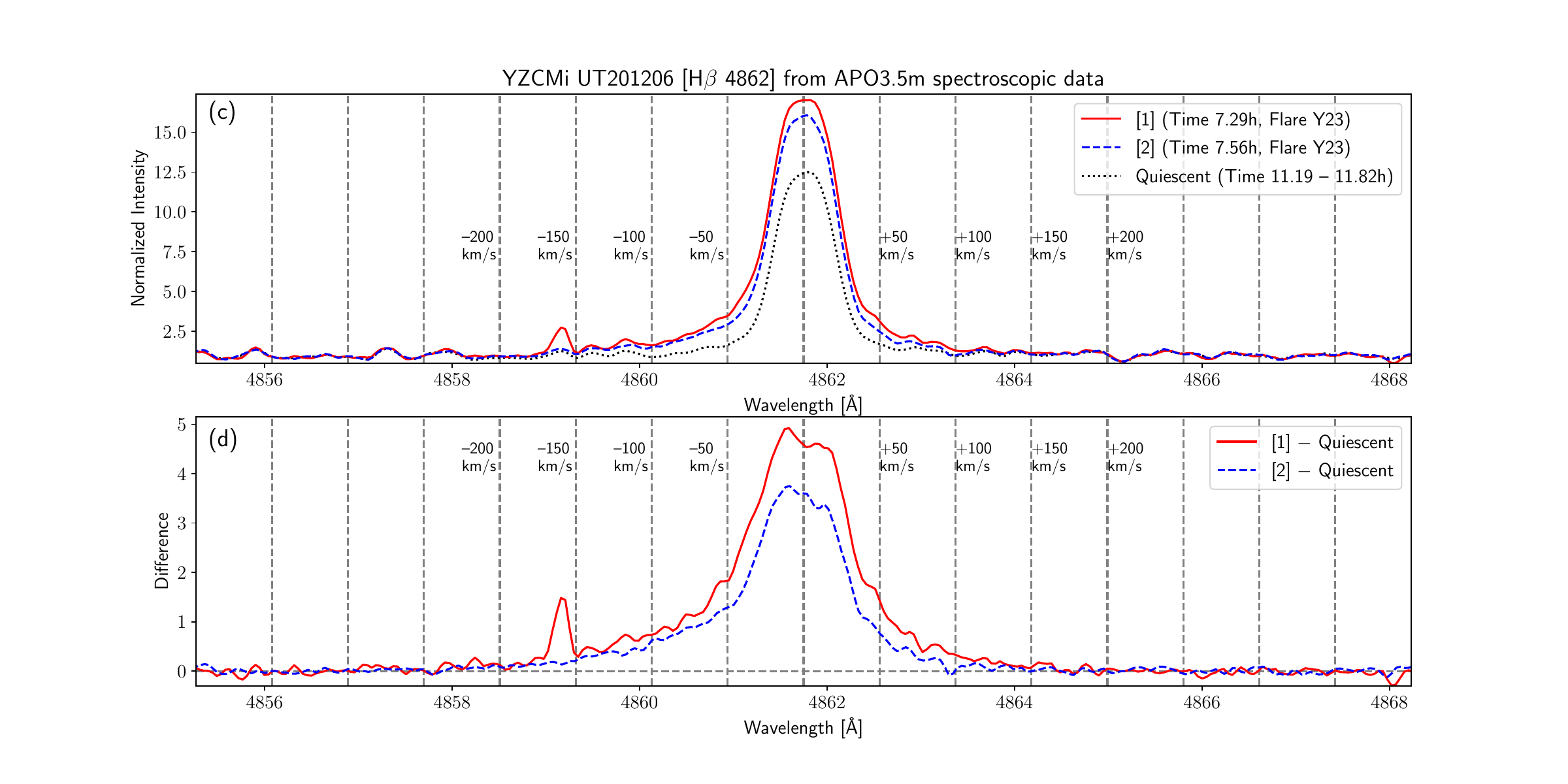}{0.58\textwidth}{\vspace{0mm}}
    }
     \vspace{-1.0cm}
            \gridline{  
     \hspace{-0.06\textwidth}
    \fig{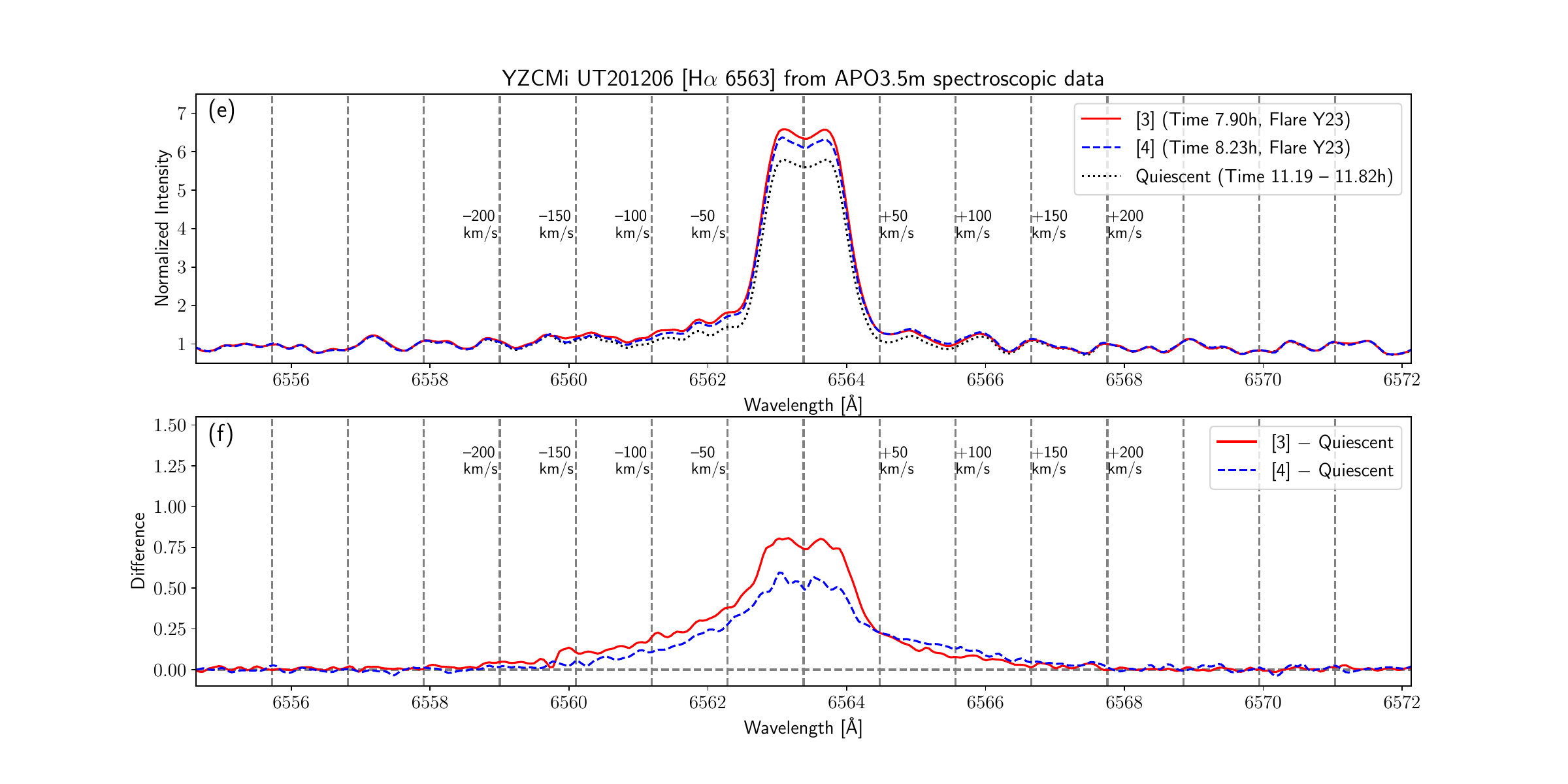}{0.58\textwidth}{\vspace{0mm}}
     \hspace{-0.06\textwidth}
       \fig{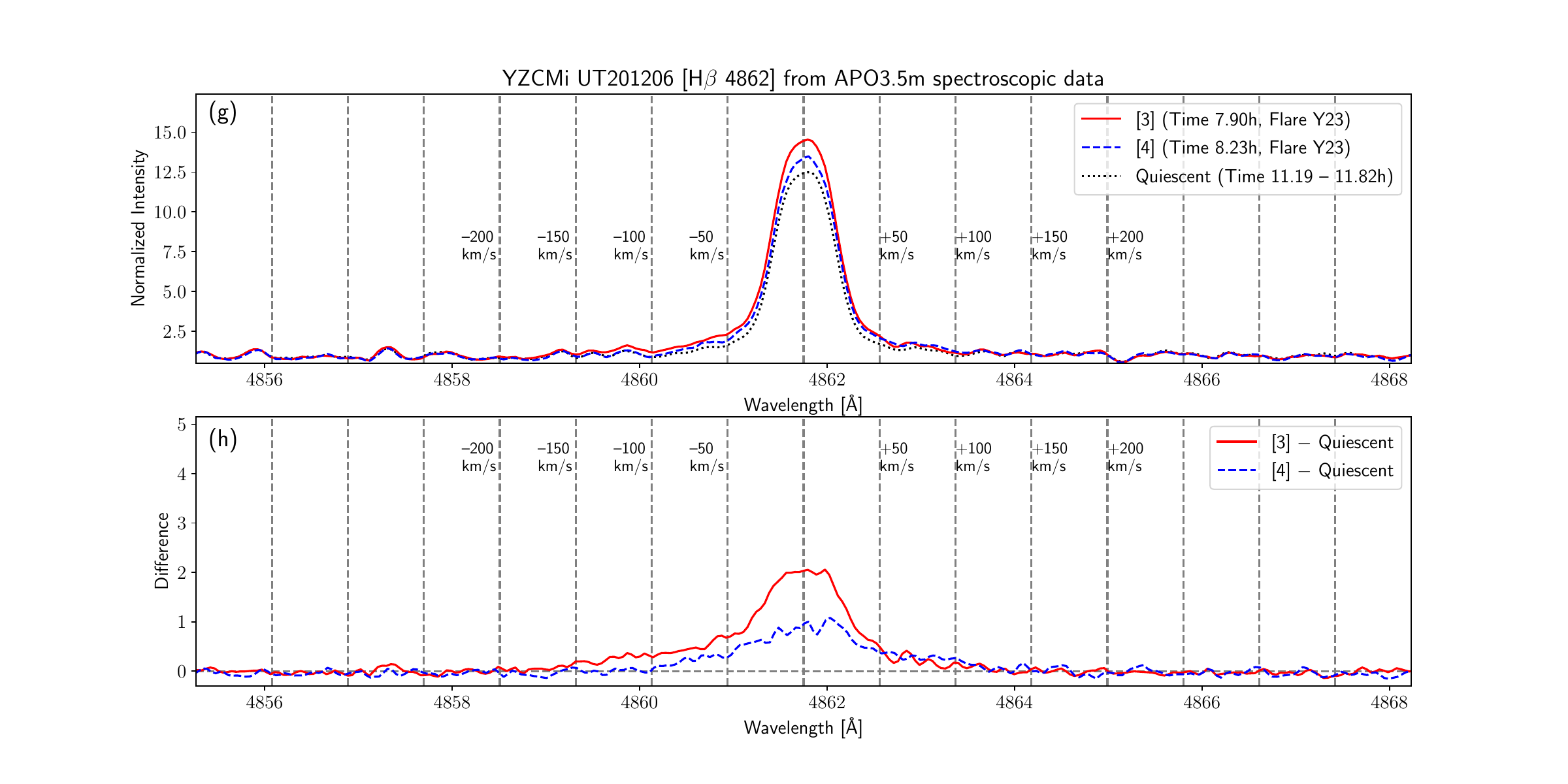}{0.58\textwidth}{\vspace{0mm}}
    }
     \vspace{-1.0cm}
            \gridline{  
     \hspace{-0.06\textwidth}
    \fig{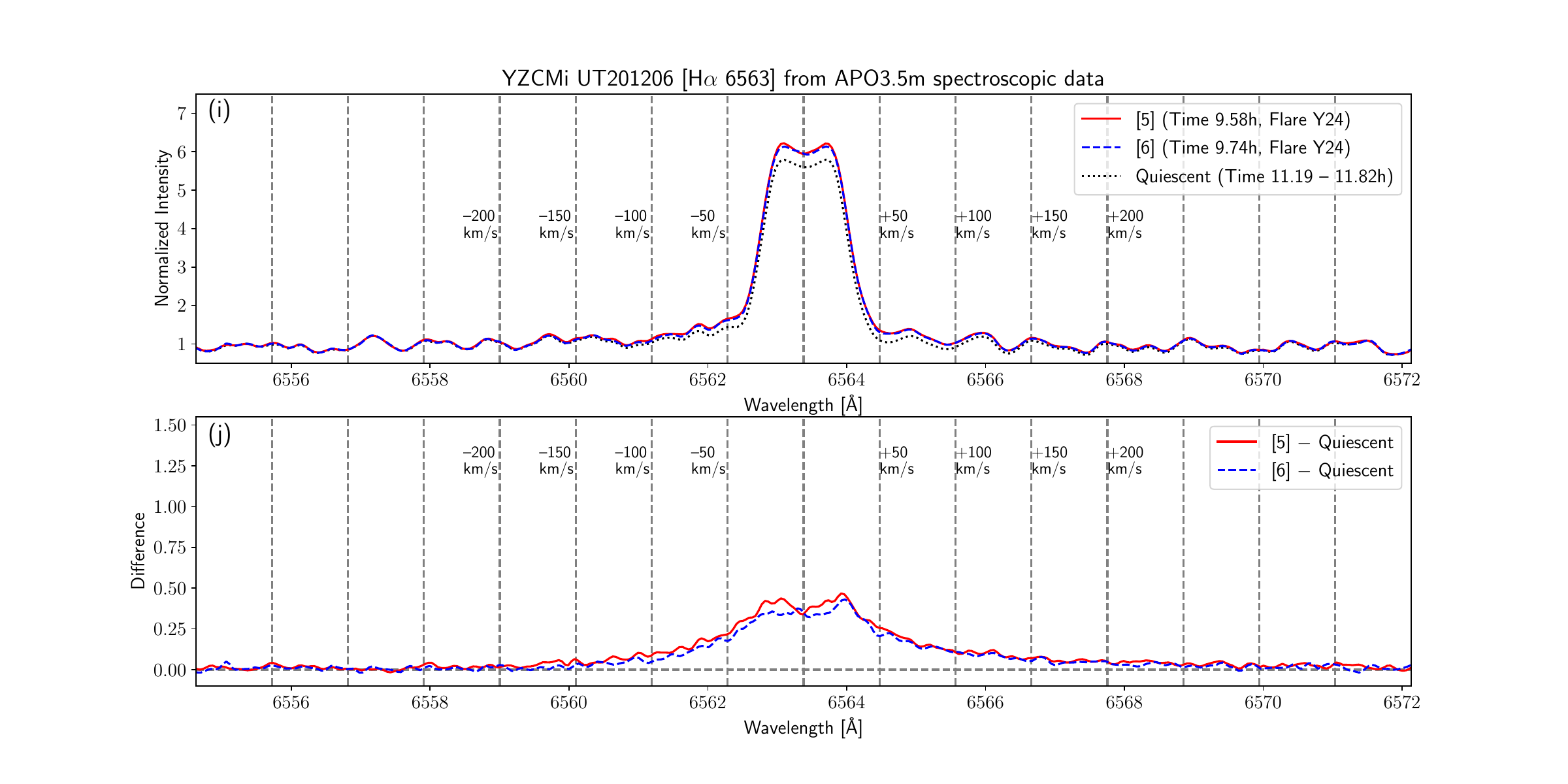}{0.58\textwidth}{\vspace{0mm}}
     \hspace{-0.06\textwidth}
       \fig{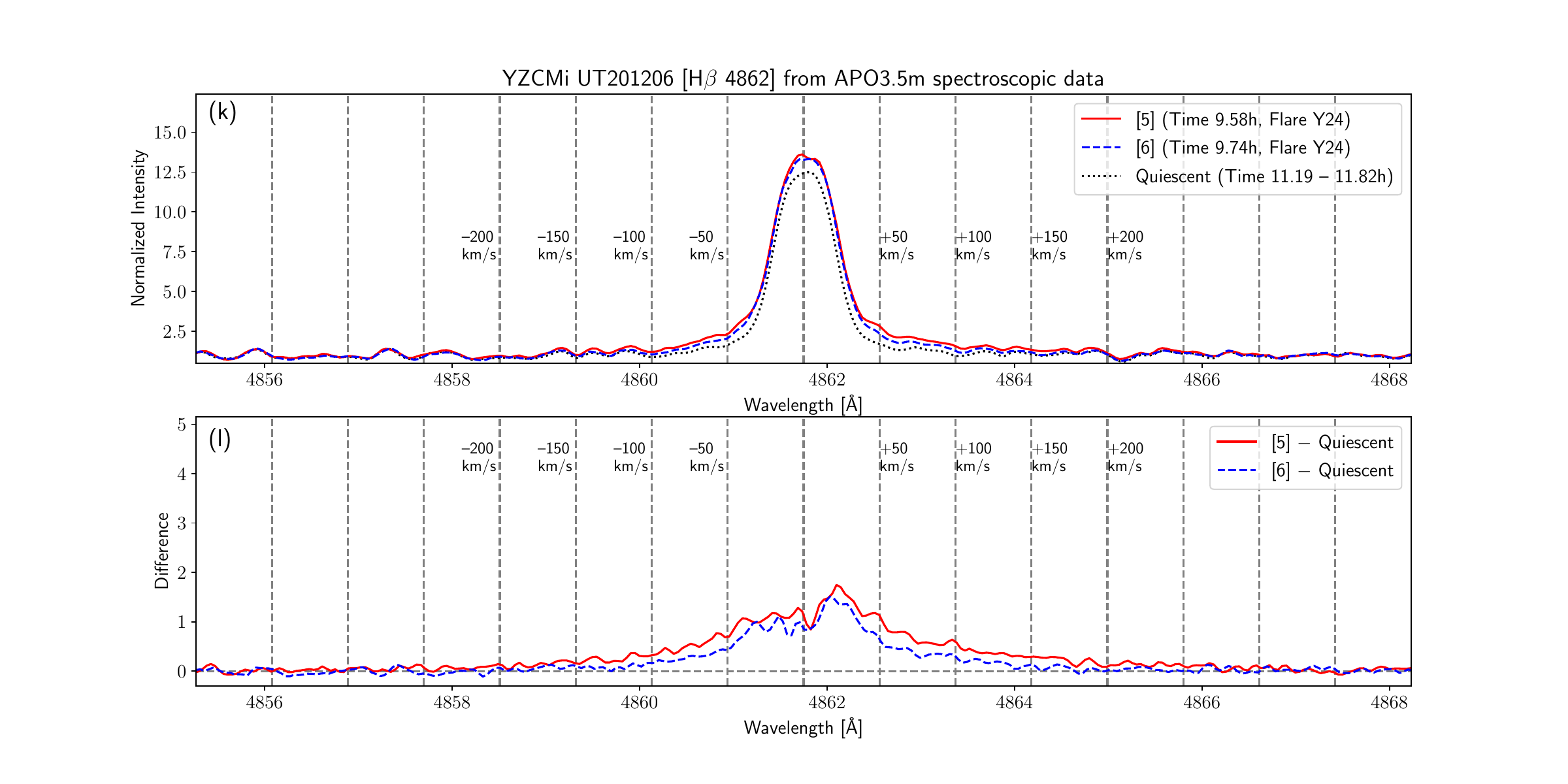}{0.58\textwidth}{\vspace{0mm}}
    }
     \vspace{-0.8cm}
     \caption{
     \color{black}\textrm{  
Line profiles of the H$\alpha$ \& H$\beta$ emission lines during Flares Y23 and Y24  (at the time [1]-[6]) on 2020 December 6 from APO3.5m spectroscopic data, which are plotted similarly with Figure \ref{fig:spec_HaHb_YZCMi_UT190127}.
 } \color{black}
     }
   \label{fig:spec_HaHb_YZCMi_UT201206}
   \end{center}
 \end{figure}

On 2020 December 6, two flares (Flares Y23 \& Y24) 
were detected in H$\alpha$ \& H$\beta$ lines as shown in Figure \ref{fig:lcEW_HaHb_YZCMi_UT201206} (a).  
Flare Y23 already started when the spectroscopic observation
started.
The H$\alpha$ \& H$\beta$ equivalent widths decreased from 10.5\AA~and 17.8\AA, respectively, and $\Delta t^{\rm{flare}}_{\rm{H}\alpha}$ is $>$1.3 hours (Table \ref{table:list1_flares}).
The photometric observation captured early phase of the flare since it started $\sim$1 hour before the spectroscopic observation started. 
During Flare Y23, the continuum brightness observed with ARCSAT $u$- \& $g$-bands increased by $\sim$1260\% and $\sim$125\%, respectively (Figure \ref{fig:lcEW_HaHb_YZCMi_UT201206} (b)). 
As for Flare Y24, the H$\alpha$ \& H$\beta$ equivalent widths increased to 8.2\AA~and 13.4\AA, respectively, and $\Delta t^{\rm{flare}}_{\rm{H}\alpha}$ is 0.7 hours (Table \ref{table:list1_flares}).
In addition to these enhancements in Balmer emission lines, 
the continuum brightness observed with ARCSAT 
$u$- \& $g$-bands increased by $\sim$ 70 -- 75\% and $\sim$4\%,  
respectively, during Flare Y24 (Figure \ref{fig:lcEW_HaHb_YZCMi_UT201206} (b)).

 \color{black}\textrm{ 
We estimated 
$L_{u}$, $L_{g}$, $E_{u}$, $E_{g}$, $L_{\rm{H}\alpha}$, $L_{\rm{H}\beta}$, $E_{\rm{H}\alpha}$, and $E_{\rm{H}\beta}$ values, and they are listed in Table \ref{table:list1_flares}.
Since the initial phase of Flare Y23 was not observed in the spectroscopic observation, the 
 H$\alpha$ \&  H$\beta$ luminosities and flare energies of Flare Y23 estimated here are only
 lower limit values.
} \color{black}
  
\begin{figure}[ht!]
   \begin{center}
      \gridline{
     \hspace{-0.07\textwidth}
      \fig{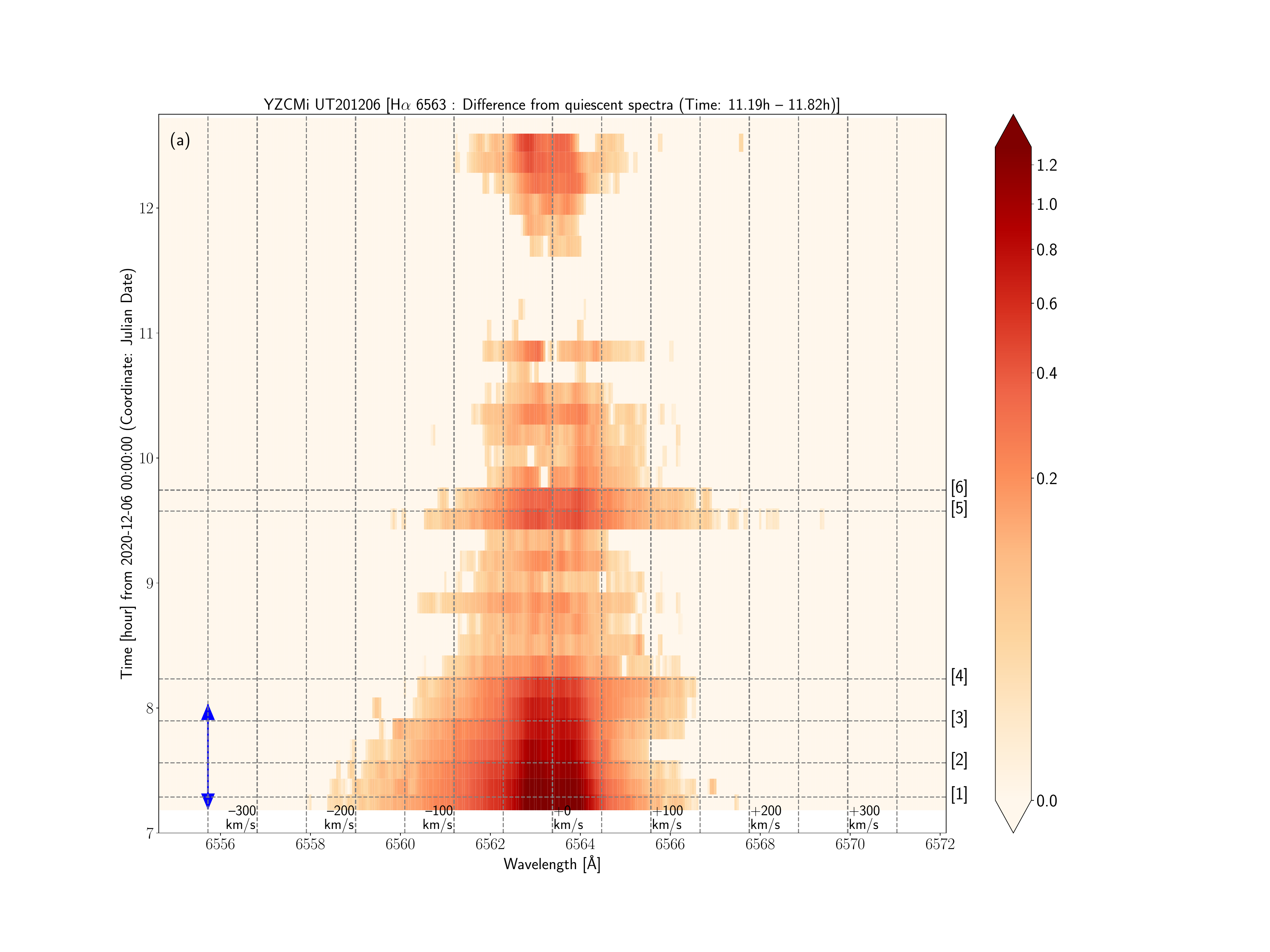}{0.63\textwidth}{\vspace{0mm}}
     \hspace{-0.11\textwidth}
    \fig{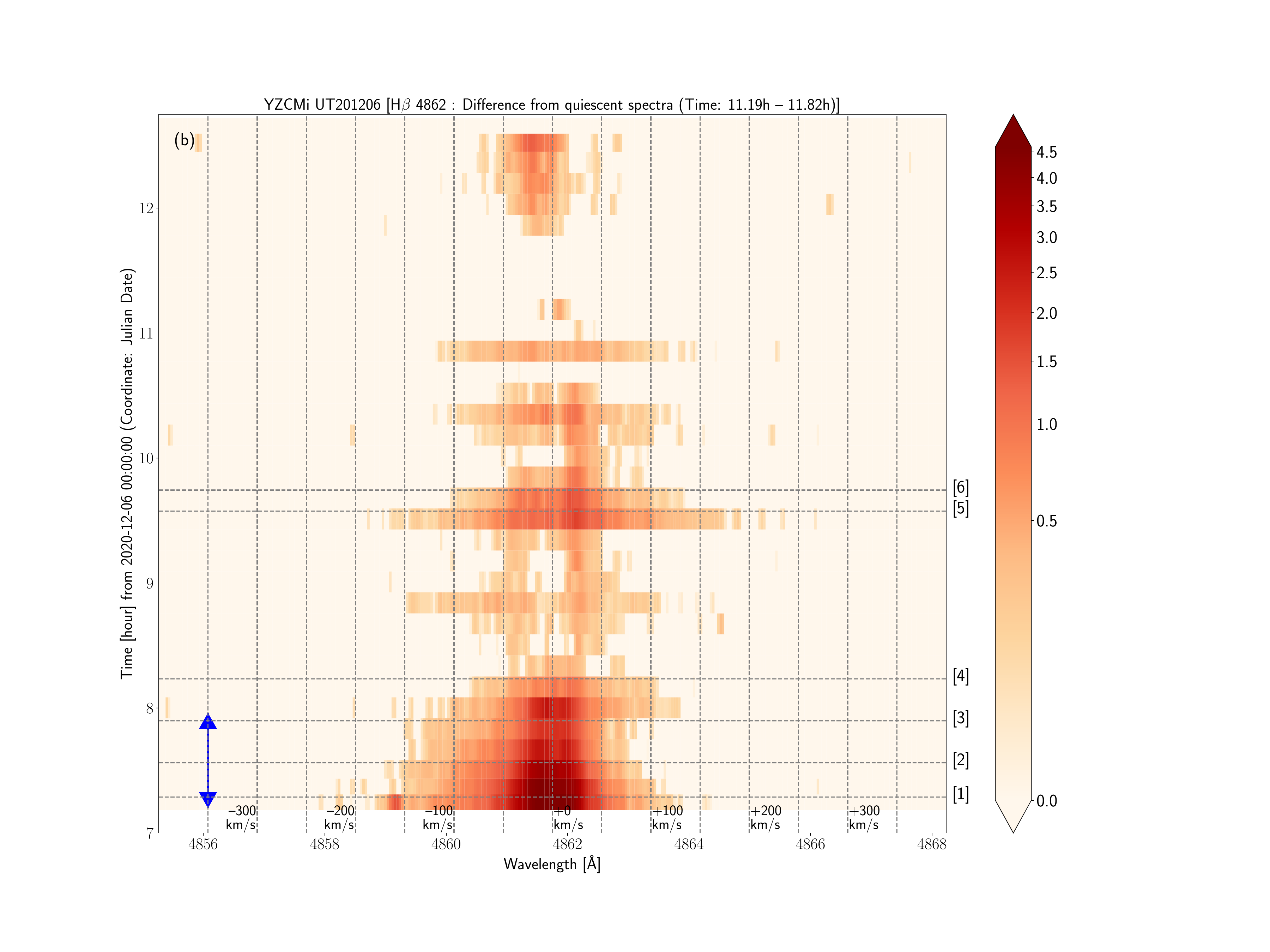}{0.63\textwidth}{\vspace{0mm}}
    }
     \vspace{-0cm}
     \caption{
          \color{black}\textrm{  
Time evolution of the H$\alpha$ \& H$\beta$ line profiles covering Flares Y23 \& Y24 on 2020 December 6, which are shown similarly with Figure \ref{fig:map_HaHb_YZCMi_UT191212}.
The grey horizontal dashed lines indicate the time [1] -- [6], which are shown in Figure \ref{fig:lcEW_HaHb_YZCMi_UT201206} (light curves) and Figure \ref{fig:spec_HaHb_YZCMi_UT201206} (line profiles).
}\color{black}
     }
   \label{fig:map_HaHb_YZCMi_UT201206}
   \end{center}
 \end{figure}

The H$\alpha$ \& H$\beta$ line profiles during Flares Y23 and Y24 are shown in
Figures \ref{fig:spec_HaHb_YZCMi_UT201206} \& \ref{fig:map_HaHb_YZCMi_UT201206}. 
The blue wing of H$\alpha$ line was
enhanced during Flare Y23 (time [1]--[4] in 
Figures \ref{fig:spec_HaHb_YZCMi_UT201206} (b), (f), \& \ref{fig:map_HaHb_YZCMi_UT201206}(a)).
This blue wing asymmetry was the largest at time [1] (up to $\sim$ -250 km s$^{-1}$) 
and continued until around the end of the flare (time [4]),
while the velocity of blue wing enhancement decayed gradually.
The similar time evolution with the blue wing asymmetry (up to $\sim$ -200 km s$^{-1}$) was seen also in the H$\beta$ line (time [1]--[4] in 
Figures \ref{fig:spec_HaHb_YZCMi_UT201206} (d), (h), \& \ref{fig:map_HaHb_YZCMi_UT201206}(b)). 
but the line wing asymmetries at around the time [1] and [2] were not as clear compared to those of H$\alpha$ line (Figures \ref{fig:spec_HaHb_YZCMi_UT201206}(b), \& \ref{fig:map_HaHb_YZCMi_UT201206}(b)). 
During Flare Y24, 
there were no clear blue or red wing asymmetries in H$\alpha$ and H$\beta$ lines 
(time [5]\&[6] in Figures \ref{fig:spec_HaHb_YZCMi_UT201206} \& \ref{fig:map_HaHb_YZCMi_UT201206}), 
and the line profiles showed
roughly symmetrical broadenings with $\sim \pm$ 100 -- 200 km s$^{-1}$ at around the peak time of the flares.

\begin{figure}[ht!]
   \begin{center}
            \gridline{  
     \hspace{-0.06\textwidth}
    \fig{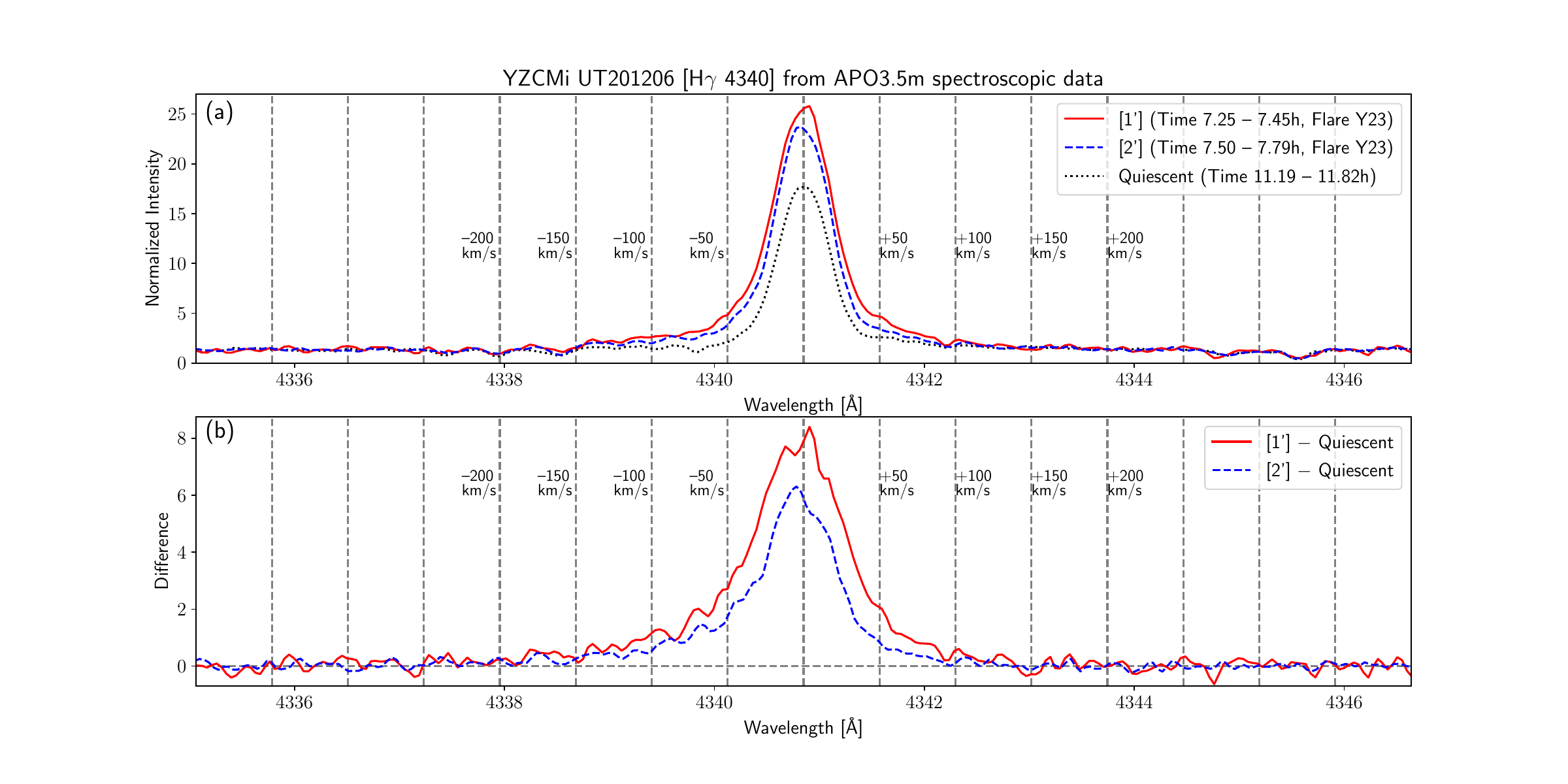}{0.58\textwidth}{\vspace{0mm}}
     \hspace{-0.06\textwidth}
       \fig{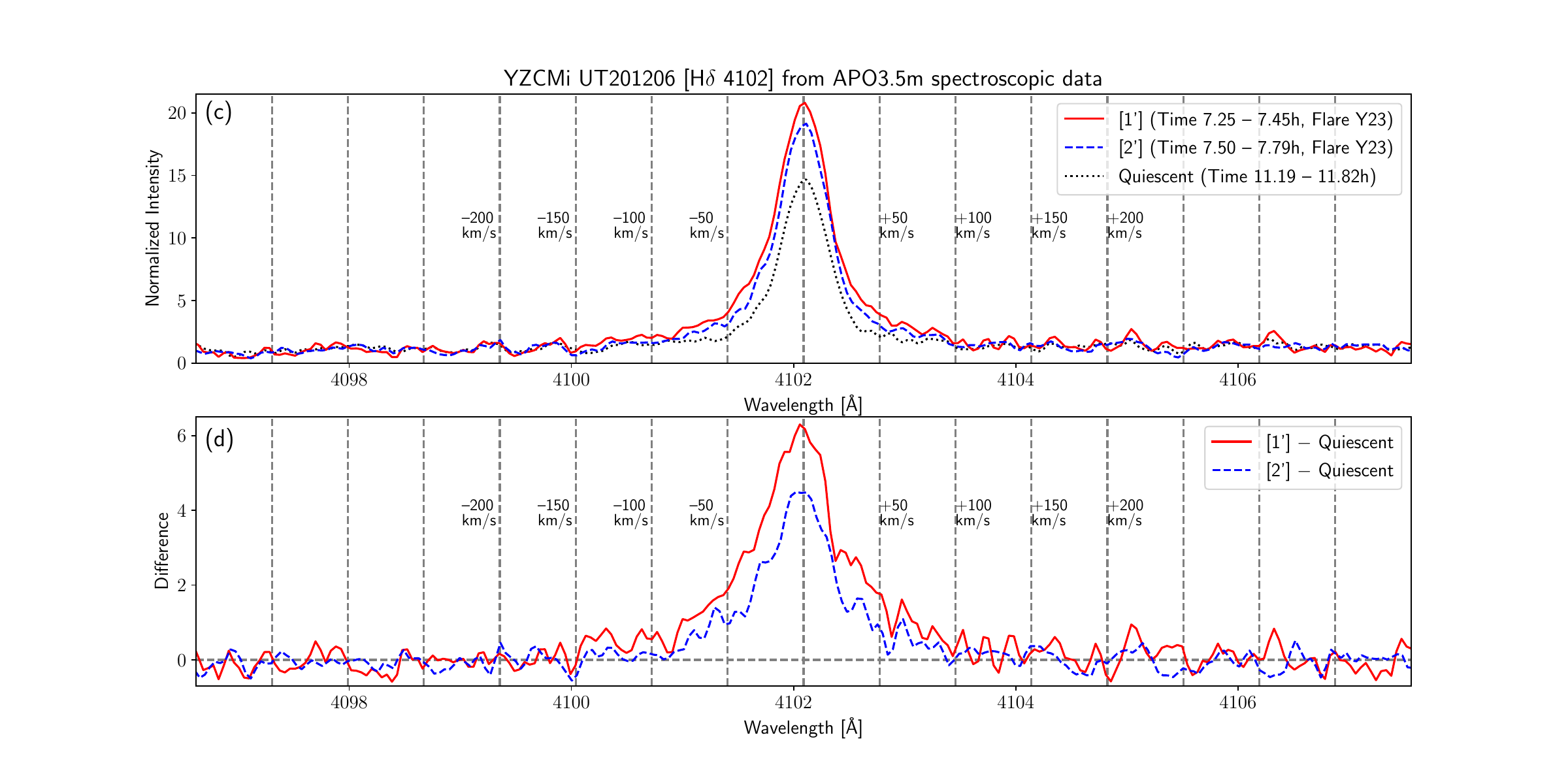}{0.58\textwidth}{\vspace{0mm}}
    }    
   \vspace{-1.0cm}
            \gridline{  
     \hspace{-0.06\textwidth}
    \fig{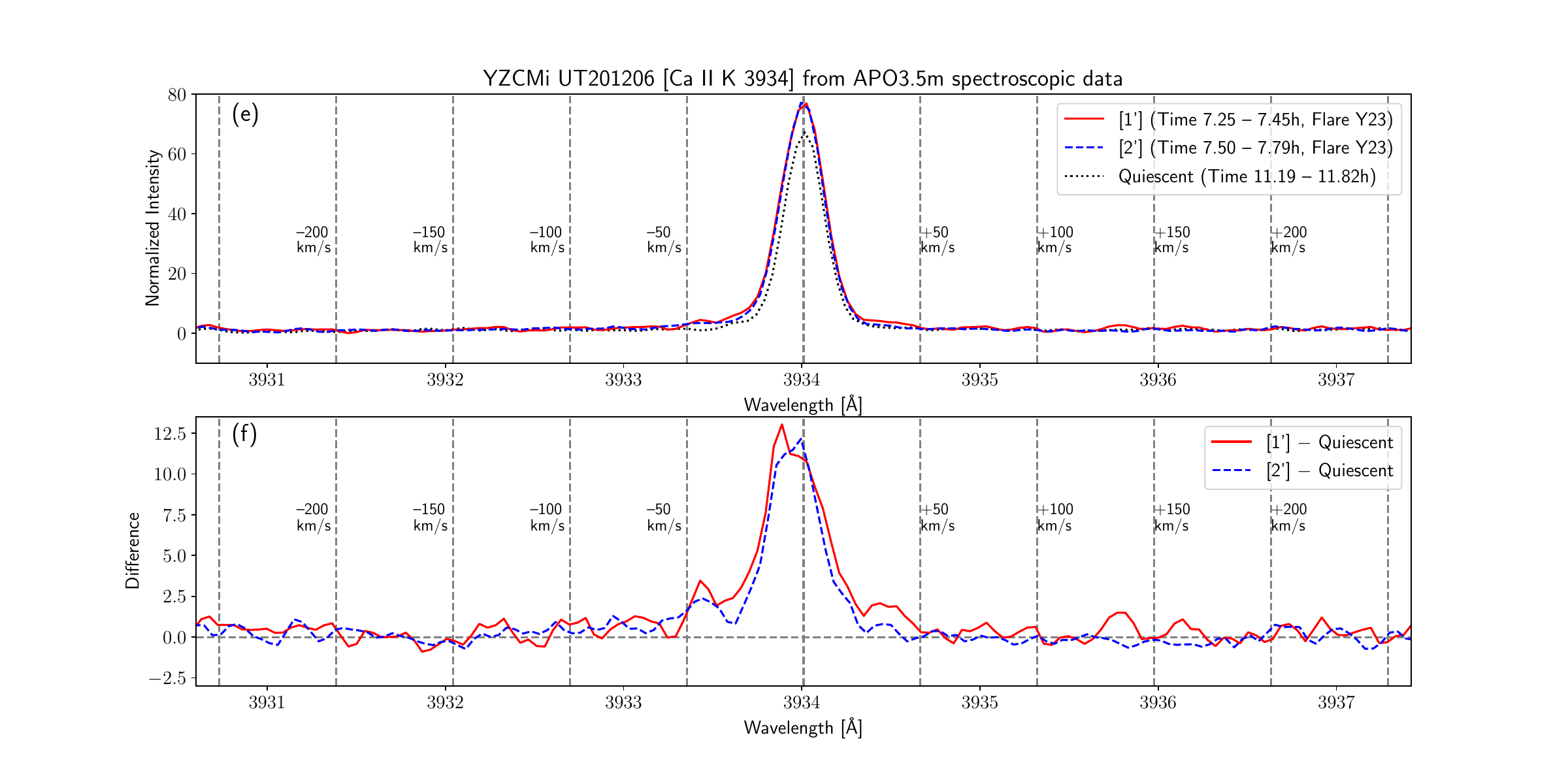}{0.58\textwidth}{\vspace{0mm}}
     \hspace{-0.06\textwidth}
         \fig{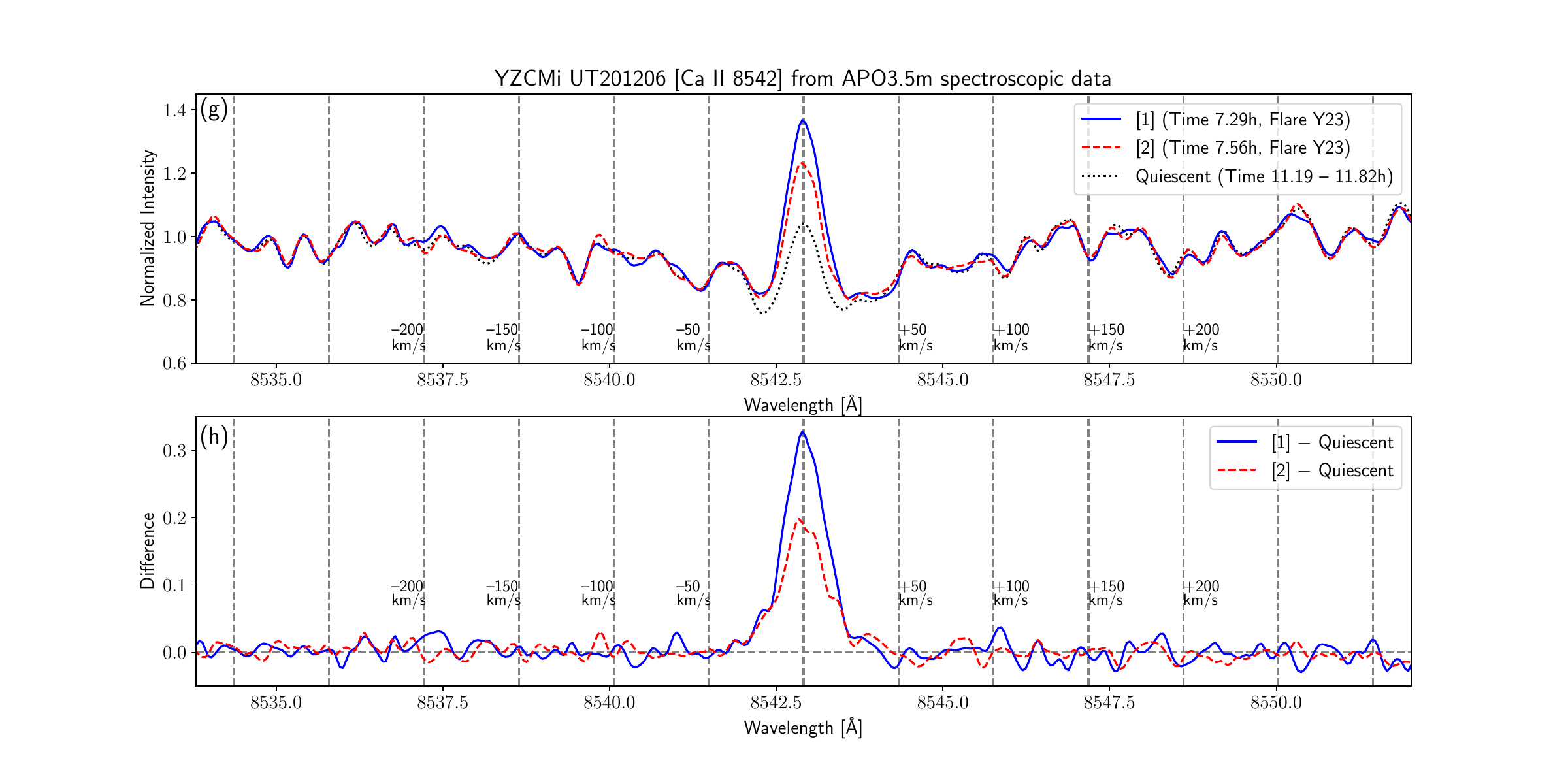}{0.58\textwidth}{\vspace{0mm}}
    }    
   \vspace{-1.0cm}
    \gridline{  
     \hspace{-0.06\textwidth}
    \fig{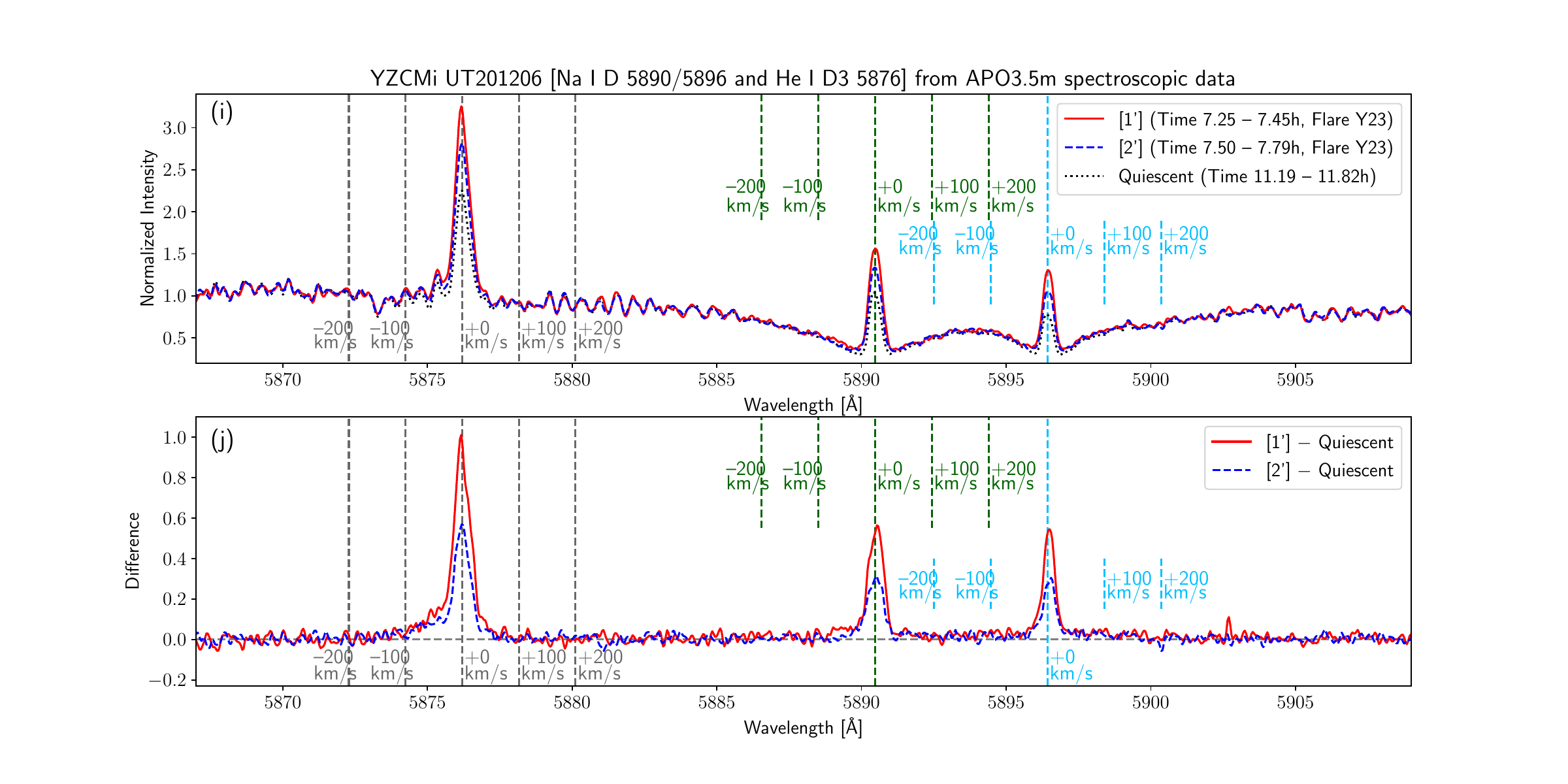}{0.58\textwidth}{\vspace{0mm}}
     \hspace{-0.06\textwidth}
         \fig{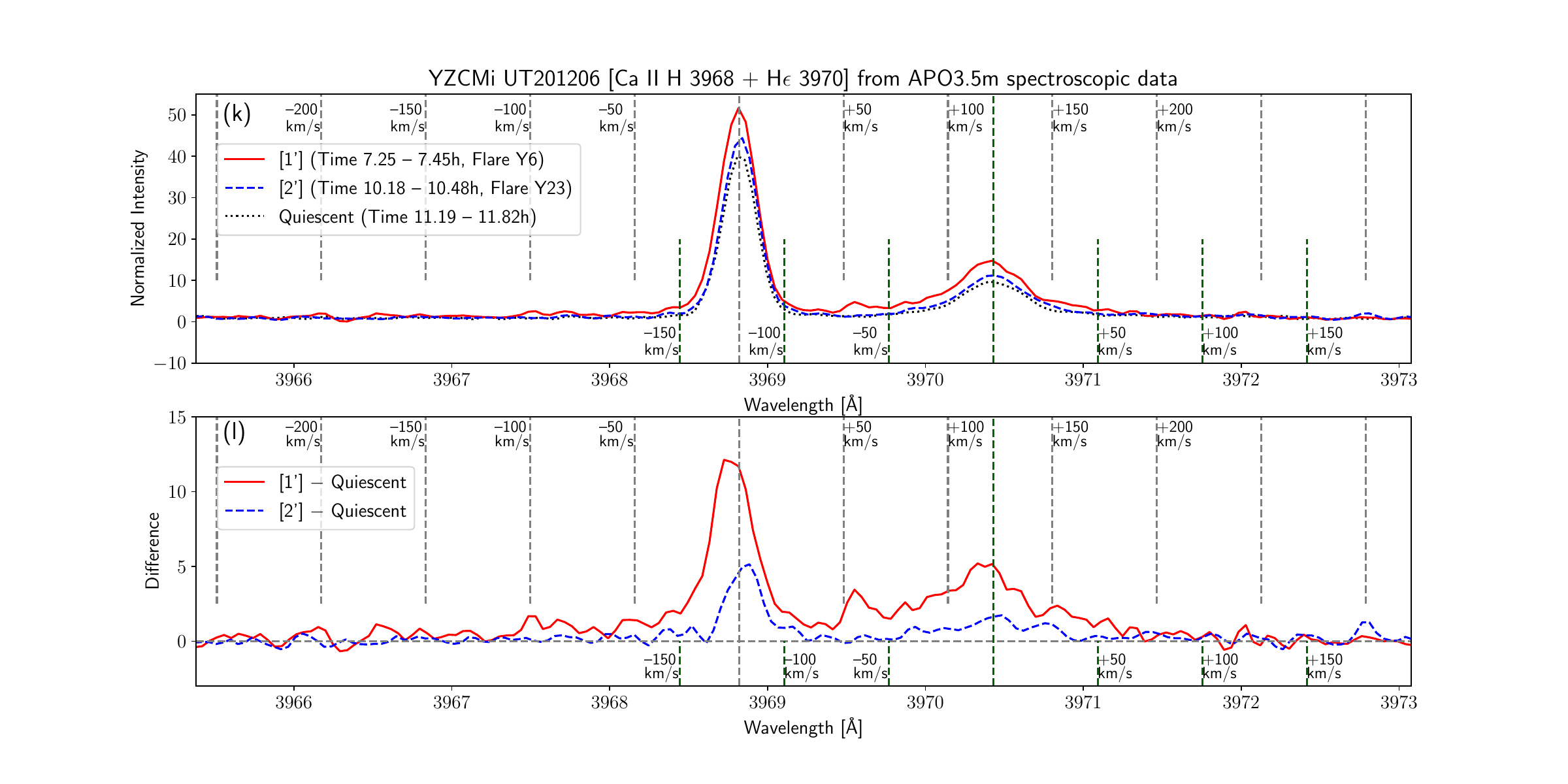}{0.58\textwidth}{\vspace{0mm}}
    }
     \vspace{-0cm}
     \caption{
\color{black}\textrm{  
(a)\&(b)
Line profiles of the H$\gamma$ emission line during Flare Y23 on 2020 December 6 from APO3.5m spectroscopic data, which are similarly plotted with Figure \ref{fig:spec_HaHb_YZCMi_UT201206}.
The blue solid and red dashed lines indicate 
the integrated line profiles over the time [1$^{\prime}$] (Time 7.25 -- 7.45h) and [2$^{\prime}$] (Time 7.50 -- 7.79h) on this date, 
which include the time [1] and [2] in Figure \ref{fig:lcEW_HaHb_YZCMi_UT201206} (light curves), respectively.
(c)\&(d), (e)\&(f), (g)\&(h), (i)\&(j), and (k)\&(l)
Same as panels (a)\&(b), but for H$\delta$, Ca II K, Ca II 8542, Na I D1 \& D2 (5890 \& 5896)$+$He I D3 5876, and H$\epsilon+$Ca II H lines, respectively.
As for the Ca II 8542 line, the data at the time [1] and [2] are plotted (not [1$^{\prime}$] and [2$^{\prime}$]).
 } \color{black}
     }
   \label{fig:spec_other_YZCMi_UT201206}
   \end{center}
 \end{figure}

The EW light curves of H$\gamma$, H$\delta$, Ca II K, Ca II 8542, Na I D1 \& D2, and He I D3 5876 lines are also shown in Figures 
\ref{fig:lcEW_HaHb_YZCMi_UT201206} (c), (d), \& (e).
The profiles of these lines and H$\epsilon+$Ca II H lines during Flare Y23 
are shown in Figure \ref{fig:spec_other_YZCMi_UT201206}.
At around time [1] or [1$^{\prime}$] (Figure \ref{fig:spec_other_YZCMi_UT201206}), 
slight blue asymmetries (slight blue wing enhancements) 
were seen in all the lines except for Ca II 8542,
while the blue wing asymmetry velocities are different among the lines. For example, 
Ca II K line shows blue asymmetry up to -100 -- -50 km s$^{-1}$ while  H$\gamma$ line up to -150 -- -100 km s$^{-1}$.

\clearpage
 
 \subsection{Flares E1 (Blue wing asymmetry) \& E2 (Blue wing asymmetry) observed on 2019 December 15} 
\label{subsec:results:2019-Dec-15} 

     \begin{figure}[ht!]
   \begin{center}
   \gridline{
    \fig{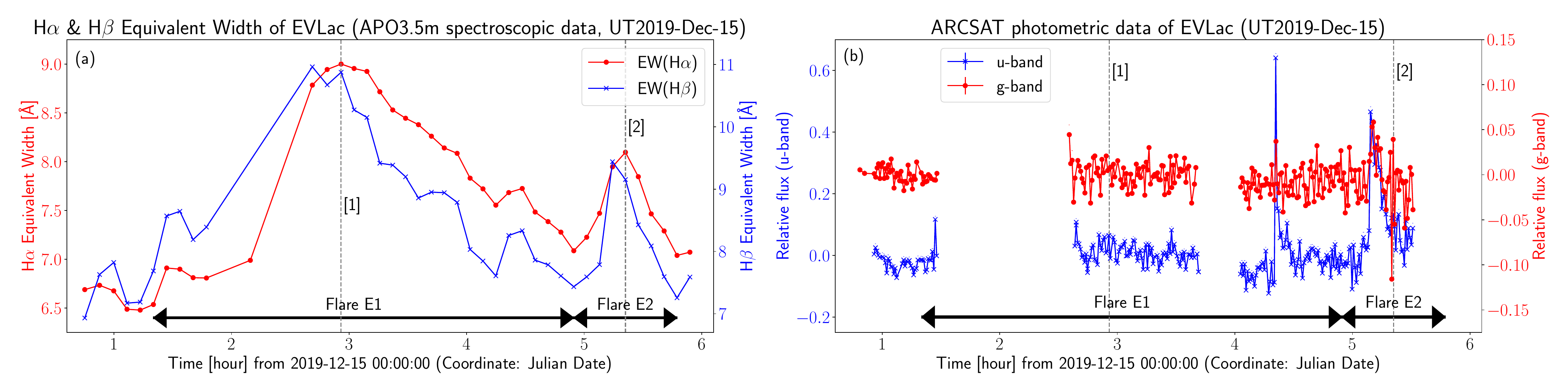}{1.0\textwidth}{\vspace{0mm}}
    }
         \vspace{-1cm}
                  \gridline{  
     \hspace{-0.02\textwidth}
    \fig{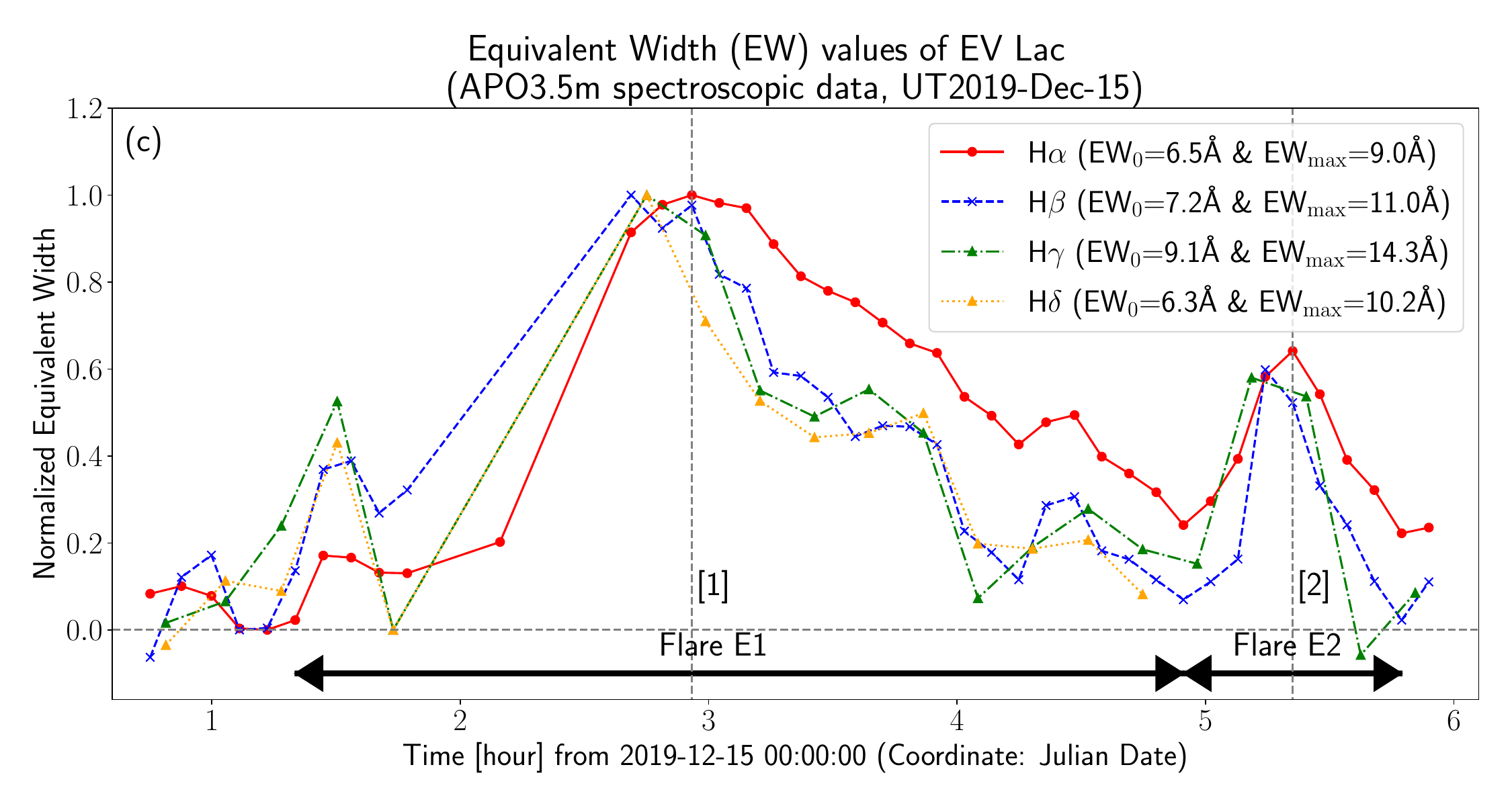}{0.5\textwidth}{\vspace{0mm}}
     \hspace{-0.02\textwidth}
    \fig{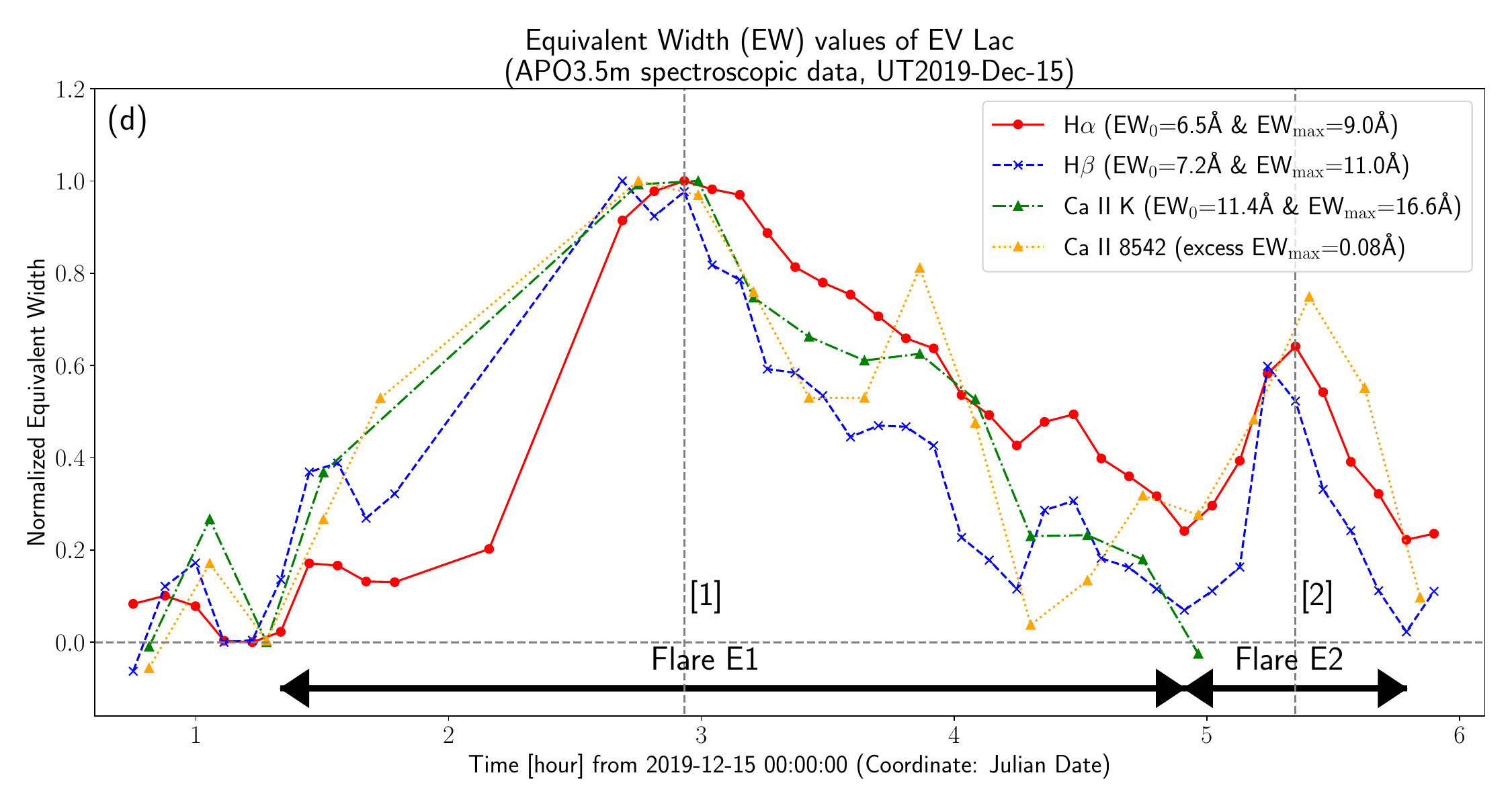}{0.5\textwidth}{\vspace{0mm}}
    }
     \vspace{-1cm}
              \gridline{  
     \hspace{-0.02\textwidth}
    \fig{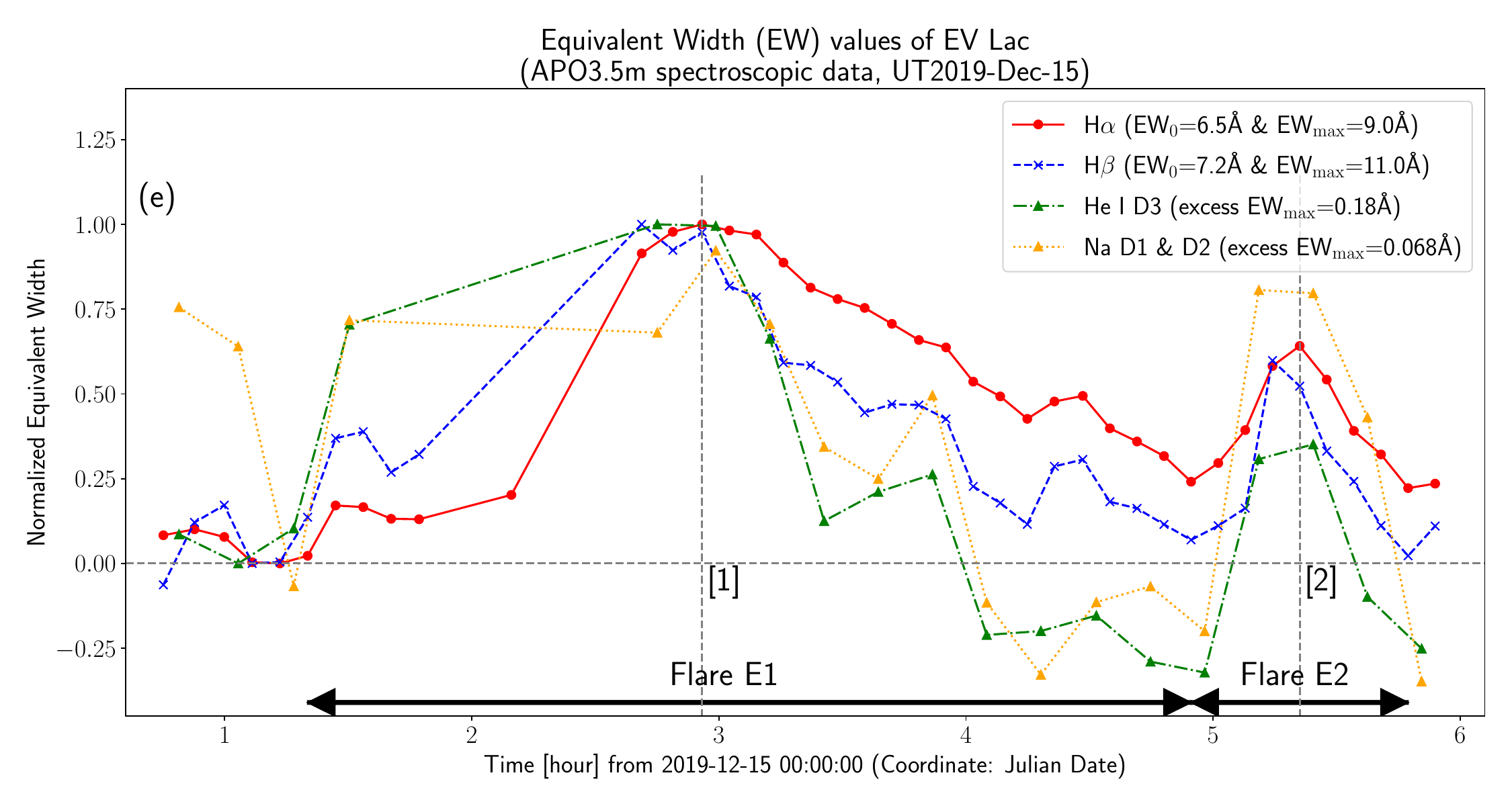}{0.5\textwidth}{\vspace{0mm}}
    }
     \vspace{-0.8cm}
     \caption{
     \color{black}\textrm{  
Light curves of EV Lac on 2019 December 15 showing Flares E1 \& E2, which are plotted 
similarly with Figure \ref{fig:lcEW_HaHb_YZCMi_UT191212}.
The grey dashed lines with numbers ([1]\&[2]) correspond to the time shown 
with the same numbers in Figures \ref{fig:spec_HaHb_EVLac_UT191215} \& \ref{fig:map_HaHb_EVLac_UT191215}.
 } \color{black}
     }
   \label{fig:lcEW_HaHb_EVLac_UT191215}
   \end{center}
 \end{figure}

On 2019 December 15, two flares (Flares E1 \& E2) 
were detected on EV Lac in H$\alpha$ \& H$\beta$ lines 
as shown in Figure \ref{fig:lcEW_HaHb_EVLac_UT191215} (a).  
As for Flare E1, the H$\alpha$ \& H$\beta$ equivalent widths increased to 9.0\AA~and 11.0\AA, respectively, and $\Delta t^{\rm{flare}}_{\rm{H}\alpha}$ is 3.6 hours (Table \ref{table:list1_flares}).
Only the late phase of Flare E1 was observed with ARCSAT $u$- \& $g$-bands 
and \color{black}\textrm{an increase of the continuum flux was observed in late phase 
at $\sim$4.3--4.4h (Figure \ref{fig:lcEW_HaHb_EVLac_UT191215} (b))}\color{black}.
It is possible there were increases of the continuum flux in the early phase of Flare E1.
As for Flare E2, the H$\alpha$ \& H$\beta$ equivalent widths increased to 8.1\AA~and 9.4\AA, respectively, and $\Delta t^{\rm{flare}}_{\rm{H}\alpha}$ is 0.9 hours (Table \ref{table:list1_flares}).
In addition to these enhancements in Balmer emission lines, the continuum brightness observed with ARCSAT $u$- \& $g$-bands 
increased by $\sim$ 60 -- 65\% and $\sim$ 5\%, 
respectively, during Flare E2 (Figure \ref{fig:lcEW_HaHb_EVLac_UT191215} (b)).

      \begin{figure}[ht!]
   \begin{center}
            \gridline{  
     \hspace{-0.06\textwidth}
    \fig{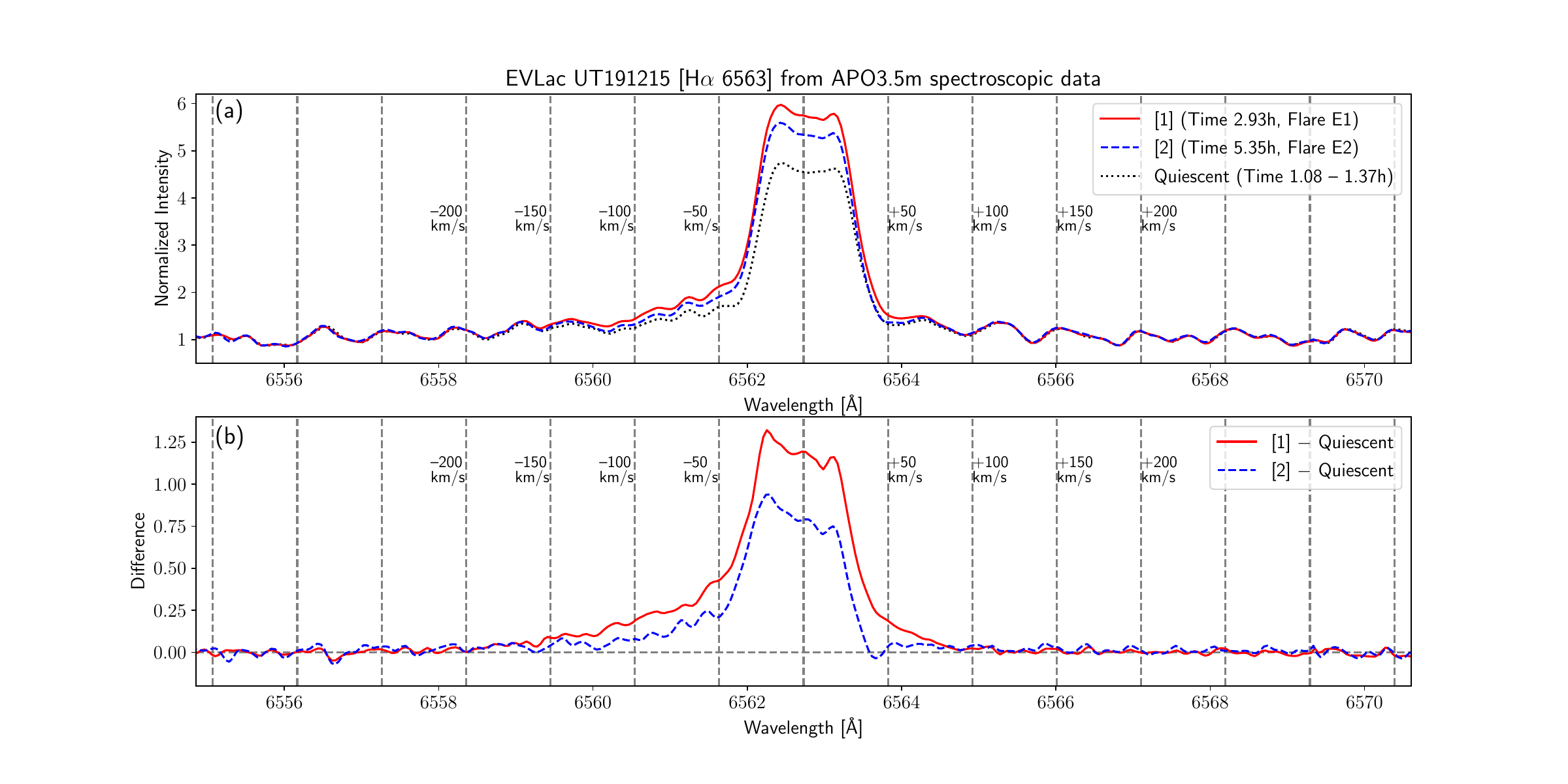}{0.58\textwidth}{\vspace{0mm}}
     \hspace{-0.06\textwidth}
       \fig{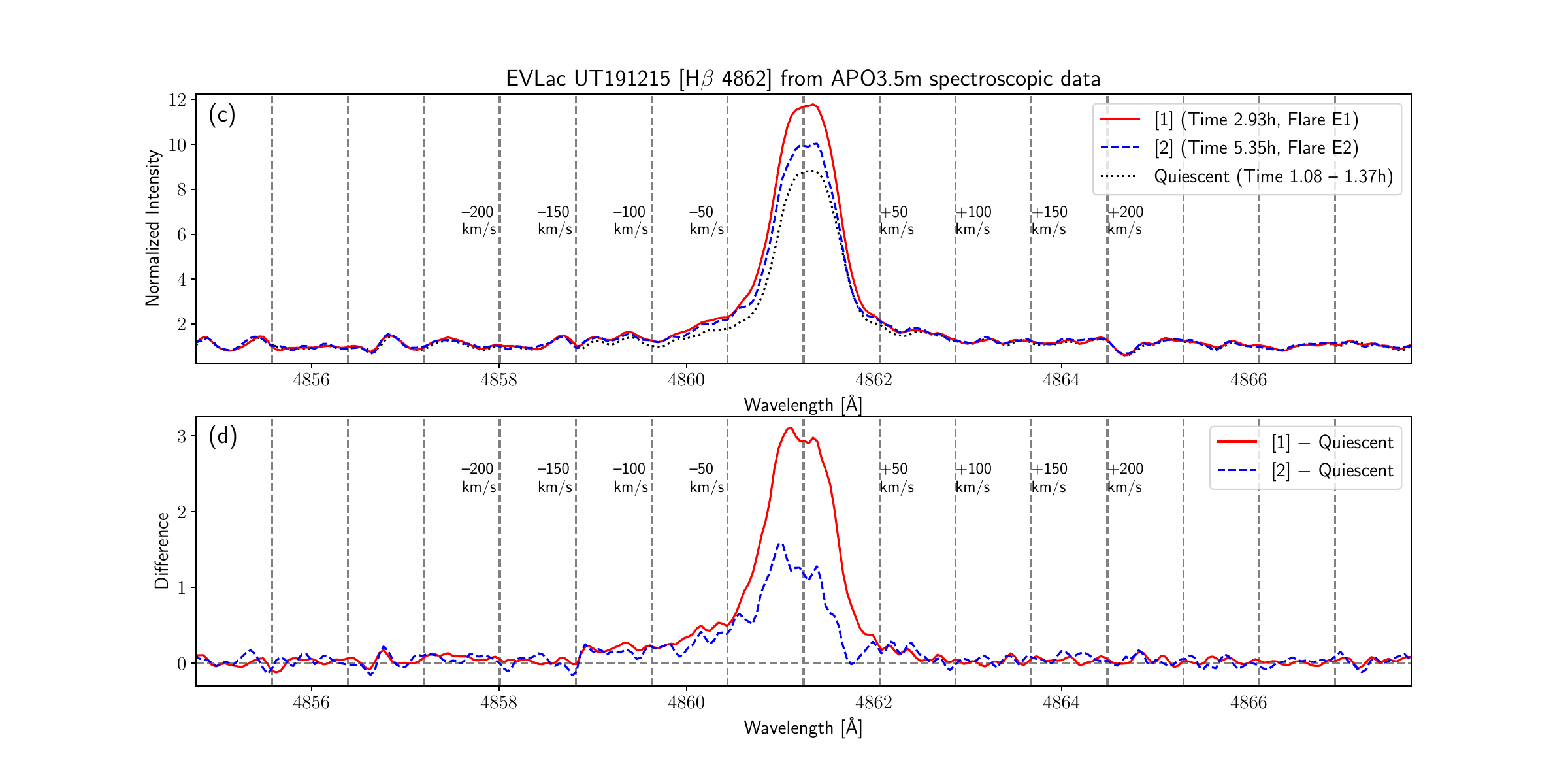}{0.58\textwidth}{\vspace{0mm}}
    }
     \vspace{-0.5cm}
     \caption{
     \color{black}\textrm{  
Line profiles of the H$\alpha$ \& H$\beta$ emission lines during Flares E1 \& E2  (at the time [1] and [2]) on 2020 December 6 from APO3.5m spectroscopic data, which are plotted similarly with Figure \ref{fig:spec_HaHb_YZCMi_UT190127}.
 } \color{black}
     }
   \label{fig:spec_HaHb_EVLac_UT191215}
   \end{center}
 \end{figure}

      \begin{figure}[ht!]
   \begin{center}
          \gridline{  
     \hspace{-0.07\textwidth}
    \fig{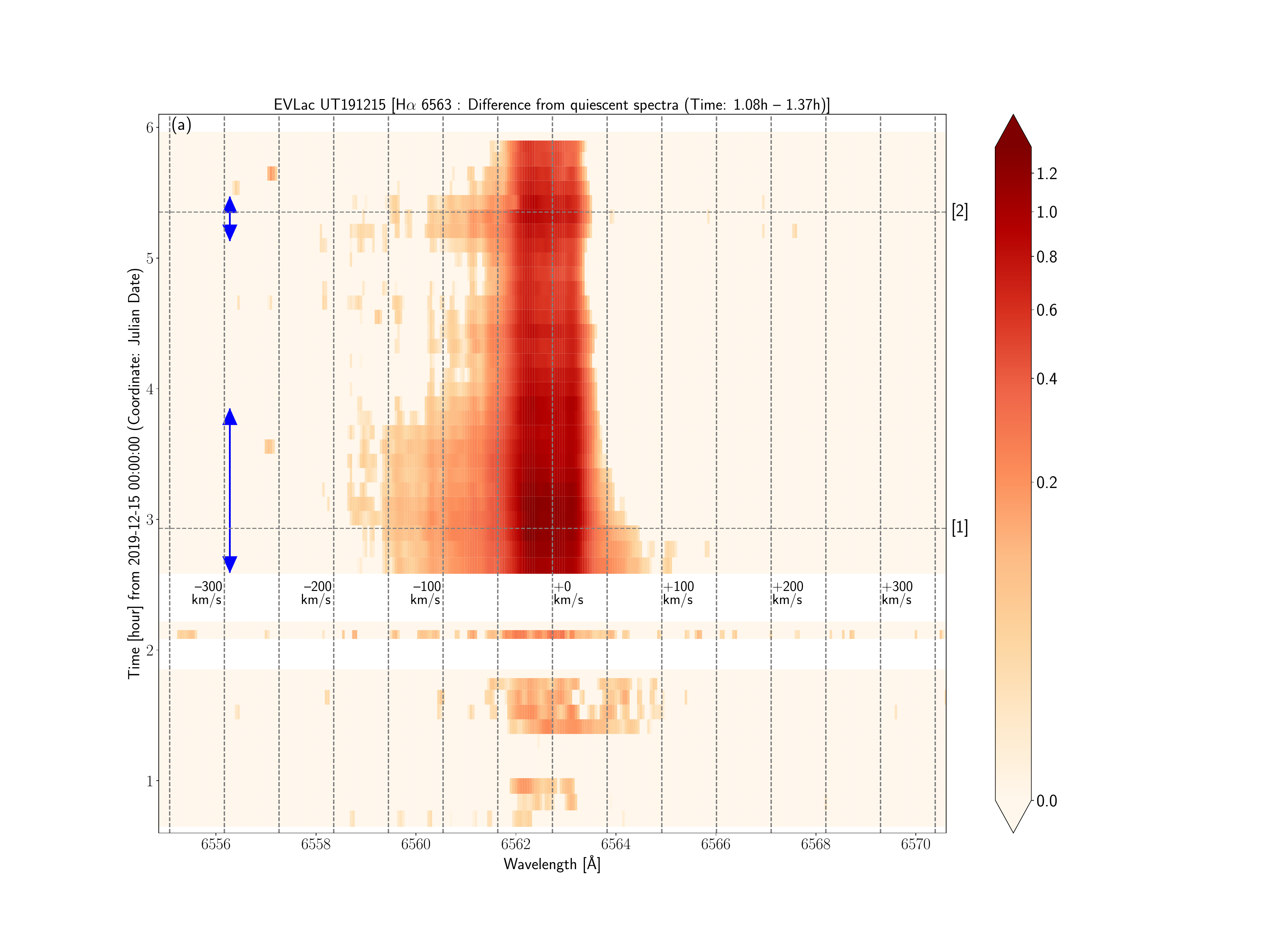}{0.63\textwidth}{\vspace{0mm}}
     \hspace{-0.11\textwidth}
    \fig{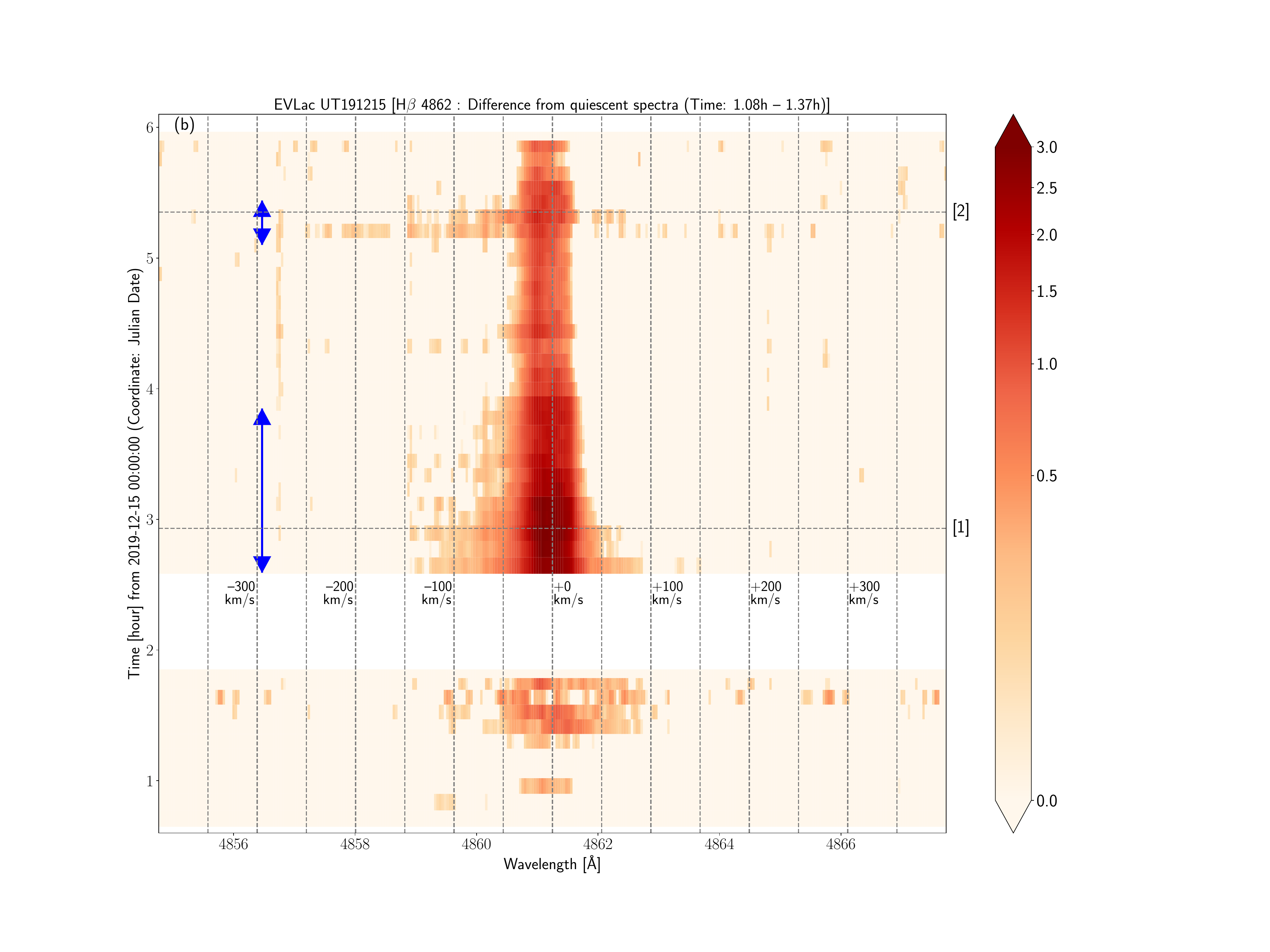}{0.63\textwidth}{\vspace{0mm}}
    }
     \vspace{-1cm}
     \caption{     
\color{black}\textrm{  
Time evolution of the H$\alpha$ \& H$\beta$ line profiles covering Flares E1 \& E2 
on 2019 December 15, which are shown similarly with Figure \ref{fig:map_HaHb_YZCMi_UT191212}.
The grey horizontal dashed lines indicate the time [1] \& [2], which are shown in Figure \ref{fig:lcEW_HaHb_EVLac_UT191215} (light curves) and Figure \ref{fig:spec_HaHb_EVLac_UT191215} (line profiles).
 } \color{black}
     }
   \label{fig:map_HaHb_EVLac_UT191215}
   \end{center}
 \end{figure}

  \color{black}\textrm{ 
The $L_{u}$, $L_{g}$, $E_{u}$, $E_{g}$, $L_{\rm{H}\alpha}$, $L_{\rm{H}\beta}$, $E_{\rm{H}\alpha}$, and $E_{\rm{H}\beta}$ values are listed in Table \ref{table:list1_flares}.
As for Flare E1, we did not estimate $L_{u}$, $L_{g}$, $E_{u}$, \& $E_{g}$ values, since only the late phase of Flare E1 was observed with ARCSAT $u$- \& $g$-bands, and no 
clear increases of the continuum brightness were observed in the late phase 
(Figure \ref{fig:lcEW_HaHb_EVLac_UT191215}(b)).
} \color{black}

The H$\alpha$ \& H$\beta$ line profiles during Flares E1 and E2 are shown in
Figures \ref{fig:spec_HaHb_EVLac_UT191215} \& \ref{fig:map_HaHb_EVLac_UT191215}. 
The blue wing of H$\alpha$ line was 
enhanced (up to -150--200 km s$^{-1}$) during the early phase of Flare E1 
(time [1] in Figures \ref{fig:spec_HaHb_EVLac_UT191215} (b) \& \ref{fig:map_HaHb_EVLac_UT191215} (a)).
The similar blue wing asymmetry (up to -150 km s$^{-1}$) was seen also in the H$\beta$ line (time [1] in Figures \ref{fig:spec_HaHb_EVLac_UT191215} (d) \& \ref{fig:map_HaHb_EVLac_UT191215} (b)), 
but the duration of the blue wing asymmetry in H$\beta$ line ($\sim$0.5 hours) 
is shorter than that of 
H$\alpha$ line ($\gtrsim$1 hours) (Figures \ref{fig:map_HaHb_EVLac_UT191215}).
The blue wing asymmetry in H$\alpha$ and H$\beta$ lines (up to -150 km s$^{-1}$) were also seen at around the peak time of Flare E2 
(time [1] in Figures \ref{fig:spec_HaHb_EVLac_UT191215} (b) \& \ref{fig:map_HaHb_EVLac_UT191215} (a)). 
The duration of the blue wing asymmetry in H$\alpha$ and H$\beta$ lines 
during Flare E2 were $\sim$20 min and $\sim$10 min, respectively.

\begin{figure}[ht!]
   \begin{center}
            \gridline{  
     \hspace{-0.06\textwidth}
    \fig{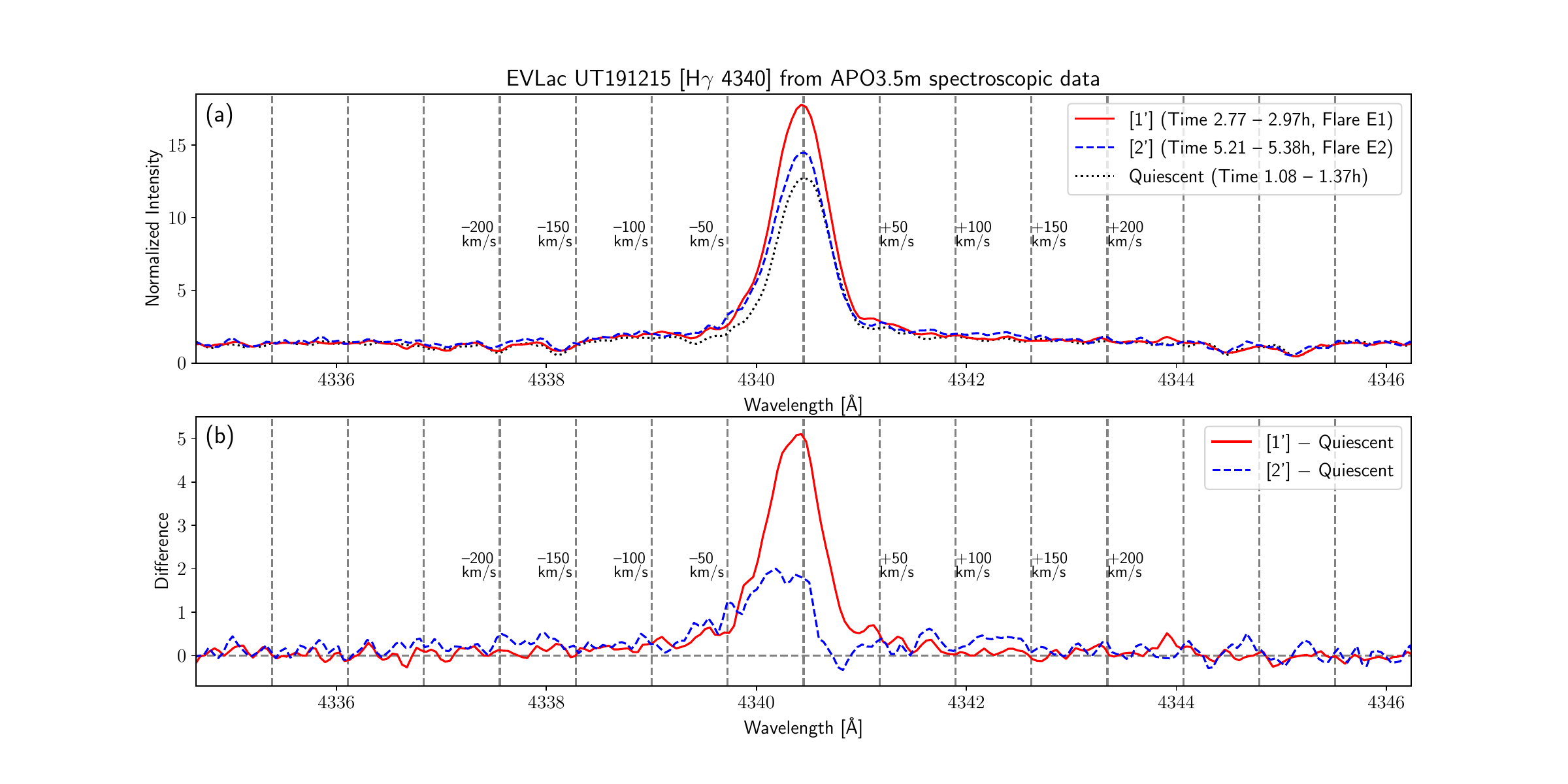}{0.58\textwidth}{\vspace{0mm}}
     \hspace{-0.06\textwidth}
       \fig{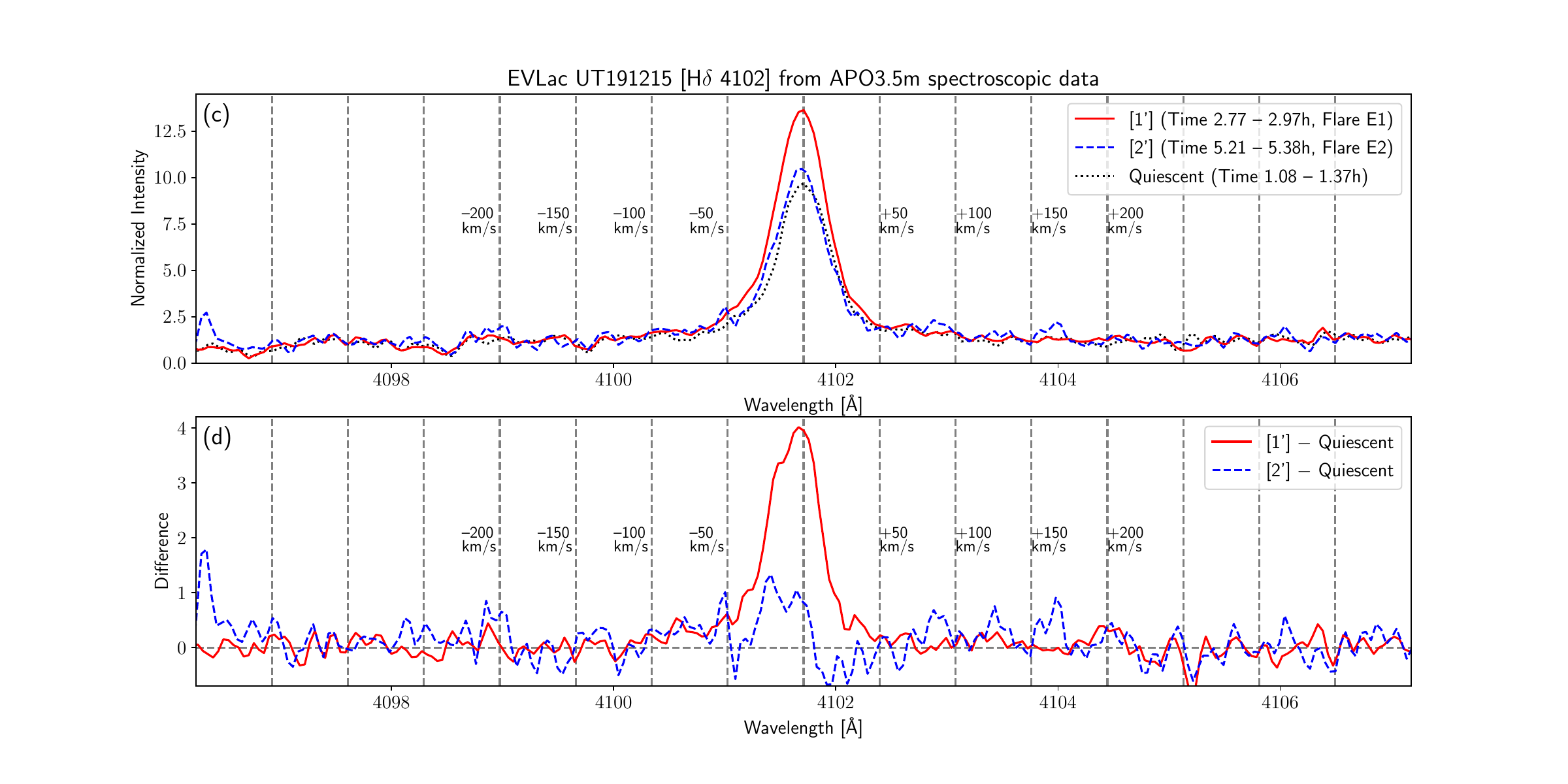}{0.58\textwidth}{\vspace{0mm}}
    }    
   \vspace{-1.0cm}
            \gridline{  
     \hspace{-0.06\textwidth}
    \fig{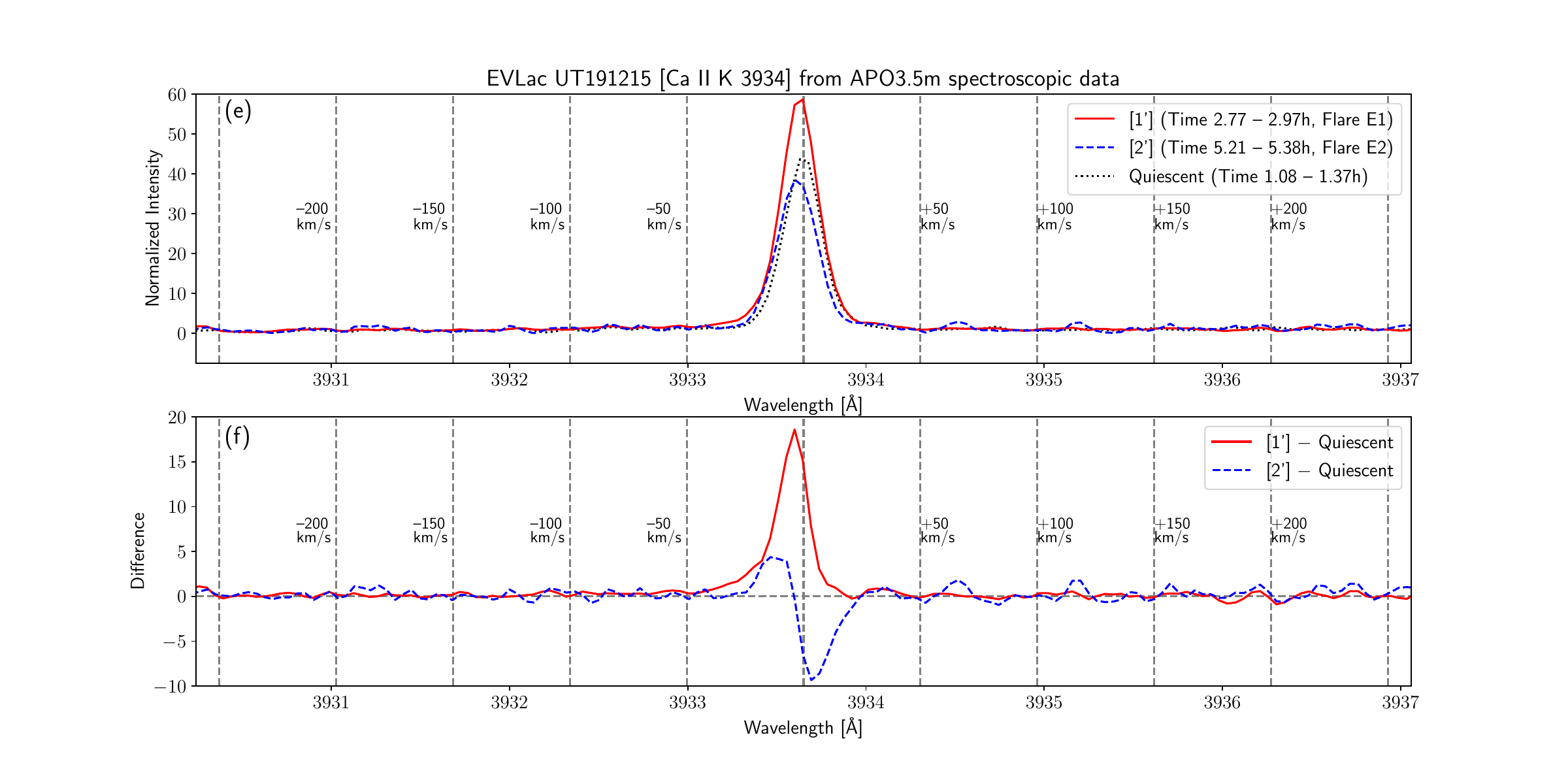}{0.58\textwidth}{\vspace{0mm}}
     \hspace{-0.06\textwidth}
        \fig{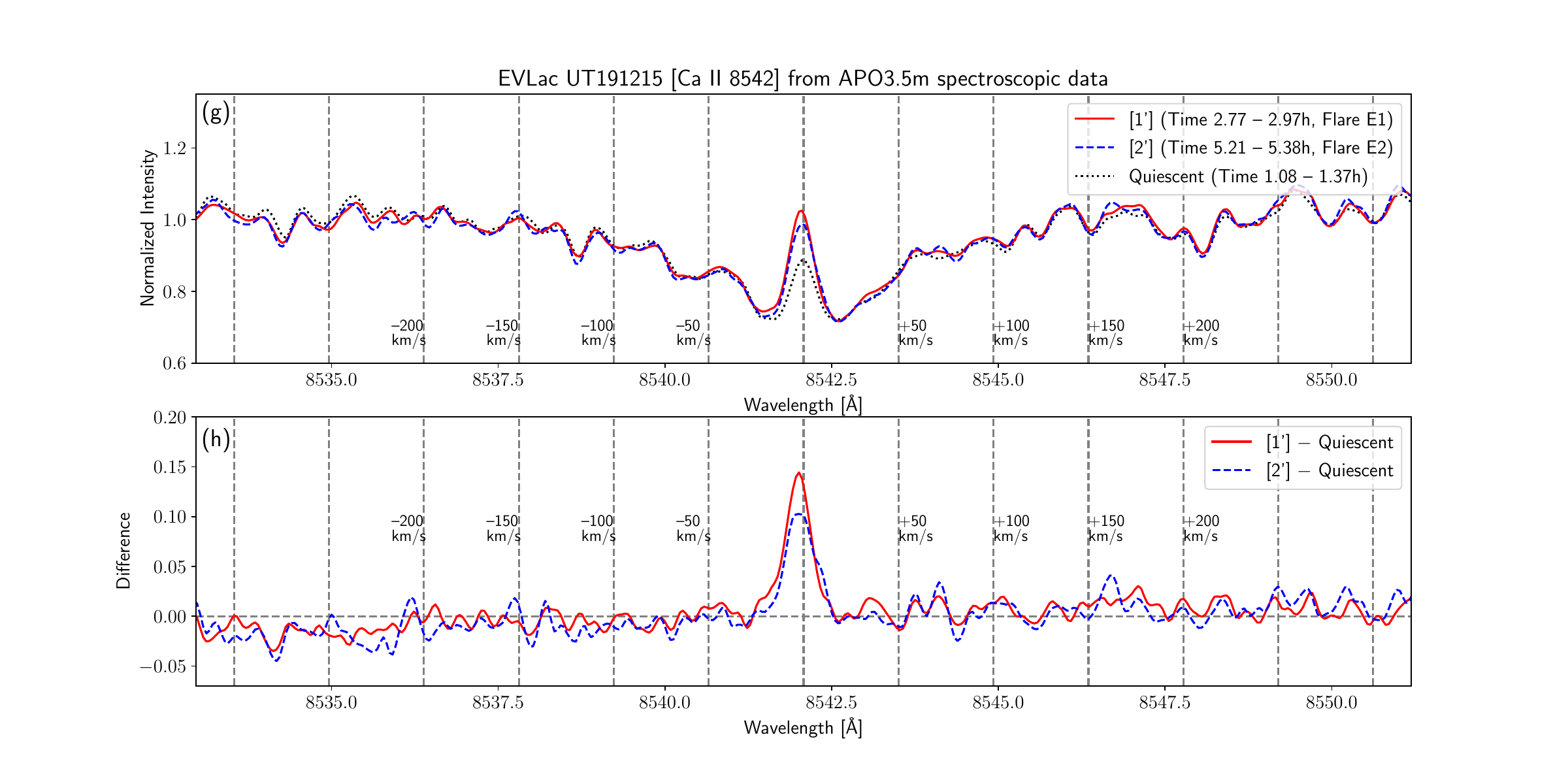}{0.58\textwidth}{\vspace{0mm}}
    }    
     \vspace{-1.0cm}
    \gridline{  
     \hspace{-0.06\textwidth}
    \fig{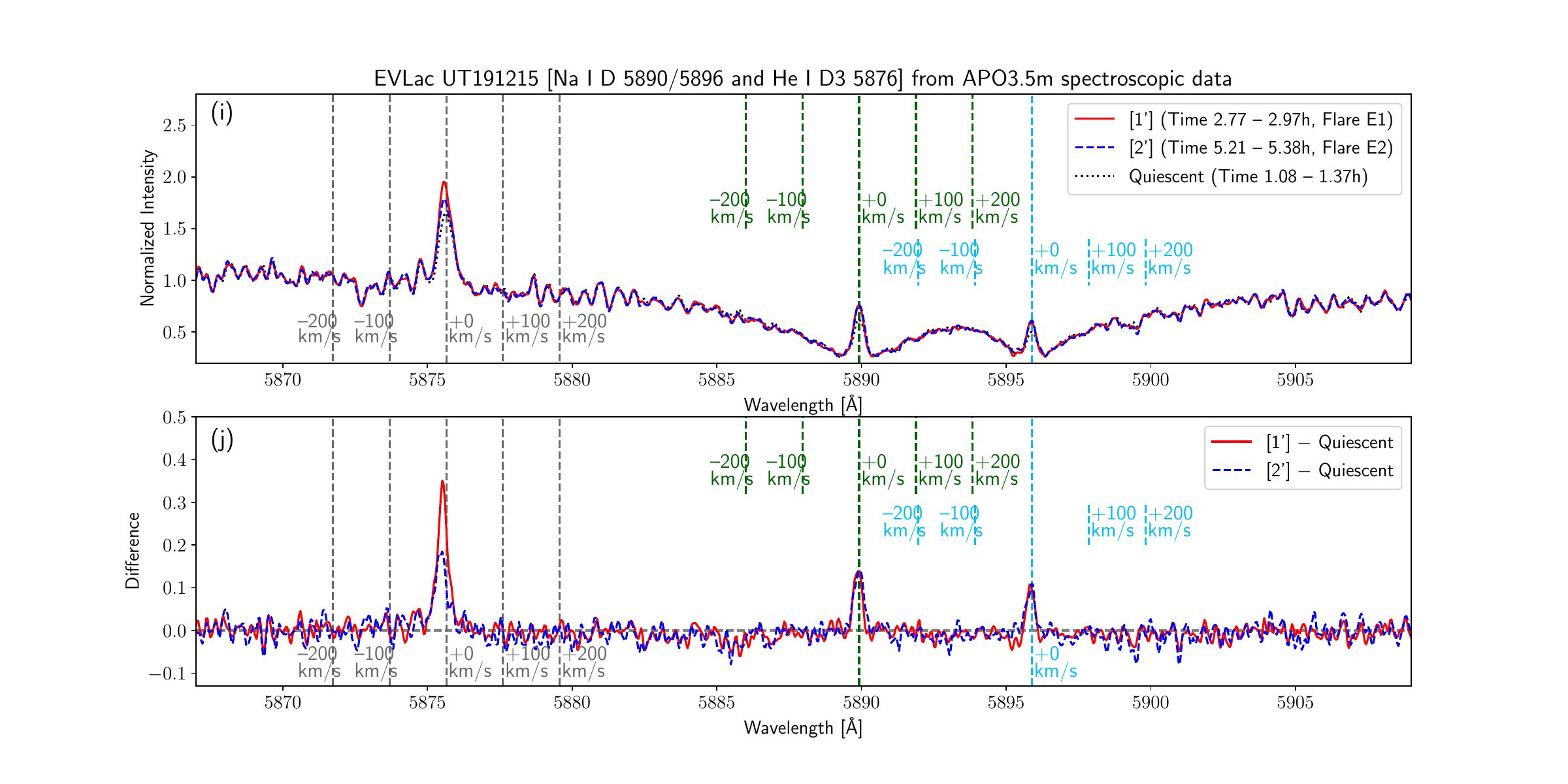}{0.58\textwidth}{\vspace{0mm}}
     \hspace{-0.06\textwidth}
    \fig{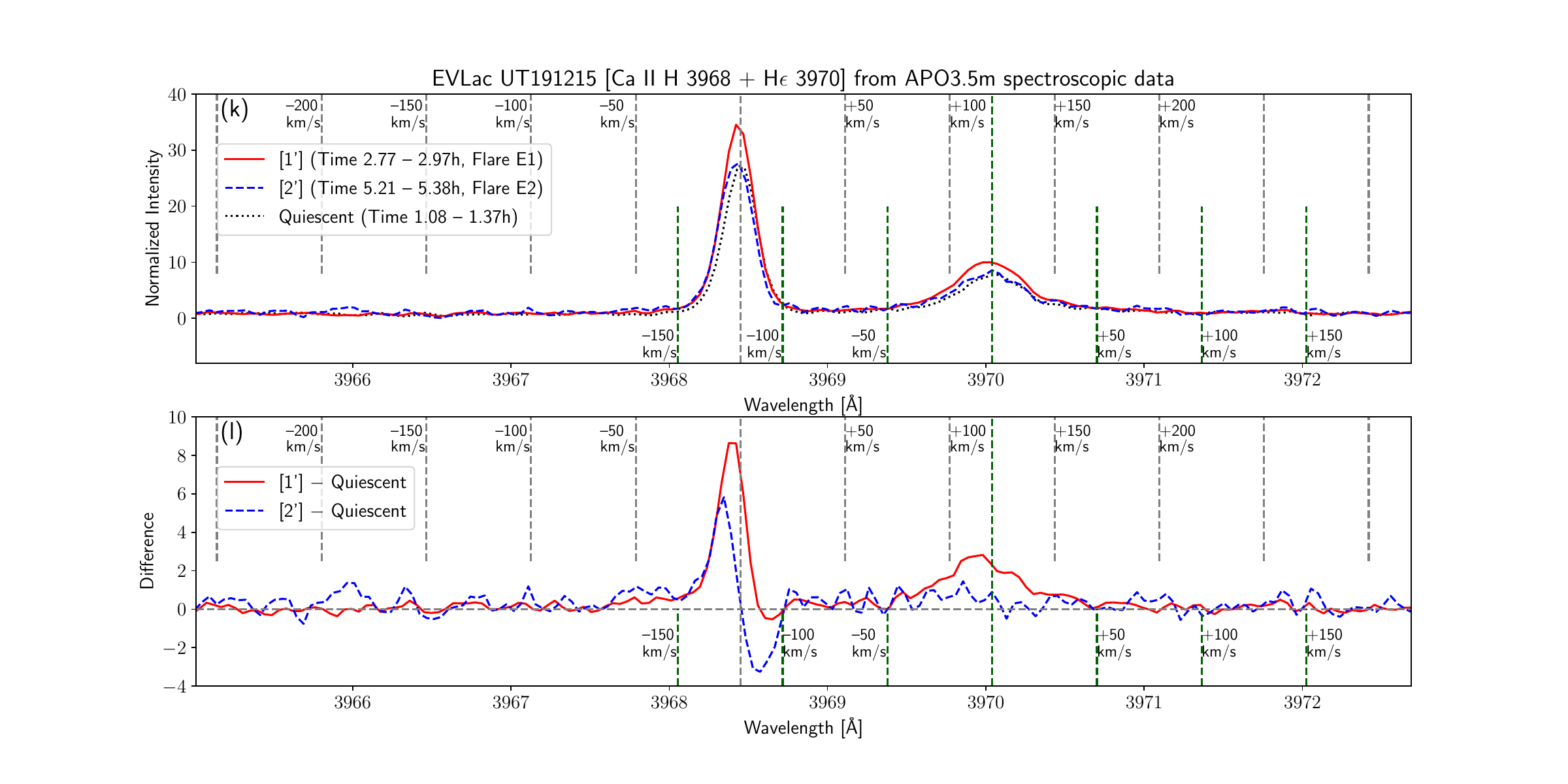}{0.58\textwidth}{\vspace{0mm}}
    }    
     \vspace{-0.5cm}
     \caption{
    \color{black}\textrm{  
(a)\&(b)
Line profiles of the H$\gamma$ emission line during Flares E1 \& E2 on 2019 December 15 from APO3.5m spectroscopic data, which are similarly plotted with Figure \ref{fig:spec_HaHb_EVLac_UT191215}.
The blue solid and red dashed lines indicate 
the integrated line profiles over the time [1$^{\prime}$] (Time 2.77 -- 2.97h) and [2$^{\prime}$] (Time 5.21 -- 5.38h) on this date, 
which include the time [1] and [2] in Figure \ref{fig:lcEW_HaHb_EVLac_UT191215} (light curves), respectively.
(c)\&(d), (e)\&(f), (g)\&(h), (i)\&(j), and (k)\&(l)
Same as panels (a)\&(b), but for H$\delta$, Ca II K, Ca II 8542, Na I D1 \& D2 (5890 \& 5896)$+$He I D3 5876, and H$\epsilon+$Ca II H lines, respectively.
 } \color{black}
}
   \label{fig:spec_other_EVLac_UT191215}
   \end{center}
 \end{figure}

The EW light curves of H$\gamma$, H$\delta$, Ca II K, Ca II 8542, Na I D1 \& D2, and He I D3 5876 lines are also shown in Figures 
\ref{fig:lcEW_HaHb_EVLac_UT191215} (c), (d), \& (e).
The profiles of these lines and H$\epsilon+$Ca II H lines during Flares E1 \& E2 
are shown in Figure \ref{fig:spec_other_EVLac_UT191215}.
As for H$\gamma$ line, 
the blue wing asymmetries up to $\sim$ -100 km s$^{-1}$ are seen during both Flares E1 \& E2.
As for H$\delta$ and Ca II H\&K lines, possible blue wing enhancements (up to $\sim$ -50 km s$^{-1}$) are seen during Flare E1, while line wing asymmetries of these lines are not clearly seen during Flare E2.
Line asymmetries of H$\epsilon$, Ca II 8542, Na I D1\&D2, and He I D3 lines are not so clear during both E1\&E2. However, we also note that there could be some slight peak blue shifts in the lines, for which we do not see clear line wing enhancements (e.g., Ca H\&K lines during Flare E2, He I D3 and Ca II 8542 lines during Flare E1).

\clearpage

 \subsection{Flare A3 (Blue wing asymmetry) observed on 2019 May 19} 
\label{subsec:results:2019-May-19} 

     \begin{figure}[ht!]
   \begin{center}
   \gridline{
    \fig{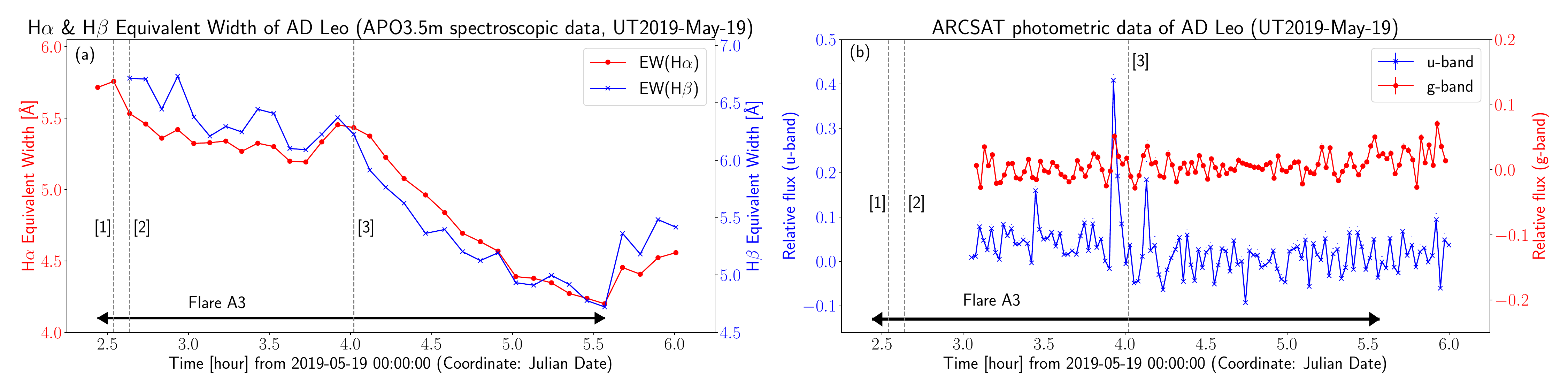}{1.0\textwidth}{\vspace{0mm}}
    }
     \vspace{-1cm}
                  \gridline{  
     \hspace{-0.02\textwidth}
    \fig{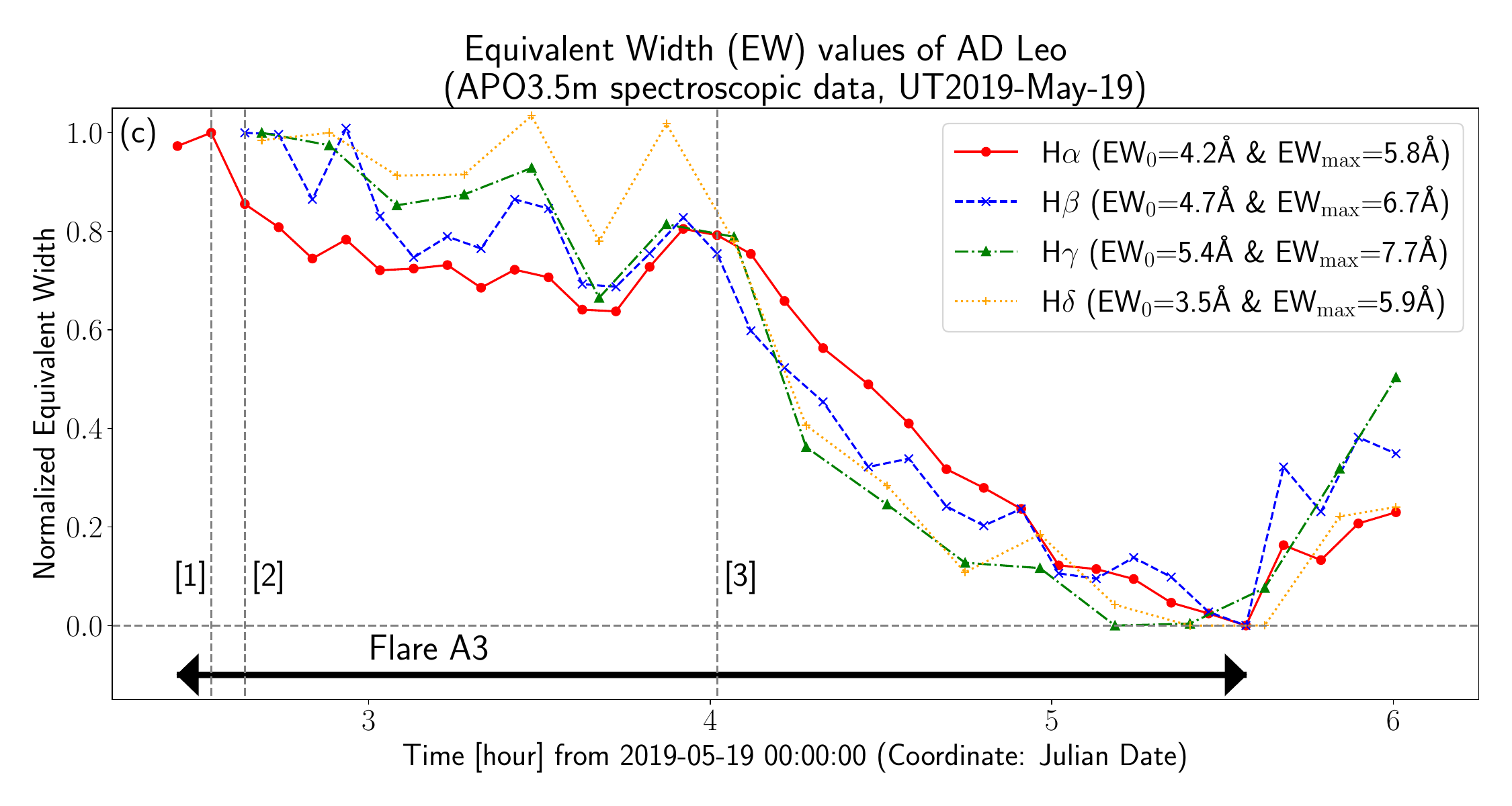}{0.5\textwidth}{\vspace{0mm}}
     \hspace{-0.02\textwidth}
    \fig{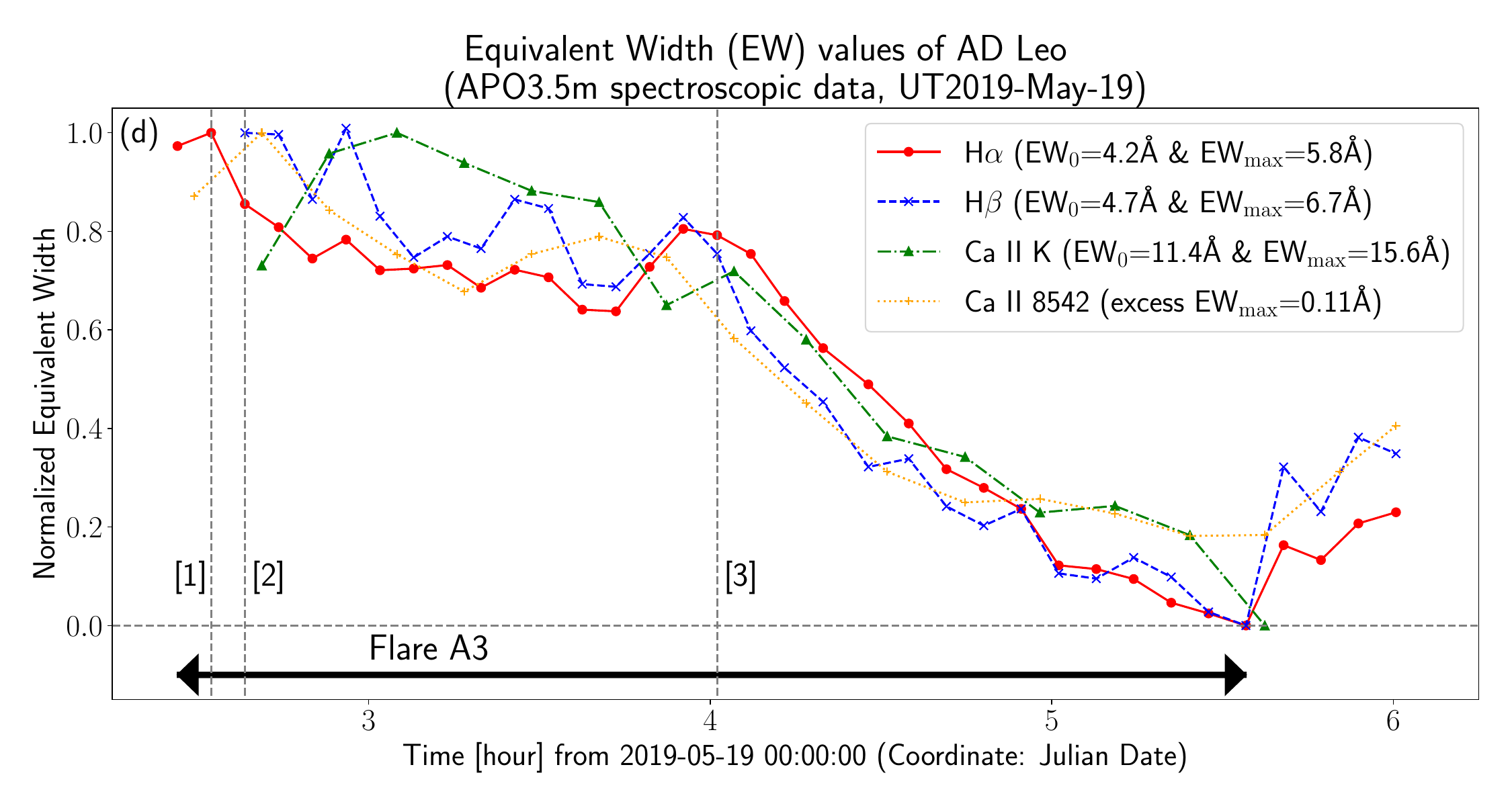}{0.5\textwidth}{\vspace{0mm}}
    }
     \vspace{-1cm}
              \gridline{  
     \hspace{-0.02\textwidth}
    \fig{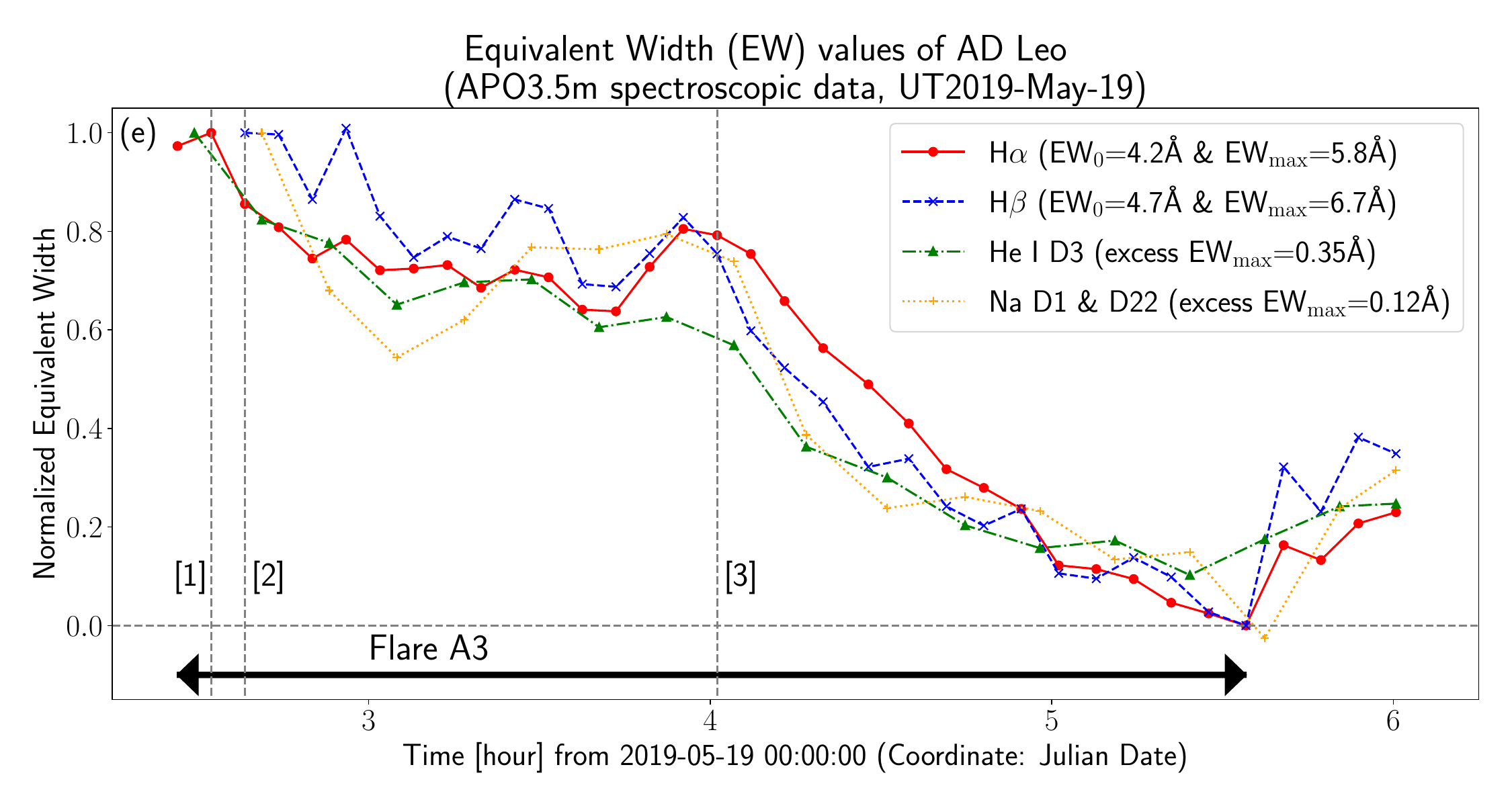}{0.5\textwidth}{\vspace{0mm}}
    }
     \vspace{-0.5cm}
     \caption{
     \color{black}\textrm{  
Light curves of AD Leo on 2019 May 19 showing Flare A3, which are plotted 
similarly with Figure \ref{fig:lcEW_HaHb_YZCMi_UT191212}.
The grey dashed lines with numbers ([1] -- [3]) correspond to the time shown 
with the same numbers in Figures \ref{fig:spec_HaHb_ADLeo_UT190519} \& \ref{fig:map_HaHb_ADLeo_UT190519}.
 } \color{black}
     }
   \label{fig:lcEW_HaHb_ADLeo_UT190519}
   \end{center}
 \end{figure}

On 2019 May 19,  one flare (Flares A3) 
was detected on AD Leo in H$\alpha$ \& H$\beta$ lines 
as shown in Figure \ref{fig:lcEW_HaHb_ADLeo_UT190519} (a).  
Flare A3 already started when the observation started.
The H$\alpha$ \& H$\beta$ equivalent widths increased to 5.8\AA~and 6.7\AA, respectively, and the $\Delta t^{\rm{flare}}_{\rm{H}\alpha}$ is $>$3.1 hour (Table \ref{table:list1_flares}).
In addition to these enhancements in Balmer emission lines, the continuum flux observed with ARCSAT $u$- \& $g$-bands increased 
by $\sim$40\% 
and $\sim$ 3 -- 4\%, respectively, 
\color{black}\textrm{
at around time $\sim$3.9--4h during Flare A3 (Figure \ref{fig:lcEW_HaHb_ADLeo_UT190519} (b)), 
However, the photomeric observation covered only the latter portion of the 
flare, and the flare already had started when the observation started.
Because of these, we cannot judge whether the main H$\alpha$ and H$\beta$ flare emission components are associated with white-light flares or not.
In these cases, we also do not list the flare peak luminosities in the continuum bands ($u$- and $g$-bands) in Table \ref{table:list1_flares}.
The lower limit of flare energies in the continuum bands ($u$- and $g$-bands) 
are estimated to be $E_{u}>2.7\times 10^{31}$erg, and $E_{g}>1.4\times 10^{31}$ erg from the existing data period.
Since Flare A3 already started when the observation started,
the $L_{\rm{H}\alpha}$, $L_{\rm{H}\beta}$, $E_{\rm{H}\alpha}$, and $E_{\rm{H}\beta}$,
which are also listed in Table \ref{table:list1_flares},
are only lower limit values.
} \color{black}

     \begin{figure}[ht!]
   \begin{center}
                   \gridline{  
     \hspace{-0.06\textwidth}
    \fig{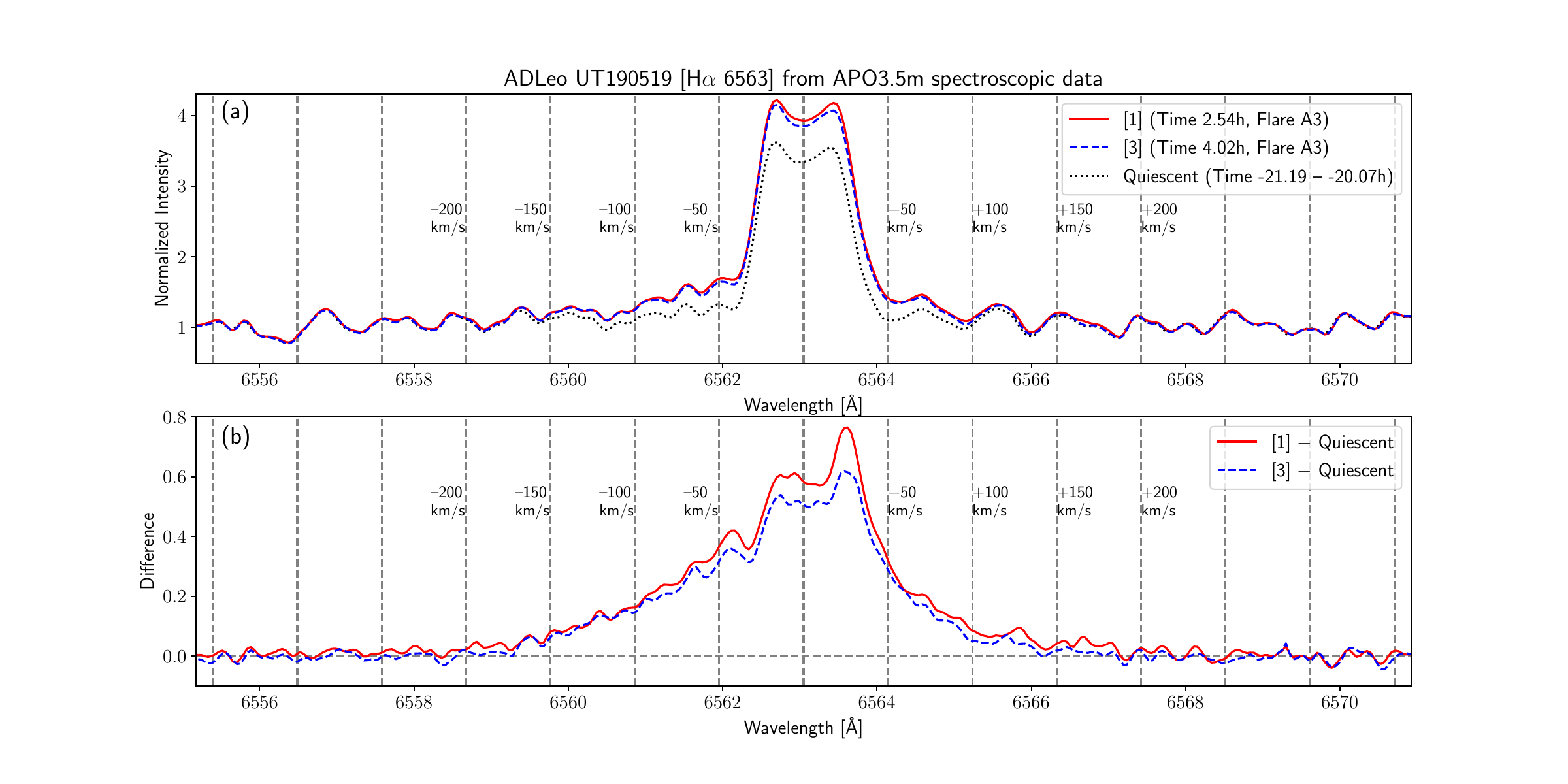}{0.58\textwidth}{\vspace{0mm}}
     \hspace{-0.06\textwidth}
       \fig{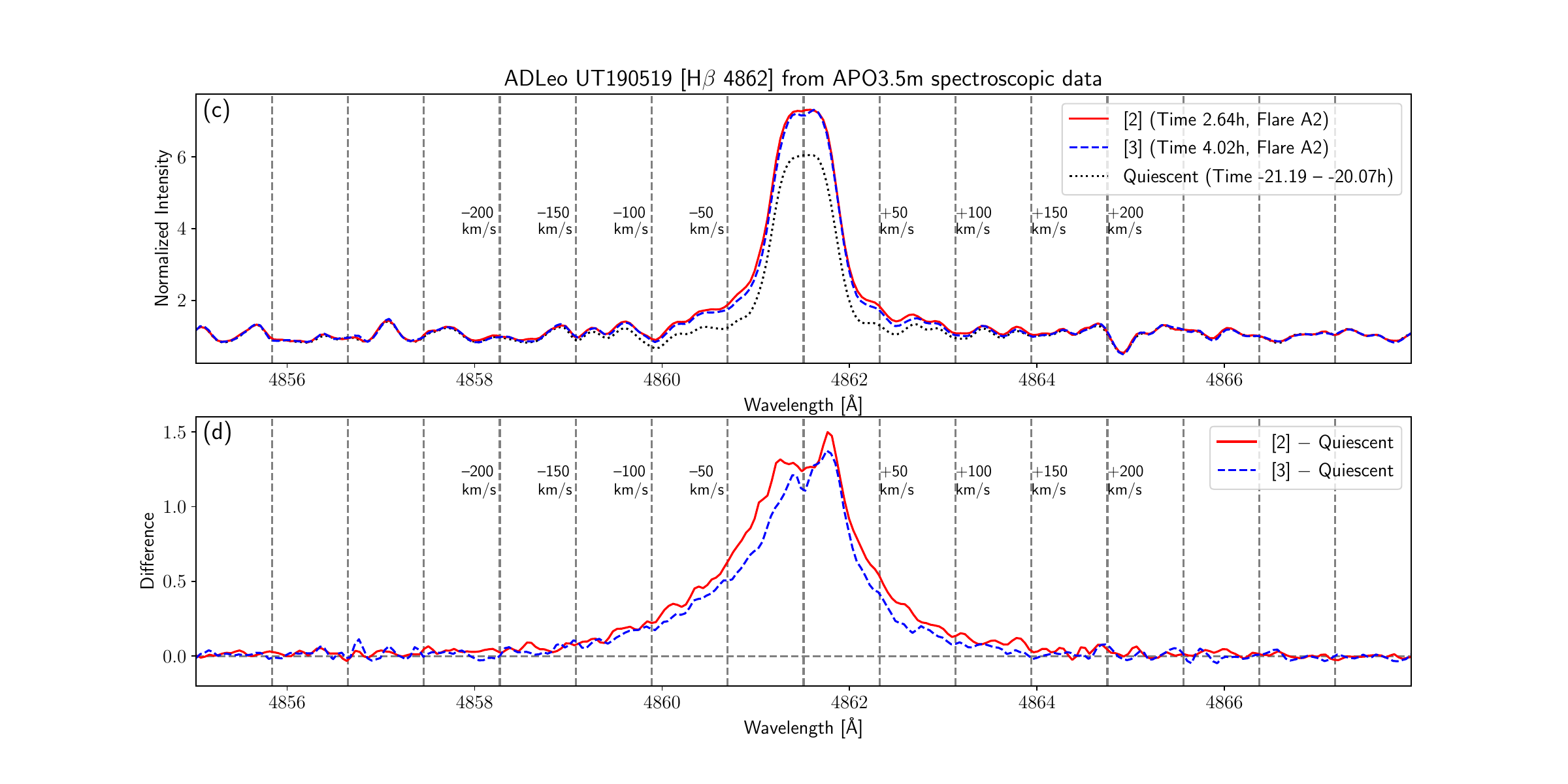}{0.58\textwidth}{\vspace{0mm}}
    }
     \vspace{-0.3cm}
     \caption{
     \color{black}\textrm{  
Line profiles of the H$\alpha$ \& H$\beta$ emission lines during Flare A3  (at the time [1] or [2], and [3]) on 2019 May 19 from APO3.5m spectroscopic data, which are plotted similarly with Figure \ref{fig:spec_HaHb_YZCMi_UT190127}.
The black dotted lines indicate the line profiles in quiescent phase, which are the average profile during -21.19h -- -20.07h from the data on 2019 May 18 
(2.81 -- 3.93h in Figure \ref{fig:lcEW_HaHb_ADLeo_UT190518} (a)).
 } \color{black}
     }
   \label{fig:spec_HaHb_ADLeo_UT190519}
   \end{center}
 \end{figure}
 
      \begin{figure}[ht!]
   \begin{center}
      \gridline{  
     \hspace{-0.07\textwidth}
   \fig{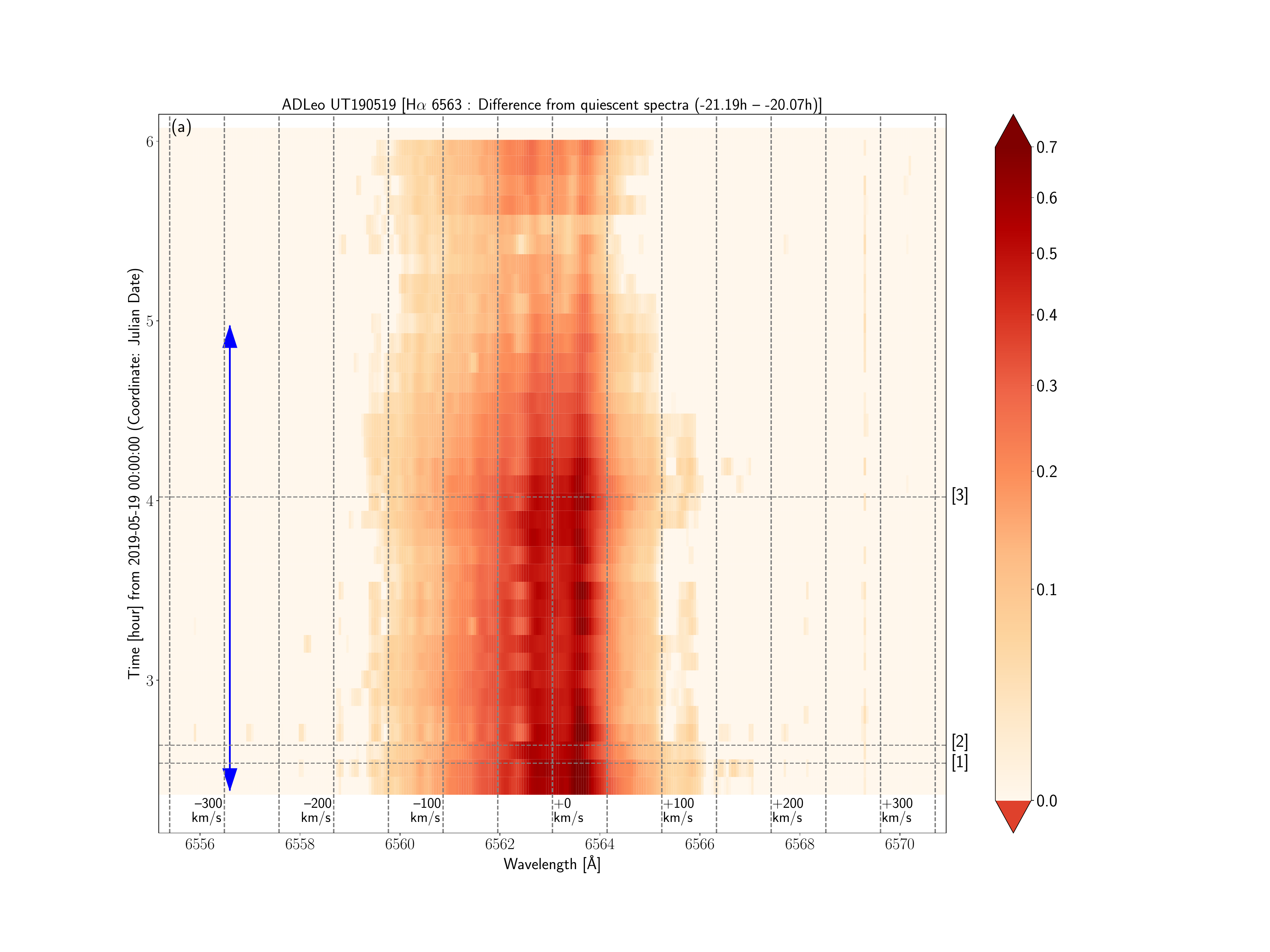}{0.63\textwidth}{\vspace{0mm}}
     \hspace{-0.11\textwidth}
    \fig{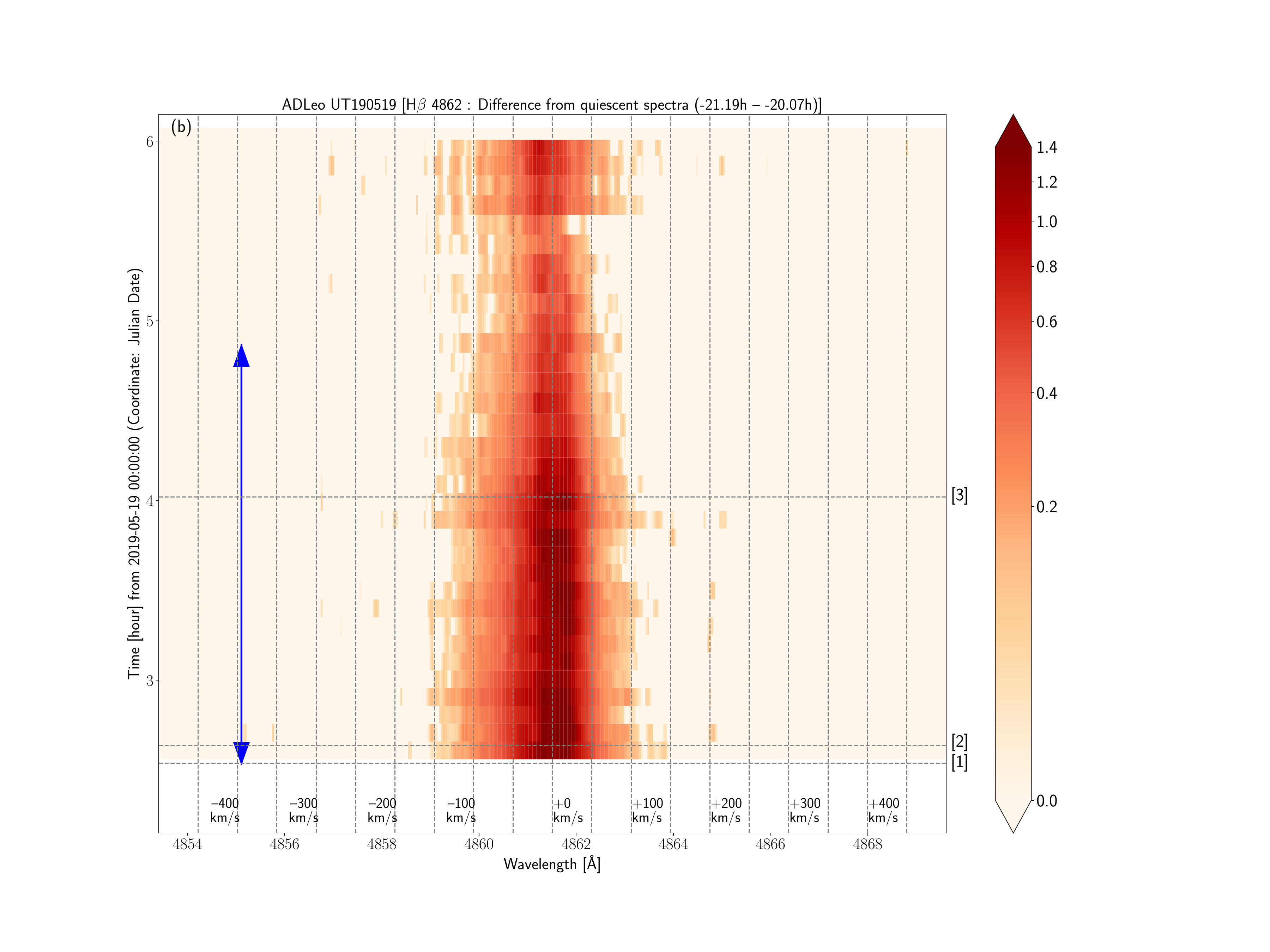}{0.63\textwidth}{\vspace{0mm}}
    }
     \vspace{-5mm}
     \caption{
\color{black}\textrm{  
Time evolution of the H$\alpha$ \& H$\beta$ line profiles covering Flare A3
on 2019 May 19, which are shown similarly with Figure \ref{fig:map_HaHb_YZCMi_UT191212}.
The grey horizontal dashed lines indicate the time [1] -- [3], 
which are shown in Figure \ref{fig:lcEW_HaHb_ADLeo_UT190519}  (light curves) and Figure \ref{fig:spec_HaHb_ADLeo_UT190519} (line profiles).
 } \color{black}
     }
   \label{fig:map_HaHb_ADLeo_UT190519}
   \end{center}
 \end{figure}

The H$\alpha$ \& H$\beta$ line profiles during Flare A3 \color{black}\textrm{are} \color{black} shown in
Figures \ref{fig:spec_HaHb_ADLeo_UT190519} \& \ref{fig:map_HaHb_ADLeo_UT190519}. 
It is noted that the data on 2019 May 18 (cf. Figures \ref{fig:lcEW_HaHb_ADLeo_UT190518} (a)) are used for quiescent profiles in these figure, 
since the quiescent phase data are limited (or there could be no quiescent phase) on 2019 May 19 as seen in Figure 
\color{black}\textrm{\ref{fig:lcEW_HaHb_ADLeo_UT190519}} \color{black}.
During Flare A3, 
the blue wings of H$\alpha$ and H$\beta$ lines were enhanced up to -150 -- -200 km s$^{-1}$ (time [1]--[3] in Figures \ref{fig:spec_HaHb_ADLeo_UT190519} (b) \& (d)).
These blue wing asymmetries continued
for more than 2 hours until the flare decayed (Figure \ref{fig:map_HaHb_ADLeo_UT190519}).

\begin{figure}[ht!]
   \begin{center}
            \gridline{  
     \hspace{-0.06\textwidth}
    \fig{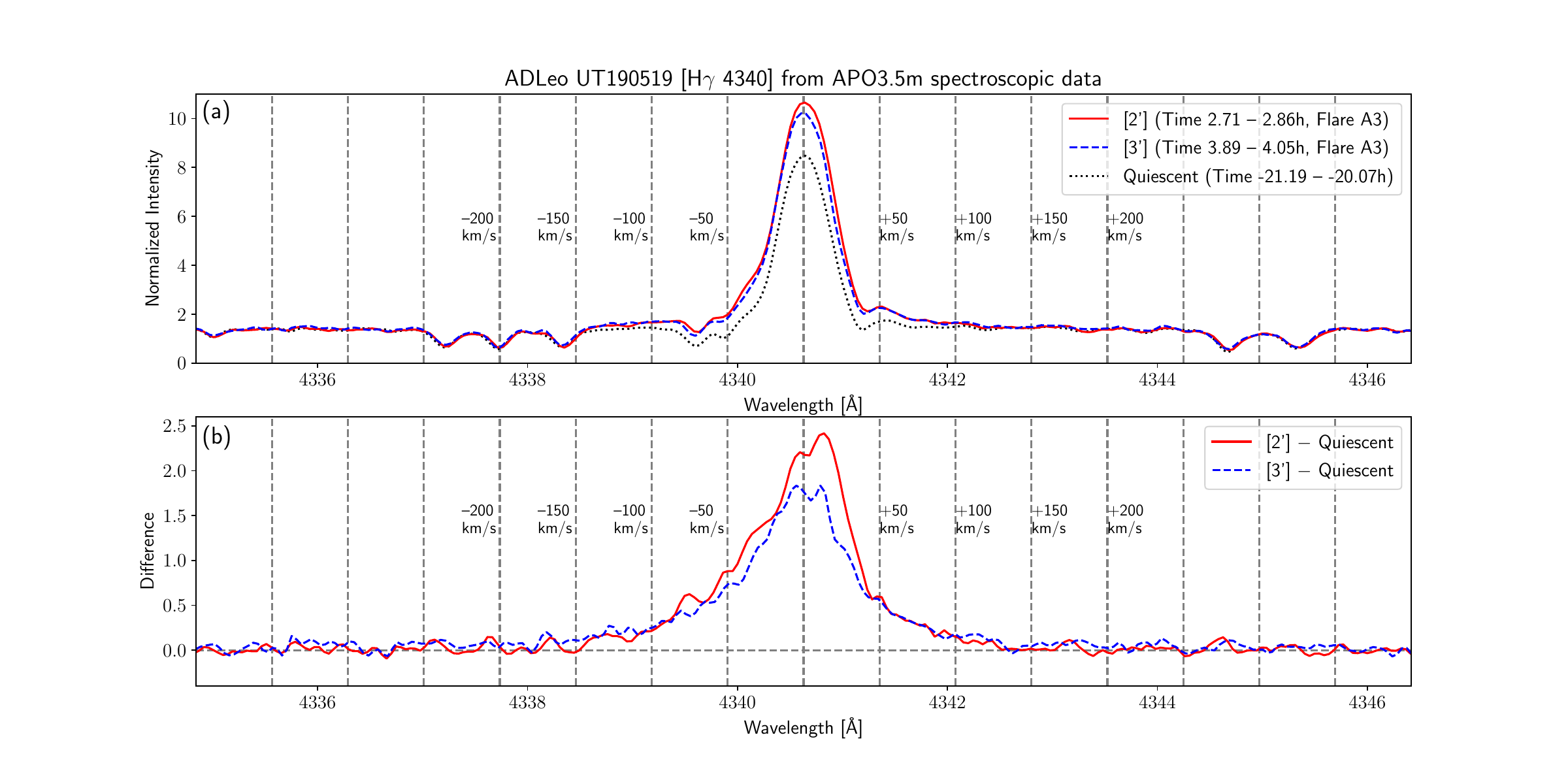}{0.58\textwidth}{\vspace{0mm}}
     \hspace{-0.06\textwidth}
       \fig{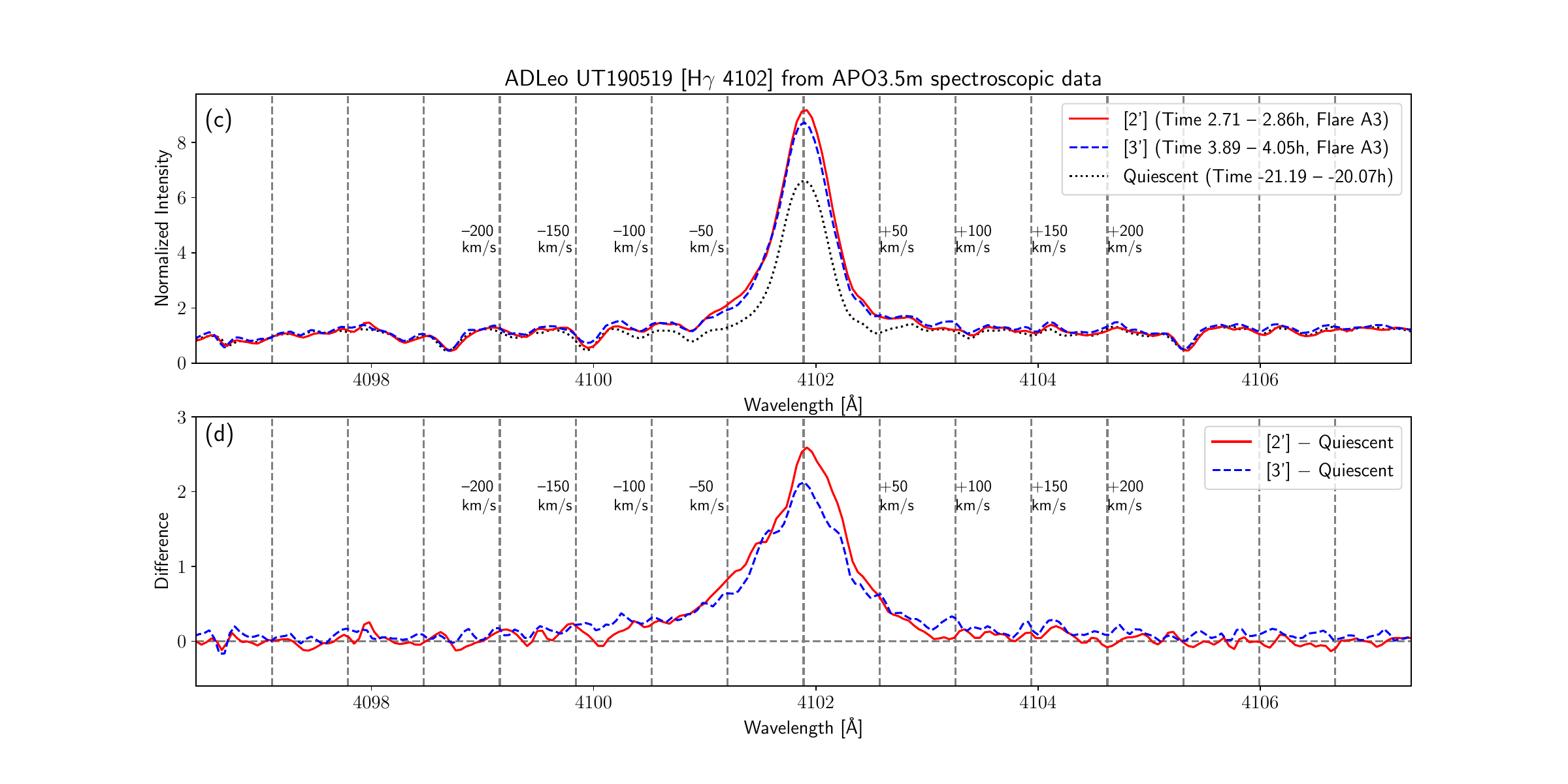}{0.58\textwidth}{\vspace{0mm}}
    }    
   \vspace{-1.0cm}
            \gridline{  
     \hspace{-0.06\textwidth}
    \fig{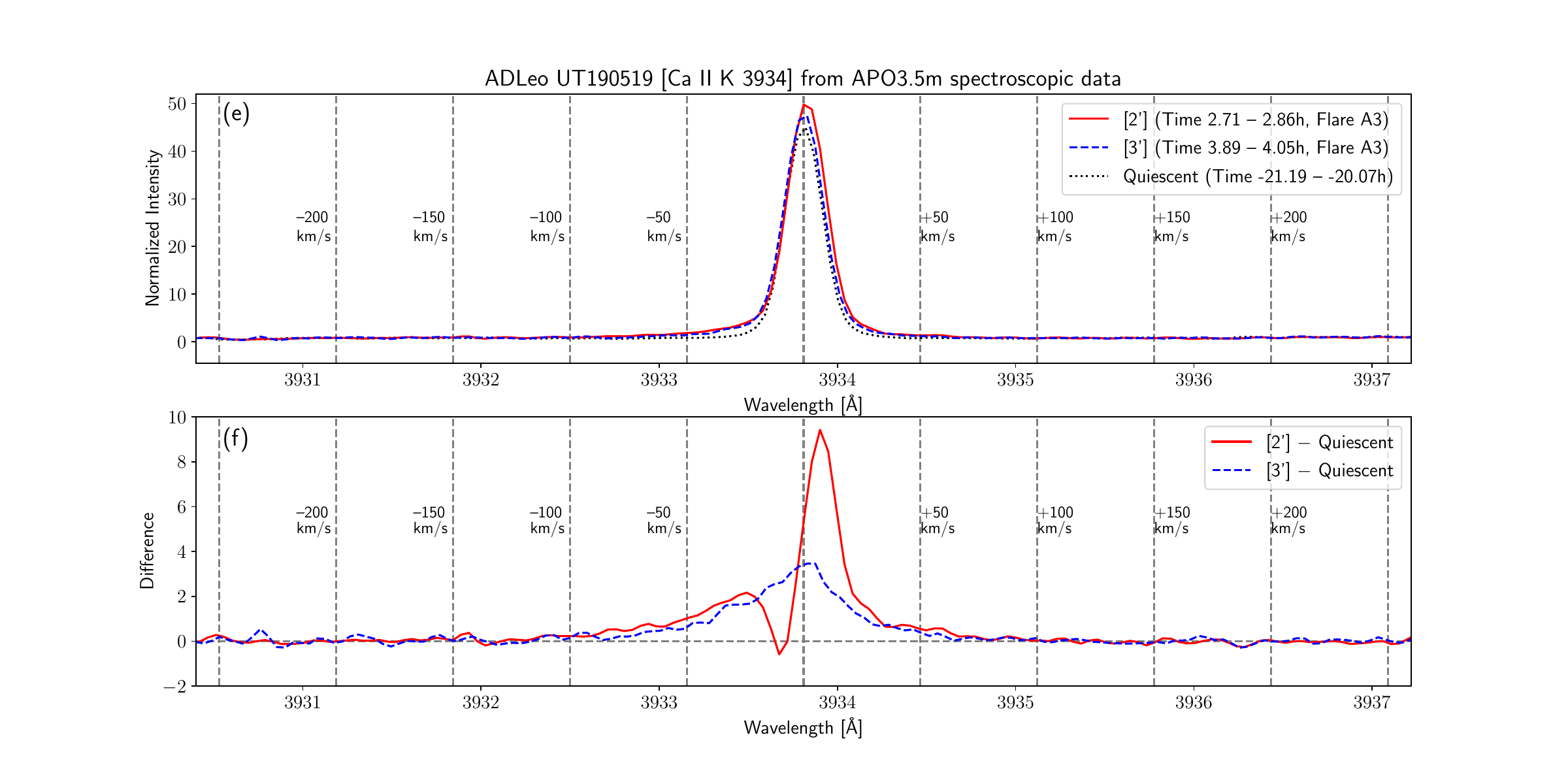}{0.58\textwidth}{\vspace{0mm}}
     \hspace{-0.06\textwidth}
         \fig{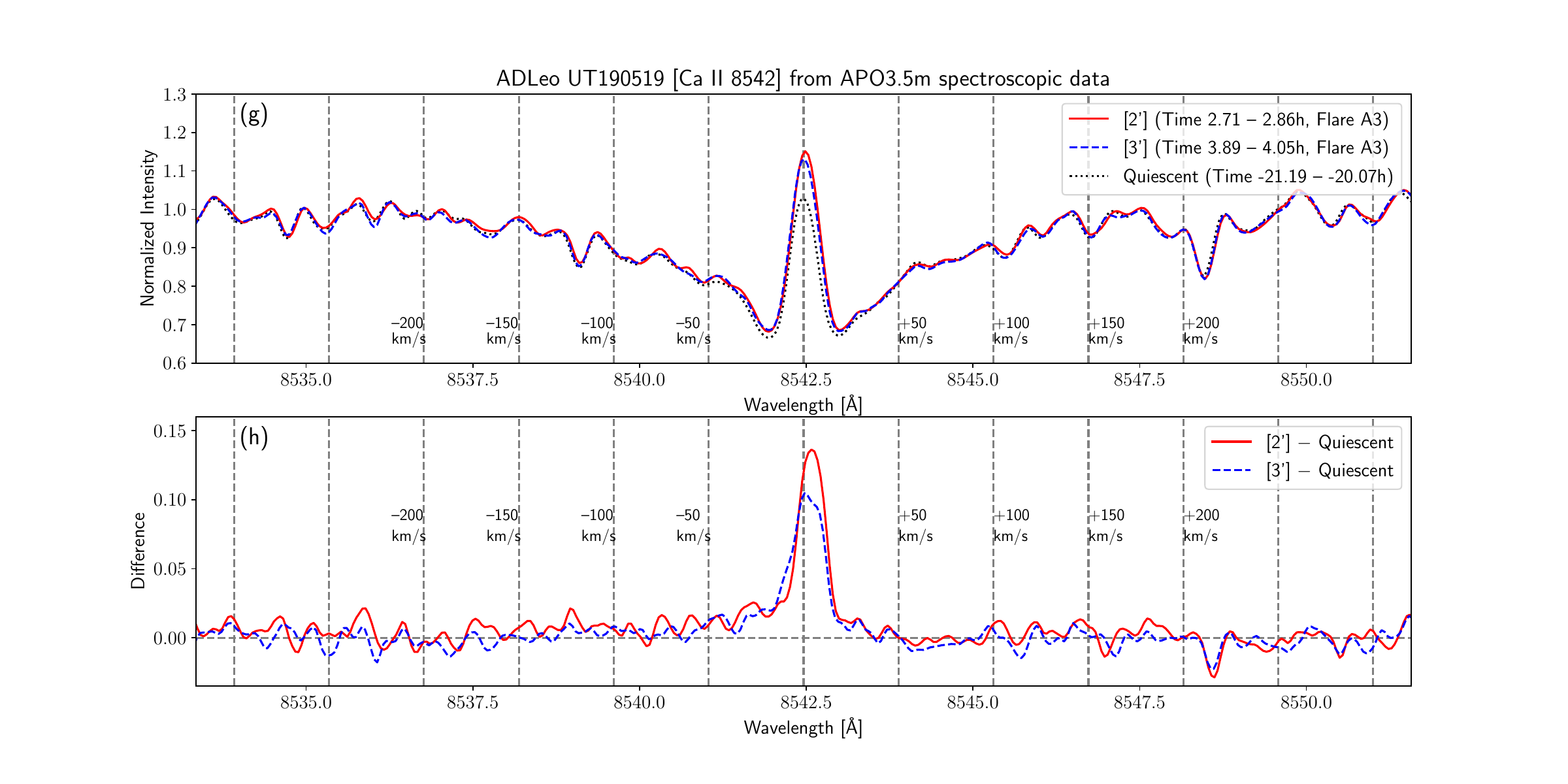}{0.58\textwidth}{\vspace{0mm}}
    }    
     \vspace{-1.0cm}
    \gridline{  
     \hspace{-0.06\textwidth}
    \fig{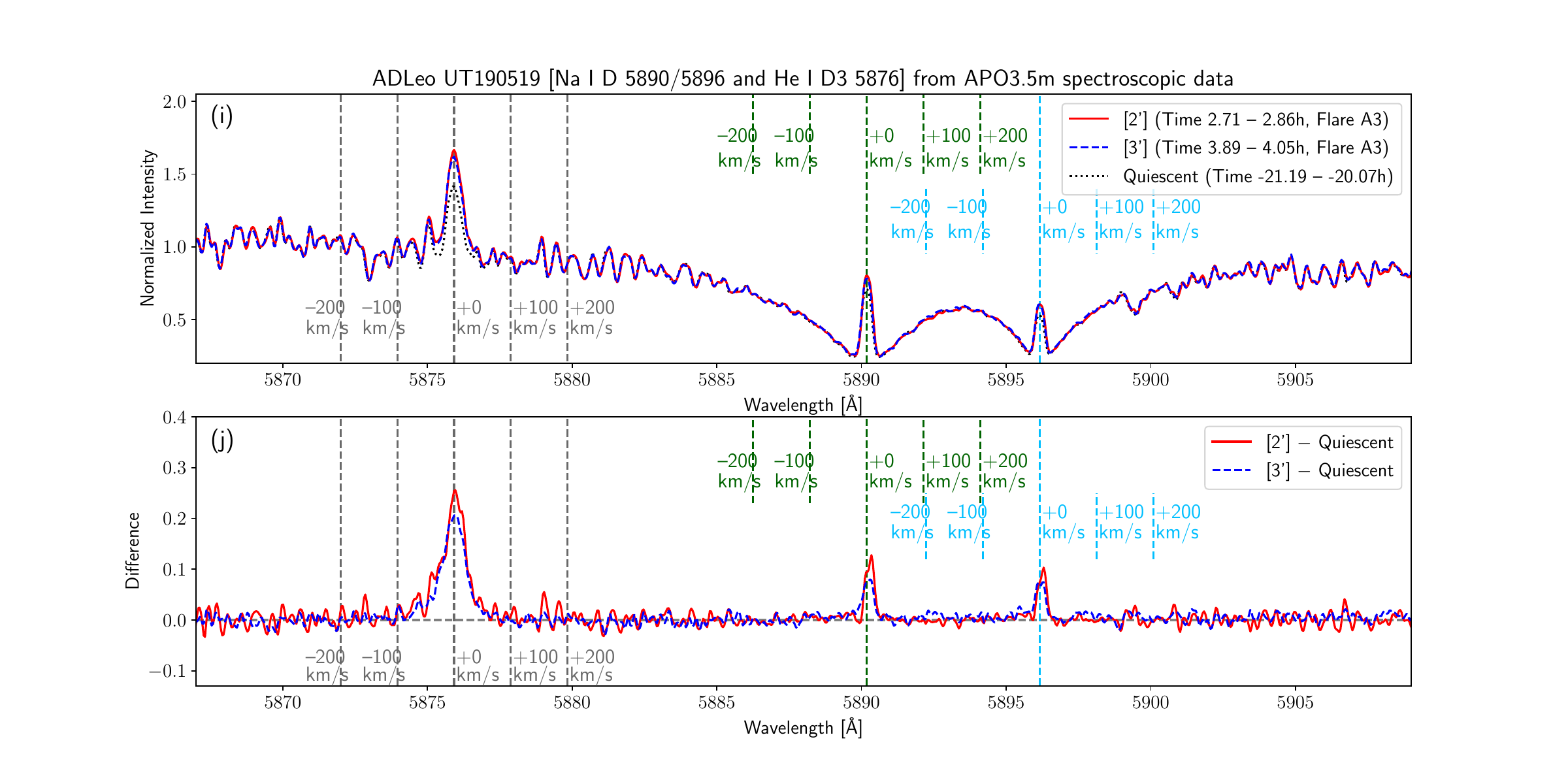}{0.58\textwidth}{\vspace{0mm}}
     \hspace{-0.06\textwidth}
         \fig{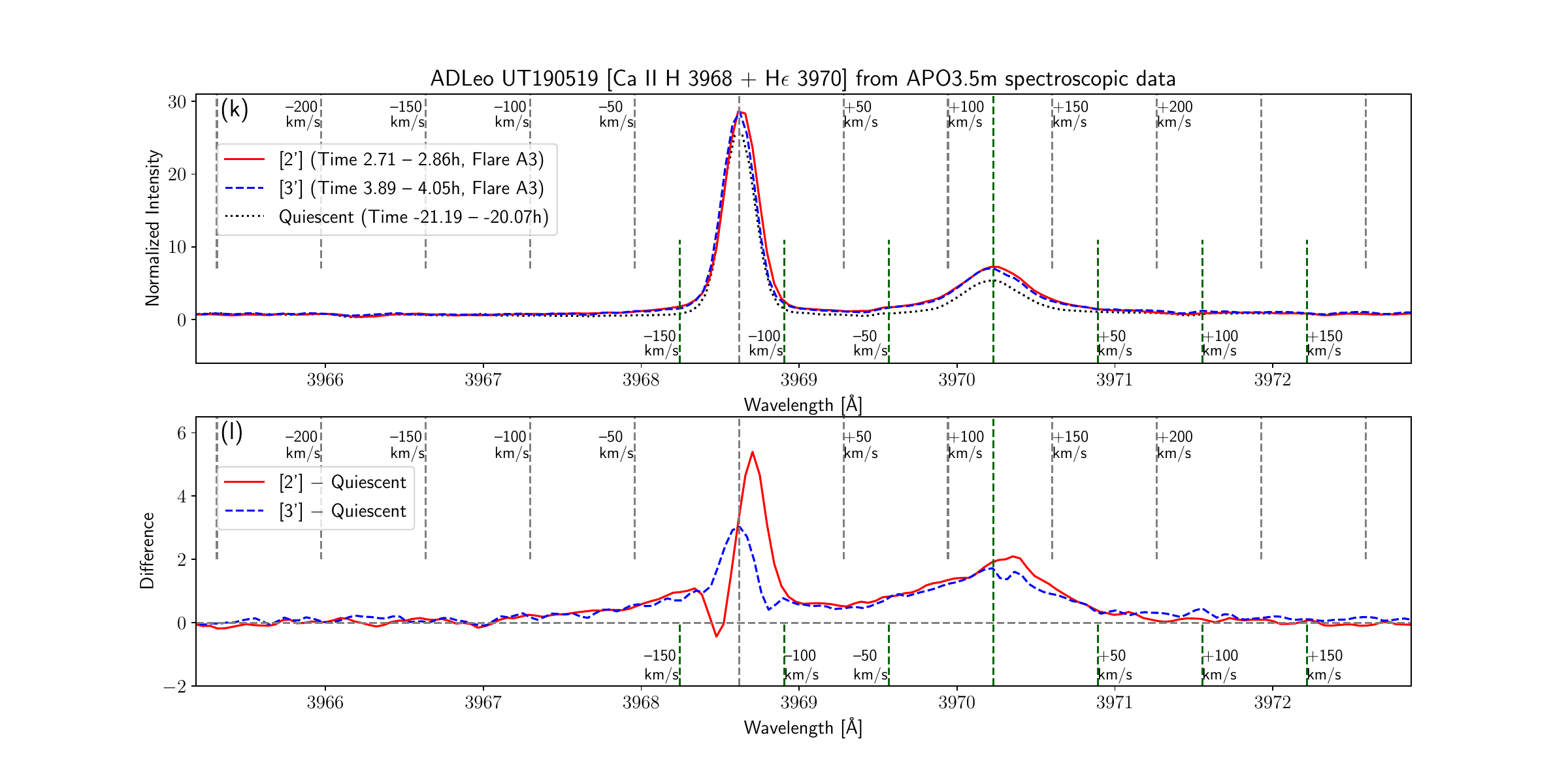}{0.58\textwidth}{\vspace{0mm}}
    }
     \vspace{-0.7cm}
     \caption{
\color{black}\textrm{  
(a)\&(b)
Line profiles of the H$\gamma$ emission line during Flare A3 on 2019 May 19 from APO3.5m spectroscopic data, which are similarly plotted with Figure \ref{fig:spec_HaHb_ADLeo_UT190519}.
The blue solid and red dashed lines indicate 
the integrated line profiles over the time [2$^{\prime}$](Time 2.71 -- 2.86h) and 
[3$^{\prime}$] (Time 3.89 -- 4.05h) on this date, which include the time [2] and [3] in Figure \ref{fig:lcEW_HaHb_ADLeo_UT190519} (light curve), respectively.
(c)\&(d), (e)\&(f), (g)\&(h), (i)\&(j), and (k)\&(l)
Same as panels (a)\&(b), but for H$\delta$, Ca II K, Ca II 8542, Na I D1 \& D2 (5890 \& 5896)$+$He I D3 5876, and H$\epsilon+$Ca II H lines, respectively.
 } \color{black}
     }
   \label{fig:spec_other_ADLeo_UT190519}
   \end{center}
 \end{figure}

The EW light curves of H$\gamma$, H$\delta$, Ca II K, Ca II 8542, Na I D1 \& D2, and He I D3 5876 lines are also shown in Figures 
\ref{fig:lcEW_HaHb_ADLeo_UT190519} (c), (d), \& (e).
The profiles of these lines and Ca II H \& H$\epsilon$ lines during Flare A3 
are shown in Figure \ref{fig:spec_other_ADLeo_UT190519}.
As for H$\gamma$, H$\delta$, H$\epsilon$, Ca II H\&K, Ca II 8542, 
and He I D3 lines, 
the blue wing asymmetries similar to H$\alpha$ \& H$\beta$ lines are seen during Flare A3, though the velocities of peak wing enhancements are different.

\clearpage

\section{Discussions} \label{sec:discussions}

\subsection{Luminosities and Energies of flares in photometric bands and H$\alpha$ line} \label{subsec:dis:flare-energy}

In this study, we observed flares in chromospheric lines (e.g., H$\alpha$ line) and white-light continuum emission bands (e.g., $u$- \& $g$-bands), as summarized in Section \ref{subsec:results:obs-summary} and Table \ref{table:list1_flares}.
\color{black}\textrm{
Among the total 41 flares observed in this paper,
6 flares (Flares Y2, Y22, E1, E7, A1, \& A3 marked with ``NEP" in Table \ref{table:list1_flares}) 
do not have appropriate data sets for 
judging 
whether the flares showed corresponding white-light continuum flux enhancements.
As for the four (Flares Y2, E7, A1, \&A3) among these six flares, the initial part of the flare time evolution 
was not observed both in the spectroscopic and photometric data, 
while the other two flares (Flares Y22 and E1) have large data gaps 
in the photomeric data.
We classified the remaining 35 flares 
into white-light (WL) flares and non white-light (NWL) flares.
The procedure is summarized as follows:
} \color{black}

\begin{enumerate} 
\renewcommand{\labelenumi}{(\roman{enumi})}

\item  \color{black}\textrm{
As also described in Section \ref{subsec:quiescent-ene}, 
if the relative flux ($\Delta f_{\rm{band, flare}}(t)$) shows the increase whose peak amplitude is larger than the photometric error ($3\sigma_{\rm{band}}$) and the associated \color{black}\textrm{flare decays }\color{black} over multiple data points, we judge that flare emission is identified in the photometric band.
} \color{black}

\item  \color{black}\textrm{ 
The M-dwarf flare amplitudes are 
generally larger in blue bands as seen for Flare Y4 in Figure \ref{fig:lcEW_HaHb_YZCMi_UT190128} 
as well as in previous studies (e.g., \citealt{Hawley+1991}; \citealt{Namekata+2020_PASJ})
since flare optical continuum spectra 
have much higher temperature than those in the quiescent phase 
(\citealt{Kowalski+2010} \& \citeyear{Kowalski+2019}; \citealt{Howard+2020_ApJ}).
Considering this point, flares are classified as white-light (WL) flares in this paper if the flare emissions in $U$- or $u$-bands are identified.
One exception is Flare Y1, which showed clear white-light emissions in $g$- \& $TESS$-bands while there are no available $u$-band observation data (Figure \ref{fig:lcEW_HaHb_YZCMi_UT190126}). 
If there are no flare emissions identified in any photometric bands with the above threshold, the flare is identified as non white-light (NWL) flares.
} \color{black}

\item \color{black}\textrm{
It is noted that the threshold $3\sigma_{\rm{band}}$ may depend on the data quality (S/N) of each night. The classification of white-light flares and non white-light flares
could be somewhat affected from this point. 
Moreover, flare colors in the optical band can include some variety among events (cf. \citealt{Kowalski+2019}), and the WL/NWL classification based on one band 
($U$- or $u$-band in this study)
could leave us some bias.
However, the purpose of the WL/NWL classification in this study
is only to show that the blue wing asymmetries can exist both in clear WL flares 
and candidate NWL flares (See Section \ref{subsec:dis:flare-blue}).
In other words, it is sufficient to investigate whether each flare shows white-light emissions within the available dataset for each flare (whose data quality has some variety among each event).
From this point of view,  
a detailed statistical classification of WL/NWL flares is beyond the scope of this paper, considering that most of the photometric data are from the ground-based observations with small ARCSAT and LCO telescopes including some data gaps.
We note here that the future studies on the WL/NWL associations during H$\alpha$\&H$\beta$ flares are necessary with more comprehensively and more well-observed dataset (e.g., $TESS$-like high precision space photometry, in blue optical wavelength band).
} \color{black}

\end{enumerate} 

\color{black}\textrm{
As a result, 31 flares showed corresponding white-light continuum flux enhancements, 
and are classified here as white-light flares (marked with ``WL" in Table \ref{table:list1_flares}). 
The remaining 4 flares (Flares Y3, Y5, Y6, and Y26 marked with ``NWL" in Table \ref{table:list1_flares}) are classified as ``candidate" non white-light
flares in this study.
It is noted three  (Flares Y3, Y5, and Y26) 
among these 4 flares
showed marginal white-light increases 
comparable to photometric errors (see Figures \ref{fig:lcEW_HaHb_YZCMi_UT190127}, \ref{fig:lcEW_HaHb_YZCMi_UT190128}, \& 
\ref{fig:lcEW_HaHb_YZCMi_UT201207}), 
while the other one Flare Y6 showed white-light emission peaks in late phase of the H$\alpha$\&H$\beta$ flare though we judged that there are no clear WL emissions that are considered to be physically associated with the early main increasing phase of the the H$\alpha$\&H$\beta$ flare (see Figure \ref{fig:lcEW_HaHb_YZCMi_UT191212}).
} \color{black}

The flare energy partition among different wavelengths 
is an important topic of stellar flares since this can have constraints on 
how flare energy release occur in the different layers of flaring atmosphere from photosphere to corona (e.g., \citealt{Osten+2015}; \citealt{Guarcello+2019}; \citealt{Stelzer+2022}).
We here briefly mention this topic on the basis of our observation data,
though the main topic of this paper is blue wing asymmetries of chromospheric lines and detailed discussions on the flare energy partition are beyond the scope of this paper.

\begin{figure}[ht!]
   \begin{center}
            \gridline{  
    \fig{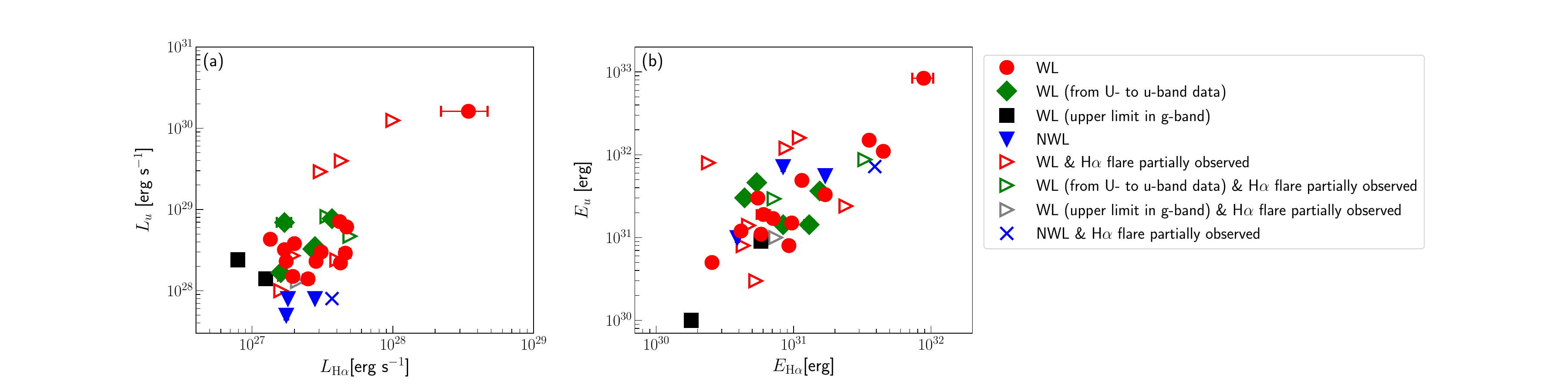}{1.12\textwidth}{\vspace{0mm}}
    }
   \vspace{-10mm}
            \gridline{  
    \fig{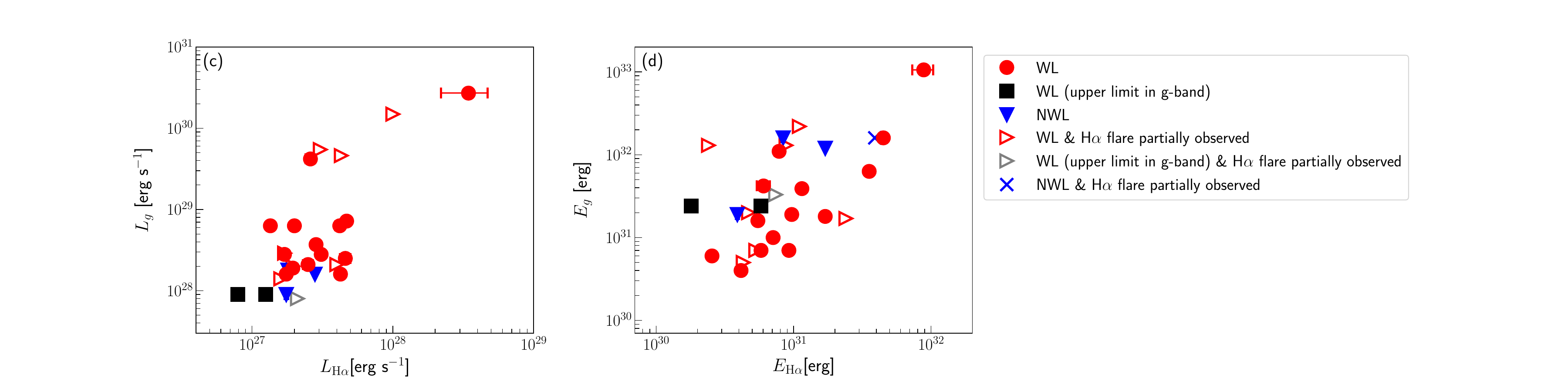}{1.12\textwidth}{\vspace{0mm}}
    }
   \vspace{-5mm}
     \caption{
     \color{black}\textrm{
Scatter plot of flare peak luminosities and energies in $u$- \& $g$- bands ($L_{u}$, $E_{u}$, $L_{g}$, and $E_{g}$) 
and H$\alpha$ line 
($L_{\rm{H}\alpha}$ and $E_{\rm{H}\alpha}$). 
Red filled circles represent the flares identified as white-light flares 
(``WL" in Table \ref{table:list1_flares}) 
and whose H$\alpha$ flare emission was observed from the flare start to end. 
Green filled diamonds represent the flares with the same properties as those marked with red filled circles, but their $L_{u}$ and $E_{u}$ values are 
converted from their LCO $U$-band values assuming the luminosity ratio between these two bands (cf. Table \ref{table:targets_quiescent_flux}), since these stars were observed not in the ARCSAT $u$- \& $g$-bands data 
but in the LCO $U$-band (The $L_{g}$ and $E_{g}$ values of these stars are not available, so these stars are not in (c) \& (d)). It is noted that because of some data gaps of the LCO $U$-band photometric observations, 
the plotted $E_{u}$ values can be lower limit values, 
although the $L_{u}$ values are less likely lower limit values since the peaks corresponding to the H$\alpha$ flare peaks are already selected (see also footnotes of Table \ref{table:list1_flares}).
Black filled squares are flares with the same properties as those marked with red filled circles, but \color{black}\textrm{the upper limit values are estimated for $L_{g}$, and $E_{g}$ values, 
while the exact $L_{u}$, and $E_{u}$ values of these flares are measured. }\color{black}
Blue downward triangles are 
the flares identified as ``candidate" non white-light flares 
(``NWL" in Table \ref{table:list1_flares}). 
As for these stars, the plotted $L_{u}$ values show the upper limit values from Table \ref{table:list1_flares}.
Red open rightward triangles represent 
the flares identified as white-light flares but whose H$\alpha$ flare phase was only partially observed (only the lower limit values for $\Delta t^{\rm{flare}}_{\rm{H}\alpha}$ are listed in Table \ref{table:list1_flares}). As for these stars, the plotted $L_{\rm{H}\alpha}$ values can be the lower limit values.
Green open rightward triangles are the flares with the same properties as those marked with red open rightward triangles but their $u$-band values are 
converted from their LCO $U$-band values as the green filled diamonds.
Gray open rightward triangles are the flares with the same properties as those marked with red open rightward triangles but \color{black}\textrm{the upper limit values are estimated for $L_{g}$, and $E_{g}$ values, while the exact $L_{u}$, and $E_{u}$ values of these flares are measured}\color{black}. 
Blue cross marks are the flares identified as ``candidate" non white-light flares 
(``NWL" in Table \ref{table:list1_flares}) and whose H$\alpha$  flare emission was only partially observed. 
The six flares (``NEP" in Table \ref{table:list1_flares}) 
without enough data for judging whether the flares are WL or NWL flares 
are not included in this figure.
} \color{black}
}
   \label{fig:Lg_LHa_Eg_EHa}
   \end{center}
 \end{figure}
 
In Figure \ref{fig:Lg_LHa_Eg_EHa}, we compare flare peak luminosities and energies 
in photometric bands ($u$- \& $g$-bands) and H$\alpha$ line. 
\color{black}\textrm{
 Figure \ref{fig:Lg_LHa_Eg_EHa}(b) suggests a rough correlation between the flare energies especially between $u$-band and 
H$\alpha$ line but detailed quantitative conclusions are beyond the scope of this paper 
considering some uncertainties of the observation data available in this study (e.g., there 
are some gaps in the photometric data and many flares only partially observed as shown with various symbols in this figure).
} \color{black}
We will come back to this point in our future paper, 
discussing in detail the differences of time evolution of various chromospheric lines during stellar flares (e.g., \citealt{Kowalski+2013}).
In addition, related with this topic, soft X-ray energy of Flare Y3 is mentioned in Section \ref{subsec:dis:Xray}.

\subsection{Flares showing blue wing asymmetries} \label{subsec:dis:flare-blue}

In this study, \color{black}\textrm{41 }\color{black} flares were detected from the total 31 night observations, 
as summarized in Section \ref{subsec:results:obs-summary} and Table \ref{table:list1_flares}.
Among these, 7 flares (Flares Y3, Y6, Y18, Y23, E1, E2, \& A3 in Section \ref{subsec:results:2019-Jan-27} -- \ref{subsec:results:2019-May-19}) showed clear blue wing asymmetries in H$\alpha$ line. 
Various notable properties, which are described in Section \ref{subsec:results:2019-Jan-27} -- \ref{subsec:results:2019-May-19}, are summarized in Table \ref{table:list_blue_flares}. 
For reference, \color{black}\textrm{three} \color{black} flares with H$\alpha$ blue wing asymmetries in \color{black}\textrm{\citet{Vida+2016}}\color{black}, \citet{Honda+2018}, and \citet{Maehara+2021} are also listed in this Table \ref{table:list_blue_flares} (\color{black}\textrm{``V2016"}\color{black}, ``H2018", and ``M2021", respectively).
In Table \ref{table:list_blue_flares}, we list $v^{\rm{H}\alpha}_{\rm{blue,max}}$ values, which are the maximum velocities of blue wing enhancements of H$\alpha$ line measured by eye. 
The same velocity values for other lines showing blue asymmetries (e.g., $v^{\rm{H}\beta}_{\rm{blue,max}}$) are also listed in Table \ref{table:velmass_blue_flares}.
We discuss blue wing asymmetry velocities more in detail by the line fitting method in Section \ref{subsec:dis:blue-ejection}. 
As summarized in Table \ref{table:list_blue_flares} and described in the following, 
there are various correspondences in flare properties (e.g., durations of blue wing asymmetries, intensities of white-light emissions, blue wing asymmetries in various chromospheric lines).

\begin{figure}[ht!]
   \begin{center}
            \gridline{  
    \fig{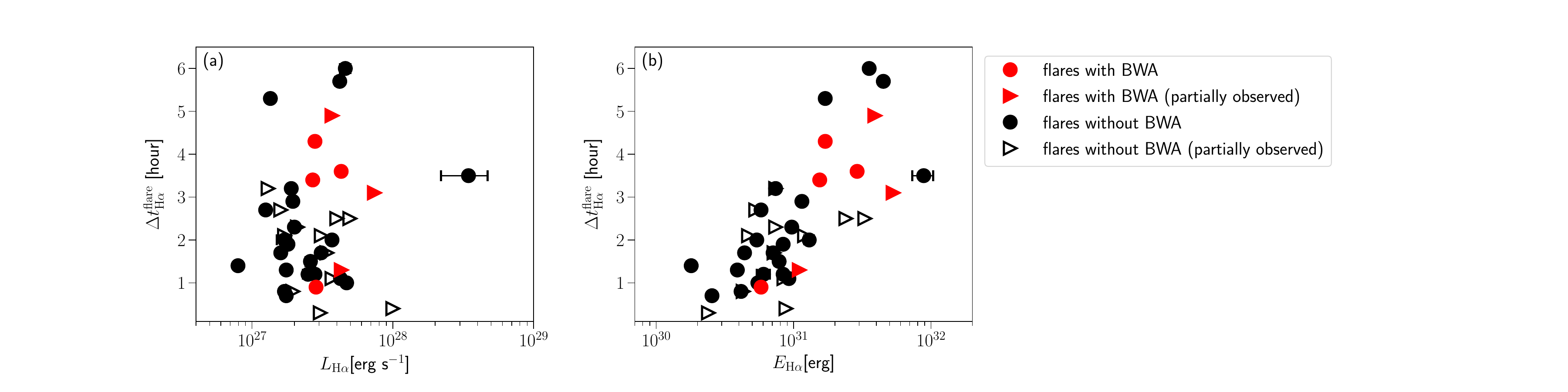}{1.12\textwidth}{\vspace{0mm}}
}
   \vspace{-5mm}
     \caption{
     \color{black}\textrm{   
(a) Scatter plot of the H$\alpha$ flare peak luminosity ($L_{\rm{H}\alpha}$) and  H$\alpha$ flare duration ($\Delta t_{\rm{H}\alpha}^{\rm{flare}}$). 
Red filled circles and red rightward filled triangles 
represent the flares with blue wing asymmetries (BWA), while 
the latter ones correspond to the flares whose H$\alpha$ flare phase was only partially observed (only the lower limit values for $\Delta t^{\rm{flare}}_{\rm{H}\alpha}$ are listed in Tables \ref{table:list1_flares} \& \ref{table:list_blue_flares}).
As for these stars, the plotted $L_{\rm{H}\alpha}$ values can be the lower limit values.
Black \color{black}\textrm{filled }\color{black} circles and black rightward open triangles are the same as the above red points but for the flares without BWAs.
(b) Same as (a), but for the H$\alpha$ flare energy ($E_{\rm{H}\alpha}$) and H$\alpha$ flare duration ($\Delta t_{\rm{H}\alpha}^{\rm{flare}}$). 
} \color{black}
}
   \label{fig:blue_LEt_all}
   \end{center}
 \end{figure}

Figure \ref{fig:blue_LEt_all} shows the scatter plots of the 
H$\alpha$ flare peak \color{black}\textrm{luminosity}\color{black}, energy, and duration values of
the 7 flares with blue wing asymmetries and the remaining 34 flares observed in this study. Blue wing asymmetries could be seen both in relatively 
large/long and small/short flares, 
although it would be difficult to statistically conclude this point
only from the limited number of observed samples in this study.

\begin{figure}[ht!]
   \begin{center}
            \gridline{  
    \fig{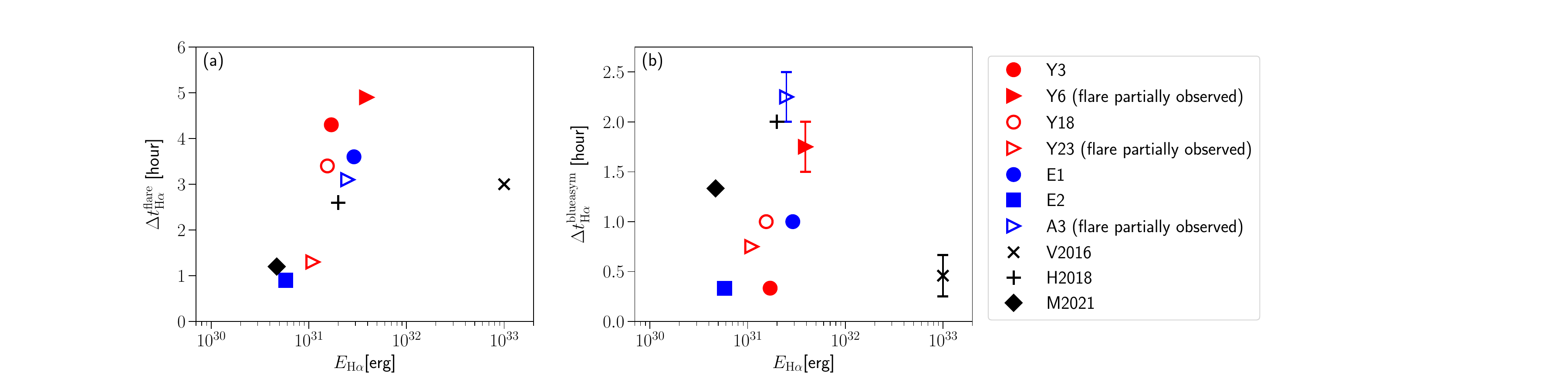}{1.12\textwidth}{\vspace{0mm}}
    }
   \vspace{-5mm}
     \caption{
\color{black}\textrm{  
Scatter plots of the H$\alpha$ flare energy ($E_{\rm{H}\alpha}$), H$\alpha$ flare duration ($\Delta t^{\rm{flare}}_{\rm{H}\alpha}$), 
and duration of H$\alpha$ blue asymmetries ($\Delta t_{\rm{H}\alpha}^{\rm{blueasym}}$) for the flares with blue wing asymmetries (cf. Table \ref{table:list_blue_flares}). 
In addition to the 7 flares reported in this study,
three events from the previous papers (V2016, H2018, and M2021 in Table \ref{table:list_blue_flares}) are also plotted.
As for Flares Y6, Y23, \& A3, \color{black}\textrm{the H$\alpha$ flare phase was only partially }\color{black} observed, and the plotted $E_{\rm{H}\alpha}$ and $\Delta t_{\rm{H}\alpha}^{\rm{blueasym}}$ values can be the lower limit values.
As for Flare Y3, $\Delta t_{\rm{H}\alpha}^{\rm{blueasym}}$=20min $\times$ 2 is listed in Table \ref{table:list_blue_flares}, but 
the single data point of $\Delta t_{\rm{H}\alpha}^{\rm{blueasym}}$= 20 min = 0.33 hour is 
only plotted here.
 } \color{black}
}
   \label{fig:blue_LEt_BWA}
   \end{center}
 \end{figure}

\color{black}\textrm{
The duration of H$\alpha$ blue wing asymmetries ($\Delta t_{\rm{H}\alpha}^{\rm{blueasym}}$ in Table \ref{table:list_blue_flares}) ranges from 20 min to 2.5 hours (Figure \ref{fig:blue_LEt_BWA}).
Comparing Figures \ref{fig:blue_LEt_BWA}(a) and (b), there is some variation
among the relations of $\Delta t^{\rm{flare}}_{\rm{H}\alpha}$ and $\Delta t_{\rm{H}\alpha}^{\rm{blueasym}}$.
} \color{black}
As a notable example, Flare Y3 showed clear short-lived H$\alpha$ blue wing asymmetries twice 
\color{black}\textrm{(20min$\times$2 at the times [3] and [5]) during the entire Flare Y3 in H$\alpha$ line lasting over 4 hours (Figures \ref{fig:lcEW_HaHb_YZCMi_UT190127} \& \ref{fig:map_HaHb_YZCMi_UT190127}).} \color{black}
Similarly, \citet{Vida+2016} also reported three distinct blue wing enhancements
spanning more than three hours \color{black}\textrm{
(``V2016" in Table \ref{table:list_blue_flares})}\color{black}.
In contrast, Flares Y23, E1, \& A3 showed H$\alpha$ blue wing asymmetries over almost all the observed phases of the flares (Figures \ref{fig:map_HaHb_YZCMi_UT201206}, \ref{fig:map_HaHb_EVLac_UT191215}, \& \ref{fig:map_HaHb_ADLeo_UT190519}), 
although initial phases of the flares were not observed during Flares Y23 \& A3.
Similarly, \citet{Honda+2018} also reported a continuous blue asymmetry of H$\alpha$ line over all phase of the flare (H2018 in Table \ref{table:list_blue_flares}). 
As another notable point, blue \color{black}\textrm{wing} \color{black} asymmetry velocities 
showed gradual decays during Flares Y6 \& Y23 (Figures \ref{fig:map_HaHb_YZCMi_UT191212} \& \ref{fig:map_HaHb_YZCMi_UT201206}).
In particular, Flare Y6 showed clear H$\alpha$ blue wing enhancement (blue wing asymmetry) up to $\sim$ -200 km s$^{-1}$ in early phase of the flare, while the line profile gradually shifted to the red wing enhancement (red wing asymmetry) up to $\sim$ +200 km s$^{-1}$, 
during the H$\alpha$ flare over 4.9 hours (Figure \ref{fig:map_HaHb_YZCMi_UT191212}).
In the middle time between blue wing asymmetry and red wing asymmetry, the H$\alpha$ line profile showed almost symmetric broadening with $\pm$150 km s$^{-1}$.
These red wing asymmetries could be caused by the chromospheric condensation, flare-driven coronal rain or post-flare loop, as summarized in Section \ref{subsec:dis:flares-redsym}.
This example (Flare Y6) may show that both blue and red wing asymmetries of H$\alpha$ line can evidently occur during the same flare of a mid M-dwarf, which suggests dynamic plasma motions upward and downward during the same flare.  
\color{black}\textrm{
It is noted that the possible change from blue wing enhancement to the red wing enhancement during a flare was also reported in \citet{Muheki+2020}.
}\color{black}
However, it can be also possible that Flare Y6 consists of different consecutive flares showing blue wing asymmetries and red wing asymmetries, respectively, 
considering that the flare light curve showed multiple peaks (cf. Figure \ref{fig:lcEW_HaHb_YZCMi_UT191212}).

There is also a notable difference 
among the intensities of red wing of the H$\alpha$ line 
when the blue wing shows an excess enhancement (blue wing asymmetry).
As for Flares Y3, Y6, \& A3, red wing of the H$\alpha$ line is broadened up to $\sim$ +150 km s$^{-1}$ when the blue wing is more enhanced up to $\sim$ -200 km s$^{-1}$ (Figures \ref{fig:map_HaHb_YZCMi_UT190127}, \ref{fig:map_HaHb_YZCMi_UT191212}, \& \ref{fig:map_HaHb_ADLeo_UT190519}).
In contrast, during Flares Y18 \& E2, the red wing of the H$\alpha$ line is broadened only up to $\sim$ +50--100 km s$^{-1}$ when the blue wing is enhanced up to $\sim$ -200 km s$^{-1}$ (Figures \ref{fig:map_HaHb_YZCMi_UT200121} \& \ref{fig:map_HaHb_EVLac_UT191215}).
These differences might suggest that
H$\alpha$ line symmetric broadening or red wing enhancements, which have been often observed during stellar flares (e.g., \citealt{Namekata+2020_PASJ}; \citealt{Wollmann+2023_A&A}), could occur to some extent simultaneously with larger H$\alpha$ blue wing enhancements (see also Section \ref{subsec:dis:flares-redsym}). 
In a relevant context, \citet{Honda+2018} reported  the possible existence of absorption components in the red wing of the H$\alpha$ line when the H$\alpha$ line showed blue wing asymmetry.

\color{black}\textrm{
Intensities of white-light continuum fluxes also showed various properties
even among these 7 flares (the ``WLF" column in Table \ref{table:list_blue_flares}).
Flare Y3 did not show clear white-light continuum flux enhancements, 
while flare emissions were observed for $\gtrsim$4 hours in various chromospheric lines 
and \textit{NICER} soft X-ray data (Figure \ref{fig:lcEW_HaHb_YZCMi_UT190127}; see also Section \ref{subsec:dis:Xray} for detailed discussions of \textit{NICER} soft X-ray data).
There were very small ``suggestive" increases in $u$- \& $g$-bands and \textit{TESS} data 
around time 6--8h in Figure \ref{fig:lcEW_HaHb_YZCMi_UT190127} (b) \& (c), although they are still a bit smaller than the white-light flare detection thresholds (see Section \ref{subsec:results:2019-Jan-27}).
We can speculate that 
these small ``suggestive" increases could be caused by the emission lines 
(e.g., Balmer lines) included in $u$-, $g$-, and $TESS$-bands 
(cf. Figure \ref{fig:filter_YZCMi}).
This flare could be possibly categorized to so-called non white-light flares, which are often seen in the case of solar flares (e.g., \citealt{Watanabe+2017}). 
\citet{Maehara+2021} also reported the H$\alpha$ blue wing asymmetry during a non white-light flare (``M2021" in Table \ref{table:list_blue_flares}).
As for Flare Y6, there are short white-light continuum flux enhancements in the middle/late phase of the flare (around time 10.0--10.5h and 12.0--12.5h in Figure \ref{fig:lcEW_HaHb_YZCMi_UT191212}), 
but there are no other clear white-light enhancements that are considered to be physically associated with the early increasing phase of the whole H$\alpha$ flare, so this flare could be categorized into non white-light flare as the whole flare event.}\color{black}

\color{black}\textrm{
As described in Section \ref{subsec:dis:flare-energy},
the 31 flares are classified as white-light flares among the 35 flares with enough data sets to judge whether the flares are white-light flares.
The remaining 4 flares (Flares Y3, Y5, Y6, and Y26) are classified as non white-light
flares in this study, while three of them (Flares Y3, Y5, and Y26) 
showed slight possible white-light increases 
almost comparable to the photometric errors and Flare Y6 showed white-light emissions in middle/late phase 
of the H$\alpha$\&H$\beta$ flare emission.
As a result, as for the 7 flares with clear blue wing asymmetries discussed here (Flares Y3, Y6, Y18, Y23, E1, E2, \& A3), 
5 flares (Flares Y3, Y6, Y18, Y23, \& E2) have enough datasets 
for judging whether they are \color{black}\textrm{white-light flares}\color{black}.
Among these 5 flares, three flares (Flares Y18, Y23, \& E2) are classified as white-light flares and two flares (Flares Y3\&Y6) are candidate non white-light flares as described in the above. 
These results can suggest that blue wing asymmetries of chromospheric lines can be seen both during ``clear" white-light and ``candidate" non white-light flares.
However, it should be noted the non white-flares in this study could actually be weak white-light flares,
since the ground-based photometry used for most of the flares in this study 
have relatively large photomeric errors and high-precision $TESS$ photometry is only available for six flares and it only observes the red wavelength range (6000--10000\AA). 
This is not the best wavelength range for stellar flare observations compared with blue optical wavelength range (e.g., $U$- \& $u$-bands), since the M-dwarf flares generally have larger amplitudes in blue optical wavelangth range than the red range (\citealt{Hawley+1991}; \citealt{Kowalski+2010}; \citealt{Brasseur+2023_ApJ}).
}\color{black}

In Table \ref{table:list_blue_flares}, 
we list which chromospheric lines showed blue wing asymmetries ([B] and [NB]\color{black}\textrm{, as explained in a footnote of the table}\color{black}) in addition to H$\alpha$ line.
Large variety is seen also for this point.
Among the seven flares with H$\alpha$ blue wing asymmetries, 
all seven flares \color{black}\textrm{in this study} \color{black} showed blue wing asymmetries also in H$\beta$ lines, though the H$\beta$ data of Flare Y18 was not so clear (Figure \ref{fig:spec_HaHb_YZCMi_UT200121_Y18}). 
Flares Y6 \& E2 showed blue wing asymmetries in higher-order Balmer lines 
up to H$\beta$ and H$\gamma$ lines, respectively, 
but they did not show blue wing asymmetries in chromospheric lines other than Balmer lines (e.g., Ca II lines, Na I D1\&D2 lines, and He I D3 lines).
Flares Y3 \& E1 showed blue wing asymmetries not only in Balmer lines 
up to H$\epsilon$ and H$\delta$ lines, respectively, but also in Ca II H\&K lines.
Moreover, Flares Y23 \& A3 showed blue wing asymmetries in almost all chromospheric lines we investigated, except for Ca II 8542 and Na I D1\&D2, respectively.
\color{black}\textrm{Blue wing asymmetries in multiple 
\color{black}\textrm{chromospheric }\color{black} lines 
have been investigated in several previous studies.
Flare V2016 on M4 dwarf V374Peg from \citet{Vida+2016} (in Table \ref{table:list_blue_flares}) showed blue asymmetries in \color{black} \textrm{
H$\alpha$, H$\beta$, and H$\gamma$ lines (see also the re-analysis results in \citealt{Leitzinger+2022}), while it is not clearly mentioned 
whether He I line showed blue asymmetries or not
in the discussions of \citet{Vida+2016} and \citet{Leitzinger+2022}}\color{black}.
A flare on EV Lac in Figure 7 of \citet{Muheki+2020_EVLac} also showed blue wing asymmetry only in H$\alpha$ line but not in H$\beta$ and He I lines. In contrast, a flare on AD Leo in Figure 6 of \citet{Muheki+2020} showed blue asymmetries both in H$\alpha$ and H$\beta$ lines.} \color{black}

The velocities of blue wing enhancements \color{black}\textrm{in} \color{black} these various chromospheric lines 
are listed in Table \ref{table:velmass_blue_flares} ($v_{\rm{blue,max}}$ in the table).
The velocities are different among different lines, and lower-order Balmer lines especially H$\alpha$ line tend to show larger velocities of blue wing asymmetries 
or wider blue wing tails, while higher-order Balmer lines, 
Ca II lines, Na I D1\&D2, and He I D3 lines show smaller velocities \footnote{
\color{black}\textrm{
Some of the data could be affected from the lower S/N ratios at bluer wavelengths (e.g., Figure \ref{fig:spec_HaHb_YZCMi_UT200121_Y18}), but most of the data have enough S/N values to determine $v_{\rm{blue,max}}$ (e.g., Figure \ref{fig:spec_HaHb_YZCMi_UT191212}). Furthermore, the velocity differences can be still seen if we integrate the data over longer time so that the data have higher S/N ratios. (e.g., integrating from Time 9.3h-10.5h in Figure \ref{fig:map_HaHb_YZCMi_UT191212}). Then it is not possible to explain all the difference trends (i.e. H$\alpha$ having largest $v_{\rm{blue,max}}$ values) only from the lower S/N at bluer wavelengths. Some noisy data (e.g., Figure \ref{fig:spec_HaHb_YZCMi_UT200121_Y18}) could be affected, but the overall trends discussed in the following of this paragraph would not be affected.}\color{black}}. 
\color{black}\textrm{
We speculate that these differences can be caused by the differences of optical depth and line wing broadening physics among other chromospheric lines as described in the following.
The differences of optical depth and line wing broadening physics (e.g., Stark effect) can 
affect these differences in the flaring atmosphere (e.g., \citealt{Kowalski+2022}), 
while those of optical depth can also affect the emission from prominences (e.g., \citealt{Okada+2020}).
}\color{black}
These differences can be clues to investigate how blue wing asymmetries occur associated with flares on mid M-dwarfs. For example, 
there is a difference of optical depth among different Balmer lines 
and H$\alpha$ line is more optically thick than other Balmer lines 
(e.g., \citealt{Drake_Ulrich_1980}; \citealt{Heinzel+1994_A&A}).
Then the visibility difference of Balmer lines could be a clue to constrain 
density and/or total emitting area values of the 
upward moving plasma that caused the blue wing enhancements during flares.
However, in order to interpret these differences more quantitatively in detail, 
it is necessary to conduct observation-based modeling studies incorporating 
radiative transfer physics of stellar (erupting) \color{black}\textrm{prominences} 
\color{black} and flaring atmospheres (e.g., \citealt{Leitzinger+2022}; \color{black}\textrm{\citealt{Kowalski+2022}}\color{black}).
Comparisons with the multi-wavelength Sun-as-a-star observation data of solar (erupting) 
\color{black}\textrm{prominences} \color{black} 
and solar flares are also very important for further quantitative discussions (e.g, \citealt{Namekata+2022_ApJ}; \citealt{Otsu+2022}; \citealt{Lynch+2022_WP}).

\begin{longrotatetable}
\begin{deluxetable*}{ccccccccl}
   \tablecaption{Flares showing blue wing asymmetries}
   \tablewidth{0pt}
   \tablehead{ 
     \colhead{Flare} & \colhead{Star name} & \colhead{UT date} &
     \colhead{WLF} &  \colhead{$E_{\rm{H}\alpha}$} &
     \colhead{$\Delta t_{\rm{H}\alpha}^{\rm{flare}}$} & 
     \colhead{$\Delta t_{\rm{H}\alpha}^{\rm{blueasym}}$} &
     \colhead{$v^{\rm{H}\alpha}_{\rm{blue,max}}$} &
      \colhead{Other lines \tablenotemark{$\dagger$}} \\
    \colhead{} & \colhead{} & \colhead{} 
    & \colhead{} & \colhead{[10$^{31}$ erg]} &
    \colhead{[h]} & \colhead{[min]} & \colhead{[km s$^{-1}$]} & \colhead{} 
    }
   \startdata
   Y3 & YZ CMi & 2019 Jan 27 & \color{black}\textrm{NWL} \color{black} & 1.7 & 4.3 & $\sim$20$\times$2 & -200 & [B] H$\beta$, H$\gamma$, H$\delta$, H$\epsilon$, Ca II H\&K  \\ 
    & &  &  &  &  & & & [NB] Ca II 8542, Na D1\&D2, He D3  \\
    & \multicolumn{8}{l}{$\cdot$ The clear short-lived H$\alpha$ blue wing asymmetries up to $\sim$ -200 km s$^{-1}$ were seen twice (20min$\times$2) during the H$\alpha$ flare over 4 hours.} \\
    & \multicolumn{8}{l}{$\cdot$ As for H$\beta$, H$\gamma$, H$\delta$, H$\epsilon$, and Ca II H\&K lines, blue wing asymmetries are not so clear at around the time of the first H$\alpha$ blue asymmetry} \\
    & \multicolumn{8}{l}{\ \ (Time [3] in 
    \color{black}\textrm{Figures \ref{fig:lcEW_HaHb_YZCMi_UT190127} -- \ref{fig:spec_HcHd_YZCMi_UT190127}}\color{black}), while they are clearly seen at around the second one (Time [5] in 
     \color{black}\textrm{Figures \ref{fig:lcEW_HaHb_YZCMi_UT190127} -- \ref{fig:spec_HcHd_YZCMi_UT190127}}\color{black}).} \\
    & \multicolumn{8}{l}{$\cdot$ Some red wing enhancements (or almost symmetric broadened wing components) were also seen for H$\alpha$, H$\beta$, H$\gamma$, H$\delta$, H$\epsilon$, and Ca II H\&K } \\
        & \multicolumn{8}{l}{\ \ lines (e.g., $\pm$150 km s$^{-1}$ for H$\alpha$ line).} \\
    & \multicolumn{8}{l}{$\cdot$ Equivalent width light curves (Figure \ref{fig:lcEW_HaHb_YZCMi_UT190127}): H$\alpha$ and Ca II K evolve similarly while other Balmer lines, Ca II 8542, Na I D1\&D2, and He I D3 lines} \\
    & \multicolumn{8}{l}{\ \ decrease faster.} \\
    & \multicolumn{8}{l}{$\cdot$ No clear white-light flux enhancements during the flare even in \textit{TESS} high precision photometric data, while there are very small ``suggestive"} \\
    & \multicolumn{8}{l}{\ \ increases (\color{black}\textrm{$E_{TESS}<2.6\times 10^{32}$erg}\color{black}).} \\
    & \multicolumn{8}{l}{$\cdot$ Flare emission was observed also in \textit{NICER} soft X-ray data ($E_{\rm{X}}$(0.5--2.0 keV)$=2.6\times 10^{32}$erg and $E_{\rm{Xray, flare}}$(\textit{GOES}-band)$=4.7\times10^{31}$erg).} \\
    \hline
   Y6 & YZ CMi & 2019 Dec 12 & \color{black}\textrm{NWL}\color{black} & \color{black}$>$4.1 & $>$4.9 & 90--120 & -200 &
    [B] H$\beta$\\
    & & & & & & & & [NB] H$\gamma$, H$\delta$, H$\epsilon$, Ca II H\&K, Ca II 8542, Na D1\&D2, \\
    & & & & & & & & \ \ He D3\\
    & \multicolumn{8}{l}{$\cdot$ The clear H$\alpha$ blue wing enhancement up to $\sim$ -200 km s$^{-1}$ was seen in early phase of the flare, while the line profile gradually shifted to} \\
    & \multicolumn{8}{l}{\ \ the red wing enhancement up to $\sim$ +200 km s$^{-1}$,
    during the H$\alpha$ flare over 4.9 hours.}  \\
    & \multicolumn{8}{l}{$\cdot$ Similar shift from blue to red wing asymmetry was also seen in H$\beta$ line (from $\sim$ -150 km s$^{-1}$ to $\sim$ +150 km s$^{-1}$).} \\
    & \multicolumn{8}{l}{$\cdot$ Late-phase red wing asymmetries were also seen clearly in H$\gamma$ \& H$\delta$ lines, and possibly in He I D3 5876 line.} \\
    & \multicolumn{8}{l}{$\cdot$ There are at least two short white-light continuum enhancements ($\lesssim$10 min each) \color{black}\textrm{in the middle/late phase of the flare, but there are no} \color{black}} \\
    & \multicolumn{8}{l}{\ \ \color{black}\textrm{other clear white-light enhancements that are considered to be physically associated with the early main increasing phase 
    of}\color{black}} \\
    & \multicolumn{8}{l}{\ \ \color{black}\textrm{the whole H$\alpha$ flare.}\color{black}} \\
    \hline
    Y18 & YZ CMi & 2020 Jan 21 & \color{black}\textrm{WL   }\color{black} & \color{black}1.5--1.6 & 3.4 & 60 & -200 &  [B] H$\beta$ \\ 
    & &  &  &  &  & & & [NB] Ca II 8542, Na D1\&D2, He D3\\
    & \multicolumn{8}{l}{$\cdot$ The blue wing enhancements up to $\sim$ -200 km s$^{-1}$ \& $\sim$ -150 km s$^{-1}$ were seen in H$\alpha$ \& H$\beta$ 
    lines, respectively, during the decay phase of the flare.} \\
    & \multicolumn{8}{l}{$\cdot$ The multiple white-light continuum flux enhancements during the H$\alpha$ flare.} \\
    \hline
    Y23 & YZ CMi & 2020 Dec 06 & \color{black}\textrm{WL   }\color{black} & color{black} $>$1.1 & $>$1.3 &  $>$45 & -250 & [B] H$\beta$, H$\gamma$, H$\delta$, H$\epsilon$, Ca II H\&K, Na D1\&D2, He D3\\
    & &  &  &  &  & & & [NB] Ca II 8542\\
    & \multicolumn{8}{l}{$\cdot$ The clear white-light continuum flux enhancements ($\sim$1260\% and $\sim$125\% in $u$\&$g$-bands, respectively) observed before the start of} \\
    & \multicolumn{8}{l}{\ \ the spectroscopic observation.}  \\
    & \multicolumn{8}{l}{$\cdot$ The H$\alpha$\&H$\beta$ blue wing enhancements up to 
    $\sim$ -250 km s$^{-1}$ and $\sim$ -200 km s$^{-1}$, respectively, were seen almost over the whole observed phase} \\
    & \multicolumn{8}{l}{\ \  of the flare, while the velocities of blue wing enhancements decayed gradually.}  \\
    & \multicolumn{8}{l}{$\cdot$ Blue wing enhancements were seen in all the lines except for Ca II 8542, while the velocities of blue wing enhancements are different} \\
    & \multicolumn{8}{l}{\ \  among the lines.}  \\
    \hline
    E1 & EV Lac & 2019 Dec 15 & \color{black}\textrm{NEP   }\color{black} & 2.9 & 3.6 & $\gtrsim$60 & -200 & [B] H$\beta$, H$\gamma$, H$\delta$, Ca II H\&K \\
    & &  &  &  &  & & & [NB] H$\epsilon$, Ca II 8542, Na D1\&D2, He D3 \\
    & \multicolumn{8}{l}{$\cdot$ Only the late-phase of the flare was observed with ARCSAT photometry, so it is possible there were increases of the continuum white-light flux } \\
    & \multicolumn{8}{l}{\ \ in the early-phase of the flare.}  \\
    & \multicolumn{8}{l}{$\cdot$ The H$\alpha$\&H$\beta$ blue wing enhancements up to $\sim$ -200 km s$^{-1}$ and $\sim$ -150 km s$^{-1}$, respectively, were seen, 
    but the duration of the blue wing } \\
    & \multicolumn{8}{l}{\ \ enhancement in H$\beta$ line ($\sim$0.5 hours) is shorter than that of H$\alpha$ line ($\gtrsim$1 hours).}  \\
    & \multicolumn{8}{l}{$\cdot$ The blue wing enhancements were also seen for H$\gamma$, H$\delta$, and Ca II H\&K lines, while possible slight blue shifts of the line peak were also seen} \\
    & \multicolumn{8}{l}{\ \ in \color{black}\textrm{Ca II 8542 and He I D3} \color{black} lines.}  \\
    \hline
    E2 & EV Lac & 2019 Dec 15 & \color{black}\textrm{WL   }\color{black} & 0.58 & 0.9 & 20 & -150 & [B] H$\beta$, H$\gamma$   \\    
    & &  &  &  &  & & & [NB] H$\delta$, H$\epsilon$, Ca II H\&K, Ca II 8542, Na D1\&D2, He D3 \\
    & \multicolumn{8}{l}{$\cdot$ The H$\alpha$\&H$\beta$ blue wing enhancements up to $\sim$ -150 km s$^{-1}$ were seen for the durations 20 and 10 min, respectively.} \\
      & \multicolumn{8}{l}{$\cdot$ The clear white-light continuum flux increase was observed almost simultaneously with the H$\alpha$\&H$\beta$ blue wing enhancements.} \\
     & \multicolumn{8}{l}{$\cdot$ The blue wing enhancements were also seen for H$\gamma$ lines, while possible slight line peak blue shifts \color{black}\textrm{were }\color{black} 
     also seen in \color{black}\textrm{Ca II H\&K} \color{black} lines.} \\
    \hline
    A3 & AD Leo & 2019 May 19 &  \color{black}\textrm{NEP   }\color{black} & $>$5.3 & $>$3.1 & 120--150 & & [B] H$\beta$, H$\gamma$, H$\delta$, \color{black}\textrm{H$\epsilon$}\color{black}, Ca II K, Ca II 8542, He D3 \\
    & &  &  &  &  & & & [NB] Na D1\&D2 \\
    & \multicolumn{8}{l}{$\cdot$ Flare already started when the observation started, so it is possible there were increases of the continuum white-light flux before the observation } \\
    & \multicolumn{8}{l}{\ \ started.}  \\
    & \multicolumn{8}{l}{$\cdot$ The H$\alpha$\&H$\beta$ blue wing enhancements up to 
   -150 -- -200 km s$^{-1}$ were seen, which continued for more than two hours until the flare decayed.} \\
    & \multicolumn{8}{l}{$\cdot$ Blue wing enhancements were seen in all the lines except for Na I D1\&D2, while the velocities of blue wing enhancements are different among } \\
    & \multicolumn{8}{l}{\ \ the lines.}  \\
        \hline
     \color{black} V2016\tablenotemark{\rm *} &  \color{black} V374 Peg  & 
      \color{black} 2005 Aug 20 &  \color{black} -- & 
       \color{black} $\sim$10$^{2}$ &  \color{black}$\sim$3 & 
        \color{black} 15-40min$\times$3 & -675 & \color{black} [B] H$\alpha$, \color{black}  H$\beta$, H$\gamma$ \\
    & &  &  &  &  & & &   [NB] He D3 (Not clearly mentioned in \citealt{Vida+2016}.) \\
    & \multicolumn{8}{l}{
$\cdot$ Three consecutive blue-wing enhancements in \color{black} Balmer \color{black} lines.
}  \\
    & \multicolumn{8}{l}{
All three blue-wing asymmetries occurred during a single flare with flaring energy $\sim$10$^{33}$ erg (cf. \citealt{Moschou+2019}), with
}  \\
    & \multicolumn{8}{l}{
the third event being the strongest, corresponding to projected velocity of 
-675 km s$^{-1}$.
}  \\
    & \multicolumn{8}{l}{
This event showed not only blue wing asymmetries, 
but also clearly separated additional emission components 
} \\
    & \multicolumn{8}{l}{ \ \ 
(see Figure 2 of \citealt{Leitzinger+2022}).
}  \\
    \hline
    H2018\tablenotemark{\rm *} & EV Lac  & 2015 Aug 15 & -- & 2.0 & 2.6 & 120 & -200 & -- \\
    & \multicolumn{8}{l}{
$\cdot$ A blue wing asymmetry in the H$\alpha$ line has been observed for $\gtrsim$2 h (almost from flare start to end).
}  \\
    & \multicolumn{8}{l}{
$\cdot$ The possible existence of absorption components in the red wing of the H$\alpha$ line was reported when the H$\alpha$ line showed blue 
} \\
    & \multicolumn{8}{l}{ \ \ 
wing asymmetry. 
} \\
    \hline
    M2021\tablenotemark{\rm *} & YZ CMi  & 2019 Jan 18 & no & 0.47 & 1.2 & 80 & -150 & --\\
    & \multicolumn{8}{l}{
$\cdot$ A H$\alpha$ flare without clear brightening in continuum,
which exhibited blue wing asymmetry lasting for $\sim$1 hour.
} \\
    \hline
   \enddata
      \tablecomments{WLF: 
      This column describes whether the flare is identified as white-light flare (WL) or non white-light flare (NWL), while
      stars with ``NEP" do not have enough photometric data
      for the WL/NWL identification 
      (from Table \ref{table:list1_flares}).
  $\Delta t_{\rm{H}\alpha}^{\rm{flare}}$: Flare duration in H$\alpha$ line
      (from Table \ref{table:list1_flares}).
      $E_{\rm{H}\alpha}$: Flare energy in H$\alpha$ line 
      (from Table \ref{table:list1_flares}).
      $\Delta t_{\rm{H}\alpha}^{\rm{blueasym}}$: Duration of H$\alpha$ blue wing asymmetry.       \color{black}\textrm{(e.g., the double-headed arrow in Figure \ref{fig:map_HaHb_YZCMi_UT190127}(a))}\color{black}.
      $v^{\rm{H}\alpha}_{\rm{blue,max}}$: the maximum velocity of blue wing enhancements estimated by eye.
      Other lines: [B] and [NB] show the lines with and without blue wing asymmetries, respectively.
   }
       \tablenotetext{\dagger}{
 There are no observation data of H$\gamma$, H$\delta$, H$\epsilon$, Ca II H\&K lines for Flare Y18, since these lines are not included in the wavelength range of SMARTS1.5m/CHIRON.
 }
      \tablenotetext{\rm *}{
  V2016, H2018, and M2021 are flares with clear blue wing asymmetries reported in \citet{Vida+2016}, \citet{Honda+2018}, and \citet{Maehara+2021}, respectively. They are listed here just for comparison.
  As for chromospheric lines, H$\alpha$ line is only observed in \citet{Honda+2018} and \citet{Maehara+2021}.  
  There were no white-light observation data in \citet{Vida+2016} and \citet{Honda+2018}. 
  The flare energy value of V2016 is from \citet{Moschou+2019}.
      }
   \label{table:list_blue_flares}
 \end{deluxetable*}
\end{longrotatetable}

\begin{longrotatetable}
\begin{deluxetable*}{l|ccccccccc}
   \tablecaption{Velocities and masses of blue wing asymmetry flares}
   \tablewidth{0pt}
   \tablehead{
    & \multicolumn{2}{c}{Y3} & \colhead{Y6} &
     \colhead{Y18} &  \colhead{Y23} &
     \colhead{E1}& \colhead{E2}&
     \colhead{A3} \\
      &  \colhead{[3]\tablenotemark{\rm a}} & \colhead{[5]\tablenotemark{\rm a}} & &
     & & & &
     }
   \startdata
  $v^{\rm{H}\alpha}_{\rm{blue,max}}$ [km s$^{-1}$] \tablenotemark{\rm b} & -250 & -250 & -200 & -200 & -250 & -200 & -150 & -200 \\
  $v^{\rm{H}\beta}_{\rm{blue,max}}$ [km s$^{-1}$] \tablenotemark{\rm b} & -200 & -200 & -150 & -150 & -200 & -150 & -150 & -150  \\
  $v^{\rm{H}\gamma}_{\rm{blue,max}}$ [km s$^{-1}$] \tablenotemark{\rm b}  & -100 & -100 & [NB] & -- & -150 & -100 & -100 & -150  \\
  $v^{\rm{H}\delta}_{\rm{blue,max}}$ [km s$^{-1}$] \tablenotemark{\rm b}  & -100 & -75 & [NB] & -- & -100 & -50 & [NB] & -150  \\
  $v^{\rm{H}\epsilon}_{\rm{blue,max}}$ [km s$^{-1}$] \tablenotemark{\rm b}  & -50 &  -50 & [NB] & -- & -75  & [NB] & [NB] & -100  \\
  $v^{\rm{CaK}}_{\rm{blue,max}}$ [km s$^{-1}$] \tablenotemark{\rm b}  & -50 & -50 & [NB] & -- & -100 & -50 & [NB] & -100 \\
  $v^{\rm{CaH}}_{\rm{blue,max}}$ [km s$^{-1}$] \tablenotemark{\rm b}  & -50 & -50 & [NB] & -- & -75 & -50 & [NB] & -100  \\
  $v^{\rm{Ca8542}}_{\rm{blue,max}}$ [km s$^{-1}$] \tablenotemark{\rm b}  & [NB]  & [NB] & [NB] & [NB] & [NB] & [NB] & [NB] & -50 \\
  $v^{\rm{NaD1\&D2}}_{\rm{blue,max}}$ [km s$^{-1}$] \tablenotemark{\rm b}  & [NB] & [NB] & [NB] & [NB] & -100 & [NB] & [NB] & [NB] \\
  $v^{\rm{He D3}}_{\rm{blue,max}}$ [km s$^{-1}$] \tablenotemark{\rm b} & [NB] & [NB] & [NB] & [NB] & -100 & [NB] & [NB] & -50 \\
    \hline
 $v^{\rm{H}\alpha}_{\rm{blue,fit}}$ [km s$^{-1}$] & 
 \color{black} -105$^{+11}_{-8}$ &
  \color{black} -88$^{+8}_{-9}$ & 
  \color{black} -122$^{+1}_{-1}$  & 
   \color{black} -115$^{+2}_{-5}$ & 
    \color{black} -106$^{+1}_{-3}$ &
     \color{black} -87$^{+7}_{-9}$ &
      \color{black} -73$^{+11}_{-18}$ & 
       \color{black} -106$^{+0}_{-3}$ \\
  $EW^{\rm{H}\alpha}_{\rm{blue,fit}}$ [\AA] & 
   \color{black} 0.23$^{+0.04}_{-0.03}$ & 
    \color{black} 0.45$^{+0.08}_{-0.08}$ & 
     \color{black} 0.32$^{+0.01}_{-0.01}$ & 
      \color{black} 0.27$^{+0.01}_{-0.02}$ & 
       \color{black} 0.45$^{+0.02}_{-0.03}$ & 
        \color{black} 0.39$^{+0.07}_{-0.08}$ & 
         \color{black} 0.19$^{+0.07}_{-0.04}$ &  
          \color{black} 0.16$^{+0.00}_{-0.01}$ \\
  $v^{\rm{H}\beta}_{\rm{blue,fit}}$ [km s$^{-1}$] & 
   \color{black} -97$^{+1}_{-6}$ &
    \color{black} -90$^{+10}_{-12}$ &
     \color{black} -107$^{+1}_{-1}$ & 
      \color{black} -116$^{+11}_{-3}$ & 
       \color{black} -97$^{+7}_{-8}$ & 
        \color{black} -86$^{+11}_{-14}$ & 
         \color{black} -86$^{+18}_{-15}$ & 
          \color{black} -85$^{+1}_{-4}$  \\
  $EW^{\rm{H}\beta}_{\rm{blue,fit}}$ [\AA] & 
   \color{black} 0.40$^{+0.01}_{-0.05}$ &
    \color{black} 0.82$^{+0.12}_{-0.13}$ &
     \color{black} 0.29$^{+0.01}_{-0.01}$ &
      \color{black} 0.24$^{+0.04}_{-0.01}$ &
       \color{black} 0.94$^{+0.13}_{-0.13}$ &
        \color{black} 0.39$^{+0.08}_{-0.09}$ & 
         \color{black} 0.28$^{+0.08}_{-0.07}$ & 
          \color{black} 0.17$^{+0.01}_{-0.02}$ \\
     \hline
  $L^{\rm{H}\alpha}_{\rm{blue}}$ [erg s$^{-1}$] & 
  \color{black} 2.5$^{+0.4}_{-0.3}\times10^{26}$ & 
  \color{black} 4.9$^{+0.9}_{-0.9}\times10^{26}$ & 
  \color{black} 3.5$^{+0.1}_{-0.1}\times10^{26}$ & 
  \color{black} 2.9$^{+0.1}_{-0.2}\times10^{26}$ & 
  \color{black} 5.0$^{+0.2}_{-0.3}\times10^{26}$ &
  \color{black} 6.8$^{+1.2}_{-1.4}\times10^{26}$ & 
  \color{black} 3.5$^{+1.3}_{-0.7}\times10^{26}$ & 
  \color{black} 6.1$^{+0.0}_{-0.4}\times10^{26}$ \\
  $L^{\rm{H}\beta}_{\rm{blue}}$ [erg s$^{-1}$] & 
  \color{black} 1.4$^{+0.1}_{-0.2}\times10^{26}$ & 
  \color{black} 2.9$^{+0.4}_{-0.5}\times10^{26}$ & 
  \color{black} 1.0$^{+0.1}_{-0.1}\times10^{26}$ &
  \color{black} 8.6$^{+0.1}_{-0.1}\times10^{25}$ & 
  \color{black} 3.5$^{+0.5}_{-0.5}\times10^{26}$ & 
  \color{black} 2.4$^{+0.5}_{-0.6}\times10^{26}$ & 
  \color{black} 1.8$^{+0.5}_{-0.5}\times10^{26}$ & 
  \color{black} 2.4$^{+0.1}_{-0.3}\times10^{26}$ \\
   $\alpha$ (=  \color{black}\textrm{$\epsilon^{\rm{H}\alpha}_{\rm{blue}}$/$\epsilon^{\rm{H}\beta}_{\rm{blue}}$} \color{black}) &
   \color{black}1.75$\pm$0.31 &
   \color{black}1.67$\pm$0.39 &
   \color{black}3.35$\pm$0.16 &
   \color{black}3.42$\pm$0.46 &
   \color{black}1.45$\pm$0.22 &
   \color{black}2.81$\pm$0.82 &
   \color{black}1.91$\pm$0.77 &
   \color{black}2.51$\pm$0.26 \\
   \color{black}\textrm{$\log \epsilon^{\rm{H}\alpha}_{\rm{blue}}$}\color{black} [erg s$^{-1}$ cm$^{-2}$ sr$^{-1}$] &
   \color{black}6.4$^{+0.2}_{-0.2}$ &
   \color{black}6.4$^{+0.2}_{-0.2}$ &
   \color{black} 5.9$^{+0.1}_{-0.1}$ &
   \color{black} 5.9$^{+0.1}_{-0.1}$ &
  \color{black} 6.5$^{+0.1}_{-0.1}$ &
  \color{black} 6.0$^{+0.3}_{-0.2}$ &
  \color{black} 6.3$^{+0.4}_{-0.3}$ &
  \color{black} 6.1$^{+0.1}_{-0.1}$  \\
   $A^{\rm{H}\alpha}_{\rm{blue}}$ [10$^{19}$ cm$^{2}$] &
   \color{black}1.6$^{+1.1}_{-0.7}$ &
   \color{black}2.8$^{+2.3}_{-1.5}$ &
   \color{black}6.9$^{+2.1}_{-1.6}$ &
   \color{black}6.1$^{+2.6}_{-2.1}$ &
   \color{black}2.3$^{+1.1}_{-0.8}$ &
   \color{black}9.9$^{+9.6}_{-5.8}$ &
   \color{black}2.5$^{+4.1}_{-1.8}$ &
   \color{black}7.3$^{+2.4}_{-2.2}$ \\
   \hline 
   \hline
  \multicolumn{9}{l}{(1) Upper limit \color{black}\textrm{$\epsilon^{\rm{H}\alpha}_{\rm{blue}}$}\color{black} case 
   (e.g., \color{black}\textrm{$\log\epsilon^{\rm{H}\alpha}_{\rm{blue}}$[erg s$^{-1}$ cm$^{-2}$ sr$^{-1}$]=6.0 for Flare Y6)}\color{black}} \\
   \hline 
   $\log EM^{(1)}_{\rm{blue}}$ [cm$^{-5}$] & 
   \color{black} 32.6 -- 33.0 & 
   \color{black} 32.7 -- 33.1 & 
   \color{black} 30.9 -- 31.2  &  
   \color{black} 30.9 -- 31.2 & 
   \color{black} 32.8 -- 33.2 & 
   \color{black} 31.5 -- 32.1 & 
   \color{black} 32.3 -- 32.7 & 
   \color{black} 31.5 -- 32.1 \\
   $\log n^{(1)}_{e}$ [cm$^{-3}$] &
   \color{black} 11.1 -- 11.5 &
   \color{black} 11.2 -- 11.5 &
   \color{black} 10.3 -- 11.5 &
   \color{black} 10.3 -- 11.5 &
   \color{black} 11.2 -- 11.5 &
   \color{black} 10.5 -- 11.5 &
   \color{black} 11.0 -- 11.5 &
   \color{black} 10.6 -- 11.5 \\
   $D^{(1)}_{\rm{blue, upp}}$ [cm] \tablenotemark{\rm c} & $R_{\rm{star}}$ & $R_{\rm{star}}$  & $R_{\rm{star}}$ & $R_{\rm{star}}$ & $R_{\rm{star}}$ & $R_{\rm{star}}$ & $R_{\rm{star}}$  & $R_{\rm{star}}$ \\
   $D^{(1)}_{\rm{blue, low}}$ [cm] & 
   \color{black} 4.0$\times$10$^{9}$ & 
   \color{black} 5.0$\times$10$^{9}$ & 
   \color{black} 7.9$\times$10$^{7}$ & 
   \color{black} 7.9$\times$10$^{7}$ & 
   \color{black} 6.3$\times$10$^{9}$ & 
   \color{black} 3.2$\times$10$^{8}$ & 
   \color{black} 2.0$\times$10$^{9}$ & 
   \color{black} 3.2$\times$10$^{8}$  \\
   $M^{(1)}_{\rm{blue}, \rm{upp}}$ [g] (\color{black}\textrm{$n_{e}/n_{H}$}\color{black}=0.17)
    &  
   \color{black}1.9$\times10^{18}$ &
   \color{black}4.2$\times10^{18}$ &  
   \color{black}7.3$\times10^{17}$ &  
   \color{black}7.0$\times10^{17}$ &  
   \color{black}3.0$\times10^{18}$ & 
   \color{black}7.5$\times10^{18}$ &  
   \color{black}3.7$\times10^{18}$ &  
   \color{black}3.5$\times10^{18}$ \\
   $M^{(1)}_{\rm{blue}, \rm{low}}$ [g] (\color{black}\textrm{$n_{e}/n_{H}$}\color{black}=0.47)& 
      \color{black}4.1$\times10^{16}$ &    
      \color{black}7.8$\times10^{16}$ &    
      \color{black}4.9$\times10^{15}$ &    
      \color{black}3.6$\times10^{15}$ &     
      \color{black}1.1$\times10^{17}$ &     
      \color{black}1.5$\times10^{16}$ &     
      \color{black}1.8$\times10^{16}$ &     
      \color{black}1.8$\times10^{16}$ \\
   \hline 
   \multicolumn{9}{l}{(2) Lower limit
   \color{black}\textrm{$\epsilon^{\rm{H}\alpha}_{\rm{blue}}$}\color{black} case 
   (e.g., \color{black}\textrm{$\log\epsilon^{\rm{H}\alpha}_{\rm{blue}}$[erg s$^{-1}$ cm$^{-2}$ sr$^{-1}$]=5.8  for Flare Y6)}\color{black}} \\
   \hline
   $\log EM^{(2)}_{\rm{blue}}$ [cm$^{-5}$]  & 
   \color{black} 31.7 -- 32.1 & 
   \color{black} 31.8 -- 32.2 & 
   \color{black} 30.5 -- 30.7  &  
   \color{black} 30.5 -- 30.7 & 
   \color{black} 32.2 -- 32.6 & 
   \color{black} 30.6 -- 31.0 & 
   \color{black} 30.9 -- 31.2 & 
   \color{black} 30.8 -- 31.2 \\
   $\log n^{(2)}_{e}$ [cm$^{-3}$] &
   \color{black} 10.7 -- 11.5 &
   \color{black} 10.7 -- 11.5 &
   \color{black} 10.1 -- 11.5 &
   \color{black} 10.1 -- 11.5 &
   \color{black} 10.9 -- 11.5 &
   \color{black} 10.1 -- 11.5 &
   \color{black} 10.3 -- 11.5 &
   \color{black} 10.2 -- 11.5 \\
   $D^{(2)}_{\rm{blue, upp}}$ [cm] \tablenotemark{\rm c} & $R_{\rm{star}}$ & $R_{\rm{star}}$  & $R_{\rm{star}}$ & $R_{\rm{star}}$ & $R_{\rm{star}}$ & $R_{\rm{star}}$ & $R_{\rm{star}}$  & $R_{\rm{star}}$ \\
   $D^{(2)}_{\rm{blue, low}}$ [cm]  &
   \color{black} 5.0$\times$10$^{8}$ & 
   \color{black} 6.3$\times$10$^{8}$ & 
   \color{black} 3.2$\times$10$^{7}$ & 
   \color{black} 3.2$\times$10$^{7}$ & 
   \color{black} 1.6$\times$10$^{9}$ & 
   \color{black} 4.0$\times$10$^{7}$ & 
   \color{black} 7.9$\times$10$^{7}$ & 
   \color{black} 6.3$\times$10$^{7}$  \\
   $M^{(2)}_{\rm{blue}, \rm{upp}}$ [g] (\color{black}\textrm{$n_{e}/n_{H}$}\color{black}=0.17) &
   \color{black}6.8$\times10^{17}$ & 
   \color{black}1.5$\times10^{18}$ &  
   \color{black}3.7$\times10^{17}$ &  
   \color{black}3.5$\times10^{17}$ &  
   \color{black}1.5$\times10^{18}$ & 
   \color{black}1.7$\times10^{18}$ &  
   \color{black}5.9$\times10^{17}$ &  
   \color{black}9.7$\times10^{17}$ \\
   $M^{(2)}_{\rm{blue}, \rm{low}}$ [g] (\color{black}\textrm{$n_{e}/n_{H}$}\color{black}=0.47) & 
      \color{black}
   \color{black} 5.1$\times10^{15}$ & 
   \color{black} 9.9$\times10^{15}$ & 
   \color{black} 1.9$\times10^{15}$ & 
   \color{black} 1.4$\times10^{15}$ &  
   \color{black} 2.7$\times10^{16}$ &  
   \color{black} 1.9$\times10^{15}$ & 
   \color{black} 7.1$\times10^{14}$ &  
   \color{black} 3.5$\times10^{15}$ \\
   \hline 
   \hline
  $M_{\rm{blue}, \rm{upp}}$ [g] & 
  \multicolumn{2}{c}{\color{black}6.1$\times10^{18}$} & 
  \color{black}7.3$\times10^{17}$ & 
  \color{black}7.0$\times10^{17}$ &  
  \color{black}3.0$\times10^{18}$ & 
  \color{black}7.5$\times10^{18}$ &  
  \color{black}3.7$\times10^{18}$ &  
  \color{black}3.5$\times10^{18}$ \\
   $M_{\rm{blue}, \rm{low}}$ [g] &
   \multicolumn{2}{c}{\color{black}1.5$\times10^{16}$} & 
   \color{black}1.9$\times10^{15}$ & 
   \color{black}1.4$\times10^{15}$ & 
   \color{black}2.7$\times10^{16}$ &  
   \color{black}1.9$\times10^{15}$  &  
   \color{black}7.1$\times10^{14}$ & 
   \color{black}3.5$\times10^{15}$ \\
   $E_{\rm{kin,upp}}$ [erg] & 
   \multicolumn{2}{c}{\color{black}2.2$\times10^{32}$} & 
   \color{black}5.4$\times10^{31}$ & 
   \color{black}4.5$\times10^{31}$ & 
   \color{black}1.7$\times10^{32}$ & 
   \color{black}2.4$\times10^{32}$ & 
   \color{black}7.2$\times10^{31}$ & 
   \color{black}1.9$\times10^{32}$\\
   $E_{\rm{kin,low}}$ [erg] & 
   \multicolumn{2}{c}{\color{black}7.9$\times10^{29}$} & 
   \color{black}1.5$\times10^{29}$ & 
   \color{black}1.0$\times10^{29}$ & 
   \color{black}1.6$\times10^{30}$ & 
   \color{black}8.7$\times10^{28}$ & 
   \color{black}2.9$\times10^{28}$ & 
   \color{black}2.1$\times10^{29}$  \\
       \hline 
   $E_{u\rm{, flare}}$  [erg] \tablenotemark{\rm d} & \multicolumn{2}{c}{\color{black}$<$5.6$\times10^{31}$} & \color{black}$<7.2\times10^{31}$ & -- & $1.6\times10^{32}$ & -- & $1.1\times10^{31}$ & \color{black}$>$2.7$\times10^{31}$ \\
  $E_{U\rm{, flare}}$ [erg] \tablenotemark{\rm d}  & \multicolumn{2}{c}{--} & -- & $4.6\times10^{31}$ & -- & -- & -- & --  \\
  $E_{g\rm{, flare}}$  [erg]  \tablenotemark{\rm d} & \multicolumn{2}{c}{\color{black}$<$1.2$\times10^{32}$} & \color{black}$<1.6\times10^{32}$ & -- & $2.2\times10^{32}$ & -- & $7\times10^{30}$ &  \color{black}$>$1.4$\times10^{31}$  \\
  $E_{V\rm{, flare}}$ [erg]  \tablenotemark{\rm d}  & \multicolumn{2}{c}{--} & -- & $2.4\times10^{31}$ & -- & -- & -- & --  \\
  $E_{TESS\rm{, flare}}$ [erg]  \tablenotemark{\rm d}  & \multicolumn{2}{c}{\color{black}$<$2.6$\times10^{32}$} & -- & -- & -- & -- & -- & --  \\
    \color{black} $E_{\rm{bol, flare}}^{(1)}$ [erg] \tablenotemark{\rm e} &
    \multicolumn{2}{c}{\color{black}$<$6.2$\times10^{32}$} &
    \color{black} $<$8.0$\times10^{32}$ & 4.2$\times10^{32}$ & 
    \color{black} 1.8$\times10^{33}$ & 
    \color{black} -- & 
    \color{black} 1.2$\times10^{32}$ & 
    \color{black} $>$3.0$\times10^{32}$\\
        \hline 
   $E_{\rm{H}\alpha\rm{, flare}}$ [erg]  \tablenotemark{\rm d}  & \multicolumn{2}{c}{1.7$\times10^{31}$} & \color{black}$>3.9\times10^{31}$ & \color{black}(1.5--1.6)$\times10^{31}$ & \color{black}$>1.1\times10^{31}$ & 2.9$\times10^{31}$ & 5.8$\times10^{30}$  & $>$5.3$\times10^{31}$ \\
  $E_{\rm{Xray, flare}}$(0.5--2.0 keV) [erg] & \multicolumn{2}{c}{2.6$\times10^{32}$} & -- & -- & -- & -- & -- & --  \\
  \tablenotemark{\rm d} & \multicolumn{2}{c}{} &  &  &  &  &  &   \\
     \color{black}  $E_{\rm{Xray, flare}}$(\textit{GOES}-band) [erg]  & 
  \multicolumn{2}{c}{\color{black}4.7$\times10^{31}$} & 
     \color{black} $>$3.6$\times10^{31}$ & 
     \color{black} (1.4--1.5)$\times10^{31}$ & 
     \color{black} $>$1.0$\times10^{31}$ & 
     \color{black} 2.7$\times10^{31}$ & 
     \color{black} 5.4$\times10^{30}$ & 
     \color{black} $>$4.9$\times10^{31}$  \\
   \tablenotemark{\rm d} & \multicolumn{2}{c}{} &  &  &  &  &  &   \\
     \color{black}$E_{\rm{bol, flare}}^{(2)}$ [erg]  \tablenotemark{\rm e}& 
    \multicolumn{2}{c}{\color{black} 7.8$\times$10$^{32}$} & 
     \color{black} $>$6.0$\times10^{32}$ & 
     \color{black} (2.3--2.5)$\times10^{32}$ & 
     \color{black} $>$1.7$\times10^{32}$ & 
     \color{black} 4.5$\times10^{32}$ & 
     \color{black} 9.0$\times10^{31}$ & 
     \color{black} $>$8.2$\times10^{32}$\\
        \hline 
     \color{black}  $E_{\rm{bol, flare}}$  [erg] \tablenotemark{\rm e}  & 
     \multicolumn{2}{c}{\color{black} 7.8$\times$10$^{32}$} & 
     \color{black} 6.0$\times10^{32}$ & 
     \color{black} 2.3$\times10^{32}$ & 
     \color{black} 1.7$\times10^{32}$ & 
     \color{black} 4.5$\times10^{32}$ & 
     \color{black} 9.0$\times10^{31}$ & 
     \color{black} 3.0$\times10^{32}$ \\  
    & \multicolumn{2}{c}{} & 
    \color{black} -- 8.0$\times10^{32}$ & 
    \color{black} -- 4.2$\times10^{32}$ & 
    \color{black} -- 1.8$\times10^{33}$  &  & 
    \color{black} -- 1.2$\times10^{32}$ & 
    \color{black} -- 1.0$\times10^{34}$\\  
   \enddata
       \tablenotetext{\rm a}{
       Blue wing asymmetries were seen twice (20min$\times$2) during Flare Y3. 
       The values for these two asymmetries, which occurred at around Time [3] and [5] in Figure \ref{fig:lcEW_HaHb_YZCMi_UT190127}, are listed separately here.
       As for the mass and kinetic energy values, the sum of the two asymmetries are listed.
       }
       \tablenotetext{\rm b}{
      $v^{\rm{H}\alpha}_{\rm{blue,max}}$, $v^{\rm{H}\beta}_{\rm{blue,max}}$, $v^{\rm{H}\gamma}_{\rm{blue,max}}$, $v^{\rm{H}\delta}_{\rm{blue,max}}$, $v^{\rm{H}\epsilon}_{\rm{blue,max}}$, $v^{\rm{CaK}}_{\rm{blue,max}}$, $v^{\rm{CaH}}_{\rm{blue,max}}$, $v^{\rm{Ca8542}}_{\rm{blue,max}}$, $v^{\rm{NaD1\&D2}}_{\rm{blue,max}}$, $v^{\rm{HeD3}}_{\rm{blue,max}}$: the maximum velocity values of blue wing enhancements measured by eye, for H$\alpha$, H$\beta$, H$\gamma$, H$\delta$, H$\epsilon$, Ca II K, Ca II H, Ca II 8542, Na I D1\&D2, and He I D3 lines. 
      [NB] means this line does not show blue \color{black}\textrm{wing} \color{black} asymmetry \color{black}\textrm{(cf. Table \ref{table:list_blue_flares})}\color{black}.  There are no observation data of H$\gamma$, H$\delta$, H$\epsilon$, Ca II H\&K lines for Flare Y18, which was observed with SMARTS1.5m/CHIRON.
   }
         \tablenotetext{\rm c}{
     Stellar radius ($R_{\rm{star}}$) of YZ CMi, EV Lac, and AD Leo are 2.1$\times$10$^{10}$ cm, 2.5$\times$10$^{10}$ cm, and 3.0$\times$10$^{10}$ cm, respectively (cf. Table \ref{table:targets_basic_para}).   
    }
         \tablenotetext{\rm d}{
         $E_{\rm{H}\alpha\rm{, flare}}$, $E_{u\rm{, flare}}$, $E_{U\rm{, flare}}$, $E_{g\rm{, flare}}$, $E_{V\rm{, flare}}$, and
         $E_{TESS\rm{, flare}}$ are flare energies in H$\alpha$ line, $u$-band, $U$-band, $g$-band, $V$-band, and $TESS$-band, respectively, which are taken from Table \ref{table:list1_flares}.
         As for Flare Y3, $E_{\rm{Xray, flare}}$(0.5--2.0 keV) and $E_{\rm{Xray, flare}}$(\textit{GOES}-band)
         are the soft X-ray energies 
         in 0.5--2.0 keV and the \textit{GOES}-band (1.5--12.4 keV) \color{black}\textrm{
         from the NICER X-ray spectra. 
         As for other flares, $E_{\rm{Xray, flare}}$(\textit{GOES}-band) are estimated from the flare energies in H$\alpha$ line using the scaling law from \citet{Haisch_1989_A+A} (cf. Section 
         \ref{subsec:dis:blue-ejection}).}\color{black}}
         \tablenotetext{\rm e}{
         $E_{\rm{bol, flare}}^{(1)}$ and $E_{\rm{bol, flare}}^{(2)}$ are the flare bolometric energy with the two different methods described in Section \ref{subsec:dis:blue-ejection}, respectively. $E_{\rm{bol, flare}}$ is the resultant bolometric flare energy used for Figure \ref{fig:ene_mass_vel_kin}.
         } 

   \label{table:velmass_blue_flares}
 \end{deluxetable*}
\end{longrotatetable}

\clearpage

\subsection{Blue wing asymmetries and possible stellar mass ejections} \label{subsec:dis:blue-ejection}

\begin{figure}[ht!]
   \begin{center}
            \gridline{  
     \hspace{-0.06\textwidth}
    \fig{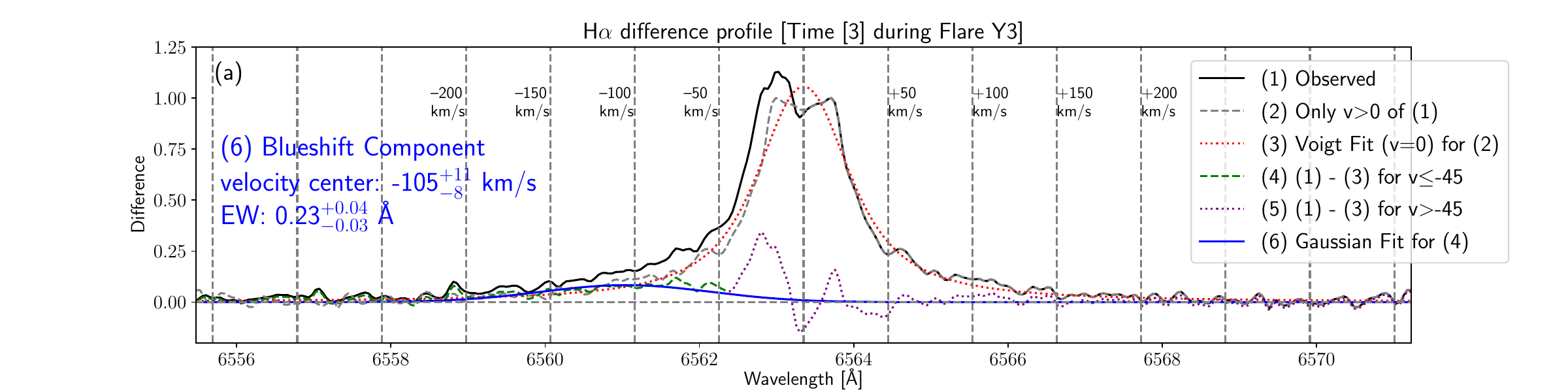}{0.56\textwidth}{\vspace{0mm}}
     \hspace{-0.06\textwidth}
    \fig{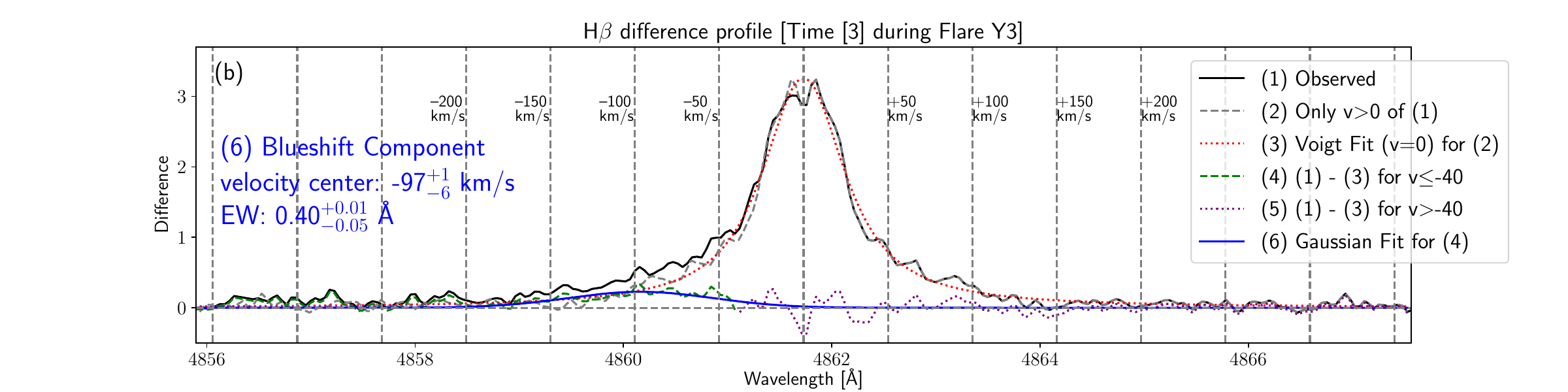}{0.56\textwidth}{\vspace{0mm}}
    }    
     \vspace{-0.04\textwidth}
            \gridline{  
     \hspace{-0.06\textwidth}
       \fig{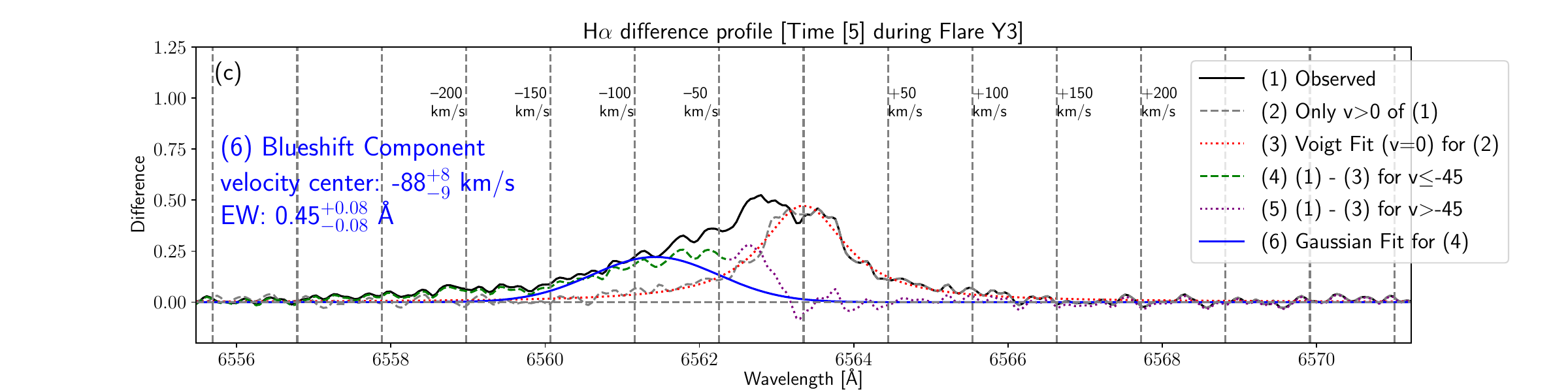}{0.56\textwidth}{\vspace{0mm}}
     \hspace{-0.06\textwidth}
       \fig{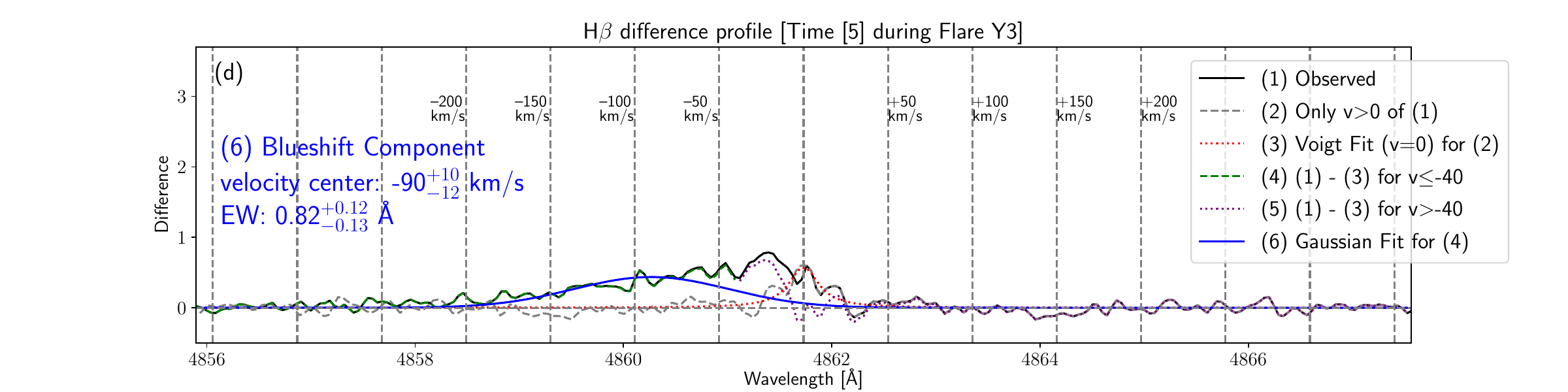}{0.56\textwidth}{\vspace{0mm}}
    }    
   \vspace{-5mm}
     \caption{
(a) Line profile change of the H$\alpha$ emission line from the quiescent phase at Time [3] during Flare Y3, which is the same as the H$\alpha$ difference profile shown in Figure \ref{fig:spec_HaHb_YZCMi_UT190127}(f). 
The horizontal and vertical axes represent the wavelength and flux normalized by the continuum. The grey vertical dashed lines with velocity values represent the Doppler velocities from the H$\alpha$ line center.
The black solid line (1) indicates the observed line profile change.
The gray dashed line (2) shows the symmetric line profile 
created by folding the red part ($>$0 km s$^{-1}$) of the original spectrum (1) 
to the blue part ($<$0 km s$^{-1}$).
The red dotted line (3) represents a Voigt function fit to the profile (2),
assuming the line-of-sight velocity of 0 km s$^{-1}$.
The blue solid line (6) shows a Gaussian fit to the residuals 
in the range shorter than the threshold velocity ($\leq$-45 km s$^{-1}$), which is shown with the green dashed line (4). 
This threshold velocity of -45 km s$^{-1}$ was determined by try-and-error and by eye
so that only the line wing component is used for the fitting. 
The range longer than the the threshold velocity ($>$-45 km s$^{-1}$) 
is plotted with the purple dotted line (5).
The result of the Gaussian fitting (6) (line-of-sight velocity and equivalent width of blue-shifted excess components) is shown in blue characters 
in the left side of the panel. 
\color{black}\textrm{The error values of the fitting results are obtained by changing the threshold velocities by $\pm$15 km s$^{-1}$ (i.e. 45$\pm$15 km s$^{-1}$ in this case).} \color{black}
(b) Same as (a) but for H$\beta$ line. The threshold velocity 
for the Gaussian fitting (6) is set to be -40 (\color{black}\textrm{$\pm$15}\color{black}) km s$^{-1}$.
(c) Same as (a) but at Time [5] during Flare Y3, which is the same as another H$\alpha$ difference profile shown in Figure \ref{fig:spec_HaHb_YZCMi_UT190127}(f).
(d) Same as (c) but for H$\beta$ line.
     }
   \label{fig:fit_blue_FlareY3}
   \end{center}
 \end{figure}

\begin{figure}[ht!]
   \begin{center}
            \gridline{  
     \hspace{-0.06\textwidth}
    \fig{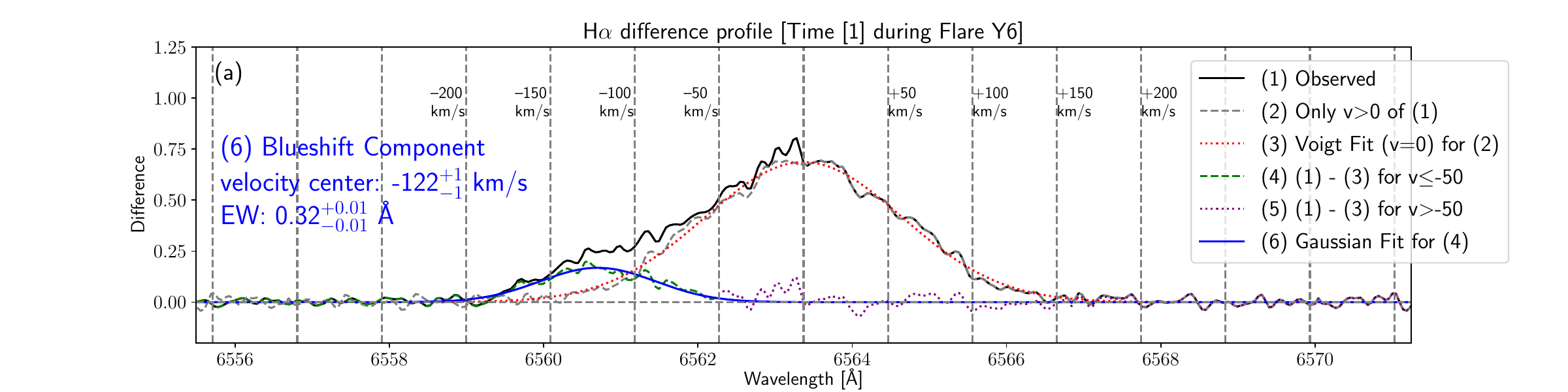}{0.56\textwidth}{\vspace{0mm}}
     \hspace{-0.06\textwidth}
       \fig{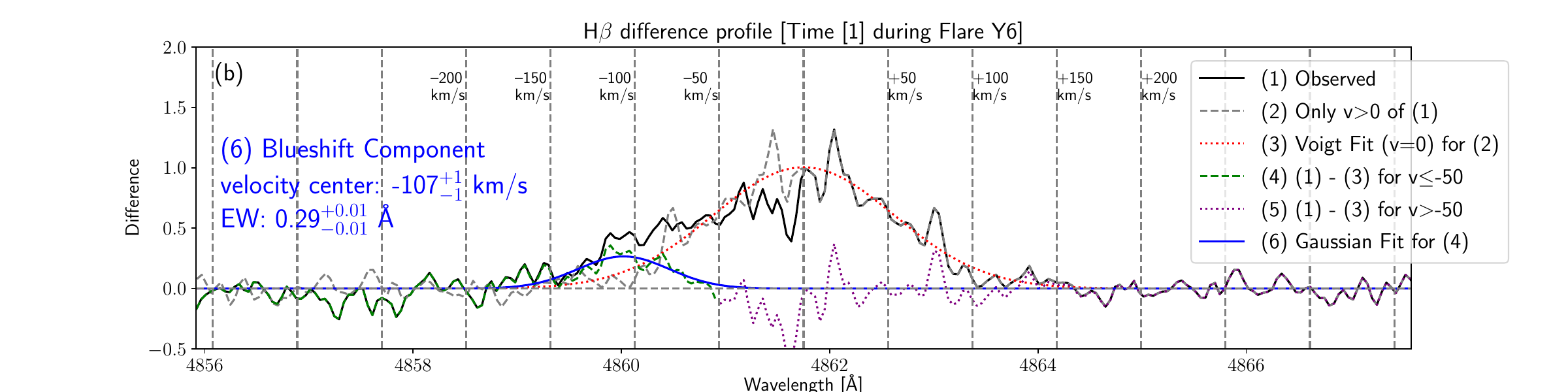}{0.56\textwidth}{\vspace{0mm}}
    }    
   \vspace{-5mm}
     \caption{
(a) Same as Figure \ref{fig:fit_blue_FlareY3}(a) but for the H$\alpha$ profile at Time [1] during Flare Y6, which is the same as the H$\alpha$ difference profile shown in Figure \ref{fig:spec_HaHb_YZCMi_UT191212}(b). 
The red dotted line (3) represents the result of Gaussian fit instead of Voigt fit. 
The threshold velocity 
for the Gaussian fitting (6) is set to be -50 (\color{black}\textrm{$\pm$15}\color{black}) km s$^{-1}$.
(b) Same as (a) but for H$\beta$ line.
     }
   \label{fig:fit_blue_FlareY6}
   \end{center}
 \end{figure}

\begin{figure}[ht!]
   \begin{center}
            \gridline{  
     \hspace{-0.06\textwidth}
    \fig{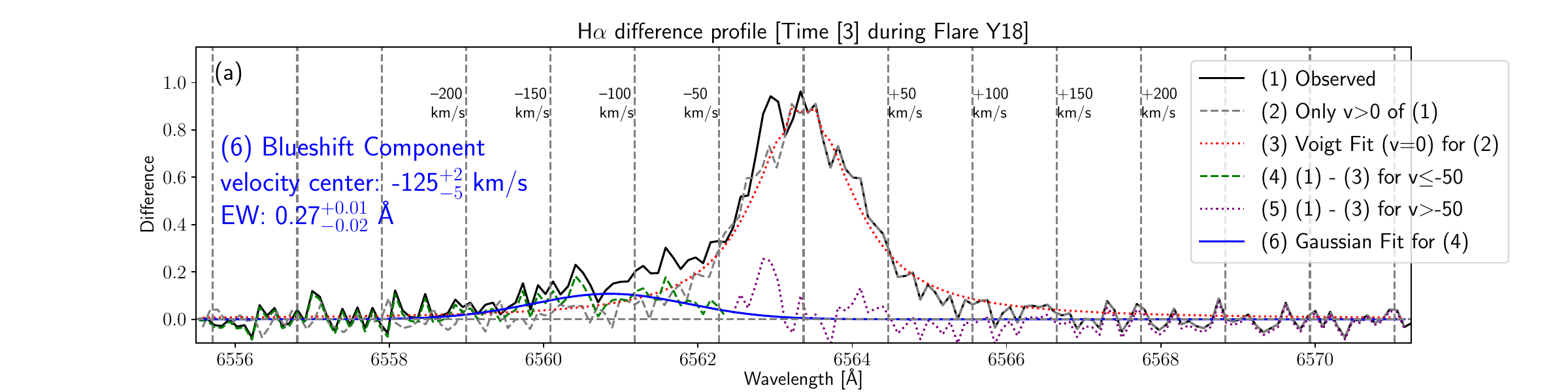}{0.56\textwidth}{\vspace{0mm}}
     \hspace{-0.06\textwidth}
       \fig{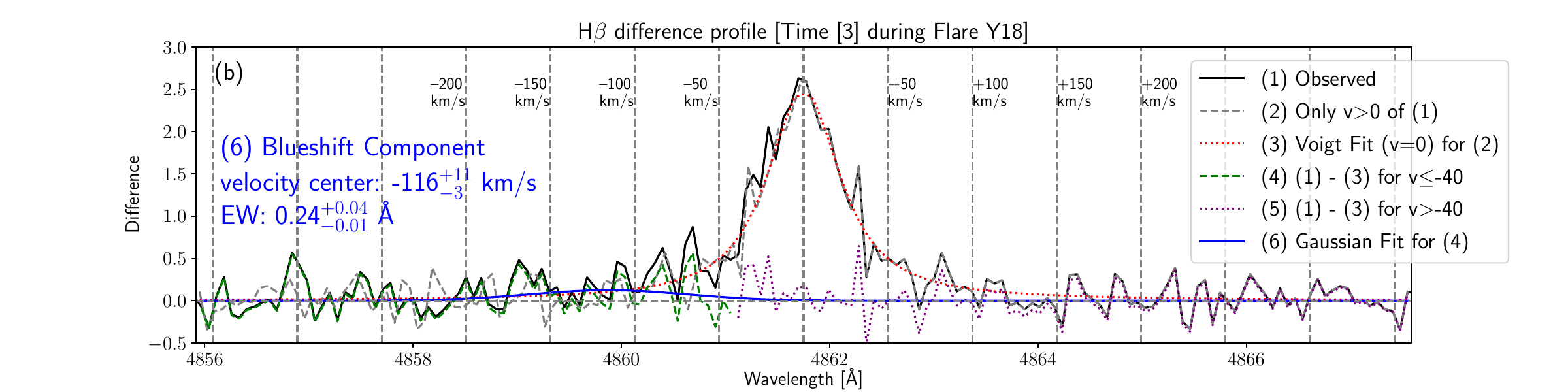}{0.56\textwidth}{\vspace{0mm}}
    }    
   \vspace{-5mm}
     \caption{
(a) Same as Figure \ref{fig:fit_blue_FlareY3}(a) but for the H$\alpha$ profile at Time [3] during Flare Y18, which is the same as the H$\alpha$ difference profile shown in Figure \ref{fig:spec_HaHb_YZCMi_UT200121_Y18}(f). 
The threshold velocity 
for the Gaussian fitting (6) is set to be -50 (\color{black}\textrm{$\pm$15}\color{black}) km s$^{-1}$.
(b) Same as (a) but for H$\beta$ line. The threshold velocity 
for the Gaussian fitting (6) is set to be -40 (\color{black}\textrm{$\pm$15}\color{black}) km s$^{-1}$.
     }
   \label{fig:fit_blue_FlareY18}
   \end{center}
 \end{figure}

\begin{figure}[ht!]
   \begin{center}
            \gridline{  
     \hspace{-0.06\textwidth}
    \fig{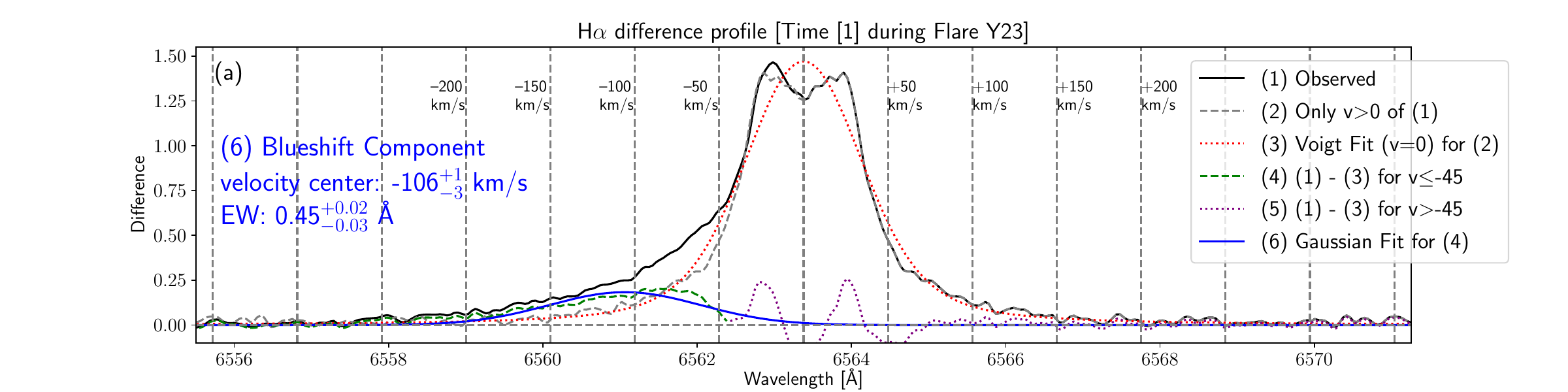}{0.56\textwidth}{\vspace{0mm}}
     \hspace{-0.06\textwidth}
       \fig{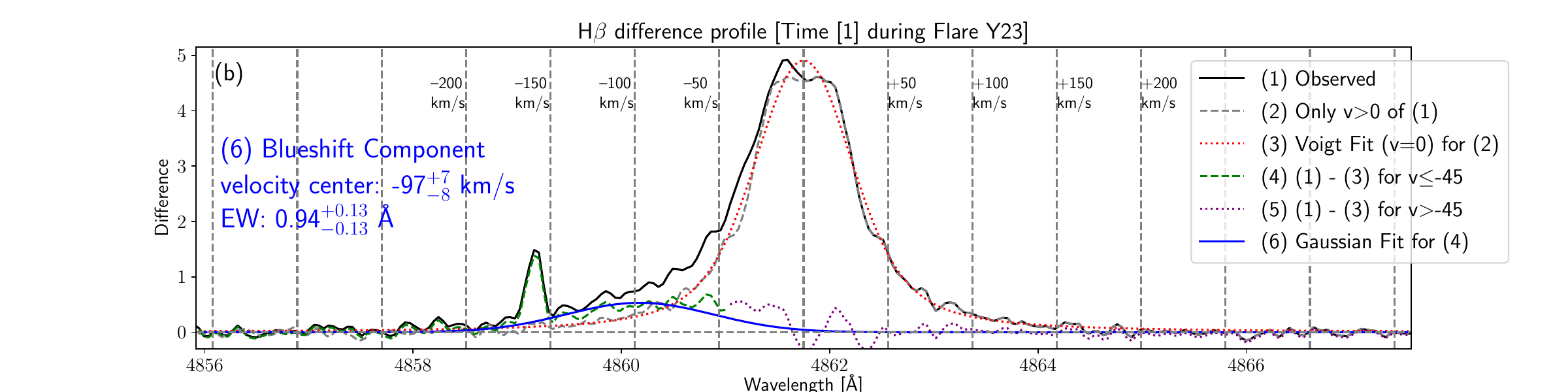}{0.56\textwidth}{\vspace{0mm}}
    }    
   \vspace{-5mm}
     \caption{
(a) Same as Figure \ref{fig:fit_blue_FlareY3}(a) but for the H$\alpha$ profile at Time [1] during Flare Y23, which is the same as the H$\alpha$ difference profile shown in Figure \ref{fig:spec_HaHb_YZCMi_UT201206}(b). 
The threshold velocity 
for the Gaussian fitting (6) is set to be -45 (\color{black}\textrm{$\pm$15}\color{black}) km s$^{-1}$.
(b) Same as (a) but for H$\beta$ line. 
     }
   \label{fig:fit_blue_FlareY23}
   \end{center}
 \end{figure}

\begin{figure}[ht!]
   \begin{center}
            \gridline{  
     \hspace{-0.06\textwidth}
    \fig{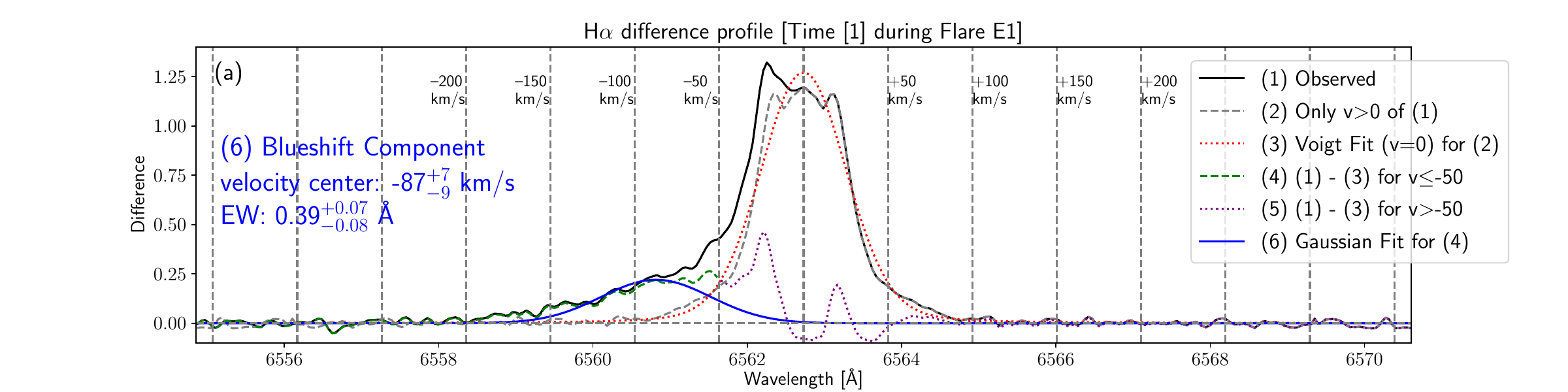}{0.56\textwidth}{\vspace{0mm}}
     \hspace{-0.06\textwidth}
       \fig{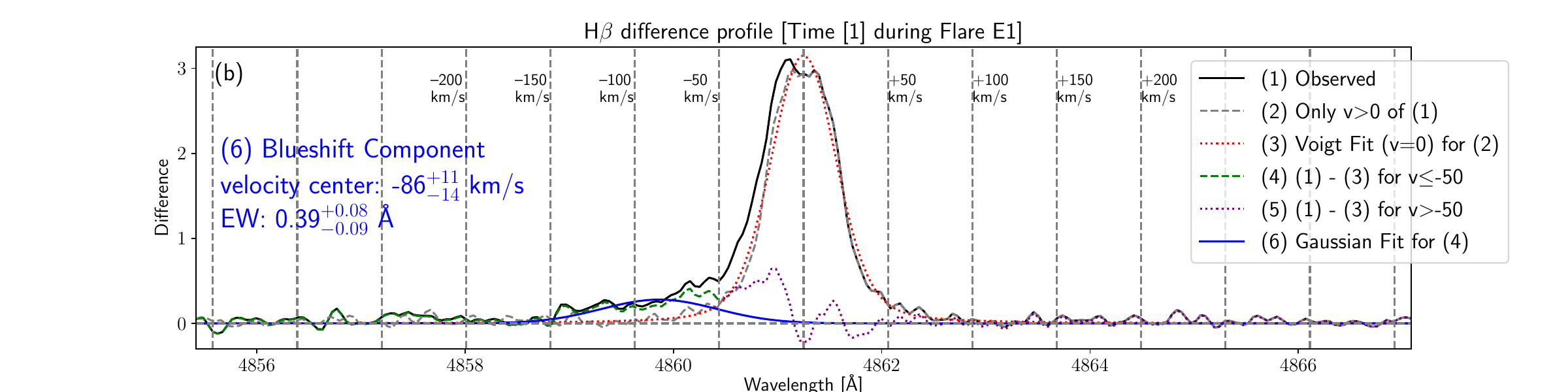}{0.56\textwidth}{\vspace{0mm}}
    }    
   \vspace{-5mm}
     \caption{
(a) Same as Figure \ref{fig:fit_blue_FlareY3}(a) but for the H$\alpha$ profile at Time [1] during Flare E1, which is the same as the H$\alpha$ difference profile shown in Figure \ref{fig:spec_HaHb_EVLac_UT191215}(b). 
The threshold velocity 
for the Gaussian fitting (6) is set to be -50 (\color{black}\textrm{$\pm$15}\color{black}) km s$^{-1}$.
(b) Same as (a) but for H$\beta$ line. 
     }
   \label{fig:fit_blue_FlareE1}
   \end{center}
 \end{figure}

\begin{figure}[ht!]
   \begin{center}
            \gridline{  
     \hspace{-0.06\textwidth}
    \fig{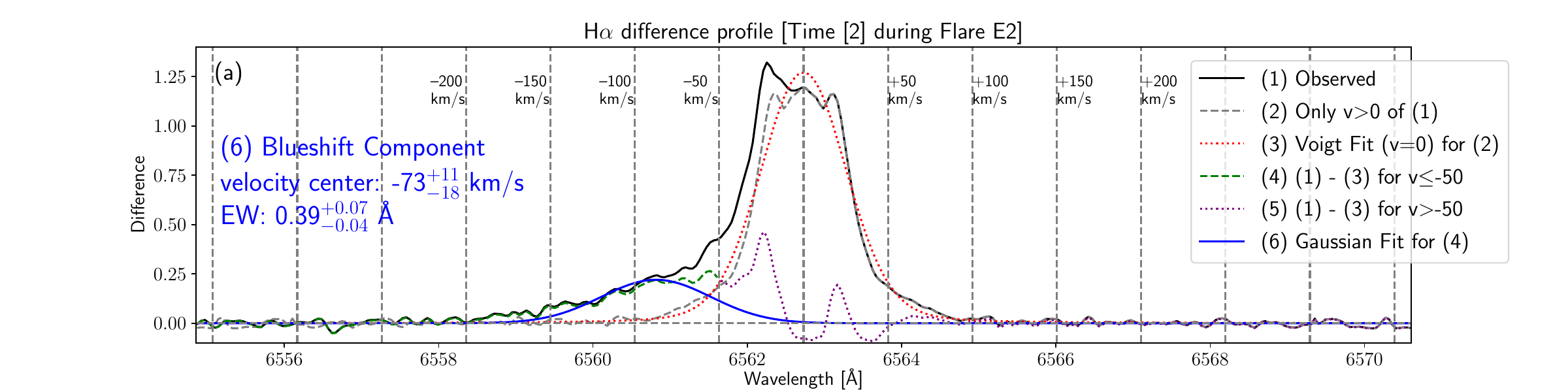}{0.56\textwidth}{\vspace{0mm}}
     \hspace{-0.06\textwidth}
       \fig{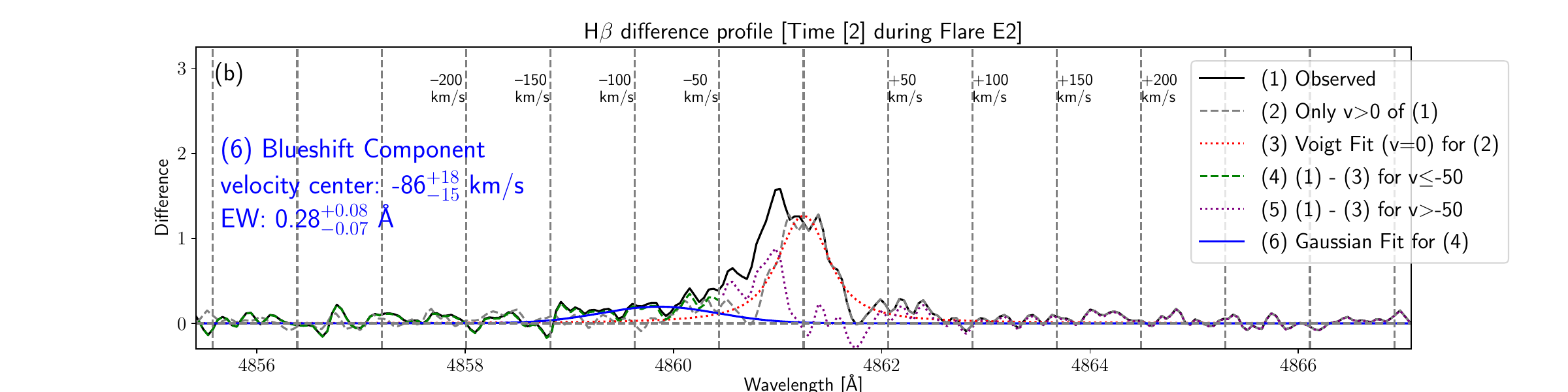}{0.56\textwidth}{\vspace{0mm}}
    }    
   \vspace{-5mm}
     \caption{
(a) Same as Figure \ref{fig:fit_blue_FlareY3}(a) but for the H$\alpha$ profile at Time [2] during Flare E2, which is the same as the H$\alpha$ difference profile shown in Figure \ref{fig:spec_HaHb_EVLac_UT191215}(b). 
The threshold velocity 
for the Gaussian fitting (6) is set to be -50 (\color{black}\textrm{$\pm$15}\color{black}) km s$^{-1}$.
(b) Same as (a) but for H$\beta$ line. 
     }
   \label{fig:fit_blue_FlareE2}
   \end{center}
 \end{figure}

\begin{figure}[ht!]
   \begin{center}
            \gridline{  
     \hspace{-0.06\textwidth}
    \fig{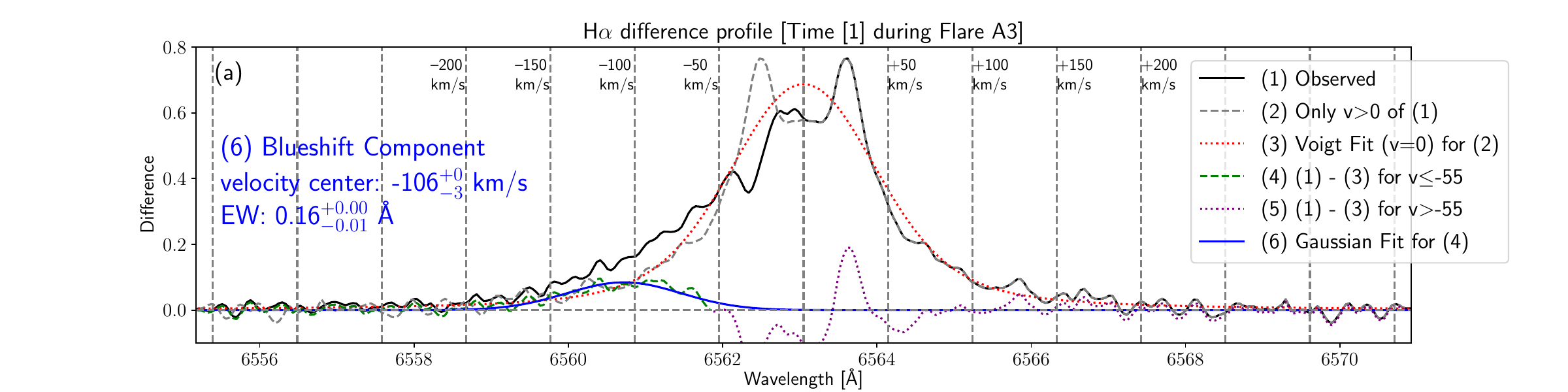}{0.56\textwidth}{\vspace{0mm}}
     \hspace{-0.06\textwidth}
       \fig{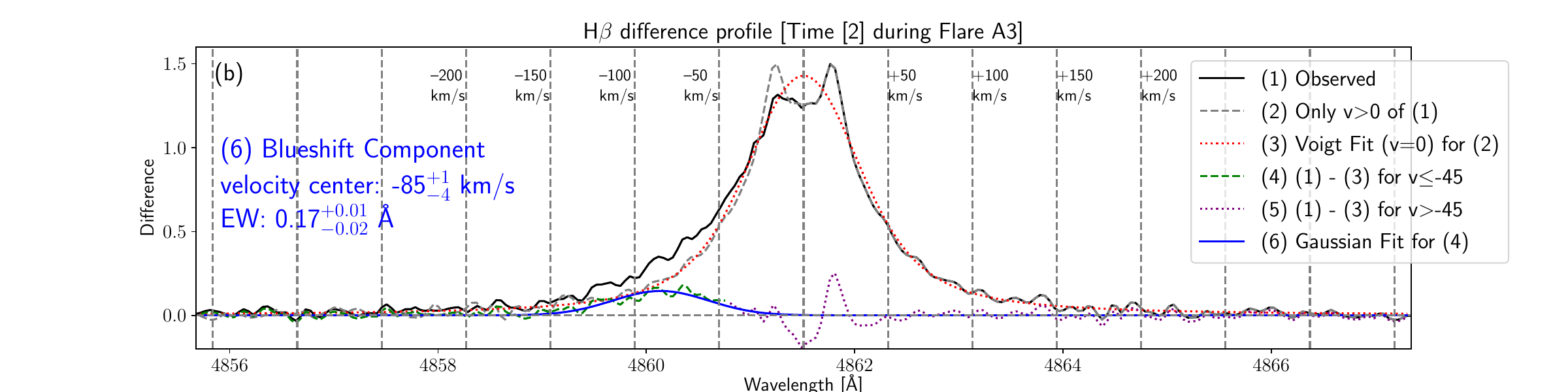}{0.56\textwidth}{\vspace{0mm}}
    }    
   \vspace{-5mm}
     \caption{
(a) Same as Figure \ref{fig:fit_blue_FlareY3}(a) but for the H$\alpha$ profile at Time [1] during Flare A3, which is the same as the H$\alpha$ difference profile shown in Figure \ref{fig:spec_HaHb_ADLeo_UT190519}(b). 
The threshold velocity 
for the Gaussian fitting (6) is set to be -55 ($\pm$15) km s$^{-1}$.
(b) Same as (a) but for H$\beta$ line at Time [2] \color{black}\textrm{during }\color{black} Flare A3, 
which is the same as the H$\beta$ difference profile shown in Figure \ref{fig:spec_HaHb_ADLeo_UT190519}(d).
The threshold velocity 
for the Gaussian fitting (6) is set to be -45 (\color{black}\textrm{$\pm$15}\color{black}) km s$^{-1}$.
\color{black}\textrm{We note that the H$\alpha$ data at Time [1] in (a) while H$\beta$ data at Time [2] in (b), since the H$\beta$ observation started later than H$\alpha$ (There are no H$\beta$ data at Time [1]) in Figure \ref{fig:lcEW_HaHb_ADLeo_UT190519}.}\color{black}
     }
   \label{fig:fit_blue_FlareA3}
   \end{center}
 \end{figure}

Using the line fitting method similar to \citet{Maehara+2021}, 
we estimated the velocities of blue wing excess components of H$\alpha$ and H$\beta$ lines 
(Figures \ref{fig:fit_blue_FlareY3}, 
\ref{fig:fit_blue_FlareY6}, 
\ref{fig:fit_blue_FlareY18}, 
\ref{fig:fit_blue_FlareY23}, 
\ref{fig:fit_blue_FlareE1}, 
\ref{fig:fit_blue_FlareE2}, 
\& \ref{fig:fit_blue_FlareA3}).
\color{black}\textrm{
We note that in this line fitting method (cf. \citealt{Maehara+2021}; \citealt{Inoue+2023_ApJ}), there is an assumption that the flare emission other than the component causing the blue wing asymmetry shows completely symmetric emission, and the red wing is not affected by any flare-related processes. If the red wing is affected by the flare-related processes simultaneously, then the measured blue wing asymmetry properties could be somewhat over-estimated or under-estimated. } \color{black}

As shown with the lines (3) in Figures \ref{fig:fit_blue_FlareY3}(a) \& (b), 
we first fitted the H$\alpha$ \& H$\beta$ difference profiles (the quiescent component subtracted profiles) with the Voigt functions, assuming the 
line-of-sight velocity of 0 km s$^{-1}$ and only using the red part 
($>$0 km s$^{-1}$) of the original spectra (lines (2)).
Next, we calculated the residuals between the fitted Voigt functions
and the observed spectra, which are shown with lines (4)\&(5) in 
Figures \ref{fig:fit_blue_FlareY3}(a) \& (b).
Finally, the residual was fitted with the Gaussian function 
to estimate the blue wing excess component (lines (6) 
in Figures \ref{fig:fit_blue_FlareY3}(a) \& (b)).
In this Gaussian fitting process, the wavelength ranges shorter than the threshold velocities (-45 and -40 km s$^{-1}$ for H$\alpha$ \& H$\beta$ profiles in Figures \ref{fig:fit_blue_FlareY3}(a) \& (b), respectively) were only used (lines (4)).
These threshold velocities were determined by trial-and-error and by eye,
so that asymmetries at the line center components (line (5))
do not affect the fitting and only the blue wing excess components
are used for the Gaussian fitting (line (6)).
Figures \ref{fig:fit_blue_FlareY3}(a) \& (b), which are described here,
show the results of the first blue asymmetry component of Flare Y3.
The fitting results of the second blue asymmetry component of Flare Y3 are shown in 
Figures \ref{fig:fit_blue_FlareY3}(c) \& (d).
The fitting results of blue asymmetry components of other six flares (Flare Y6, Y18, Y23, E1, E2, \& A3) are shown in Figures 
\color{black}\textrm{\ref{fig:fit_blue_FlareY6} -- \ref{fig:fit_blue_FlareA3}.} \color{black}As for Flare Y6 shown in Figure \ref{fig:fit_blue_FlareY6},
the Gaussian fitting was conducted instead of the initial Voigt fitting (lines (3)),
considering the line profile of the original spectra (lines (1)).
The threshold velocities of the Gaussian fitting\color{black}\textrm{(6)} \color{black} (e.g., -45 and -40 km s$^{-1}$ for H$\alpha$ \& H$\beta$ profiles in Figure \ref{fig:fit_blue_FlareY3}(a) \& (b)) are different among the events and lines, and the values are shown in the figures.
The results of Gaussian fitting (6) (line-of-sight velocity and equivalent width of blue wing enhancement components of H$\alpha$ and H$\beta$ lines) 
are shown with blue characters in Figures 
\color{black}\textrm{\ref{fig:fit_blue_FlareY3} -- \ref{fig:fit_blue_FlareA3}}\color{black},
and these values are listed in Table \ref{table:velmass_blue_flares} ($v^{\rm{H}\alpha}_{\rm{blue,fit}}$, $EW^{\rm{H}\alpha}_{\rm{blue,fit}}$, $v^{\rm{H}\beta}_{\rm{blue,fit}}$, $EW^{\rm{H}\beta}_{\rm{blue,fit}}$).
The error values of these fitting results are roughly obtained by changing the threshold velocities by $\pm$15 km s$^{-1}$. 
This range ``$\pm$15 km s$^{-1}$" is roughly assumed by considering the accuracy of the ``bye-eye" determination of the threshold velocity.
For example, in the case of the first asymmetry component (Time [3]) of Flare Y3, 
$v^{\rm{H}\alpha}_{\rm{blue,fit}}$=-105$^{+11}_{-8}$ km s$^{-1}$, 
and $EW^{\rm{H}\alpha}_{\rm{blue,fit}}$=0.23$^{+0.04}_{-0.03}$ \AA, 
$v^{\rm{H}\beta}_{\rm{blue,fit}}$=-97$^{+1}_{-6}$ km s$^{-1}$, 
and $EW^{\rm{H}\beta}_{\rm{blue,fit}}$=0.40$^{+0.01}_{-0.05}$ \AA, 
by considering the threshold velocities of -45$\pm$15 and -40$\pm$15 km s$^{-1}$ for H$\alpha$ \& H$\beta$ profiles in Figure \ref{fig:fit_blue_FlareY3}(a) \& (b).
In addition, it is noted that 
the H$\beta$ profile of Flare Y18 in Figure \ref{fig:fit_blue_FlareY18}(b) are particularly noisy, the error values for this event can be larger than those estimated here, and we have to \color{black}\textrm{keep this in mind }\color{black} in the following analyses.

The estimated Doppler velocities of the 7 blue-shift (blue wing asymmetry) events 
($v^{\rm{H}\alpha}_{\rm{blue,fit}}$) range from -73 to -122 km s$^{-1}$
(Table \ref{table:velmass_blue_flares}).
\color{black}\textrm{
Since the asymmetries do not recur periodically both in the blue and red wings independently of flares, these 7 blue wing asymmetries should be more likely to be related to flares, and cannot be explained by the rotationally modulated emission from the co-rotating prominence (e.g., \citealt{Jardine+2020}).} \color{black}
These velocities (73 -- 122 km s$^{-1}$) are also a bit larger than the upward velocities of blue asymmetries observed 
\color{black}\textrm{in H$\alpha$ line mainly in the early phase of solar flares (e.g., \citealt{Canfield+1990}; \citealt{Heinzel+1994_SoPh}). For reference, 
such blue asymmetries of solar flares have been also observed in other chromospheric lines (e.g., Mg II lines)  mainly in early phase (e.g., \citealt{Tei+2018}; \citealt{Huang+2019}; \citealt{Li+2019}).} \color{black}
The durations of these solar blue asymmetries (a few min) are
one or two orders of magnitude shorter than those of the blue wing asymmetries 
in this study ($\Delta t_{\rm{H}\alpha}^{\rm{blueasym}}=$20 min -- 2.5 hours in Table \ref{table:list_blue_flares}).
In contrast, the velocities of the \color{black}\textrm{blue wing asymmetries in this study} \color{black} (73--122 km s$^{-1}$) 
are in the same range of solar prominence/filament eruptions (e.g., 10--400 km s$^{-1}$ according to \citealt{Gopalswamy+2003}).
The timescale of solar prominence/filament eruptions observed in H$\alpha$ line 
is roughly 20 min -- 1 hour (\citealt{Namekata+2022_NatAst}; \citealt{Otsu+2022}),
and this could be comparable or a bit shorter than the durations of 
the blue wing asymmetries in this study 
($\Delta t_{\rm{H}\alpha}^{\rm{blueasym}}=$ 20 min -- 2.5 hours).
In the case of stellar flares, since we cannot obtain spatial information 
of the stellar surface, such prominence/filament \color{black}\textrm{eruption} \color{black} may be \color{black}\textrm{a} \color{black} possible cause
of the blue wing asymmetries in chromospheric lines associated with flares.
We note that \citet{Leitzinger+2022} suggested that unlike the Sun, 
the “filament” can be visible \color{black}\textrm{in} \color{black} emission even on the stellar disk in the case of M-dwarfs, 
since the stellar background emission components are quite weak.
In the following of this subsection, 
we discuss the blue wing asymmetries detected in this study,
from the viewpoint of stellar \color{black}\textrm{prominence} \color{black} eruptions.

\begin{figure}[ht!]
   \begin{center}
            \gridline{  
     \hspace{-0.06\textwidth}
    \fig{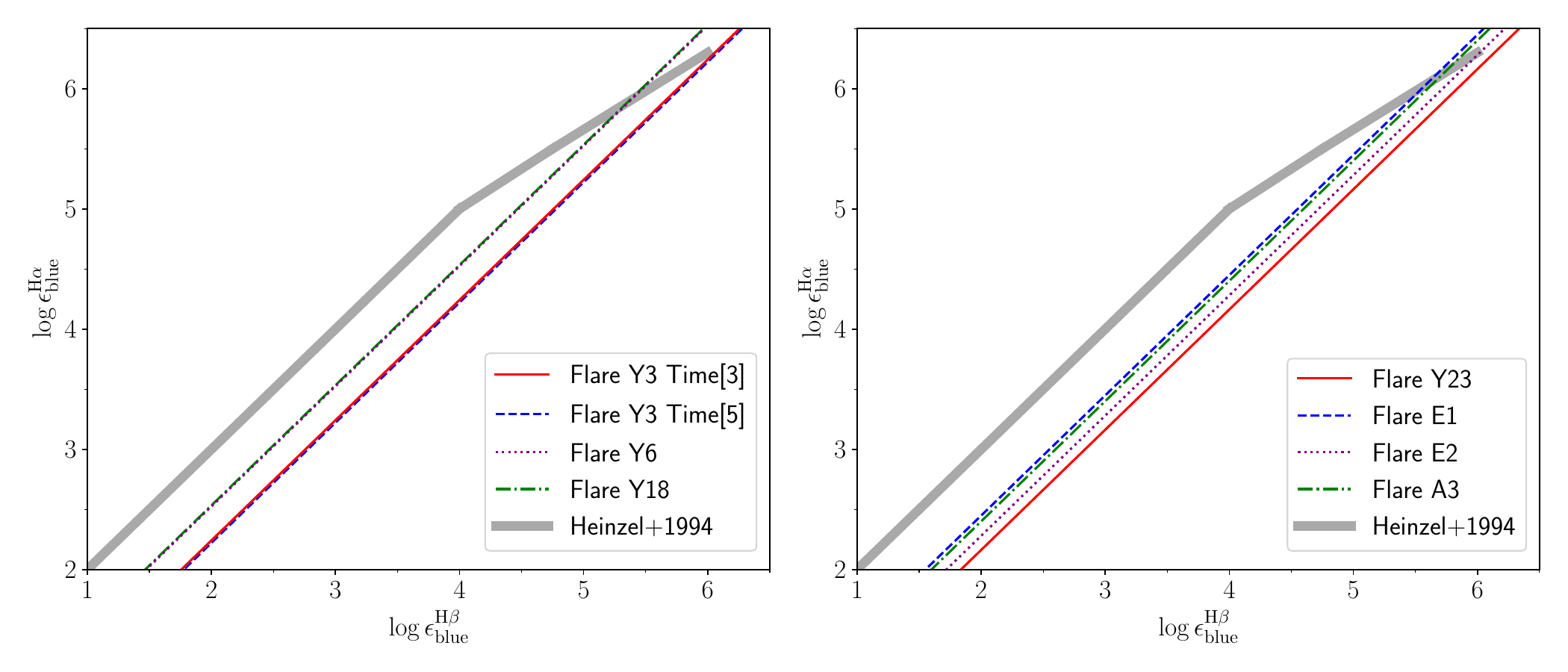}{0.82\textwidth}{\vspace{0mm}}
    }
   \vspace{-10mm}
     \caption{
Relations between H$\alpha$ and H$\beta$ 
\color{black}\textrm{line integrated intensities} \color{black}
emitted from the prominence.
Observed relations between \color{black}\textrm{$\epsilon^{\rm{H}\alpha}_{\rm{blue}}$ and  $\epsilon^{\rm{H}\beta}_{\rm{blue}}$} \color{black} (Equation (\ref{eq:log_Fha_Fhb})) of the blue wing asymmetry events are plotted with red solid lines, blue dashed lines, purple dotted lines and green dash dotted lines. 
The gray shaded area represent the result of the theoretical calculation of the NLTE slab model of solar prominence (taken from Figure 1 of \citealt{Heinzel+1994_A&A}).
     }
   \label{fig:Fha_Fhb_blue}
   \end{center}
 \end{figure}

The estimated equivalent width values of the H$\alpha$ and H$\beta$ emissions from 
blue-shifted excess components ($EW^{\rm{H}\alpha}_{\rm{blue,fit}}$ and $EW^{\rm{H}\beta}_{\rm{blue,fit}}$ in Table \ref{table:velmass_blue_flares}) can be converted to the luminosities of
H$\alpha$ and H$\beta$ emissions ($L^{\rm{H}\alpha}_{\rm{blue}}$ and  $L^{\rm{H}\beta}_{\rm{blue}}$ in Table \ref{table:velmass_blue_flares}),
by applying $EW^{\rm{H}\alpha}_{\rm{blue,fit}}$ and $EW^{\rm{H}\beta}_{\rm{blue,fit}}$ values into Equation (\ref{eq:L_line_flare}).
As done in \citet{Maehara+2021} and \citet{Inoue+2023_ApJ}, if we assume the simple slab NLTE emission model of solar prominences (e.g., \citealt{Heinzel+1994_A&A}) can be applied to the upward moving plasma (possible \color{black}\textrm{prominence} \color{black} eruptions) 
showing the blue-shifted excess components (blue wing asymmetries) on the M-dwarfs,
the luminosities of H$\alpha$ and H$\beta$ emissions ($L^{\rm{H}\alpha}_{\rm{blue}}$ and  $L^{\rm{H}\beta}_{\rm{blue}}$) can be calculated as
\color{black}\textrm{
  \begin{eqnarray}   \label{eq:Halpha_L_blue_F_blue_Ablue}
  L^{\rm{H}\alpha}_{\rm{blue}} = \int\int \epsilon^{\rm{H}\alpha}_{\rm{blue}} dAd\Omega
  = 2\pi A^{\rm{H}\alpha}_{\rm{blue}} \epsilon^{\rm{H}\alpha}_{\rm{blue}} \ , 
   \end{eqnarray}
   and 
   \begin{eqnarray}   \label{eq:Hbeta_L_blue_F_blue_Ablue}
  L^{\rm{H}\beta}_{\rm{blue}} = \int\int \epsilon^{\rm{H}\beta}_{\rm{blue}} dAd\Omega
  = 2\pi A^{\rm{H}\beta}_{\rm{blue}} \epsilon^{\rm{H}\beta}_{\rm{blue}} \ ,
  \end{eqnarray}
} \color{black}
\noindent
where $\epsilon^{\rm{H}\alpha}_{\rm{blue}}$ \& $\epsilon^{\rm{H}\beta}_{\rm{blue}}$
are \color{black}\textrm{the H$\alpha$ \& H$\beta$ line integrated intensities (cf. Table 1 of \citealt{Heinzel+1994_A&A})\footnote{\citet{Heinzel+1994_A&A} used the symbol ``$E$" for the line integrated intensities but ``$\epsilon$" is used in this study so that this cannot be confused with flare energies.}}  \color{black}, 
and $A^{\rm{H}\alpha}_{\rm{blue}}$ \& $A^{\rm{H}\beta}_{\rm{blue}}$ are the area of the region emitting H$\alpha$ and H$\beta$ lines.
If we assume H$\alpha$ and H$\beta$ emissions originate from the same area
($A^{\rm{H}\alpha}_{\rm{blue}}=A^{\rm{H}\beta}_{\rm{blue}}$), 
these two Equations (\ref{eq:Halpha_L_blue_F_blue_Ablue}) and (\ref{eq:Hbeta_L_blue_F_blue_Ablue}) are combined into one equation:
\color{black}\textrm{
   \begin{eqnarray}   \label{eq:alpha_F_Halpha_Hbeta}
  \frac{\epsilon^{\rm{H}\alpha}_{\rm{blue}}}{\epsilon^{\rm{H}\beta}_{\rm{blue}}} 
  = \frac{L^{\rm{H}\alpha}_{\rm{blue}}}{L^{\rm{H}\beta}_{\rm{blue}}} \equiv \alpha \ .
  \end{eqnarray}
} \color{black}
The $\alpha$ values calculated from $L^{\rm{H}\alpha}_{\rm{blue}}$ and  $L^{\rm{H}\beta}_{\rm{blue}}$ values are listed 
in Table \ref{table:velmass_blue_flares} 
(\color{black}\textrm{Note: The error values of $L^{\rm{H}\alpha}_{\rm{blue}}$, $L^{\rm{H}\beta}_{\rm{blue}}$, and $\alpha$ values in Table \ref{table:velmass_blue_flares} are from those of $EW^{\rm{H}\alpha}_{\rm{blue,fit}}$ and $EW^{\rm{H}\beta}_{\rm{blue,fit}}$ values}\color{black}). 
Then we get linear relations between logarithms of H$\alpha$ and H$\beta$
\color{black}\textrm{
line integrated intensities:
   \begin{eqnarray}   \label{eq:log_Fha_Fhb}
\log \epsilon^{\rm{H}\alpha}_{\rm{blue}}= \log\alpha + \log \epsilon^{\rm{H}\beta}_{\rm{blue}}
  \end{eqnarray}
} \color{black}
and these relations are plotted in Figure \ref{fig:Fha_Fhb_blue}.

\citet{Heinzel+1994_A&A} conducted the theoretical calculation of the NLTE slab model of solar prominence, and estimated the relation between H$\alpha$ and H$\beta$ \color{black}\textrm{line integrated intensities} \color{black}
(Figure 1 therein). If we assume this relation can be applied to 
the upward moving plasma showing the present blue-shifted excess components,
we can roughly determine the values of \color{black}\textrm{$\epsilon^{\rm{H}\alpha}_{\rm{blue}}$,} \color{black}
by comparing Equation (\ref{eq:log_Fha_Fhb}) with the theoretical relation as in Figure \ref{fig:Fha_Fhb_blue}. 
The resultant values of \color{black}\textrm{$\epsilon^{\rm{H}\alpha}_{\rm{blue}}$ } \color{black}
are listed in Table \ref{table:velmass_blue_flares}.
\color{black}\textrm{The error values of $\epsilon^{\rm{H}\alpha}_{\rm{blue}}$
listed in Table  \ref{table:velmass_blue_flares} are from 
the errors of $\alpha$ values and the scatter of the data points in Figure 1 of \citet{Heinzel+1994_A&A} ($\approx$ the width of the gray shaded area in Figure \ref{fig:Fha_Fhb_blue}). 
} \color{black}
By adapting Figure 5 of \citet{Heinzel+1994_A&A}, 
these values of
\color{black}\textrm{$\log \epsilon^{\rm{H}\alpha}_{\rm{blue}}$}\color{black}
[erg s$^{-1}$ cm$^{-2}$ sr$^{-1}$] = 5.9 -- 6.4 
correspond to the optical thicknesses of the H$\alpha$ line ($\tau_{\rm{H}\alpha}$) 
roughly ranging from 10 to 300 ($\log\tau_{\rm{H}\alpha}\sim 1.0-2.5$).
Using the resultant \color{black}\textrm{$\epsilon^{\rm{H}\alpha}_{\rm{blue}}$ }\color{black} 
values and Equation (\ref{eq:Halpha_L_blue_F_blue_Ablue}), the $A^{\rm{H}\alpha}_{\rm{blue}}$ values are obtained as listed in Table \ref{table:velmass_blue_flares}.
The resultant values of $A^{\rm{H}\alpha}_{\rm{blue}}\sim 10^{19} - 10^{20}$cm$^{2}$ roughly correspond to \color{black}\textrm{0.5--4}\color{black}\% of the visible stellar surface of the target stars (YZ CMi, EV Lac, and AD Leo). 
This value can be a bit smaller than or comparable to the area of starspots estimated from 
the amplitude of rotational modulations 
(e.g., the total spot coverage $\sim$ 6 -- 17\% for YZ CMi in \citealt{Maehara+2021}).

In Figure 15 of \citet{Heinzel+1994_A&A}, the correlation between H$\alpha$ \color{black}\textrm{line integrated intensity ($\epsilon^{\rm{H}\alpha}_{\rm{blue}}$) }\color{black} 
and emission measure $EM_{\rm{blue}}$=$n_{e}^{2}D$ 
is provided, where $n_{e}$ and $D$ are the electron density and geometrical thickness of the prominence, respectively.\footnote{
It is noted that the definition of emission measure 
for the H$\alpha$ emission here ($EM_{\rm{blue}}$=$n_{e}^{2}D$, from \citealt{Heinzel+1994_A&A}) 
is different from that for the X-ray emission in Section \ref{subsec:dis:Xray} 
(EM$=n^{2}V$, where $n$ is the electron density and $V$ is the volume,
from \citealt{Shibata_Yokoyama+2002}).
}
By adapting this correlation, the $EM_{\rm blue}$ values are obtained as listed in Table \ref{table:velmass_blue_flares}.
In this table, the $EM_{\rm blue}$ values are separately listed for two cases (e.g., $EM_{\rm blue}^{(1)}$ and $EM_{\rm blue}^{(2)}$). 
\color{black}\textrm{These two cases correspond to upper and lower limit values of $\epsilon^{\rm{H}\alpha}_{\rm{blue}}$ values (e.g., $\log \epsilon^{\rm{H}\alpha}_{\rm{blue}}$= 6.0 and 5.8 [erg s$^{-1}$ cm$^{-2}$ sr$^{-1}$] for Flare Y6), respectively,
which come 
from the error range of $\epsilon^{\rm{H}\alpha}_{\rm{blue}}$ in Table \ref{table:velmass_blue_flares} (e.g., $\log \epsilon^{\rm{H}\alpha}_{\rm{blue}}$[erg s$^{-1}$ cm$^{-2}$ sr$^{-1}$]=$5.9^{+0.1}_{-0.1}$ for Flare Y6). }\color{black}
Assuming the observed electron density range of solar prominences 
($\log n_{\rm{e}}$[cm$^{-3}$] = 10 -- 11.5  from \citealt{Hirayama+1986}), 
the geometrical thickness $D_{\rm{blue}}$ ($=EM_{\rm blue}/n_{\rm{e}}^{2}$) can be estimated from the emission measure $EM_{\rm blue}$. 
Since this assumed range of $n_{\rm{e}}$ could be wide, 
here we have another rough constraint that the prominence geometrical thickness is no larger than the stellar radius ($D_{\rm{blue}}\leq R_{\rm{star}}$). 
From this constraint, the lower limit of $n_{\rm{e}}$ can be determined as
$n_{\rm{e}}\geq \sqrt{EM_{\rm blue}/R_{\rm{star}}}$. 
The resultant estimated range of $n_{\rm{e}}$ and $D_{\rm{blue}}$ are listed in Table \ref{table:velmass_blue_flares}. For example, 
$\log n_{\rm{e}}^{(1)}$[cm$^{-3}$]$=10.3-11.5$ 
and $D_{\rm blue}^{(1)}=7.9\times10^{7}\rm{cm}- R_{\rm{star}}$ (=$2.0\times10^{10}$cm) for the upper limit \color{black}\textrm{$\epsilon^{\rm{H}\alpha}_{\rm{blue}}$} \color{black} case of Flare Y6.

With the estimated surface area ($A^{\rm{H}\alpha}_{\rm{blue}}$) and geometrical thickness ($D_{\rm{blue}}$) values, we can estimate the mass of the upward moving plasma showing the blue-shifted excess components ($M_{\rm{blue}}$): 
  \begin{eqnarray} \label{eq:Mass_estimate}
        M_{\rm{blue}} &\sim& A^{\rm{H}\alpha}_{\rm{blue}} D_{\rm{blue}} n_{\mathrm{H}} m_{\mathrm{H}} \\
        &=& A^{\rm{H}\alpha}_{\rm{blue}}\bigg(\frac{EM_{\rm blue}}{n_{\rm{e}}^{2}}\bigg) n_{\mathrm{H}} m_{\mathrm{H}} \\
        \label{eq:Mass_estimate_3}
        &=& A^{\rm{H}\alpha}_{\rm{blue}}\bigg(\frac{EM_{\rm blue}}{n_{\rm{e}}}\bigg)\bigg(\frac{n_{\rm{e}}}{n_{\rm{H}}}\bigg)^{-1} m_{\mathrm{H}}
        \ , 
  \end{eqnarray} 
where $n_{\mathrm{H}}$ is the total hydrogen density and $m_{\mathrm{H}}$ is the mass of hydrogen atom. 
Here we roughly assume the prominence ionization fraction 
from Table 1 of \citet{Labrosse+2010}, and $i=n_{\mathrm{e}}/n(\mathrm{H}^{0}) \approx n(\mathrm{H}^{+})/n(\mathrm{H}^{0}) = 0.2 - 0.9$, where 
$n(\mathrm{H}^{+})$ and $n(\mathrm{H}^{0})$ are the proton density and \color{black}\textrm{neutral }\color{black} hydrogen density, respectively.
From this, 
  \begin{eqnarray} \label{eq:ne_nH}
\frac{n_{\mathrm{e}}}{n_{\mathrm{H}}} &=&
\bigg(\frac{n_{\mathrm{e}}}{n(\mathrm{H}^{0})}\bigg)
\bigg(\frac{n(\mathrm{H}^{0})}{n_{\mathrm{H}}}\bigg) \\
&=&
\bigg(\frac{n_{\mathrm{e}}}{n(\mathrm{H}^{0})}\bigg)
\bigg(\frac{n(\mathrm{H}^{0})}{n(\mathrm{H}^{0})+n(\mathrm{H}^{+})}\bigg)\\
&\approx&
\bigg(\frac{n_{\mathrm{e}}}{n(\mathrm{H}^{0})}\bigg)
\bigg(\frac{n(\mathrm{H}^{0})}{n(\mathrm{H}^{0})+n_{\mathrm{e}}}\bigg)\\
&=&i/(i+1)\\
\label{eq:ne_nH_last}
&=&0.17-0.47 \ .
  \end{eqnarray} 
Then for example, 
the upper and lower limit of the prominence mass in the case of Flare Y6 
are estimated as follows:
   \begin{eqnarray} \label{eq:Mass_upp_FlareY6}
        M_{\rm{blue}, \rm{upp}} 
        &\sim& A^{\rm{H}\alpha}_{\rm{blue, upp}}\bigg(\frac{EM_{\rm{blue, upp}}^{(1)}}{n_{\rm{e, low}}^{(1)}}\bigg)\bigg(\bigg(\frac{n_{\rm{e}}}{n_{\rm{H}}}\bigg)_{\rm{low}}\bigg)^{-1} m_{\mathrm{H}}      \\
        &\sim& ((\color{black}\textrm{6.9+1.6}\color{black})\times 10^{19})
        \bigg(\color{black}\frac{10^{31.2}}{10^{10.3}}\color{black}\bigg)(\color{black}\textrm{0.17}\color{black})^{-1} (1.7\times10^{-24}) \\
        &\sim& \color{black}7.3\times 10^{17}\color{black} \rm{g} \ , 
  \end{eqnarray}  
and 
   \begin{eqnarray} \label{eq:Mass_low_FlareY6}
        M_{\rm{blue}, \rm{low}} 
        &\sim& A^{\rm{H}\alpha}_{\rm{blue, low}}\bigg(\frac{EM_{\rm{blue, low}}^{(2)}}{n_{\rm{e, upp}}^{(2)}}\bigg)\bigg(\bigg(\frac{n_{\rm{e}}}{n_{\rm{H}}}\bigg)_{\rm{upp}}\bigg)^{-1} m_{\mathrm{H}}      \\
        &\sim& ((\color{black}\textrm{6.9-1.6}\color{black})\times 10^{19})
        \bigg(\color{black}\frac{10^{30.5}}{10^{11.5}}\color{black} \bigg) (\color{black}\textrm{0.47}\color{black})^{-1} (1.7\times10^{-24}) \\
        &\sim& \color{black}1.9\times 10^{15}\color{black} \rm{g}  \ .
  \end{eqnarray}  

\begin{figure}[ht!]
   \begin{center}
            \gridline{  
    \fig{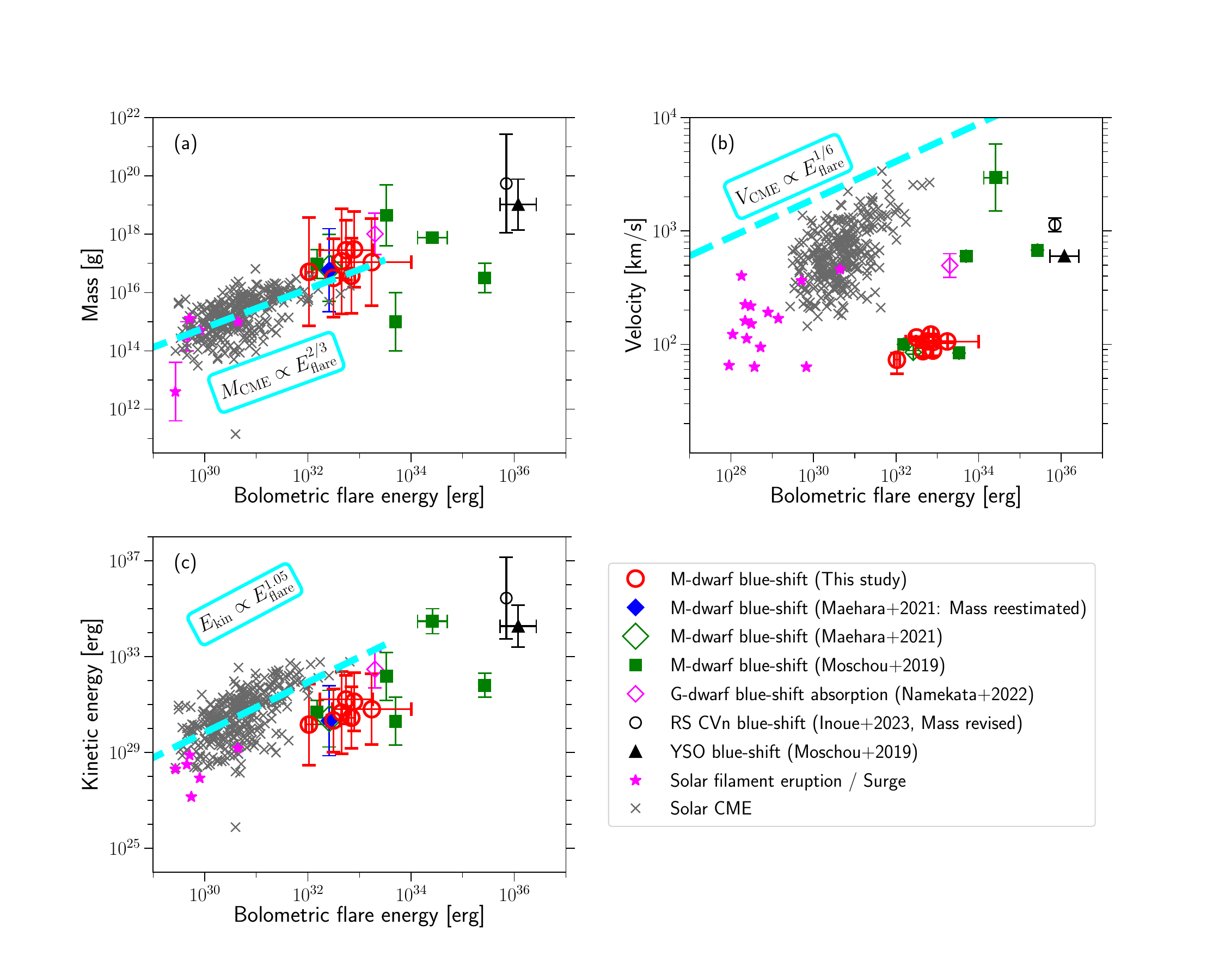}{1.0\textwidth}{\vspace{0mm}}
    }
   \vspace{-10mm}
     \caption{
Mass, velocity, and kinetic energy as a function of flare bolometric energy, for solar and stellar flares and prominence eruptions/CMEs. 
(a) Mass of prominence eruptions/CMEs as a function of flare bolometric energy. 
Red open circles represent the seven blue-shift (blue wing asymmetry) events on M-dwarfs (YZ CMi, EV Lac, ADLeo) reported in this study (Table \ref{table:velmass_blue_flares}).
The blue filled diamond and the green open diamond are the blue-shift event on M-dwarf YZ CMi reported in \citet{Maehara+2021}. The latter is the original datapoint, 
while the former is the datapoint with the mass value reestimated in this study.
Green filled squares and the black filled upward triangle are blue-shift events on M-dwarfs and a young stellar object (YSO), respectively, 
which are reported in \citet{Moschou+2019}. The pink open diamond represents 
the blue-shift absorption event on the young Sun-like star EK Dra reported in 
\citet{Namekata+2022_NatAst}, while the black open filled circle denotes the 
blue-shift event on the RS CVn type binary star reported in \citet{Inoue+2023_ApJ}. \color{black}
\textrm{It is noted that the mass value of \citet{Inoue+2023_ApJ} is slightly revised in this study as described in Section \ref{subsec:dis:blue-ejection}.}
\color{black}
Pink filled star marks correspond to filament eruptions / surges on the Sun taken 
from \citet{Namekata+2022_NatAst}, while gray crosses are CME
events on the Sun \color{black}\textrm{taken from \citet{Yashiro+2009} (see also \citealt{Drake+2013}). We acknowledge that the data of \citet{Yashiro+2009} were provided  thorough private communication with Dr. Seiji Yashiro.} \color{black}
The cyan dashed line represents the relation: $M_{\rm{CME}}\propto E^{2/3}_{\rm{flare}}$ shown by \citet{Takahashi+2016}, which is fitted to the solar CME data points in this figure.
(b) Velocity of prominence eruptions/CMEs as a function of flare bolometric energy. 
Pink filled star marks represent the filament eruptions on the Sun from \citet{Seki+2019}, and the other symbols are the same as in (a).
The scaling law denoted by the cyan dashed line ($V_{\rm{CME}}\propto E^{1/6}_{\rm{flare}}$) 
\color{black}\textrm{taken from \citet{Takahashi+2016} was plotted 
to show the the upper limit of CME speeds as a function of flare magnitude (cf. Equation (9) of \citealt{Takahashi+2016}).
} \color{black}
(c) Kinetic energy of prominence eruptions/CMEs as a function of flare bolometric energy. Symbols are plotted in the same way as in (a).
The scaling relation denoted by the cyan dashed line ($E_{\rm{kin}}\propto E^{1.05}_{\rm{flare}}$) is obtained from \citet{Namekata+2022_NatAst}, 
which is also fitted to the solar CME data points in this figure.
     }
   \label{fig:ene_mass_vel_kin}
   \end{center}
 \end{figure}

\color{black} 
\textrm{
It must be noted that there is an important assumption that we simply 
applied the solar prominence model of \citet{Heinzel+1994_A&A} for estimating the 
parameters (e.g., mass) of the upward moving plasma (prominence eruptions) of M-dwarfs.
It is assumed that the parameter space of the prominence plasma  
(e.g., density, temperature) is the same for the Sun and M dwarfs.
The models of \citet{Heinzel+1994_A&A} 
are computed for solar incident radiation 
(solar intensity and spectral energy distribution, and line profile shapes) 
and for a fixed height of 10,000km above the solar surface.
These model setups can be different between the Sun and M dwarfs.
The resulting prominence parameters can also depend crucially on the scattering of the incident radiation
since the emission of solar prominence is dominated by this scattering process (cf. Section 2 of  \citet{Heinzel+1994_A&A}).
Moreover, \citet{Heinzel+1994_A&A} do not calculate the models of erupting prominences 
but those of static prominences. 
This can also affect the calculation results 
because of the Doppler dimming and brightening effects
(e.g. \citealt{Heinzel+1987_SolPhys}; \citealt{Gontikakis+1997_A&A}).
These effects affect different lines in different ways, and for example, 
the parameter $\alpha$ in Equation (\ref{eq:alpha_F_Halpha_Hbeta}) can be affected.
}\color{black}
It is then important to assess the reliability of the obtained values by conducting the calculation of erupting prominences of M-dwarfs.
However, the discussions on prominence parameters (e.g., mass) in this study already include errors with two or three order-of-magnitude with various other assumptions (e.g., prominence shapes), and we only conduct broad discussions over many order-of-magnitude in the following (cf. Figure \ref{fig:ene_mass_vel_kin}). 
So in this study, we only use the simple assumption of solar prominence model from \citet{Heinzel+1994_A&A}, and the calculation of erupting prominences of M-dwarfs is beyond the scope of this study, considering the main focus of this paper is reporting the blue wing asymmetries from the huge campaign observations. We demonstrate that as a next study, 
it is important to conduct the NLTE model calculations of eruption prominences in the M-dwarf stellar atmosphere (cf. \citealt{Leitzinger+2022}) and reevaluate the prominence parameters (e.g., mass) more accurately .

\textrm{
We also estimated the kinetic energy
of the upward moving plasma showing the blue-shifted excess components 
($E_{\rm{kin,upp}}$ and $E_{\rm{kin,low}}$ in Table \ref{table:velmass_blue_flares})
from the velocity values of H$\alpha$ blue asymmetry components ($v^{\rm{H}\alpha}_{\rm{blue,fit}}$) and these mass values ($M_{\rm{blue}, \rm{upp}}$ and $M_{\rm{blue}, \rm{low}}$) listed in Table \ref{table:velmass_blue_flares}.
When we estimate $E_{\rm{kin,upp}}$ and $E_{\rm{kin,low}}$ values here, we simply used 
the line-of-sight velocity value $v^{\rm{H}\alpha}_{\rm{blue,fit}}$ as we use the same method with our previous studies estimating kinetic energies using Doppler shift velocities (e.g., \citealt{Maehara+2021}; \citealt{Namekata+2022_NatAst}; \citealt{Inoue+2023_ApJ}).
We must keep in mind the effects that the line-of-sight velocity is always smaller than or equal to the true velocity and the $E_{\rm{kin,upp}}$ values here cannot be true ``upper limit" values, when we discuss the kinetic energy values in the following.
}
\color{black}

As also done in the previous studies (e.g., \citealt{Moschou+2019}; \citealt{Maehara+2021}; \citealt{Namekata+2022_NatAst}; \citealt{Inoue+2023_ApJ}), 
these estimated mass ($M^{\rm{H}\alpha}_{\rm{blue}}$), velocity 
($v^{\rm{H}\alpha}_{\rm{blue,fit}}$), and kinetic energy ($E_{\rm{kin}}$) of the upward moving plasma (or \color{black}\textrm{prominence} \color{black}
eruptions) showing blue wing enhancements (blue wing asymmetries) can be discussed as a function of flare energy (Figure \ref{fig:ene_mass_vel_kin}).
In Figure \ref{fig:ene_mass_vel_kin}, 
we use flare bolometric energy ($E_{\rm{bol, flare}}$: cf. \citealt{Osten+2015})
for more general discussions, instead of \textit{GOES}-band X-ray energy used in some previous studies (e.g., \citealt{Moschou+2019}; \citealt{Maehara+2021}).
\color{black}\textrm{
We estimated the flare bolometric energies with the following two methods.
We simply used the both values to estimate the value ranges of bolometric energies. 
We note that the both methods include several assumptions/ambiguities. 
For example, the earlier method relies on the ground-based (ARCSAT and LCO) photometric data
including some data gaps in this study. The latter relies on the scaling law between H$\alpha$ and GOES X-ray band flare energies from \citet{Haisch_1989_A+A}, which only used small number of stellar flares. Moreover, the H$\alpha$ and GOES X-ray emission components consist of roughly up to a few percent of the total flare energy, and they are emitted upper part of the stellar atmosphere (chromosphere and corona) while the dominant part of bolometric energy is emitted as white-light emission from lower atmosphere (e.g., \citealt{Emslie+2012}; \citealt{Osten+2015}). 
For more precise and accurate estimations of flare bolometric energies, more comprehensive multi-wavelength observation data to estimate flare energies from X-rays to optical are important. 
This point should be kept in mind when discussing flare bolometric energies in the following,
while we only conduct the order-of-magnitude discussions for flare energy values in Figure \ref{fig:ene_mass_vel_kin}.
} \color{black}

In the first method, we
convert the flare energies in the $u$- and $U$-bands into 
bolometric energies assuming the energy partitions of \citet{Osten+2015}, since most of the flares in this study were observed in either $u$- or $U$-bands and  
the flare amplitude signal-to-noise ratios in $u$- or $U$-bands are generally 
better than those in $g$- or $V$-bands in this study (see light curve figures in Section \ref{sec:results}).
The fraction of $U$-band flare energy to the the bolometric energy 
is $E_{U, \rm{ flare}}/E_{\rm{bol, flare}} \sim 0.11$ (Table 2 of \citealt{Osten+2015}).
Then, assuming the luminosity ratio of $U$- and $u$-bands 
in Table \ref{table:targets_quiescent_flux},
the fraction of $u$-band to the the bolometric energy 
is $E_{u, \rm{ flare}}/E_{\rm{bol, flare}} \sim 0.09$.
\color{black}\textrm{
The resultant energy values from this first method 
are listed as $E_{\rm{bol, flare}}^{(1)}$ in Table \ref{table:velmass_blue_flares}.
$E_{\rm{bol, flare}}^{(1)}$ values of Flares Y23 \& E2 were estimated to be $E_{\rm{bol, flare}}^{(1)}$=1.8$\times$10$^{33}$ and 1.2$\times$10$^{32}$ erg, respectively, from the observed $E_{u, \rm{ flare}}$ values, and that of Flare Y18 was to be $E_{\rm{bol, flare}}^{(1)}$=4.2$\times$10$^{32}$ erg from the observed $E_{U, \rm{ flare}}$ value. 
As for Flares Y3 and Y6, the upper limit of bolometric energies could be estimated to be $E_{\rm{bol, flare}}^{(1)}<6.2\times10^{32}$ and $<8.0\times10^{32}$  erg from the observed upper limit of $E_{u, \rm{ flare}}$ value. 
As for Flare A3, only the lower limit of bolometric energy was estimated to be $E_{\rm{bol, flare}}^{(1)}>3.0\times10^{32}$ erg from the observed lower limit of $E_{u, \rm{ flare}}$, since the flare already started when the observation started.
Flare E1 did not have simultaneous photometric data. 
} \color{black}

\color{black}\textrm{
In the second method, we convert the \textit{GOES}-band X-ray (1.5--12.4 keV = 1--8 \AA~range: $E_{\rm{Xray, flare}}$(\textit{GOES}-band))
into bolometric energies ($E_{\rm{bol, flare}}$) assuming the energy partitions of \citet{Osten+2015} ($E_{\rm{Xray, flare}}$(\textit{GOES}-band)$=0.06E_{\rm{bol, flare}}$ in Table 2 therein). 
As for Flare Y3, the \textit{GOES}-band X-ray energy estimated from NICER X-ray spectra in Section \ref{subsec:results:2019-Jan-27} are used here ($E_{\rm{Xray, flare}}$(\textit{GOES}-band)= $4.7 \times 10^{31}$ erg).
As for the other flares without NICER X-ray data in this study, 
the \textit{GOES}-band X-ray energy was converted from the H$\alpha$ flare energy ($E_{\rm{H}\alpha\rm{, flare}}$), using the empirical relationship between H$\alpha$ and \textit{GOES}-band soft X-ray flare energies in Figure 2 and Equation (1) 
of \citet{Haisch_1989_A+A}.
The resultant energy values from this first method 
are listed as $E_{\rm{bol, flare}}^{(2)}$ in Table \ref{table:velmass_blue_flares}.
} \color{black}

\color{black}\textrm{
Using the bolometric energies estimated with these two methods
($E_{\rm{bol, flare}}^{(1)}$ and $E_{\rm{bol, flare}}^{(2)}$), 
the resultant $E_{\rm{bol, flare}}$ values are estimated 
and shown in Table \ref{table:velmass_blue_flares} and in Figure \ref{fig:ene_mass_vel_kin}.
As for Flares Y6, Y18, Y23, \& E2, the ranges of $E_{\rm{bol, flare}}$ values are estimated 
by simply taking the value differences of $E_{\rm{bol, flare}}^{(1)}$ and $E_{\rm{bol, flare}}^{(2)}$.
As for Flare Y3, we only used $E_{\rm{bol, flare}}^{(2)}$ for $E_{\rm{bol, flare}}$ value, 
since the estimated upper limit value of $E_{\rm{bol, flare}}^{(1)}$ is smaller than $E_{\rm{bol, flare}}^{(2)}$. As for Flare E1, only $E_{\rm{bol, flare}}^{(2)}$ is used for $E_{\rm{bol, flare}}$ value
since there were no photometric data. As for Flare A3, $E_{\rm{bol, flare}}^{(1)}$ 
is used for the lower limit of $E_{\rm{bol, flare}}$ value.
The upper limit of $E_{\rm{bol, flare}}$ is set to be $10^{34}$ erg,
very roughly assuming that the total flare energy is not larger than one order of magnitude larger than the limit $E_{\rm{bol, flare}}^{(2)}>8.2\times10^{32}$ erg in the second method from the available H$\alpha$ observation data.  
} \color{black}

In Figure \ref{fig:ene_mass_vel_kin}, the bolometric energies of solar events (the pink filled star marks and gray crosses) are plotted, assuming the energy conversion from \textit{GOES} X-ray band to bolometric energy for solar flares : $E_{GOES}=0.01E_{\rm{bol, flare}}$ (\citealt{Emslie+2012}; \citealt{Osten+2015}).
As for the data from \citet{Namekata+2022_NatAst} and \citet{Inoue+2023_ApJ}, the bolometric energy values estimated in these papers are used in Figure \ref{fig:ene_mass_vel_kin}.
\color{black}\textrm{
As for the event from \citet{Maehara+2021} (listed as ``M2021" in Table \ref{table:list_blue_flares}),  
we estimated to be $E_{\rm{bol, flare}}=2.6\times 10^{32}$ erg from their reported 
H$\alpha$ flare energy ($E_{\rm{H}\alpha}=4.7\times 10^{30}$ erg) 
on the basis of the above second method using the scaling relation of \citet{Haisch_1989_A+A} \footnote{\color{black}\textrm{
\citet{Maehara+2021} reported $E_{GOES}=8\times 10^{31}$ erg (see also Figure 10 therein).
However, the X-ray band luminosity in the 0.04--2.0 keV band
is not a proper GOES bandpass energy (see Equation (5) of \citealt{Maehara+2021} and Equation (1) of \citealt{Moschou+2019}).
The correct value is $E_{GOES}=4.4\times 10^{30}$ erg using the $E_{GOES}-E_{\rm{H}\alpha}$ scaling relation of \citet{Haisch_1989_A+A}. 
It is then noted that the relative location in the x-axis 
between the data of \citealt{Maehara+2021} and 
those of other events (e.g., the events from \citealt{Moschou+2019}) is a bit changed in 
Figure \ref{fig:ene_mass_vel_kin} from Figure 10 of \citet{Maehara+2021}, although
overall order-of-magnitude discussions are not affected.
}}. 
It is noted \citet{Maehara+2021}
included the TESS data, but this event did not show clear white-light emission and can be categorized as a non white-light flare.
} \color{black}
The bolometric energies of the stellar blue-shift events from \citet{Moschou+2019}
\color{black}\textrm{(including the event ``V2016" in Table \ref{table:list_blue_flares}) } \color{black}
are plotted in Figure \ref{fig:ene_mass_vel_kin}, assuming the 
energy conversion relation from \textit{GOES} X-ray band energy to bolometric energy 
$E_{GOES}=0.06E_{\rm{bol, flare}}$, which is the scaling relation for active stars in \citet{Osten+2015}.
\color{black}
\textrm{
\citet{Moschou+2019} originally included the events observed from X-ray absorptions, 
but only the events detected by blue-shifts of chromospheric lines are plotted in Figure 
\ref{fig:ene_mass_vel_kin} for simple comparison with blue-shift events reported in this study.
}
\color{black}

In addition, the mass value of the M-dwarf blue-shift event of \citet{Maehara+2021} 
is reestimated to be \color{black}\textrm{$2.2\times10^{15} - 1.5\times10^{18}$} \color{black}g (blue filled diamonds in Figure  \ref{fig:ene_mass_vel_kin}(a)\&(c)), 
by assuming the $F_{\rm{blue}}^{\rm{H}\alpha}$ range from the 7 M-dwarf events in this study ($\log F_{\rm{blue}}^{\rm{H}\alpha}$[erg s$^{-1}$ cm$^{-2}$ sr$^{-1}$] = 5.9 -- 6.4) and using the almost the same estimation method as in this study. 
Only the difference of the method with this study is that we assume the $F_{\rm{blue}}^{\rm{H}\alpha}$ range, 
since \citet{Maehara+2021} only had H$\alpha$ data and we cannot determine
$F_{\rm{blue}}^{\rm{H}\alpha}$ value from the relation 
of $F_{\rm{blue}}^{\rm{H}\alpha}$ 
and $F_{\rm{blue}}^{\rm{H}\beta}$ 
(cf. Figure \ref{fig:Fha_Fhb_blue}).
The mass value of the blue-shift event of the RS CVn-type star from our previous paper \citet{Inoue+2023_ApJ} is also slightly revised from $9.5\times10^{17}$--$1.4\times10^{21}$ g to $1.1\times10^{18}$--$2.7\times10^{21}$ g. 
This is because in \citet{Inoue+2023_ApJ}, although we used the same basic equations 
with this study (cf. Eq. \ref{eq:Mass_estimate_3}), 
we mistakenly assumed 
$n_{\mathrm{e}}/n_{\mathrm{H}}=n_{\mathrm{e}}/n(\mathrm{H}^{0})=0.2-0.9$, which is incorrect. 
The correct value is $n_{\mathrm{e}}/n_{\mathrm{H}}=0.17-0.47$ (cf. Eq. (\ref{eq:ne_nH_last})), and the resultant mass value range is slightly affected (the lower limit value becomes double). Overall discussions does not change since there were already larger range of values.

As we can see in Figure \ref{fig:ene_mass_vel_kin}(b),
the maximum observed line-of-sight velocities 
of the 7 blue-shift (blue wing asymmetry) events reported in this study 
($v^{\rm{H}\alpha}_{\rm{blue,fit}}$) range from 73 to 122 km s$^{-1}$.
These values are in the same range of solar filament/prominence eruptions 
associated with CMEs (10--400 km s$^{-1}$ in \citealt{Gopalswamy+2003}; see also the pink star marks in  Figure \ref{fig:ene_mass_vel_kin}(b)).
\color{black}\textrm{These values are also roughly comparable to the M-dwarf blueshift event 
from \citet{Maehara+2021} (the green open diamond mark) and some of the events from \citet{Moschou+2019} (the green filled square marks). 
In addition, the velocities of M-dwarf blue wing asymmetries from the other papers 
(\citealt{Vida+2019}; \citeauthor{Muheki+2020} \citeyear{Muheki+2020} \& \citeyear{Muheki+2020_EVLac}) are also in the similar ranges (e.g., the observed maximum velocities of M-dwarf blue wing asymmetries are 100--300 km s$^{-1}$ in \citet{Vida+2019}).
} \color{black}

It has been discussed whether blue wing asymmetries on M-dwarfs cause stellar CMEs (\citealt{Vida+2016} \& \citeyear{Vida+2019}; \color{black}\textrm{\citealt{Moschou+2019}; \citealt{Muheki+2020_EVLac}; } \color{black} \citealt{Maehara+2021}).
\color{black}\textrm{
The blue wing velocities have been compared with escape velocities, as one potential interpretation
that the observed velocities are relatively slow (e.g., \citealt{Moschou+2019}; \citealt{Vida+2019}; \citealt{Muheki+2020_EVLac}).
For example, the velocities of blue-shift events (73--122 km s$^{-1}$) in this study are smaller than the escape velocities at the stellar surface ($\sim$600 km s$^{-1}$ for YZ CMi, EV Lac, and AD Leo).
However, this cannot simply lead to the conclusion that 
the plasma is not ejected from the star, 
as the blue-shift events only provide the lower limit of the velocities and as summarized in the following, based on the relevant discussions and similar interpretations in previous papers (e.g., \citealt{Moschou+2019}; \citealt{Vida+2019}; \citealt{Maehara+2021}; \citealt{Namekata+2022_NatAst}).
\citet{Gopalswamy+2003} showed 
} \color{black} 
that the average CME core velocity ($\sim$350 km s$^{-1}$) and average CME velocity ($\sim$610 km s$^{-1}$) 
are $\sim$4 and $\sim$8 times larger than that of the associated prominence eruptions 
($\sim$80 km s$^{-1}$). 
This indicates that prominences with initial slow speeds 
are accelerated as they are lifted up and they evolve into CMEs. 
However, this indicates that if we assume similar acceleration mechanism would work,\footnote{
We note that although this assumption on the acceleration can be certainly possible on the basis of 
the solar observations/models of prominence eruptions and CMEs (e.g., \citealt{Otsu+2022}),
this can be also only speculation, 
considering that the acceleration is not (or cannot be) observed 
within the available observational dataset of Balmer lines in this paper.
Future observations of blue-shifts simultaneously with other CME detection methods may help more understanding (see the brief remark in the later part of this subsection).
} these prominence eruptions would be accelerated into $\sim$ 300 -- 1000 km s$^{-1}$.
This value is generally larger than the escape velocities at $\sim$ 2 -- 3$R_{\rm{star}}$ ($\sim$ 300 -- 450 km s$^{-1}$), and the prominence eruptions with the velocity 
of $\sim$100 km s$^{-1}$ could evolve into CMEs.
Moreover, the observed blue-shift velocities are line-of-sight velocities,
and the radial velocities of prominence eruptions 
can be larger considering the projection angle effect, 
which suggests that these prominence eruptions could evolve into CMEs 
with faster velocities. 
In addition, it is noted that red wing enhancements were observed during some flares with blue wing asymmetries (especially late-phase red wing asymmetry during Flare Y6) as summarized in Section \ref{subsec:dis:flare-blue}, 
which indicates that some of the materials fell back to the stellar surface. This phenomenon is often observed in the case of solar filament/prominence
eruptions even in the case that they evolve into CMEs (\citealt{Wood+2016}; \citealt{Namekata+2022_NatAst}; \citealt{Otsu+2022}).

The erupted masses of the 7 blue-shift events are estimated to be $M^{\rm{H}\alpha}_{\rm{blue}}\sim
10^{15}-10^{19}$ g (Table \ref{table:velmass_blue_flares}).
We note that some blue-shift events have long durations 
(e.g., $\Delta t_{\rm{H}\alpha}^{\rm{blueasym}}\sim$ 2 hours in the case of Flares Y6\&A3),
and it could be speculated that 
\color{black}\textrm{
these events were observed as superpositions of multiple consecutive flare events (cf. models of sympathetic eruptions as in \citealt{Torok+2011_ApJ}, \citealt{Lynch+2013_ApJ}, \citealt{Lynch+2016_ApJ}).
}\color{black}
This might cause the underestimate of mass since we only used the data at the peak of the continuous blue wing asymmetry events, and more detailed studies are necessary in the future.
In another point, we assumed the theoretical calculation results of \citet{Heinzel+1994_A&A} for the mass estimation process (e.g., Figure \ref{fig:Fha_Fhb_blue}), but this is only the calculation for solar prominences.
\color{black} 
\textrm{
As described in the earlier part of this subsection, 
this could significantly affect the reliability of the results presented here, 
and it is } \color{black}
important to conduct the NLTE model calculations of prominences in the M-dwarf stellar atmosphere for more accurate mass estimations in the future
(cf. \citealt{Leitzinger+2022}). 
Although there is a very large range of uncertainty of the mass estimation method,
Figure \ref{fig:ene_mass_vel_kin}(a) shows that 
these estimated mass of the 7 blue-shift events are roughly on the relation
expected from solar CMEs (the cyan line in Figure \ref{fig:ene_mass_vel_kin}(a)), and are roughly on the same relation with other stellar events in the previous studies (\citealt{Moschou+2019}; \citealt{Maehara+2021}; \citealt{Namekata+2022_NatAst}).
\color{black}\textrm{In addition, \citet{Vida+2019} reported the masses from M-dwarf blue wing asymmetries are 10$^{15}$--10$^{18}$ g, and this range is roughly the same as that of the 7 events in this study.} \color{black}
These results might suggest that these possible prominence eruptions on M-dwarfs
could share a common underlying mechanism with solar \color{black}\textrm{filament/prominence }\color{black}
eruptions/CMEs
(i.e. magnetic energy release) (\citealt{Aarnio+2012}; \citealt{Drake+2013}; \citealt{Takahashi+2016}; \citealt{Kotani+2023_ApJ}), 
although the large uncertainty of the mass estimation method should be considered.

In contrast, Figure \ref{fig:ene_mass_vel_kin}(c) shows that 
kinetic energies of the 7 blue-shift events ($E_{\rm{kin}}\sim 10^{29} - 10^{32}$ erg in Table \ref{table:velmass_blue_flares}) are roughly two orders of
magnitude smaller than the the relation
expected from solar CMEs (the cyan line in Figure \ref{fig:ene_mass_vel_kin}(c)),
as also indicated in the previous studies \color{black}\textrm{(\citealt{Maehara+2021}; \citealt{Namekata+2022_NatAst})}\color{black}.
\color{black}
\textrm{
First, it is noted that these small kinetic energies can be at least partly affected by the fact that the Doppler velocities measured from spectra are always the lower limits of real velocities because of projection effects. Moreover, these}
\color{black}
small kinetic energies can be also understood through a solar analogy.
As described above, the velocities of filament/prominence eruptions are 4--8 times
lower than the corresponding CMEs (e.g., \citealt{Gopalswamy+2003}),
and the kinetic energies of \color{black}\textrm{filament/prominence }\color{black} 
eruptions are typically smaller 
(the pink filled star marks in Figure \ref{fig:ene_mass_vel_kin}(c)).
Therefore the kinetic energy for stellar events estimated from the velocity of 
M-dwarf blue-shift events would be 1--2 orders of
magnitude smaller than the solar CME trend (\citealt{Maehara+2021}; \citealt{Namekata+2022_NatAst}).

However, it is still not clear
whether the prominence eruptions on M-dwarfs can really cause CMEs.
Recent numerical studies (e.g., \citealt{Drake+2016}; \citealt{Alvarado-Gomez+2018}; \citealt{Sun+2022}) have discussed that CMEs would be suppressed by the strong overlying magnetic fields. 
Zeeman Doppler Imaging (ZDI) observations in \citet{Morin+2008} 
suggested that the mid M-dwarf flare stars investigated in this study (YZ CMi, EV Lac, and AD Leo) have mainly axisymmetric large-scale poloidal fields.
In the case of these three stars, the magnetic energy in dipole mode accounts for 56--75\% of the whole magnetic energy, 
and such large-scale and strong dipole magnetic fields may cause the suppression or deceleration of CMEs.
In one possibility, the small kinetic energies the 7 blue-shift events shown
in Figure \ref{fig:ene_mass_vel_kin}(c) could be explained by the deceleration by the overlying magnetic fields (e.g., \citealt{Alvarado-Gomez+2018}; \citealt{Moschou+2019}).
\color{black}\textrm{The recent paper \citet{Bellotti+2023_A+A}
reported that AD Leo still showed mainly axisymmetric large-scale poloidal fields in April -- June 2019, when Flare A3 was observed,
while the numerical CME modeling incorporating the ZDI results (cf. \citealt{Alvarado-Gomez+2018}) is beyond the scope of this paper (a future research topic).
There were no reported ZDI magnetic field observations during our campaign for the other 6 blue-shift events (on YZ CMi and EV Lac), and we
}\color{black}
do not know how the real magnetic topologies were when we observed these 6 blue-shift events, since magnetic field topologies can change with time (\citealt{Morin+2008}\color{black}\textrm{; \citealt{{Bellotti+2023_A+A}}} \color{black}). Then in the future, it is important to conduct 
\color{black}\textrm{more} \color{black}
simultaneous flare campaign and magnetic field observations. In addition, future observations 
of blue-shifts simultaneously with other CME detection methods
(e.g., UV/X-ray dimmings as in \citealt{Veronig+2021}; \citealt{Loyd+2022}, radio bursts as in \citealt{Zic+2020}) 
may help whether and how prominence eruptions detected as blue-shifts of chromospheric lines could be evolved into CMEs, 
since different methods could be sensitive to different phases of the CME evolution (e.g., Figure 1 of \citealt{Namekata+2022_IAUS}).

Mass, velocity, and kinetic energy of the possible prominence eruptions of M-dwarfs
shown in Figure \ref{fig:ene_mass_vel_kin} could eventually lead to understanding the  statistical properties of M-dwarf CMEs with more observational samples in the future,
although it is still not clear whether they can really cause CMEs.
This would help us to evaluate the effects of CMEs on exoplanets 
orbiting around M-dwarfs (e.g., loss of atmosphere, atmospheric chemistry, radiation dose; cf. \citealt{Lammer+2007}; \citealt{Segura+2010}; \citealt{Scheucher+2018}; \citealt{Tilley+2019}; \citealt{Yamashiki+2019}; \citealt{Airapetian+2020}; \citealt{Chen+2021}; \citealt{Grayver+2022}). 
Furthermore, it has been discussed that
stellar mass loss from \color{black}\textrm{filament/prominence }\color{black} eruptions/CMEs could significantly affect 
the evolution of stellar mass and angular momentum loss (\citealt{Osten+2015}; \citealt{Cranmer2017}; \citealt{Odert+2017}; \citealt{Vidotto_2021}; \citealt{Wood+2021}),
and more observational samples of prominence eruptions would provide more insights 
in the case of M-dwarfs.

\subsection{Coronal parameters from \textit{NICER} soft X-ray data and implications for flare emission process} \label{subsec:dis:Xray}

Soft X-ray emission during a stellar flare is 
caused by the chromospheric evaporation process, which is coronal plasma filling of coronal magnetic loops (e.g., \citealt{Guedel+2004}; \citealt{Shibata+2011}). Soft X-ray spectroscopic and photometric data can help us to investigate the physical parameters of coronal plasma and magnetic loops such as temperature, loop length, electron density, magnetic field strength 
(e.g., \citealt{Shibata_Yokoyama+2002}; \citealt{Osten+2006}; \citealt{Raassen+2007};  \citealt{Pillitteri+2022}).

Flare Y3, which showed blue wing asymmetry of Balmer lines, was observed also in \textit{NICER} soft X-ray data as described in Section \ref{subsec:results:2019-Jan-27}.
The temperature ($T$) and emission measure (EM$=n^{2}V$) values of the quiescent (non-flaring) and flare components are estimated from the model fitting of X-ray spectra (Figure \ref{fig:NICER_BBlc_190127}(e) \& Figure \ref{fig:NICER_190127_spec1}), 
and the resultant values are listed in Table \ref{table:corona_para}. 
Here $n$ is the electron density and $V$ is the volume.
\citeauthor{Shibata_Yokoyama_1999} (\citeyear{Shibata_Yokoyama_1999}, \citeyear{Shibata_Yokoyama+2002}) discussed the scaling laws of $T$ and EM for solar/stellar flares on the basis of the magnetic reconnection model, which considers the energy balance between conduction cooling and reconnection heating (cf. \citealt{Shibata+2011} for review). 
The scaling laws derived by \citet{Shibata_Yokoyama+2002} show that the 
flare magnetic field strength ($B$) and characteristic length of the flare loop ($L$) can be expressed in terms of the
the flare emission measure (EM$=n^{2}V$), the pre-flare coronal electron density ($n_{0}$), and flare temperature ($T$):

  \begin{eqnarray}
    B &=& 50\left(\frac{\rm{EM}}{10^{48}\rm{cm}^{-3}}\right)^{-1/5}\left(\frac{n_{0}}{10^{9}\rm{cm}^{-3}}\right)^{3/10}\left(\frac{T}{10^{7}\rm{K}}\right)^{17/10} \rm{G}\ ,  \label{eq:B-EMT}\\
    L &=& 10^{9}\left(\frac{\rm{EM}}{10^{48}\rm{cm}^{-3}}\right)^{3/5}\left(\frac{n_{0}}{10^{9}\rm{cm}^{-3}}\right)^{-2/5}\left(\frac{T}{10^{7}\rm{K}}\right)^{-8/5} \rm{cm}\ .  \label{eq:L-EMT}
  \end{eqnarray}

Here simple order-of-magnitude estimates are used and the emitting volume is give by $V=L^{3}$. This simple method derived by \citet{Shibata_Yokoyama+2002} was validated 
with Sun-as-a-star observations and can estimate the loop length and magnetic field strength with an accuracy of a factor of 3 \citep{Namekata+2017_PASJ}.
As we have shown in Figure \ref{fig:NICER_BBlc_190127}, the temperature and emission measure of the flare component at the peak of Flare Y3 are $T=1.1\times 10^{7}$ K and EM$=2.2\times 10^{51}$cm$^{-3}$ (Table \ref{table:corona_para}). 
The X-ray spectrum of quiescent (preflare) phase was well fitted with the two temperature components: $T_{1}=3.1\times 10^{6}$ K and $T_{2}=1.1\times 10^{7}$ K.
\color{black}
\textrm{
The hot quiescent plasma temperature ($T_{2}$) 
is close to the Y3 flare peak temperature. 
It is also higher than the hot quiescent plasma temperature reported in the earlier 
XMM-Newton observation of YZ CMi ($\sim$0.64 keV = 7.4 $\times$ 10$^{6}$ K), 
while the EM is similar to the quiescent emission
during the XMM-Newton observation (\citealt{Raassen+2007}). 
This result may suggest 
that the preflare phase
contains emission from the decay of a previous flare.
}
\color{black}
Since there were no simultaneous grating X-ray line observations that can be used for estimating preflare densities in our \textit{NICER} data, we use the previous measurements of quiescent electron densities $n_{0}$ of a dMe flare star similar to the target star YZCMi (dM4e flare star).
\citet{Osten+2006} measured electron densities of the quiescent atmosphere of the d3.5Me flare star EV Lac using transition region and coronal lines.
Their measurements indicate nearly constant electron densities ($n\sim 10^{11}$cm$^{-3}$) between $T = 10^{5.2}$ and $10^{6.4}$K, while at higher coronal temperatures, there is a sharp increase of 2 orders of magnitude in density ($n\sim 10^{13}$cm$^{-3}$ at $T = 10^{6.9}-10^{7.0}$K) (see Figure 9 therein). Taking into account the measured preflare temperature values ($T_{1}=3.1\times 10^{6}$ K and $T_{2}=1.1\times 10^{7}$ K in the above) and the results of \citet{Osten+2006}, we consider three cases of preflare densities of $n_{0}=10^{11}$, $10^{12}$, and $10^{13}$cm$^{-3}$  
when we estimate magnetic field $B$ and loop length $L$ values from Equations (\ref{eq:B-EMT}) \& (\ref{eq:L-EMT}).
With the flare peak temperature and emission measure $T=1.1\times 10^{7}$ K and EM$=2.2\times 10^{51}$cm$^{-3}$ in the above, $B$ and $L$ values are estimated to be $B\sim 50$G and $L\sim  \color{black}\textrm{1.4}\color{black}\times 10^{10}$cm $=0.66R_{\rm{star}}$ if $n_{0}=10^{11}$cm$^{-3}$.  $R_{\rm{star}}$ is the radius of the target star YZ CMi (Table \ref{table:targets_basic_para}). 
$B\sim 100$G and $L\sim 5.5\times 10^{9}$cm $=0.26R_{\rm{star}}$ if $n_{0}=10^{12}$cm$^{-3}$, and $B\sim 200$G and $L\sim 2.2\times 10^{9}$cm $=0.10R_{\rm{star}}$ if $n_{0}=10^{13}$cm$^{-3}$. They are listed in Table \ref{table:corona_para}.

\begin{deluxetable*}{l|cc}[htbp]
   \tablecaption{Coronal parameters of Flare Y3 \color{black}\textrm{(on YZ CMi)} \color{black} from \textit{NICER} soft X-ray data}
   \tablewidth{0pt}
   \tablehead{
  \multicolumn{3}{l}{Coronal temperature ($T$) \& Emission Measure (EM$=n^{2}V$)} 
     }
   \startdata
   Quiescent (preflare) phase & $T_{1}=3.1\times 10^{6}$ K & EM$_{1}=1.8\times 10^{51}$cm$^{-3}$ \\
    & $T_{2}=1.1\times 10^{7}$ K & EM$_{2}=9.4\times 10^{50}$cm$^{-3}$ \\
    \hline
   Flare peak & $T=1.1\times 10^{7}$ K & EM$=2.2\times 10^{51}$cm$^{-3}$ \\
     \hline
    \hline
  \multicolumn{3}{l}{Coronal magnetic field ($B$) \& loop length ($L$)} \\
      \hline
   $n_{0}=10^{11}$cm$^{-3}$ & $B\sim 50$G & $L\sim 
   \color{black}\textrm{1.4}\color{black}\times 10^{10}$cm $=0.66R_{\rm{star}}$ \\
   $n_{0}=10^{12}$cm$^{-3}$ & $B\sim 100$G & $L\sim 5.5\times 10^{9}$cm $=0.26R_{\rm{star}}$ \\
   $n_{0}=10^{13}$cm$^{-3}$ & $B\sim 200$G & $L\sim 2.2\times 10^{9}$cm $=0.10R_{\rm{star}}$ \\
    \enddata
      \tablecomments{$n_{0}$: preflare coronal density. $R_{\rm{star}}=0.30R_{\odot}$ (Table \ref{table:targets_basic_para}).
   }
   \label{table:corona_para}
 \end{deluxetable*}

These estimated values ($B=50-200$G and $L= 2.2\times 10^{9}-  \color{black}\textrm{1.4}\color{black}\times 10^{10}$cm$=0.10-0.66R_{\rm{star}}$; Table \ref{table:corona_para}) 
can be compared with the estimation results of previous studies.
First, our result \color{black}\textrm{suggests} \color{black} that flare loop length is at least larger than $L\sim0.10R_{\rm{star}}$, and this is roughly consistent with the result of \citet{Maehara+2021}.
They estimated that at least 10--20\% of stellar surface of YZ CMi 
would be covered by starspots on the basis of the rotational modulations of \textit{TESS} and ground-based photometric data. 
Moreover, \citet{Maehara+2021} also discussed the statistical relation of flare energy and duration from optical flares observed by \textit{TESS}, 
and estimated the $B$ and $L$ values of YZ CMi (Figure 14 therein), 
by using the method based on the magnetic reconnection model proposed by \citet{Namekata+2017_ApJ}.
As a result, the $B$ and $L$ values estimated from \textit{NICER} X-ray data ($B=50-200$G and $L= 2.2\times 10^{9}-  \color{black}\textrm{1.4}\color{black}\times 10^{10}$cm) in this study 
are roughly in the range of those from flare duration statistics of \textit{TESS} data in \citet{Maehara+2021} \footnote{
\color{black}\textrm{
\citet{Bicz+2022_ApJ} derived relatively larger loop length values ($L= 10^{10} - 10^{11}$cm) and smaller magnetic field values ($B=15-45$G) from the duration statistics of \textit{TESS} data, although they also used the same scaling relation proposed by \citet{Namekata+2017_ApJ} as done in \citet{Maehara+2021}. This difference can be caused by the definition of flare duration: they defined total duration as flare duration and applied it into the same scaling relation of \citet{Namekata+2017_ApJ}, but the 
coefficient of the original scaling relation of \citet{Namekata+2017_ApJ} is determined with
the e-folding decay time (not flare ``total" duration) of solar flares. 
We note that because of this, the larger loop length and smaller magnetic field values were estimated in \citet{Bicz+2022_ApJ}.
}\color{black}
}.
These consistency among different methods 
can support the validity of the method used in this study.
\color{black}
\textrm{In addition, 
the derived loop length is similar to the estimated length of a flare observed from YZ CMi with the EUVE satellite 
in 1994 (0.14--0.50 $R_{\rm{star}}$, \citealt{Mullan+2006_ApJS}).
This paper also reported a gigantic flare with a loop length of 1.1--1.5 $R_{\rm{star}}$ from the star.
}\color{black}

The $B$ and $L$ values estimated in this study would be helpful for future modeling studies
discussing blue wing asymmetries of M-dwarf flares.
\color{black}
\textrm{
These values can be helpful for modeling 
how the prominences, 
erupt associated with flares (cf. \citealt{Shibata+2011}; \citealt{Yuhong+2018_ApJ}) and cause blue wing asymmetries of Balmer lines. 
For example, the loop lengths can be closely related with the timescale of flares (\citealt{Maehara+2015}; \citealt{Reep+2023_arXiv}), 
and coronal magnetic field strength can be a factor that determines the CME evolution (e.g., \citealt{Alvarado-Gomez+2018}; \citealt{Sun+2022}).
}
\color{black}
Moreover, this kind of X-ray observation has been \color{black}\textrm{still very limited for blue wing asymmetry events: for example, a flare on M5.5 dwarf CN Leo
in \citealt{Fuhrmeister+2008} \& \citealt{Liefke+2010_A&A}, 
that on M5.5 dwarf Proxima Centauri reported in \citealt{Fuhrmeister+2011},  
and that on K-dwarf AB Dor in \citet{Lalitha+2013_A&A}. }\color{black}
It is necessary to increase the number of X-ray observation of blue asymmetry flares 
for further statistical discussions.

In this subsection, we assumed preflare coronal electron density 
of YZ CMi $n_{0}=10^{11} - 10^{13}$cm$^{-3}$.
This is orders of magnitude larger than that of the Sun $n_{0, \odot}\sim10^{9} - 10^{10}$cm$^{-3}$ (e.g., \citealt{Shibata_Yokoyama+2002}; \citealt{Allred+2005}). 

Such higher preflare coronal density is also expected for M-dwarfs from the theoretical point of view because \color{black}
\textrm{photospheric}
\color{black} density of M-dwarfs is higher than that of the Sun  (\citealt{Sakaue-Shibata_2021}).
In order to predict the preflare coronal density more consistently, we need to develop multi-coronal loop model extending the method considered by \citet{Takasao+2020}.
Higher preflare coronal density can be discussed with the propagation of nonthermal electron beam along the coronal loop, which are important to understand strong white-light emission of M-dwarf flares (e.g., \citealt{Allred+2006}; \citealt{Namekata+2020_PASJ}).

\begin{figure}[ht!]
   \begin{center}
   \gridline{
    \fig{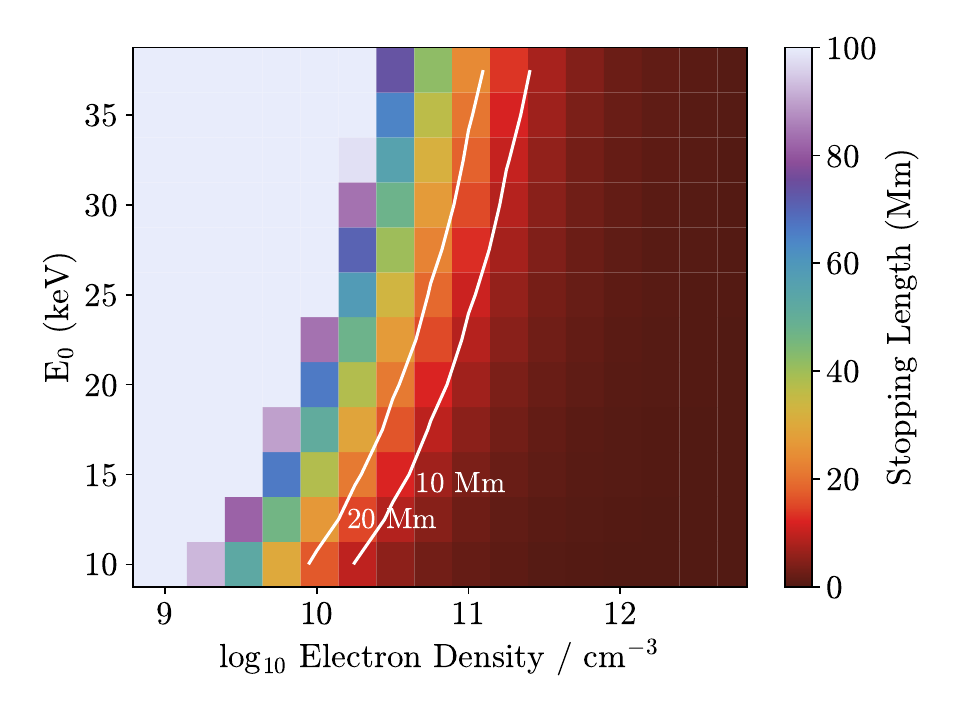}{0.65\textwidth}{\vspace{0mm}}
    }
     \vspace{-5mm}
     \caption{
Contour of stopping lengths for mono-energetic electron beams with initial kinetic energy $E_{0}$ in constant density slabs (fully ionized). Two white lines show the stopping lengths of 10 and 20 Mm.
     }
   \label{fig:stopping_length_v1}
   \end{center}
 \end{figure}

Then we conducted simple calculations to determine the stopping lengths for high energy electrons in M-dwarf preflare corona of various electron densities. 
\color{black}\textrm{
In these calculations, we integrated the analytic formula from 
\citet{Holman+2011_SSRv} and \citet{Holman_2012_ApJ}:
 \begin{eqnarray}
 \label{eq:electron_distribution_1}
\frac{dE}{dl} = - 0.150 \bigg(\frac{\Lambda_{ee}}{23}\bigg) n_{10} \bigg( \frac{20}{E} \bigg) [\mathrm{keV}~\mathrm{Mm}^{-1}]
 \end{eqnarray}
for initial energy $E_0$ (see Figure \ref{fig:stopping_length_v1}), where $l$ is the path length, $E$ is the kinetic energy of the electron in keV, $n_{10}$ is the ambient electron density in units of $10^{10}$ cm$^{-3}$, and $\Lambda_{ee}$ is the Coulomb logarithm 
(e.g., \citealt{Allred+2015_ApJ}).
} \color{black}
Figure \ref{fig:stopping_length_v1} shows the contour of stopping lengths for mono-energetic electron beams in constant density fully ionized slabs.
This simple calculation result shown in Figure \ref{fig:stopping_length_v1} 
suggests that 
significant (or at least some) fraction of the electron beams
can be stopped in high density (e.g., $n_{0}\gtrsim10^{11}$cm$^{-3}$) preflare coronal loop, especially for soft power-law distributions of the electron beams with $\delta\gtrsim 7$. 
We note that these simple mono-energetic calculations very roughly appropriate very ``soft" power-law distributions of the electron beams with the spectral index $\delta\gtrsim 7$, 
which are often determined in solar flare hard X-ray observations (e.g., \citealt{Milligan+2014}; \citealt{Thalmann+2015}; \citealt{Warmuth+2016}; \citealt{Kowalski+2019}). 
For hard power-law case of $\delta \sim 3 - 5$ and/or low-energy cutoff $\gg$ 10 keV,
there would be enough high energy electrons to penetrate the dense coronae and produce continuum radiation in the chromosphere / photosphere.
This calculation result in Figure \ref{fig:stopping_length_v1} can be consistent with the fact that Flare Y3 does not show clear white-light emission (cf. Table \ref{table:list_blue_flares}), 
and may suggest absence of strong electron beams (e.g., F13 beam in \citealt{Kowalski2016}) for this non white-light flare.
This might also suggest that thermal conduction heating can largely contribute to causing 
chromospheric line emissions and soft X-ray emission during Flare Y3 (e.g., \citealt{Hori+1997}).
However, there could be other explanations of the cause of non white-light flares (cf. \citealt{Watanabe+2017}), and more detailed calculations (e.g., density stratification along the loop, power law distributions of electron energies) as done in recent radiative hydrodynamic calculations (e.g., RADYN calculations: \citealt{Allred+2006}; \citealt{Namekata+2020_PASJ}; \citealt{Kowalski+2022}) are necessary for detailed quantitative discussions. The brief discussion in this section also suggests detailed investigations of coronal densities using X-ray high resolution spectra (e.g., \citealt{Guedel2004_A&ARv}; \citealt{Osten+2006}; \citealt{Pillitteri+2022}) are important as a next-step study for understanding white-light emission of M-dwarf flares with/without chromospheric line wing asymmetries. 

In addition, we also note that stellar flare energy partitions
among different wavelengths (e.g., white-light, X-ray, H$\alpha$) 
have been discussed in several recent observational studies 
(\citealt{Osten+2015}; \citealt{Guarcello+2019}; \citealt{Paudel+2021_ApJ}; \citealt{Stelzer+2022}).
In particular, it is interesting to compare optical white-light energy partitions in the case of stellar flares to big solar flares (\citealt{Emslie+2012}; \citealt{Cliver+2022}).
\color{black}\textrm{From} \color{black} this point of view, non white-light flares like this Flare Y3 
can be interesting, since they were not incorporated well in the discussions of the above previous studies. For example, in the case of Flare Y3, X-ray energy
\color{black}\textrm{is } \color{black}
larger than $TESS$-band white-light and H$\alpha$ energies ($E_{\rm{X}}$(0.5--2.0 keV)$=2.6\times 10^{32}$erg, $E_{TESS}<2.6\times 10^{32}$erg, and $E_{\rm{H}\alpha}=1.7\times 10^{31}$erg). 

It could also be interesting to note that the location of the flares (e.g., limb darkening effect) could also affect the observed energy partitions (\citealt{Woods+2006}).
In future studies, 
it is important to discuss this point statistically 
with much larger number of 
multi-wavelength data of white-light flares and non white-light flares.

\subsection{Additional notable properties other than blue asymmetries} \label{subsec:dis:flares-redsym}

In addition to blue wing asymmetries, the \color{black}\textrm{41 }\color{black} flares detected in this study 
also showed various notable properties.
Clear red wing asymmetries (enhancements of red wing of H$\alpha$ line) were also observed in at least 11 flares among the total \color{black}\textrm{41 }\color{black} flares (These flares are marked with ``R" in Table \ref{table:list1_flares}). 
Flares Y6, Y17, \& E5 in Sections \ref{subsec:results:2019-Dec-12}, \ref{subsec:results:2020-Jan-20}, \& \ref{subsec:results:2020-Aug-27} 
are remarkable examples of flares with red wing asymmetries among these 11 flares.
One possible cause of the red wing asymmetries is the process called 
chromospheric condensations, which is the downward flow of cool plasma in the chromosphere (e.g., \citealt{Ichimoto+1984}; \citealt{Longcope_2014}; \citealt{Graham+2015}; \citealt{Kowalski+2017}). 
Another possible cause is the flare-driven coronal rain or the post-flare loop (e.g., \citealt{Antorin2020}; \citealt{Wu+2022}; \citealt{Wollmann+2023_A&A}).
Flares Y8, E5, and A2 showed H$\alpha$ and H$\beta$ symmetric line broadenings with $\gtrsim 300-400$ km s$^{-1}$, accompanied by large white-light flares (See Appendix \ref{subsec:results:2020-Jan-14}, \ref{subsec:results:2020-Aug-27}, \& \ref{subsec:results:2019-May-18}). 
These broadenings can be caused by 
high-energy non-thermal electron beams penetrating into the lower atmosphere (e.g., 
\citealt{Oks+2016}; \citealt{Namekata+2020_PASJ}; \citealt{Kowalski+2022}).
In particular, Flare E5 (in Appendix \ref{subsec:results:2020-Aug-27}) can be the most interesting since this flare showed both red wing asymmetries and broad symmetric broadenings accompanied by large white-light flares.
These additional notable properties are important topics of stellar flares,
and these flares will be discussed in detail in our future papers.

As described in the above, our flare data also include 
a large number of light curves of various chromospheric lines (e.g., H$\alpha$, H$\beta$, H$\gamma$, H$\delta$, Ca K, Ca II 8542, Na I D1\&D2, He I D3) whose line formation heights are different (e.g., \citealt{Vernazza+1981}; \citealt{Heinzel+2019}).
In some (or many) cases, different chromospheric lines evolve differently.
For example, during Flare Y3, which also showed blue wing asymmetries, 
H$\alpha$ and Ca II K evolved similarly 
while other Balmer lines, Ca II 8542, Na I D1\&D2, and He I D3 lines decayed faster (Figure \ref{fig:lcEW_HaHb_YZCMi_UT190127}).
These differences can provide us clues to investigate temperature and density 
evolution of the chromosphere during flares (e.g., time decrement of Balmer lines: e.g., \citealt{Hawley+1991}; \citealt{Kowalski+2013}), 
and it is important to compare with radiative hydro-dynamic modeling results as well as solar flare observation results.
These points will also be investigated more in our future papers.

In addition to flares, 
Figures 
\color{black}\textrm{\ref{fig:rot1_YZCMi_2019Q1} -- \ref{fig:rot_all1_ADLeo_2020Q2}} \color{black} show
that the H$\alpha$ \& H$\beta$ equivalent width values of the quiescent phase (non-flare phase) exhibit some variabilities among the observation dates. 
In particular, Figure \ref{fig:Ha_rot_YZCMi_2019Q4to2020Q1} (a)\&(b) show some quasi-periodic modulations of H$\alpha$ \& H$\beta$ EW values, 
and this could be related with the rotational modulations (\citealt{Toriumi+2020}; \citealt{Maehara+2021}; \citealt{Namekata+2022_ApJL}; \citealt{Schofer+2022}), 
considering the YZ CMi's rotation period of 2.77 days (Table \ref{table:targets_basic_para}).
This topic is being highlighted in recent studies, including possible relations with flare activities.
\citet{Maehara+2021} suggested that 
the amplitude of rotational modulations of YZ CMi in the H$\alpha$ line 
can change depending on the difference in flare activity (flare frequency) 
during each observation run. 
In contrast, \citet{Schofer+2022} showed there were no clear periodic rotational modulations in H$\alpha$ line of YZ CMi and EV Lac, while the photometric indexes (e.g., TiO \color{black}\textrm{7050 \AA } \color{black} 
index, \textit{TESS} photometry) of them showed clear periodic modulations.
These modulations of chromospheric lines in the quiescent phase (non-flare phase) 
will also be discussed more in detail in our future papers possibly with more dataset.

\section{Summary and Conclusions} \label{sec:summary}

We conducted the time-resolved simultaneous optical spectroscopic and photometric observations of mid M dwarf flare stars YZ CMi, EV Lac, and AD Leo.
High-dispersion spectroscopic observations were obtained using APO 3.5m and CTIO/SMARTS 1.5m telescopes, and various chromospheric lines (H$\alpha$, H$\beta$, H$\gamma$, H$\delta$, H$\epsilon$,
Ca II H\&K, Ca II 8542, He I D3, and Na I D1\&D2 lines) were investigated. 
As a result, \color{black}\textrm{41 }\color{black} flares (Flares Y1--\color{black}\textrm{Y29 }\color{black} on YZ CMi, Flares E1--
\color{black}\textrm{E9 }\color{black} on EV Lac, and Flares A1--A3 on AD Leo) were detected (Table \ref{table:list1_flares}) during the 31 nights over two years (2019 January -- 2021 February). The energy ranges of the observed \color{black}\textrm{41 }\color{black} flares are $10^{30}-10^{32}$ erg in the H$\alpha$ line, and $10^{30}-10^{33}$ erg in $u$- \& $g$-band continuum bands (Figure \ref{fig:Lg_LHa_Eg_EHa}).
Among the \color{black}\textrm{41 }\color{black} flares, seven flares (Flares Y3, Y6, Y18, Y23, E1, E2, \& A3) 
showed clear blue wing asymmetries in H$\alpha$ line.
There are various correspondences in flare properties (e.g., durations of blue wing asymmetries, intensities of white-light emissions, blue wing \color{black}\textrm{asymmetries }\color{black} in various chromospheric lines) as listed in Tables \ref{table:list_blue_flares} \& \ref{table:velmass_blue_flares}, and the key findings of this study are as follows.

\begin{enumerate} 
\renewcommand{\labelenumi}{(\roman{enumi})}

\item The duration of the H$\alpha$ blue wing asymmetries range from 20 min to 2.5 hours (Table \ref{table:list_blue_flares}).
As a notable example, Flare Y3 showed short-lived H$\alpha$ blue wing asymmetries  twice (20min$\times$2) during the H$\alpha$ flare over 4 hours (Figure \ref{fig:map_HaHb_YZCMi_UT190127}). 
In contrast, Flares Y23, E1, \& A3 showed continuous H$\alpha$ blue wing asymmetries over almost all the observed phases of the flares (Figures \ref{fig:map_HaHb_YZCMi_UT201206}, \ref{fig:map_HaHb_EVLac_UT191215}, \& \ref{fig:map_HaHb_ADLeo_UT190519}). As another notable point, \color{black}\textrm{the velocities of }\color{black} blue wing asymmetries showed gradual decays during Flares Y6 \& Y23 (Figures \ref{fig:map_HaHb_YZCMi_UT191212} \& \ref{fig:map_HaHb_YZCMi_UT201206}). In particular, Flares Y6 showed the gradual shift from blue wing asymmetry to red wing asymmetry, during the H$\alpha$ flare over 4.9 hours (Figure \ref{fig:map_HaHb_YZCMi_UT191212}).

\item \color{black}\textrm{
Among the seven flares with blue wing asymmetries, two flares (Flare Y3 \& Y6) are categorized as candidate non white-light (NWL) flares and three flares (Flares Y18, Y23, \& E2) are clearly white-light (WL) flares (Table \ref{table:list_blue_flares}), 
while the remaining two flares (Flares E1 \& A3) do not have enough data coverage of simultaneous 
spectroscopic and photometric data to judge whether they are white-light or non white-light flares.
For reference, among all the 41 flares, 4 flares are categorized as candidate NWL flares and 31 flares are clear WL flares (Table \ref{table:list_blue_flares}), 
while the remaining 6 flares do not have enough data coverage of simultaneous 
spectroscopic and photometric data.
These results can suggest that blue wing asymmetries of chromospheric lines can be commonly seen both during white-light and candidate non white-light flares.
} \color{black}

\item All of the seven flares showed blue wing asymmetries also in the H$\beta$ line,
but there is a large variety in which other chromospheric lines showed blue wing asymmetries ([B] and [NB] in Table \ref{table:list_blue_flares}). 
For example, two flares (Flares Y6 \& E2) showed blue wing asymmetries only in lower-order Balmer lines (up to H$\beta$ and H$\gamma$ lines, respectively). 
In contrast, the other two flares (Flares Y23 \& A3) showed blue wing asymmetries in almost all the chromospheric lines (except for Ca II 8542 and Na I D1\&D2, respectively). The velocities of blue wing enhancements are different among different lines, and lower-order Balmer lines especially the H$\alpha$ line tend to show larger velocities of blue wing asymmetries, while higher-order Balmer lines, Ca II lines, Na I D1\&D2, and He I D3 lines show smaller velocities ($v_{\rm{blue,max}}$ in Table \ref{table:velmass_blue_flares}).
It is speculated that these differences can be caused by the differences of optical depth and line wing broadening physics, 
but observation-based modeling studies incorporating radiative transfer physics (e.g., \citealt{Leitzinger+2022}) and comparison with solar flare data are necessary for further quantitative discussions.

\item The line-of-sight velocities of the blue wing excess components (blue wing asymmetries) are estimated to range from -73 to -122 km s$^{-1}$ ($v^{\rm{H}\alpha}_{\rm{blue,fit}}$ in Table \ref{table:velmass_blue_flares}), 
and these are in the same range of solar prominence/filament eruptions 
(Figure \ref{fig:ene_mass_vel_kin}(b)).
These velocity values (73--122 km s$^{-1}$) represent possible prominence eruptions 
of M-dwarfs and they are smaller than the escape velocities at the stellar surface ($\sim$600 km s$^{-1}$ for YZ CMi, EV Lac, and AD Leo).
\color{black}\textrm{
The prominence eruptions could evolve into CMEs, assuming that the similar acceleration mechanism from prominence eruptions to CMEs on the Sun would work also in these M-dwarf cases (See also (vii) for the necessity of further investigations).
}\color{black}

\item Assuming the relation from the NLTE slab model calculation of solar prominences (\citealt{Heinzel+1994_A&A}), the surface flux densities of the upward moving plasma causing blue-shifts are estimated from the luminosity ratio of blue wing asymmetry components
in H$\alpha$ and H$\beta$ lines (cf. Figure \ref{fig:Fha_Fhb_blue}).
Using these values, the erupted mass of the seven blue-shift (blue wing asymmetry) events
are estimated to be $M^{\rm{H}\alpha}_{\rm{blue}}\sim
10^{15}-10^{19}$ g (Table \ref{table:velmass_blue_flares}).
These estimated mass of the seven blue-shift events are roughly on the relation
expected from solar CMEs, and are roughly on the same relation with other stellar events in the previous studies (Figure \ref{fig:ene_mass_vel_kin}(a)).
This might suggest that these possible prominence eruptions on M-dwarfs
could share a common underlying mechanism with solar \color{black}\textrm{filament/prominence }\color{black} eruptions/CMEs
(i.e. magnetic energy release), although the large uncertainty of the mass estimation method should be considered.

\item 
In contrast, the kinetic energies of the seven blue-shift events ($E_{\rm{kin}}\sim10^{29}-10^{32}$ erg in Table \ref{table:velmass_blue_flares}) are roughly two orders of
magnitude smaller than the the relation expected from solar CMEs (Figure \ref{fig:ene_mass_vel_kin}(c)), as also shown in previous studies.
These small kinetic energies can be understood if we assume
the velocity difference/evolution of prominence eruptions and CMEs.

\item
The mass, velocity, and kinetic energy of the possible prominence eruptions of M-dwarfs
in this study (Figure \ref{fig:ene_mass_vel_kin}) could eventually lead to understanding the statistical properties of M-dwarf CMEs with more observational samples in the future.
However, it is still not clear whether the prominence eruptions on M-dwarfs can really cause CMEs (e.g., possible suppression by overlying magnetic field), 
as discussed in Section \ref{subsec:dis:blue-ejection}. 
Further investigations are also necessary 
for understanding the observed various properties of blue wing asymmetries.
Future observations of blue-shifts simultaneously with other CME detection methods (e.g., UV/X-ray dimmings as in \citealt{Veronig+2021}, \citealt{Loyd+2022}; radio bursts as in \citealt{Zic+2020}) are important to investigate whether and how prominence eruptions detected as blue-shifts of chromospheric lines could be evolved into CMEs.

\item One flare (Flare Y3) was also observed with \textit{NICER} soft X-ray data, 
which enabled us to estimate the flare magnetic field and length of the
flare loop of a flare with blue wing asymmetry in chromospheric lines.
Coronal temperature ($T$) \& Emission Measure (EM) values are 
estimated from the model fitting of soft X-ray spectra (Table \ref{table:corona_para}).
Using the simple scaling law of $T$ and EM (\citealt{Shibata_Yokoyama+2002}), 
the flare magnetic field strength ($B$) and characteristic length of the flare loop ($L$)
are estimated to be $B=50-200$G and $L= 2.2\times 10^{9}-  \color{black}\textrm{1.4}\color{black}\times 10^{10}$cm$=0.10-0.66R_{\rm{star}}$ (Table \ref{table:corona_para}).
The $B$ and $L$ values estimated in this study would be helpful for future modeling studies
discussing blue wing asymmetries of M-dwarf flares.

\item 
The preflare coronal density value of $n_{0}=10^{11} - 10^{13}$cm$^{-3}$ is assumed 
to interpret the soft X-ray data of this Flare Y3.
A significant (or at least some) fraction of the electron beams can be stopped in
such high density (e.g., $n_{0}>10^{11}$cm$^{-3}$) preflare coronal loop (Figure \ref{fig:stopping_length_v1}), especially for soft power-law distributions of the electron beams with $\delta\gtrsim 7$.
This could be consistent with the fact that this Flare Y3 did not show clear white-light emission.
It should be also noted that in the case of this Flare Y3, soft X-ray energy dominates white-light and H$\alpha$ energies ($E_{\rm{X}}$(0.5--2.0 keV)$=2.6\times 10^{32}$erg, $E_{TESS}<1.8\times 10^{31}$erg, and $E_{\rm{H}\alpha}=1.7\times 10^{31}$erg).

\item In addition to blue wing asymmetries, our flare data of this study 
also showed various notable properties, 
as summarized in Section \ref{subsec:dis:flares-redsym}.
For example, clear red wing asymmetries (enhancements of red wing of H$\alpha$ line) were also observed in at least 11 flares among the total \color{black}\textrm{41 }\color{black} flares, while three flares showed symmetric line broadenings with $> 300-400$ km s$^{-1}$ accompanied by large white-light flares. These topics will be discussed in detail
in our future papers.
 
\end{enumerate}

\acknowledgments

\color{black}
\textrm{
The authors thank the anonymous referee for the detailed \& constructive comments, which improved the whole paper very much. 
We acknowledge Dr. Seiji Yashiro for providing the solar CME data.}
\color{black}
This work was supported by JSPS (Japan Society for the Promotion of Science) KAKENHI Grant Numbers 21J00106 (Y.N.), 20K04032, 20H05643 (H.M.), 21J00316 (K.N.), and 21H01131 (H.M., S.H., K.I., D.N., K.S.), and JST CREST Grant Number JPMJCR1761 (K.I.).
A.F.K. and Y.N. acknowledge support from NASA ADAP award program Number 80NSSC21K0632.
Y.N. was supported by JSPS Overseas Research Fellowship Program, and JSPS Postdoctoral Research Fellowship Program. K.N. was supported by JSPS Postdoctoral Research Fellowship Program.
K.H. was supported by NASA under award number 80GSFC21M0002.
We also acknowledge the International Space Science Institute and the supported International Teams 464: ``The Role Of Solar And Stellar Energetic Particles On (Exo)Planetary Habitability (ETERNAL, \url{http://www.issibern.ch/teams/exoeternal/})" and  510: ``Solar Extreme Events: Setting Up a Paradigm (SEESUP, \url{https://www.issibern.ch/teams/solextremevent/})'.

In this study, we used the observation data obtained with the Apache Point Observatory (APO) 3.5 m \& 0.5m telescopes, which are owned and operated by the Astrophysical Research Consortium. We are grateful to APO 3.5m Observing Specialists (Candace Gray, Jack Dembicky, Russet McMillan, Theodore Rudyk, and David DeColibus) and other staff members of Apache Point Observatory for their contributions in carrying out our observations. We used APO observation time allocated to the University of Colorado (CU) and the University of Washington (UW). We appreciate CU and UW members for their help to plan our observations. This research also used data from the Small and Moderate Aperture Research Telescope System (SMARTS) 1.5m telescope at Cerro Tololo Inter-American Observatory (CTIO), and this telescope is operated as part of the SMARTS Consortium. We appreciate the SMARTS Principal Scientist Todd Henry and the other staff members for their large contributions in carrying out our observations. 
This research makes use of observations from the Las Cumbres Observatory (LCO) global telescope network. We used the LCO observation time allocated to the University of Colorado. 
\color{black}
\textrm{
This paper includes data collected with the \textit{TESS} mission, obtained from the Mikulski Archive for Space Telescopes (MAST) at the Space Telescope Science Institute (STScI). 
The specific observations analyzed can be accessed via \dataset[DOI: 10.17909/rv7c-zn98]{https://doi.org/10.17909/rv7c-zn98}. }
\color{black}
Funding for the \textit{TESS} mission is provided by the NASA Explorer Program. STScI is operated by the Association of Universities for Research in Astronomy, Inc., under NASA contract NAS 5-26555. \textit{NICER} analysis software and data calibration were provided by the NASA \textit{NICER} mission and the Astrophysics Explorers Program. This research also has made use of the SVO Filter Profile Service
(\url{http://svo2.cab.inta-csic.es/theory/fps/}) supported from the Spanish MINECO through grant AYA2017-84089.

%

\facilities{APO/ARC 3.5m (ARCES), APO/ARCSAT 0.5m (flarecam), CTIO/SMARTS 1.5m (CHIRON), LCO/1m (Sinistro), LCO/0.4m (SBIG STL-6303), \textit{TESS}, \textit{NICER}}




\appendix

\section{Flare light curves and H$\alpha$ \& H$\beta$ spectra of the observation dates when blue wing asymmetries were not detected}\label{sec:app:non-blue-lcspec}

In this appendix, we describe
the detailed flare light curve and flare H$\alpha$ \& H$\beta$ spectra 
from the observation dates when blue wing asymmetries were not detected (cf. Section \ref{subsec:results:obs-summary} \& Table \ref{table:list1_flares}). 
Additional notable properties seen in these flares other than blue wing asymmetries are briefly summarized in Section \ref{subsec:dis:flares-redsym} and will be discussed in detail \color{black}\textrm{in} \color{black} future papers.

\subsection{Flare Y1 observed on 2019 January 26} \label{subsec:results:2019-Jan-26}

     \begin{figure}[ht!]
   \begin{center}
   \gridline{
    \fig{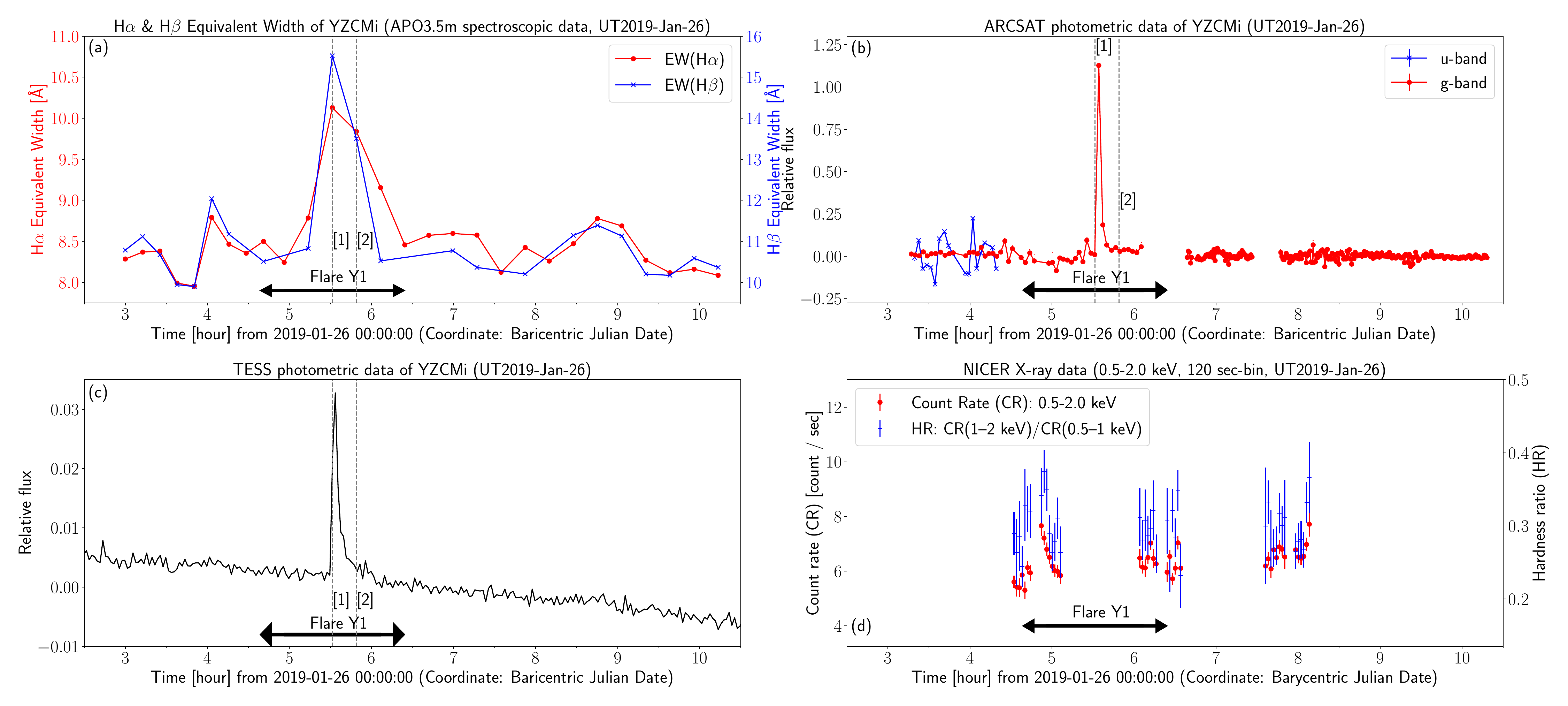}{1.0\textwidth}{\vspace{0mm}}}
     \caption{
     \color{black}\textrm{  
Light curves of YZ CMi on 2019 January 26 showing Flares Y1, which are plotted 
similarly with Figures \ref{fig:lcEW_HaHb_YZCMi_UT190127}(a)--(d).
The grey dashed lines with numbers ([1],[2]) correspond to the time shown with the same numbers in Figures \ref{fig:spec_HaHb_YZCMi_UT190126} \& \ref{fig:map_HaHb_YZCMi_UT190126}.
 } \color{black}
     }
   \label{fig:lcEW_HaHb_YZCMi_UT190126}
   \end{center}
 \end{figure}

On 2019 January 26, a flare (Flare Y1) was detected in H$\alpha$ \& H$\beta$ lines as shown in Figure \ref{fig:lcEW_HaHb_YZCMi_UT190126} (a).  During this Flare Y1,
the H$\alpha$ \& H$\beta$ equivalent widths increased up to 10.1\AA~and 15.5\AA, respectively, and the flare duration in H$\alpha$ ($\Delta t^{\rm{flare}}_{\rm{H}\alpha}$) is 1.5 hours (Table \ref{table:list1_flares}). 
In addition to the enhancements in Balmer emission lines, the continuum brightness observed by ARCSAT $g$-band and \textit{TESS} increased by $\sim$100\% and $\sim$3\%, respectively (Figures \ref{fig:lcEW_HaHb_YZCMi_UT190126} (b) \& (c)). 
We note that there were no ARCSAT $u$-band data during Flare Y1 since we only took $g$-band data for most of the time because of unstable weather on that date. Flare Y1 was not \color{black}\textrm{identified }\color{black} in \textit{NICER} X-ray data since the flare occurred during the observation gap caused by the orbital period of ISS (Figure \ref{fig:lcEW_HaHb_YZCMi_UT190126} (d)).
 \color{black}\textrm{ 
We estimated 
$L_{g}$, $L_{TESS}$, $E_{g}$, $E_{TESS}$, $L_{\rm{H}\alpha}$, $L_{\rm{H}\beta}$, $E_{\rm{H}\alpha}$, and $E_{\rm{H}\beta}$ values, and they are listed in Table \ref{table:list1_flares} (see Section \ref{subsec:quiescent-ene} for the estimation method).
} \color{black}
 
\begin{figure}[ht!]
   \begin{center}
     \gridline{  
     \hspace{-0.06\textwidth}
    \fig{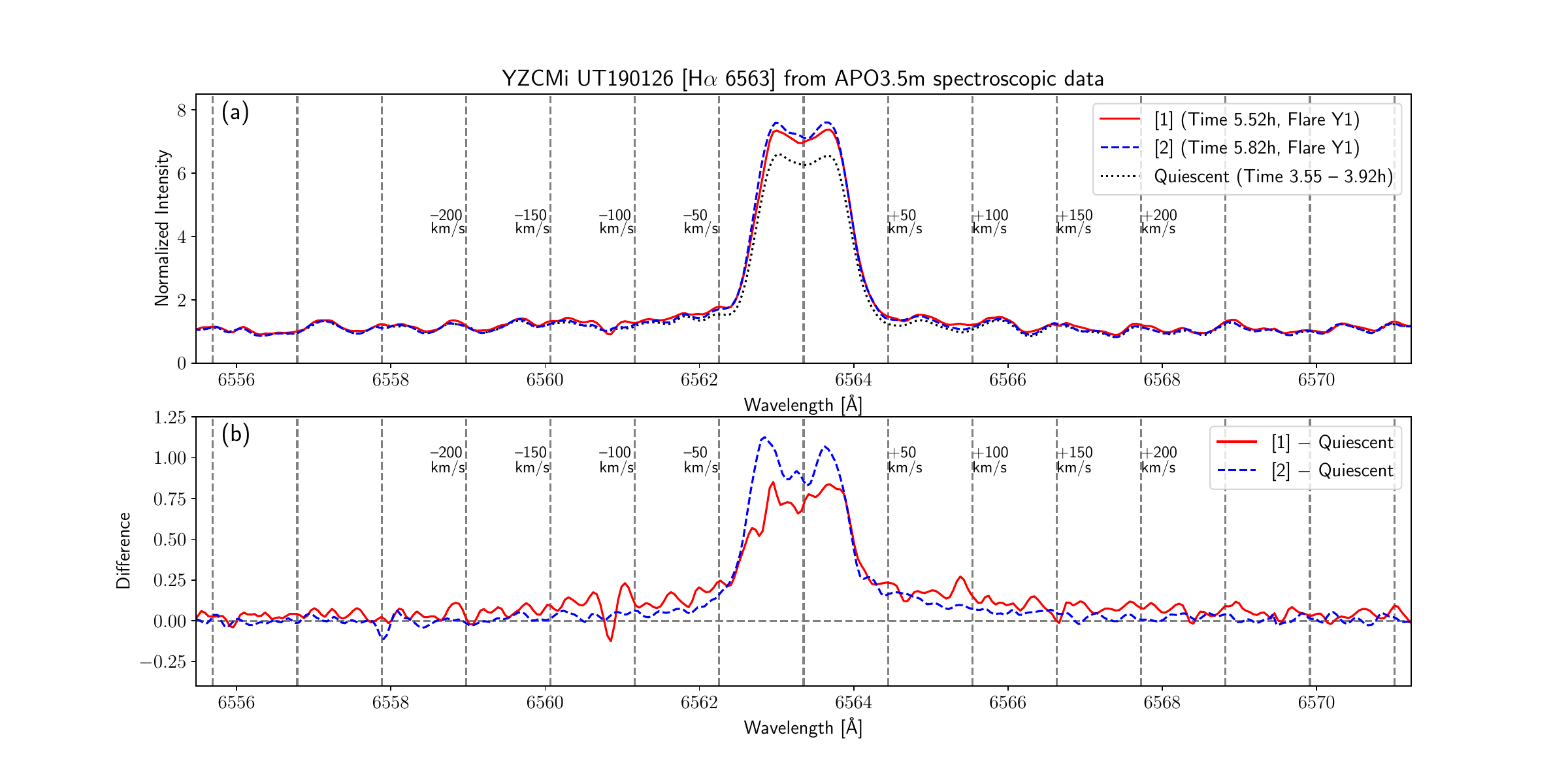}{0.58\textwidth}{\vspace{0mm}}
     \hspace{-0.06\textwidth}
       \fig{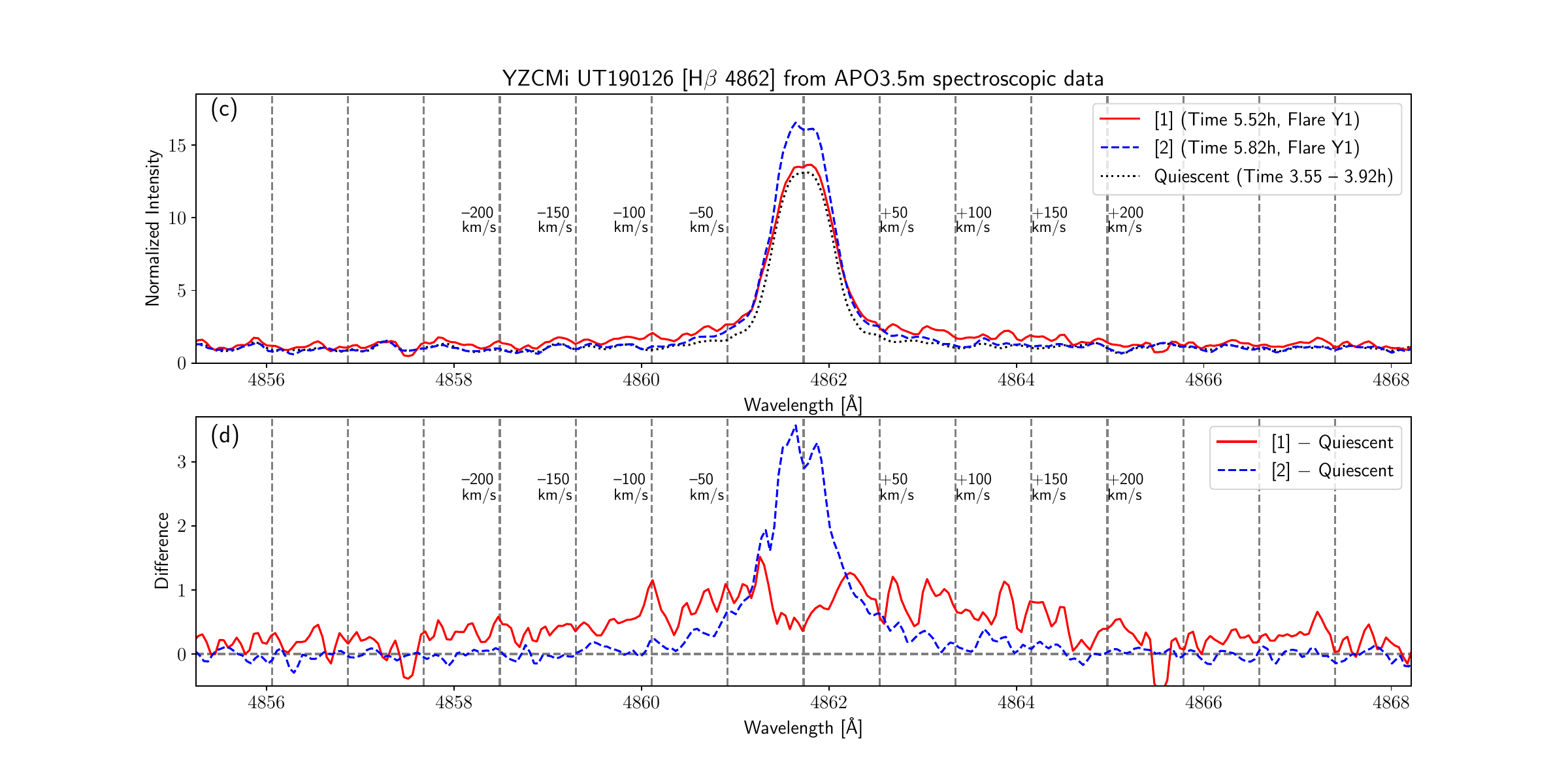}{0.58\textwidth}{\vspace{0mm}}
    }
     \vspace{-1cm}
     \caption{
    \color{black}\textrm{  
Line profiles of the H$\alpha$ \& H$\beta$ emission lines during Flare Y1 (at the time [1] and [2]) on 2019 January 26 from APO3.5m spectroscopic data, which are plotted similarly with Figure \ref{fig:spec_HaHb_YZCMi_UT190127}.
 } \color{black}
     }
   \label{fig:spec_HaHb_YZCMi_UT190126}
   \end{center}
 \end{figure}

\begin{figure}[ht!]
   \begin{center}
     \gridline{  
     \hspace{-0.07\textwidth}
    \fig{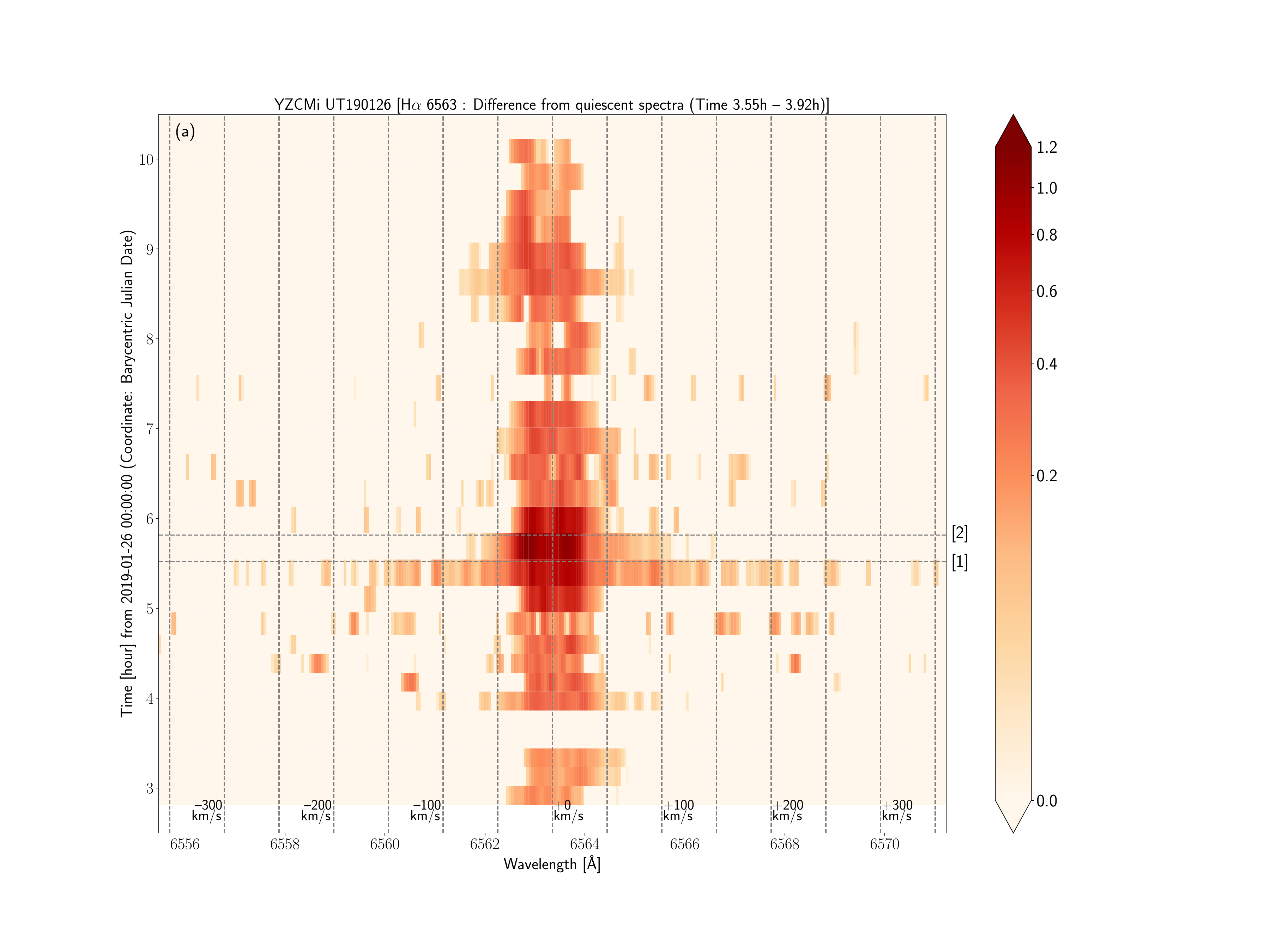}{0.62\textwidth}{\vspace{0mm}}
     \hspace{-0.11\textwidth}
    \fig{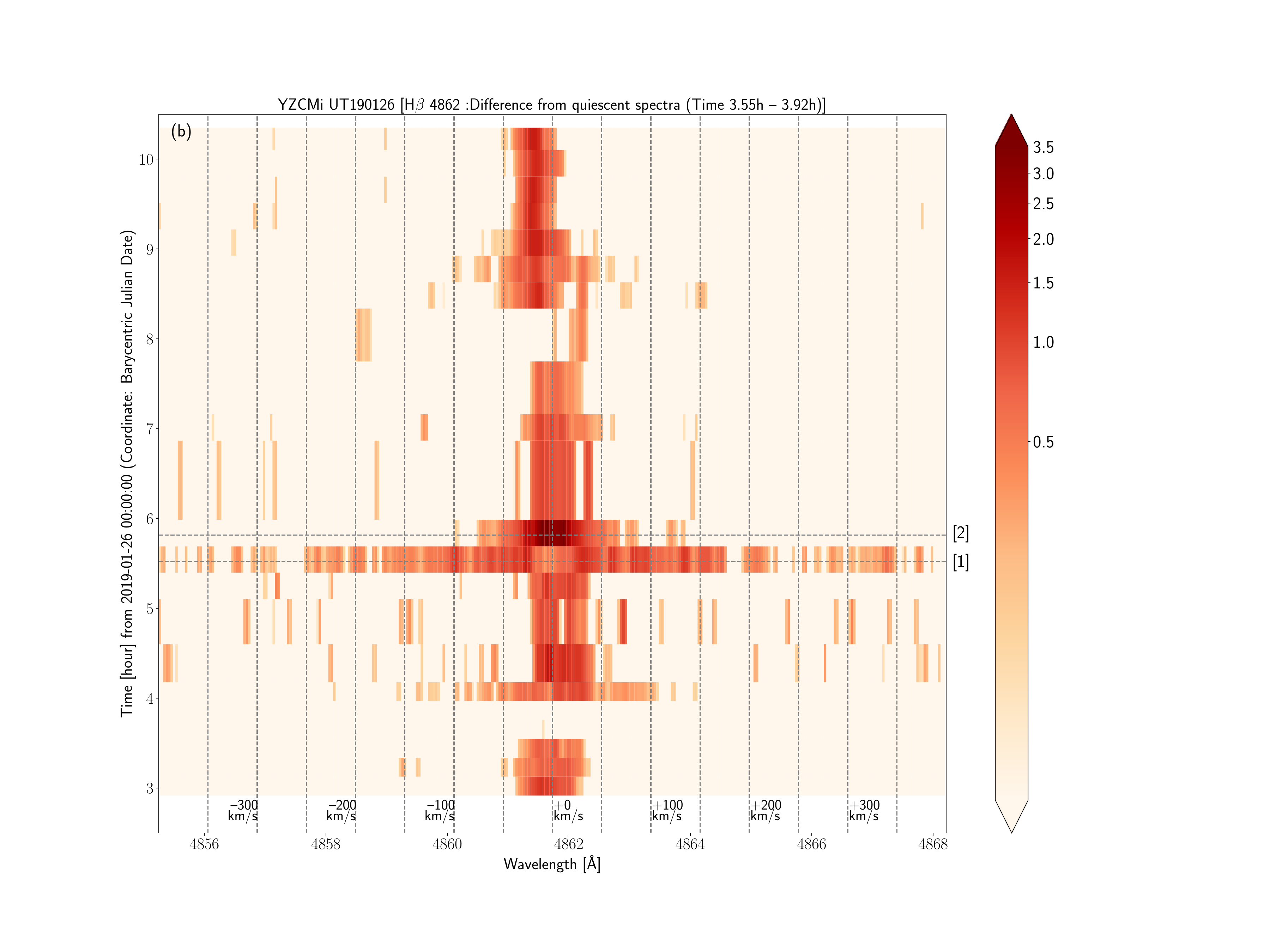}{0.62\textwidth}{\vspace{0mm}}
    }
     \vspace{-0.5cm}
     \caption{
         \color{black}\textrm{  
Time evolution of the H$\alpha$ \& H$\beta$ line profiles covering Flares Y1 on 2019 January 26, which are shown similarly with Figure \ref{fig:map_HaHb_YZCMi_UT190127}.
The grey horizontal dashed lines indicate the time [1] and [2], which are shown in Figure \ref{fig:lcEW_HaHb_YZCMi_UT190126} (light curves) and Figure \ref{fig:spec_HaHb_YZCMi_UT190126} (line profiles).
}\color{black}
     }
   \label{fig:map_HaHb_YZCMi_UT190126}
   \end{center}
 \end{figure}

Figures \ref{fig:spec_HaHb_YZCMi_UT190126} \& \ref{fig:map_HaHb_YZCMi_UT190126}
show the H$\alpha$ \& H$\beta$ line profiles during Flare Y1. 
We could not see any significant line wing asymmetries during this flare.
We can see clear line-wing broadening of the H$\alpha$ \& H$\beta$ line profiles (H$\alpha$: $\pm$150 km s$^{-1}$, H$\beta$: $\pm$200--250 km s$^{-1}$), 
which is especially seen around the flare peak time 
(see the time [1] in Figures \ref{fig:lcEW_HaHb_YZCMi_UT190126},
\ref{fig:spec_HaHb_YZCMi_UT190126}, \&
\ref{fig:map_HaHb_YZCMi_UT190126}).
 \color{black}\textrm{ 
We note that as for the H$\alpha$ and H$\beta$ lines, the larger enhancements in the line wings contributed to a bigger total equivalent widths at [1] than [2], while the peak intensities at the line centers are smaller at [1] than at [2] (Figures \ref{fig:lcEW_HaHb_YZCMi_UT190126} \&
\ref{fig:spec_HaHb_YZCMi_UT190126}).
} \color{black}

\subsection{Flares Y4 \& Y5 observed on 2019 January 28} \label{subsec:results:2019-Jan-28} 

\begin{figure}[ht!]
   \begin{center}
   \gridline{
    \fig{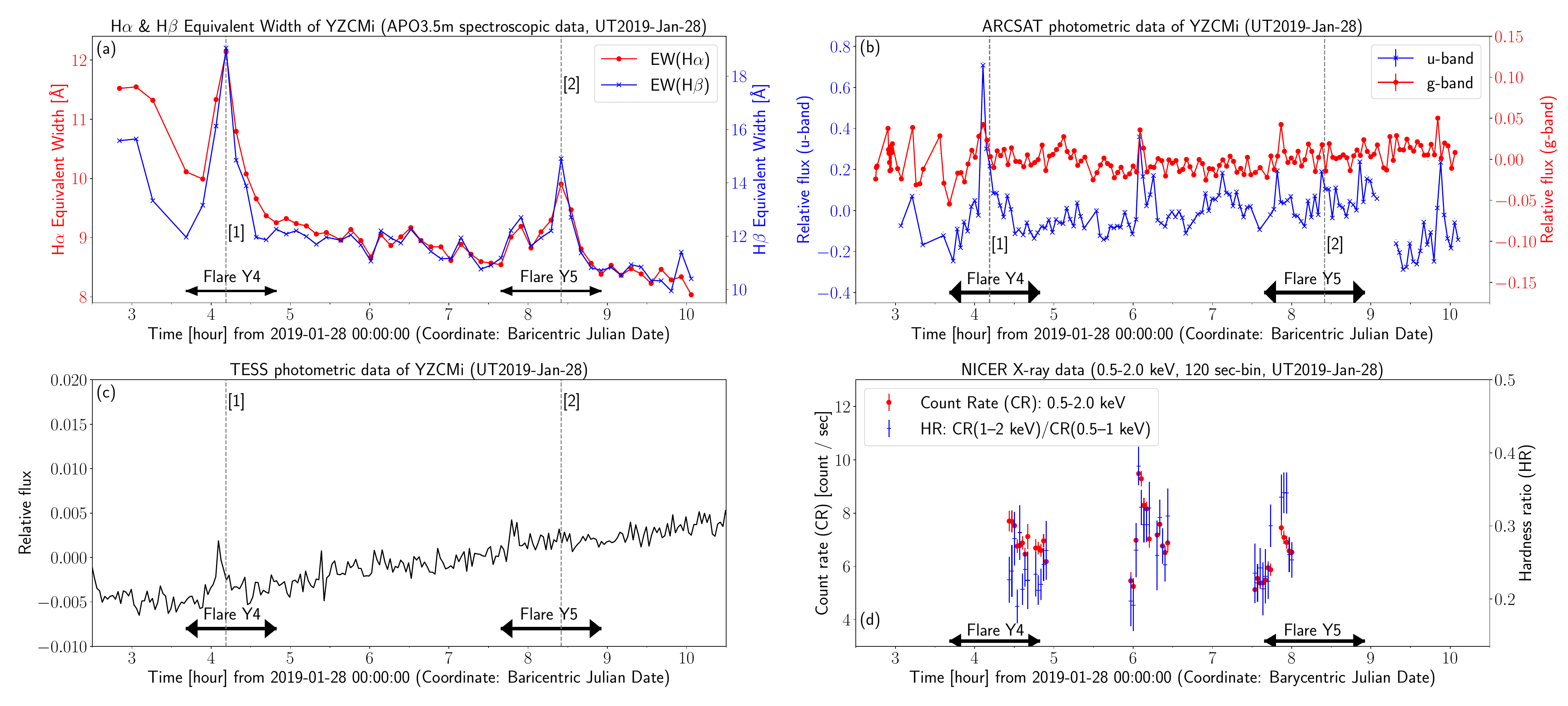}{1.0\textwidth}{\vspace{0mm}}}
     \vspace{0mm}
     \caption{
     \color{black}\textrm{  
Light curves of YZ CMi on  2019 January 28 showing Flares Y4 \& Y5, which are plotted 
similarly with Figures \ref{fig:lcEW_HaHb_YZCMi_UT190127}(a)--(d).
 The grey dashed lines with numbers ([1],[2]) correspond to the time shown with the same numbers in Figures \ref{fig:spec_HaHb_YZCMi_UT190128} \& \ref{fig:map_HaHb_YZCMi_UT190128}.
 } \color{black}
     }
   \label{fig:lcEW_HaHb_YZCMi_UT190128}
   \end{center}
 \end{figure}

On 2019 January 28, two flares (Flares Y4 \& Y5) were detected in H$\alpha$ \& H$\beta$ lines as shown in Figure \ref{fig:lcEW_HaHb_YZCMi_UT190128} (a).  During Flare Y4,
the H$\alpha$ \& H$\beta$ equivalent widths increased up to 12.1\AA~and 19.1\AA, respectively, and $\Delta t^{\rm{flare}}_{\rm{H}\alpha}$ is 1.0 hour (Table \ref{table:list1_flares}). 
In addition to the enhancements in Balmer emission lines, the continuum brightness observed by ARCSAT $u$- \& $g$-bands and \textit{TESS} increased by $\sim$70\%, $\sim$4--5\%, and $\sim$0.5\%, respectively, during Flare Y4 (Figures \ref{fig:lcEW_HaHb_YZCMi_UT190128} (b) \& (c)). 
During Flare Y5, the H$\alpha$ \& H$\beta$ equivalent widths increased up to 9.9\AA~and 14.9\AA, respectively, and $\Delta t^{\rm{flare}}_{\rm{H}\alpha}$ is 1.3 hours (Table \ref{table:list1_flares}).
The continuum brightness increase observed by ARCSAT $u$- \& $g$-bands and \textit{TESS} during Flare Y5 are not clear compared with Flare Y4 \color{black}\textrm{
($3\sigma_{u}$=15\%, $3\sigma_{u}$=2.3\%, and $3\sigma_{TESS}$=0.29\%)} \color{black},
although there might exist small increase around the time 7.8h -- 8.0h (Figures \ref{fig:lcEW_HaHb_YZCMi_UT190128} (b) \& (c)).  
\color{black}\textrm{
Both the peaks of Flares Y4 \& Y5 in H$\alpha$ \& H$\beta$ lines were 
in the gaps of the \textit{NICER} X-ray observation, 
and we cannot identify the X-ray emission from these flares in \textit{NICER} X-ray data (Figure \ref{fig:lcEW_HaHb_YZCMi_UT190128} (d)).
$L_{g}$, $L_{TESS}$, $E_{g}$, $E_{TESS}$, $L_{\rm{H}\alpha}$, $L_{\rm{H}\beta}$, $E_{\rm{H}\alpha}$, and $E_{\rm{H}\beta}$ values are estimated and listed in Table \ref{table:list1_flares}.
} \color{black}

      \begin{figure}[ht!]
   \begin{center}
        \gridline{  
     \hspace{-0.06\textwidth}
    \fig{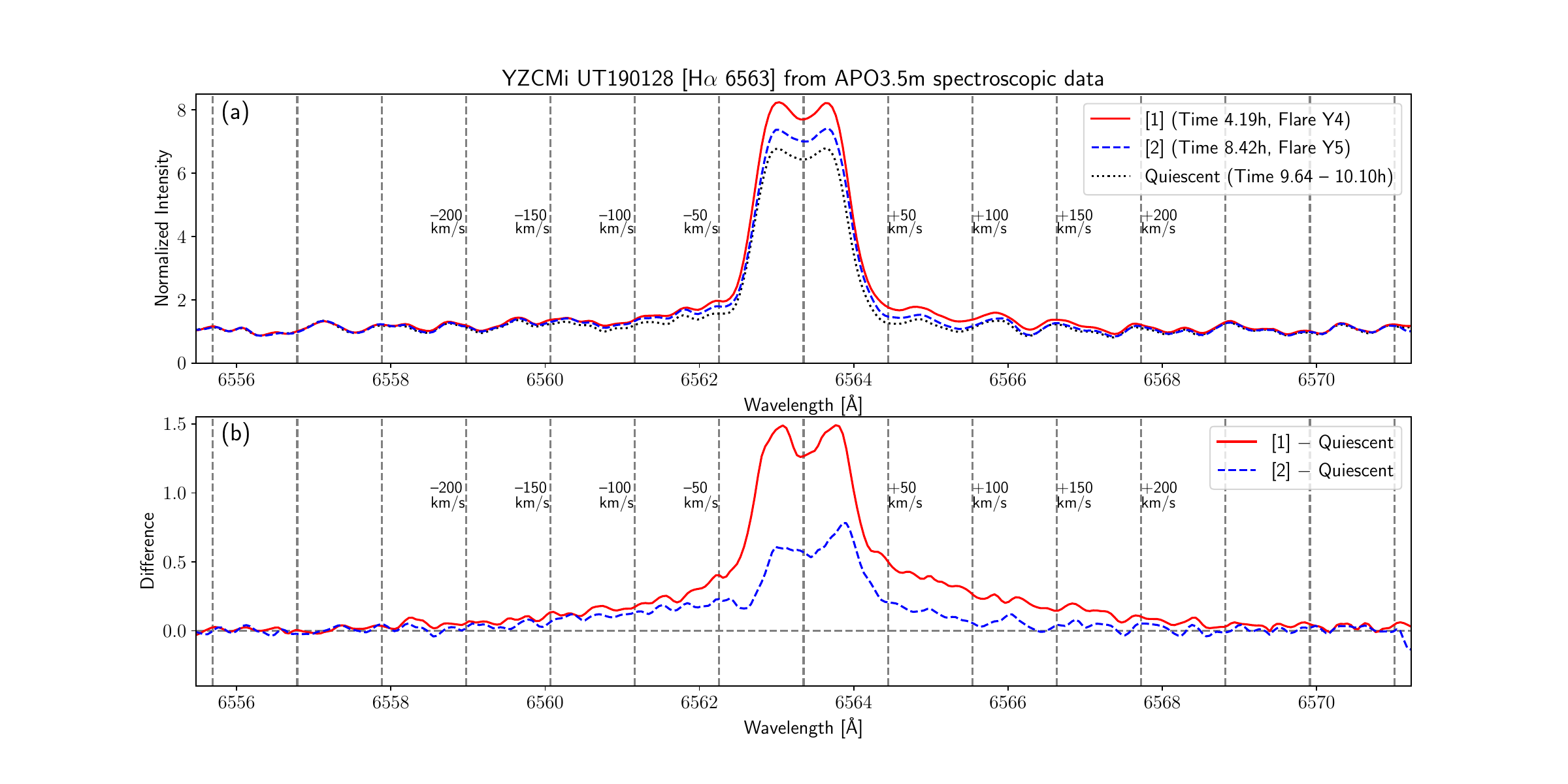}{0.58\textwidth}{\vspace{0mm}}
     \hspace{-0.06\textwidth}
       \fig{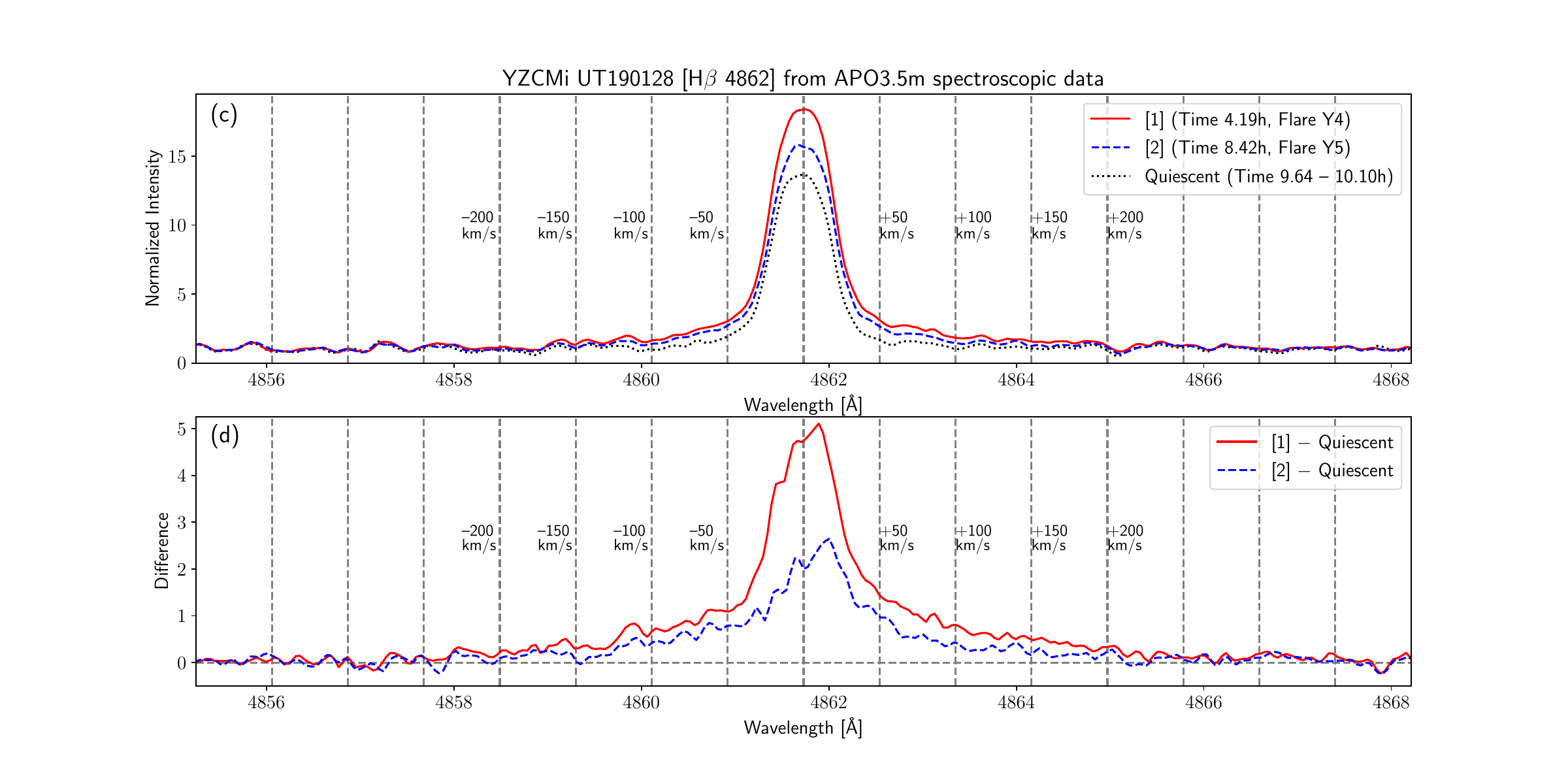}{0.58\textwidth}{\vspace{0mm}}
    }
     \vspace{-1cm}
     \caption{
    \color{black}\textrm{  
Line profiles of the H$\alpha$ \& H$\beta$ emission lines during Flares Y4 \& Y5 (at the time [1] and [2]) on 2019 January 28 from APO3.5m spectroscopic data, which are plotted similarly with Figure \ref{fig:spec_HaHb_YZCMi_UT190127}.
 } \color{black}
     }
   \label{fig:spec_HaHb_YZCMi_UT190128}
   \end{center}
 \end{figure}

      \begin{figure}[ht!]
   \begin{center}
      \gridline{  
     \hspace{-0.07\textwidth}
    \fig{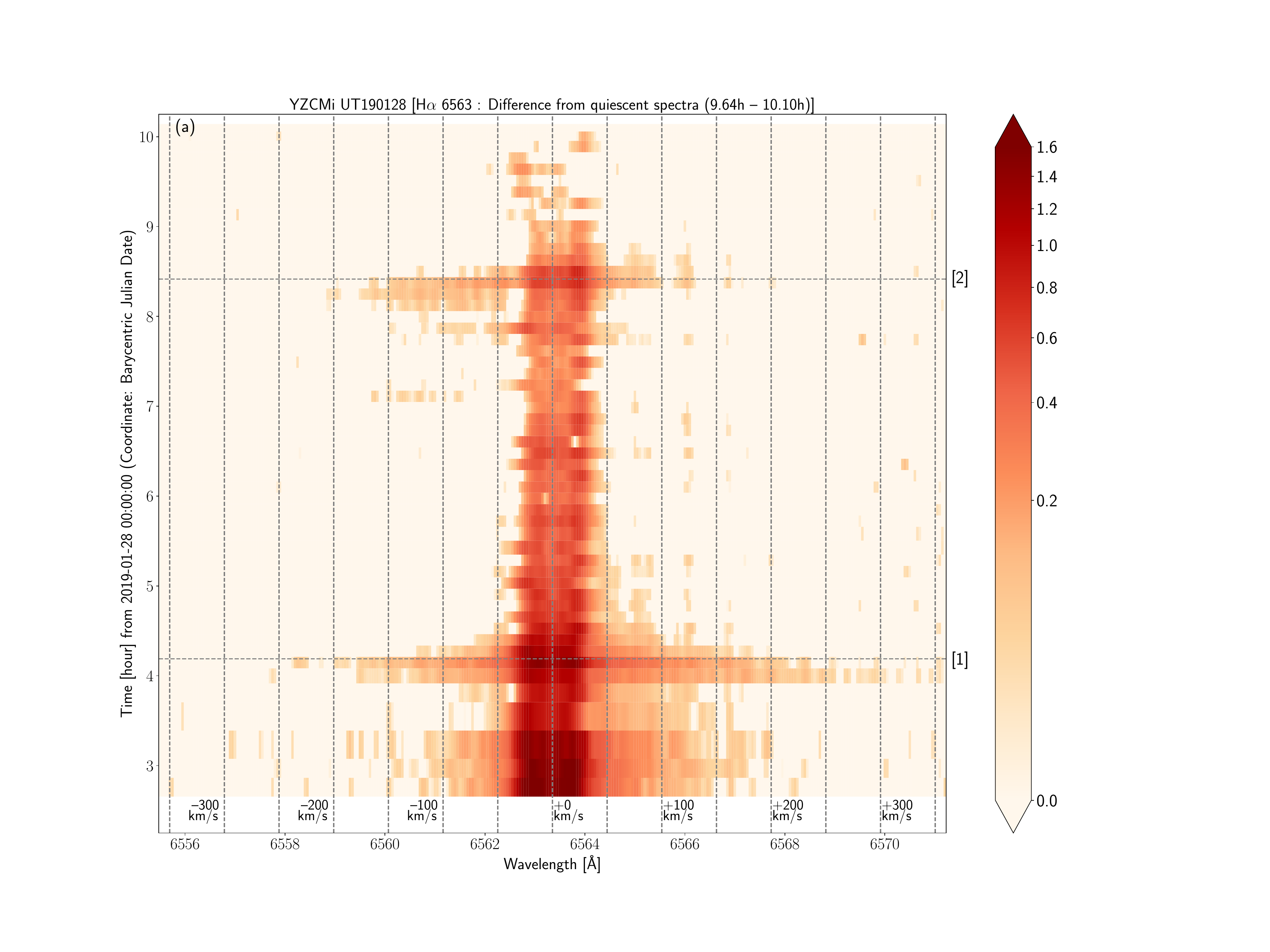}{0.63\textwidth}{\vspace{0mm}}
     \hspace{-0.11\textwidth}
    \fig{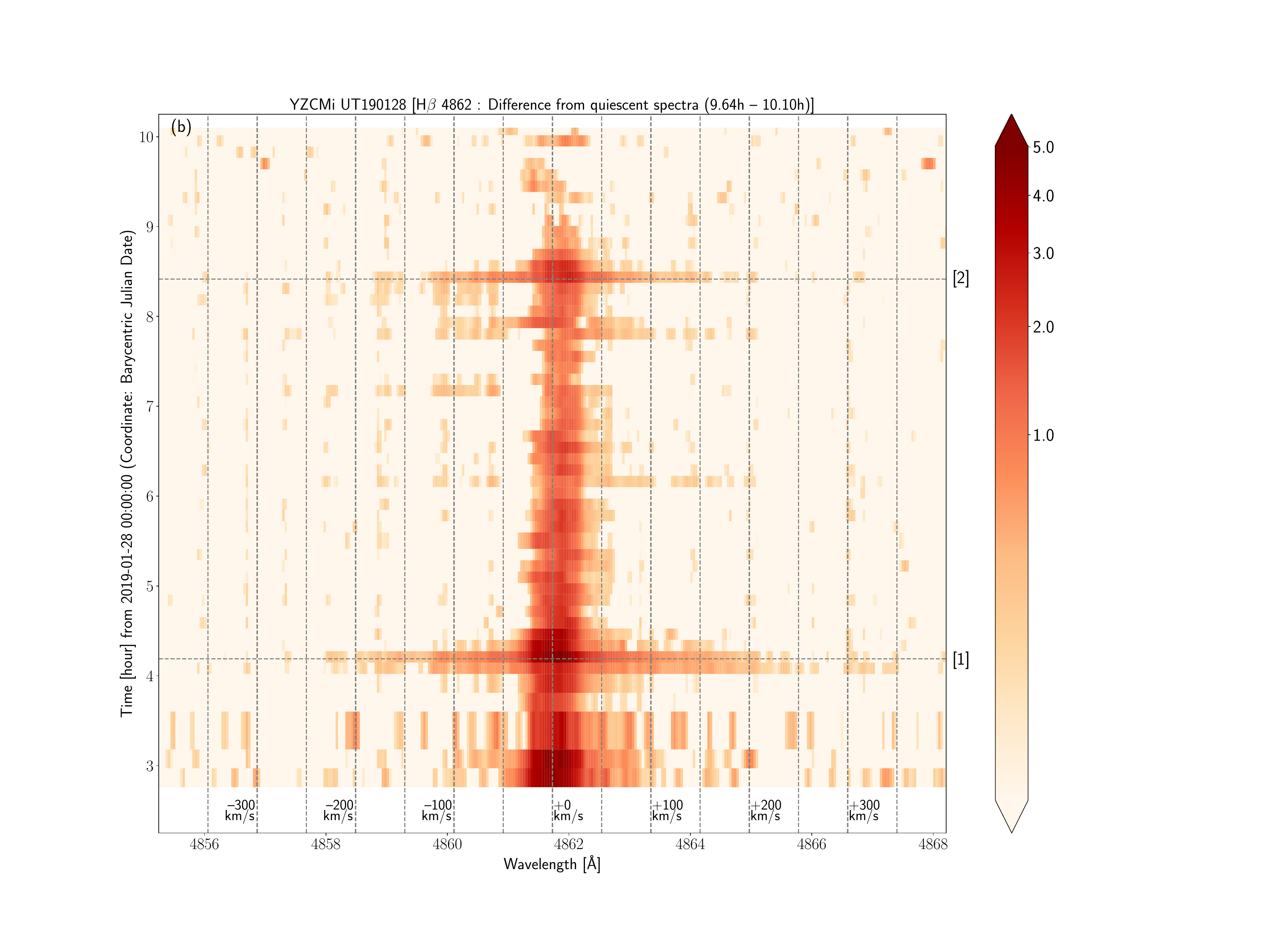}{0.63\textwidth}{\vspace{0mm}}
    }
     \vspace{-0.5cm}
     \caption{
    \color{black}\textrm{  
Time evolution of the H$\alpha$ \& H$\beta$ line profiles covering Flares Y4 \& Y5 on 2019 January 28, which are shown similarly with Figure \ref{fig:map_HaHb_YZCMi_UT190127}.
The grey horizontal dashed lines indicate the time [1] and [2], which are shown in Figure \ref{fig:lcEW_HaHb_YZCMi_UT190128} (light curves) and Figure \ref{fig:spec_HaHb_YZCMi_UT190128} (line profiles).
}\color{black}
     }
   \label{fig:map_HaHb_YZCMi_UT190128}
   \end{center}
 \end{figure}

 The H$\alpha$ \& H$\beta$ line profiles during Flares Y4 and Y5 are shown in
Figures \ref{fig:spec_HaHb_YZCMi_UT190128} \& \ref{fig:map_HaHb_YZCMi_UT190128}. 
At around the peak time of Flare Y4, we can see line-wing broadening (from $-$150--200 km s$^{-1}$ to $+$200--250 km s$^{-1}$) and the red wing was slightly enhanced compared with the blue wing (red wing asymmetry) (the time [1] in Figures \ref{fig:lcEW_HaHb_YZCMi_UT190128},
\ref{fig:spec_HaHb_YZCMi_UT190128}, \&
\ref{fig:map_HaHb_YZCMi_UT190128}). 
This slight red wing asymmetry is seen in both H$\alpha$ and H$\beta$ lines.

 \subsection{Flares Y7 \& Y8 observed on 2020 January 14} \label{subsec:results:2020-Jan-14}

\begin{figure}[ht!]
   \begin{center}
 \gridline{
    \fig{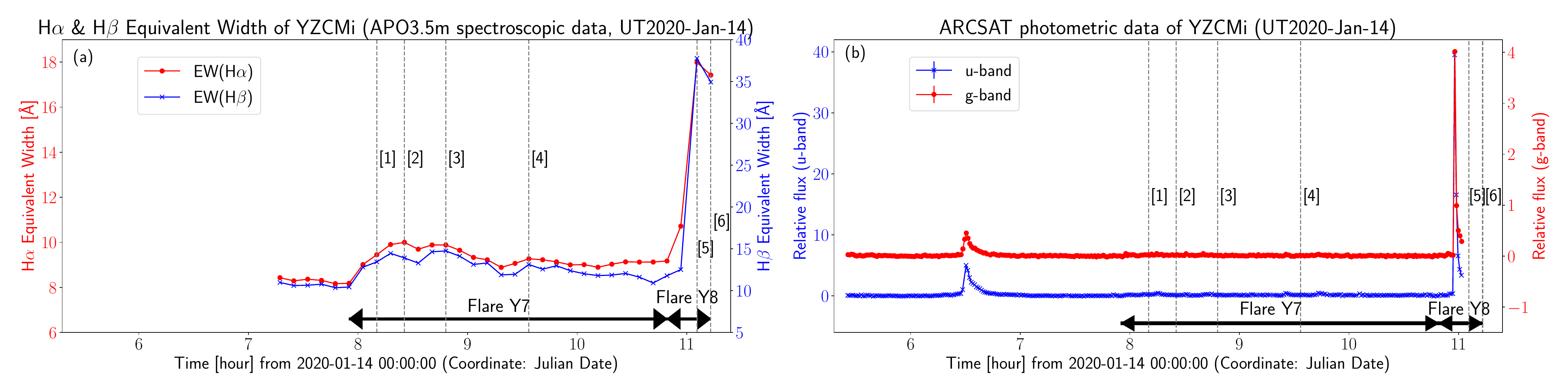}{1.0\textwidth}{\vspace{0mm}}
    }
     \vspace{-1cm}
 \gridline{
    \fig{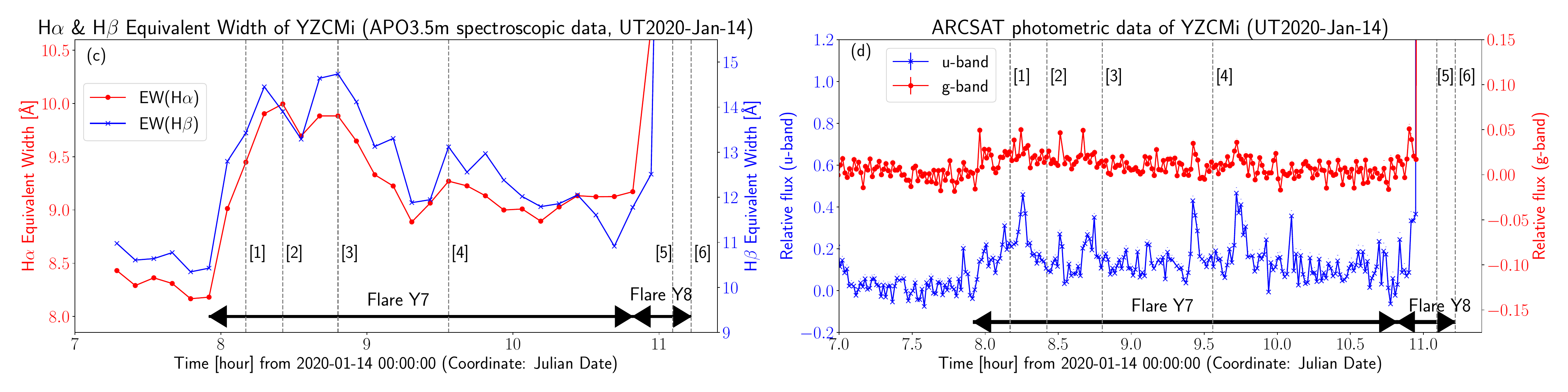}{1.0\textwidth}{\vspace{0mm}}
    }
     \vspace{-0.5cm}
     \caption{
     \color{black}\textrm{  
(a)\&(b)
Light curves of YZ CMi on 2020 January 14 showing Flares Y7 \& Y8, which are plotted 
similarly with Figures \ref{fig:lcEW_HaHb_YZCMi_UT191212} (a)\&(b).
The grey dashed lines with numbers ([1]--[6]) correspond to the time shown with the same numbers in Figures \ref{fig:spec_HaHb_YZCMi_UT200114} \& \ref{fig:map_HaHb_YZCMi_UT200114}.
(c)\&(d)
Enlarged panels of (a)\&(b).
 } \color{black}
     }
   \label{fig:lcEW_HaHb_YZCMi_UT200114}
   \end{center}
 \end{figure}

On 2020 January 14, two flares (Flares Y7 \& Y8) were detected in H$\alpha$ \& H$\beta$ lines as shown in Figures \ref{fig:lcEW_HaHb_YZCMi_UT200114} (a) \& (c).  During Flare Y7,
the H$\alpha$ \& H$\beta$ equivalent widths increased up to 10.0\AA~and 14.7\AA, respectively, and the flare duration in H$\alpha$ $\Delta t^{\rm{flare}}_{\rm{H}\alpha}$ is 2.9 hours (Table \ref{table:list1_flares}). 
In addition to the enhancements in Balmer emission lines, the continuum brightness observed with ARCSAT $u$- \& $g$-bands increased by $\sim$40\% and $\sim$4\%, respectively, \color{black}\textrm{associated with the H$\alpha$ and H$\beta$ emissions 
of }\color{black} 
Flare Y7 (Figures \ref{fig:lcEW_HaHb_YZCMi_UT200114} (b) \& (d)). 
As for Flare Y8, the H$\alpha$ \& H$\beta$ equivalent widths increased up to 18.0\AA~and 37.8\AA, respectively, but only the initial 0.4 \color{black}\textrm{hour }\color{black} of the flare were observed (Figures \ref{fig:lcEW_HaHb_YZCMi_UT200114} (a) \& (c)).
In addition to these enhancements in Balmer emission lines, the continuum brightness observed with ARCSAT $u$- \& $g$-bands increased by $\sim$4000\% and $\sim$400\%, respectively (Figures \ref{fig:lcEW_HaHb_YZCMi_UT200114} (b) \& (d)) during the observed intial phase of the flare. 
\color{black}\textrm{ 
$L_{u}$, $L_{g}$, $E_{u}$, $E_{g}$, $L_{\rm{H}\alpha}$, $L_{\rm{H}\beta}$, $E_{\rm{H}\alpha}$, and $E_{\rm{H}\beta}$ values are estimated and listed in Table \ref{table:list1_flares}.
We note that 
the flare energy values of Flare Y8 listed here is only the lower limit and is expected to be much smaller than the real total energy values, since only the initial 0.4 hour data of the flare were observed.} \color{black}

\begin{figure}[ht!]
   \begin{center}
            \gridline{  
     \hspace{-0.06\textwidth}
    \fig{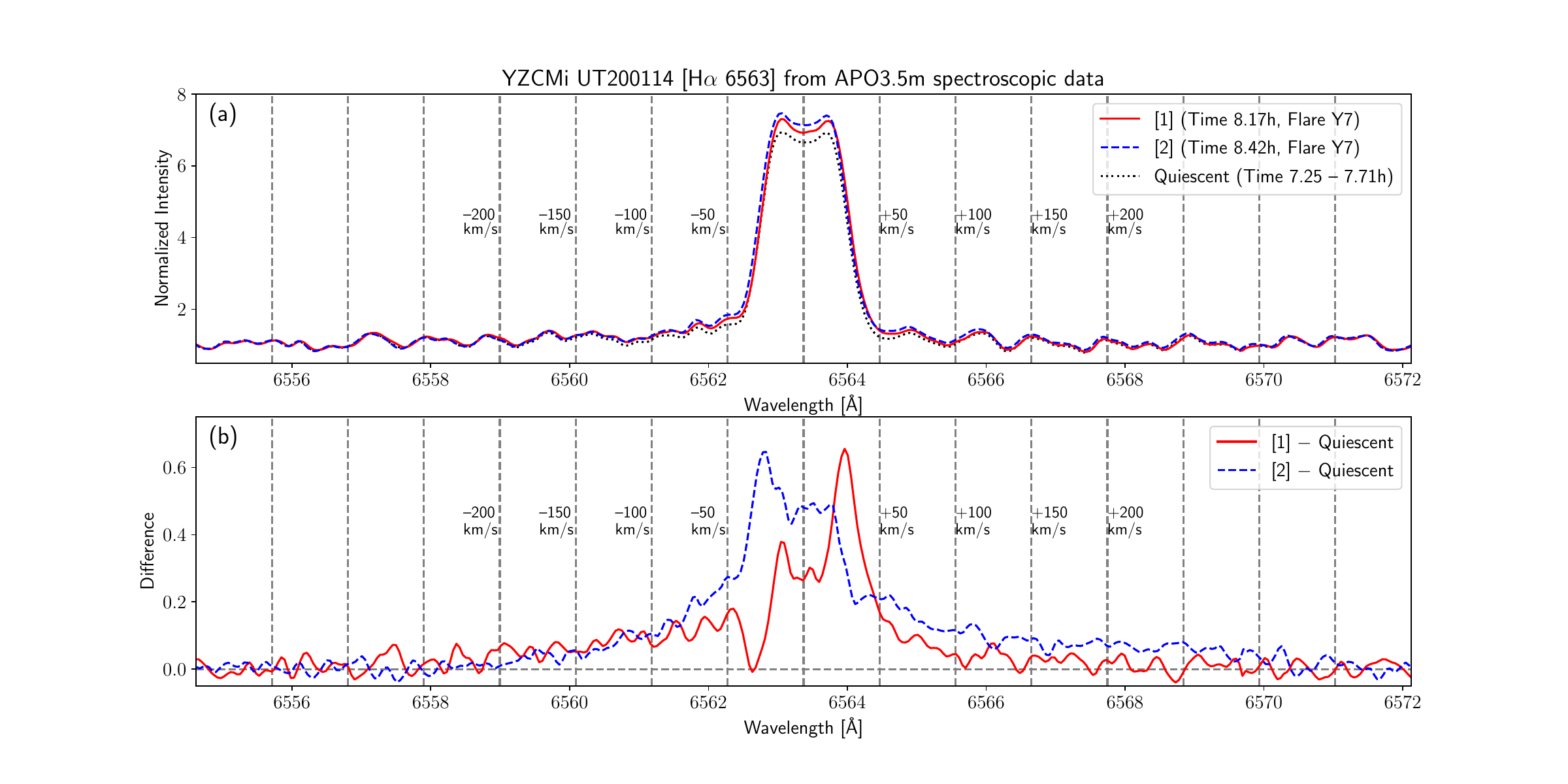}{0.58\textwidth}{\vspace{0mm}}
     \hspace{-0.06\textwidth}
       \fig{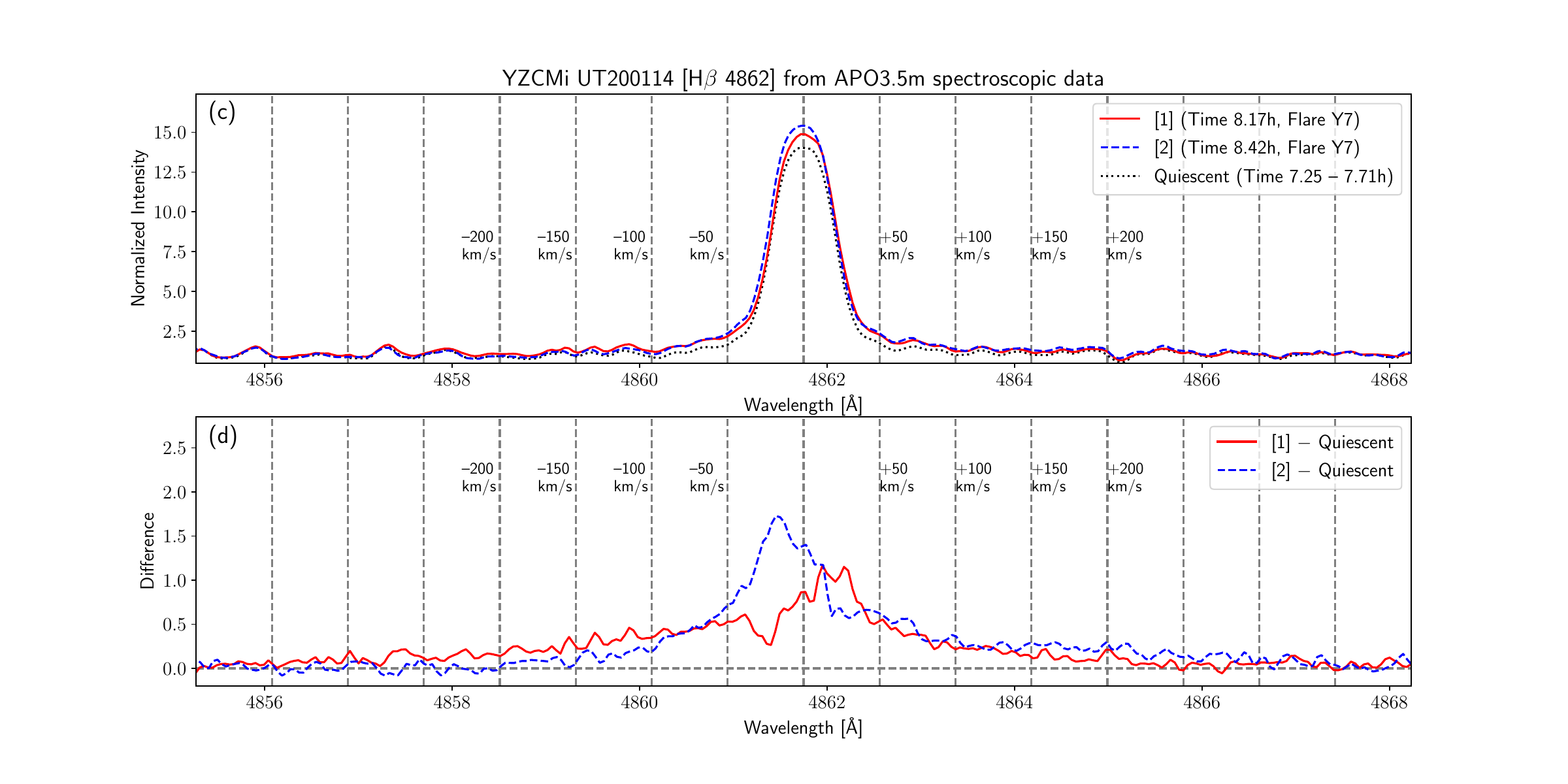}{0.58\textwidth}{\vspace{0mm}}
    }
     \vspace{-1.0cm}
            \gridline{  
     \hspace{-0.06\textwidth}
    \fig{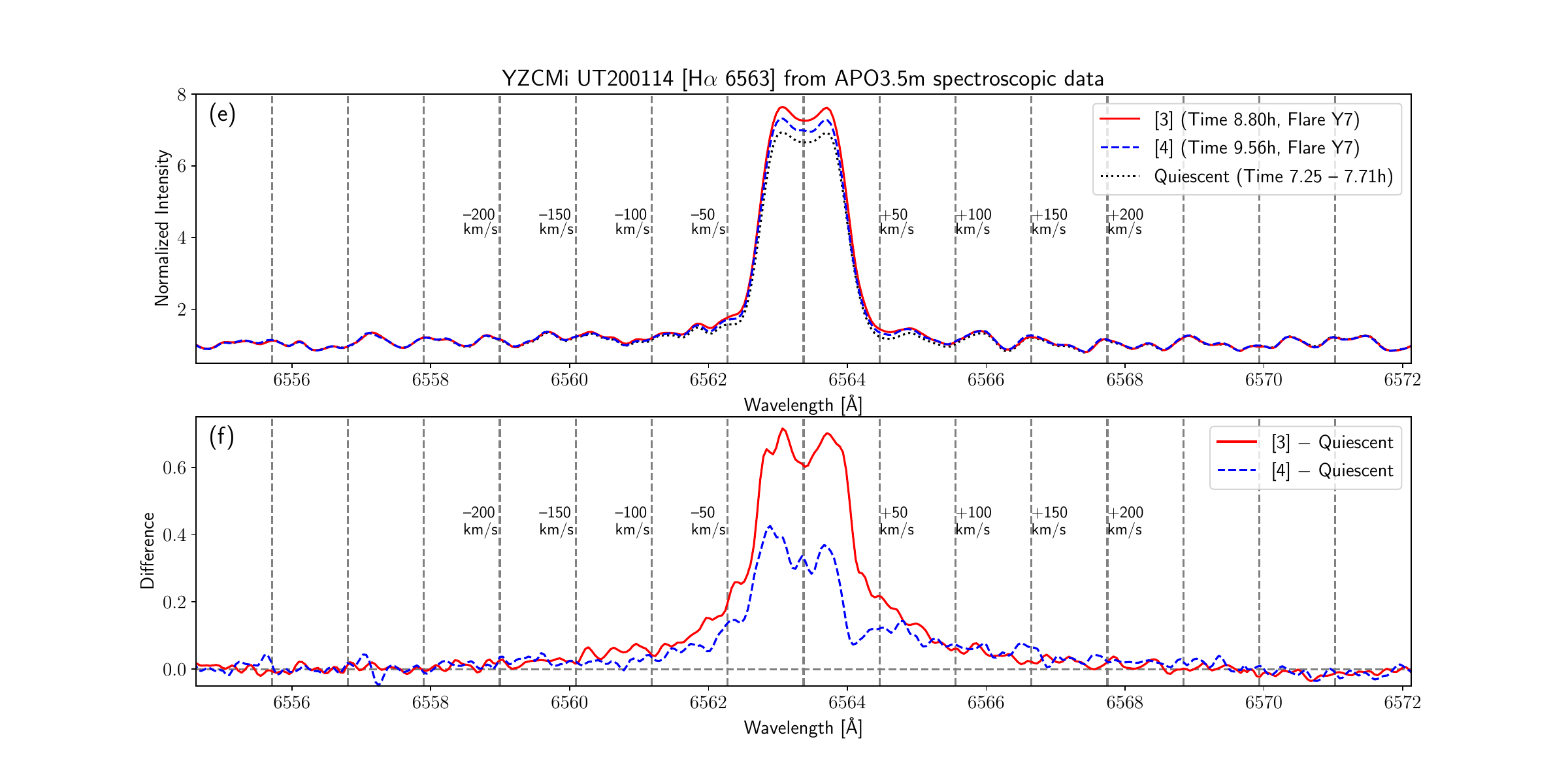}{0.58\textwidth}{\vspace{0mm}}
     \hspace{-0.06\textwidth}
   \fig{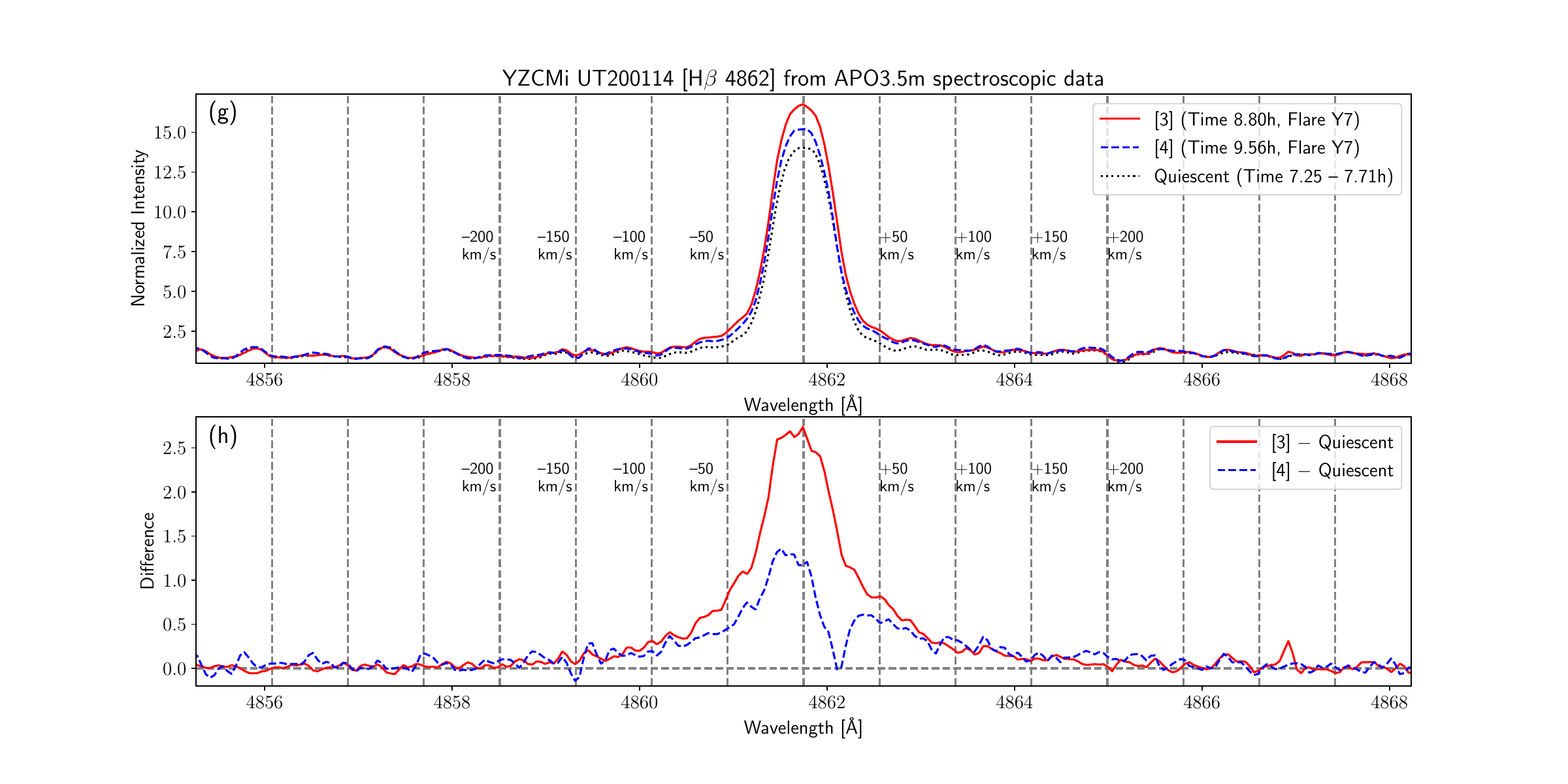}{0.58\textwidth}{\vspace{0mm}}
    }
     \vspace{-1.0cm}
             \gridline{  
     \hspace{-0.06\textwidth}
    \fig{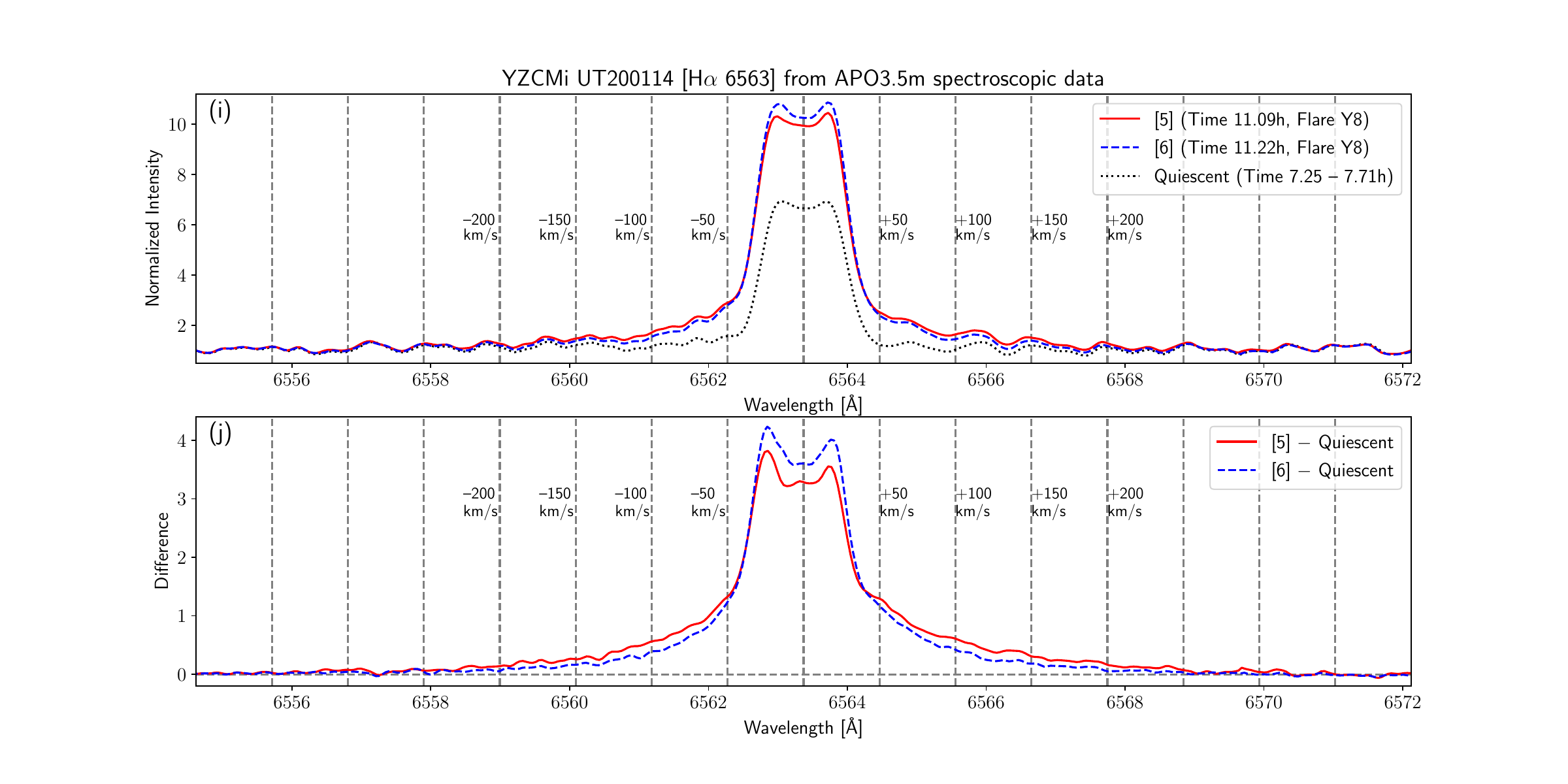}{0.58\textwidth}{\vspace{0mm}}
     \hspace{-0.06\textwidth}
       \fig{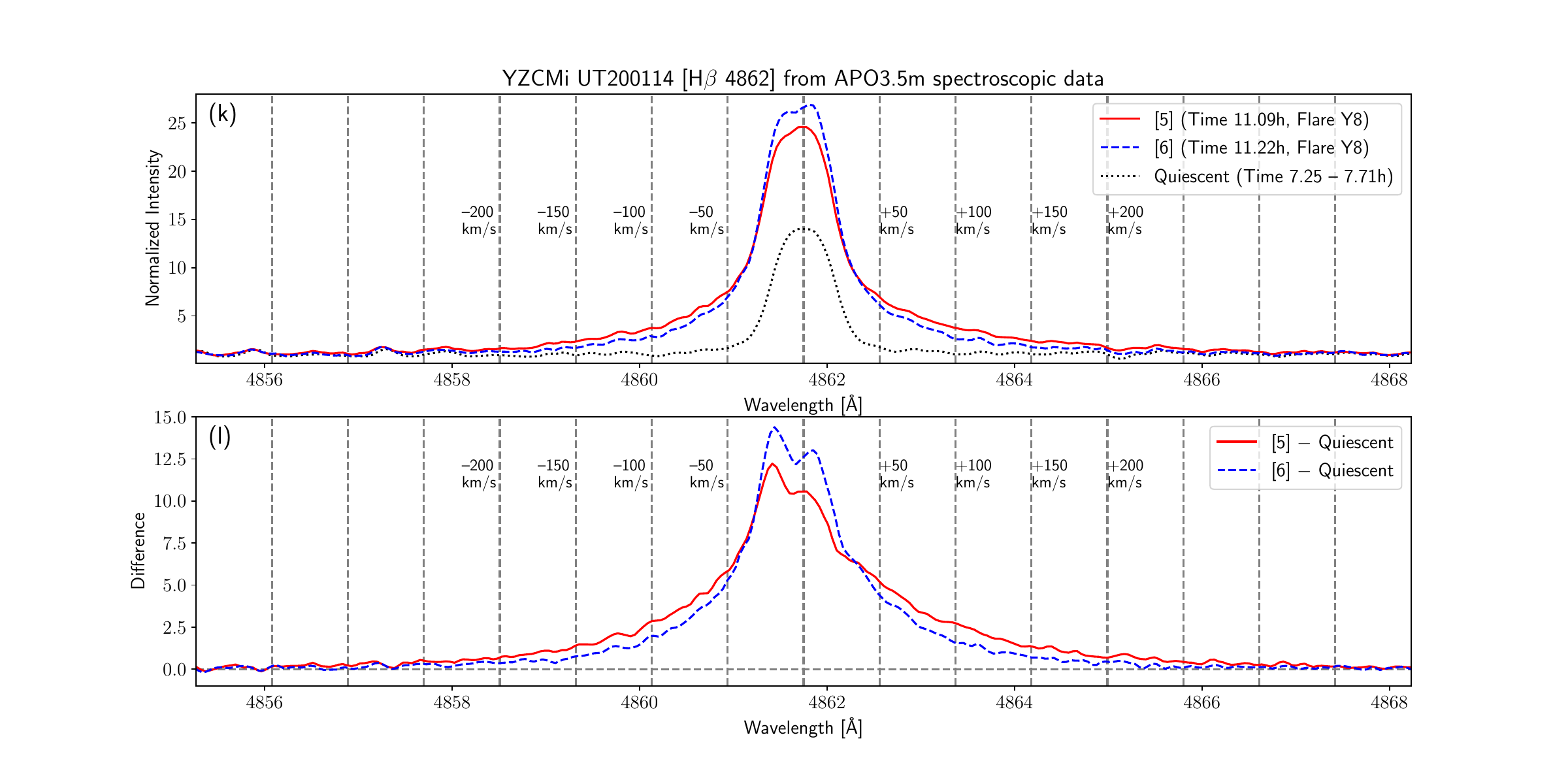}{0.58\textwidth}{\vspace{0mm}}
    }
     \vspace{-0.5cm}
     \caption{
   \color{black}\textrm{  
Line profiles of the H$\alpha$ \& H$\beta$ emission lines during Flares Y7\&Y8 on 2020 January 14 (at the time [1]-[6]) from APO3.5m spectroscopic data, which are plotted similarly with Figure \ref{fig:spec_HaHb_YZCMi_UT190127}.
 } \color{black}
     }
   \label{fig:spec_HaHb_YZCMi_UT200114}
   \end{center}
 \end{figure}

\begin{figure}[ht!]
   \begin{center}
          \gridline{  
     \hspace{-0.07\textwidth}
    \fig{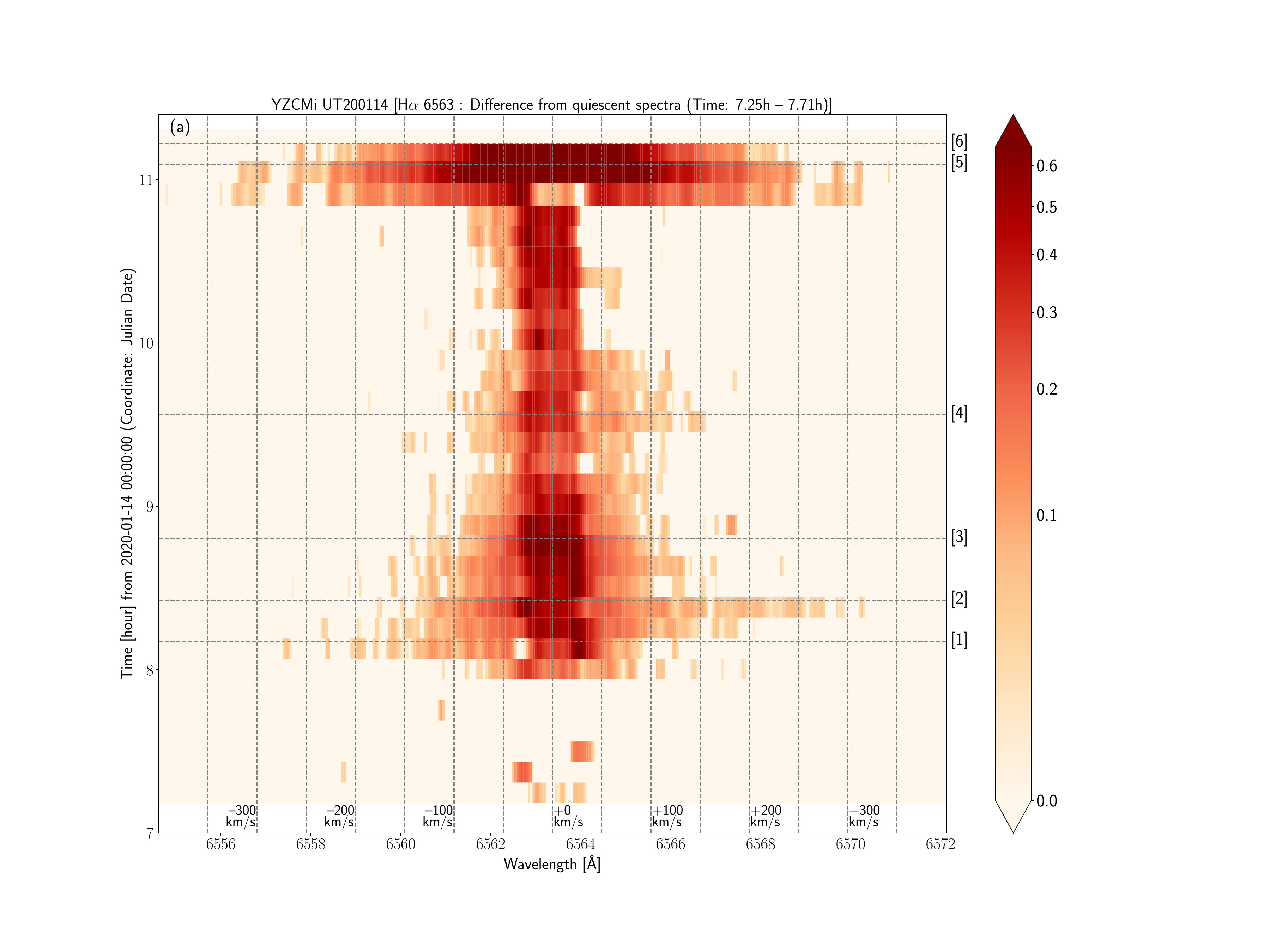}{0.63\textwidth}{\vspace{0mm}}
     \hspace{-0.11\textwidth}
    \fig{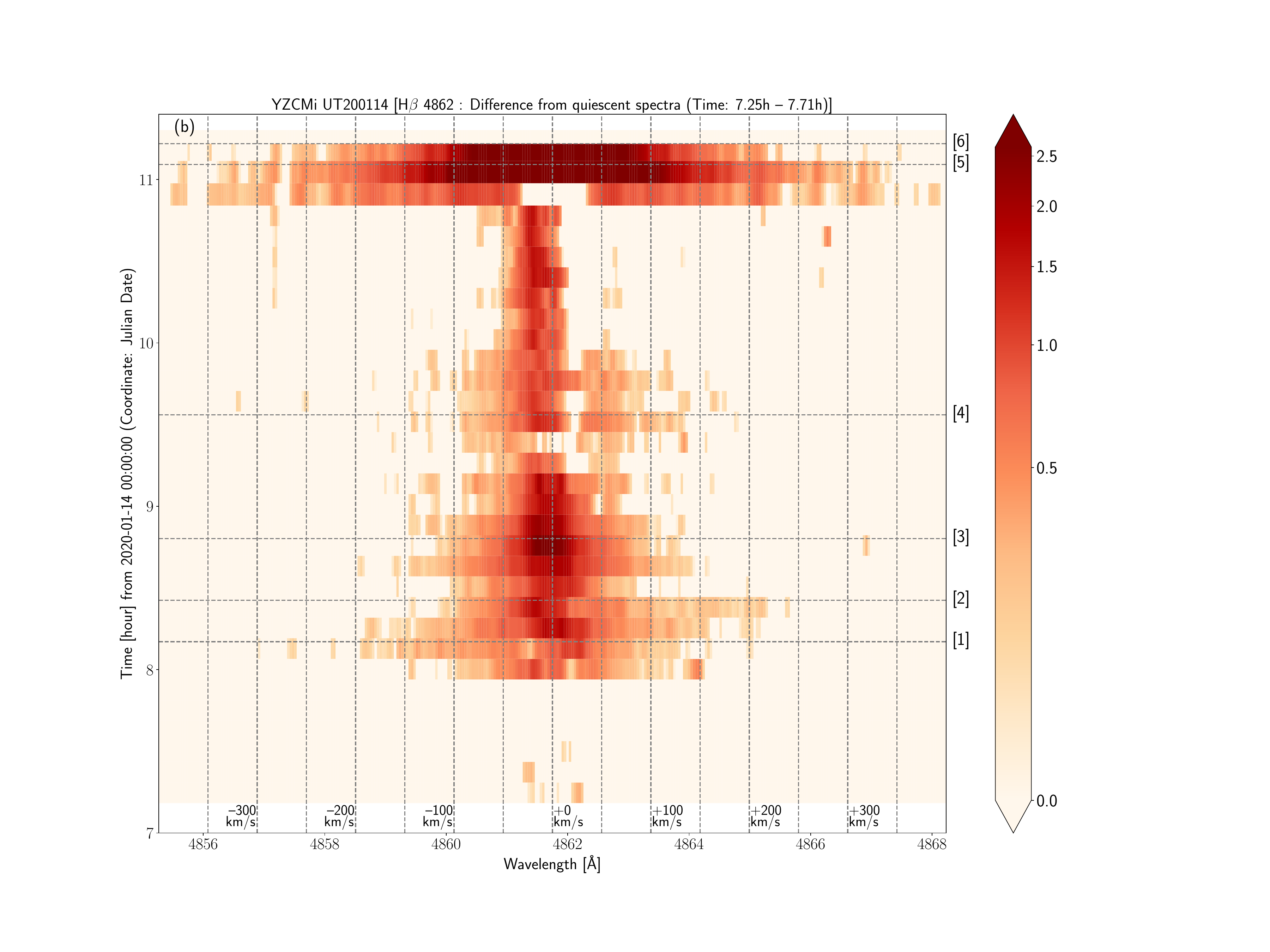}{0.63\textwidth}{\vspace{0mm}}
    }
      \vspace{-1.0cm}
          \gridline{  
     \hspace{-0.07\textwidth}
    \fig{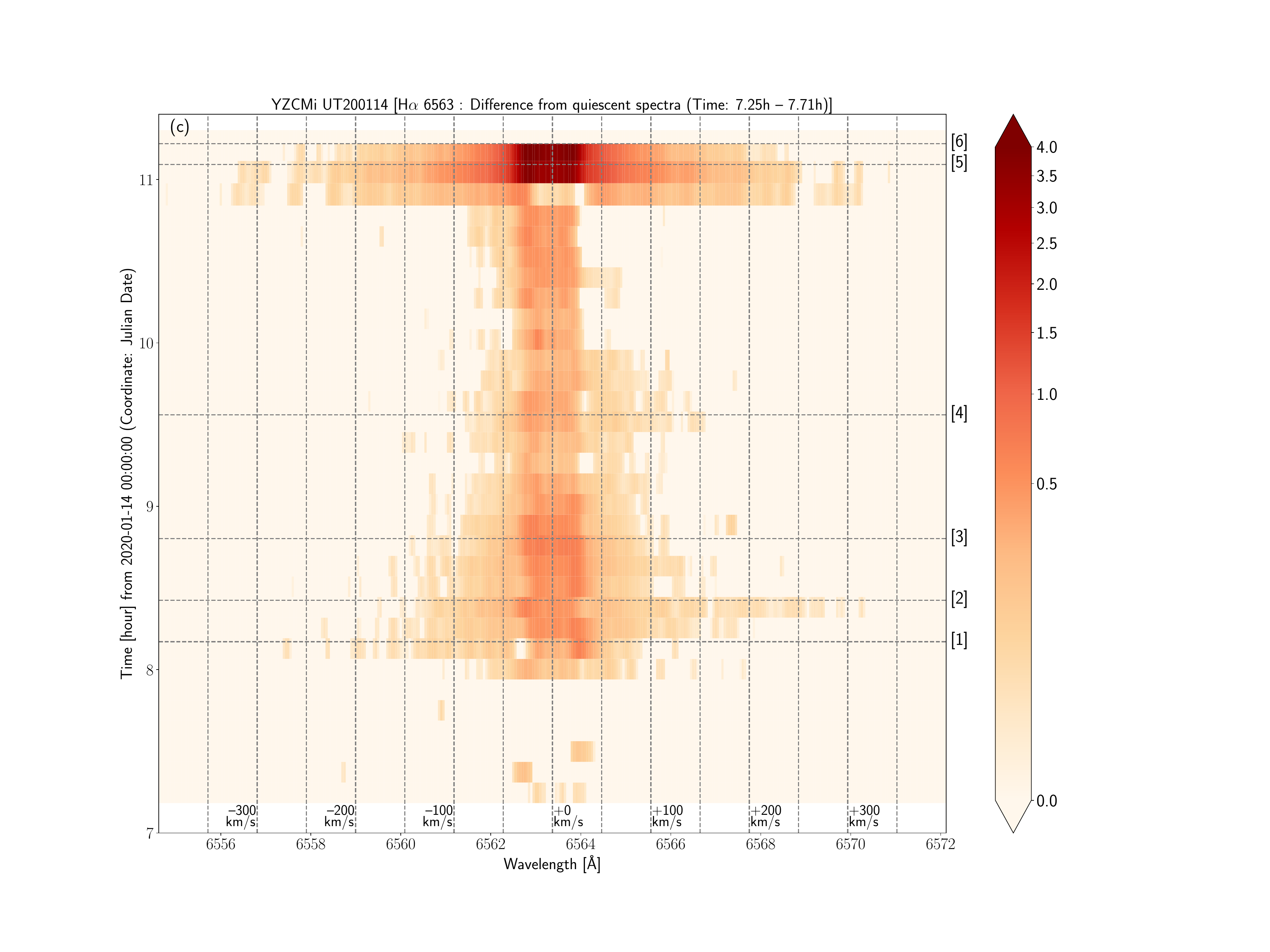}{0.63\textwidth}{\vspace{0mm}}
     \hspace{-0.11\textwidth}
    \fig{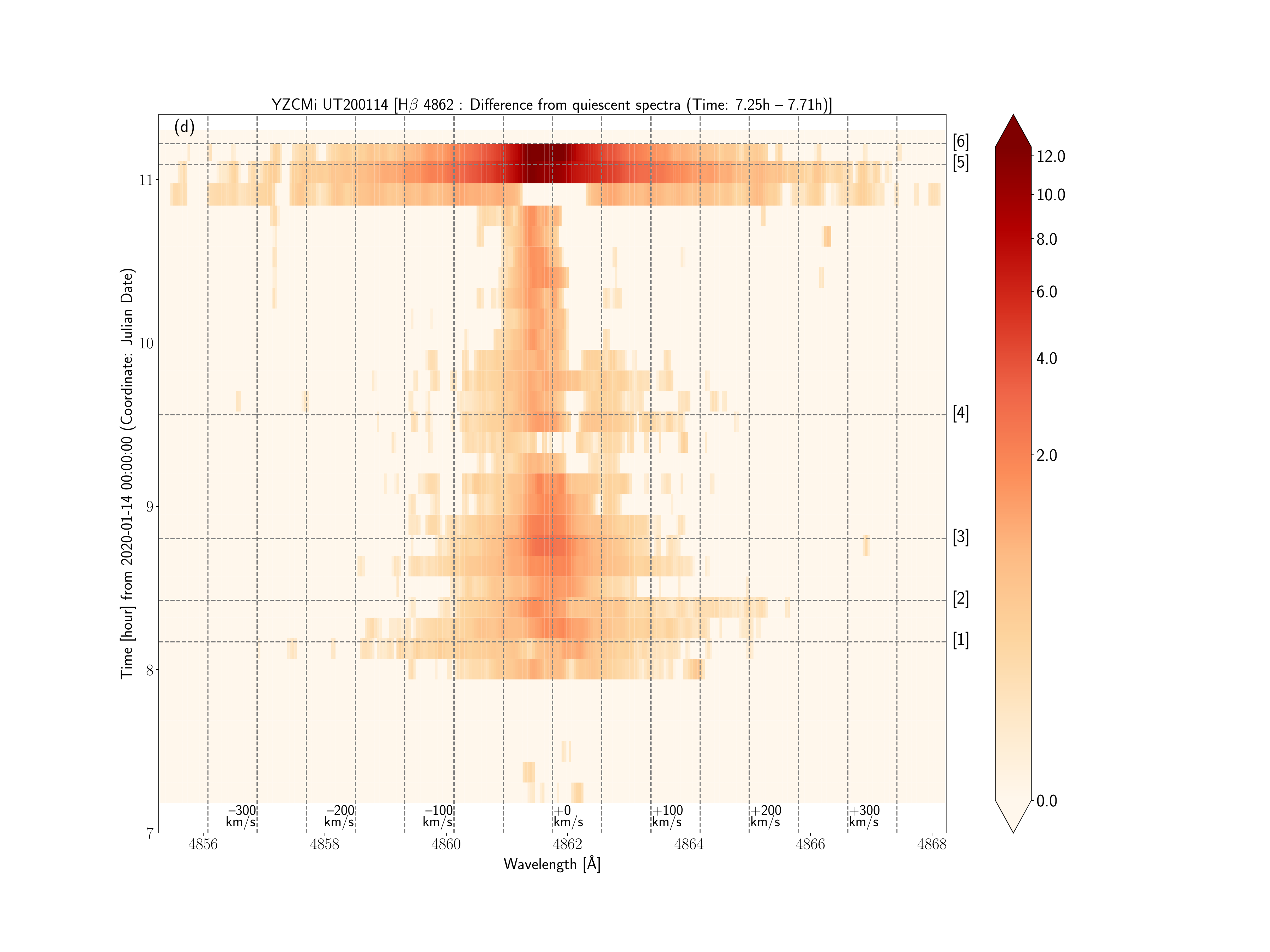}{0.63\textwidth}{\vspace{0mm}}
    }
      \vspace{-1cm}
     \caption{
         \color{black}\textrm{  
(a) \& (b)         
Time evolution of the H$\alpha$ \& H$\beta$ line profiles covering Flares Y7 \& Y8 on 2020 January 14, which are plotted similarly with Figure \ref{fig:map_HaHb_YZCMi_UT191212}.
The grey horizontal dashed lines indicate the time [1] -- [6], which are shown in Figures \ref{fig:lcEW_HaHb_YZCMi_UT200114}(a) \& (c) (light curves) and Figure \ref{fig:spec_HaHb_YZCMi_UT200114} (line profiles).
(c) \& (d)
Same as panels (a) \& (b), but the ranges of the color map contours are different.
}\color{black}
     }
   \label{fig:map_HaHb_YZCMi_UT200114}
   \end{center}
 \end{figure}

The H$\alpha$ \& H$\beta$ line profiles during Flares Y7 and Y8 are shown in
Figures \ref{fig:spec_HaHb_YZCMi_UT200114} \& \ref{fig:map_HaHb_YZCMi_UT200114}. 
During Flare Y7, the blue wing of H$\alpha$ line could be 
slightly enhanced (up to -200km s$^{-1}$) only at around the time [1], while the red wing 
could be slightly enhanced (up to +250--300 km s$^{-1}$) at around the time [2] (Figures \ref{fig:spec_HaHb_YZCMi_UT200114}(b), \& \ref{fig:map_HaHb_YZCMi_UT200114}(a)).
Since, the blue wing enhancement was so small,
we cannot judge that this flare shows clear blue wing asymmetry.  
In the later phase of Flare Y7 (around time [3] and [4]), the wing enhancements of the H$\alpha$ line profile was
weaker while the line center enhancement continued over two hours (Figures \ref{fig:spec_HaHb_YZCMi_UT200114}(f), \& \ref{fig:map_HaHb_YZCMi_UT200114}(a)).
Similar time evolution were seen also in the H$\beta$ line, but the line wing asymmetries at around the time [1] and [2] were unclear compared with those of H$\alpha$ line (Figures \ref{fig:spec_HaHb_YZCMi_UT200114}(d), \& \ref{fig:map_HaHb_YZCMi_UT200114}(b)). 
As for Flare Y8, only the initial phase of the flare was observed but probably the flare peak in H$\alpha$ \& H$\beta$ lines was observed. 
Both H$\alpha$ \& H$\beta$ line profiles show remarkable and symmetric line wing enhancements ($\pm$250--300 km s$^{-1}$ for H$\alpha$ line and $\pm$300--350 km s$^{-1}$ for H$\beta$ line) (Figures \ref{fig:spec_HaHb_YZCMi_UT200114}(j),(l) \& \ref{fig:map_HaHb_YZCMi_UT200114}). There was the continuum intensity enhancement during the flare, but the peak of the continuum intensity could be a few minutes earlier than the peak of H$\alpha$ \& H$\beta$ line equivalent widths (Figures \ref{fig:lcEW_HaHb_YZCMi_UT200114} (a) \& (b)). 
 \color{black}\textrm{ 
We also note that as for the H$\alpha$ and H$\beta$ lines, the larger enhancements in the line wings contributed to a bigger total equivalent widths at [5] than [6], while the peak intensities at the line centers are smaller at [5] than at [6] (Figures \ref{fig:lcEW_HaHb_YZCMi_UT200114} \&
\ref{fig:spec_HaHb_YZCMi_UT200114}).
} \color{black}

\clearpage

 \subsection{Flares Y9, Y10, \& Y11 observed on 2020 January 16} \label{subsec:results:2020-Jan-16} 

\begin{figure}[ht!]
   \begin{center}
   \gridline{
    \fig{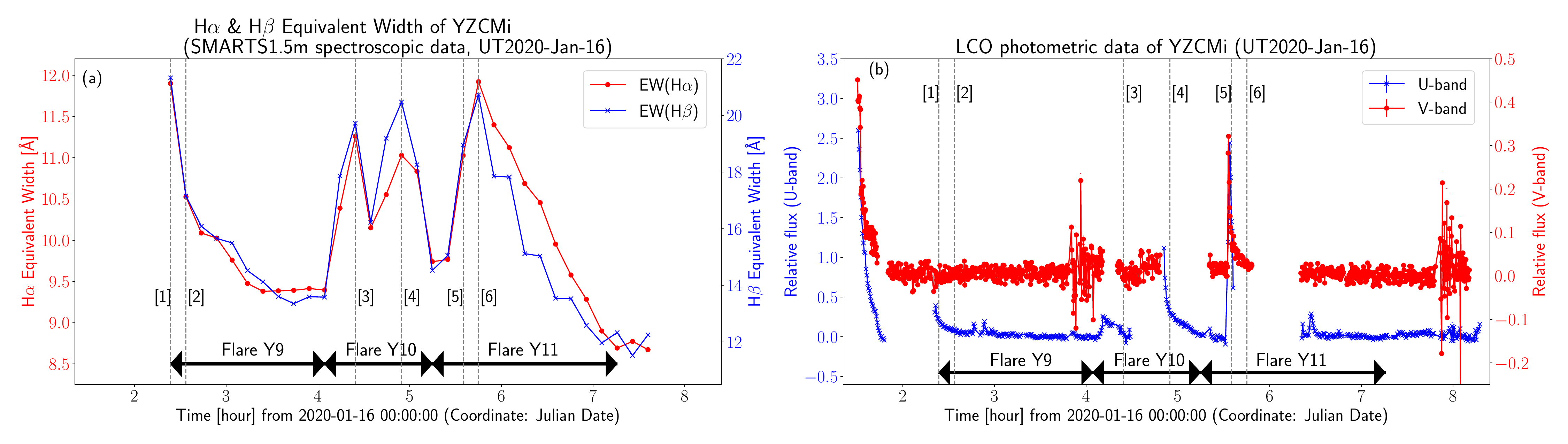}{1.0\textwidth}{\vspace{0mm}}}   
     \vspace{-5mm}
     \caption{
   \color{black}\textrm{  
Light curves of YZ CMi on 2020 January 16 showing Flares Y9, Y10, \& Y11, which are plotted 
similarly with Figures \ref{fig:lcEW_HaHb_YZCMi_UT200121} (a)\&(b).
 The grey dashed lines with numbers ([1]--[6]) correspond to the time shown with the same numbers in Figures \ref{fig:spec_HaHb_YZCMi_UT200116} \& \ref{fig:map_HaHb_YZCMi_UT200116}.
  } \color{black}
     }
   \label{fig:lcEW_HaHb_YZCMi_UT200116}
   \end{center}
   
 \end{figure}

On 2020 January 16, three flares (Flares Y9, Y10, \& Y11) were detected in H$\alpha$ \& H$\beta$ lines as shown in Figure \ref{fig:lcEW_HaHb_YZCMi_UT200116} (a).  
Flare Y9 already started when the observation started.
The H$\alpha$ \& H$\beta$ equivalent widths decreased from 11.9\AA~and 21.3\AA, respectively, and the flare duration in H$\alpha$ $\Delta t^{\rm{flare}}_{\rm{H}\alpha}$ is $>$1.7 hours (Table \ref{table:list1_flares}). 
In addition to the enhancements in Balmer emission lines, the continuum brightness observed with LCO $U$- \& $V$-bands increased at least by $\sim$250\% and $\sim$45\%, respectively, associated with Flare Y9 before the H$\alpha$ \& H$\beta$ observation started (Figure \ref{fig:lcEW_HaHb_YZCMi_UT200116} (b)). 
We note the continuum brightness increase already started even before the LCO observation started, \color{black}\textrm{and} \color{black} the amplitude values described here ($\sim$250\% and $\sim$45\%) can be only lower limit values.
Flare Y10 occurred soon after Flare Y9.
The H$\alpha$ \& H$\beta$ equivalent widths increased up to 11.3\AA~and 20.5\AA, respectively, and $\Delta t^{\rm{flare}}_{\rm{H}\alpha}$ is 1.2 hours (Table \ref{table:list1_flares}).
In addition to these enhancements in Balmer emission lines, the continuum brightness observed with LCO $U$-band increased at least by $\sim$100\% during Flare Y10 (Figure \ref{fig:lcEW_HaHb_YZCMi_UT200116} (b)). Since LCO photometric data have some gaps during the flare, it is difficult to measure the brightness increase amplitude in $V$-band data, 
and the amplitude value in $U$-band described here can be also only the lower limit value.
As for Flare Y11, 
the H$\alpha$ \& H$\beta$ equivalent widths increased up to 11.9\AA~and 20.8\AA, respectively, and the flare duration in H$\alpha$ $\Delta t^{\rm{flare}}_{\rm{H}\alpha}$ is 2.0 hours (Table \ref{table:list1_flares}). 
In addition to these enhancements in Balmer emission lines, the continuum brightness observed with LCO $U$- \& $V$-bands increased at least by $\sim$200--250\% and $\sim$30--35\%, respectively, during Flare Y11 (Figure \ref{fig:lcEW_HaHb_YZCMi_UT200116}).

          \begin{figure}[ht!]
   \begin{center}
            \gridline{  
     \hspace{-0.06\textwidth}
    \fig{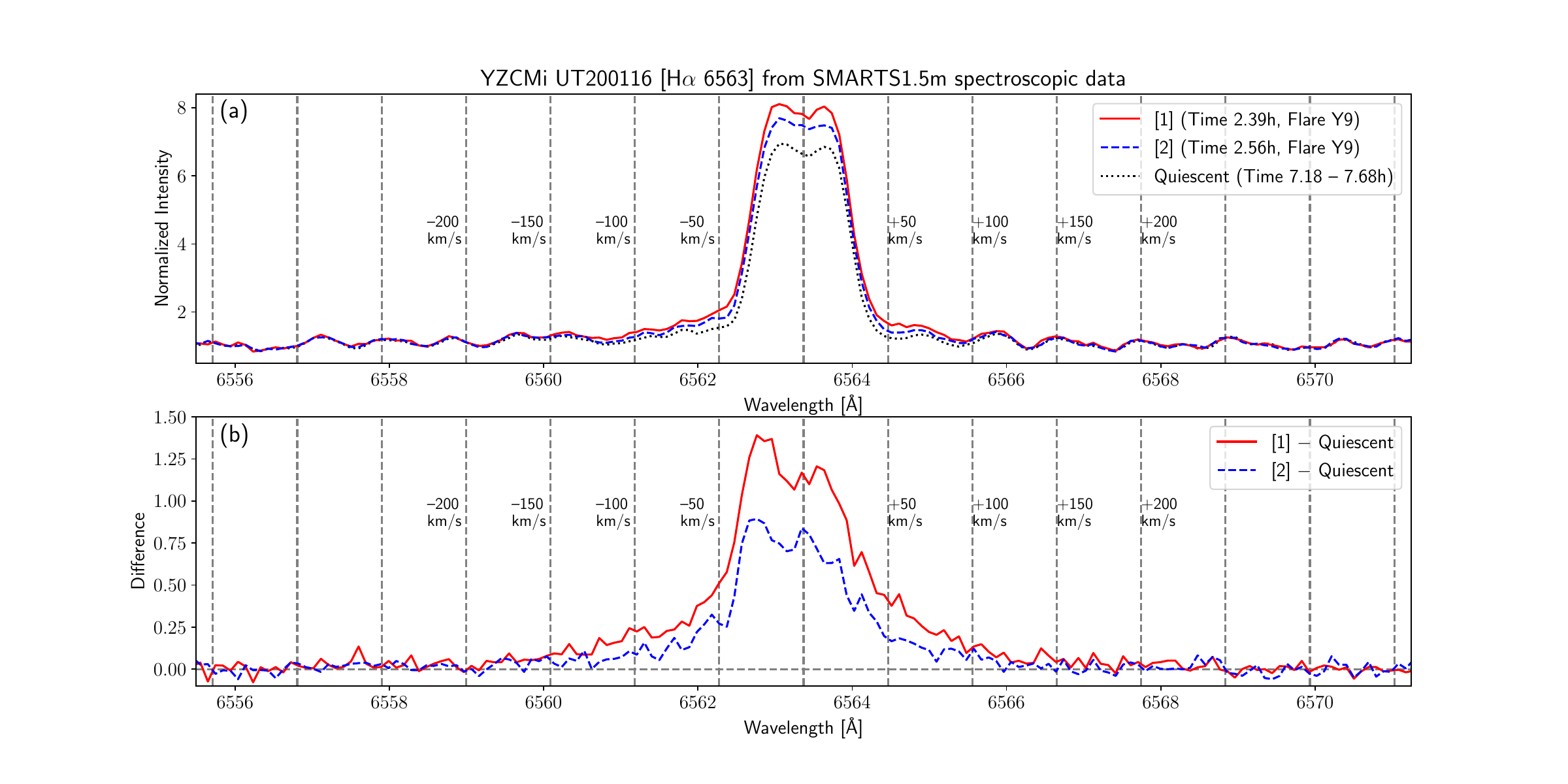}{0.58\textwidth}{\vspace{0mm}}
     \hspace{-0.06\textwidth}
       \fig{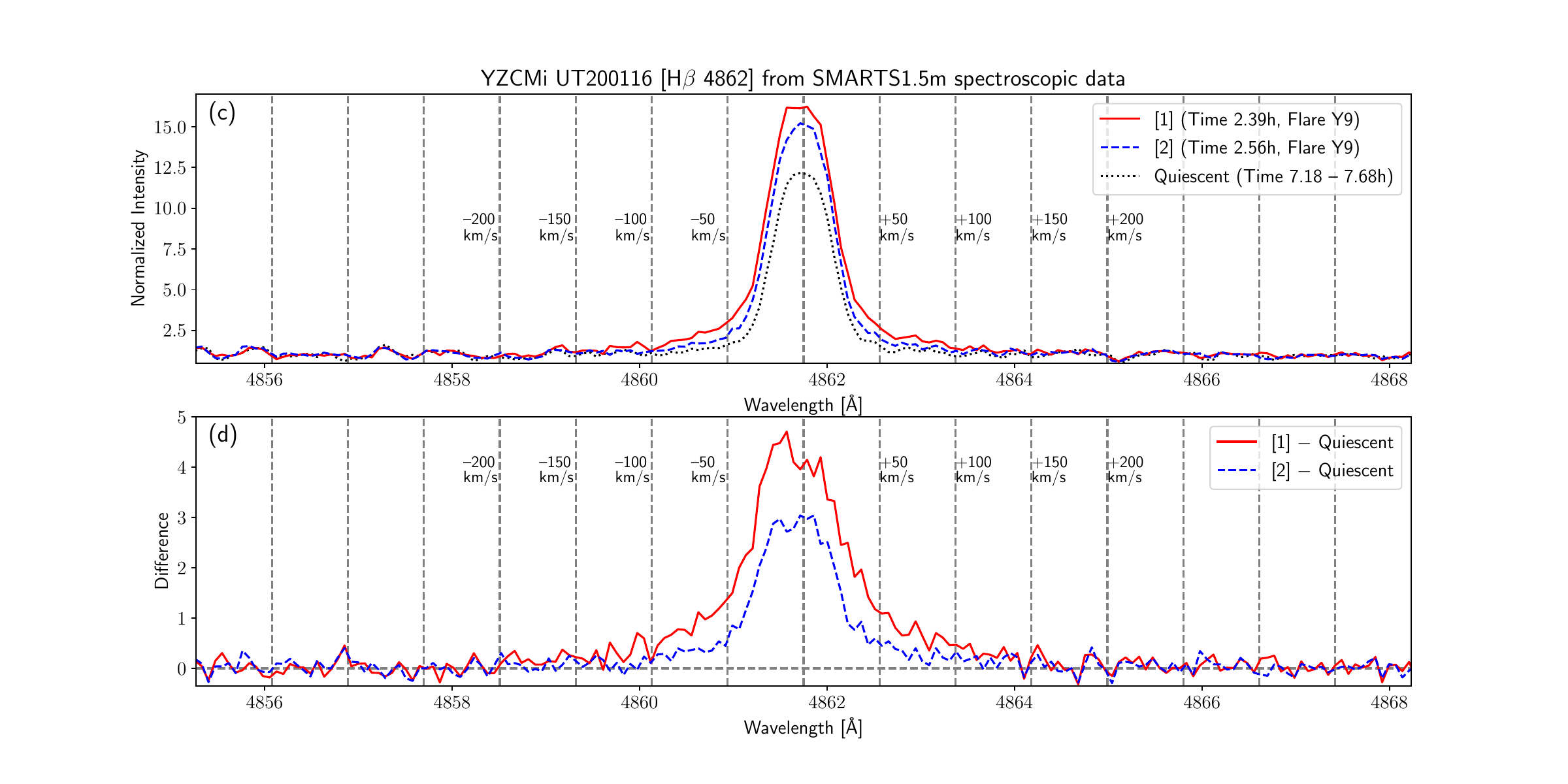}{0.58\textwidth}{\vspace{0mm}}
    }
     \vspace{-1.0cm}
            \gridline{  
     \hspace{-0.06\textwidth}
    \fig{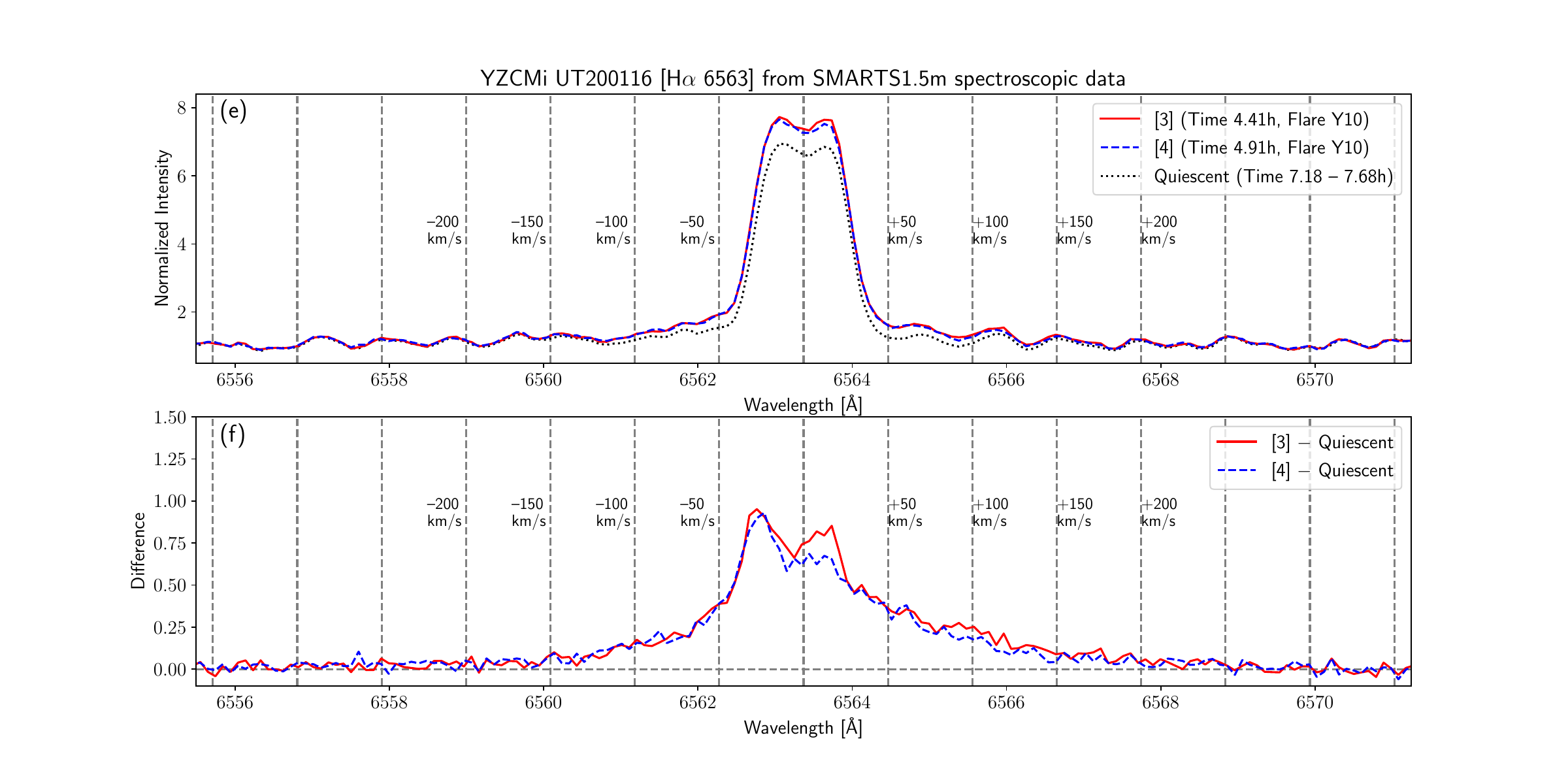}{0.58\textwidth}{\vspace{0mm}}
     \hspace{-0.06\textwidth}
       \fig{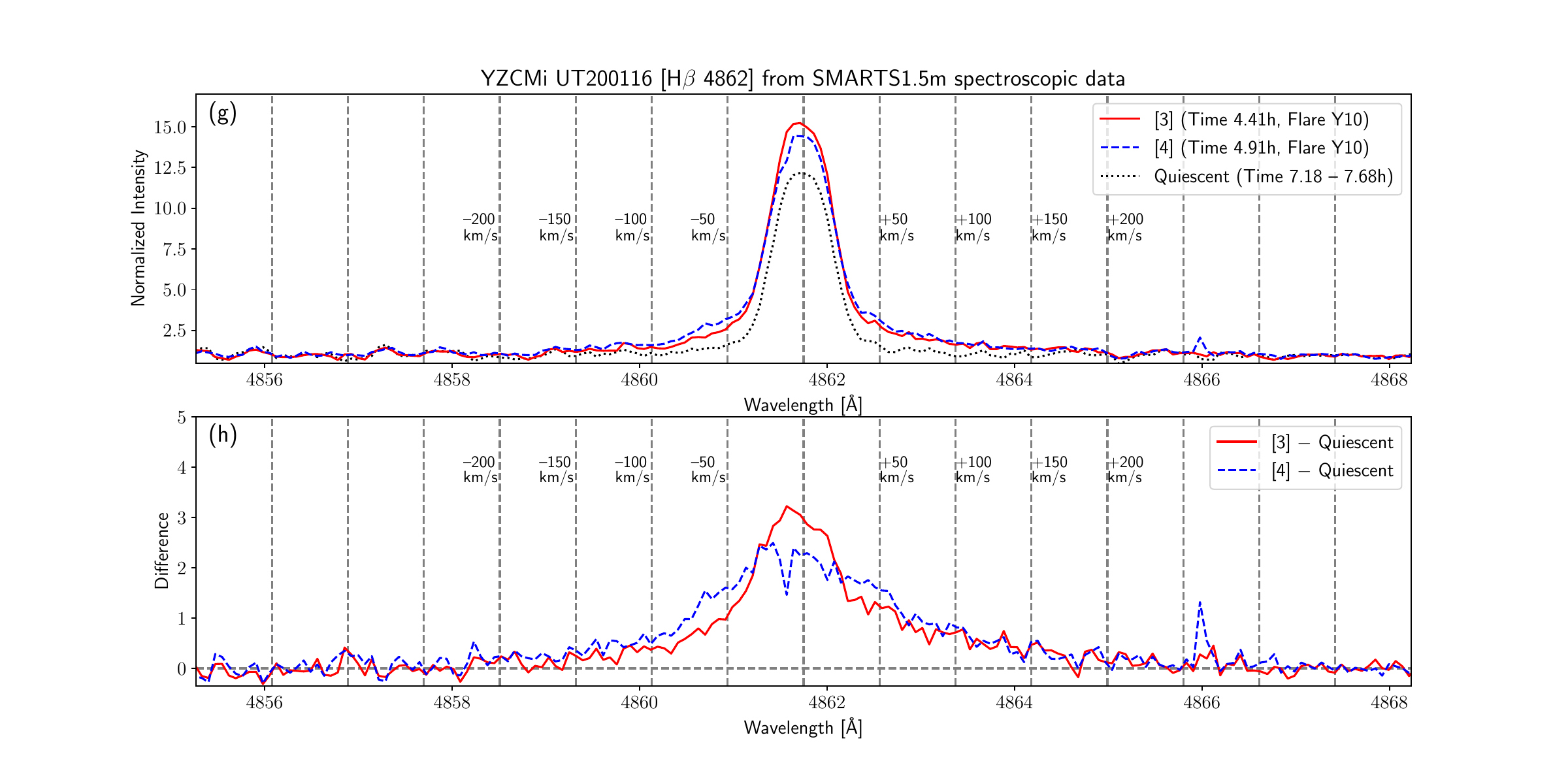}{0.58\textwidth}{\vspace{0mm}}
    }
     \vspace{-1.0cm}
            \gridline{  
     \hspace{-0.06\textwidth}
    \fig{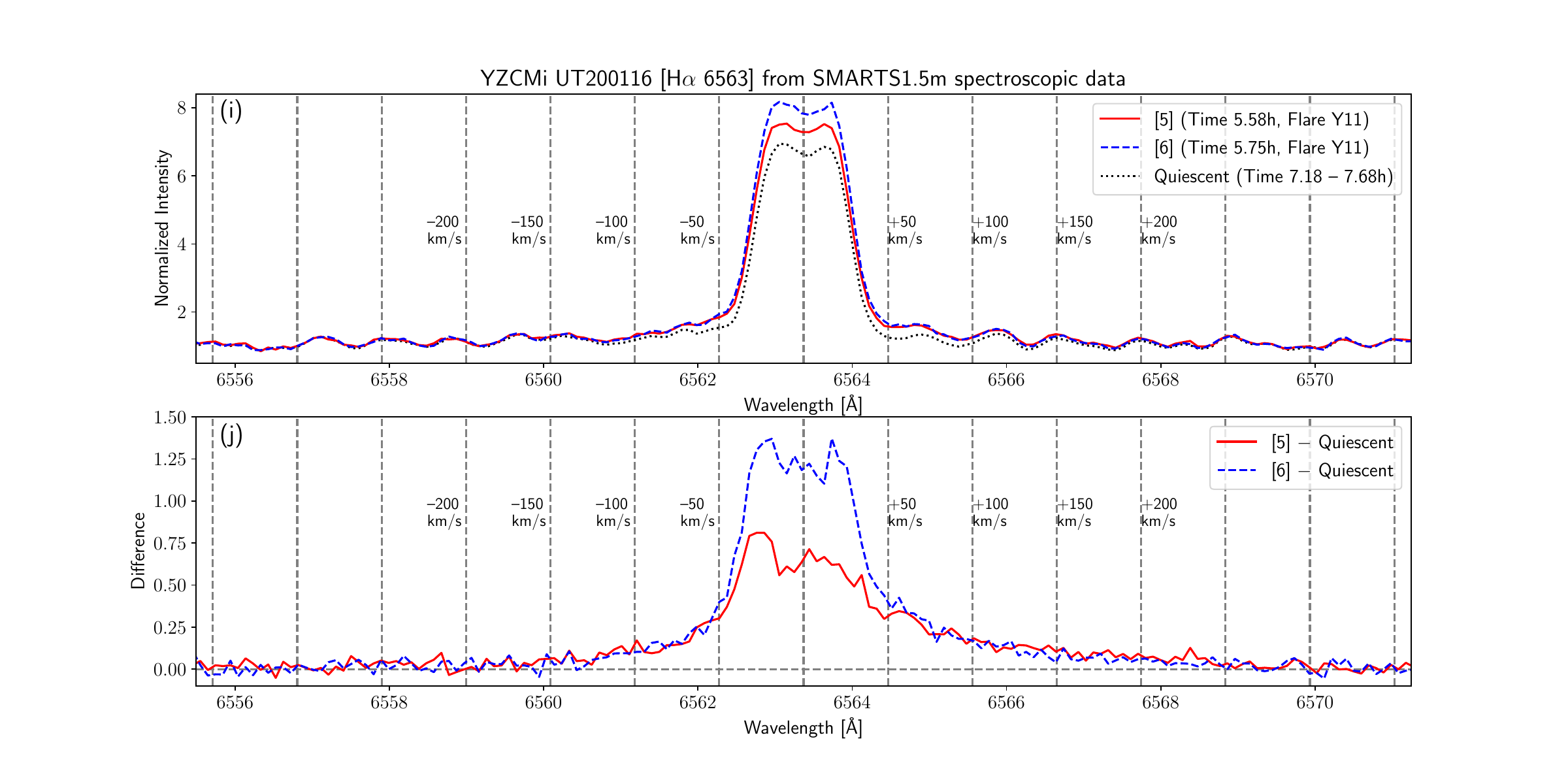}{0.58\textwidth}{\vspace{0mm}}
     \hspace{-0.06\textwidth}
       \fig{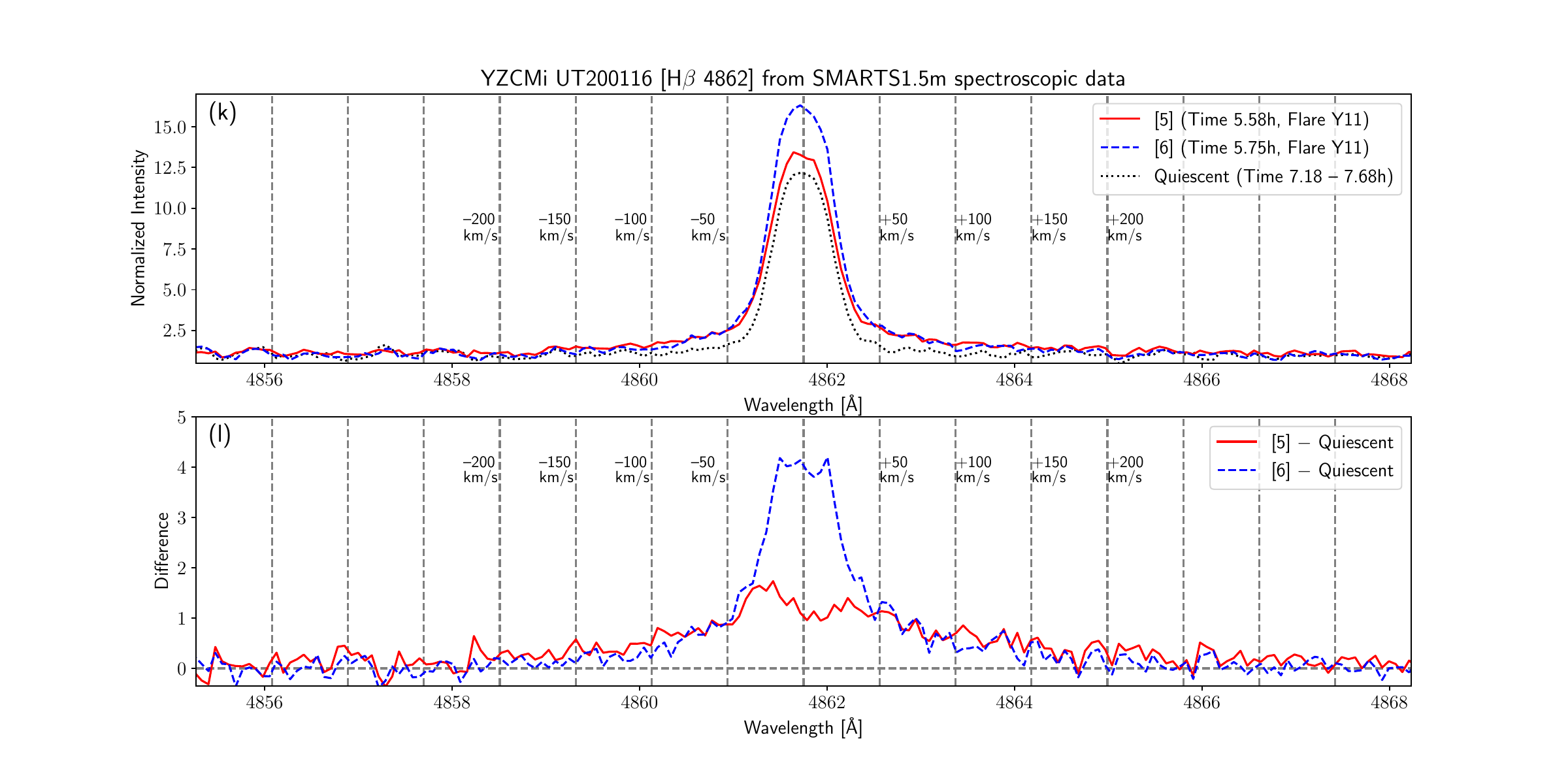}{0.58\textwidth}{\vspace{0mm}}
    }
     \vspace{-0.5cm}
     \caption{
  \color{black}\textrm{  
Line profiles of the H$\alpha$ \& H$\beta$ emission lines during Flares Y9, Y10, \& Y11 on 2020 January 16 (at the time [1]-[6]) from SMARTS 1.5m spectroscopic data, which are plotted similarly with Figure \ref{fig:spec_HaHb_YZCMi_UT190127}.
 } \color{black}
     }
   \label{fig:spec_HaHb_YZCMi_UT200116}
   \end{center}
 \end{figure}

 \color{black}\textrm{ 
$L_{U}$, $L_{V}$, $E_{U}$, $E_{V}$, $L_{\rm{H}\alpha}$, $L_{\rm{H}\beta}$, $E_{\rm{H}\alpha}$, and $E_{\rm{H}\beta}$ values are estimated and listed in Table \ref{table:list1_flares}.
We note that the $L_{U}$, $L_{V}$, $E_{U}$, and $E_{V}$ values of the three flares (Flares Y9, Y10, \& Y11) described here can be only the lower limit values, since Flare Y9 already started before the observation started, and LCO photometric observations have some gaps during all the three flares (We do not calculate $L_{V}$ and $E_{V}$ values because of the large gaps during Flare Y10 in $V$-band). 
It is noted that the main peaks of white-light emissions corresponding to the H$\alpha$ and H$\beta$ emissions \color{black}\textrm{are covered }\color{black} in $U$-band observations without any effects from the gaps.
In addition, 
$L_{\rm{H}\alpha}$, $L_{\rm{H}\beta}$, $E_{\rm{H}\alpha}$, and $E_{\rm{H}\beta}$ values of Flare Y9 are only the lower limit values,
since the flare already started when the observation started, and it can be possible the flare peak time in H$\alpha$ \& H$\beta$ lines was before the observation started.
} \color{black}

\begin{figure}[ht!]
   \begin{center}
           \gridline{  
     \hspace{-0.07\textwidth}
    \fig{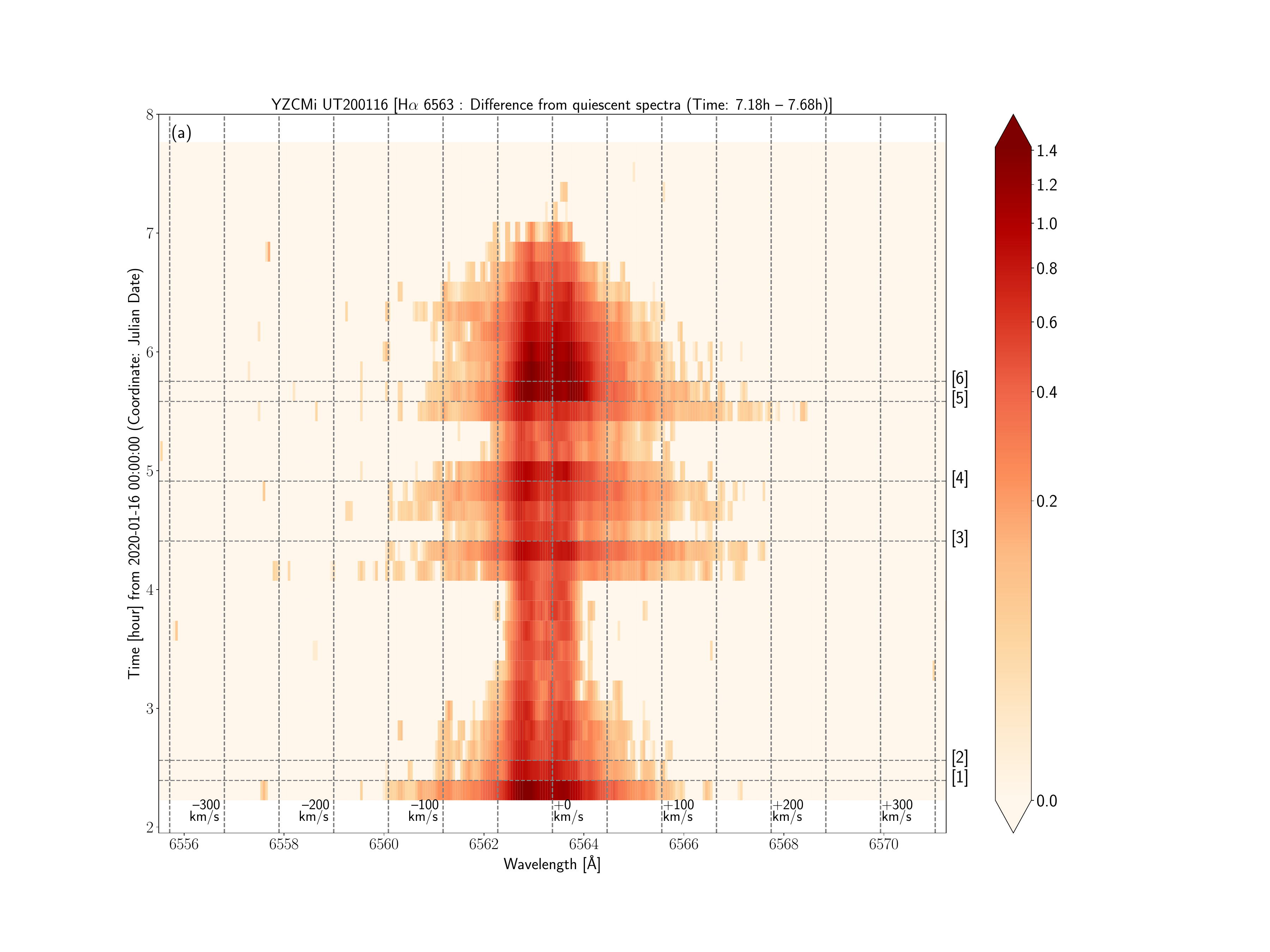}{0.63\textwidth}{\vspace{0mm}}
     \hspace{-0.11\textwidth}
    \fig{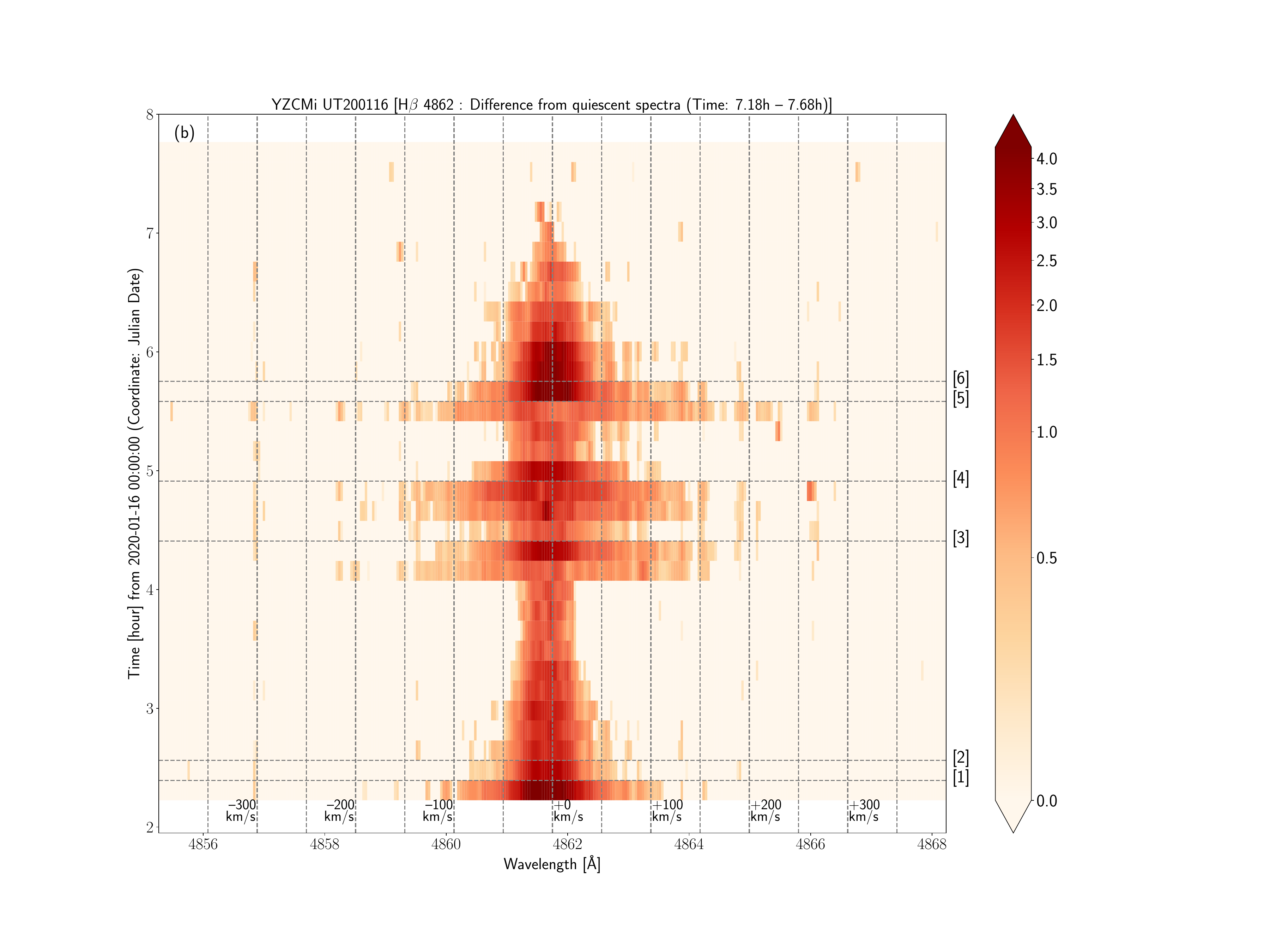}{0.63\textwidth}{\vspace{0mm}}
    }
     \vspace{-0.5cm}
     \caption{
            \color{black}\textrm{  
Time evolution of the H$\alpha$ \& H$\beta$ line profiles covering Flares Y9, Y10, \& Y11 on 2020 January 16, which are plotted similarly with Figure \ref{fig:map_HaHb_YZCMi_UT191212}.
The grey horizontal dashed lines indicate the time [1] -- [6], which are shown in Figure \ref{fig:lcEW_HaHb_YZCMi_UT200116} (light curves) and Figure \ref{fig:spec_HaHb_YZCMi_UT200116} (line profiles).
}\color{black}
     }
   \label{fig:map_HaHb_YZCMi_UT200116}
   \end{center}
 \end{figure}

The H$\alpha$ \& H$\beta$ line profiles during Flares Y9, Y10, and Y11 are shown in
Figures \ref{fig:spec_HaHb_YZCMi_UT200116} \& \ref{fig:map_HaHb_YZCMi_UT200116}. 
During Flare Y9, there are no clear line \color{black}\textrm{wing }\color{black} asymmetries in H$\alpha$ \& H$\beta$ lines,
while there are slight blue part enhancements at -20 -- -25 km s$^{-1}$) from the line center of H$\alpha$ \& H$\beta$ lines (the time [1]\&[2] in Figures \ref{fig:spec_HaHb_YZCMi_UT200116} \& \ref{fig:map_HaHb_YZCMi_UT200116}). 
Around the peak time of Flare Y10 and Flare Y11 (the time [3]--[6] in Figures \ref{fig:spec_HaHb_YZCMi_UT200116} \& \ref{fig:map_HaHb_YZCMi_UT200116}), 
we can see the red wing enhancements up to $\sim +200$ km s$^{-1}$ in H$\alpha$ line and those up to +150 -- +200 km s$^{-1}$ in H$\beta$ line.
The red wing asymmetry in H$\alpha$ line can be seen during most of the decay phase in the case of Flare Y11. 
The peak time of red wing asymmetries roughly correspond to the flare peak time in continuum brightness, comparing Figures \ref{fig:lcEW_HaHb_YZCMi_UT200116} and \ref{fig:map_HaHb_YZCMi_UT200116}.

\clearpage

\subsection{Flares Y12 \& Y13 observed on 2020 January 18} 
\label{subsec:results:2020-Jan-18} 

        \begin{figure}[ht!]
   \begin{center}
   \gridline{
    \fig{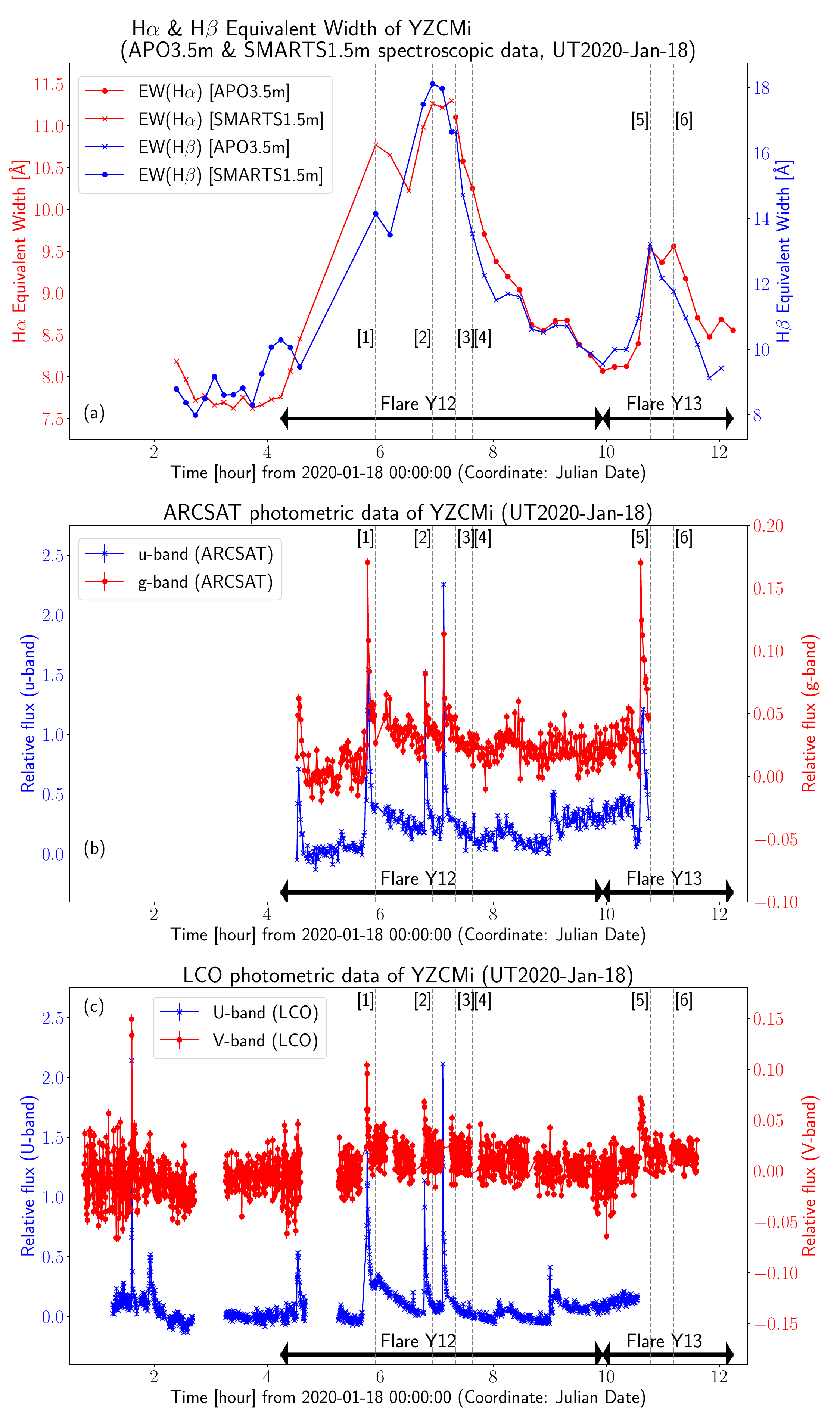}{0.5\textwidth}{\vspace{0mm}}}   
     \vspace{-5mm}
     \caption{
Light curves of YZ CMi on 2020 January 18 showing Flares Y12 \& Y13, which are plotted 
similarly with \color{black}\textrm{Figures \ref{fig:lcEW_HaHb_YZCMi_UT191212} (a)\&(b)}\color{black}.
In panel (a), Red circles and blue asterisks correspond to H$\alpha$ \& H$\beta$ EWs from APO 3.5m data, respectively, while red asterisks and blue circles correspond to H$\alpha$ \& H$\beta$ EWs from SMARTS 1.5m data, respectively. 
The ARCSAT $u$- \& $g$-band photometric data are plotted in \color{black}\textrm{(b)} \color{black}, 
while LCO $U$- \& $V$-band photometric data are plotted in \color{black}\textrm{(c)} \color{black}.
The grey dashed lines with numbers ([1]--[6]) correspond to the time shown with the same numbers in Figures \ref{fig:spec_HaHb_YZCMi_UT200118} \& \ref{fig:map_HaHb_YZCMi_UT200118}.
     }
   \label{fig:lcEW_HaHb_YZCMi_UT200118}
   \end{center}
 \end{figure}

On 2020 January 18, two flares (Flares Y12 \& Y13) were detected in H$\alpha$ \& H$\beta$ lines as shown in Figures \ref{fig:lcEW_HaHb_YZCMi_UT200118} (a) \& (c).  
As for Flare Y12, the H$\alpha$ \& H$\beta$ equivalent widths increased up to 11.3\AA~and 18.1\AA, respectively, and $\Delta t^{\rm{flare}}_{\rm{H}\alpha}$ is 5.7 hours (Table \ref{table:list1_flares}).
In addition to these enhancements in Balmer emission lines, the continuum brightness observed with ARCSAT $u$- \& $g$-band and LCO $U$- \& $V$-band increased at least by
$\sim$ 220 -- 230\%, $\sim$17\%, $\sim$ 210 -- 220\%, and $\sim$ 10 -- 11\%, 
respectively, during Flare Y12 (Figures \ref{fig:lcEW_HaHb_YZCMi_UT200118} (b) \& (d)). 
Since LCO photometric data have some observation gaps during the flare, 
the amplitude value in $U$- \& $V$-band described here can be only the lower limit values.
As for Flare Y13, the H$\alpha$ \& H$\beta$ equivalent widths increased up to 9.6\AA~and 13.2\AA, respectively, and $\Delta t^{\rm{flare}}_{\rm{H}\alpha}$ is 2.3 hours (Table \ref{table:list1_flares}).
In addition to these enhancements in Balmer emission lines, the continuum brightness observed with ARCSAT $u$- \& $g$-band and LCO $V$-band increased at least by
$\sim$120\%, $\sim$17\%, and $\sim$7--8\%, 
respectively, during Flare Y13 (Figures \ref{fig:lcEW_HaHb_YZCMi_UT200118} (b) \& (d)). 
There are no LCO $U$-band observation over the most phases of Flare Y13.
We also note that since ARCSAT photometric observation ended before Flare Y13 in H$\alpha$ \& H$\beta$ lines ended, some additional brightness changes in $u$- \& $g$-band might exist.

 \color{black}\textrm{ 
$L_{U}$, $L_{u}$, $L_{g}$, $L_{V}$, $E_{U}$, $E_{u}$, $E_{g}$, $E_{V}$, $L_{\rm{H}\alpha}$, $L_{\rm{H}\beta}$, $E_{\rm{H}\alpha}$, and $E_{\rm{H}\beta}$ values are estimated and listed in Table \ref{table:list1_flares}.
We note that the $L_{U}$, $L_{V}$, $E_{U}$, and $E_{V}$ values
can be only the lower limit values since the LCO observation has gaps during Flare Y12.
We note that since ARCSAT photometric observation ended before Flare Y13 (in Balmer lines) ended, the $L_{u}$, $L_{g}$, $E_{u}$, and $E_{g}$ values
can be only the lower limit values.
} \color{black}
 
           \begin{figure}[ht!]
   \begin{center}
            \gridline{  
     \hspace{-0.06\textwidth}
    \fig{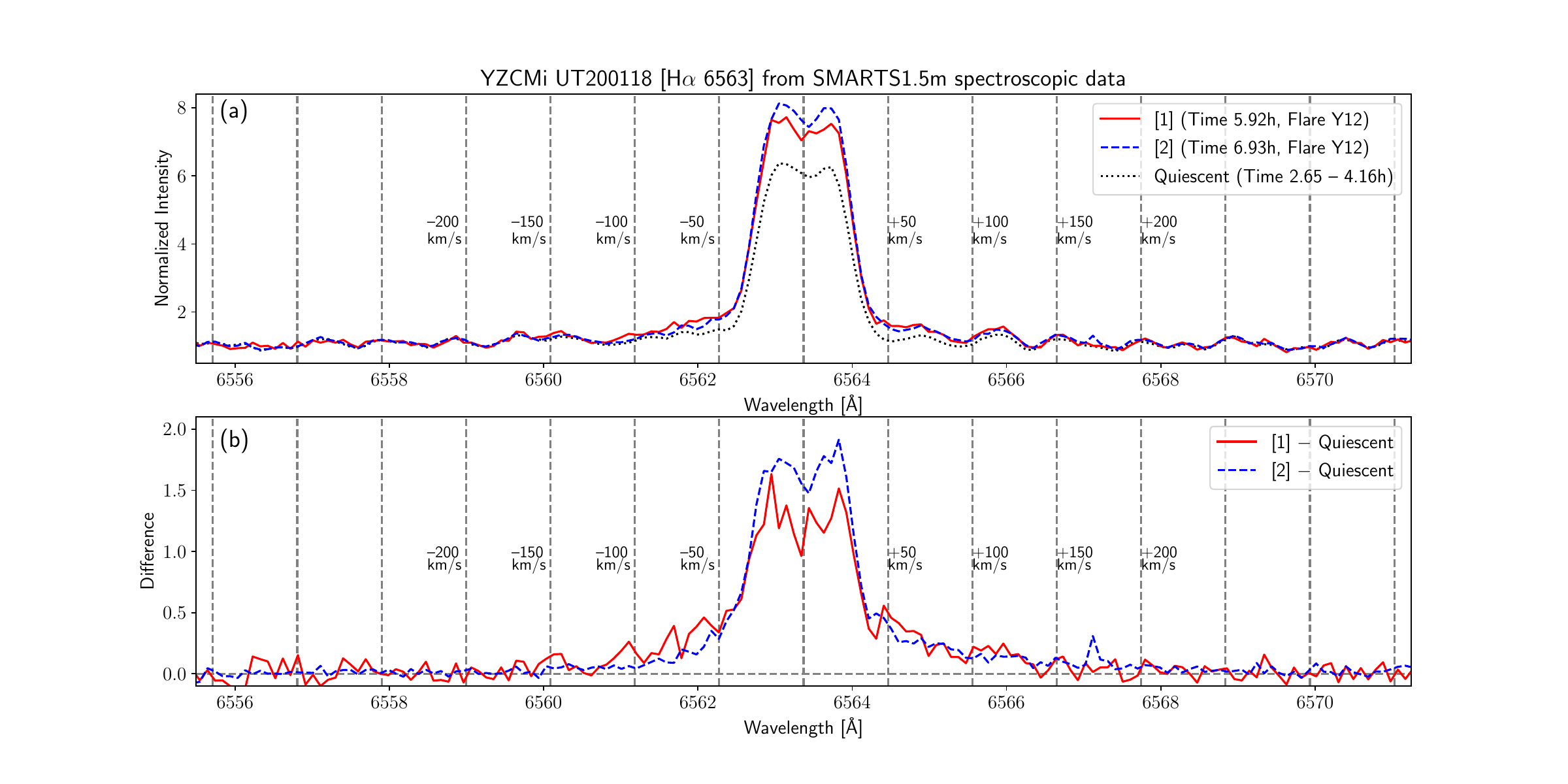}{0.58\textwidth}{\vspace{0mm}}
     \hspace{-0.06\textwidth}
       \fig{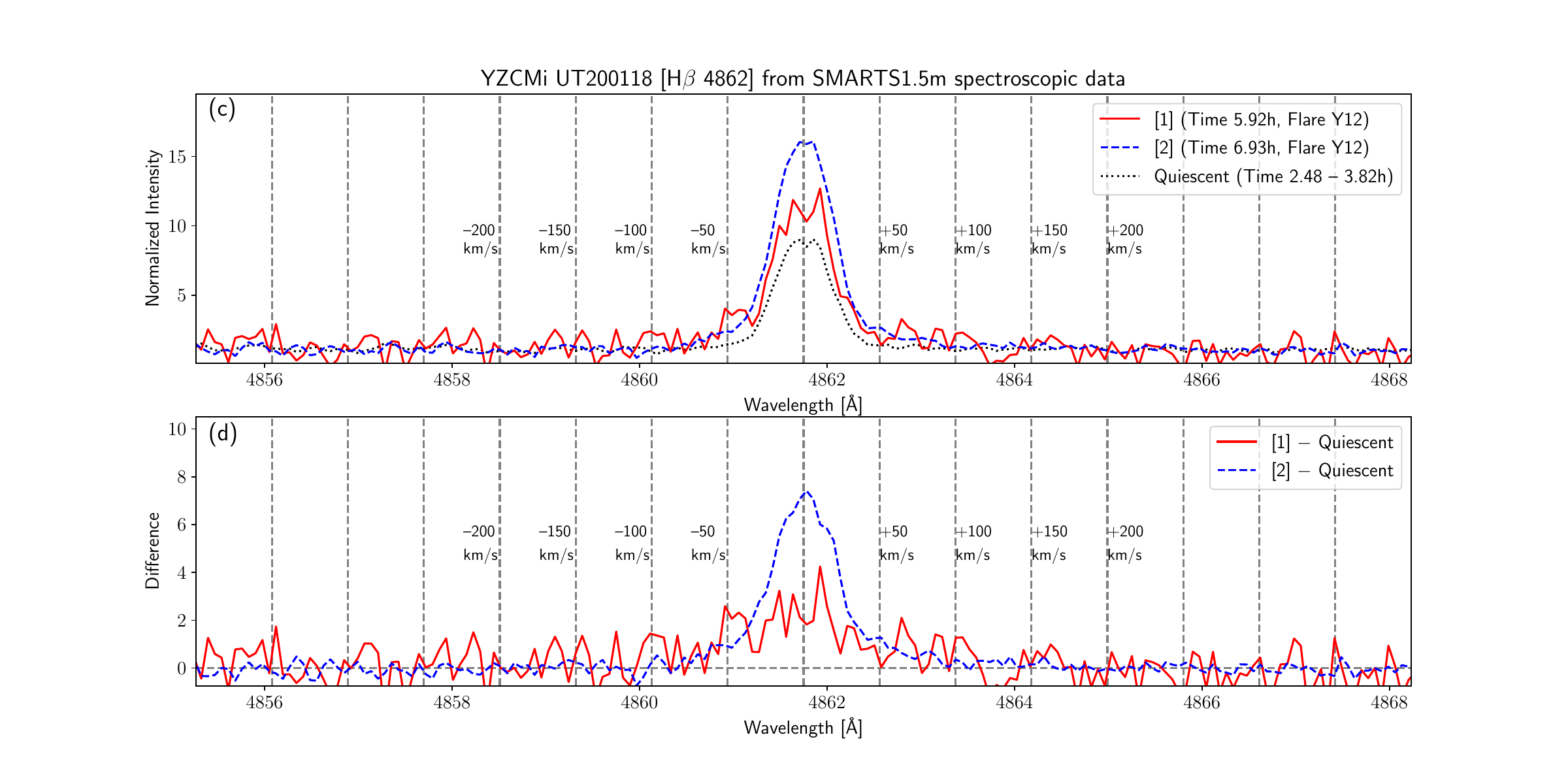}{0.58\textwidth}{\vspace{0mm}}
    }
     \vspace{-1.0cm}
            \gridline{  
     \hspace{-0.06\textwidth}
    \fig{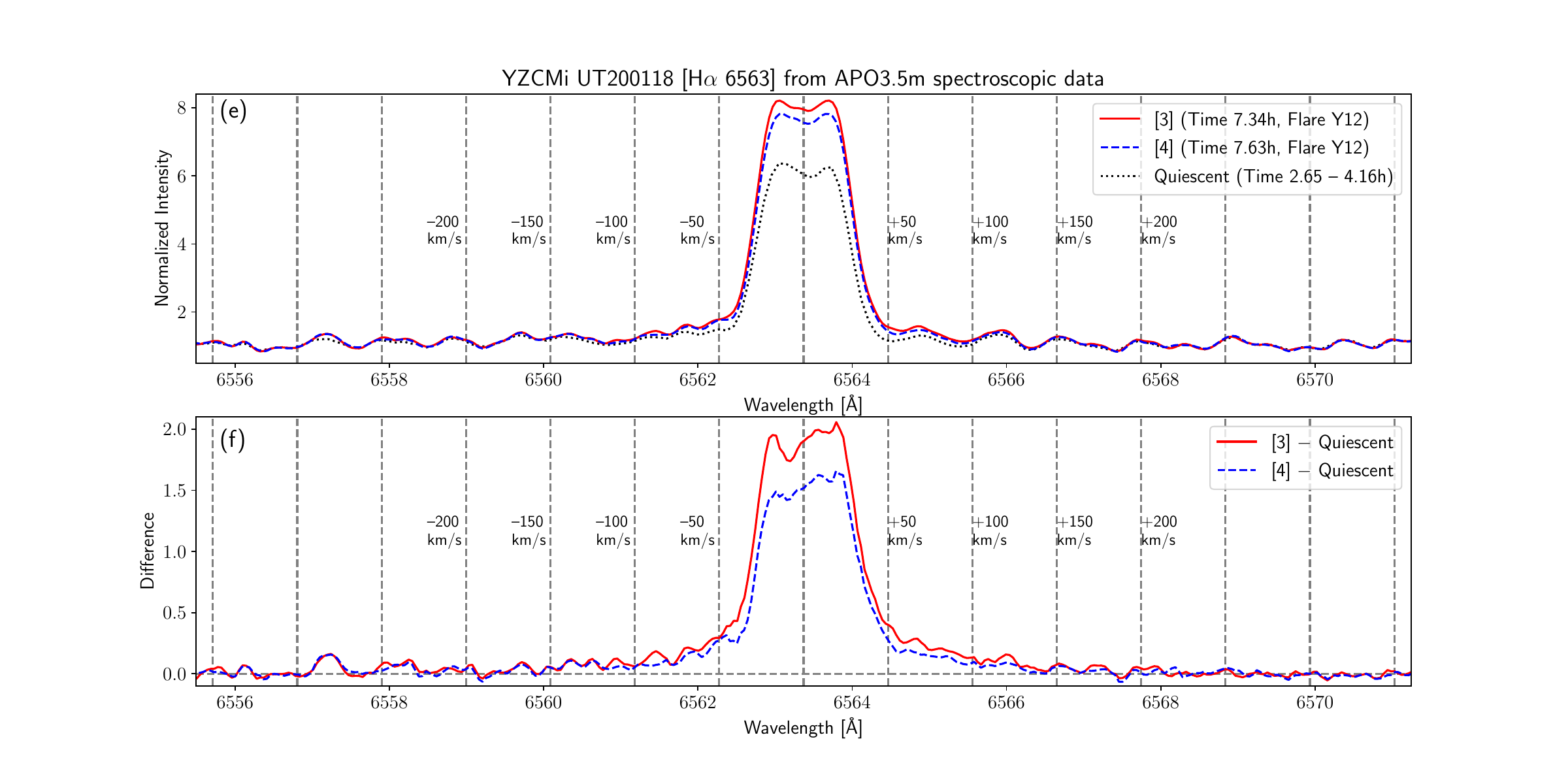}{0.58\textwidth}{\vspace{0mm}}
     \hspace{-0.06\textwidth}
       \fig{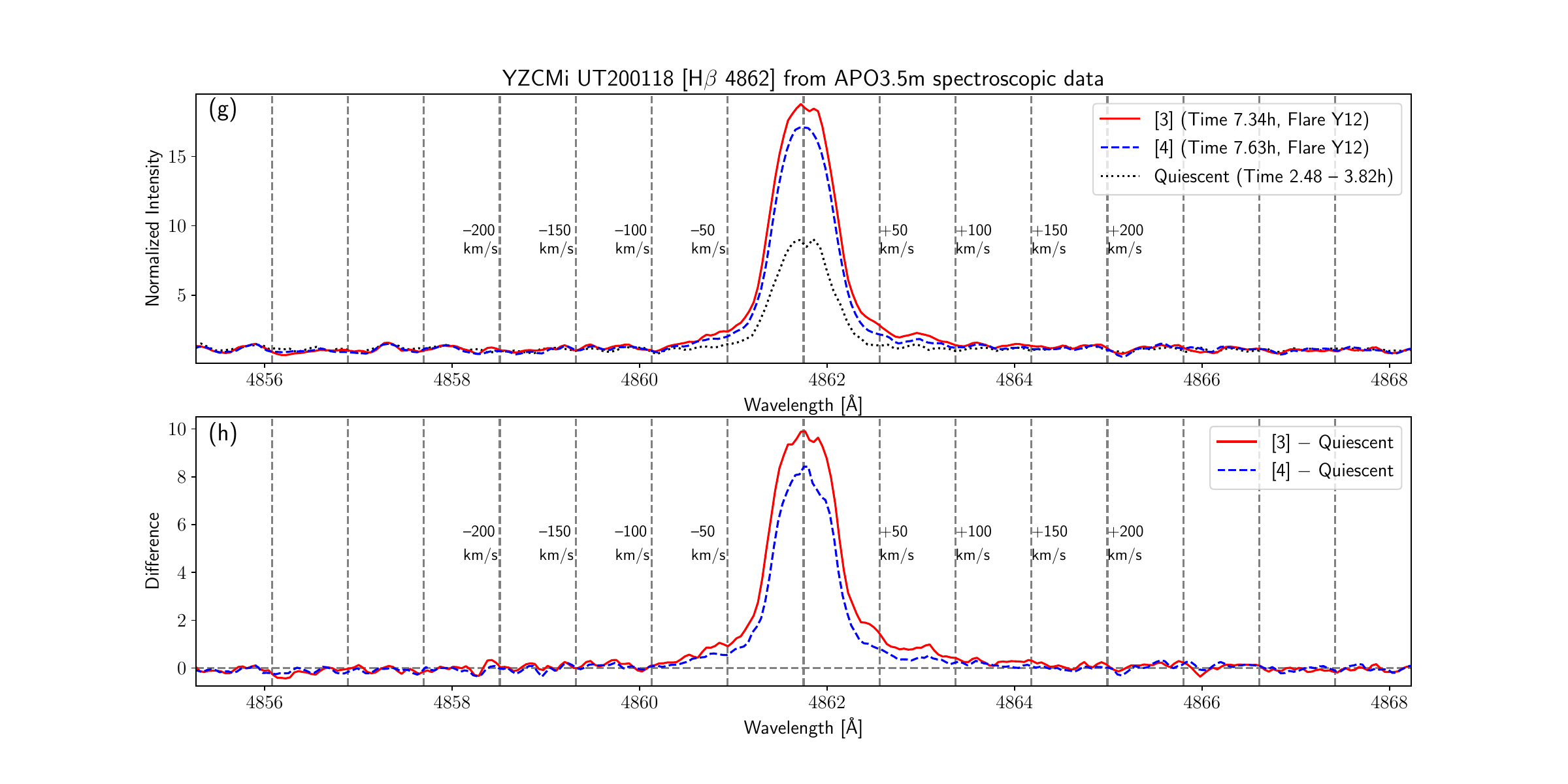}{0.58\textwidth}{\vspace{0mm}}
    }
     \vspace{-1.0cm}
            \gridline{  
     \hspace{-0.06\textwidth}
    \fig{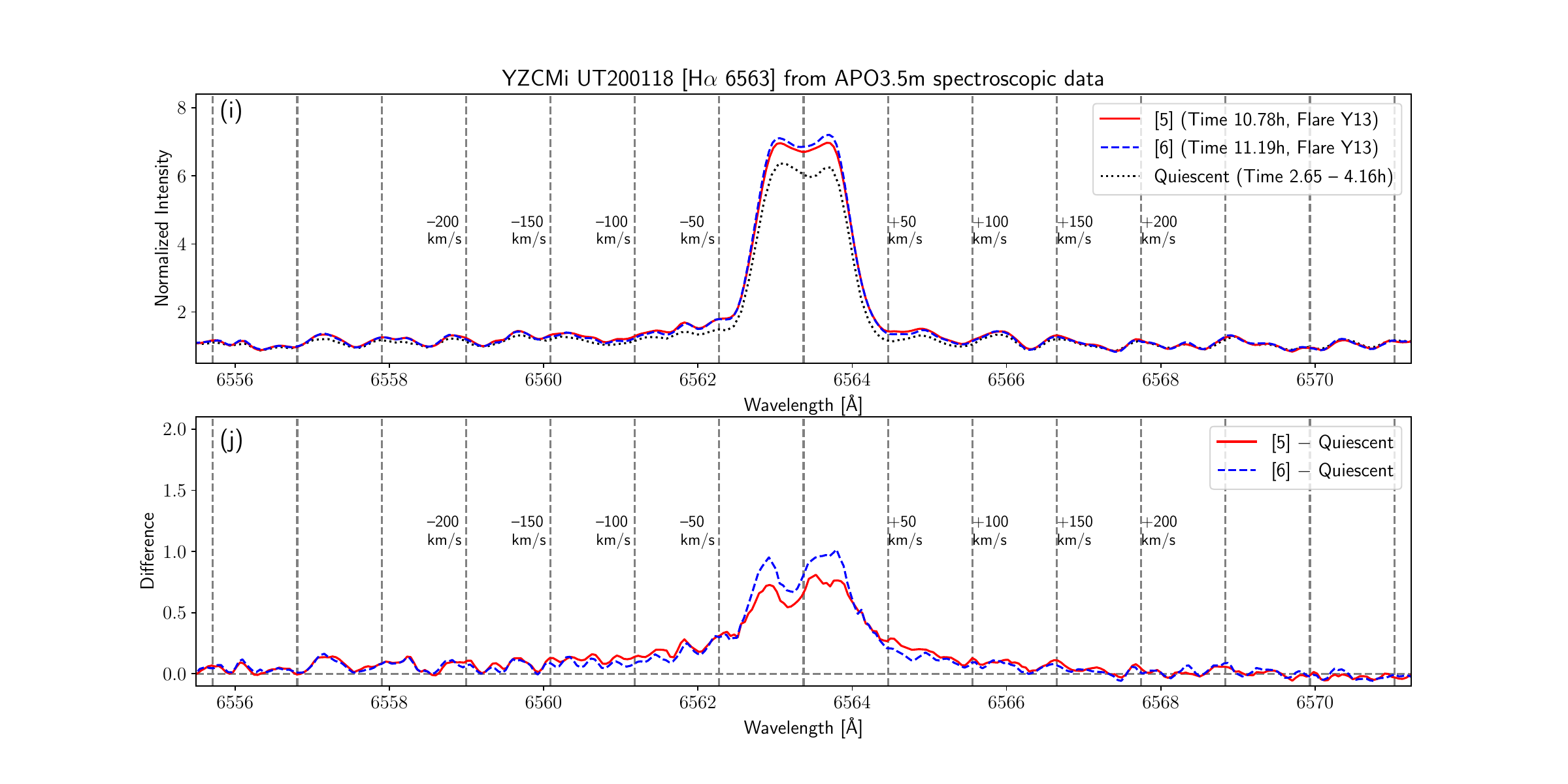}{0.58\textwidth}{\vspace{0mm}}
     \hspace{-0.06\textwidth}
       \fig{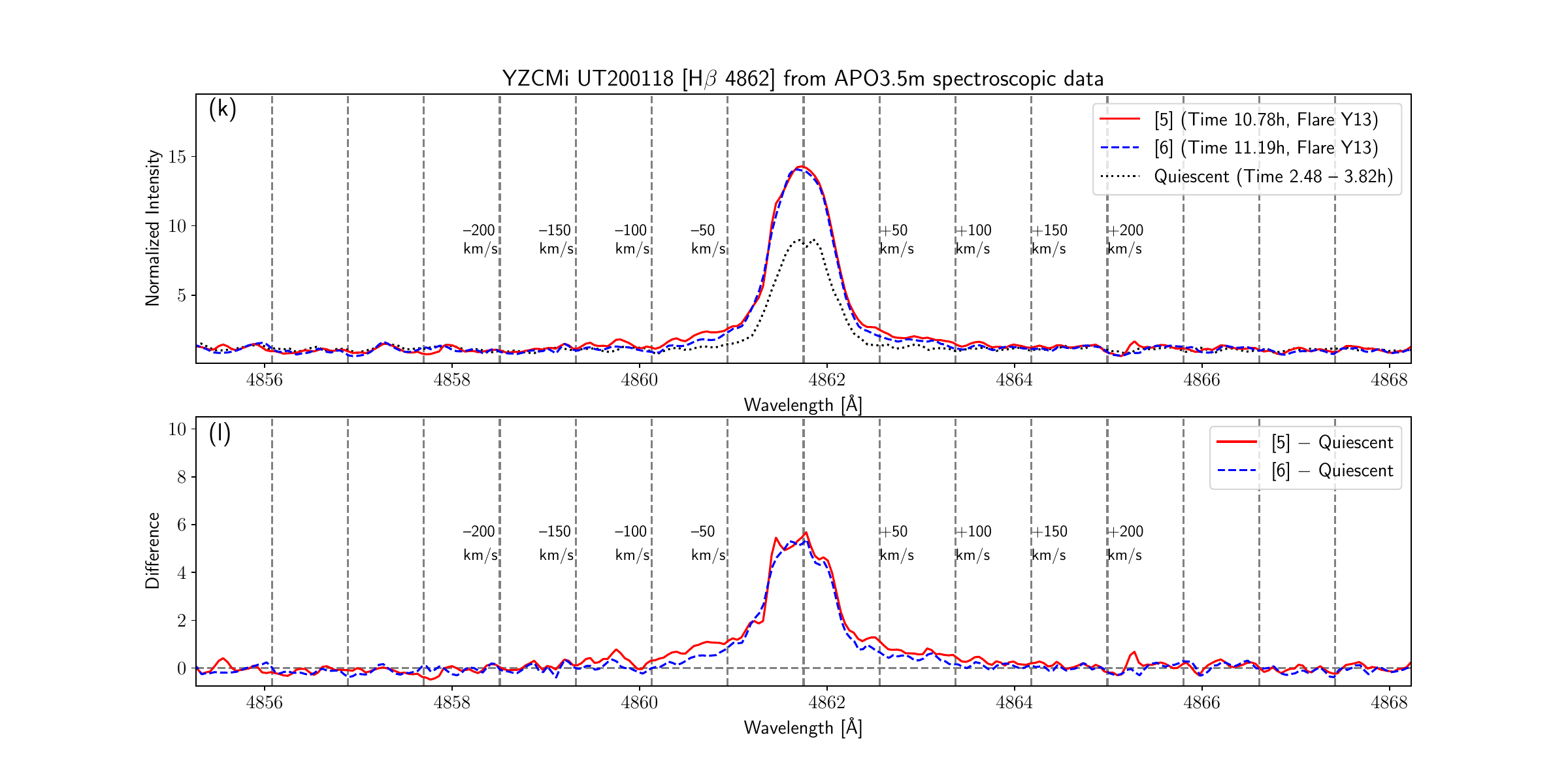}{0.58\textwidth}{\vspace{0mm}}
    }
     \vspace{-0.5cm}
     \caption{
   \color{black}\textrm{  
(a)--(d)
Line profiles of the H$\alpha$ \& H$\beta$ emission lines at the time [1] and [2] during Flare Y12 on 2020 January 18  from SMARTS1.5m spectroscopic data, which are plotted similarly with Figure \ref{fig:spec_HaHb_YZCMi_UT190127}.
(e)--(h)
Same as (a)--(d), but those at the time [3] and [4] during Flares Y12
from APO3.5m spectroscopic data.
(i)--(l)
Same as (a)--(d), but those at the time [5] and [6] during Flares Y13.
 } \color{black}
}
   \label{fig:spec_HaHb_YZCMi_UT200118}
   \end{center}
 \end{figure}

\begin{figure}[ht!]
   \begin{center}
           \gridline{  
     \hspace{-0.07\textwidth}
    \fig{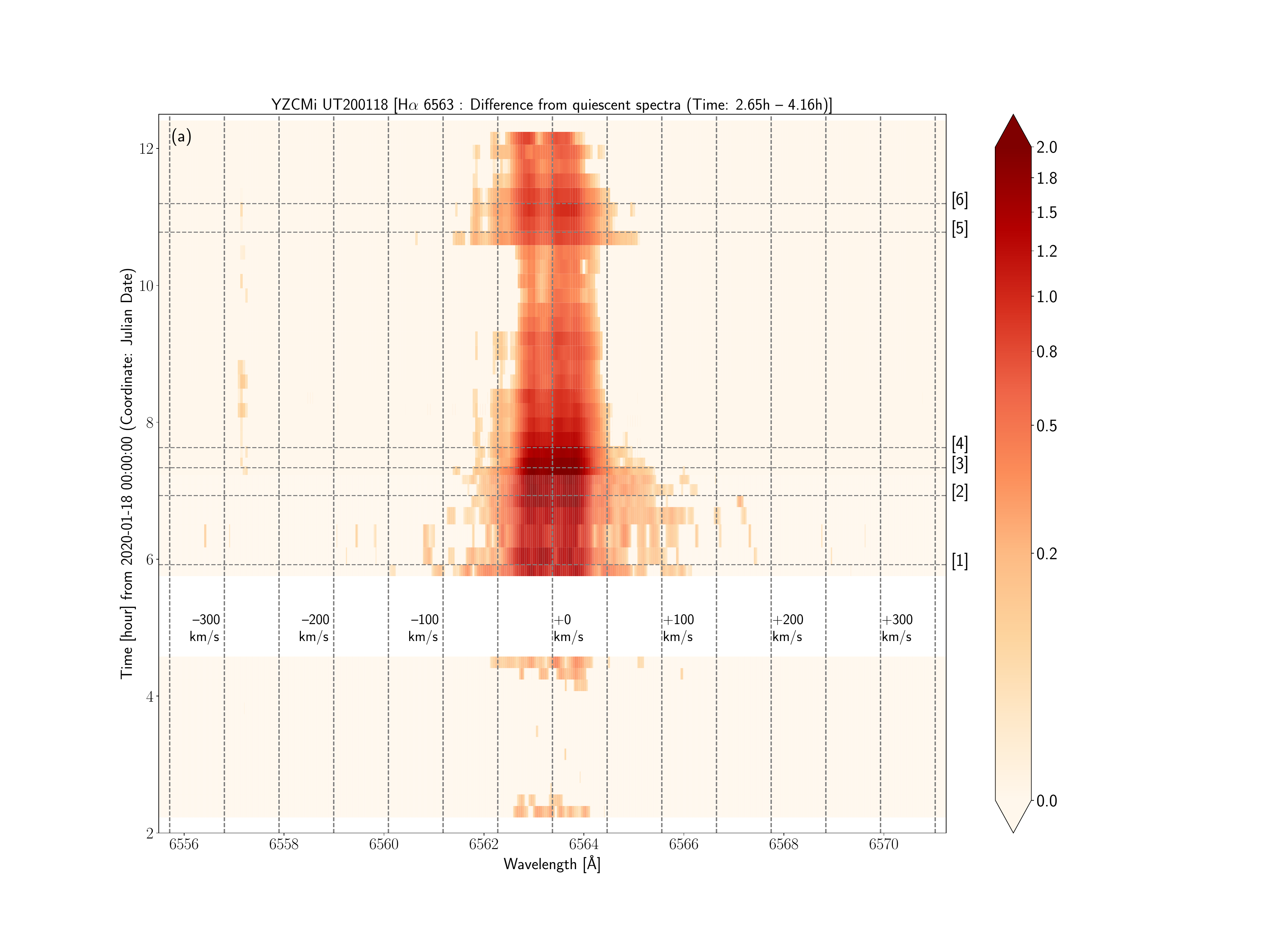}{0.63\textwidth}{\vspace{0mm}}
     \hspace{-0.11\textwidth}
    \fig{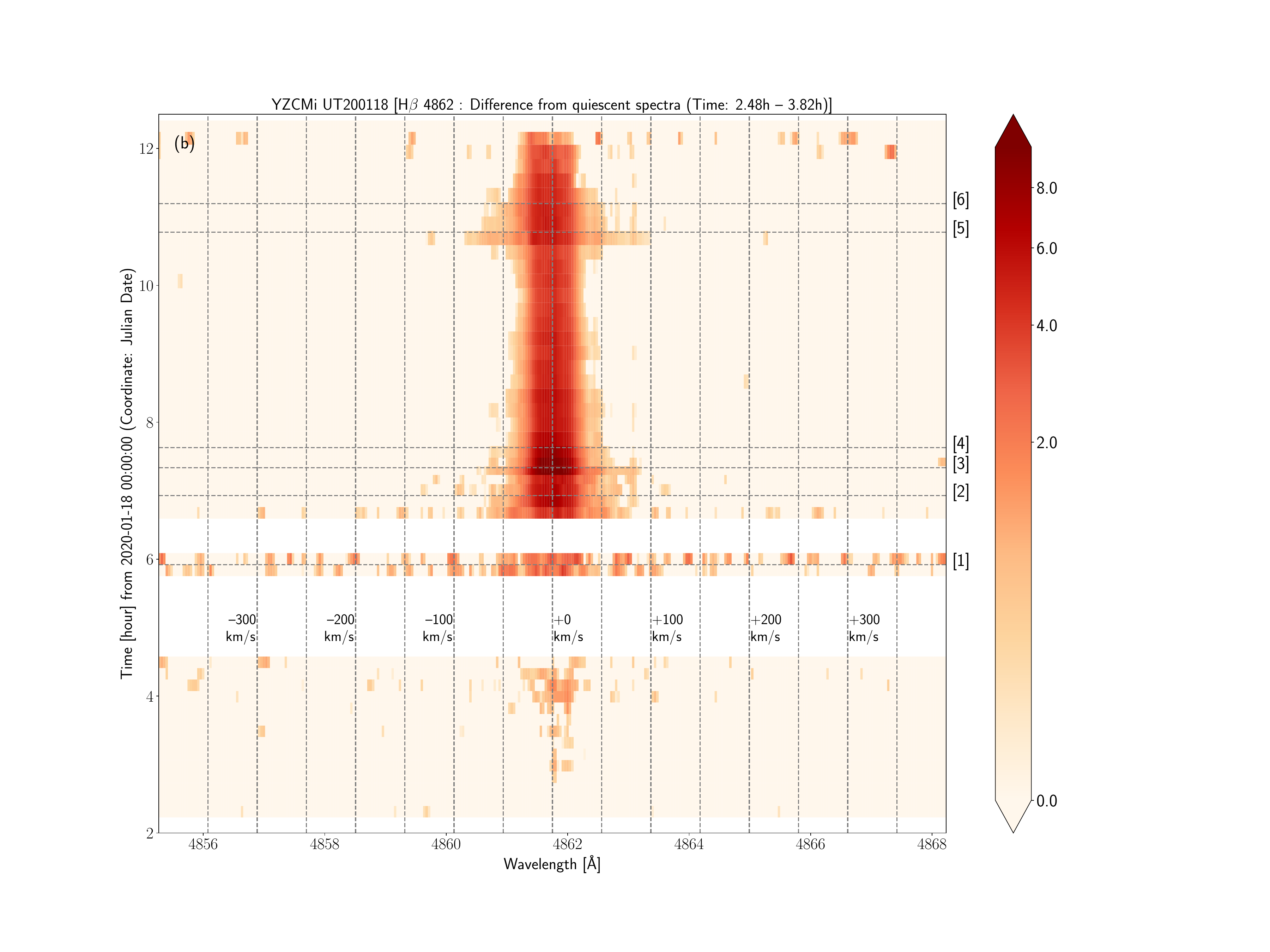}{0.63\textwidth}{\vspace{0mm}}
    }
     \vspace{-0.5cm}
     \caption{
         \color{black}\textrm{  
Time evolution of the H$\alpha$ \& H$\beta$ line profiles covering Flares Y12 \& Y13 on 2020 January 18, which are plotted similarly with Figure \ref{fig:map_HaHb_YZCMi_UT191212}.
The grey horizontal dashed lines indicate the time [1] -- [6], which are shown in Figure \ref{fig:lcEW_HaHb_YZCMi_UT200118} (light curves) and Figure \ref{fig:spec_HaHb_YZCMi_UT200118} (line profiles).
}\color{black}
     }
   \label{fig:map_HaHb_YZCMi_UT200118}
   \end{center}
 \end{figure}

 The H$\alpha$ \& H$\beta$ line profiles during Flares Y12 and Y13 are shown in
Figures \ref{fig:spec_HaHb_YZCMi_UT200118} \& \ref{fig:map_HaHb_YZCMi_UT200118}. 
During Flare Y12, the red wing of H$\alpha$ \& H$\beta$ lines 
could be slightly enhanced (up to $\sim +100$km s$^{-1}$) for $\sim$2 hours (time [1]--[4] in Figures \ref{fig:spec_HaHb_YZCMi_UT200118} \& \ref{fig:map_HaHb_YZCMi_UT200118}).
During Flare Y13, the line profiles of H$\alpha$ \& H$\beta$ lines did not show
clear wing asymmetries.

\clearpage
 
 \subsection{Flares Y14 \& Y15 observed on 2020 January 19} 
\label{subsec:results:2020-Jan-19} 

 \begin{figure}[ht!]
   \begin{center}
   \gridline{
    \fig{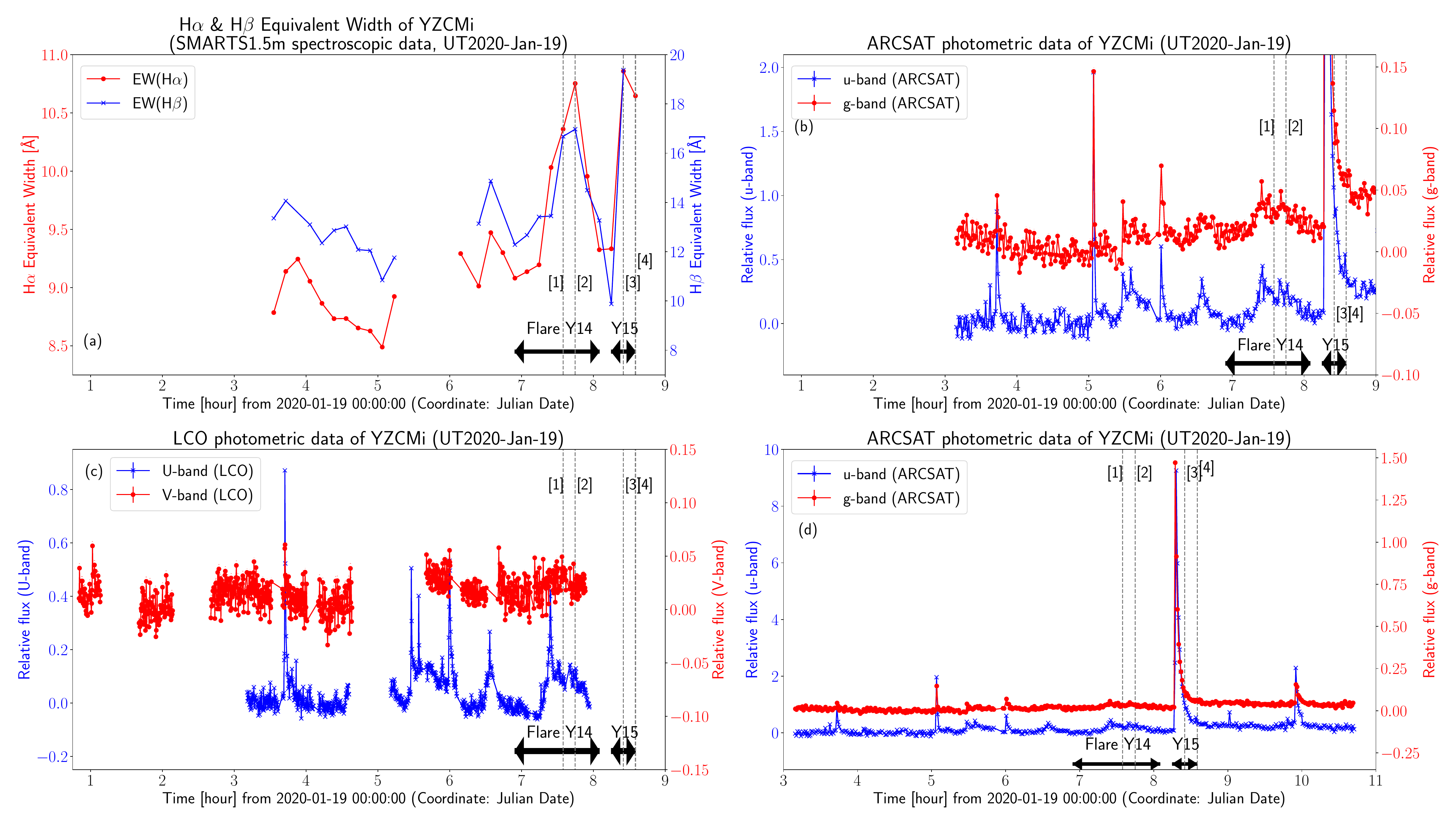}{1.0\textwidth}{\vspace{0mm}}}   
     \vspace{-5mm}
     \caption{
     \color{black}\textrm{  
Light curves of YZ CMi on 2020 January 19 showing Flares Y14 \& Y15, which are plotted 
similarly with Figure \ref{fig:lcEW_HaHb_YZCMi_UT200118}.
In panel (a), red circles and blue asterisks correspond 
to H$\alpha$ \& H$\beta$ EWs from SMARTS 1.5m data. 
The ARCSAT $u$- \& $g$-band photometric data are plotted in (b) \& (d), while the LCO $U$- \& $V$-band photometric data are plotted in (c).
The ranges of horizontal and vertical axes are different between (b) and (d), 
while the range of horizontal axis in (b) is the same as those of (a) and (c).
The grey dashed lines with numbers ([1]--[4]) correspond to the time shown with the same numbers in Figures \ref{fig:spec_HaHb_YZCMi_UT200119} \& \ref{fig:map_HaHb_YZCMi_UT200119}.
} \color{black}
}
   \label{fig:lcEW_HaHb_YZCMi_UT200119}
   \end{center}
 \end{figure}

On 2020 January 19, two flares (Flares Y14 \& Y15) were detected in H$\alpha$ \& H$\beta$ lines as shown in Figure \ref{fig:lcEW_HaHb_YZCMi_UT200119} (a).  
There could be another flare at around the time 5.0h -- 6.5h (Figure \ref{fig:lcEW_HaHb_YZCMi_UT200119} (a)) considering the brightness increases in continuum bands (Figures \ref{fig:lcEW_HaHb_YZCMi_UT200119} (b)\&(c)), but 
the spectroscopic data have observation gap at around the time 5.0h -- 6.5h because of the relatively bad S/N ratio of the data.
As for Flare Y14, the H$\alpha$ \& H$\beta$ equivalent widths increased up to 10.8\AA~and 17.0\AA, respectively, and $\Delta t^{\rm{flare}}_{\rm{H}\alpha}$ is 1.2 hours (Table \ref{table:list1_flares}).
In addition to these enhancements in Balmer emission lines, the continuum brightness observed with ARCSAT $u$- \& $g$-band and LCO $U$- \& $V$-band increased at least by
$\sim$40\%, $\sim$4\%, $\sim$40\%, and $\sim$2--3\%, 
respectively, during Flare Y14 (Figures \ref{fig:lcEW_HaHb_YZCMi_UT200119} (b) \& (c)). 
As for Flare Y15, the H$\alpha$ \& H$\beta$ equivalent widths increased up to 10.9\AA~and 19.4\AA, respectively. 
Only the initial phase ($\sim$0.3 hours) of Flare Y15 was observed.
In addition to these enhancements in Balmer emission lines, the continuum brightness observed with ARCSAT $u$- \& $g$-band increased by
$\sim$925\%, and $\sim$147\%, 
respectively, during Flare Y15 (Figures \ref{fig:lcEW_HaHb_YZCMi_UT200119} (b) \& (d)). 
The LCO observation ended before Flare Y15.

            \begin{figure}[ht!]
   \begin{center}
            \gridline{  
     \hspace{-0.06\textwidth}
    \fig{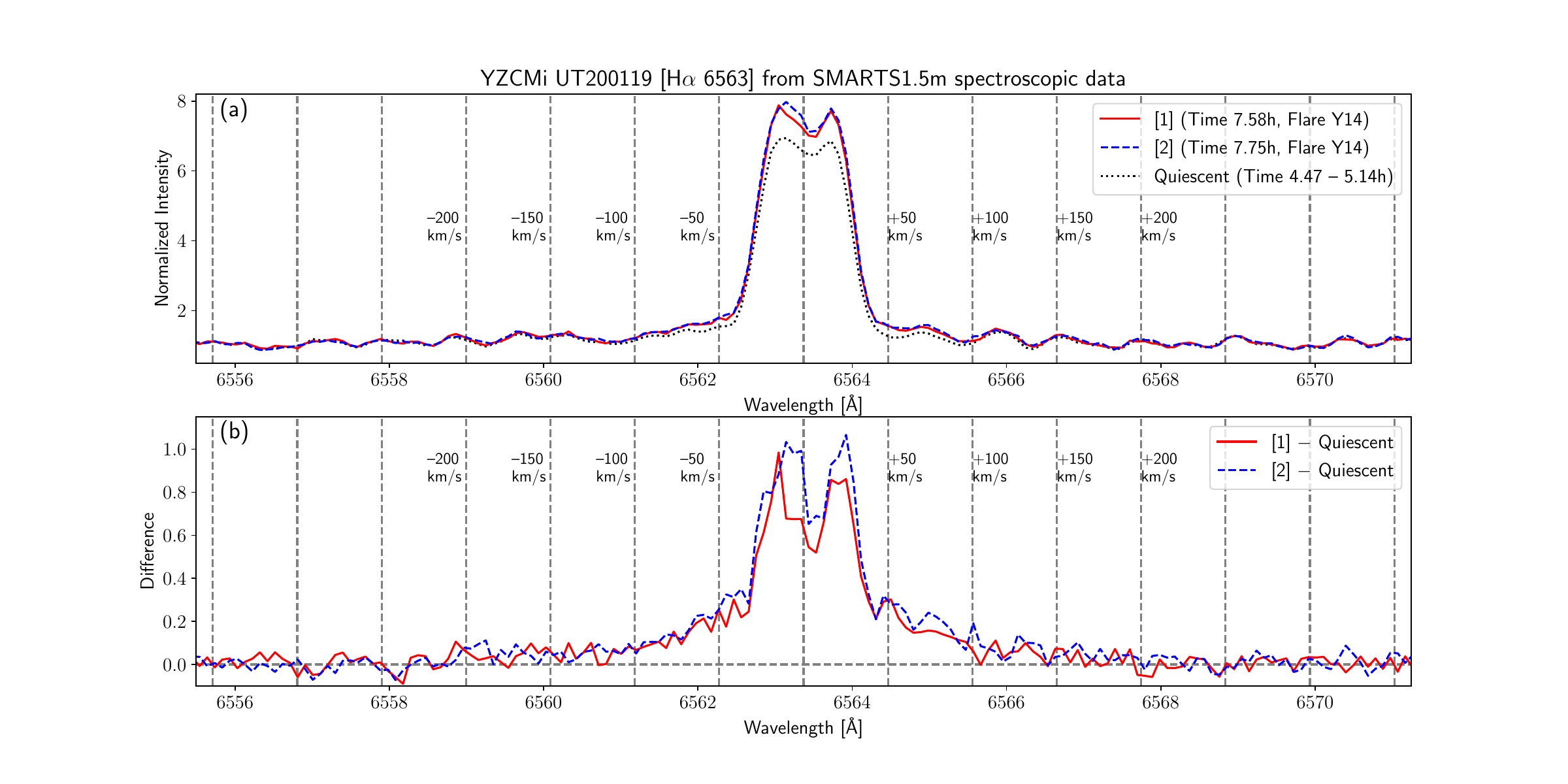}{0.58\textwidth}{\vspace{0mm}}
     \hspace{-0.06\textwidth}
       \fig{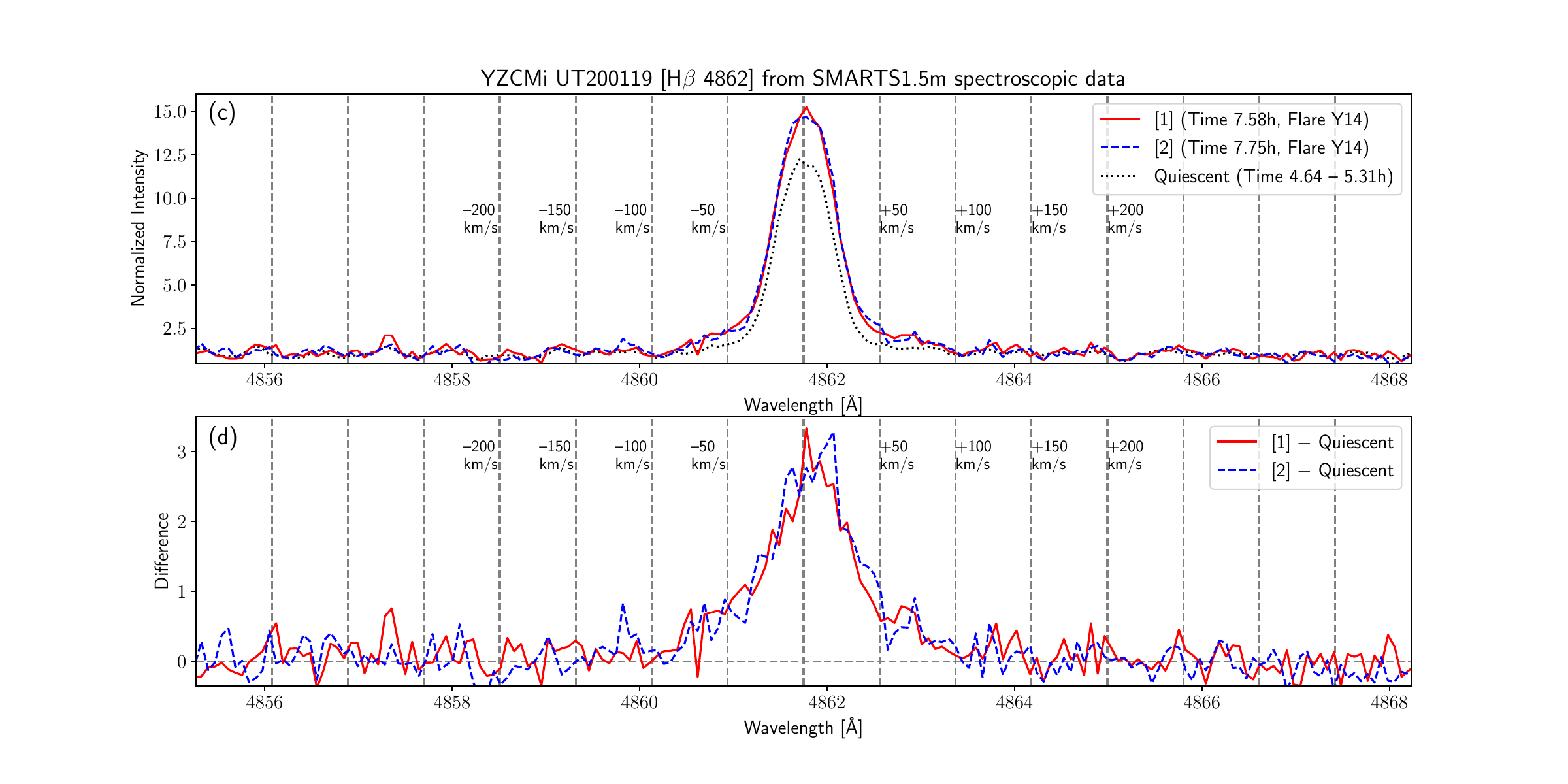}{0.58\textwidth}{\vspace{0mm}}
    }
     \vspace{-1.0cm}
            \gridline{  
     \hspace{-0.06\textwidth}
    \fig{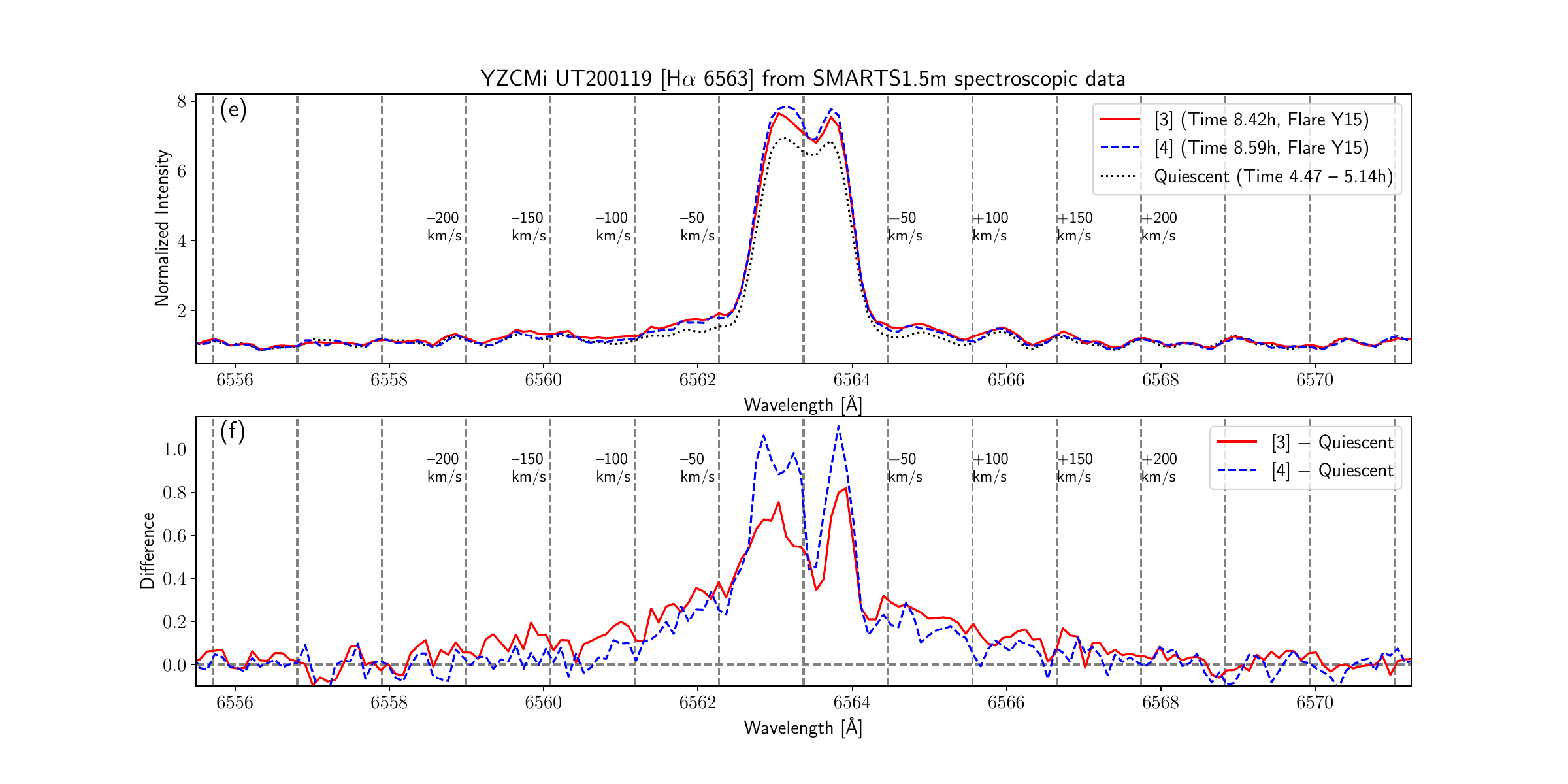}{0.58\textwidth}{\vspace{0mm}}
     \hspace{-0.06\textwidth}
       \fig{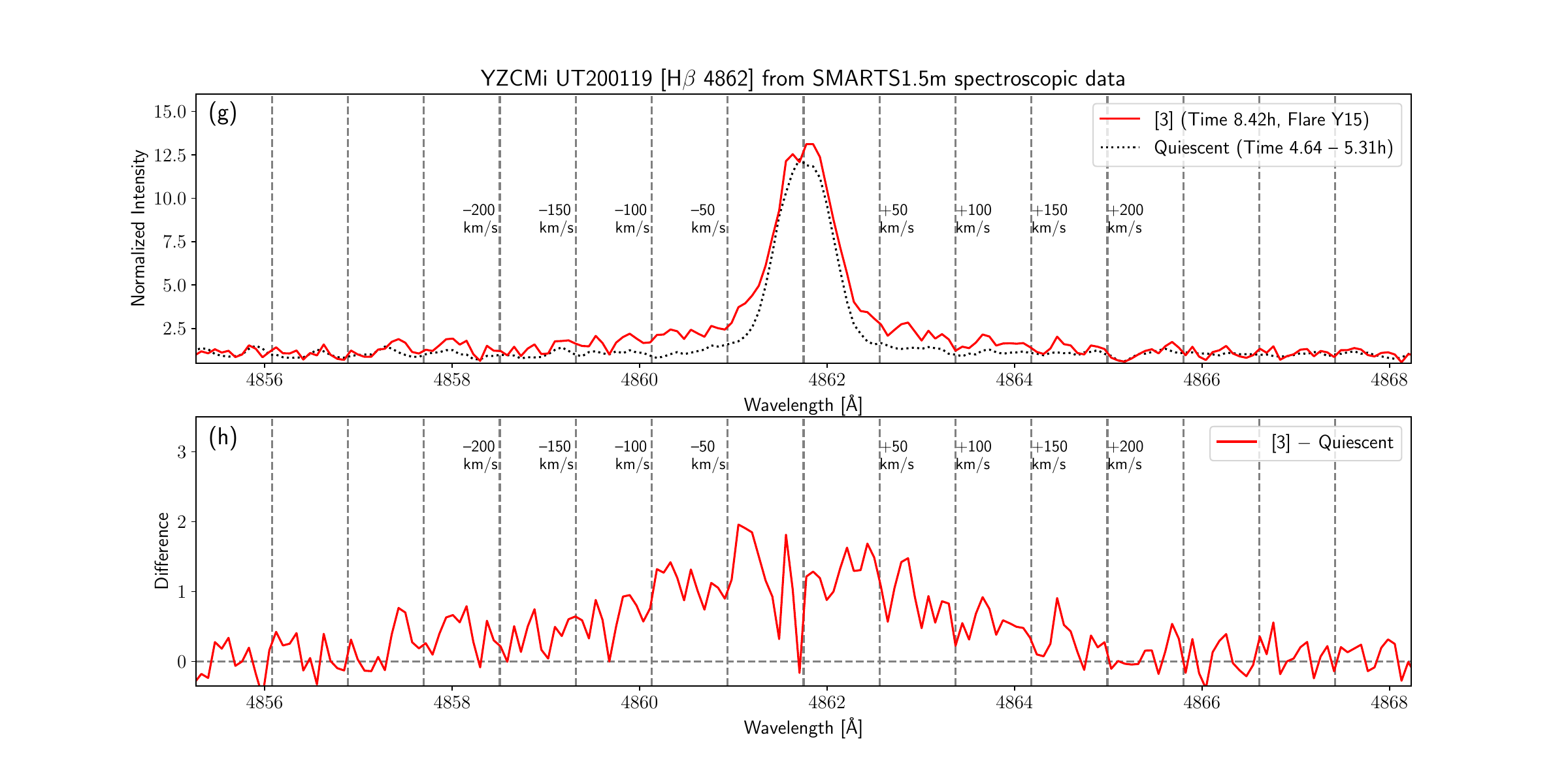}{0.58\textwidth}{\vspace{0mm}}
    }
     \vspace{-0.5cm}
     \caption{
   \color{black}\textrm{  
Line profiles of the H$\alpha$ \& H$\beta$ emission lines during Flares Y14\&Y15 on 2020 January 19 (at the time [1]-[4]) from SMARTS1.5m spectroscopic data, which are plotted similarly with Figure \ref{fig:spec_HaHb_YZCMi_UT190127}.
The H$\beta$ data at the time [4] were not plotted in (g)\&(h), 
because of the bad S/N ratio of the spectroscopic data.
 } \color{black}
     }
   \label{fig:spec_HaHb_YZCMi_UT200119}
   \end{center}
 \end{figure}

\begin{figure}[ht!]
   \begin{center}
      \gridline{
     \hspace{-0.07\textwidth}
      \fig{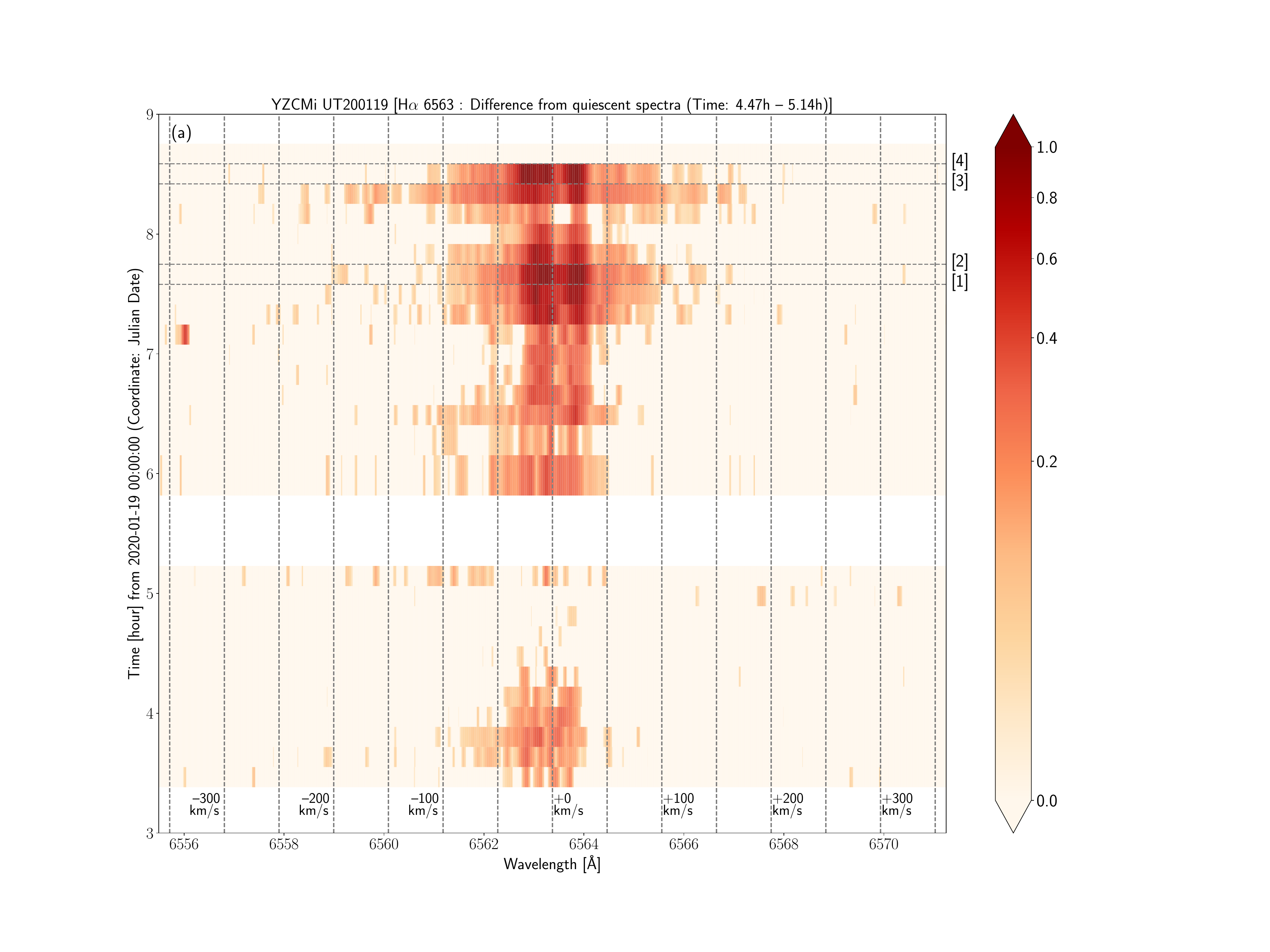}{0.63\textwidth}{\vspace{0mm}}
     \hspace{-0.11\textwidth}
    \fig{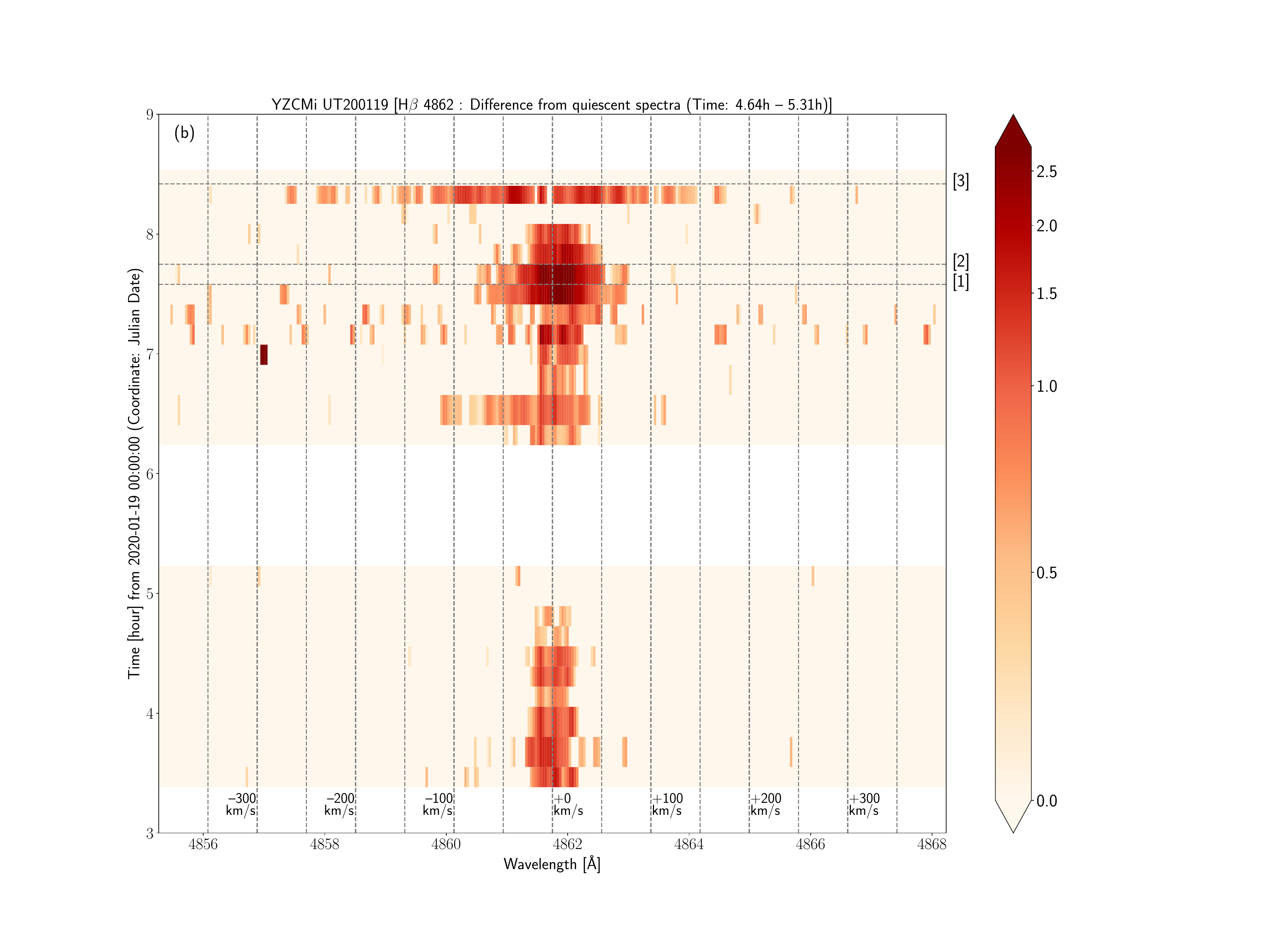}{0.63\textwidth}{\vspace{0mm}}
    }
     \vspace{-0.5cm}
     \caption{
         \color{black}\textrm{  
Time evolution of the H$\alpha$ \& H$\beta$ line profiles covering Flares Y14 \& Y15 on 2020 January 19, which are plotted similarly with Figure \ref{fig:map_HaHb_YZCMi_UT191212}.
The grey horizontal dashed lines indicate the time [1] -- [4], which are shown in Figure \ref{fig:lcEW_HaHb_YZCMi_UT200119} (light curves) and Figure \ref{fig:spec_HaHb_YZCMi_UT200119} (line profiles).
}\color{black}
     }
   \label{fig:map_HaHb_YZCMi_UT200119}
   \end{center}
 \end{figure}

 \color{black}\textrm{ 
$L_{U}$, $L_{u}$, $L_{g}$, $L_{V}$, $E_{U}$, $E_{u}$, $E_{g}$, $E_{V}$, $L_{\rm{H}\alpha}$, $L_{\rm{H}\beta}$, $E_{\rm{H}\alpha}$, and $E_{\rm{H}\beta}$ values are estimated and listed in Table \ref{table:list1_flares}.
We note that since Flare Y15 was partially observed (only the intital $\sim$0.3 hours) in Balmer lines (Figure \ref{fig:lcEW_HaHb_YZCMi_UT200119} (a)), 
the peak luminosity and energy values of Flare Y15 listed here 
could be only the lower limit values.
} \color{black}
The H$\alpha$ \& H$\beta$ line profiles during Flares Y14 and Y15 are shown in
Figures \ref{fig:spec_HaHb_YZCMi_UT200119} \& \ref{fig:map_HaHb_YZCMi_UT200119}. 
During both flares, there were no clear line wing asymmetries.
The H$\alpha$ line wings at around the peak time of Flares Y14 and Y15 showed almost symmetric line broadenings with $\pm$100--150 km s$^{-1}$ and $\pm$150--200 km s$^{-1}$, respectively.

\clearpage
 
 \subsection{Flares Y16 \& Y17 observed on 2020 January 20} 
\label{subsec:results:2020-Jan-20} 
 
         \begin{figure}[ht!]
   \begin{center}
   \gridline{
    \fig{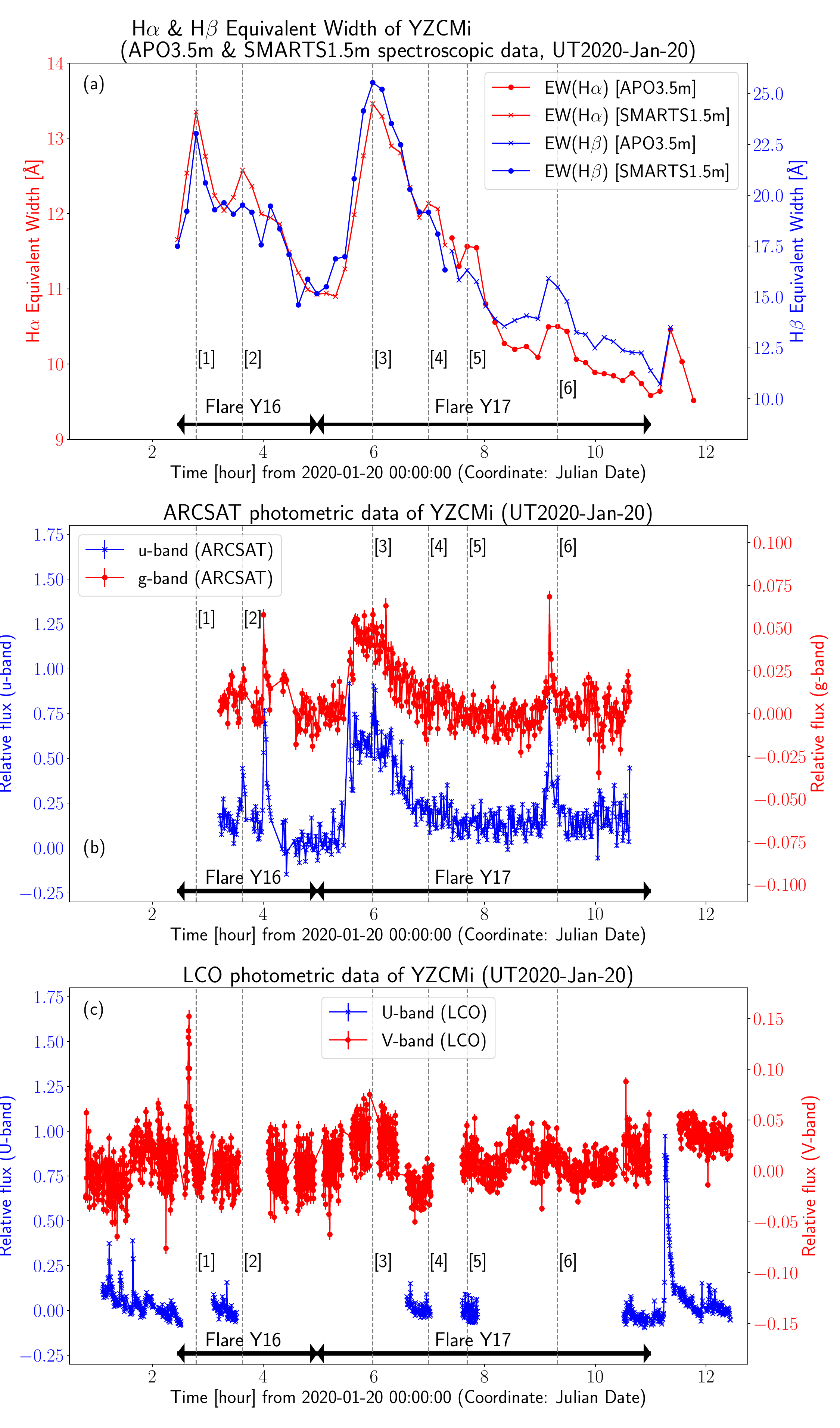}{0.5\textwidth}{\vspace{0mm}}}   
     \vspace{-5mm}
     \caption{
     \color{black}\textrm{  
Light curves of YZ CMi on 2020 January 20 showing Flares Y16 \& Y17, which are plotted 
similarly with Figure \ref{fig:lcEW_HaHb_YZCMi_UT200118}.
The grey dashed lines with numbers ([1]--[6]) correspond to the time shown with the same numbers in Figures \ref{fig:spec_HaHb_YZCMi_UT200120} \& \ref{fig:map_HaHb_YZCMi_UT200120}.
 } \color{black}
     }
   \label{fig:lcEW_HaHb_YZCMi_UT200120}
   \end{center}
 \end{figure}

On 2020 January 20, two flares (Flares Y16 \& Y17) were detected in H$\alpha$ \& H$\beta$ lines as shown in Figures \ref{fig:lcEW_HaHb_YZCMi_UT200120} (a) \& (c).  
Flare Y16 already started when the observation started.
As for Flare Y16, the H$\alpha$ \& H$\beta$ equivalent widths increased up to 13.3\AA~and 23.0\AA, respectively, and $\Delta t^{\rm{flare}}_{\rm{H}\alpha}$ is $>$2.5 hours (Table \ref{table:list1_flares}).
In addition to these enhancements in Balmer emission lines, the continuum brightness observed with ARCSAT $u$- \& $g$-band  increased at least by
$\sim$75\%, and $\sim$6\% in the later phase (at around Time 4h), 
respectively, during Flare Y16 (Figure \ref{fig:lcEW_HaHb_YZCMi_UT200120} (b)).
We note that the start time of ARCSAT photometric observation is later than that of spectroscopic observations and the continuum brightness increases with larger amplitude can be missed.  We also note that there is a continuum brightness increase with $\sim$10--15\% in LCO $V$-band just before the spectroscopic observation started (Figure \ref{fig:lcEW_HaHb_YZCMi_UT200120} (d)).
As for Flare Y17, the H$\alpha$ \& H$\beta$ equivalent widths increased up to 13.5\AA~and 25.5\AA, respectively, and $\Delta t^{\rm{flare}}_{\rm{H}\alpha}$ is 6.0 hours (Table \ref{table:list1_flares}).
In addition to these enhancements in Balmer emission lines, the continuum brightness observed with ARCSAT $u$- \& $g$-band and LCO $V$-band increased by
$\sim$75\%, $\sim$7\%, and $\sim$5\%, 
respectively, during Flare Y17 (Figures \ref{fig:lcEW_HaHb_YZCMi_UT200120} (b) \& (d)). 
We note that there are gaps of the LCO photometric observations also during Y17 and the continuum brightness increases with larger amplitude might be missed.

 \color{black}\textrm{ 
$L_{u}$, $L_{g}$, $L_{V}$, $E_{u}$, $E_{g}$, $E_{V}$, $L_{\rm{H}\alpha}$, $L_{\rm{H}\beta}$, $E_{\rm{H}\alpha}$, and $E_{\rm{H}\beta}$ values are estimated and listed in Table \ref{table:list1_flares}.
The $L_{u}$, $L_{g}$, $L_{V}$, $E_{u}$, $E_{g}$, $E_{V}$ values of Flare Y16 
are only the lower limit values,
since only the later phase in $u$- \& $g$-bands 
and only the earlier phase in $V$-band was observed, respectively. 
(Figure \ref{fig:lcEW_HaHb_YZCMi_UT200120}).
The $L_{\rm{H}\alpha}$, $L_{\rm{H}\beta}$, $E_{\rm{H}\alpha}$, and $E_{\rm{H}\beta}$ values of Flare Y16 described here are also only the lower limit values, since
the initial phase of Flare Y16 was not observed.
} \color{black}

            \begin{figure}[ht!]
   \begin{center}
            \gridline{  
     \hspace{-0.06\textwidth}
    \fig{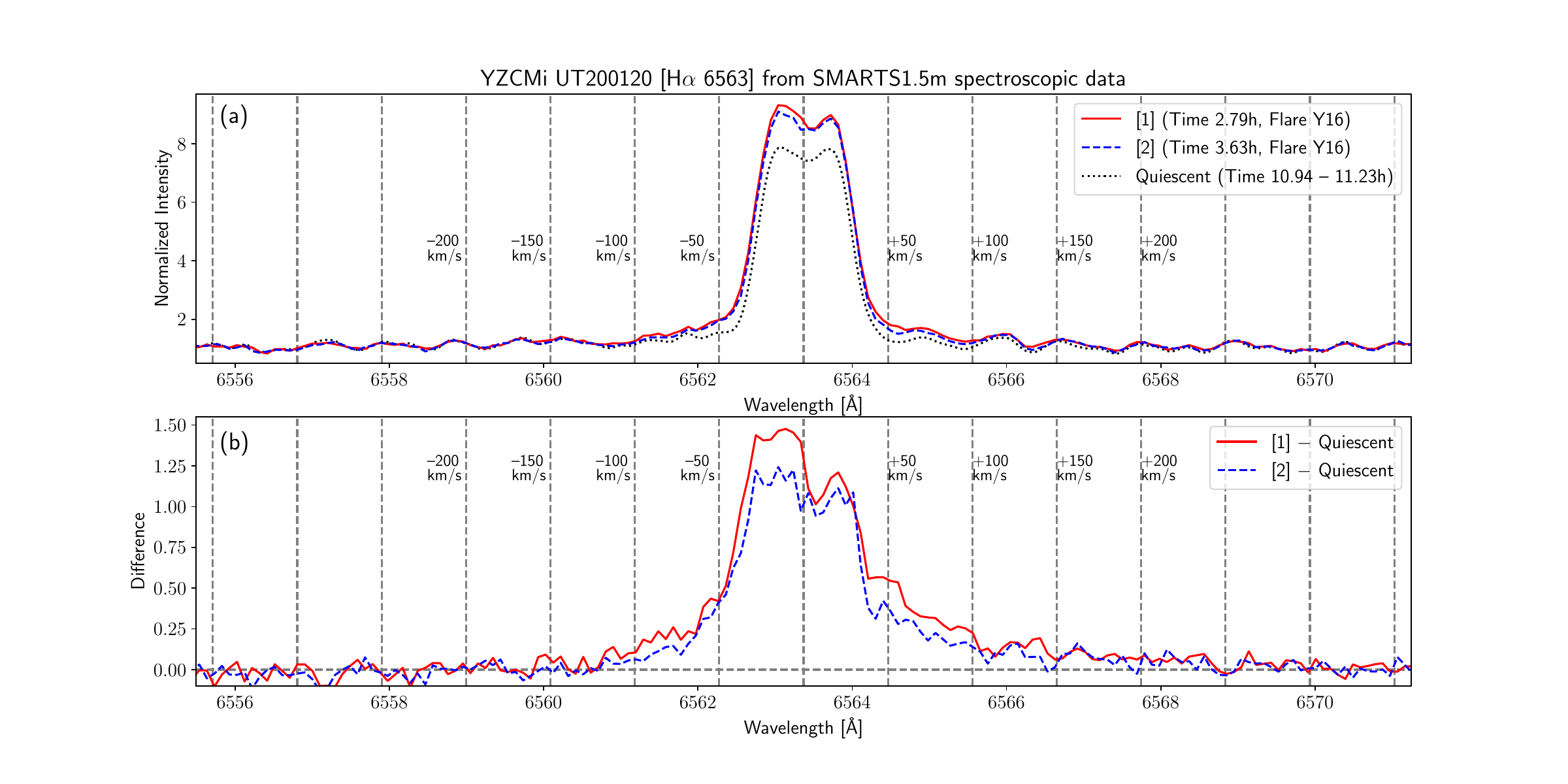}{0.58\textwidth}{\vspace{0mm}}
     \hspace{-0.06\textwidth}
       \fig{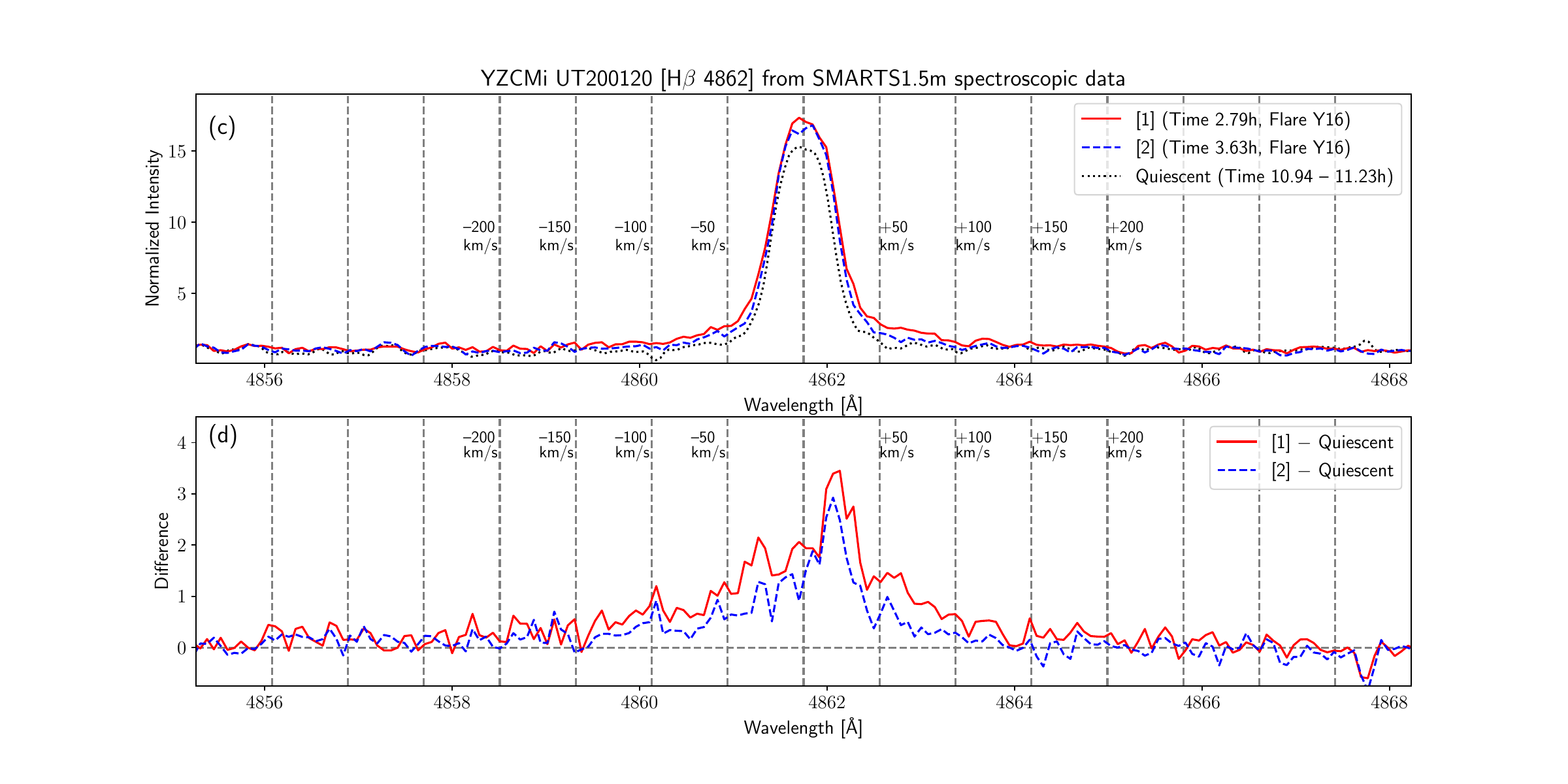}{0.58\textwidth}{\vspace{0mm}}
    }
     \vspace{-1.0cm}
            \gridline{  
     \hspace{-0.06\textwidth}
    \fig{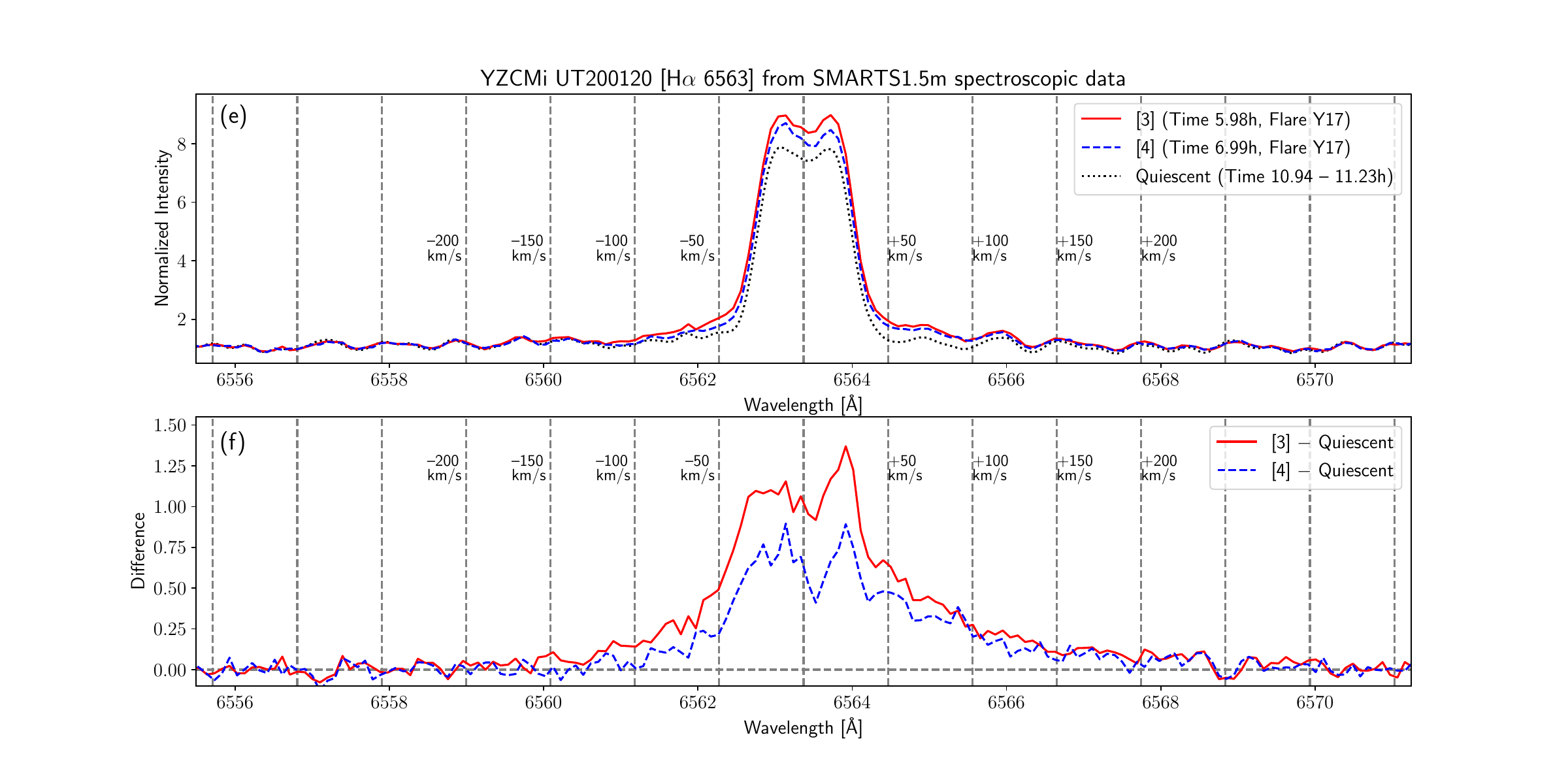}{0.58\textwidth}{\vspace{0mm}}
     \hspace{-0.06\textwidth}
       \fig{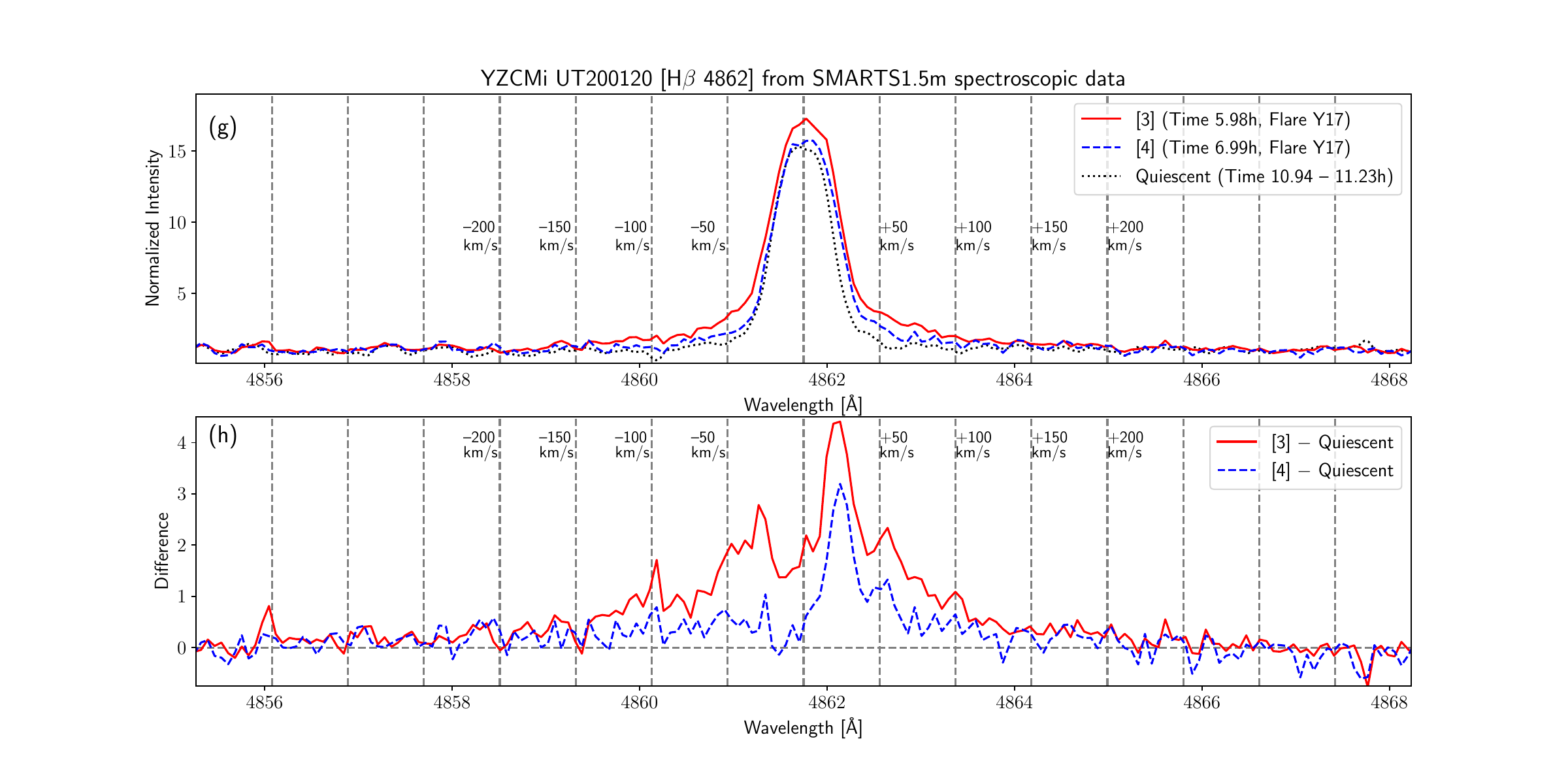}{0.58\textwidth}{\vspace{0mm}}
    }
     \vspace{-1.0cm}
            \gridline{  
     \hspace{-0.06\textwidth}
    \fig{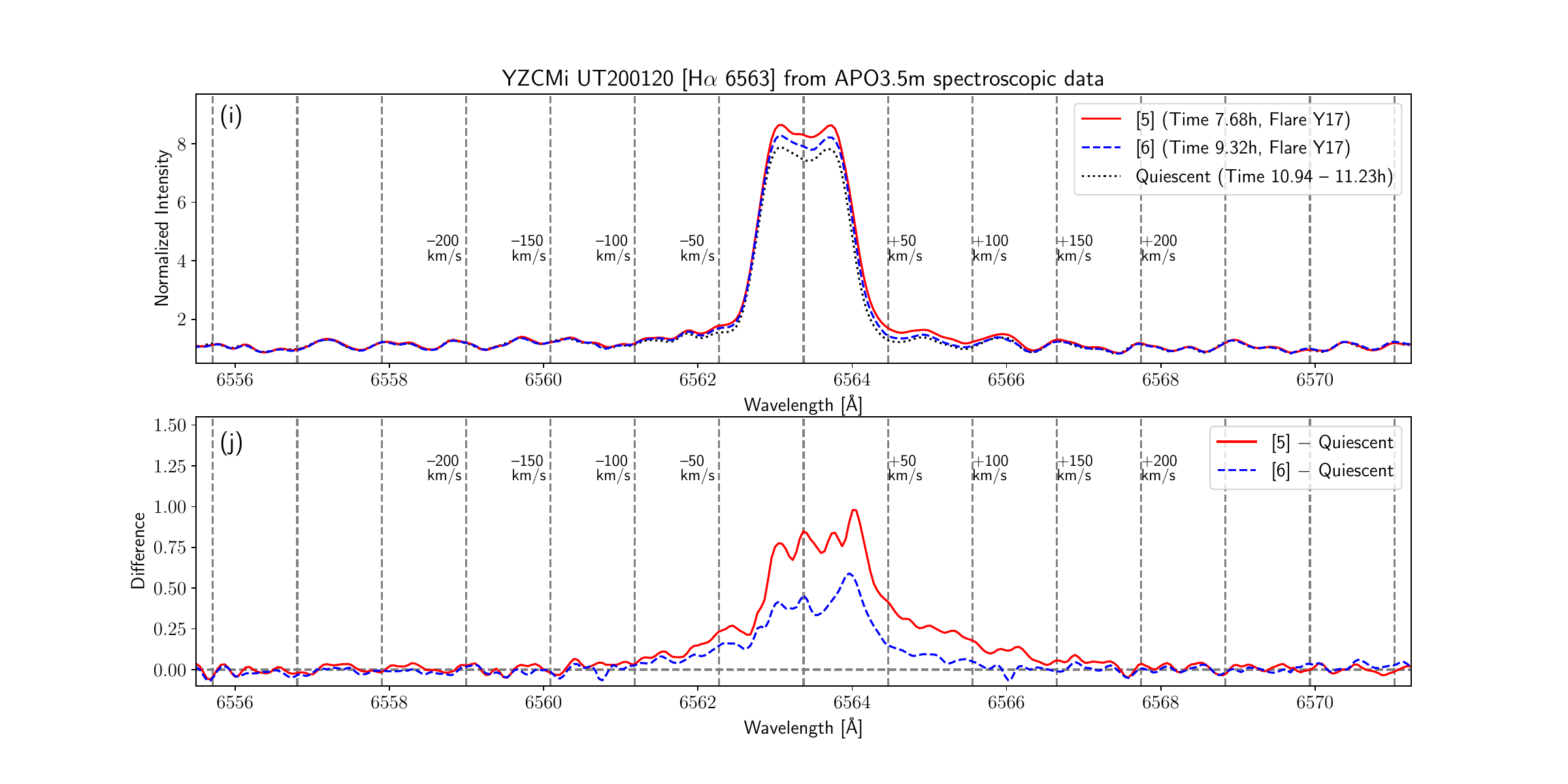}{0.58\textwidth}{\vspace{0mm}}
     \hspace{-0.06\textwidth}
       \fig{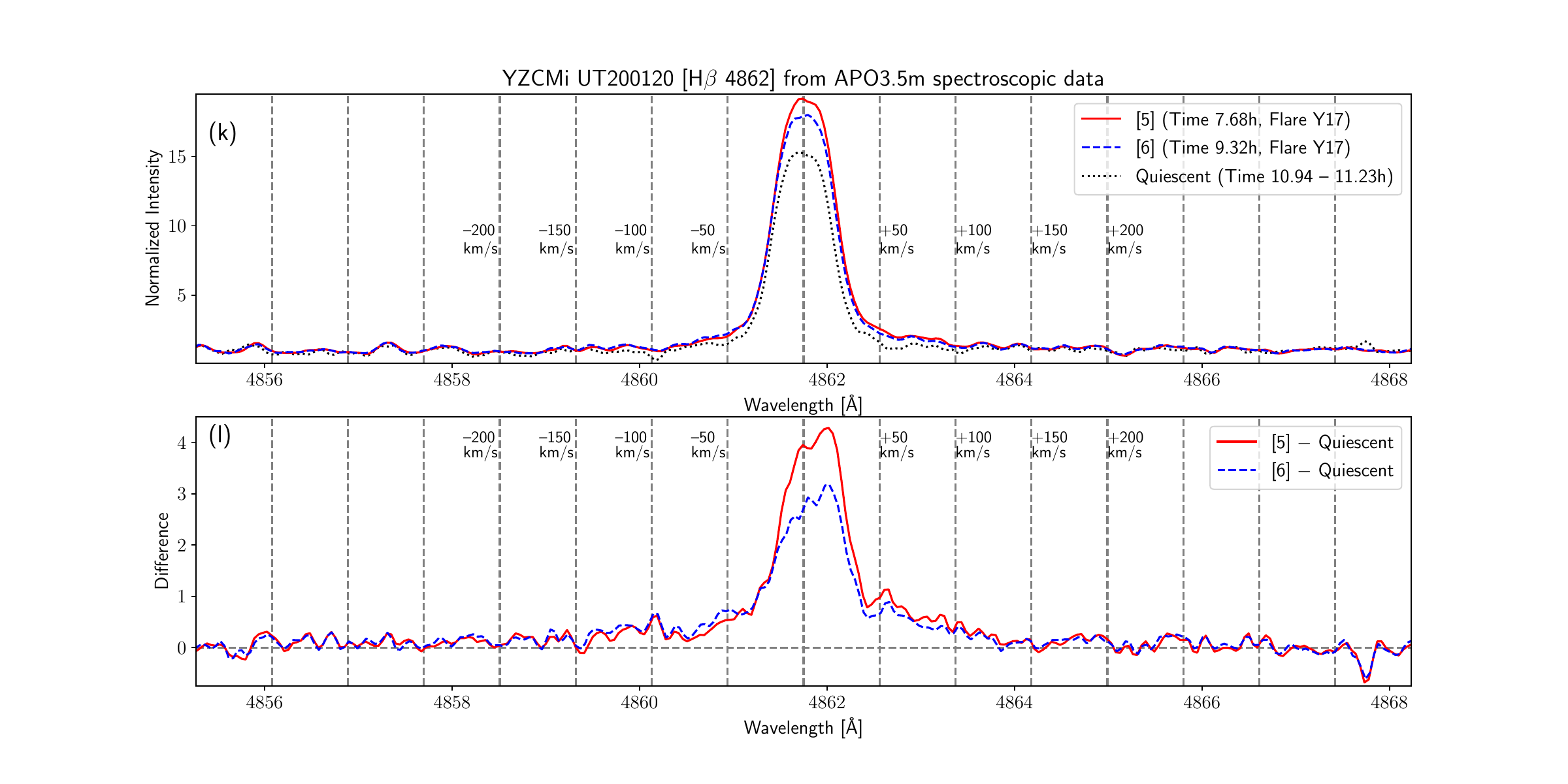}{0.58\textwidth}{\vspace{0mm}}
    }
     \vspace{-0.5cm}
     \caption{
   \color{black}\textrm{  
(a)--(d)
Line profiles of the H$\alpha$ \& H$\beta$ emission lines at the time [1] and [2] during Flares Y16 \& Y17 on 2020 January 20 from SMARTS 1.5m spectroscopic data, which are plotted similarly with Figure \ref{fig:spec_HaHb_YZCMi_UT190127}.
(e)--(h)
Same as (a)--(d), but those at the time [3] and [4] during Flares Y17
from SMARTS 1.5m spectroscopic data.
(i)--(l)
Same as (a)--(d), but those at the time [5] and [6] during Flares Y17
from APO 3.5m spectroscopic data.
 } \color{black}
     }
   \label{fig:spec_HaHb_YZCMi_UT200120}
   \end{center}
 \end{figure}
 
\begin{figure}[ht!]
   \begin{center}
      \gridline{
     \hspace{-0.07\textwidth}
      \fig{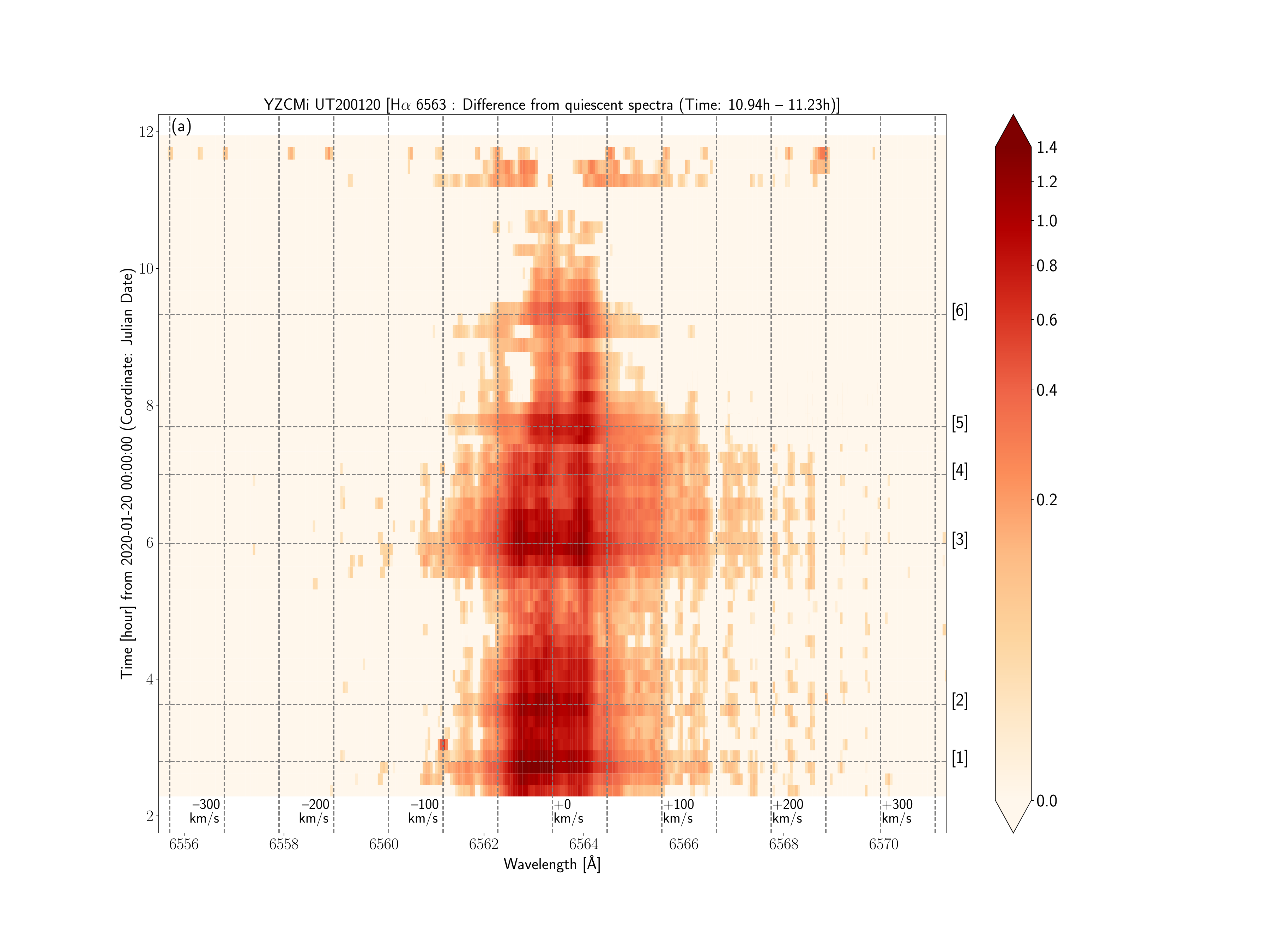}{0.63\textwidth}{\vspace{0mm}}
     \hspace{-0.11\textwidth}
    \fig{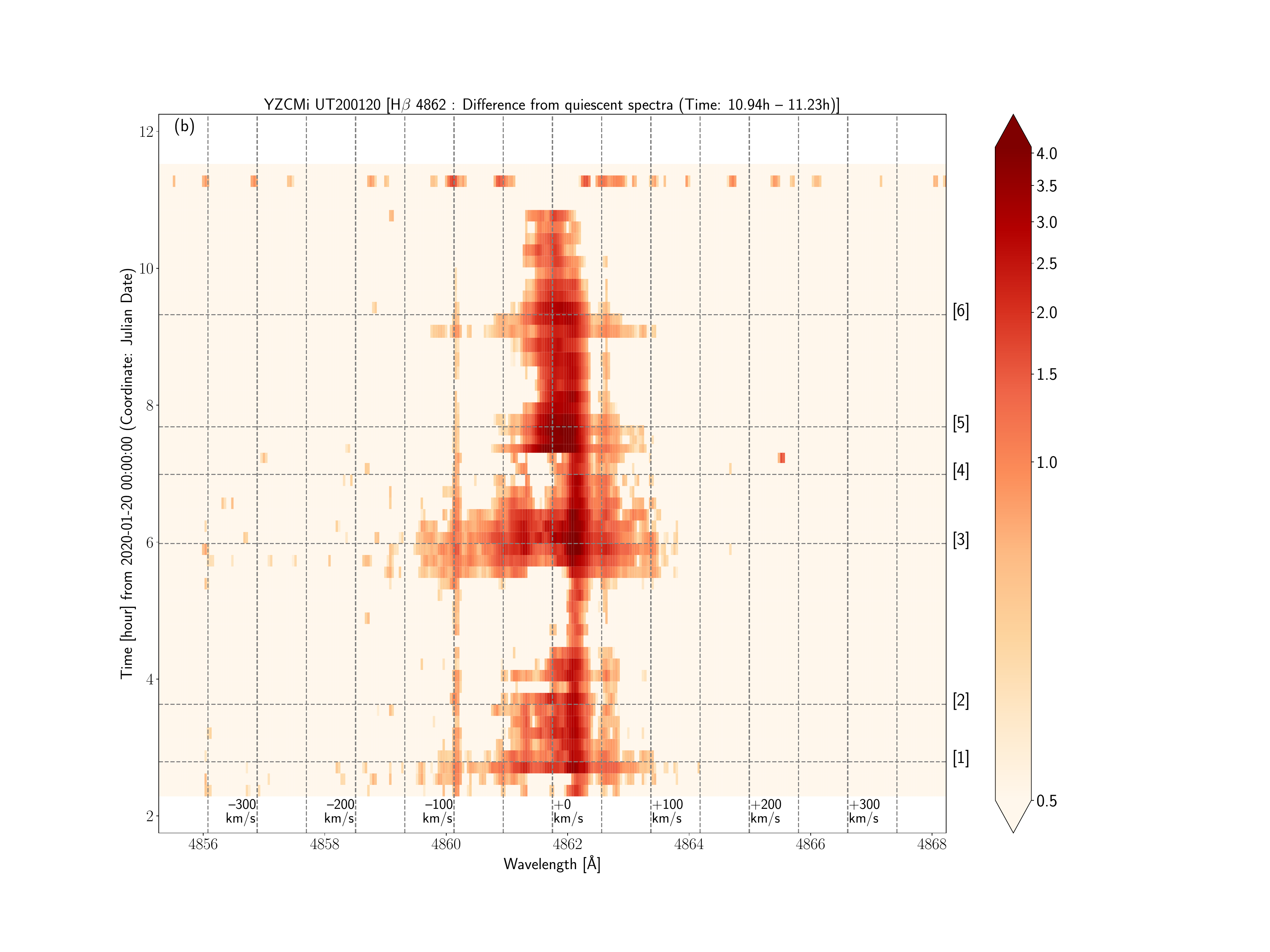}{0.63\textwidth}{\vspace{0mm}}
    }
     \vspace{-0.5cm}
     \caption{
         \color{black}\textrm{  
Time evolution of the H$\alpha$ \& H$\beta$ line profiles  covering Flares Y16 \& Y17 on 2020 January 20, which are plotted similarly with Figure \ref{fig:map_HaHb_YZCMi_UT191212}.
The grey horizontal dashed lines indicate the time [1] -- [6], which are shown in Figure \ref{fig:lcEW_HaHb_YZCMi_UT200120} (light curves) and Figure \ref{fig:spec_HaHb_YZCMi_UT200120} (line profiles).
}\color{black}
     }
   \label{fig:map_HaHb_YZCMi_UT200120}
   \end{center}
 \end{figure}

The H$\alpha$ \& H$\beta$ line profiles during Flares Y16 and Y17 are shown in
Figures \ref{fig:spec_HaHb_YZCMi_UT200120} \& \ref{fig:map_HaHb_YZCMi_UT200120}. 
During Flare Y16, the red wing of H$\alpha$ line (up to $\sim$ +150 km s$^{-1}$) was slightly enhanced (time [1], [2] in Figures \ref{fig:spec_HaHb_YZCMi_UT200120}(b)), while around the H$\alpha$ line center, the blue part ($\sim$ -20 -- -30 km s$^{-1}$) was slightly enhanced (time [1] in Figures \ref{fig:spec_HaHb_YZCMi_UT200120}(b)).
This slight enhancement of red wing of H$\alpha$ line continued almost until the end of Flare Y16 (Figure \ref{fig:map_HaHb_YZCMi_UT200120} (a)).
The H$\beta$ line profile change during Flare Y16 was a bit different from that of H$\alpha$ line. There were no red wing enhancements in the H$\beta$ line profile, and 
it could be possible there was slight blue wing enhancement ($\sim$ -100 km s$^{-1}$).
The H$\alpha$ line profile during Flare Y17 showed 
the properties similar to Flare Y16.
The red wing of H$\alpha$ line (up to $\sim$ +200 km s$^{-1}$) was slightly enhanced 
over the early phase of the flare (time [3]--[5] in Figures \ref{fig:spec_HaHb_YZCMi_UT200120}(f), (j), \& \ref{fig:map_HaHb_YZCMi_UT200120}(a)).
However, this red wing enhancement was not clear in H$\beta$ line, and 
it could be possible there was slight blue wing enhancement ($\sim$ -100 km s$^{-1}$) in the early phase of the flare (time [3] in Figures \ref{fig:spec_HaHb_YZCMi_UT200120}(h) \& \ref{fig:map_HaHb_YZCMi_UT200120}(b)).

\clearpage

\subsection{Flare Y20 observed on 2020 January 22} 
\label{subsec:results:2020-Jan-22}

       \begin{figure}[ht!]
   \begin{center}
   \gridline{
    \fig{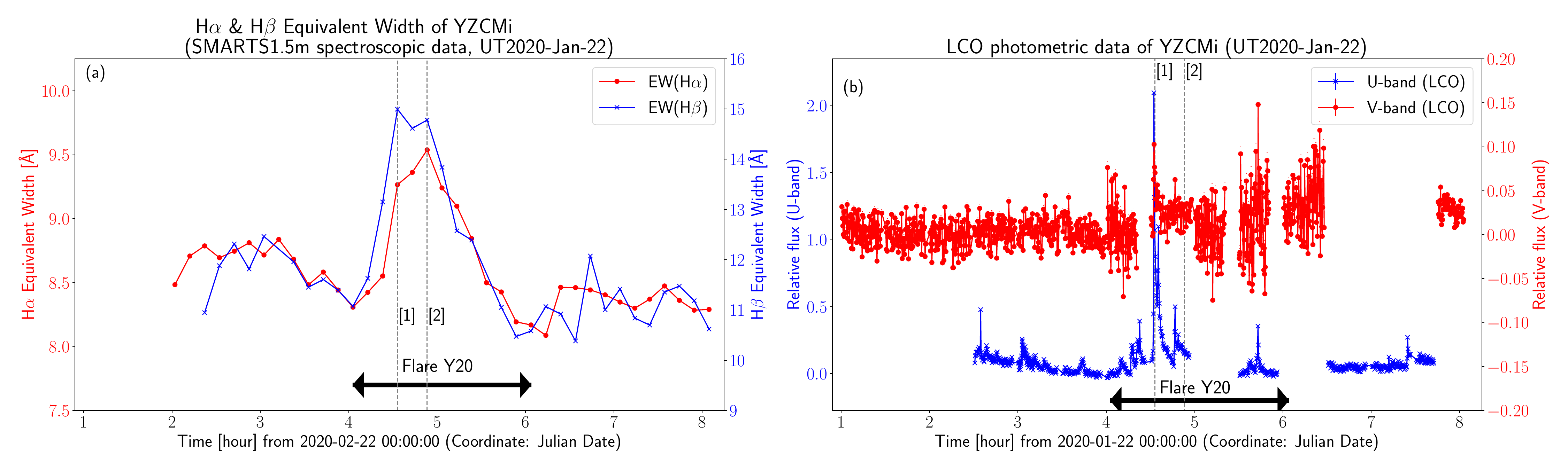}{1.0\textwidth}{\vspace{0mm}}}  
     \vspace{-5mm}
     \caption{
     \color{black}\textrm{  
Light curves of YZ CMi on 2020 January 22 showing Flare Y20, which are plotted 
similarly with Figures \ref{fig:lcEW_HaHb_YZCMi_UT200121} (a)\&(b).
The grey dashed lines with numbers ([1],[2]) correspond to the time shown with the same numbers in Figures \ref{fig:spec_HaHb_YZCMi_UT200122}, \& \ref{fig:map_HaHb_YZCMi_UT200122}.
 } \color{black}
     }
   \label{fig:lcEW_HaHb_YZCMi_UT200122}
   \end{center}
 \end{figure}

 On 2020 January 22, one flare (Flare Y20) was detected in H$\alpha$ \& H$\beta$ lines as shown in Figure \ref{fig:lcEW_HaHb_YZCMi_UT200122} (a).  
As for Flare Y20, the H$\alpha$ \& H$\beta$ equivalent widths increased up to 9.5\AA~and 11.2\AA, respectively, and $\Delta t^{\rm{flare}}_{\rm{H}\alpha}$ is 2.0 hours (Table \ref{table:list1_flares}).
In addition to these enhancements in Balmer emission lines, the continuum brightness observed with LCO $U$- \& $V$-band increased at least by
$\sim$ 200 -- 210\%, and $\sim$5\%, 
respectively, during Flare Y20 (Figure \ref{fig:lcEW_HaHb_YZCMi_UT200122} (b)). 
We note that the LCO photometric observation has gaps during Flare Y20, 
and it could be possible that we missed the continuum brightness increases during the gap time.
  \color{black}\textrm{ 
$L_{U}$, $L_{V}$, $E_{U}$, $E_{V}$, $L_{\rm{H}\alpha}$, $L_{\rm{H}\beta}$, $E_{\rm{H}\alpha}$, and $E_{\rm{H}\beta}$ values are estimated and listed in Table \ref{table:list1_flares}.
Since the LCO observation has gaps during Flare Y20 
(Figure \ref{fig:lcEW_HaHb_YZCMi_UT200122} (b)), 
the energy values in $U$- \& $V$-bands
could be only the lower limit values.
} \color{black}

       \begin{figure}[ht!]
   \begin{center}
           \gridline{  
     \hspace{-0.06\textwidth}
    \fig{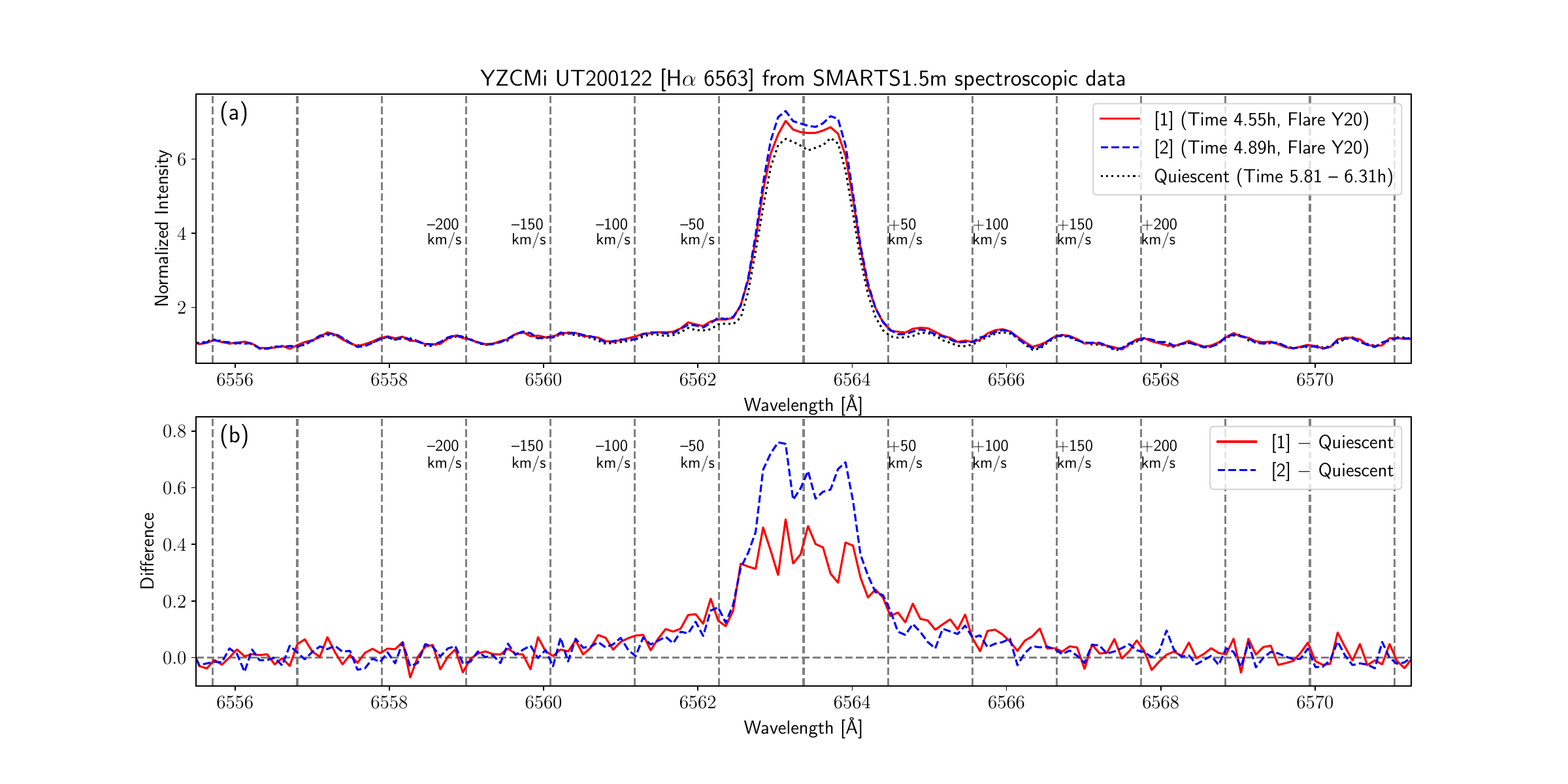}{0.58\textwidth}{\vspace{0mm}}
     \hspace{-0.06\textwidth}
       \fig{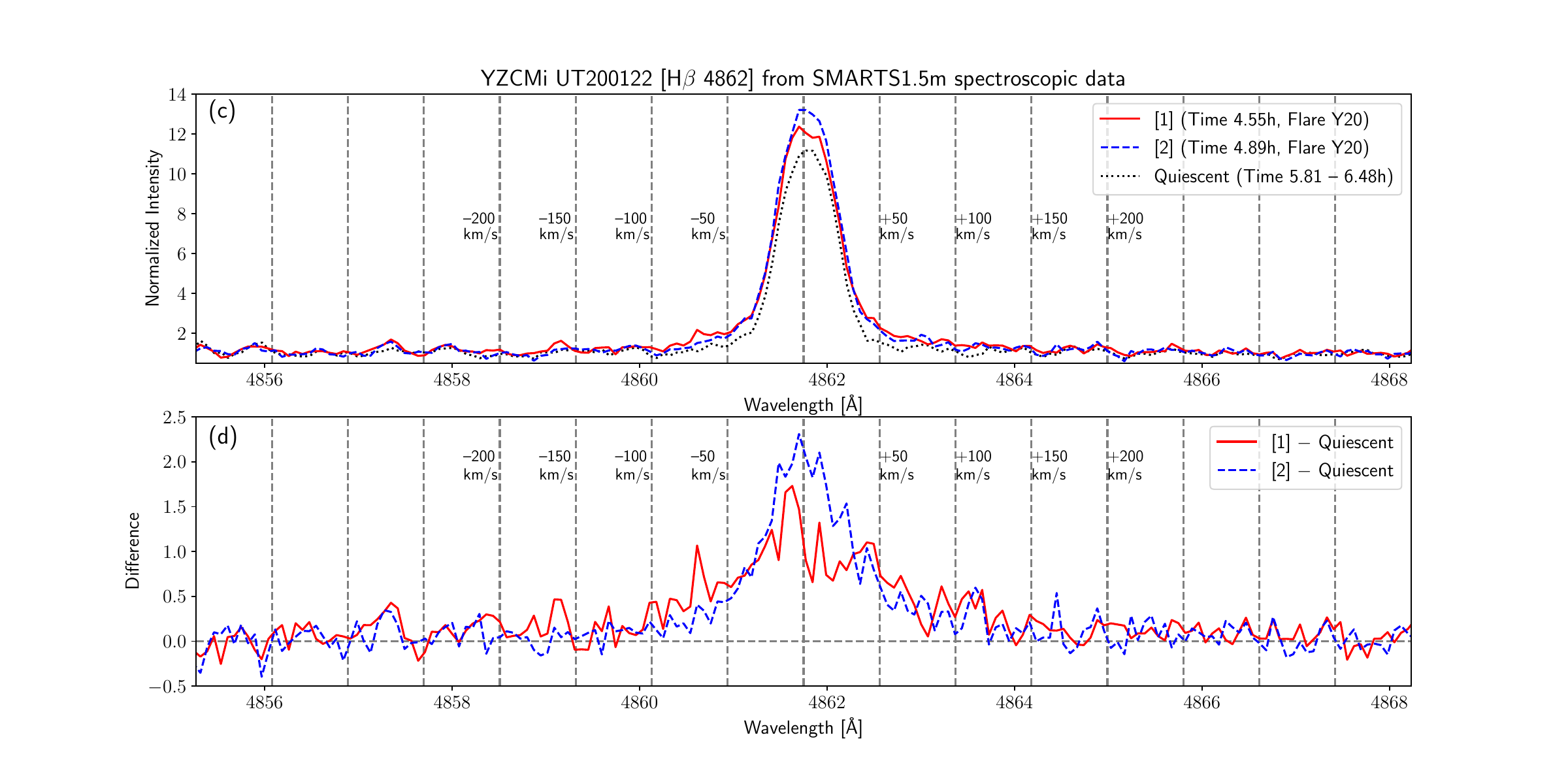}{0.58\textwidth}{\vspace{0mm}}
    }
     \vspace{-0.5cm}
     \caption{
   \color{black}\textrm{  
Line profiles of the H$\alpha$ \& H$\beta$ emission lines during Flare Y20 on 2020 January 22 (at the time [1] and [2]) from SMARTS1.5m spectroscopic data, which are plotted similarly with Figure \ref{fig:spec_HaHb_YZCMi_UT190127}.
 } \color{black}
     }
   \label{fig:spec_HaHb_YZCMi_UT200122}
   \end{center}
 \end{figure}

\begin{figure}[ht!]
   \begin{center}
      \gridline{
     \hspace{-0.07\textwidth}
      \fig{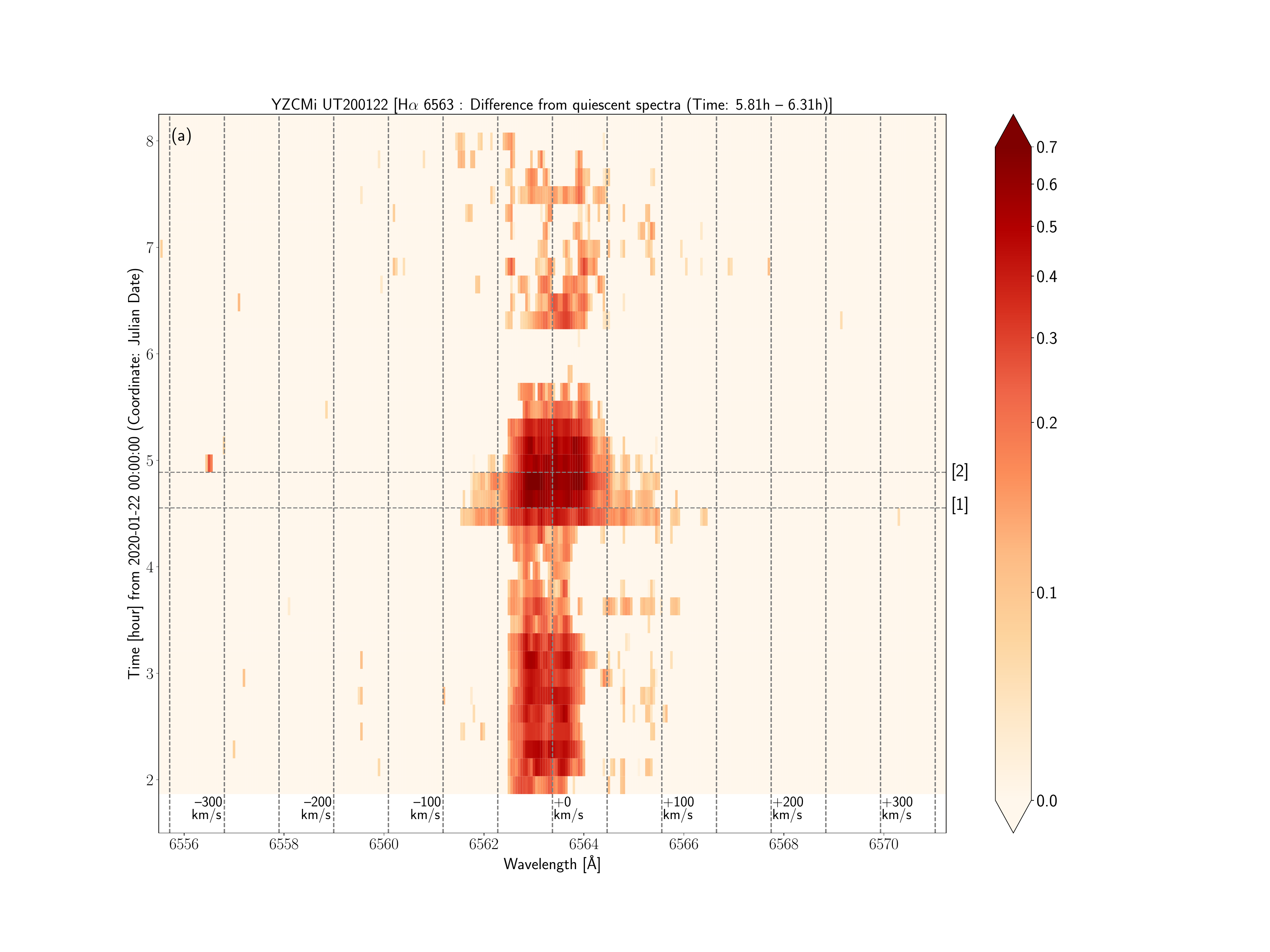}{0.63\textwidth}{\vspace{0mm}}
     \hspace{-0.11\textwidth}
    \fig{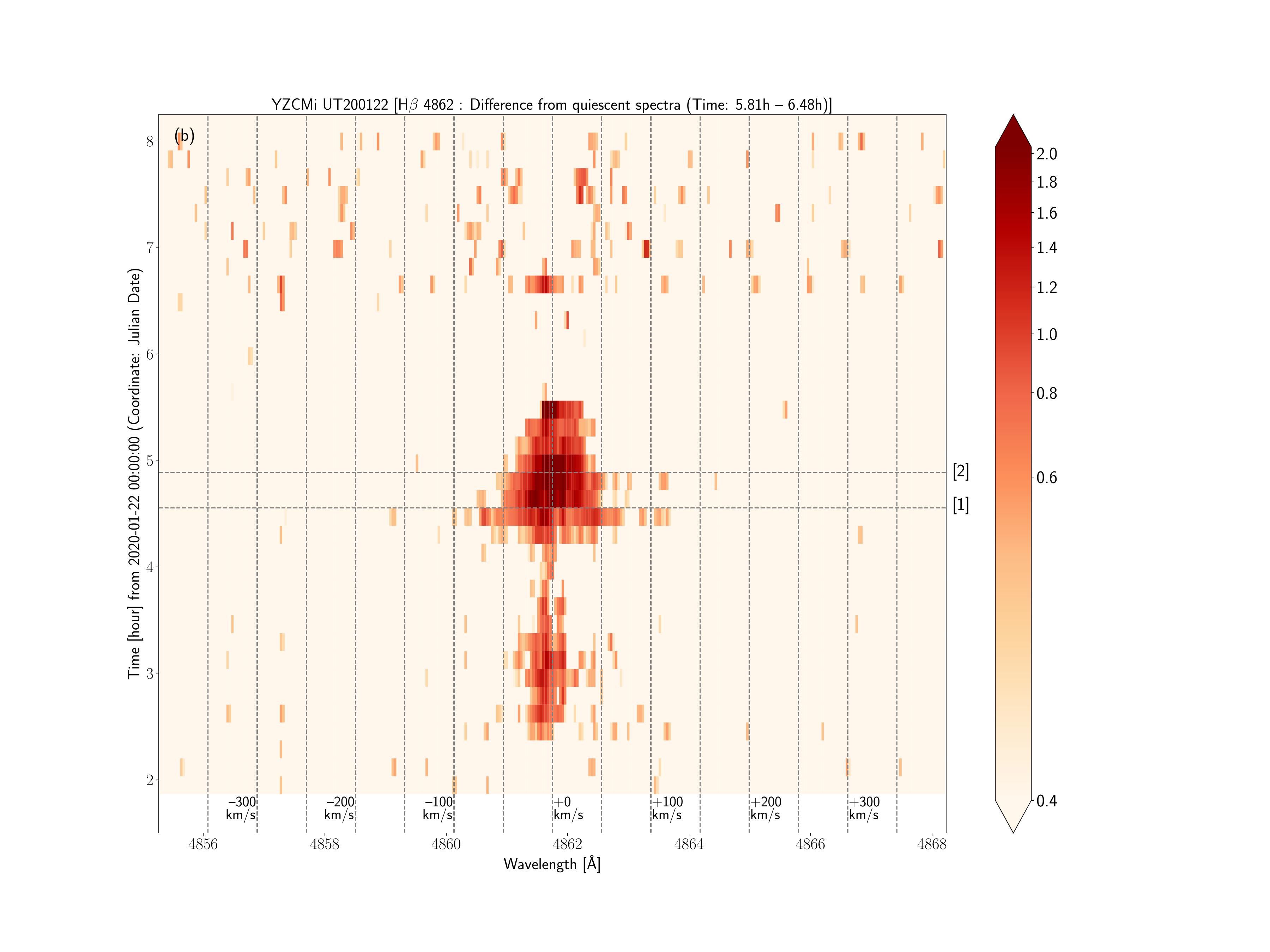}{0.63\textwidth}{\vspace{0mm}}
    }
     \vspace{-0.5cm}
     \caption{
        \color{black}\textrm{  
Time evolution of the H$\alpha$ \& H$\beta$ line profiles covering Flare Y20 on 2020 January 22, which are plotted similarly with Figure \ref{fig:map_HaHb_YZCMi_UT191212}.
The grey horizontal dashed lines indicate the time [1] \& [2], which are shown in Figures \ref{fig:lcEW_HaHb_YZCMi_UT200122}  (light curves) and 
\ref{fig:spec_HaHb_YZCMi_UT200122} (line profiles).
}\color{black}
     }
   \label{fig:map_HaHb_YZCMi_UT200122}
   \end{center}
 \end{figure}

 The H$\alpha$ \& H$\beta$ line profiles during Flare Y20 are shown in
Figures \ref{fig:spec_HaHb_YZCMi_UT200122} \& \ref{fig:map_HaHb_YZCMi_UT200122}. 
During Flare Y20, there were no clear blue or red wing asymmetries in H$\alpha$ and H$\beta$ lines (time [1],[2] in Figures \ref{fig:spec_HaHb_YZCMi_UT200122}(b) \& (d)), and the line profiles showed
roughly symmetrical broadenings with $\sim \pm$150 km s$^{-1}$.

 \clearpage

\subsection{Flares Y21 \& Y22 observed on 2020 January 23} 
\label{subsec:results:2020-Jan-23}

\begin{figure}[ht!]
   \begin{center}
   \gridline{
    \fig{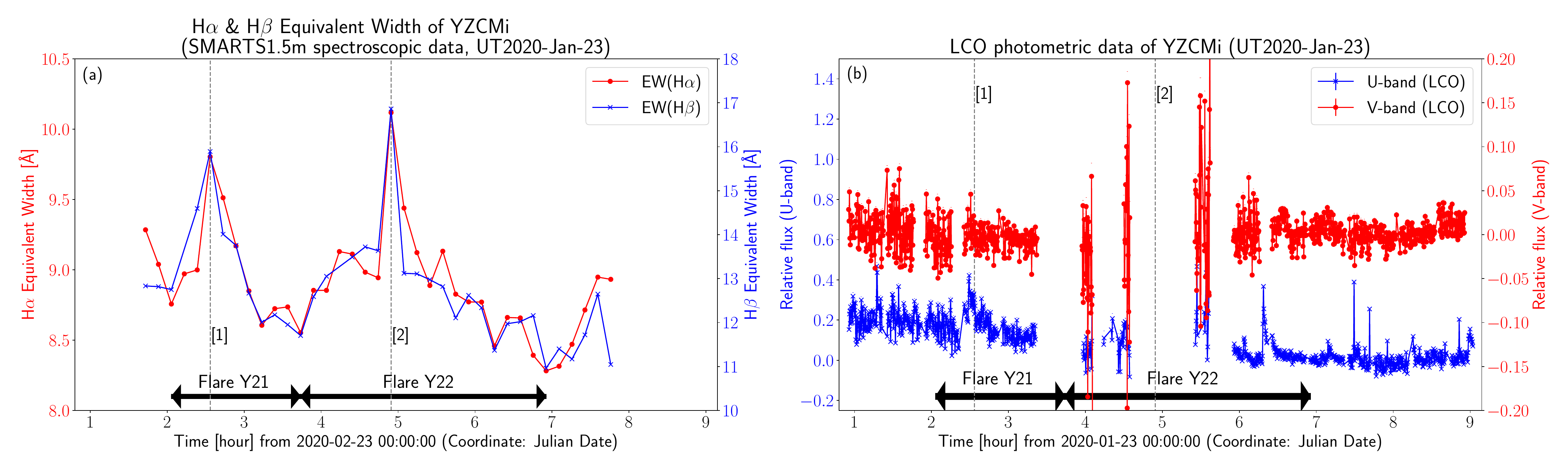}{1.0\textwidth}{\vspace{0mm}}}  
     \vspace{-5mm}
     \caption{
    \color{black}\textrm{  
Light curves of YZ CMi on 2020 January 23 showing Flares Y21 \& Y22, which are plotted 
similarly with Figures \ref{fig:lcEW_HaHb_YZCMi_UT200121} (a)\&(b).
The grey dashed lines with numbers ([1],[2]) correspond to the time shown with the same numbers in Figures \ref{fig:spec_HaHb_YZCMi_UT200123} \& \ref{fig:map_HaHb_YZCMi_UT200123}.
 } \color{black}
     }
   \label{fig:lcEW_HaHb_YZCMi_UT200123}
   \end{center}
 \end{figure}

On 2020 January 23, two flares (Flares Y21\& Y22) were detected in H$\alpha$ \& H$\beta$ lines as shown in Figure \ref{fig:lcEW_HaHb_YZCMi_UT200123} (a).  
As for Flare Y21, the H$\alpha$ \& H$\beta$ equivalent widths increased up to 9.8\AA~and 15.9\AA, respectively, and $\Delta t^{\rm{flare}}_{\rm{H}\alpha}$ is 1.7 hours (Table \ref{table:list1_flares}).
The continuum brightness observed with LCO $U$-band by
$\sim$10--20\% during Flare Y21 (Figure \ref{fig:lcEW_HaHb_YZCMi_UT200123} (b)). 
The brightness increase in $V$-band is not larger than the photometric error in $V$-band (\color{black}\textrm{$3\sigma_{V}$}\color{black}=3.7\%).
As for Flare Y22, the H$\alpha$ \& H$\beta$ equivalent widths increased up to 10.1\AA~and 16.9\AA, respectively, and $\Delta t^{\rm{flare}}_{\rm{H}\alpha}$ is 3.2 hours (Table \ref{table:list1_flares}).
For most of the time during Flare Y22, there were no LCO photometric observation data (Figure \ref{fig:lcEW_HaHb_YZCMi_UT200123} (b)), and so we cannot know whether there was the increase of the continuum brightness.
   \color{black}\textrm{ 
$L_{U}$, $L_{V}$, $E_{U}$, $E_{V}$, $L_{\rm{H}\alpha}$, $L_{\rm{H}\beta}$, $E_{\rm{H}\alpha}$, and $E_{\rm{H}\beta}$ values are estimated and listed in Table \ref{table:list1_flares}.
As for Flare Y22, no $L_{U}$, $L_{V}$, $E_{U}$, and $E_{V}$ values are estimated because of 
no LCO photometric observation data.
} \color{black}

\begin{figure}[ht!]
   \begin{center}
           \gridline{  
     \hspace{-0.06\textwidth}
    \fig{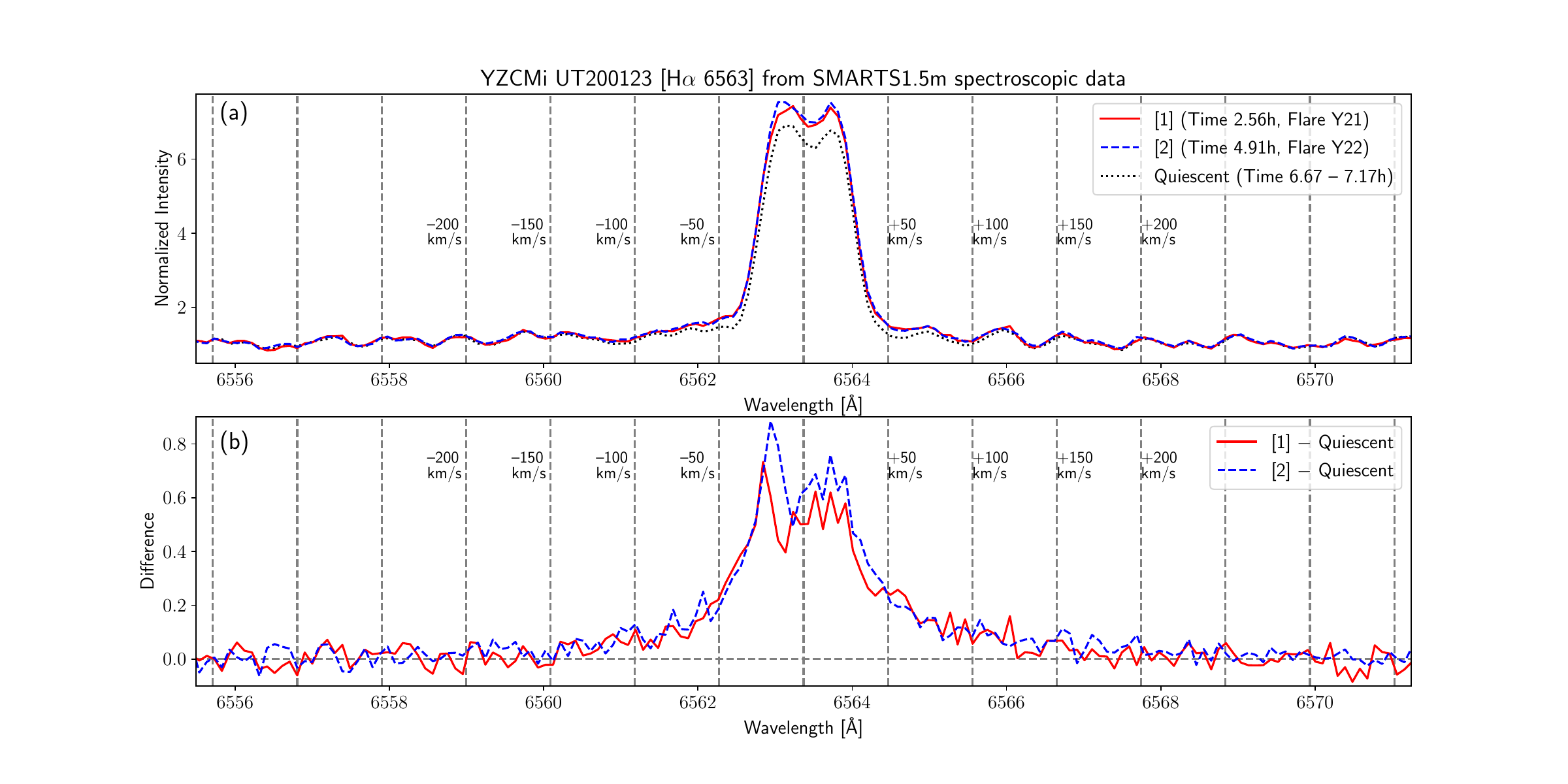}{0.58\textwidth}{\vspace{0mm}}
     \hspace{-0.06\textwidth}
       \fig{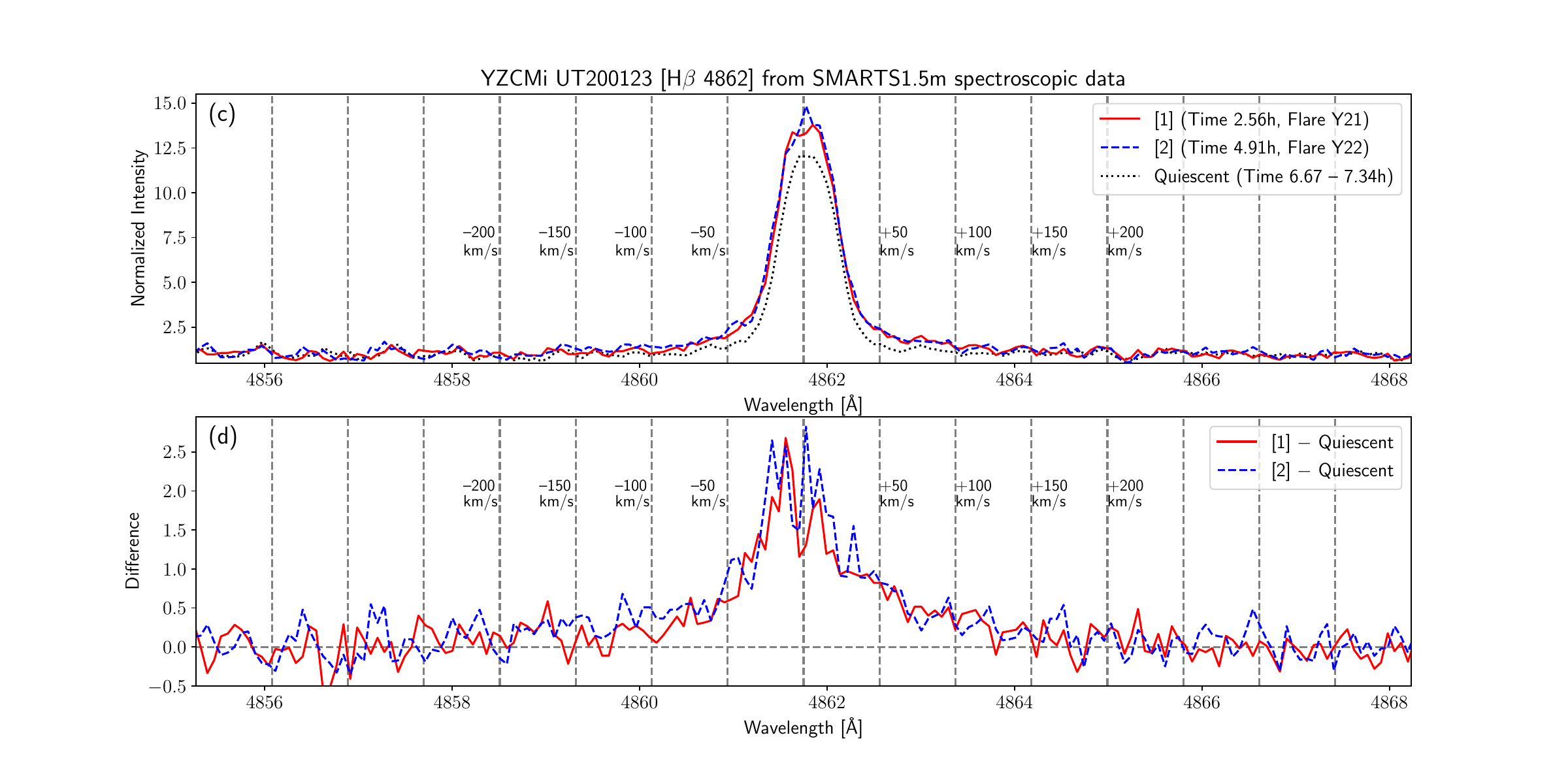}{0.58\textwidth}{\vspace{0mm}}
    }
     \vspace{-0.55cm}
     \caption{
   \color{black}\textrm{  
Line profiles of the H$\alpha$ \& H$\beta$ emission lines during Flares Y21 \& Y22 on 2020 January 23 (at the time [1] and [2]) from SMARTS1.5m spectroscopic data, which are plotted similarly with Figure \ref{fig:spec_HaHb_YZCMi_UT190127}.
 } \color{black}
     }
   \label{fig:spec_HaHb_YZCMi_UT200123}
   \end{center}
 \end{figure}

\begin{figure}[ht!]
   \begin{center}
      \gridline{
     \hspace{-0.07\textwidth}
      \fig{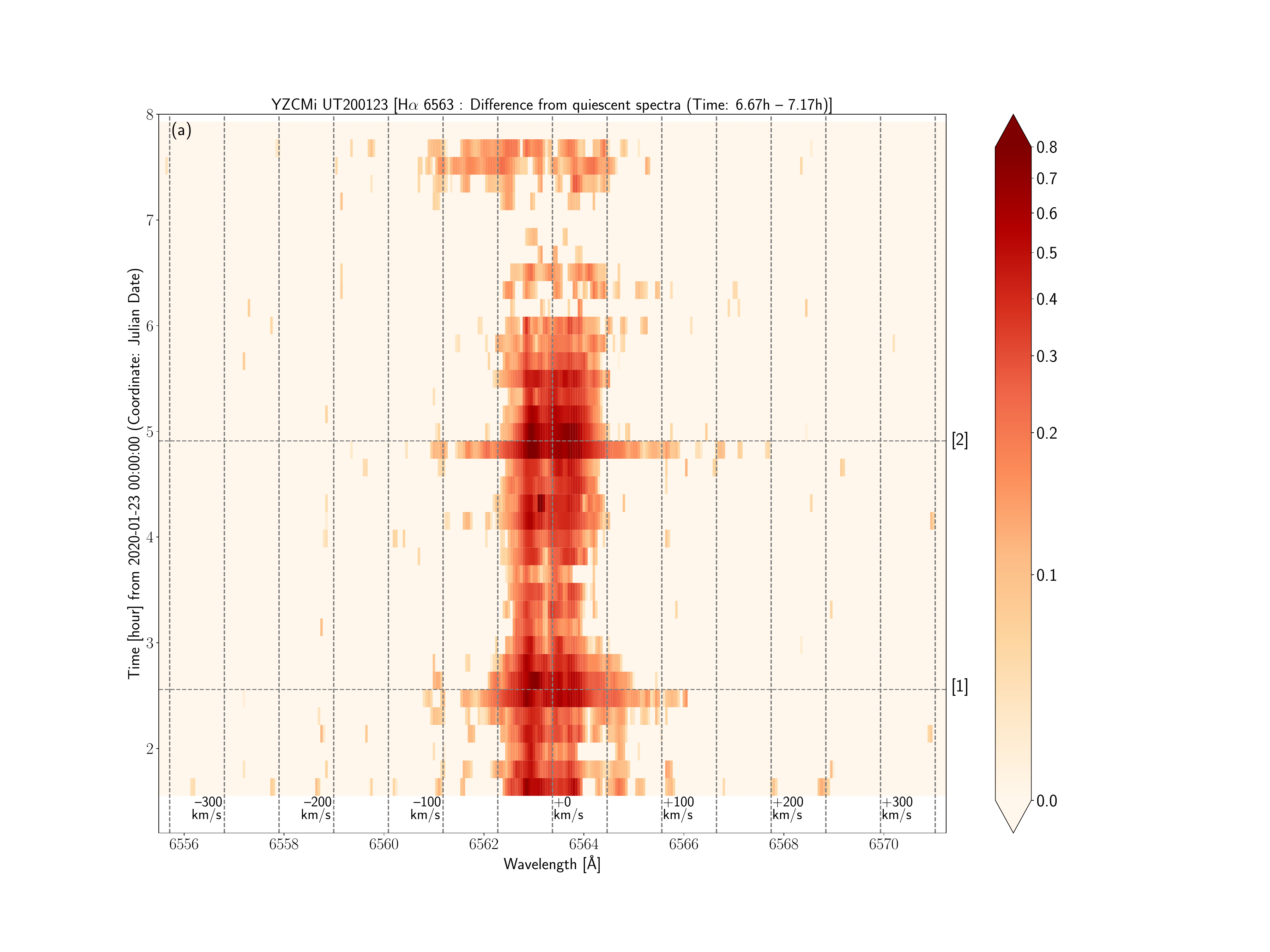}{0.63\textwidth}{\vspace{0mm}}
     \hspace{-0.11\textwidth}
    \fig{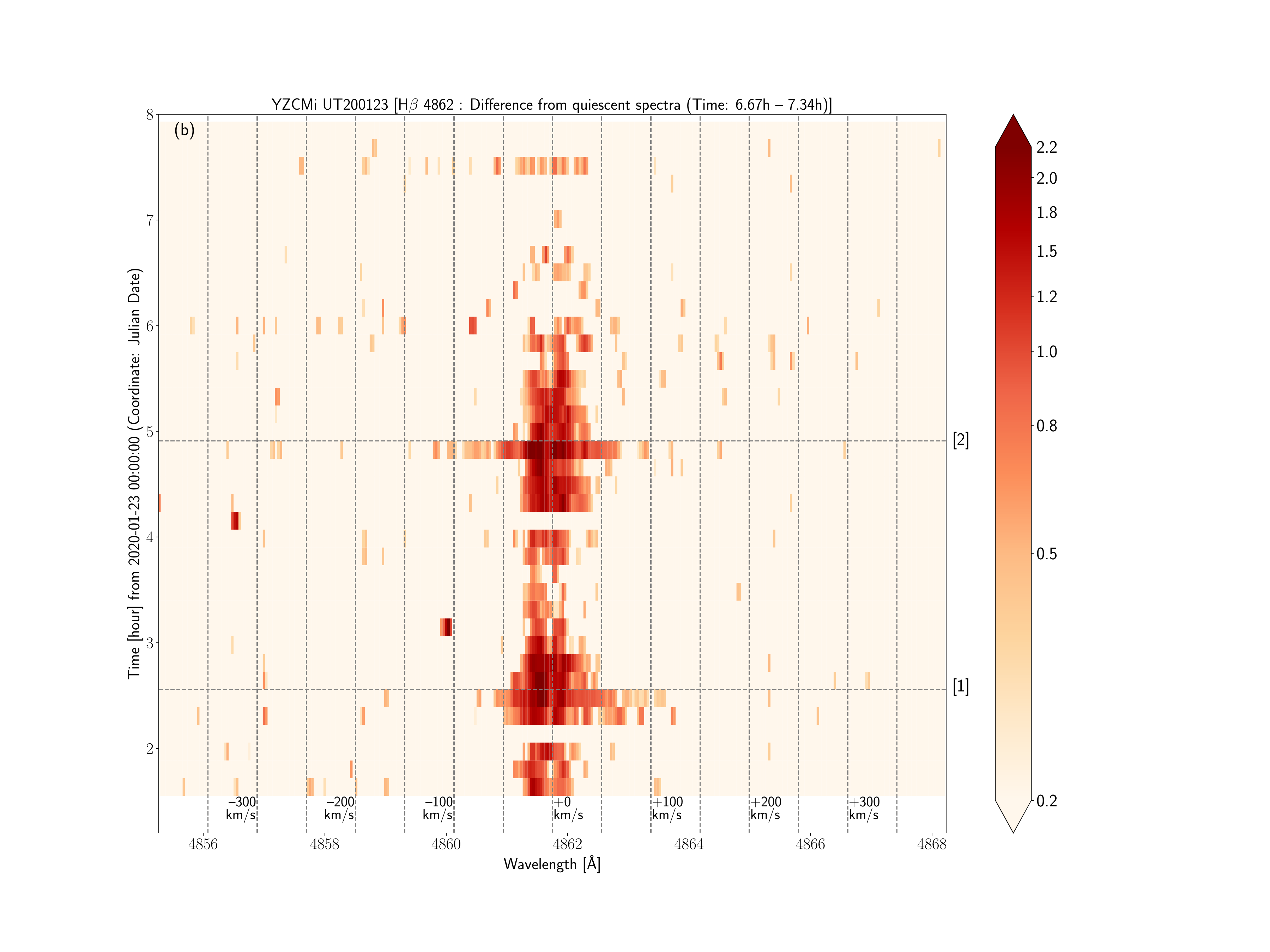}{0.63\textwidth}{\vspace{0mm}}
    }
     \vspace{-0.5cm}
     \caption{
         \color{black}\textrm{  
Time evolution of the H$\alpha$ \& H$\beta$ line profiles overing Flares Y21 \& Y22 
on 2020 January 23, which are plotted similarly with Figure \ref{fig:map_HaHb_YZCMi_UT191212}.
The grey horizontal dashed lines indicate the time [1] \& [2], which are shown in Figures \ref{fig:lcEW_HaHb_YZCMi_UT200123}  (light curves) and 
\ref{fig:spec_HaHb_YZCMi_UT200123} (line profiles).
}\color{black}
     }
   \label{fig:map_HaHb_YZCMi_UT200123}
   \end{center}
 \end{figure}

The H$\alpha$ \& H$\beta$ line profiles during Flares Y21 \& Y22 are shown in
Figures \ref{fig:spec_HaHb_YZCMi_UT200123} \& \ref{fig:map_HaHb_YZCMi_UT200123}. 
During Flares Y21 \& Y22, 
there were no clear blue or red wing asymmetries in H$\alpha$ and H$\beta$ lines 
(time [1],[2] in Figures \ref{fig:spec_HaHb_YZCMi_UT200123}(b) \& (d)), and the line profiles showed
roughly symmetrical broadenings with $\sim \pm$150--200 km s$^{-1}$ at around the peak time of the flares.

 \clearpage

\subsection{Flares Y25, Y26, Y27, \& Y28 observed on 2020 December 7} 
\label{subsec:results:2020-Dec-07} 

\begin{figure}[ht!]
   \begin{center}
   \gridline{
    \fig{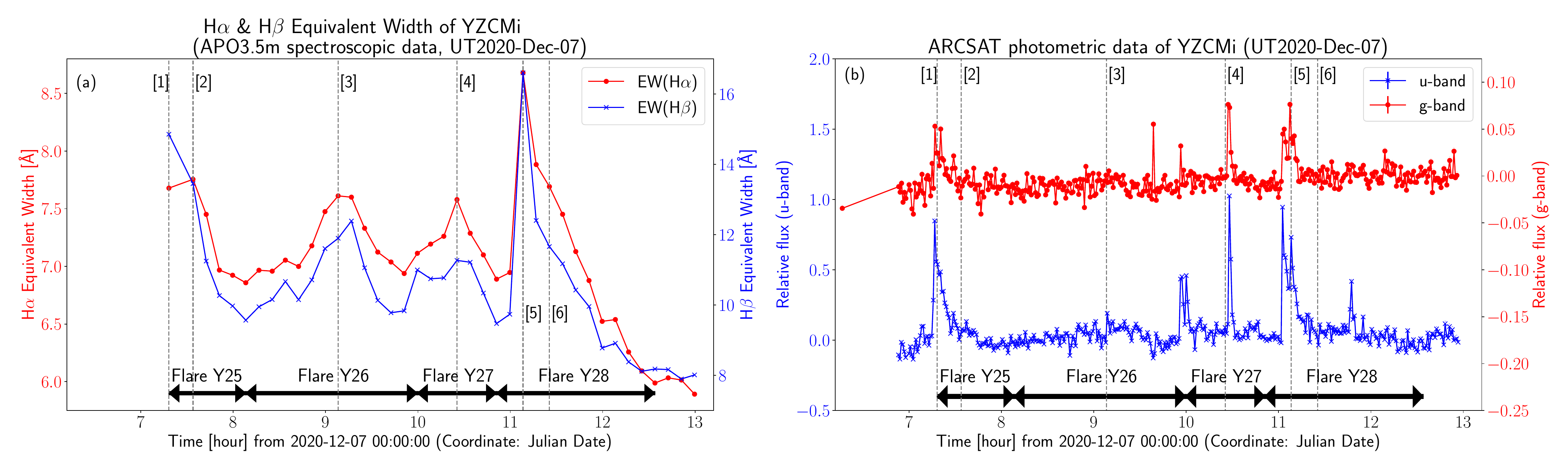}{1.0\textwidth}{\vspace{0mm}}}   
     \vspace{-5mm}
     \caption{
     \color{black}\textrm{  
Light curves of YZ CMi on 2020 December 7 showing Flares Y25, Y26, Y27, \& Y28,
which are plotted 
similarly with Figures \ref{fig:lcEW_HaHb_YZCMi_UT191212} (a)\&(b).
The grey dashed lines with numbers ([1]--[6]) 
correspond to the time shown with the same numbers in Figures \ref{fig:spec_HaHb_YZCMi_UT201207} \& \ref{fig:map_HaHb_YZCMi_UT201207}.
 } \color{black}
     }
   \label{fig:lcEW_HaHb_YZCMi_UT201207}
   \end{center}
 \end{figure}

On 2020 December 7, four flares (Flares Y25, Y26, Y27, \& Y28) 
were detected in H$\alpha$ \& H$\beta$ lines as shown in Figure \ref{fig:lcEW_HaHb_YZCMi_UT201207} (a).  
Flare Y25 already started when the spectroscopic observation
started.
The H$\alpha$ \& H$\beta$ equivalent widths decreased from 7.8\AA~and 14.9\AA, respectively, and $\Delta t^{\rm{flare}}_{\rm{H}\alpha}$
is $>$0.8 hours (Table \ref{table:list1_flares}).
The photometric observation captured a bit earlier phase of the flare since it started $\sim$0.5 hour before the spectroscopic observation started. 
During Flare Y25, the continuum brightness observed with ARCSAT $u$- \& $g$-bands increased by $\sim$80\% and $\sim$5--6\%, respectively 
(Figure \ref{fig:lcEW_HaHb_YZCMi_UT201207} (b)). 
As for Flare Y26, the H$\alpha$ \& H$\beta$ equivalent widths increased up to 7.6\AA~and 12.4\AA, respectively, and $\Delta t^{\rm{flare}}_{\rm{H}\alpha}$ is 1.9 hours (Table \ref{table:list1_flares}).
\color{black}\textrm{ 
The continuum brightness increases with ARCSAT $u$- \& $g$-bands are not enough larger than the photometric error (3$\sigma_{u}$=9.2\% and 3$\sigma_{g}$=2.3\%) and it is judged that there are no clear white-light emissions, although there are some slight possible increase in $u$-band at around 9.1-9.2h
(Figure \ref{fig:lcEW_HaHb_YZCMi_UT201207} (b)).
 } \color{black}
As for Flare Y27, the H$\alpha$ \& H$\beta$ equivalent widths increased up to 7.6\AA~and 11.3\AA, respectively, and $\Delta t^{\rm{flare}}_{\rm{H}\alpha}$ is 0.8 hours (Table \ref{table:list1_flares}).
In addition to these enhancements in Balmer emission lines, the continuum brightness observed with ARCSAT $u$- \& $g$-bands increased by $\sim$100\% and $\sim$7--8\%, respectively, during Flare Y27 (Figure \ref{fig:lcEW_HaHb_YZCMi_UT201207} (b))
As for Flare Y28, the H$\alpha$ \& H$\beta$ equivalent widths increased up to 8.7\AA~and 16.6\AA, respectively, and $\Delta t^{\rm{flare}}_{\rm{H}\alpha}$ is 1.7 hours (Table \ref{table:list1_flares}).
In addition to these enhancements in Balmer emission lines, the continuum brightness observed with ARCSAT $u$- \& $g$-bands increased by $\sim$90--100\% and $\sim$7--8\%, respectively, during Flare Y28 (Figure \ref{fig:lcEW_HaHb_YZCMi_UT201207} (b)).

 \color{black}\textrm{ 
$L_{u}$, $L_{g}$, $E_{u}$, $E_{g}$, $L_{\rm{H}\alpha}$, $L_{\rm{H}\beta}$, $E_{\rm{H}\alpha}$, and $E_{\rm{H}\beta}$ values (including upper limit values) 
are estimated and listed in Table \ref{table:list1_flares}.
Since the initial phase of Flare Y25 was not observed in the spectroscopic observation, the $L_{\rm{H}\alpha}$, $L_{\rm{H}\beta}$, $E_{\rm{H}\alpha}$, and $E_{\rm{H}\beta}$ values of Flare Y25 estimated here are only lower limit values.
} \color{black}

\begin{figure}[ht!]
   \begin{center}
            \gridline{  
     \hspace{-0.06\textwidth}
    \fig{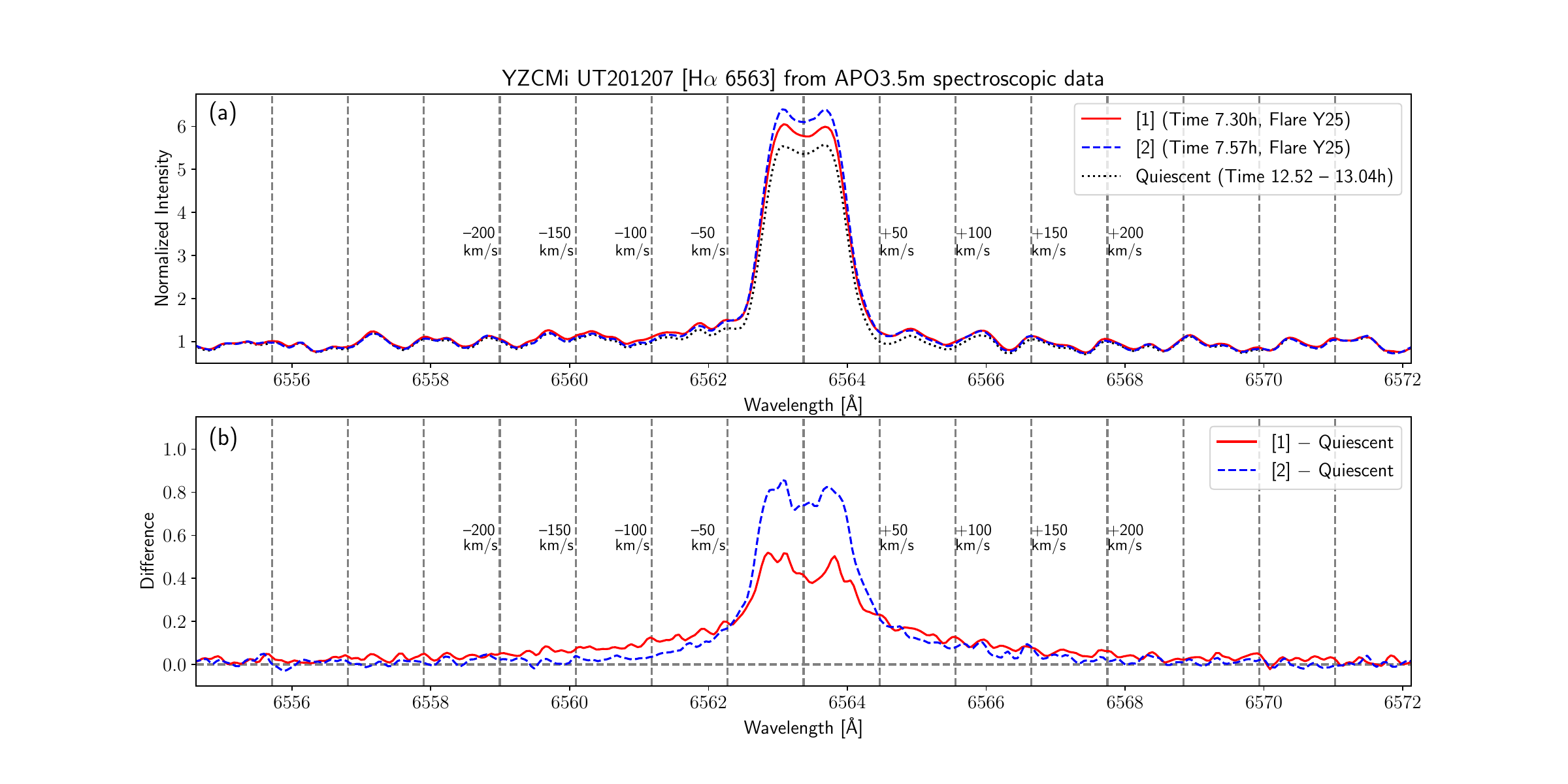}{0.58\textwidth}{\vspace{0mm}}
     \hspace{-0.06\textwidth}
       \fig{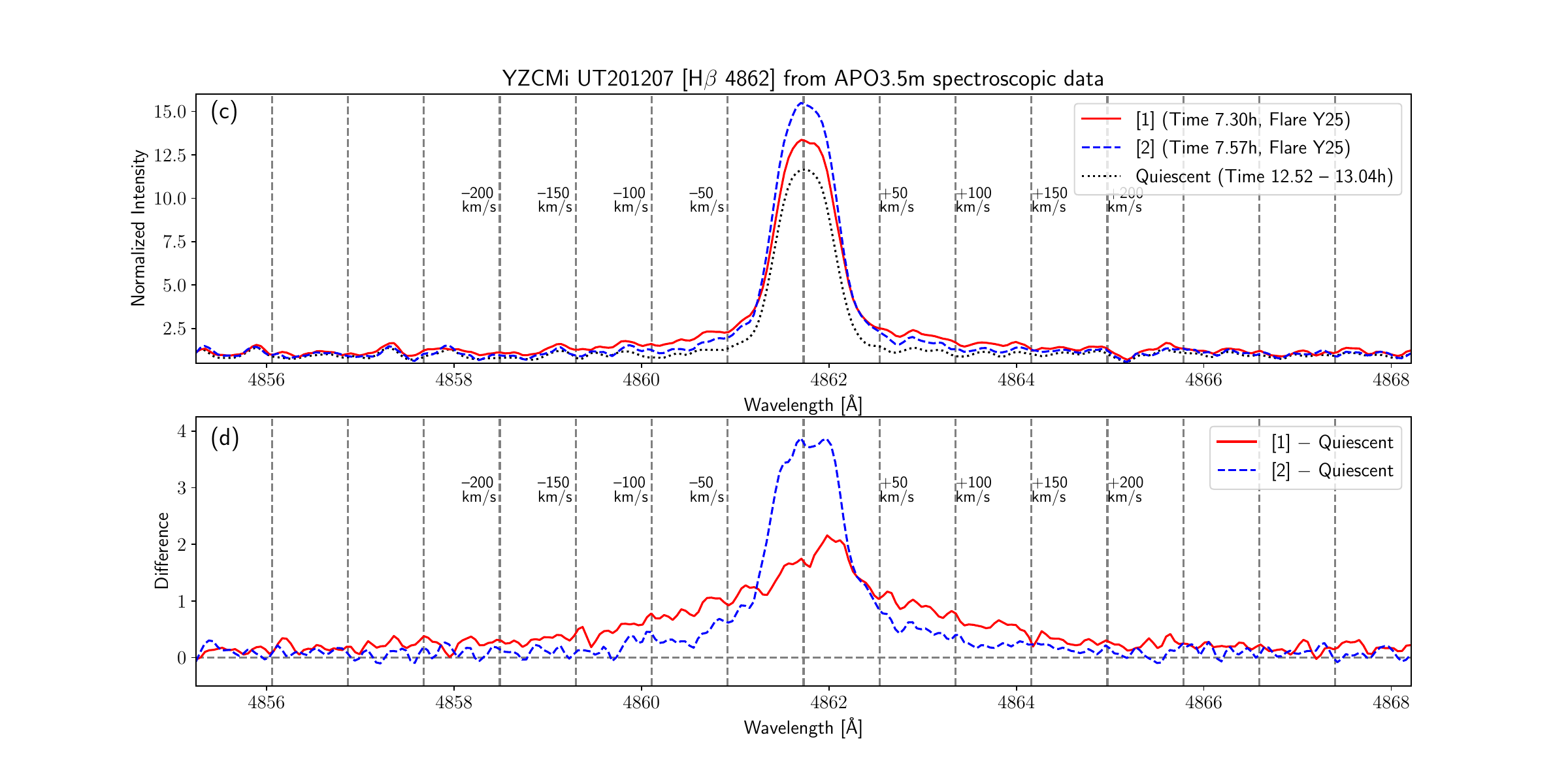}{0.58\textwidth}{\vspace{0mm}}
    }
     \vspace{-1.0cm}
            \gridline{  
     \hspace{-0.06\textwidth}
    \fig{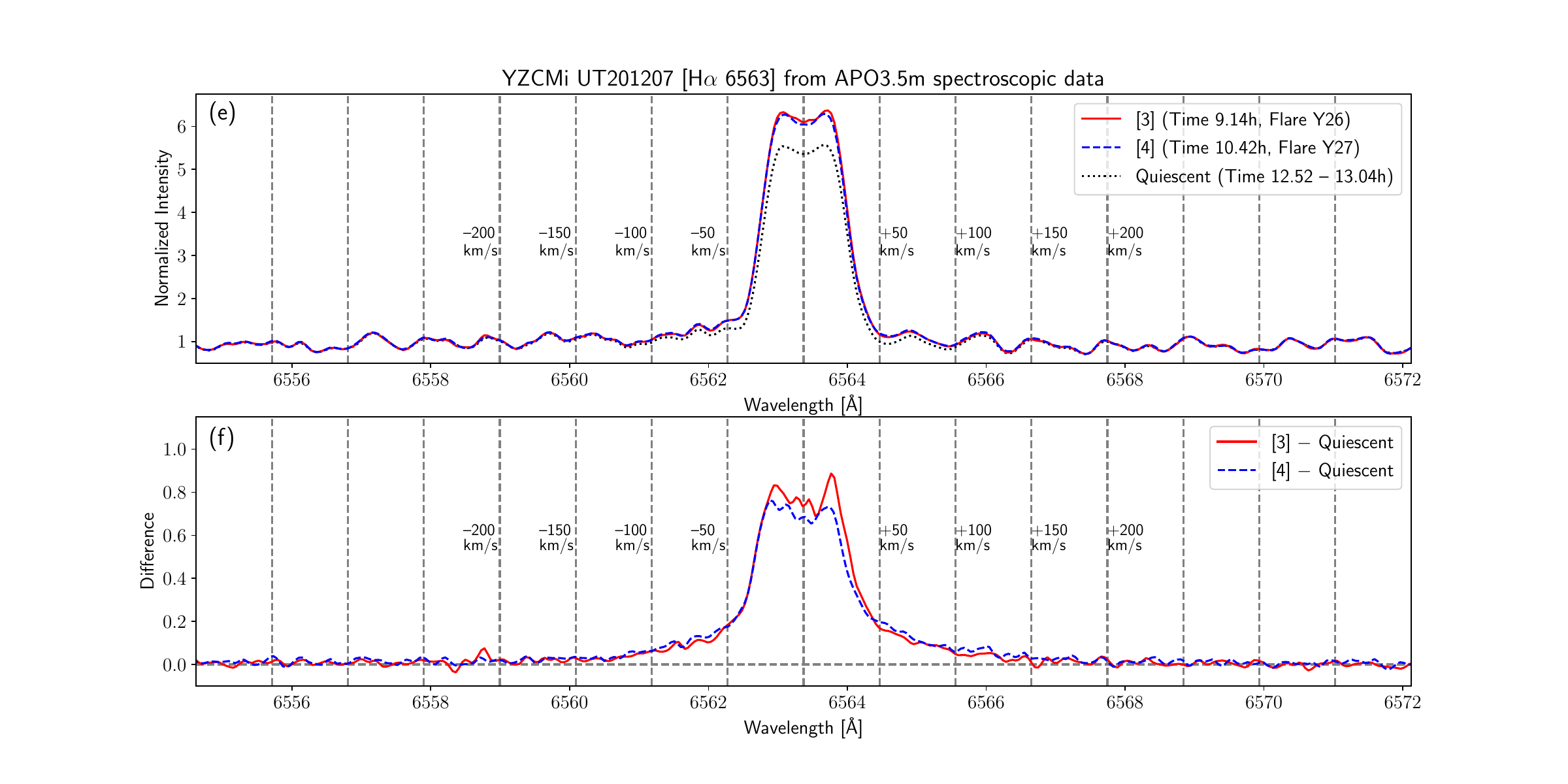}{0.58\textwidth}{\vspace{0mm}}
     \hspace{-0.06\textwidth}
       \fig{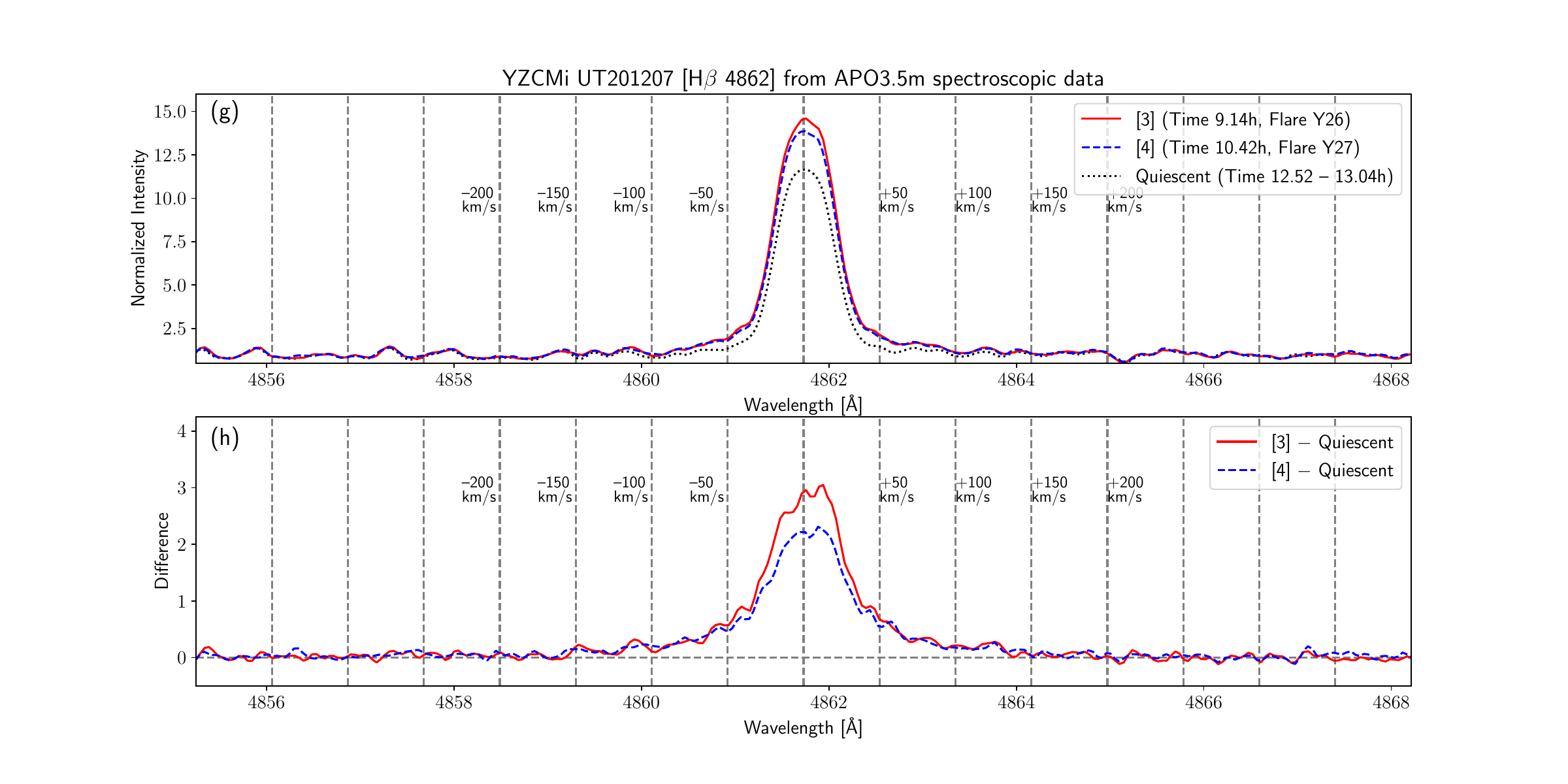}{0.58\textwidth}{\vspace{0mm}}
    }
     \vspace{-1.0cm}
            \gridline{  
     \hspace{-0.06\textwidth}
    \fig{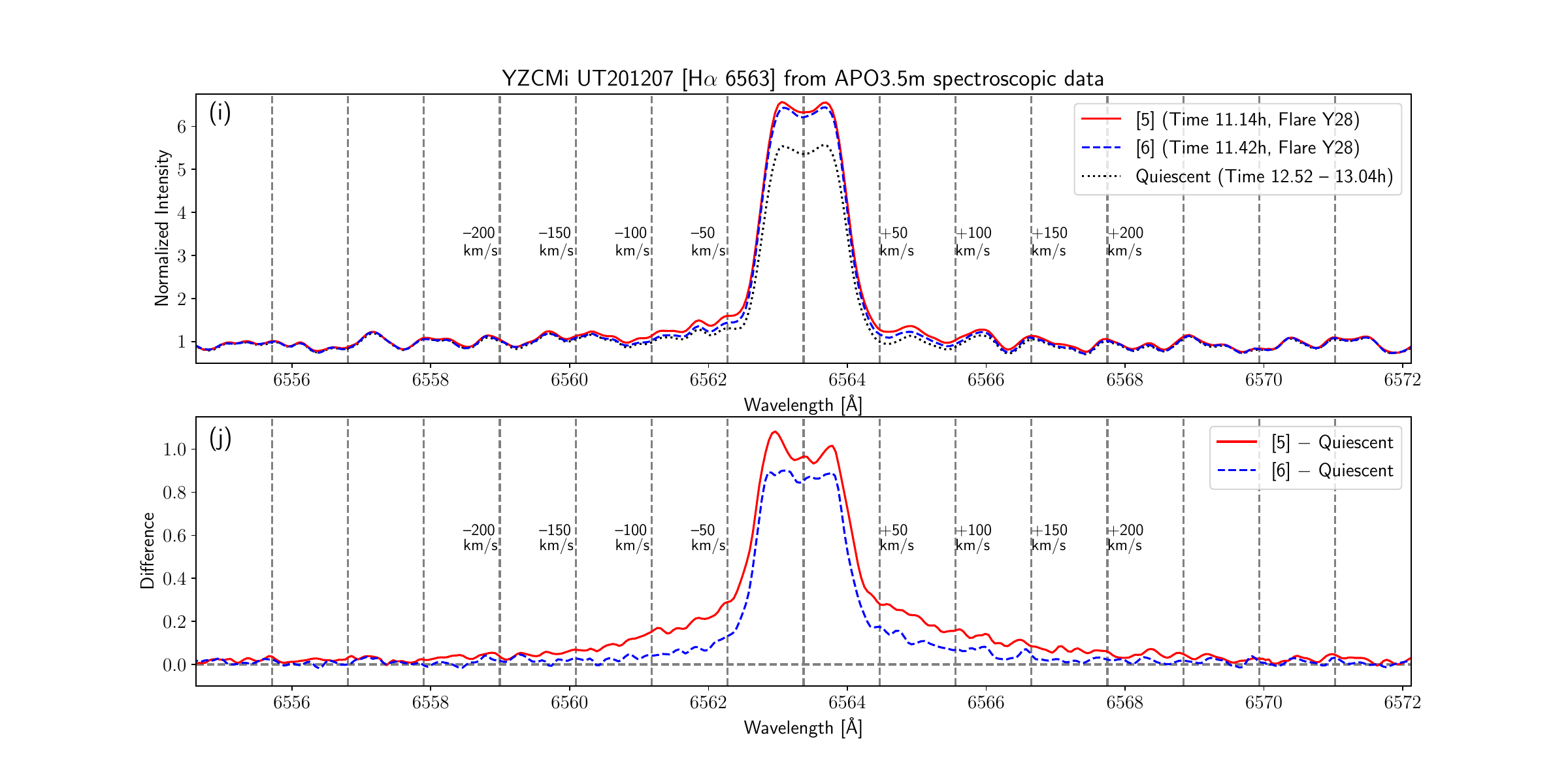}{0.58\textwidth}{\vspace{0mm}}
     \hspace{-0.06\textwidth}
       \fig{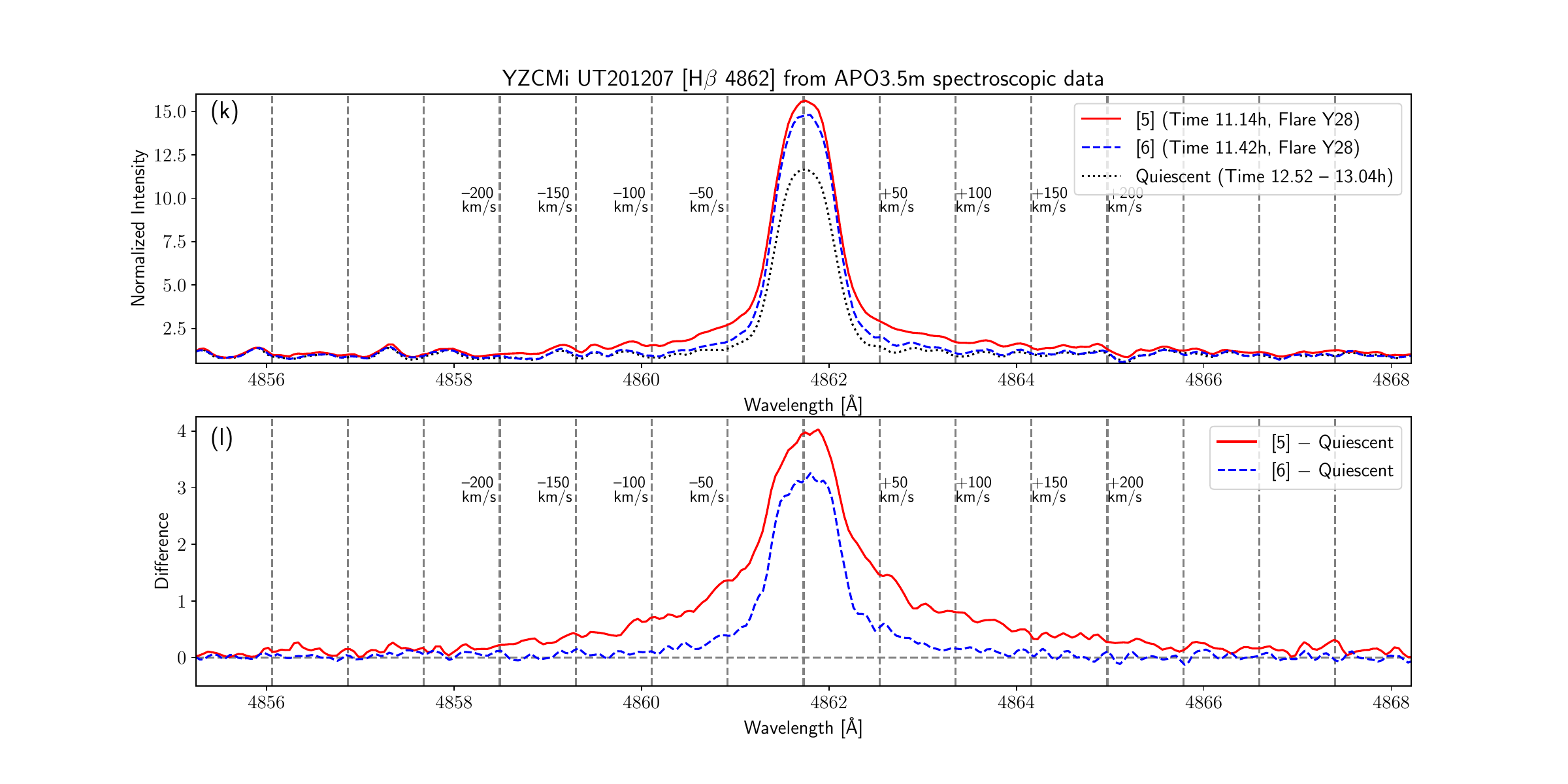}{0.58\textwidth}{\vspace{0mm}}
    }
     \vspace{-0.5cm}
     \caption{
   \color{black}\textrm{  
Line profiles of the H$\alpha$ \& H$\beta$ emission lines during Flares Y25, Y26, Y27, \& Y28 on 2020 December 7 (at the time [1]--[6]) from APO3.5m spectroscopic data, which are plotted similarly with Figure \ref{fig:spec_HaHb_YZCMi_UT190127}.
 } \color{black}
     }
   \label{fig:spec_HaHb_YZCMi_UT201207}
   \end{center}
 \end{figure}

\begin{figure}[ht!]
   \begin{center}
      \gridline{
     \hspace{-0.07\textwidth}
      \fig{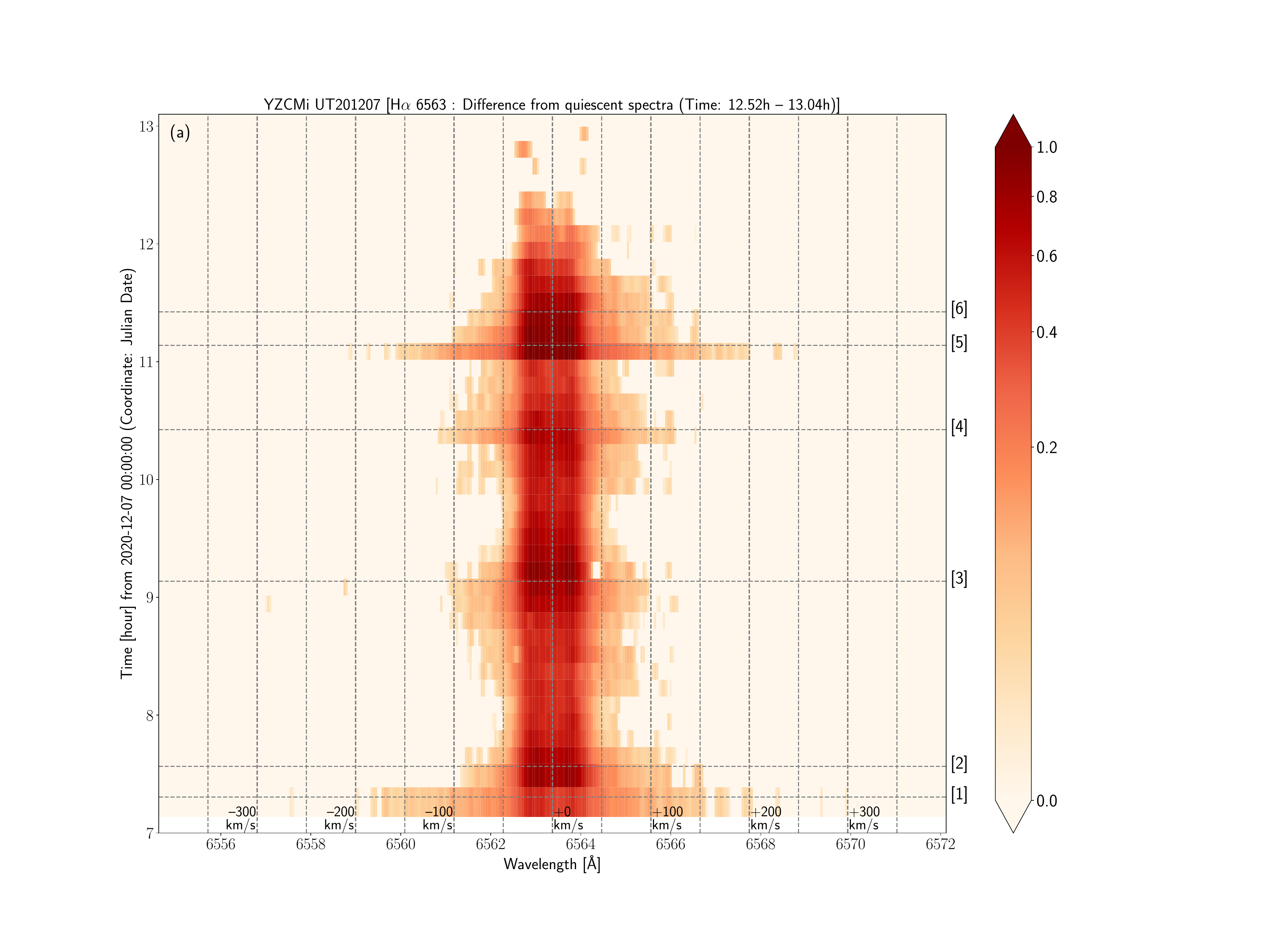}{0.63\textwidth}{\vspace{0mm}}
     \hspace{-0.11\textwidth}
    \fig{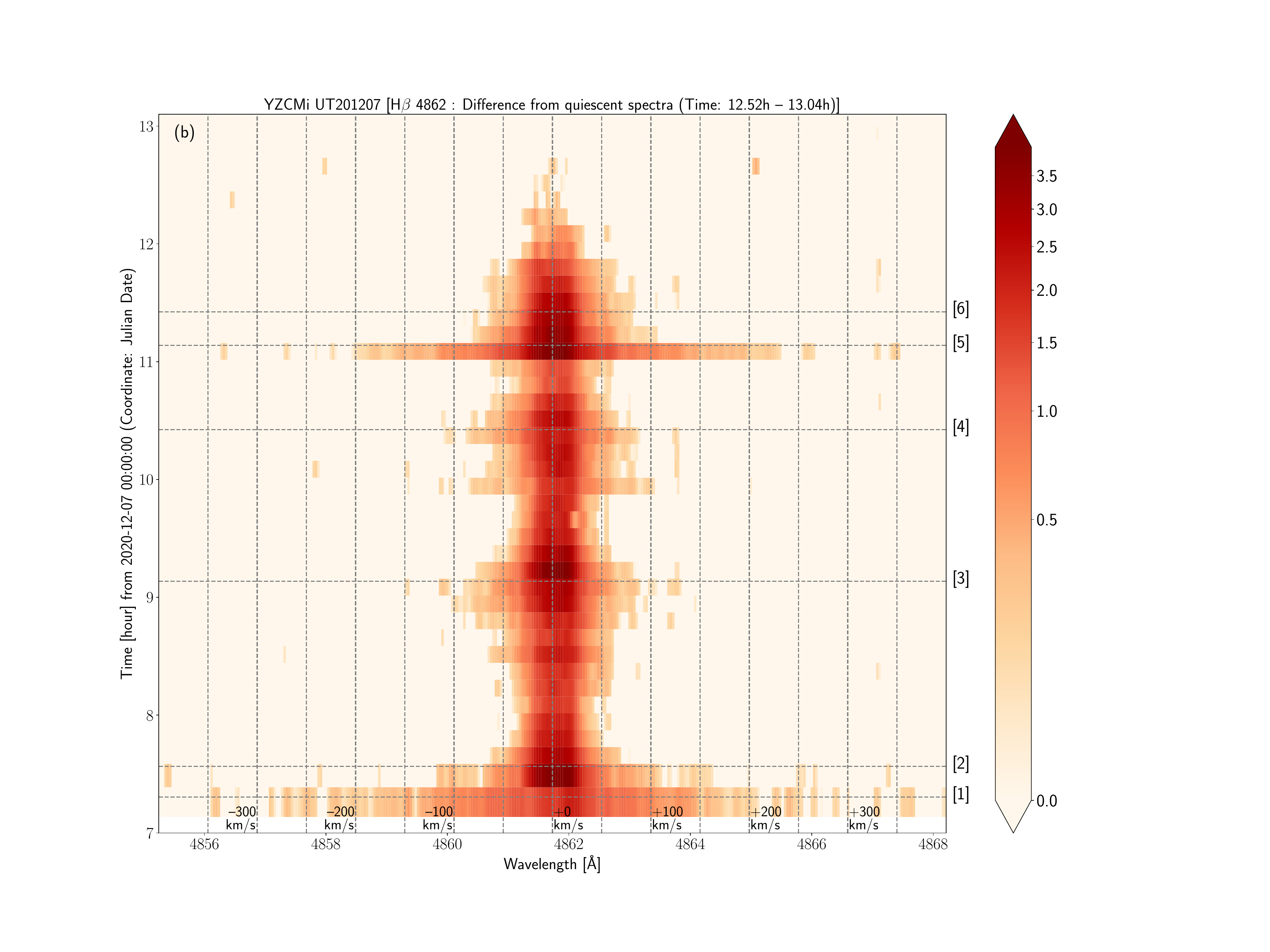}{0.63\textwidth}{\vspace{0mm}}
    }
     \vspace{-0.5cm}
     \caption{
         \color{black}\textrm{  
Time evolution of the H$\alpha$ \& H$\beta$ line profiles covering Flares Y25, Y26, Y27, \& Y28 on 2020 December 7, which are plotted similarly with Figure \ref{fig:map_HaHb_YZCMi_UT191212}.
The grey horizontal dashed lines indicate the time [1] -- [6], which are shown in Figure \ref{fig:lcEW_HaHb_YZCMi_UT201207} (light curves) and Figure \ref{fig:spec_HaHb_YZCMi_UT201207} (line profiles).
}\color{black}
     }
   \label{fig:map_HaHb_YZCMi_UT201207}
   \end{center}
 \end{figure}

The H$\alpha$ \& H$\beta$ line profiles during 
Flares Y25, Y26, Y27, \& Y28 are shown in
Figures \ref{fig:spec_HaHb_YZCMi_UT201207} \& \ref{fig:map_HaHb_YZCMi_UT201207}.
During Flare Y25, 
there were no clear blue or red wing asymmetries in H$\alpha$ and H$\beta$ lines 
(time [1] \& [2] in Figures \ref{fig:spec_HaHb_YZCMi_UT201207} \& \ref{fig:map_HaHb_YZCMi_UT201207}), 
and the line profiles showed
roughly symmetrical broadenings with $\sim\pm$150--200 km s$^{-1}$ (H$\alpha$) and 
$\sim\pm$250--300 km s$^{-1}$ (H$\beta$) at around the peak time of the flares (time [1] in Figures \ref{fig:spec_HaHb_YZCMi_UT201207} \& \ref{fig:map_HaHb_YZCMi_UT201207}).
During Flares Y26 \& Y27, 
there were no clear blue or red wing asymmetries in H$\alpha$ and H$\beta$ lines 
(time [3] \& [4] in Figures \ref{fig:spec_HaHb_YZCMi_UT201207} \& \ref{fig:map_HaHb_YZCMi_UT201207}), 
and the line profiles showed
roughly symmetrical broadenings with $\sim\pm$100--150 km s$^{-1}$ at around the peak time of the flares.
During Flare Y28, 
there were no clear blue wing asymmetries in H$\alpha$ and H$\beta$ lines 
(time [5] \& [6] in Figures \ref{fig:spec_HaHb_YZCMi_UT201207} \& \ref{fig:map_HaHb_YZCMi_UT201207}), 
and the line profiles showed
broadenings with $\sim\pm$150--200 km s$^{-1}$ (H$\alpha$) and 
$\sim\pm$200--250 km s$^{-1}$ (H$\beta$) at around the peak time of the flares (time [5] in Figures \ref{fig:spec_HaHb_YZCMi_UT201207} \& \ref{fig:map_HaHb_YZCMi_UT201207}).

\clearpage
 
 \subsection{Flare Y29 observed on 2021 January 31} 
\label{subsec:results:2021-Jan-31} 

       \begin{figure}[ht!]
   \begin{center}
   \gridline{
    \fig{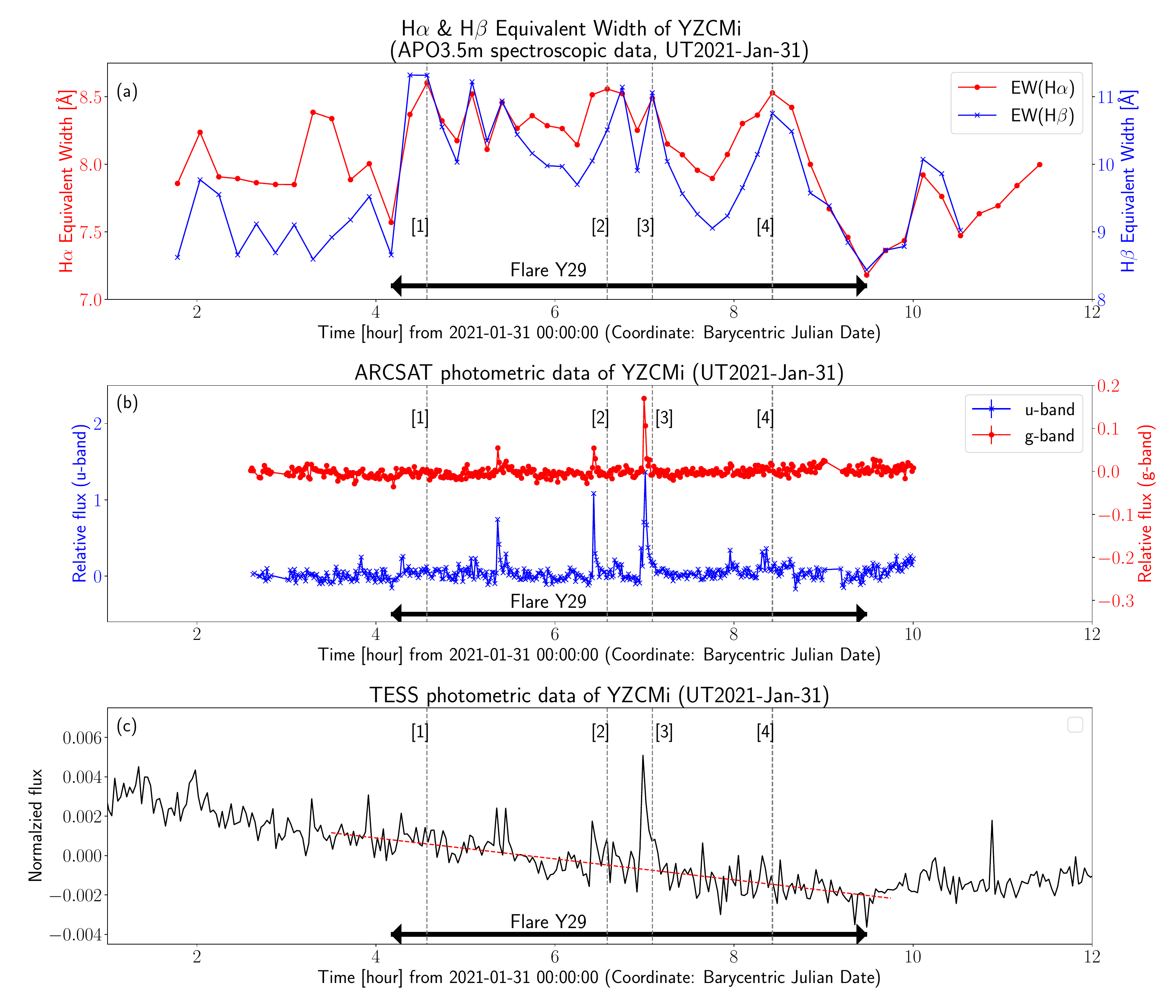}{0.85\textwidth}{\vspace{0mm}}}   
     \vspace{-0.5cm}
     \caption{
     \color{black}\textrm{  
Light curves of YZ CMi on 2021 January 31 showing Flare Y29, which are plotted 
similarly with Figures \ref{fig:lcEW_HaHb_YZCMi_UT190127} (a)--(c).
The grey dashed lines with numbers ([1]--[4]) correspond to the time shown with the same numbers in Figures \ref{fig:spec_HaHb_YZCMi_UT210131} \& \ref{fig:map_HaHb_YZCMi_UT210131}.
In (c), the red solid line is plotted to identify quiescent level of the TESS data.
 } \color{black}
     }
   \label{fig:lcEW_HaHb_YZCMi_UT210131}
   \end{center}
 \end{figure}

\color{black}\textrm{ 
On 2021 January 31,
one flare (Flare Y29) were detected in H$\alpha$ \& H$\beta$ lines as shown in Figure \ref{fig:lcEW_HaHb_YZCMi_UT210131} (a).  
During Flare Y29,
the H$\alpha$ \& H$\beta$ equivalent widths increased up to 8.6\AA~and 11.3\AA, respectively, and $\Delta t^{\rm{flare}}_{\rm{H}\alpha}$ is 5.3 hours (Table \ref{table:list1_flares}). 
In addition to the enhancements in Balmer emission lines, the continuum brightness observed by ARCSAT $u$- \& $g$-bands and \textit{TESS} increased by $\sim$140\%, $\sim$17--18\%, and $\sim$0.5\%, respectively, during Flare Y29 (Figures \ref{fig:lcEW_HaHb_YZCMi_UT210131} (b) \& (c)). 
$L_{u}$, $L_{g}$, $L_{TESS}$, $E_{u}$, $E_{g}$, $E_{TESS}$, $L_{\rm{H}\alpha}$, $L_{\rm{H}\beta}$, $E_{\rm{H}\alpha}$, and $E_{\rm{H}\beta}$ values are estimated and listed in Table \ref{table:list1_flares}.
} \color{black}

\begin{figure}[ht!]
   \begin{center}
            \gridline{  
     \hspace{-0.06\textwidth}
    \fig{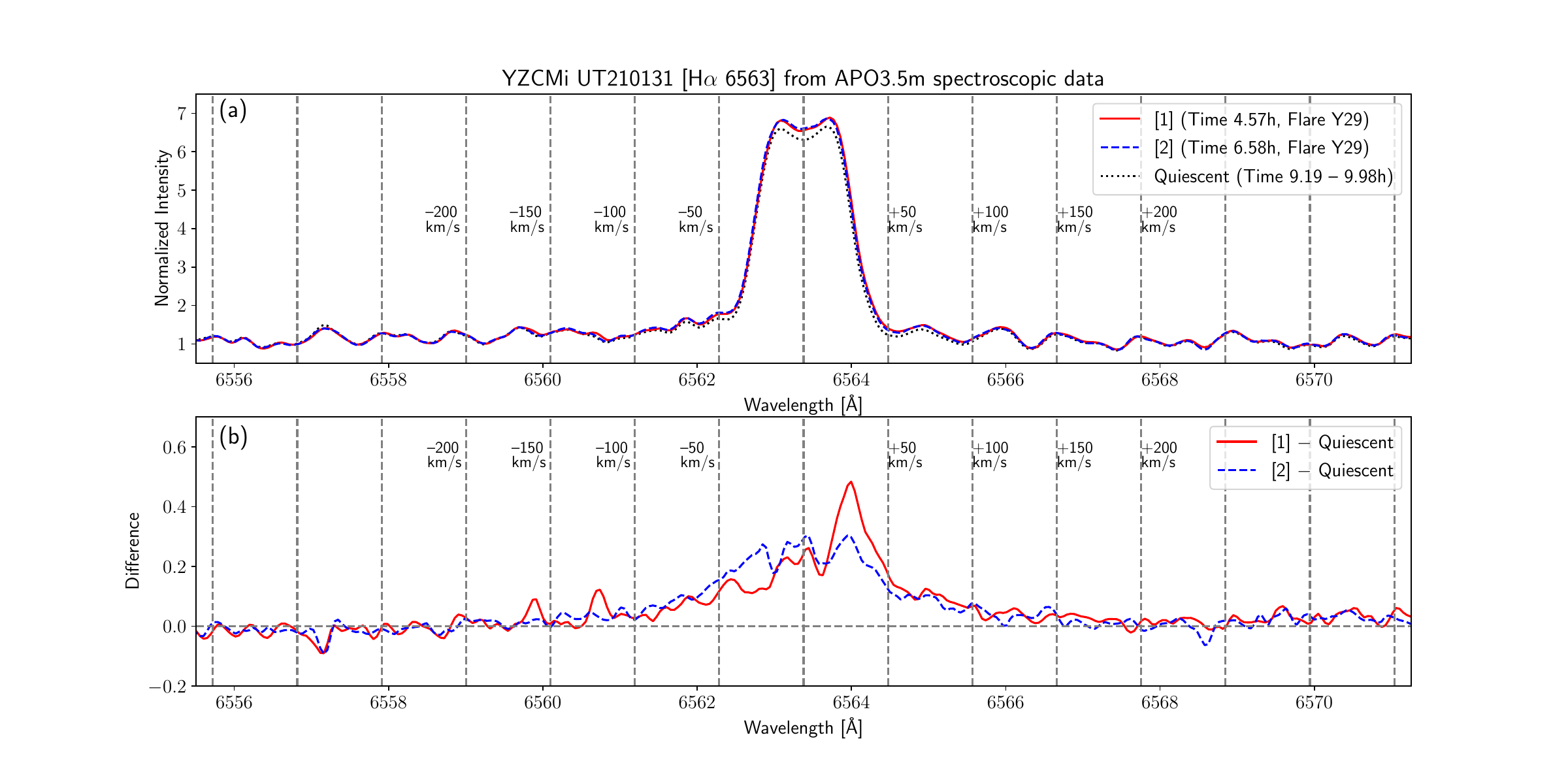}{0.58\textwidth}{\vspace{0mm}}
     \hspace{-0.06\textwidth}
       \fig{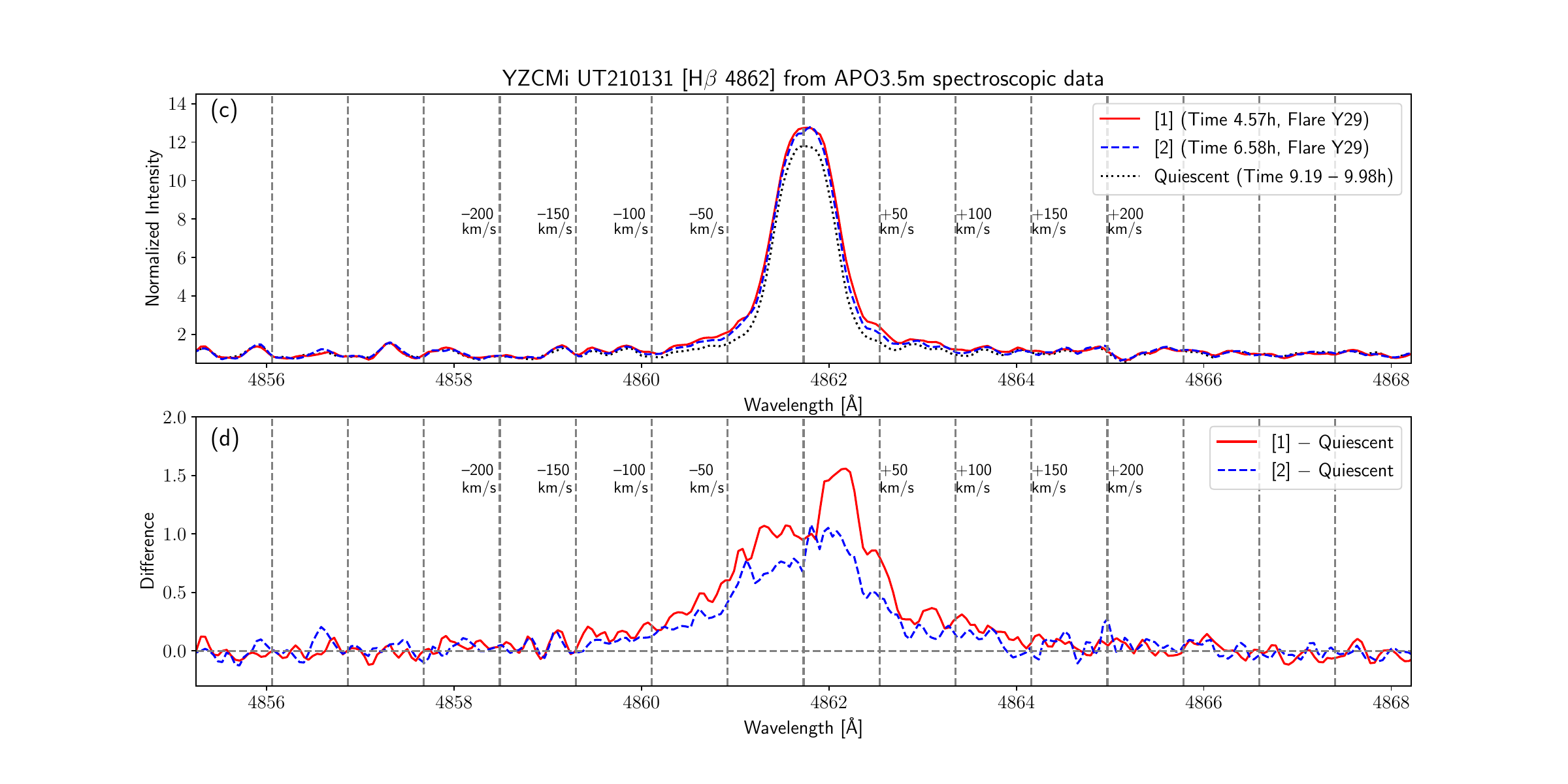}{0.58\textwidth}{\vspace{0mm}}
    }
     \vspace{-1.0cm}
            \gridline{  
     \hspace{-0.06\textwidth}
    \fig{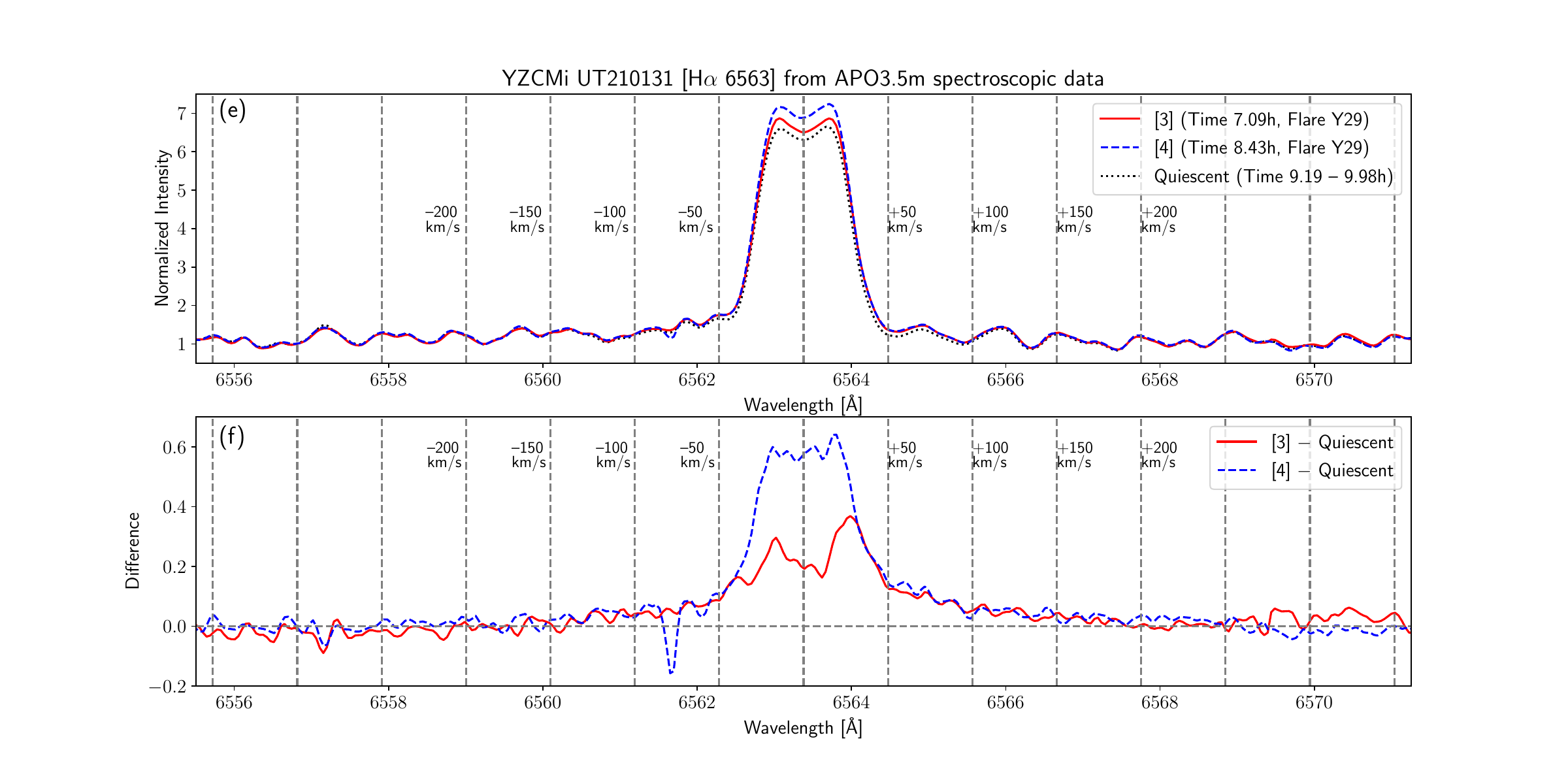}{0.58\textwidth}{\vspace{0mm}}
     \hspace{-0.06\textwidth}
       \fig{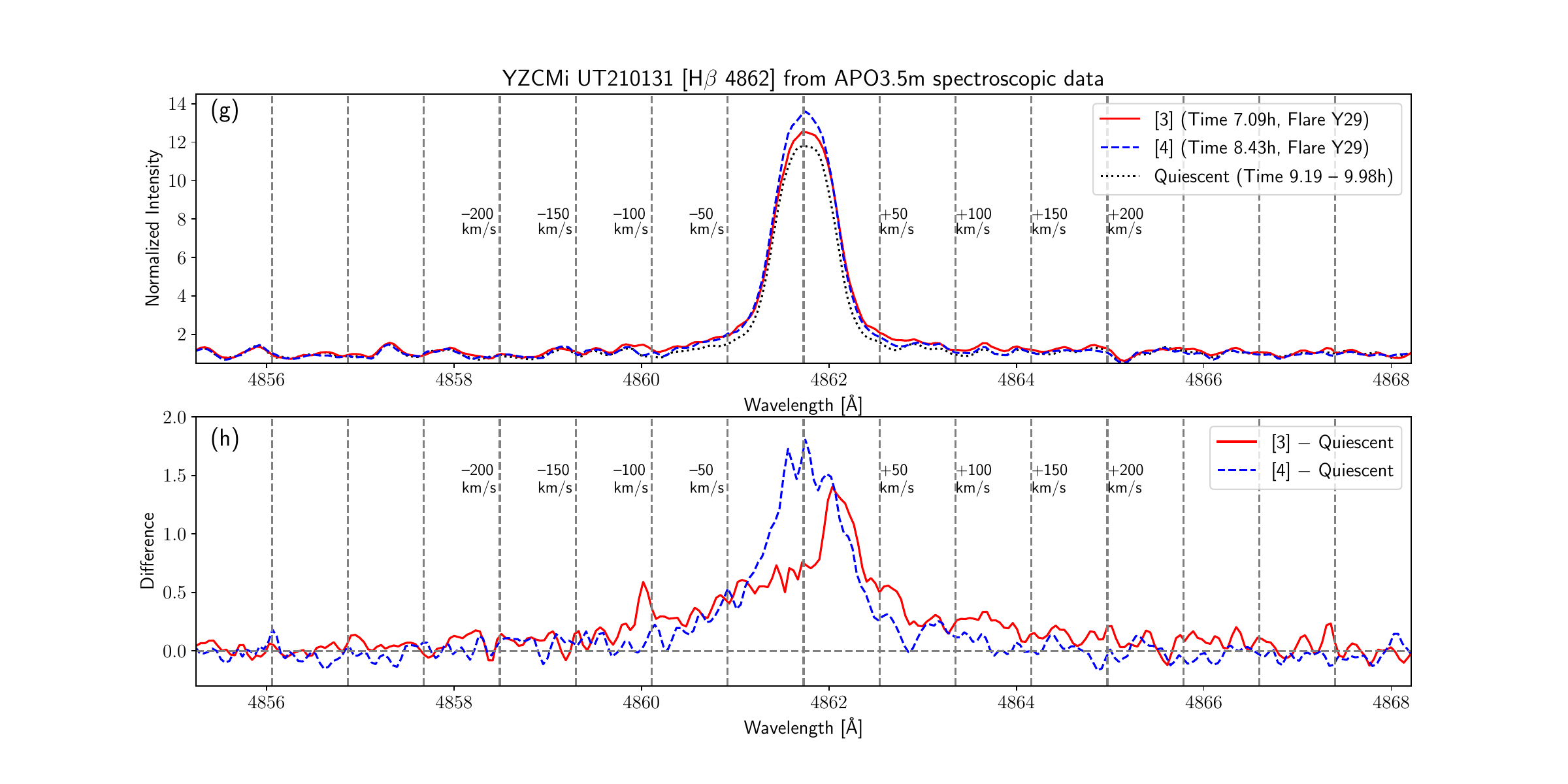}{0.58\textwidth}{\vspace{0mm}}
    }
     \vspace{-0.5cm}
     \caption{
   \color{black}\textrm{  
Line profiles of the H$\alpha$ \& H$\beta$ emission lines during Flare Y29 on 2021 January 31 (at the time [1]--[4]) from APO3.5m spectroscopic data, which are plotted similarly with Figure \ref{fig:spec_HaHb_YZCMi_UT190127}.
 } \color{black}
     }
   \label{fig:spec_HaHb_YZCMi_UT210131}
   \end{center}
 \end{figure}

\begin{figure}[ht!]
   \begin{center}
      \gridline{
     \hspace{-0.07\textwidth}
      \fig{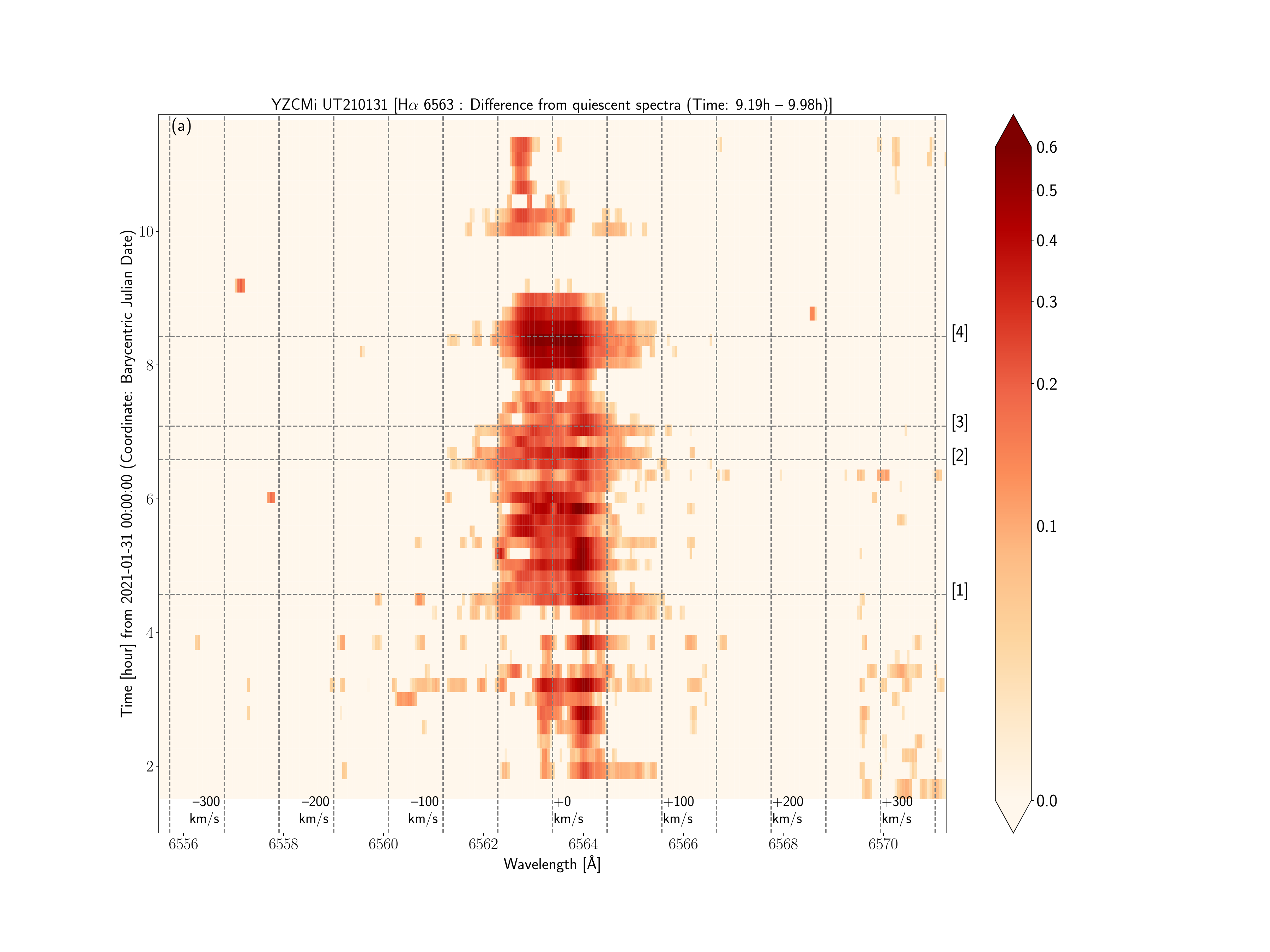}{0.63\textwidth}{\vspace{0mm}}
     \hspace{-0.11\textwidth}
    \fig{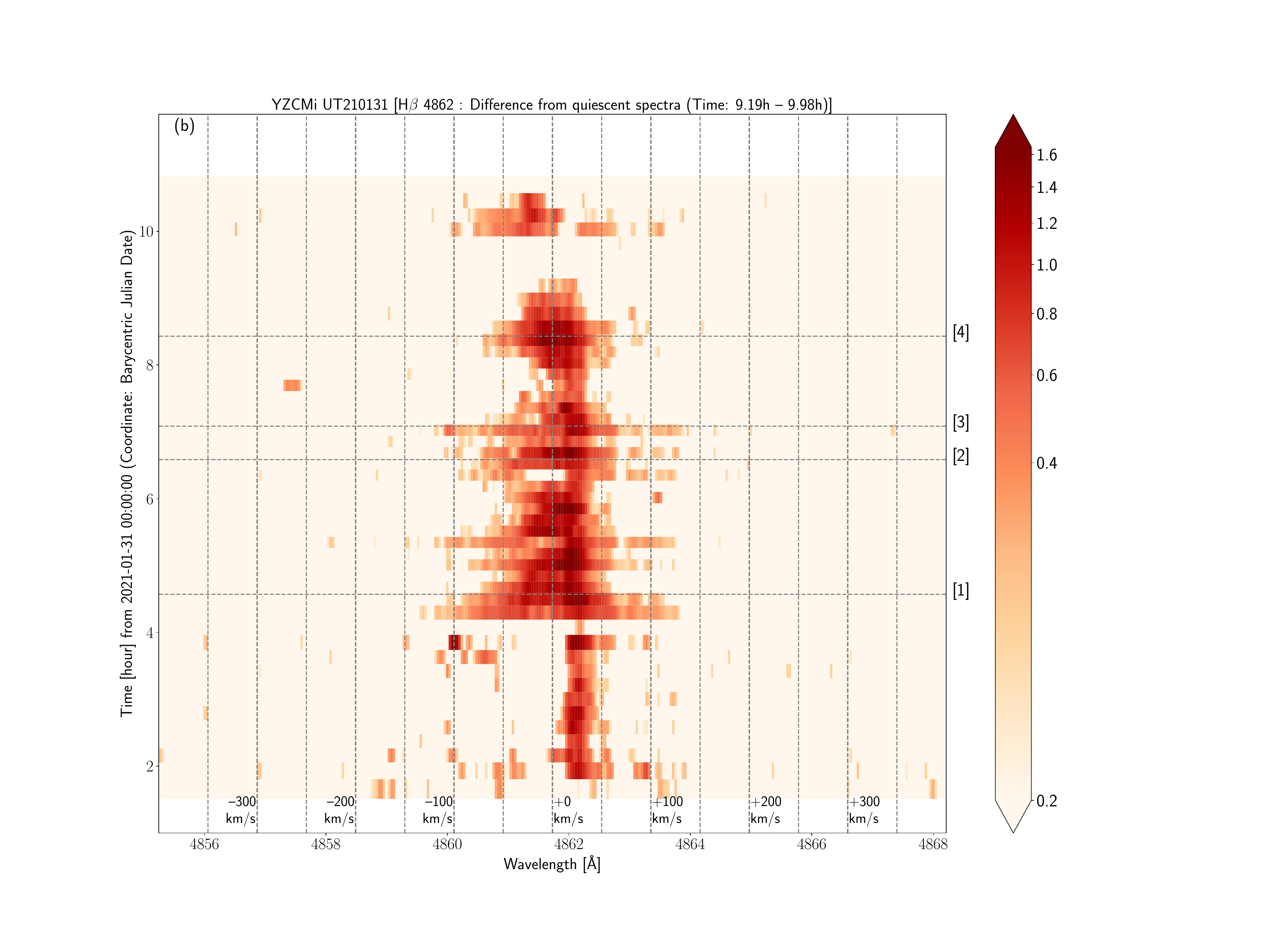}{0.63\textwidth}{\vspace{0mm}}
    }
     \vspace{-0.5cm}
     \caption{
         \color{black}\textrm{  
Time evolution of the H$\alpha$ \& H$\beta$ line profiles covering Flare Y29 
on 2021 January 31, which are plotted similarly with Figure \ref{fig:map_HaHb_YZCMi_UT190127}.
The grey horizontal dashed lines indicate the time [1]--[4], which are shown in Figure \ref{fig:lcEW_HaHb_YZCMi_UT210131} (light curves) and Figure \ref{fig:spec_HaHb_YZCMi_UT210131} (line profiles).
}\color{black}
     }
   \label{fig:map_HaHb_YZCMi_UT210131}
   \end{center}
 \end{figure}

The H$\alpha$ \& H$\beta$ line profiles during 
\color{black}\textrm{Flare Y29 }\color{black} are shown in
Figures \ref{fig:spec_HaHb_YZCMi_UT210131} \& \ref{fig:map_HaHb_YZCMi_UT210131}.
During \color{black}\textrm{Flare Y29}\color{black}, 
there were no clear blue or red wing asymmetries in H$\alpha$ and H$\beta$ lines 
(time [1] - [4] in Figures \ref{fig:spec_HaHb_YZCMi_UT210131} \& \ref{fig:map_HaHb_YZCMi_UT210131}), 
and the line profiles showed
roughly symmetrical broadenings with $\sim\pm$100--150 km s$^{-1}$ at around the peak time of the emission changes.

  \clearpage
  
 \subsection{Flares E3 \& E4 observed on 2020 August 26} 
\label{subsec:results:2020-Aug-26}

\begin{figure}[ht!]
   \begin{center}
    \gridline{
      \fig{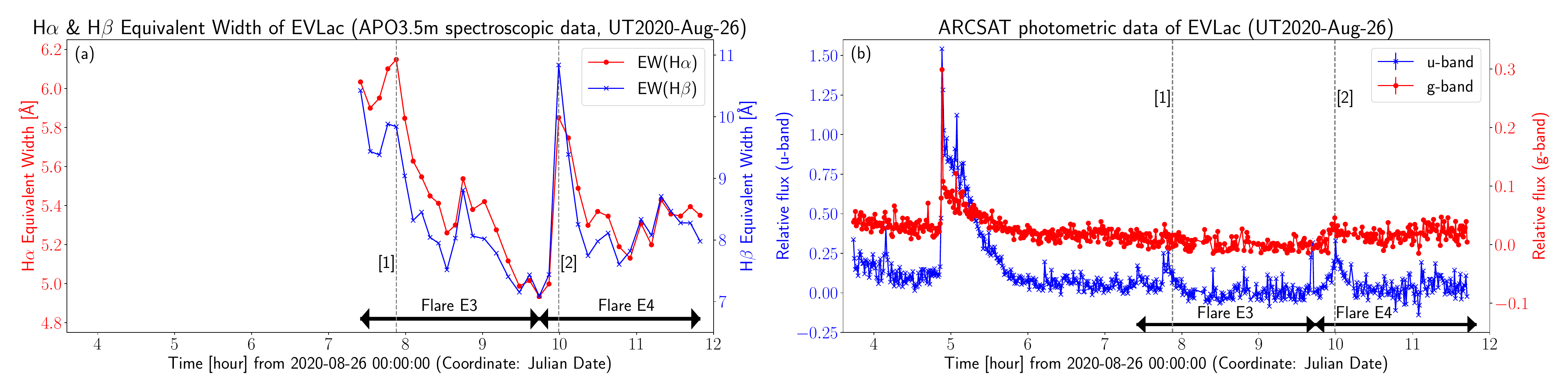}{1.0\textwidth}{\vspace{0mm}}
    }
     \vspace{-5mm}
     \caption{
     \color{black}\textrm{  
Light curves of EV Lac on 2020 August 26 showing Flares E3 \& E4, which are plotted 
similarly with Figures \ref{fig:lcEW_HaHb_YZCMi_UT191212} (a)\&(b).
The grey dashed lines with numbers ([1]\&[2]) correspond to the time shown 
with the same numbers in Figures \ref{fig:spec_HaHb_EVLac_UT200826} \& \ref{fig:map_HaHb_EVLac_UT200826}.
 } \color{black}
     }
   \label{fig:lcEW_HaHb_EVLac_UT200826}
   \end{center}
 \end{figure}

On 2020 August 26, two flares (Flares E3 \& E4) 
were detected on EV Lac in H$\alpha$ \& H$\beta$ lines 
as shown in Figure \ref{fig:lcEW_HaHb_EVLac_UT200826} (a).  
Flare E3 already started before the spectroscopic observation started.
As for Flare E3, the H$\alpha$ \& H$\beta$ equivalent widths decreased from 6.1\AA~and 10.4\AA, respectively, and $\Delta t^{\rm{flare}}_{\rm{H}\alpha}$ is $>$2.3 hours (Table \ref{table:list1_flares}).
In addition to these enhancements in Balmer emission lines, the continuum brightness observed with 
ARCSAT $u$-band increased by $\sim$20--25\% during Flare E3 (Figure \ref{fig:lcEW_HaHb_EVLac_UT200826} (b)),
while the increase in $g$-band is not so clear and comparable to the photometric error 
(3$\sigma_{g}$=2.1\%).\color{black}\textrm{
It is noted that the clear brightness increase in $u$\&$g$-bands at around Time 5h
(Figure \ref{fig:lcEW_HaHb_EVLac_UT200826} (b)) could be related with Flare E3, since the flare already started when the observation started (Figure \ref{fig:lcEW_HaHb_EVLac_UT200826} (a)).
  } \color{black}
As for Flare \color{black}\textrm{E4} \color{black}, the H$\alpha$ \& H$\beta$ equivalent widths increased up to 5.9\AA~and 10.8\AA, respectively, and $\Delta t^{\rm{flare}}_{\rm{H}\alpha}$ is $>$2.1 hours (Table \ref{table:list1_flares}).
Flare E4 did not end before the spectroscopic observation finished.
In addition to these enhancements in Balmer emission lines, the continuum brightness observed with ARCSAT $u$- \& $g$-bands increased by $\sim$25--30\% and $\sim$3--4\%, respectively, during Flare E4 (Figure \ref{fig:lcEW_HaHb_EVLac_UT200826} (b)).
\color{black}\textrm{ 
$L_{u}$, $L_{g}$, $E_{u}$, $E_{g}$, $L_{\rm{H}\alpha}$, $L_{\rm{H}\beta}$, $E_{\rm{H}\alpha}$, and $E_{\rm{H}\beta}$ values are estimated and listed in Table \ref{table:list1_flares}.
Since the flare already started before the spectroscopic observation began, 
the $L_{\rm{H}\alpha}$, $L_{\rm{H}\beta}$, $E_{\rm{H}\alpha}$, and $E_{\rm{H}\beta}$ values 
of Flare E3 listed here can be only lower limit values.
Since the flare did not end before the spectroscopic observation finished, 
the $L_{\rm{H}\alpha}$, $L_{\rm{H}\beta}$, $E_{\rm{H}\alpha}$, and $E_{\rm{H}\beta}$ values 
of Flare E4 can be also only lower limit values.
} \color{black}

\begin{figure}[ht!]
   \begin{center}
            \gridline{  
     \hspace{-0.06\textwidth}
    \fig{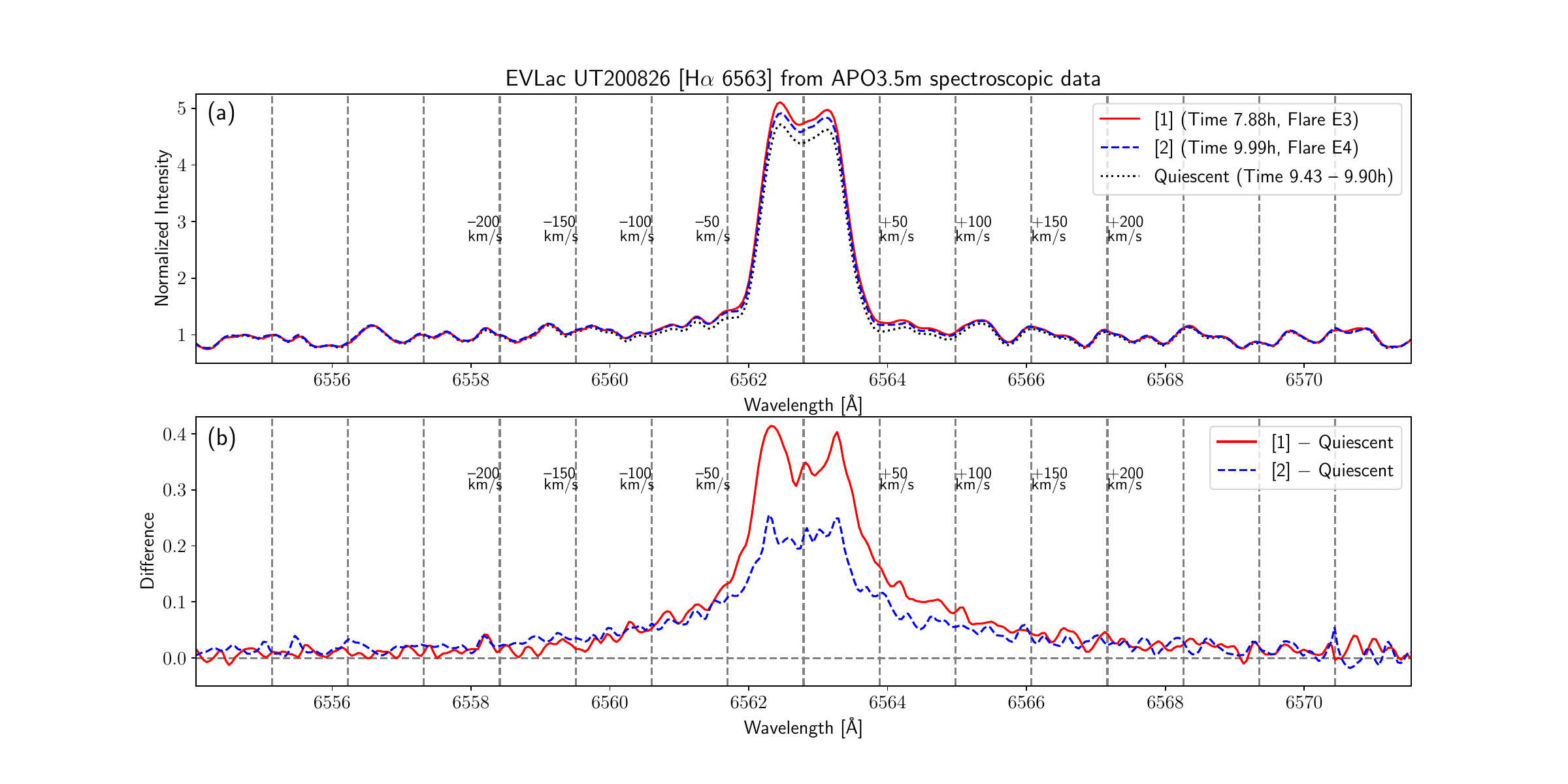}{0.58\textwidth}{\vspace{0mm}}
     \hspace{-0.06\textwidth}
       \fig{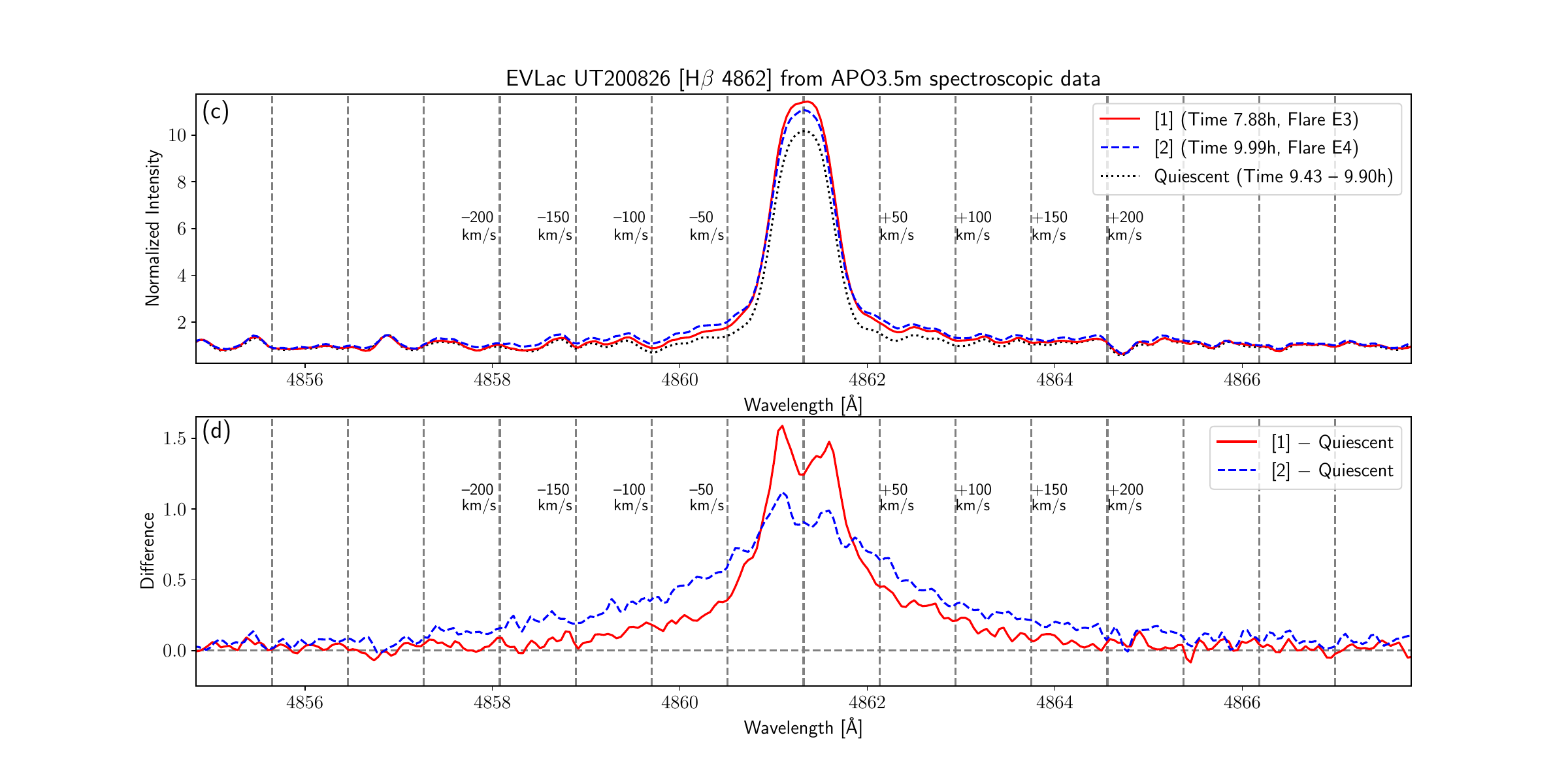}{0.58\textwidth}{\vspace{0mm}}
    }
     \vspace{-0.5cm}
     \caption{
   \color{black}\textrm{  
Line profiles of the H$\alpha$ \& H$\beta$ emission lines during Flares E3 \& E4 
on 2020 August 26 (at the time [1] and [2]) from APO3.5m spectroscopic data, which are plotted similarly with Figure \ref{fig:spec_HaHb_YZCMi_UT190127}.
 } \color{black}
     }
   \label{fig:spec_HaHb_EVLac_UT200826}
   \end{center}
 \end{figure}

\begin{figure}[ht!]
   \begin{center}
          \gridline{  
     \hspace{-0.07\textwidth}
    \fig{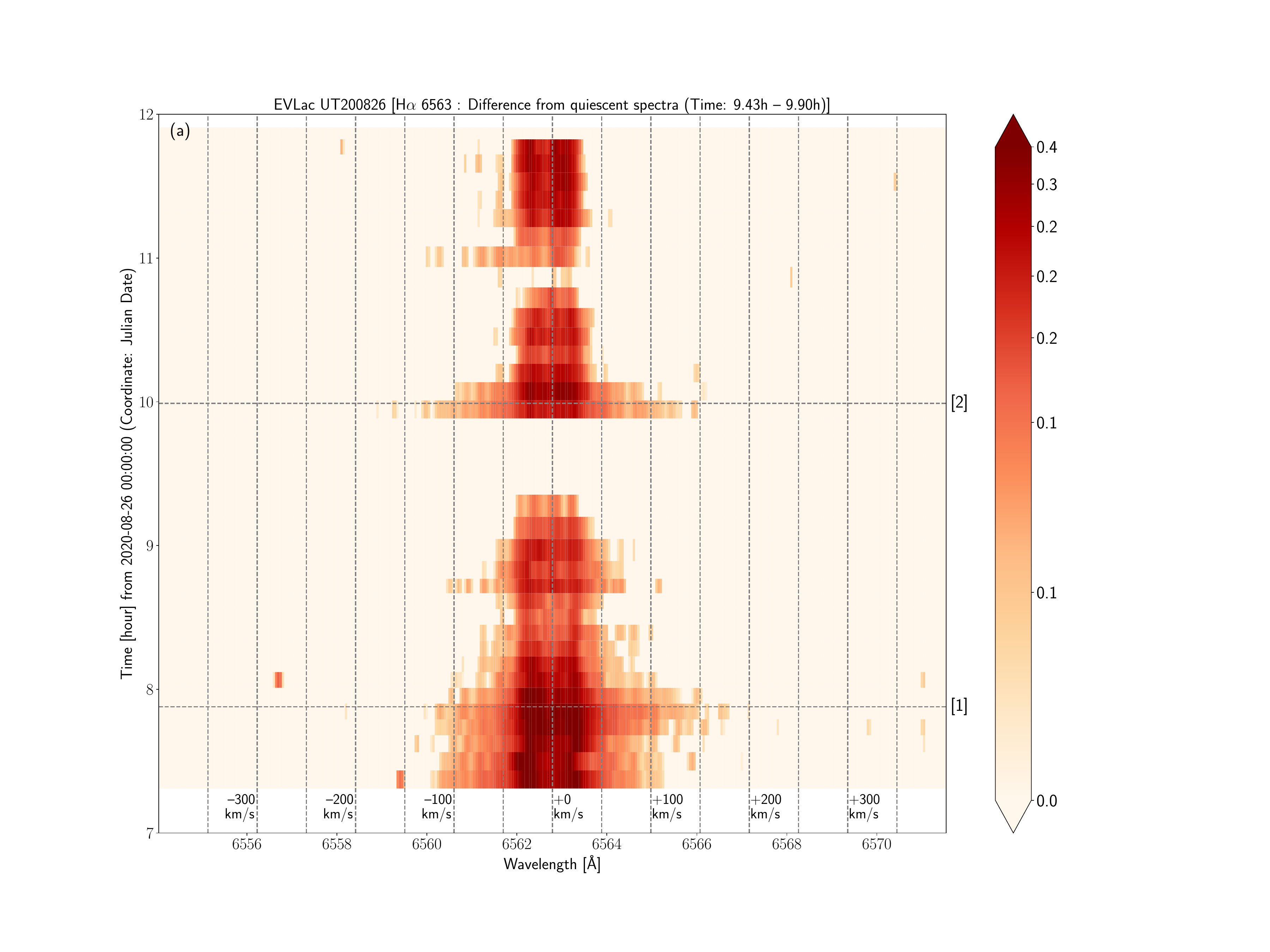}{0.63\textwidth}{\vspace{0mm}}
     \hspace{-0.11\textwidth}
    \fig{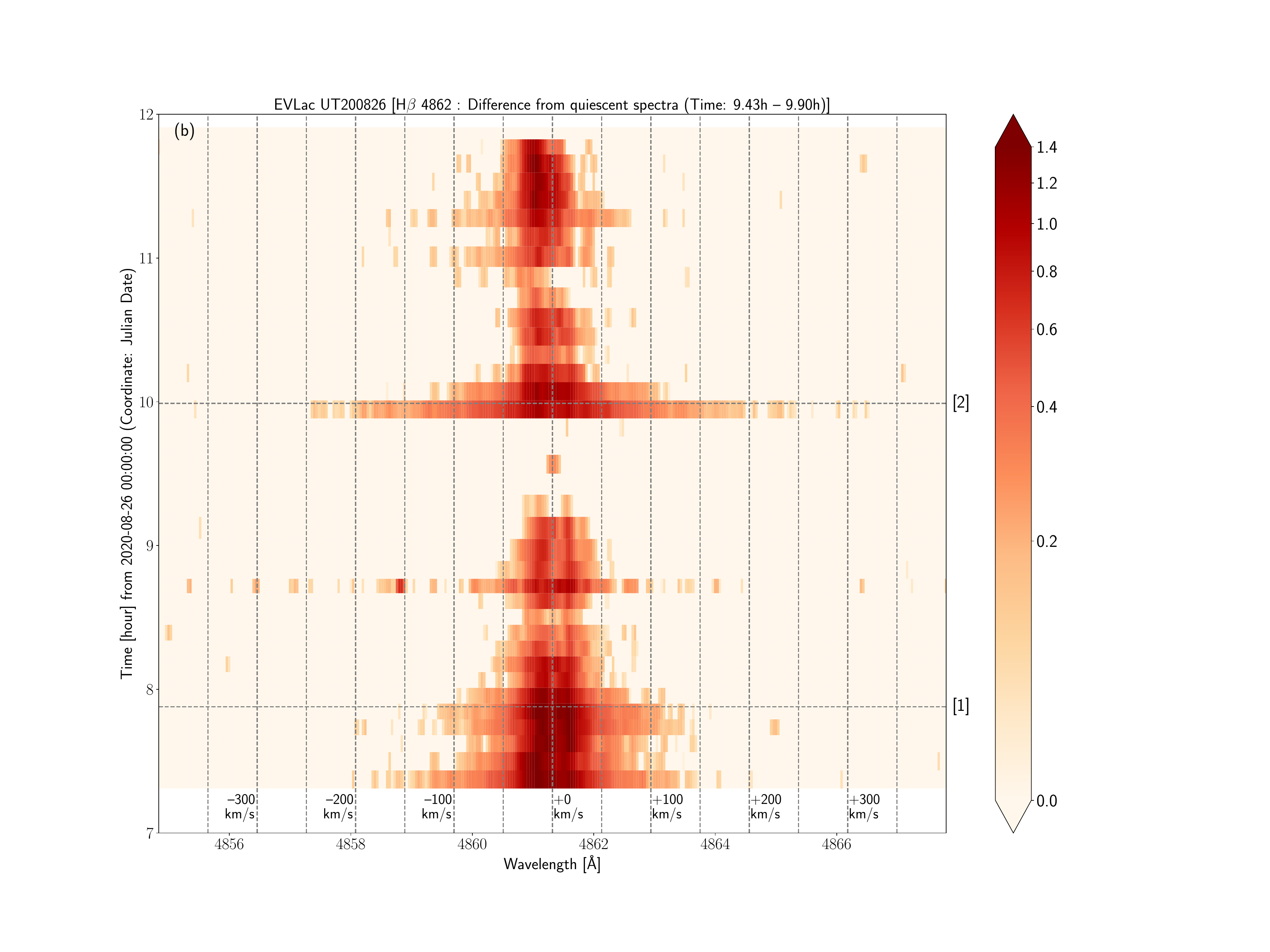}{0.63\textwidth}{\vspace{0mm}}
    }
     \vspace{-0.5cm}
     \caption{
         \color{black}\textrm{  
Time evolution of the H$\alpha$ \& H$\beta$ line profiles covering Flares E3 \& E4
on 2020 August 26, which are plotted similarly with Figure \ref{fig:map_HaHb_YZCMi_UT191212}.
The grey horizontal dashed lines indicate the time [1] \& [2], which are shown in Figure \ref{fig:lcEW_HaHb_EVLac_UT200826} (light curves) and Figure \ref{fig:spec_HaHb_EVLac_UT200826} (line profiles).
}\color{black}
     }
   \label{fig:map_HaHb_EVLac_UT200826}
   \end{center}
 \end{figure}

The H$\alpha$ \& H$\beta$ line profiles during Flares E3 and E4 are shown in
Figures \ref{fig:spec_HaHb_EVLac_UT200826} \& \ref{fig:map_HaHb_EVLac_UT200826}. 
During Flares E3, there were no clear blue or red wing asymmetries in H$\alpha$ and H$\beta$ lines (time [1] in Figures \ref{fig:spec_HaHb_EVLac_UT200826} \& \ref{fig:map_HaHb_EVLac_UT200826}), 
and the line profiles showed
roughly symmetrical broadenings with $\sim\pm$150--200 km s$^{-1}$ at around the peak time of the flares (time [1] in Figures \ref{fig:spec_HaHb_EVLac_UT200826} \& \ref{fig:map_HaHb_EVLac_UT200826}).
During Flares E4, there were also no clear blue or red wing asymmetries in H$\alpha$ and H$\beta$ lines (time [2] in Figures \ref{fig:spec_HaHb_EVLac_UT200826} \& \ref{fig:map_HaHb_EVLac_UT200826}), 
and the line profiles showed
roughly symmetrical broadenings with $\sim\pm$150--200 km s$^{-1}$ (H$\alpha$) and $\sim\pm$200--250 km s$^{-1}$ (H$\beta$) at around the peak time of the flares (time [2] in Figures \ref{fig:spec_HaHb_EVLac_UT200826} \& \ref{fig:map_HaHb_EVLac_UT200826}).
 
\clearpage

\subsection{Flare E5 observed on 2020 August 27} 
\label{subsec:results:2020-Aug-27} 
 
  \begin{figure}[ht!]
   \begin{center}
   \gridline{
    \fig{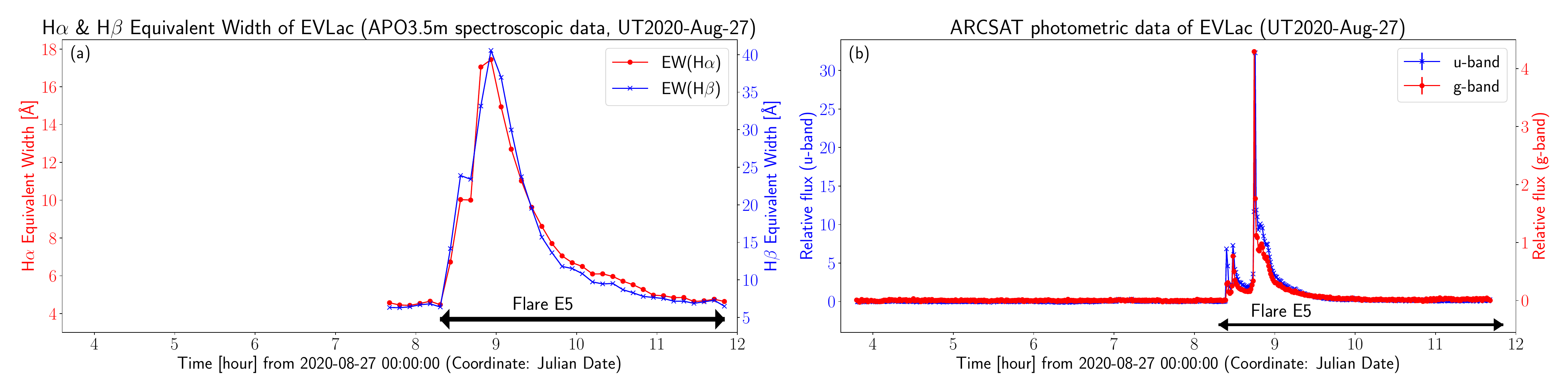}{1.0\textwidth}{\vspace{0mm}}}  
     \vspace{-0.5cm}
   \gridline{
    \fig{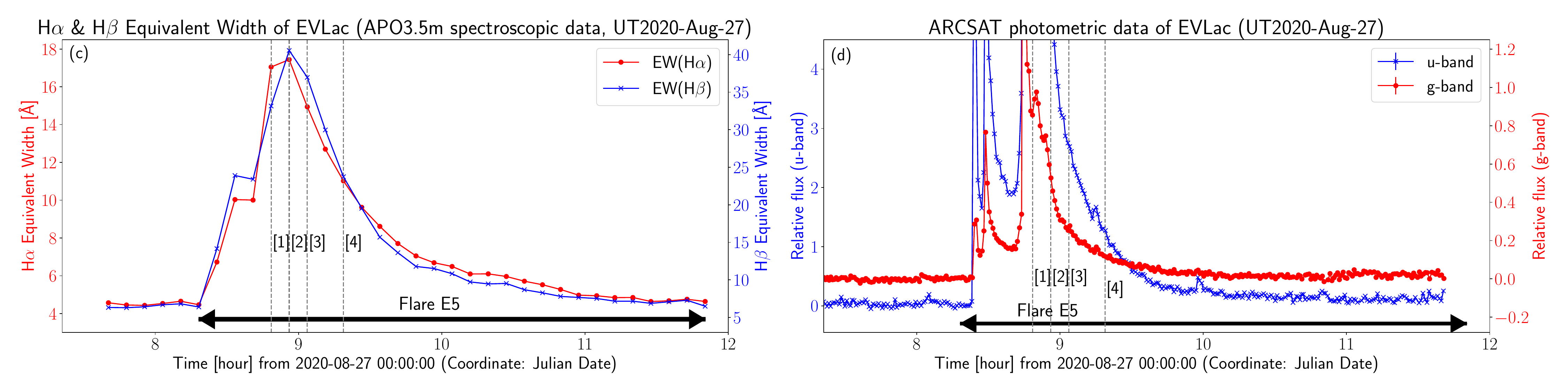}{1.0\textwidth}{\vspace{0mm}}
    } 
     \vspace{-0.5cm}
     \caption{
     \color{black}\textrm{  
Light curves of EV Lac on 2020 August 27 showing Flare E5, which are plotted 
similarly with Figures \ref{fig:lcEW_HaHb_YZCMi_UT191212} (a)\&(b).
(c) \& (d) are enlarged panels of (a) \& (b).
The grey dashed lines with numbers ([1]--[4]) in (c)\&(d) correspond to the time shown 
with the same numbers in Figures \ref{fig:spec_HaHb_EVLac_UT200827} \& \ref{fig:map_HaHb_EVLac_UT200827}.
 } \color{black}
     }
   \label{fig:lcEW_HaHb_EVLac_UT200827}
   \end{center}
 \end{figure}

On 2020 August 27, one flare (Flare E5) 
were detected on EV Lac in H$\alpha$ \& H$\beta$ lines 
as shown in Figures \ref{fig:lcEW_HaHb_EVLac_UT200827} (a) \& (c).  
As for Flare E5, the H$\alpha$ \& H$\beta$ equivalent widths increased up to 17.4\AA~and 40.6\AA, respectively, and $\Delta t^{\rm{flare}}_{\rm{H}\alpha}$ is 3.5 hours (Table \ref{table:list1_flares}).
In addition to these enhancements in Balmer emission lines, the continuum brightness observed with ARCSAT $u$- \& $g$-bands increased by $\sim$3230\% and $\sim$430\%, respectively, during Flare E5 (Figures \ref{fig:lcEW_HaHb_EVLac_UT200826} (b)).
\color{black}\textrm{ 
$L_{u}$, $L_{g}$, $E_{u}$, $E_{g}$, $L_{\rm{H}\alpha}$, $L_{\rm{H}\beta}$, $E_{\rm{H}\alpha}$, and $E_{\rm{H}\beta}$ values are estimated and listed in Table \ref{table:list1_flares}.
} \color{black}
 
\begin{figure}[ht!]
   \begin{center}
                \gridline{  
     \hspace{-0.06\textwidth}
    \fig{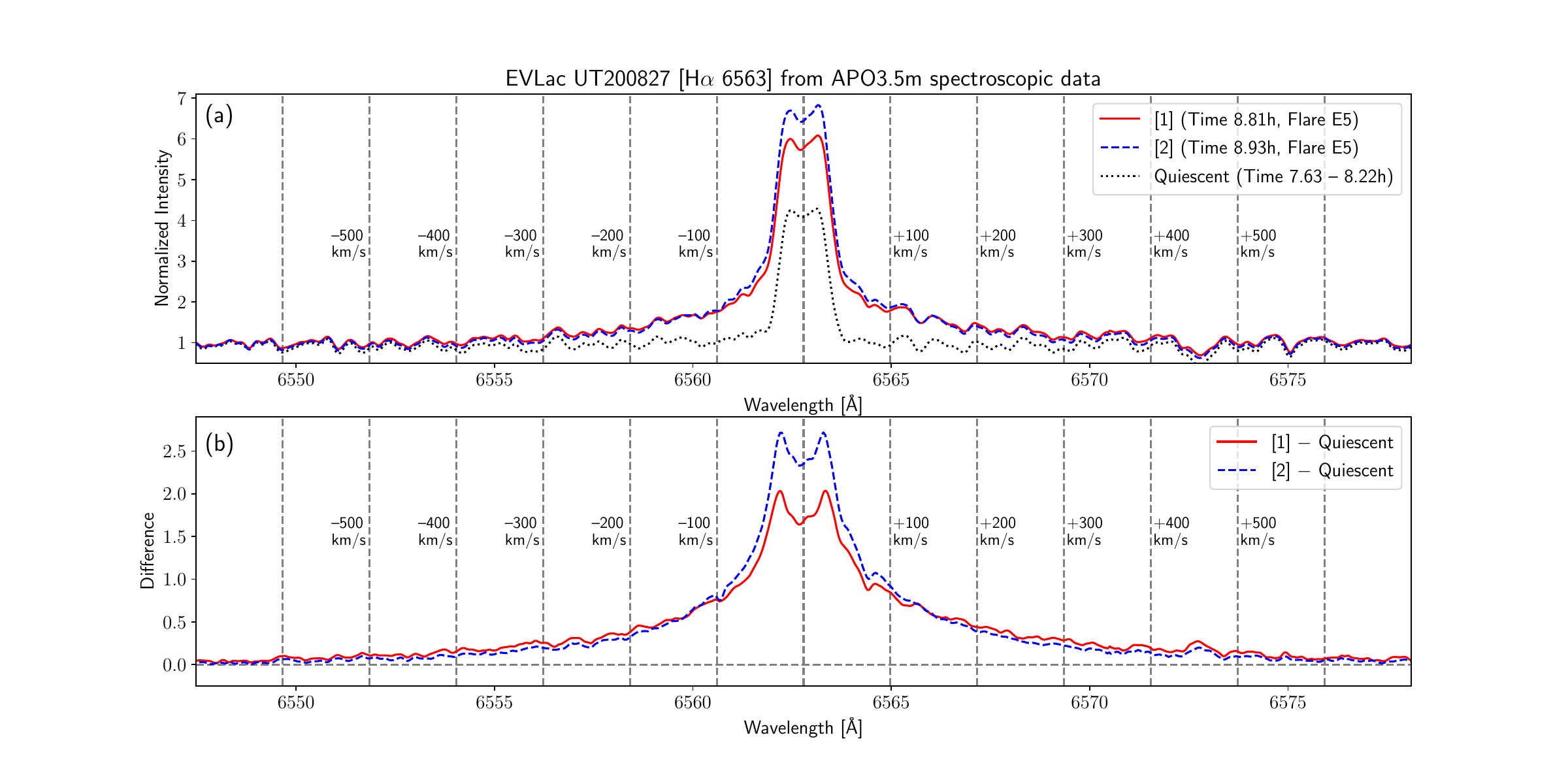}{0.58\textwidth}{\vspace{0mm}}
     \hspace{-0.06\textwidth}
       \fig{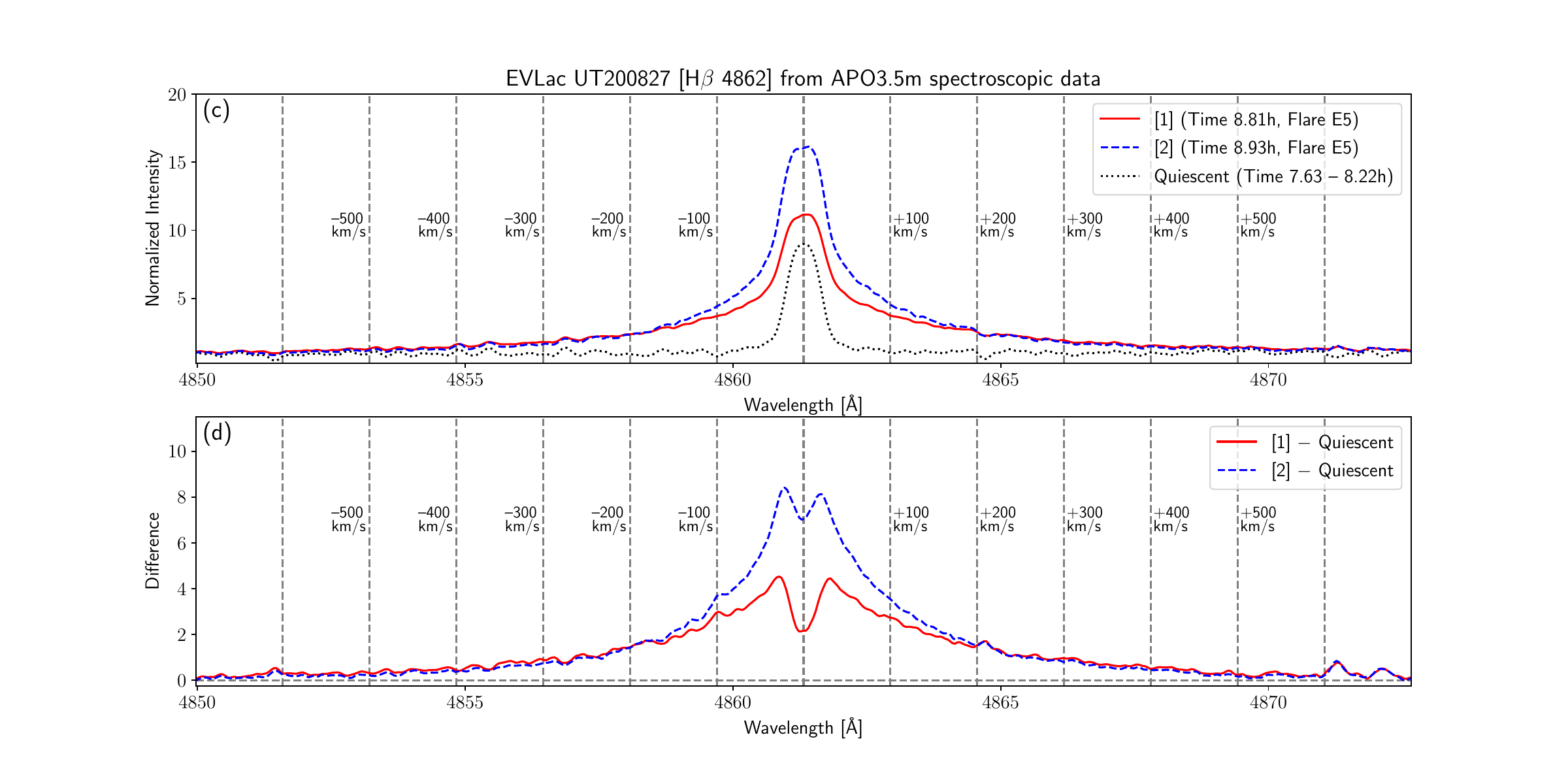}{0.58\textwidth}{\vspace{0mm}}
    }
         \vspace{-0.5cm}
                \gridline{  
     \hspace{-0.06\textwidth}
    \fig{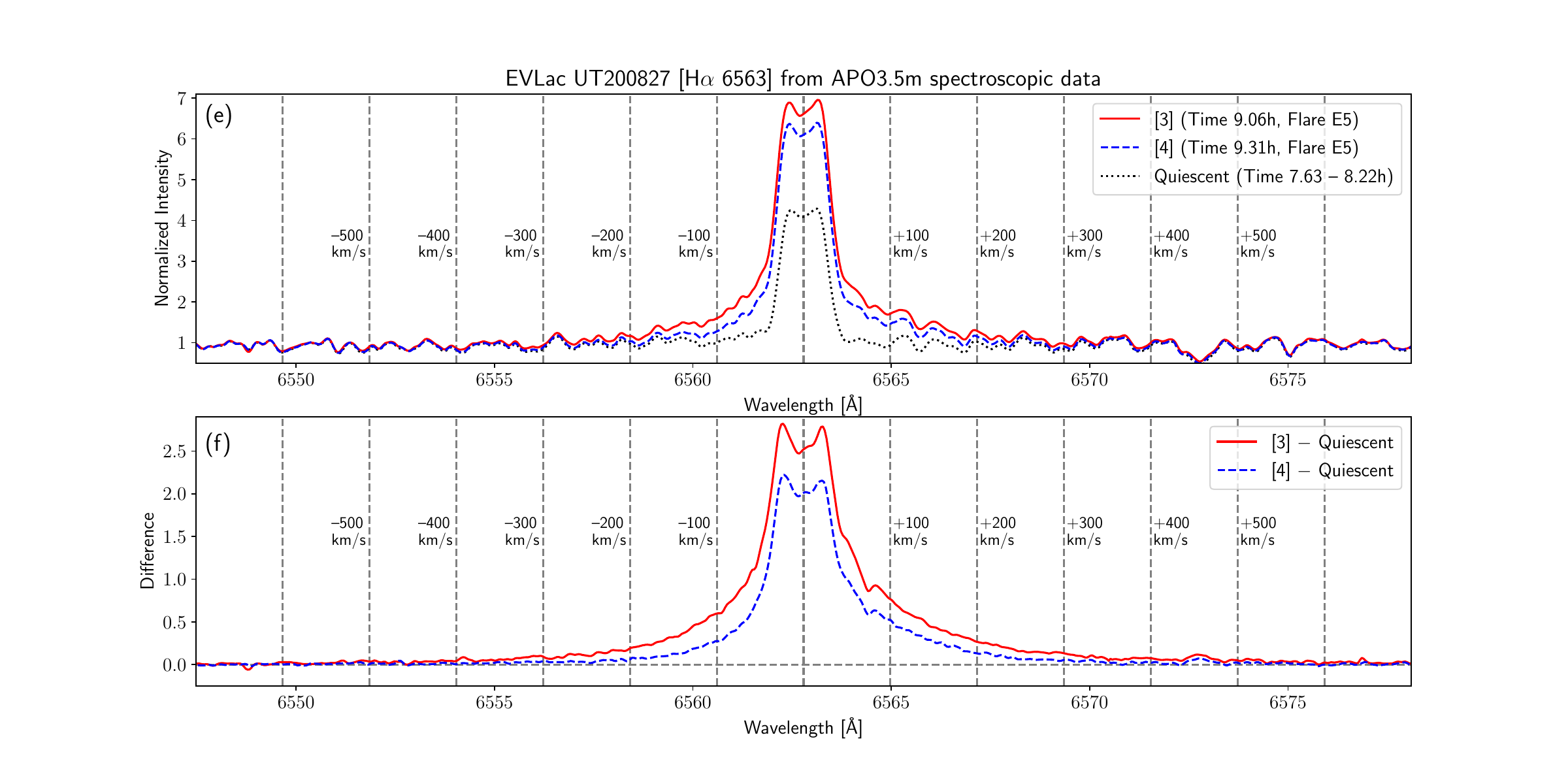}{0.58\textwidth}{\vspace{0mm}}
     \hspace{-0.06\textwidth}
       \fig{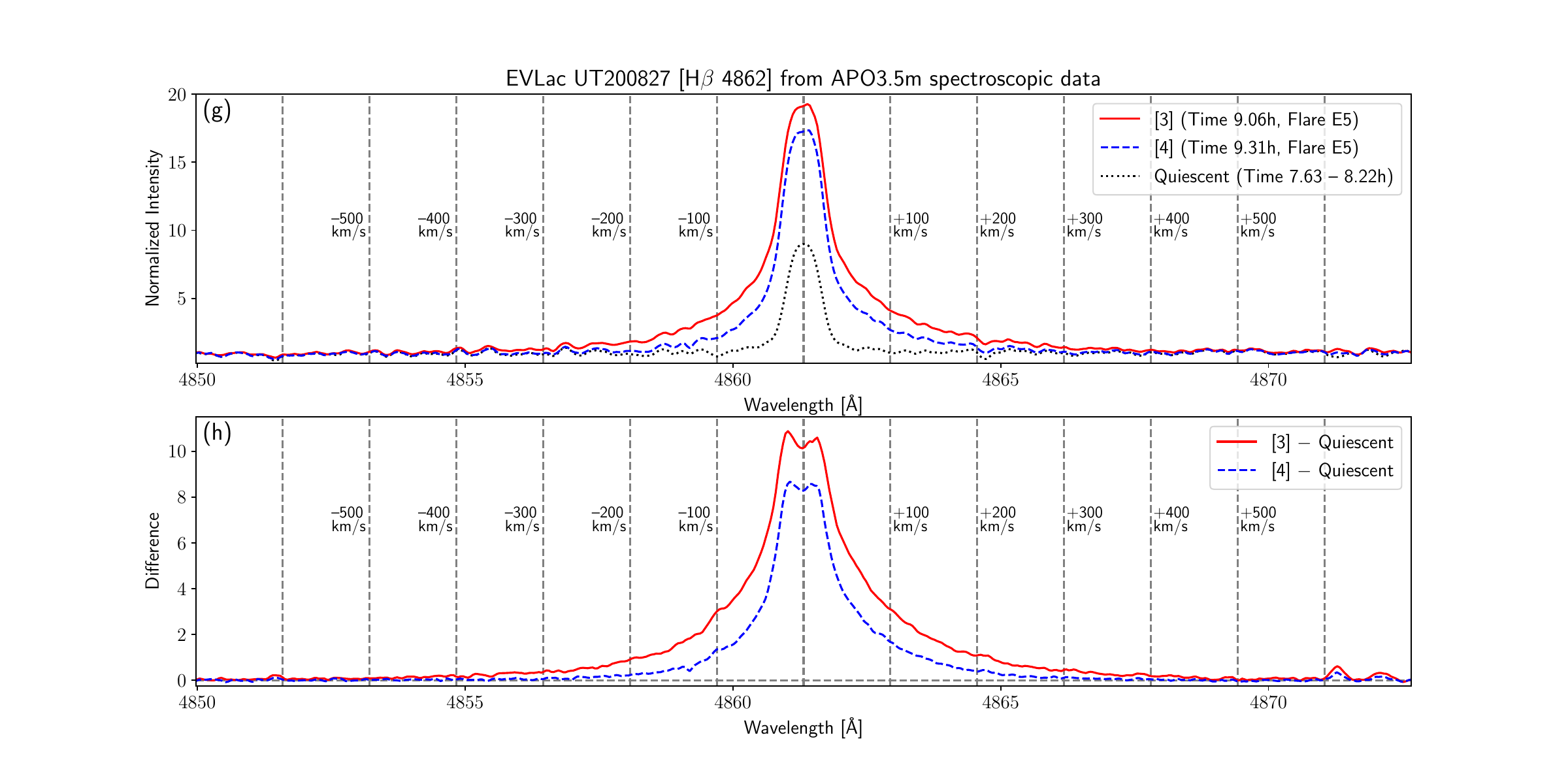}{0.58\textwidth}{\vspace{0mm}}
    }
     \vspace{-0.5cm}
     \caption{
   \color{black}\textrm{  
Line profiles of the H$\alpha$ \& H$\beta$ emission lines during Flare E5 
on 2020 August 27 (at the time [1]--[4]) from APO3.5m spectroscopic data, which are plotted similarly with Figure \ref{fig:spec_HaHb_YZCMi_UT190127}.
 } \color{black}
     }
   \label{fig:spec_HaHb_EVLac_UT200827}
   \end{center}
 \end{figure}

\begin{figure}[ht!]
   \begin{center}
      \gridline{  
     \hspace{-0.07\textwidth}
    \fig{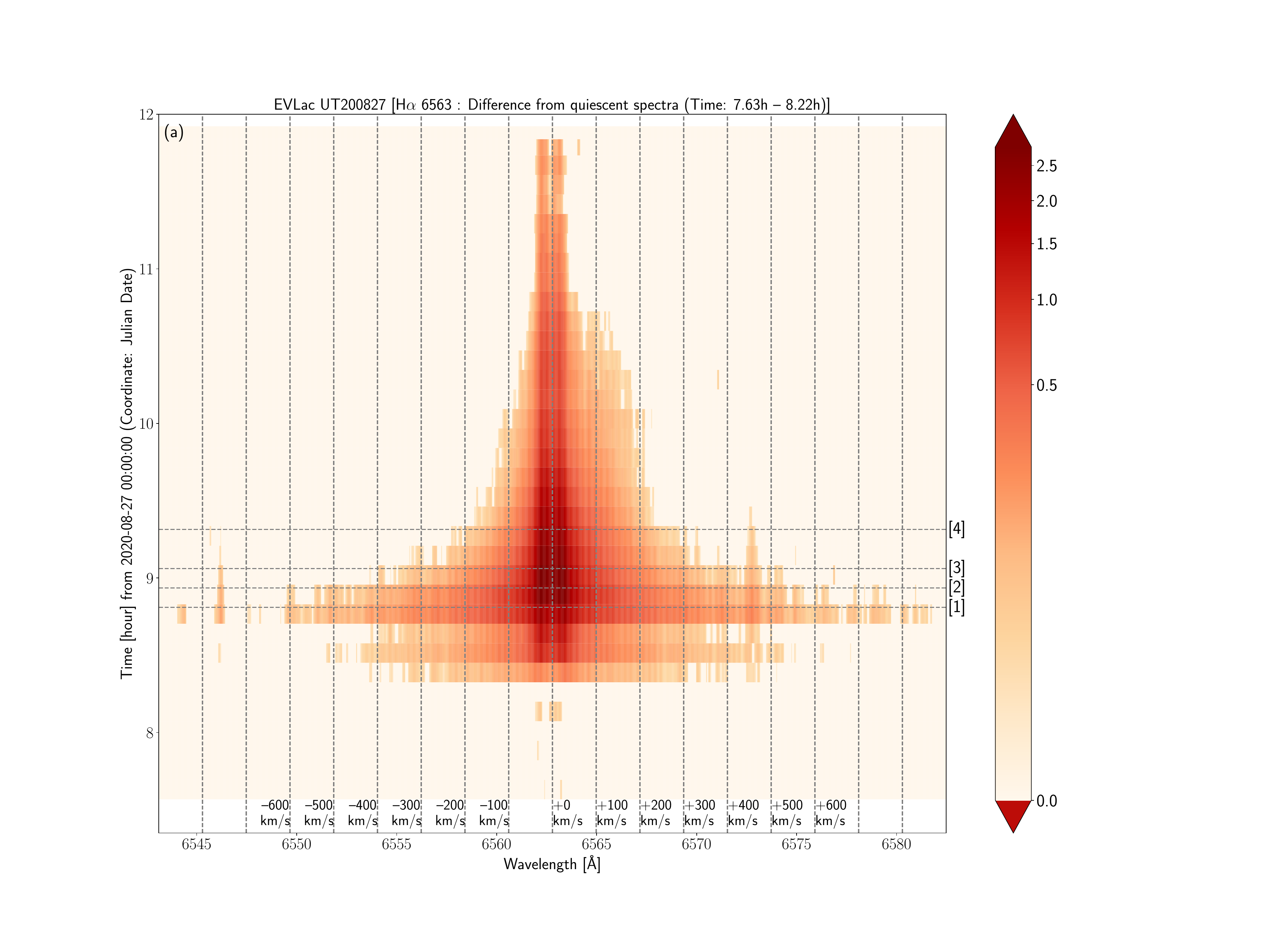}{0.63\textwidth}{\vspace{0mm}}
     \hspace{-0.11\textwidth}
    \fig{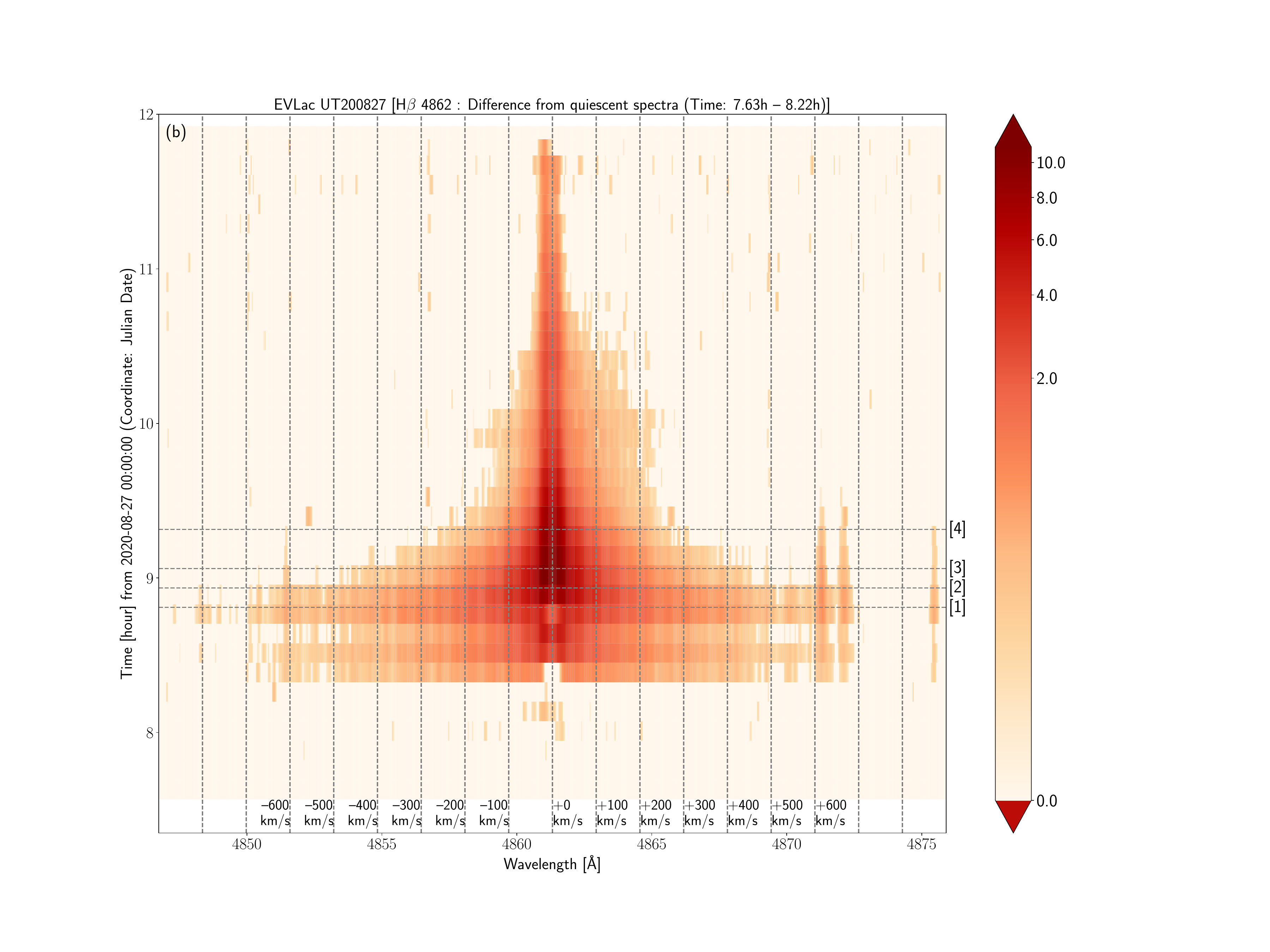}{0.63\textwidth}{\vspace{0mm}}
    }
     \vspace{-0.5cm}
     \caption{
         \color{black}\textrm{  
Time evolution of the H$\alpha$ \& H$\beta$ line profiles covering Flare E5
on 2020 August 27, which are plotted similarly with Figure \ref{fig:map_HaHb_YZCMi_UT191212}.
The grey horizontal dashed lines indicate the time [1] -- [4], 
which are shown in Figure \ref{fig:lcEW_HaHb_EVLac_UT200827} (light curves) and Figure \ref{fig:spec_HaHb_EVLac_UT200827} (line profiles).
}\color{black}
     }
   \label{fig:map_HaHb_EVLac_UT200827}
   \end{center}
   \end{figure}

The H$\alpha$ \& H$\beta$ line profiles during Flare E5 are shown in
Figures \ref{fig:spec_HaHb_EVLac_UT200827} \& \ref{fig:map_HaHb_EVLac_UT200827}. 
At around the peak time of Flares E5 (e.g., time [1]\&[2] in Figures \ref{fig:spec_HaHb_EVLac_UT200827} \& \ref{fig:map_HaHb_EVLac_UT200827}), 
the line profiles of H$\alpha$ and H$\beta$ lines 
show roughly symmetrical broadenings or possibly slight red wing asymmetries 
with $\sim\pm$600--800 km s$^{-1}$ (H$\alpha$) and 
$\sim\pm$600--700 km s$^{-1}$ (H$\beta$).
During the decay phase of Flare E5 (e.g., time [3]\&[4] in Figures \ref{fig:spec_HaHb_EVLac_UT200827} \& \ref{fig:map_HaHb_EVLac_UT200827}),
the line profiles of H$\alpha$ and H$\beta$ lines 
show clear red wing asymmetries for two hours (Figure \ref{fig:map_HaHb_EVLac_UT200827}).

 \clearpage
 
\subsection{Flare E6 observed on 2020 August 29} 
\label{subsec:results:2020-Aug-29}

\begin{figure}[ht!]
   \begin{center}
    \gridline{
      \fig{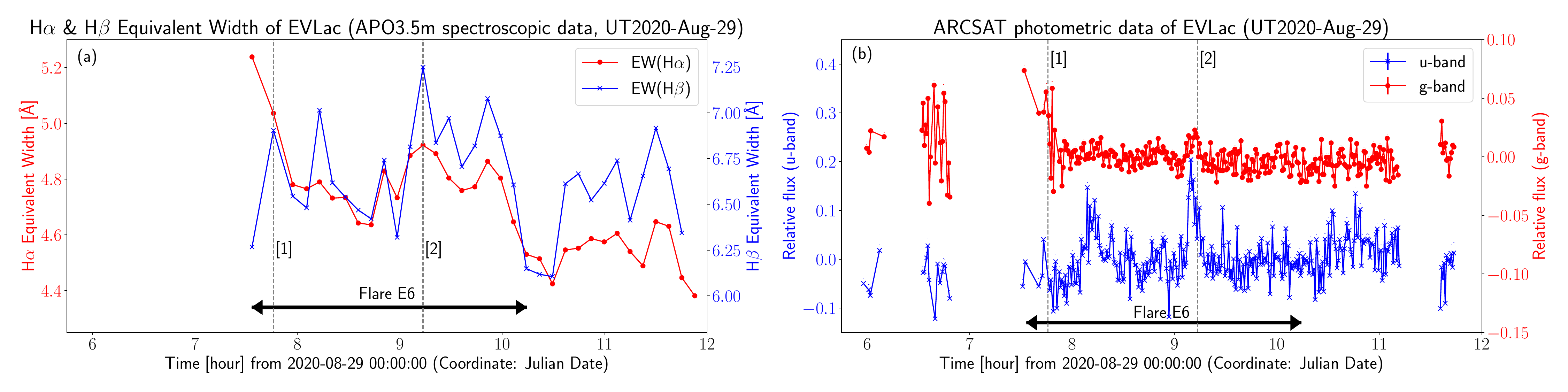}{1.0\textwidth}{\vspace{0mm}}
    }
     \vspace{-5mm}
     \caption{
     \color{black}\textrm{  
Light curves of EV Lac on 2020 August 29 showing Flare E6, which are plotted 
similarly with Figures \ref{fig:lcEW_HaHb_YZCMi_UT191212} (a)\&(b).
The grey dashed lines with numbers ([1] \& [2]) correspond to the time shown 
with the same numbers in Figures \ref{fig:spec_HaHb_EVLac_UT200829} \& \ref{fig:map_HaHb_EVLac_UT200829}.
 } \color{black}
     }
   \label{fig:lcEW_HaHb_EVLac_UT200829}
   \end{center}
 \end{figure}

On 2020 August 29, one flare (Flares E6) 
was detected on EV Lac in H$\alpha$ \& H$\beta$ lines 
as shown in Figure \ref{fig:lcEW_HaHb_EVLac_UT200829} (a).  
Flare E6 already started before the spectroscopic observation started.
As for Flare E6, the H$\alpha$ \& H$\beta$ equivalent widths increased up to 5.2\AA~and 7.2\AA, respectively, and $\Delta t^{\rm{flare}}_{\rm{H}\alpha}$ is $>$2.7 hours (Table \ref{table:list1_flares}).
In addition to these enhancements in Balmer emission lines, the continuum brightness observed with ARCSAT $u$- \& $g$-bands increased by $\sim$20\% and $\sim$2\%, respectively, during Flare E6 (Figure \ref{fig:lcEW_HaHb_EVLac_UT200829} (b)).
\color{black}\textrm{ 
$L_{u}$, $L_{g}$, $E_{u}$, $E_{g}$, $L_{\rm{H}\alpha}$, $L_{\rm{H}\beta}$, $E_{\rm{H}\alpha}$, and $E_{\rm{H}\beta}$ values are estimated and listed in Table \ref{table:list1_flares}.
Since the flare already started before the spectroscopic observation began, 
the luminosity and energy values of Flare E6 
estimated here can be only lower limit values.
} \color{black}

\begin{figure}[ht!]
   \begin{center}
            \gridline{  
     \hspace{-0.06\textwidth}
    \fig{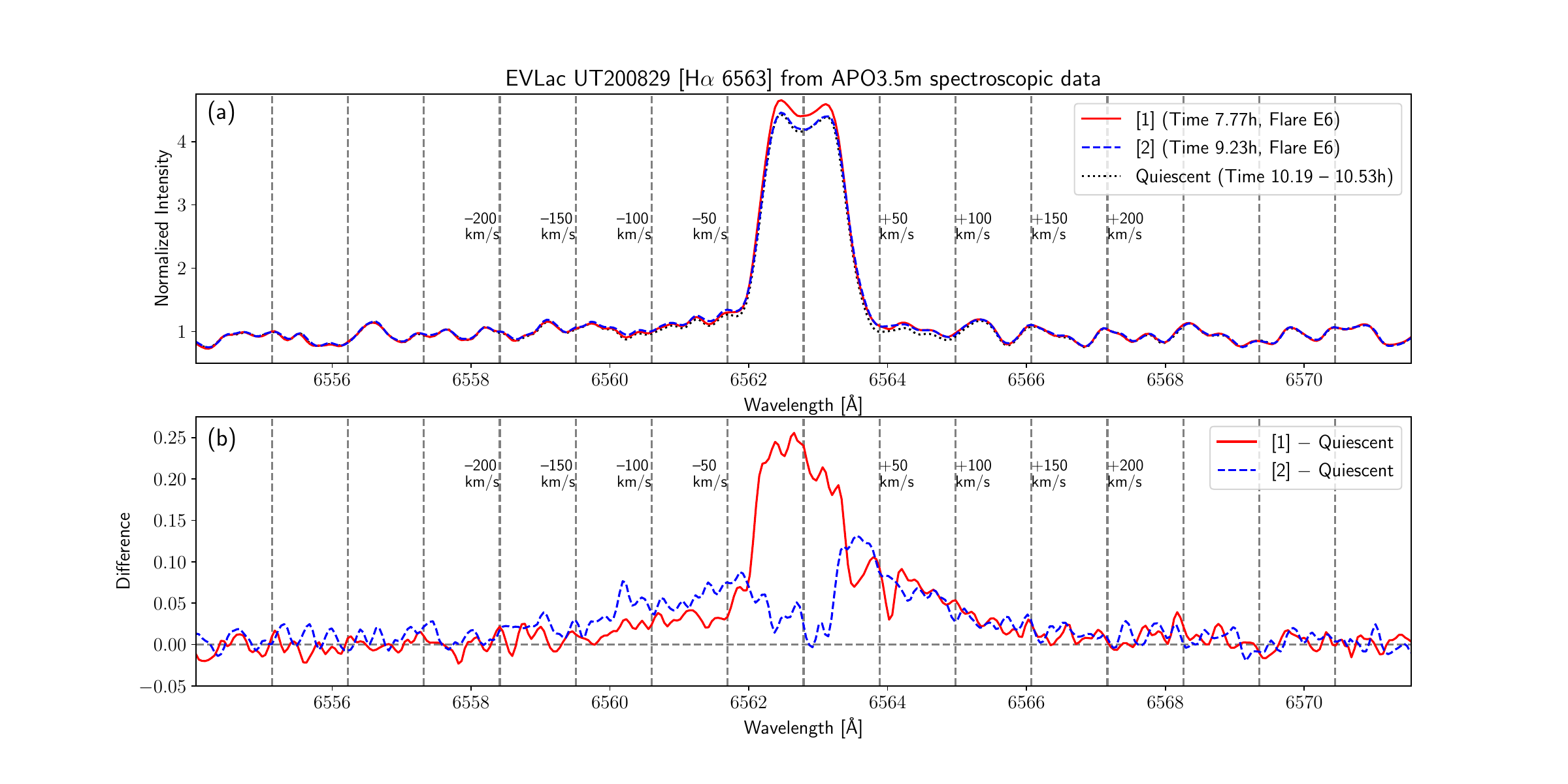}{0.58\textwidth}{\vspace{0mm}}
     \hspace{-0.06\textwidth}
       \fig{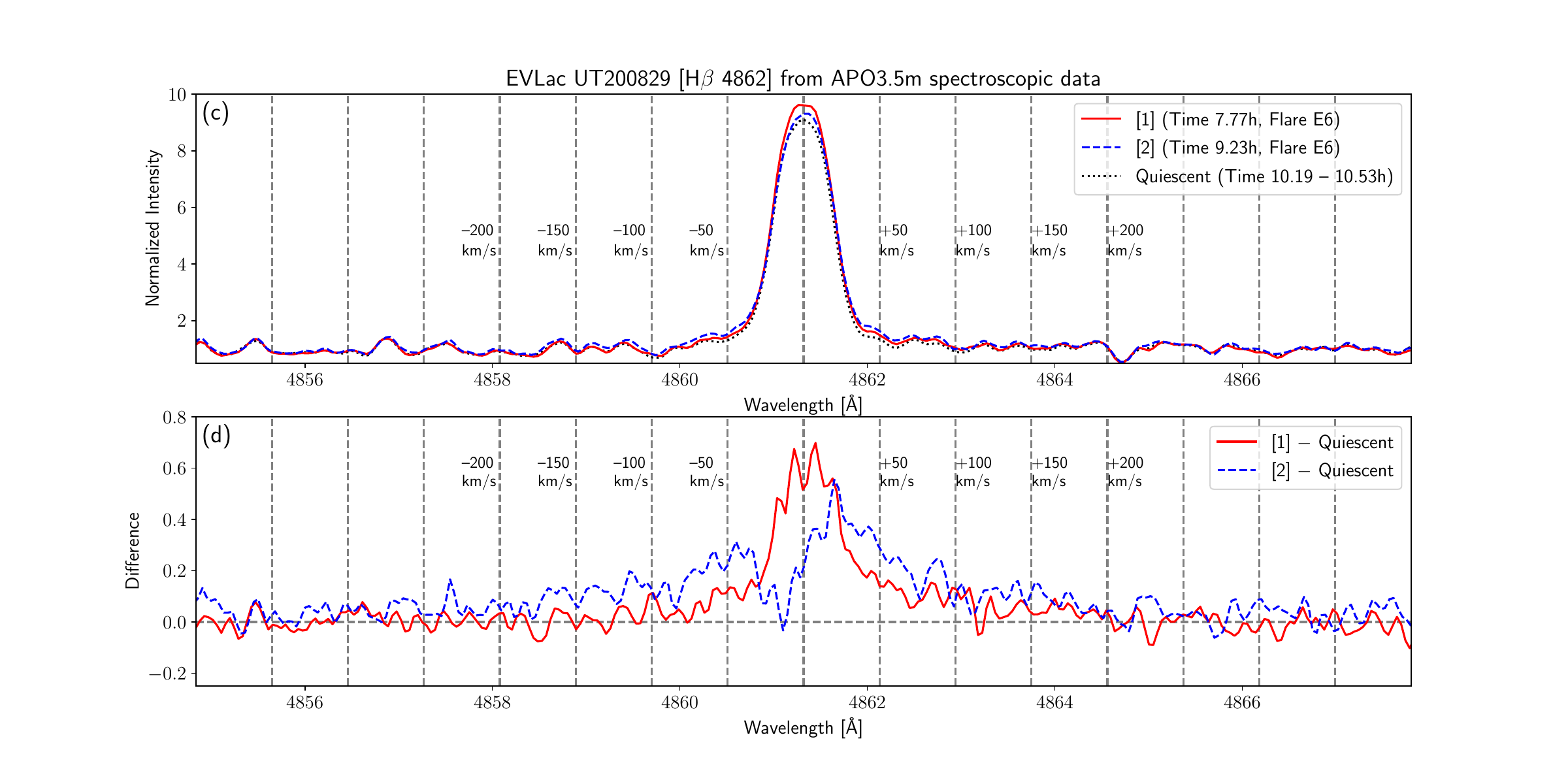}{0.58\textwidth}{\vspace{0mm}}
    }
     \vspace{-0.5cm}
     \caption{
   \color{black}\textrm{  
Line profiles of the H$\alpha$ \& H$\beta$ emission lines during Flare E6 
on 2020 August 29 (at the time [1] and [2]) from APO3.5m spectroscopic data, which are plotted similarly with Figure \ref{fig:spec_HaHb_YZCMi_UT190127}.
 } \color{black}
     }
   \label{fig:spec_HaHb_EVLac_UT200829}
   \end{center}
 \end{figure}

\begin{figure}[ht!]
   \begin{center}
          \gridline{  
     \hspace{-0.07\textwidth}
    \fig{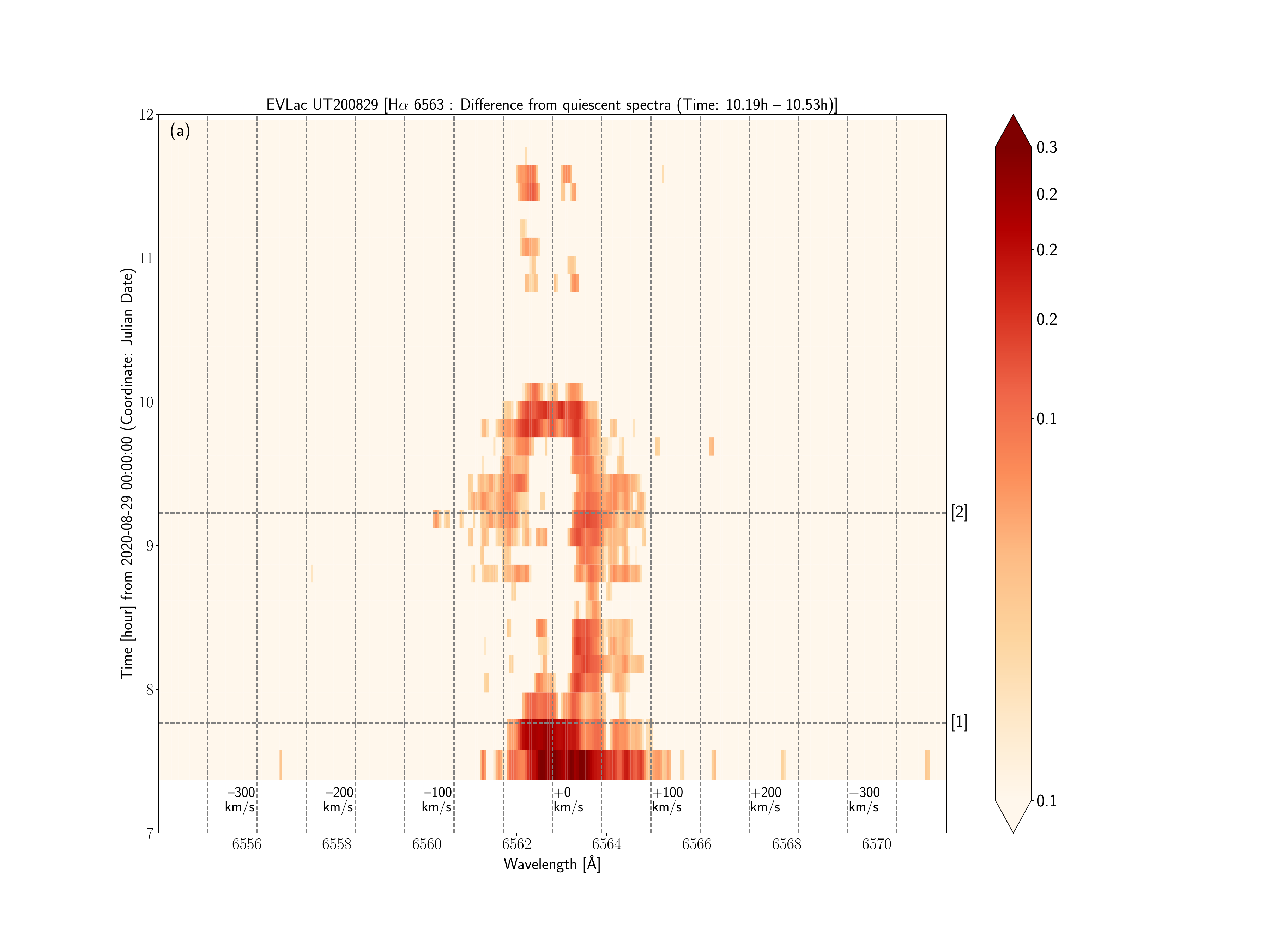}{0.63\textwidth}{\vspace{0mm}}
     \hspace{-0.11\textwidth}
    \fig{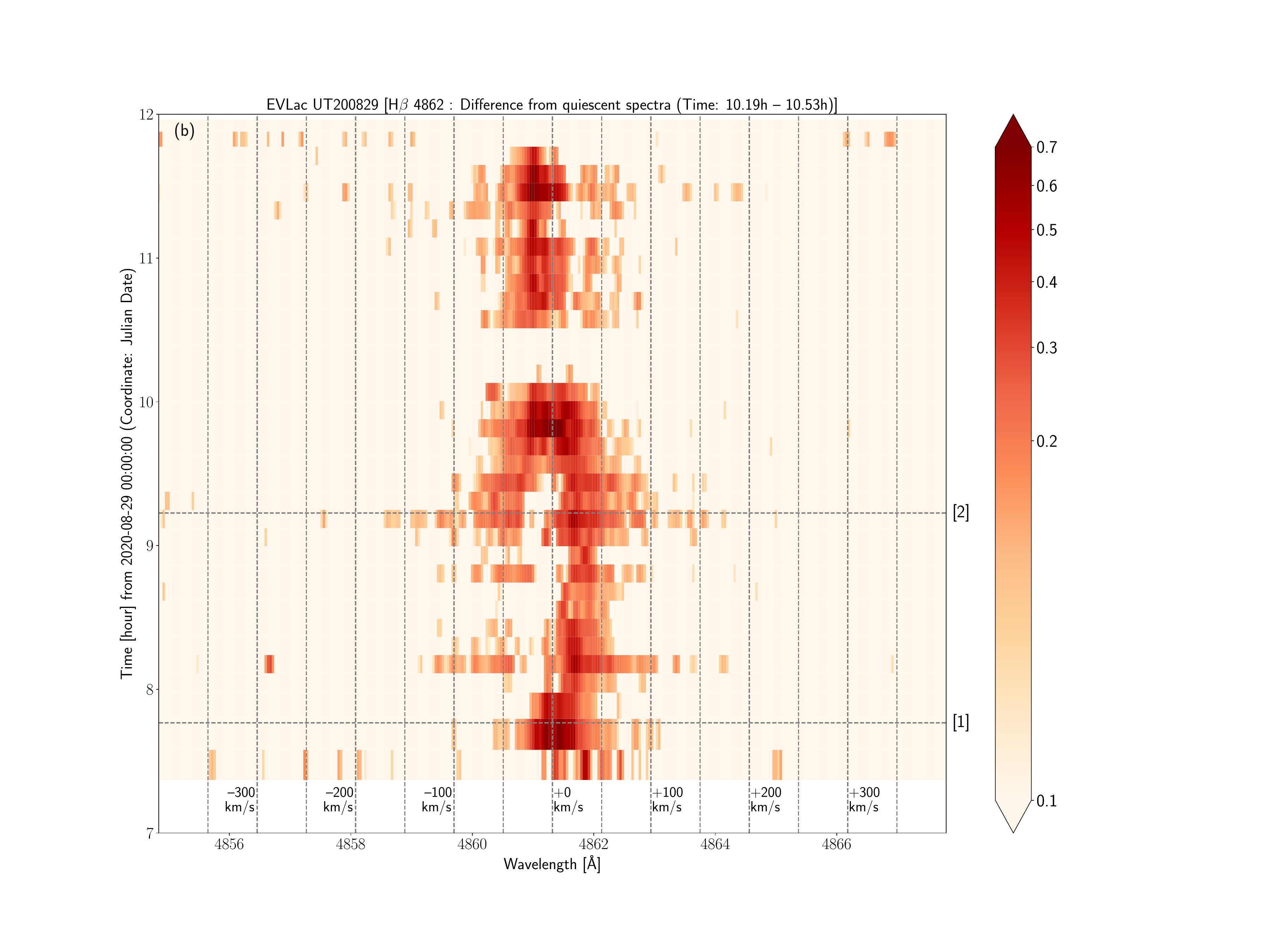}{0.63\textwidth}{\vspace{0mm}}
    }
     \vspace{-0.5cm}
     \caption{
        \color{black}\textrm{  
Time evolution of the H$\alpha$ \& H$\beta$ line profiles covering Flare E6
on 2020 August 29, which are plotted similarly with Figure \ref{fig:map_HaHb_YZCMi_UT191212}.
The grey horizontal dashed lines indicate the time [1] \& [2], 
which are shown in Figure \ref{fig:lcEW_HaHb_EVLac_UT200829} (light curves) and Figure \ref{fig:spec_HaHb_EVLac_UT200829} (line profiles).
}\color{black}
     }
   \label{fig:map_HaHb_EVLac_UT200829}
   \end{center}
 \end{figure}

The H$\alpha$ \& H$\beta$ line profiles during Flare E6 are shown in
Figures \ref{fig:spec_HaHb_EVLac_UT200829} \& \ref{fig:map_HaHb_EVLac_UT200829}. 
There were no clear blue or red wing asymmetries during Flare E6.

\clearpage
\subsection{Flare E7 observed on 2020 September 1} 
\label{subsec:results:2020-Sep-01}

\begin{figure}[ht!]
   \begin{center}
    \gridline{
      \fig{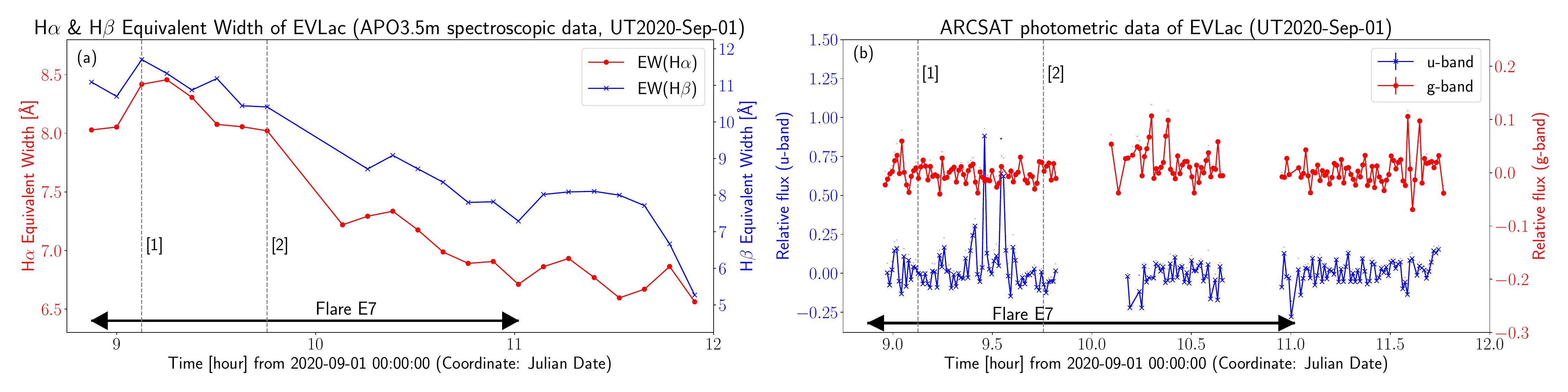}{1.0\textwidth}{\vspace{0mm}}
    }
     \vspace{-5mm}
     \caption{
     \color{black}\textrm{  
Light curves of EV Lac on 2020 September 1 showing Flare E7, which are plotted 
similarly with Figures \ref{fig:lcEW_HaHb_YZCMi_UT191212} (a)\&(b).
The grey dashed lines with numbers ([1] \& [2]) correspond to the time shown 
with the same numbers in Figures \ref{fig:spec_HaHb_EVLac_UT200901} \& \ref{fig:map_HaHb_EVLac_UT200901}.
 } \color{black}
     }
   \label{fig:lcEW_HaHb_EVLac_UT200901}
   \end{center}
 \end{figure}

On 2020 September 1, one flare (Flares E7) 
was detected on EV Lac in H$\alpha$ \& H$\beta$ lines 
as shown in Figure \ref{fig:lcEW_HaHb_EVLac_UT200901} (a).  
Flare E7 already started before the spectroscopic observation started.
The H$\alpha$ \& H$\beta$ equivalent widths increased up to 8.5\AA~and 11.7\AA, respectively, and $\Delta t^{\rm{flare}}_{\rm{H}\alpha}$ is $>$2.1 hours (Table \ref{table:list1_flares}).
There are some gaps of ARCSAT photometric observation data during Flare E7, and the flare itself had already started when the observation started (Figure \ref{fig:lcEW_HaHb_YZCMi_UT200123} (b)). 
Because of these, we cannot know whether this flare is a white-light flare or not, 
and we also do not \color{black}\textrm{estimate }\color{black} luminosities and energies in photometric bands for this flare.
$L_{\rm{H}\alpha}$, $L_{\rm{H}\beta}$, $E_{\rm{H}\alpha}$, and $E_{\rm{H}\beta}$ values are estimated and listed in Table \ref{table:list1_flares}.
Since the flare already started before the spectroscopic observation began, 
the luminosity and energy values of Flare E7 
estimated here can be only lower limit values.

\begin{figure}[ht!]
   \begin{center}
            \gridline{  
     \hspace{-0.06\textwidth}
    \fig{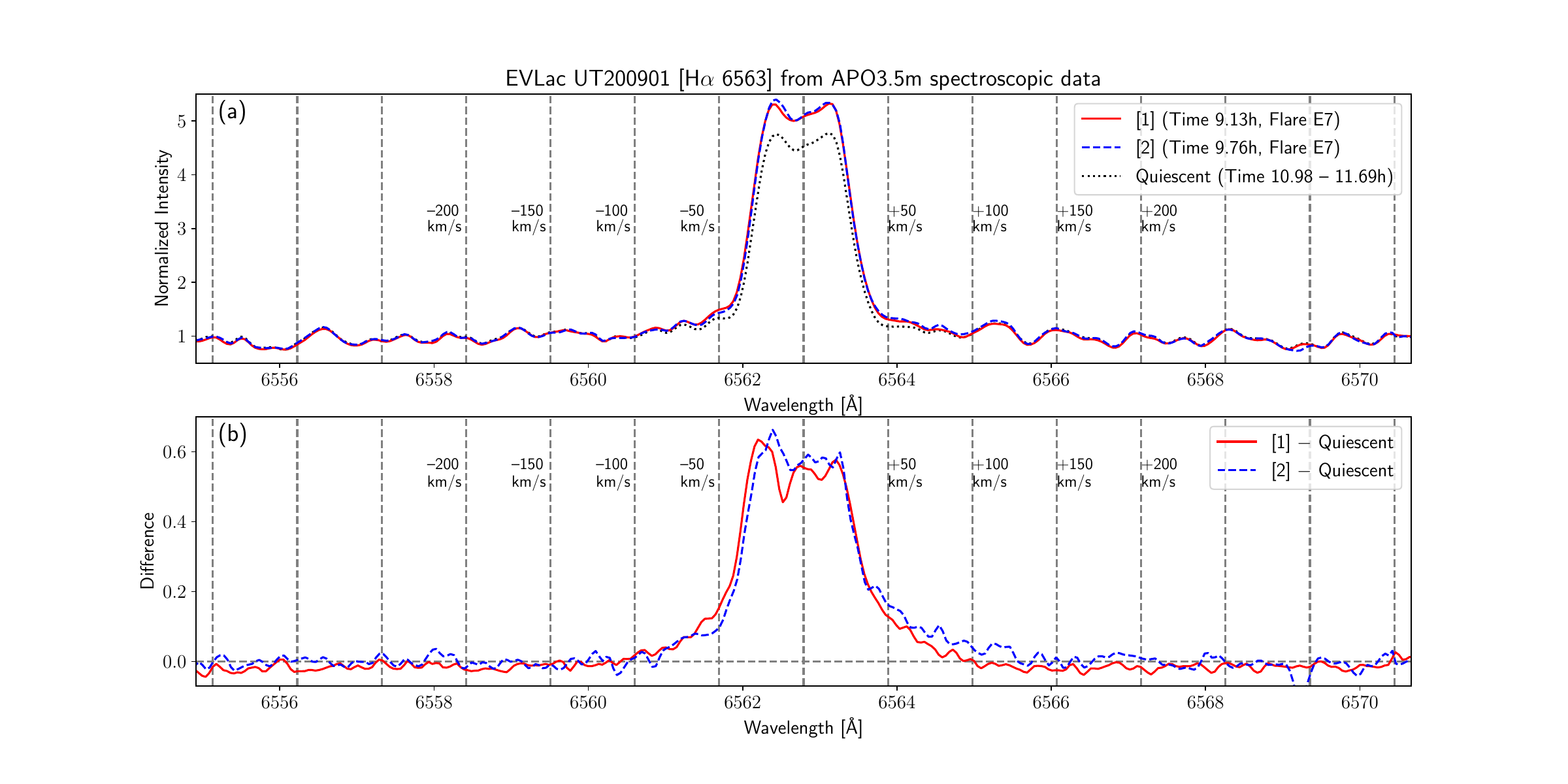}{0.58\textwidth}{\vspace{0mm}}
     \hspace{-0.06\textwidth}
       \fig{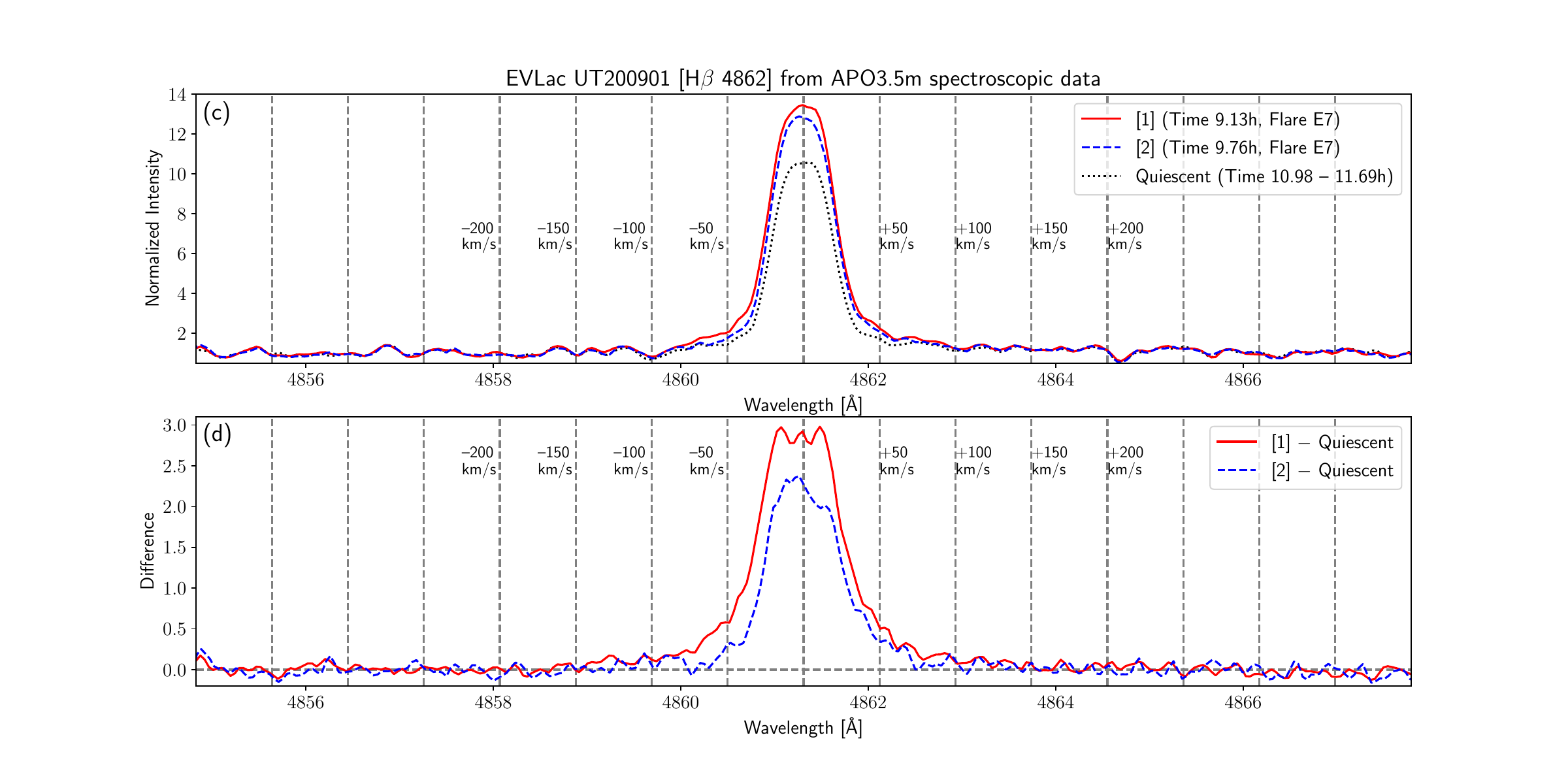}{0.58\textwidth}{\vspace{0mm}}
    }
     \vspace{-0.5cm}
     \caption{
   \color{black}\textrm{  
Line profiles of the H$\alpha$ \& H$\beta$ emission lines during Flare E7
on 2020 September 1 (at the time [1] and [2]) from APO3.5m spectroscopic data, which are plotted similarly with Figure \ref{fig:spec_HaHb_YZCMi_UT190127}.
 } \color{black}
     }
   \label{fig:spec_HaHb_EVLac_UT200901}
   \end{center}
 \end{figure}

\begin{figure}[ht!]
   \begin{center}
          \gridline{  
     \hspace{-0.07\textwidth}
    \fig{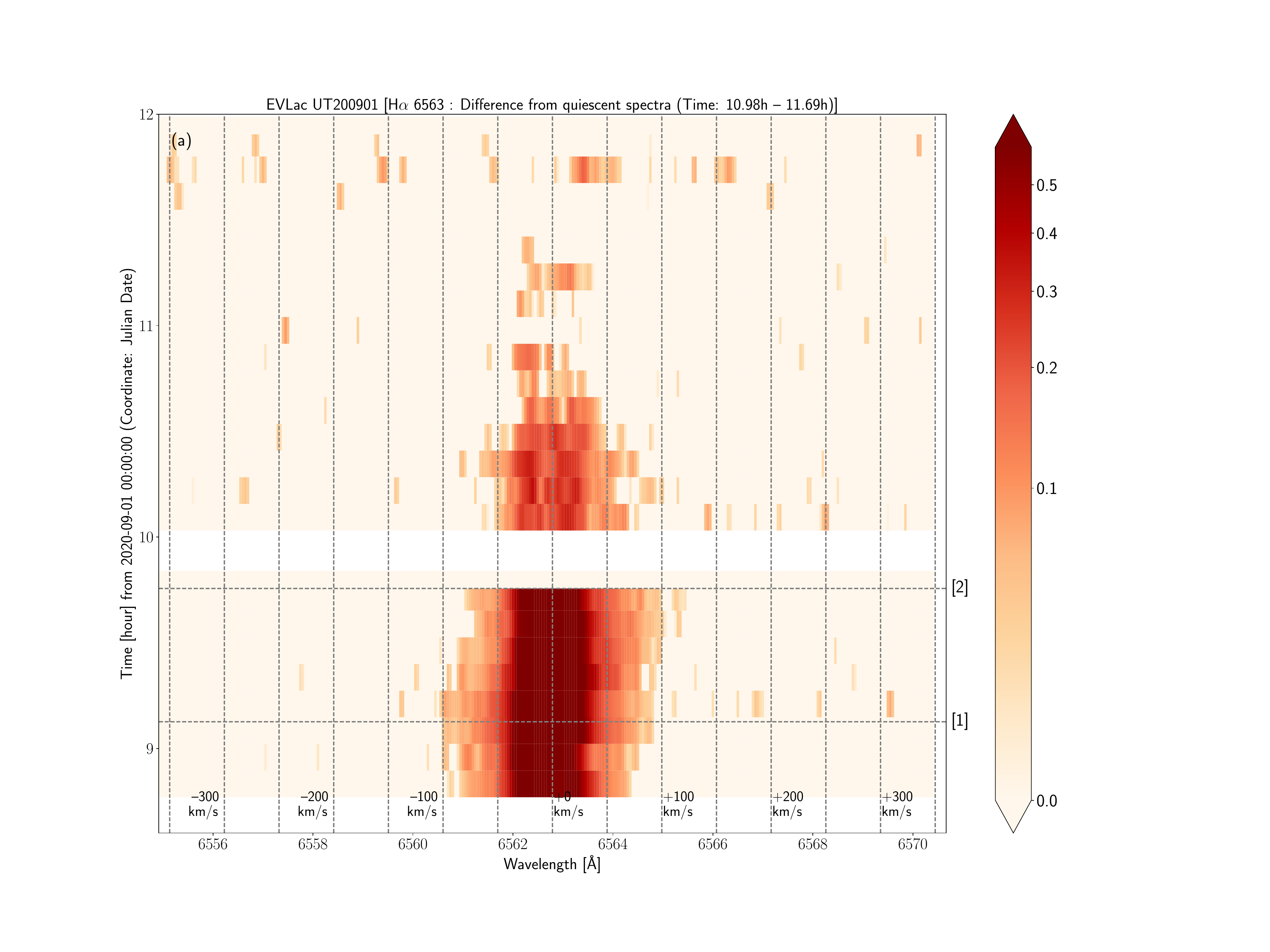}{0.63\textwidth}{\vspace{0mm}}
     \hspace{-0.11\textwidth}
    \fig{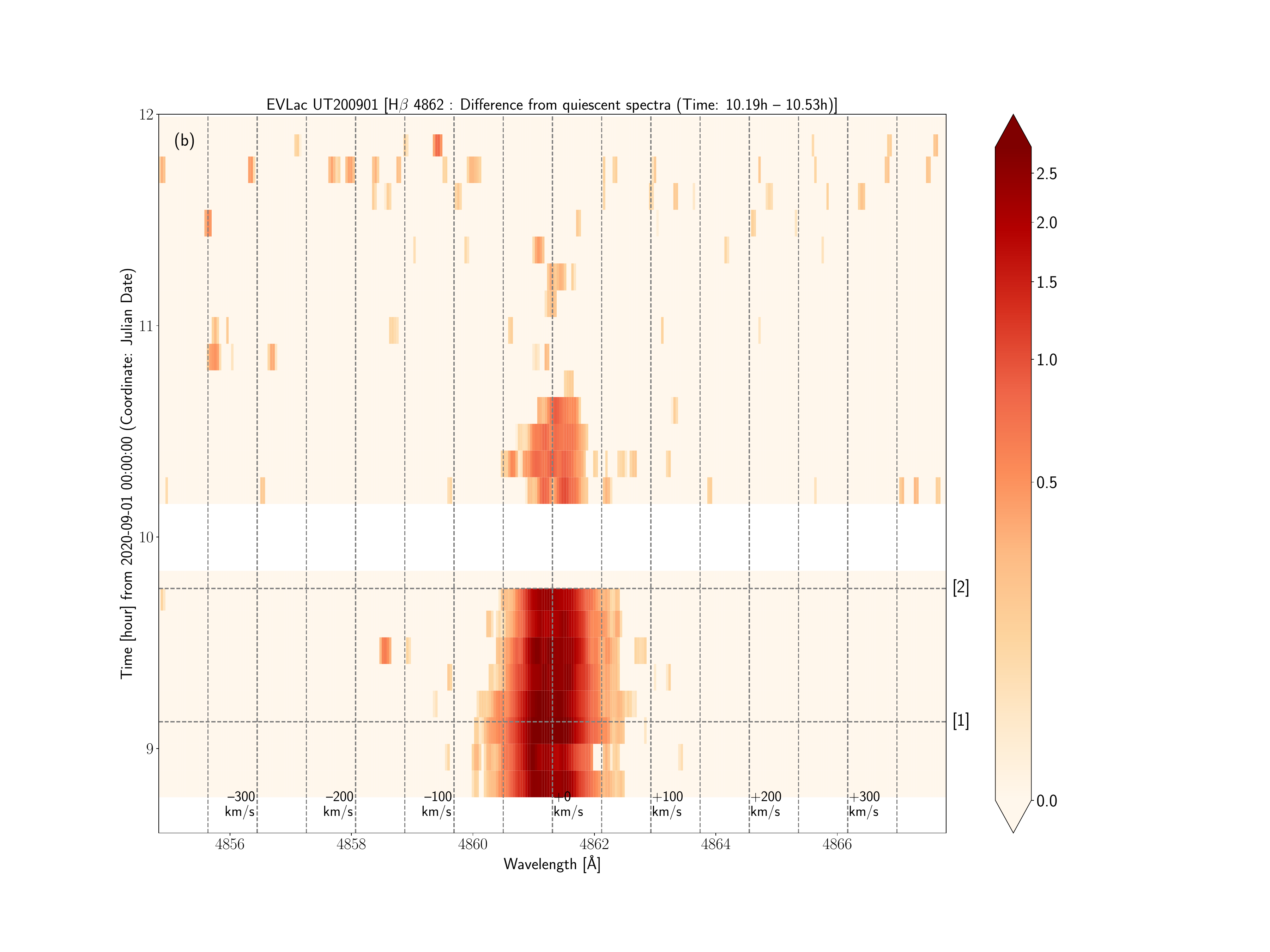}{0.63\textwidth}{\vspace{0mm}}
    }   
     \vspace{-0.5cm}
     \caption{
         \color{black}\textrm{  
Time evolution of the H$\alpha$ \& H$\beta$ line profiles covering Flare E7
on 2020 September 1, which are plotted similarly with Figure \ref{fig:map_HaHb_YZCMi_UT191212}.
The grey horizontal dashed lines indicate the time [1] \& [2], 
which are shown in Figure \ref{fig:lcEW_HaHb_EVLac_UT200901} (light curves) and Figure \ref{fig:spec_HaHb_EVLac_UT200901} (line profiles).
}\color{black}
     }
   \label{fig:map_HaHb_EVLac_UT200901}
   \end{center}
 \end{figure}

The H$\alpha$ \& H$\beta$ line profiles during Flare E7 are shown in
Figures \ref{fig:spec_HaHb_EVLac_UT200901} \& \ref{fig:map_HaHb_EVLac_UT200901}. 
There were no clear red or blue wing asymmetries in the H$\alpha$ and H$\beta$ lines during Flare E7.
 
\clearpage
 
\subsection{Flares E8 \& E9 observed on 2020 September 2} 
\label{subsec:results:2020-Sep-02}

\begin{figure}[ht!]
   \begin{center}
    \gridline{
      \fig{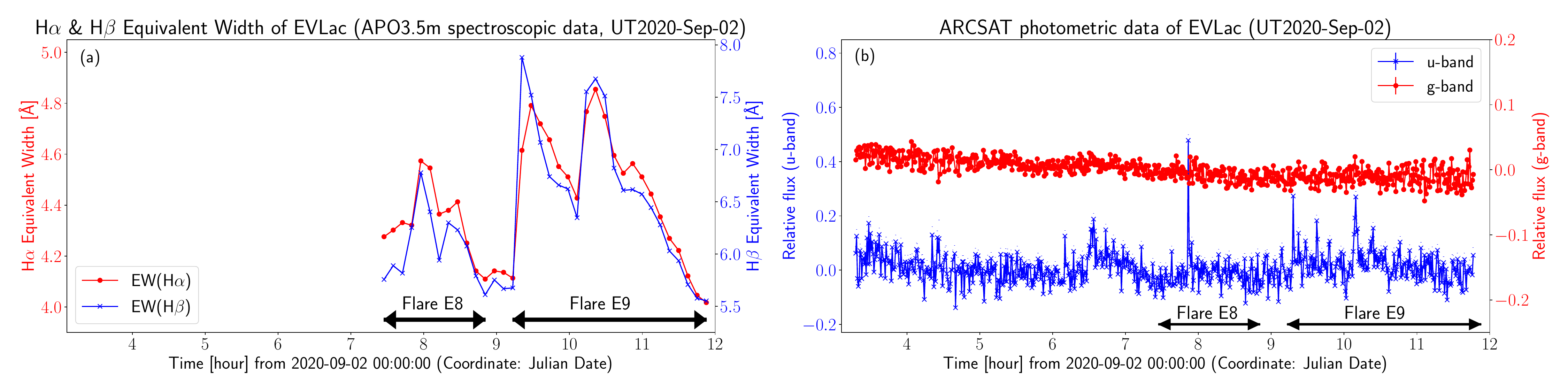}{1.0\textwidth}{\vspace{0mm}}
    }
     \vspace{-0.5cm}
        \gridline{
      \fig{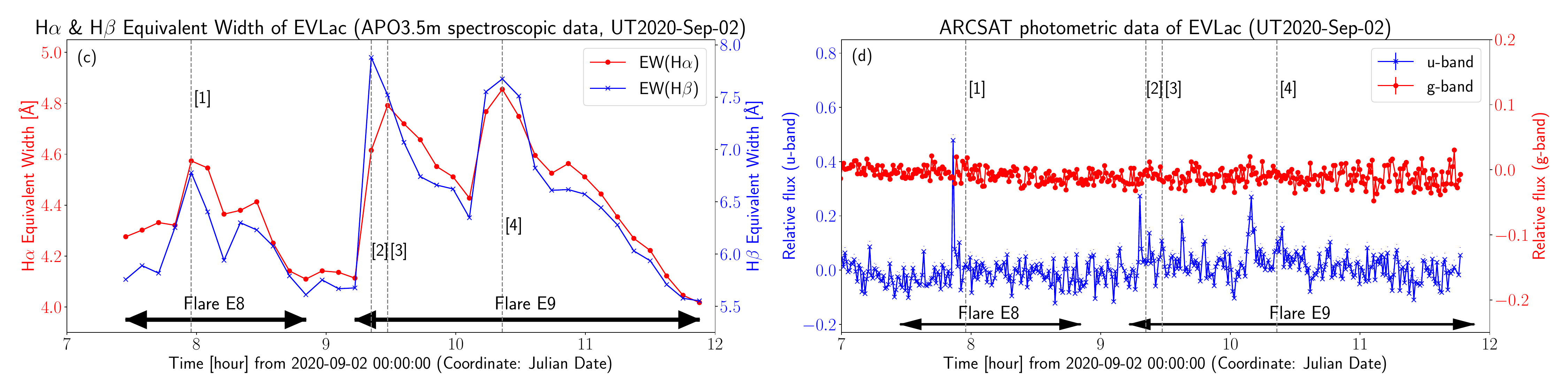}{1.0\textwidth}{\vspace{0mm}}
    }
     \vspace{-0.5cm}
     \caption{
     \color{black}\textrm{  
Light curves of EV Lac on 2020 September 2 showing Flares E8 \& E9, which are plotted 
similarly with Figures \ref{fig:lcEW_HaHb_YZCMi_UT191212} (a)\&(b).
(c)\&(d) are enlarged panels of (a)\&(b).
The grey dashed lines with numbers ([1]-[4]) in (a)\&(b) correspond to the time shown 
with the same numbers in Figures \ref{fig:spec_HaHb_EVLac_UT200902} \& \ref{fig:map_HaHb_EVLac_UT200902}.
 } \color{black}
     }
   \label{fig:lcEW_HaHb_EVLac_UT200902}
   \end{center}
 \end{figure}

On 2020 September 2, two flares (Flares E8 \& E9) 
were detected on EV Lac in H$\alpha$ \& H$\beta$ lines 
as shown in Figure \ref{fig:lcEW_HaHb_EVLac_UT200902} (c).  
As for Flare E8, the H$\alpha$ \& H$\beta$ equivalent widths increased up to 4.6\AA~and 6.8\AA, respectively, and $\Delta t^{\rm{flare}}_{\rm{H}\alpha}$ is 1.4 hours (Table \ref{table:list1_flares}).
In addition to these enhancements in Balmer emission lines, the continuum brightness observed with ARCSAT $u$-band increased by $\sim$45--50\% while that with $g$-band did not show clear increases \color{black}\textrm{compared with photometric error (\color{black}\textrm{$3\sigma_{g}$}\color{black}=2.5\%)} \color{black}, during Flare E8 (Figure \ref{fig:lcEW_HaHb_EVLac_UT200902} (d)).
As for Flare E9, the H$\alpha$ \& H$\beta$ equivalent widths increased up to 4.9\AA~and 7.9\AA, respectively, and $\Delta t^{\rm{flare}}_{\rm{H}\alpha}$ is 2.7 hours (Table \ref{table:list1_flares}).
In addition to these enhancements in Balmer emission lines, the continuum brightness observed with ARCSAT $u$-band increased by $\sim$25\% while that with $g$-band did not show clear increases \color{black}\textrm{compared with photometric error (\color{black}\textrm{$3\sigma_{g}$}\color{black}=2.5\%)} \color{black}, during Flare E9 (Figure \ref{fig:lcEW_HaHb_EVLac_UT200902} (d)).
 \color{black}\textrm{ 
$L_{u}$, $L_{g}$, $E_{u}$, $E_{g}$, $L_{\rm{H}\alpha}$, $L_{\rm{H}\beta}$, $E_{\rm{H}\alpha}$, and $E_{\rm{H}\beta}$ values are estimated and listed in Table \ref{table:list1_flares}.
} \color{black}

\begin{figure}[ht!]
   \begin{center}
            \gridline{  
     \hspace{-0.06\textwidth}
    \fig{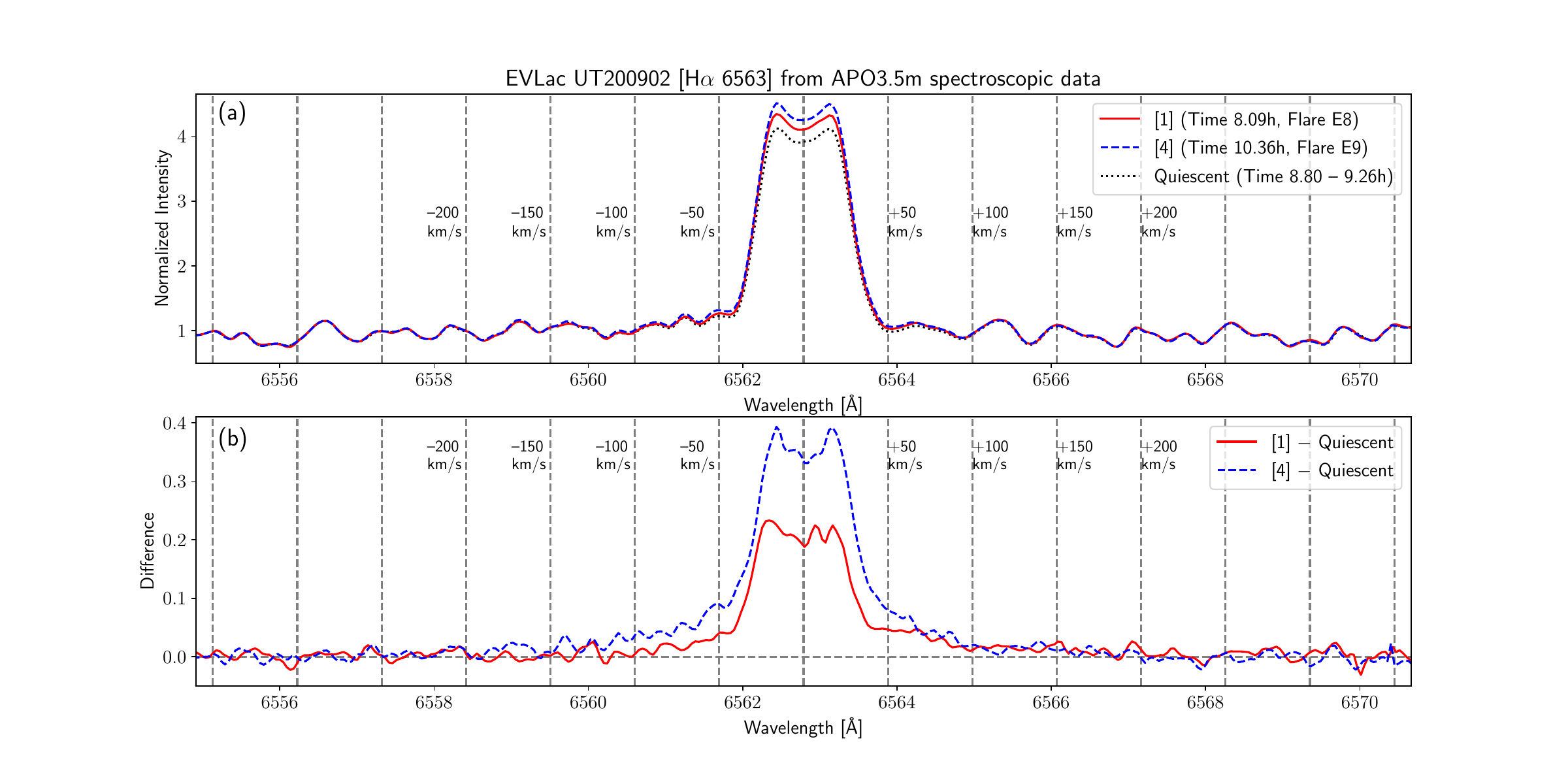}{0.58\textwidth}{\vspace{0mm}}
     \hspace{-0.06\textwidth}
       \fig{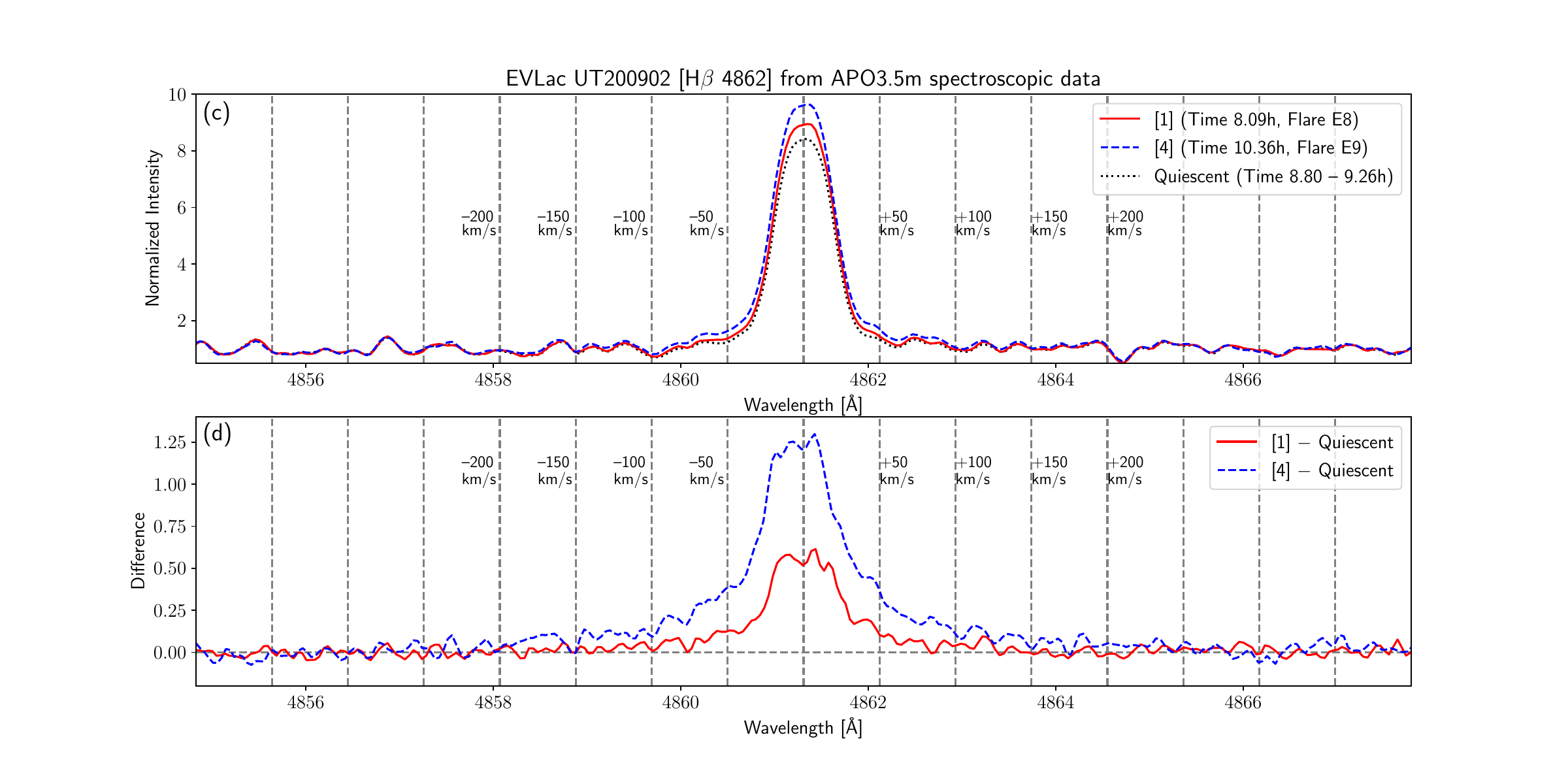}{0.58\textwidth}{\vspace{0mm}}
    }
     \vspace{-1.0cm}
            \gridline{  
     \hspace{-0.06\textwidth}
    \fig{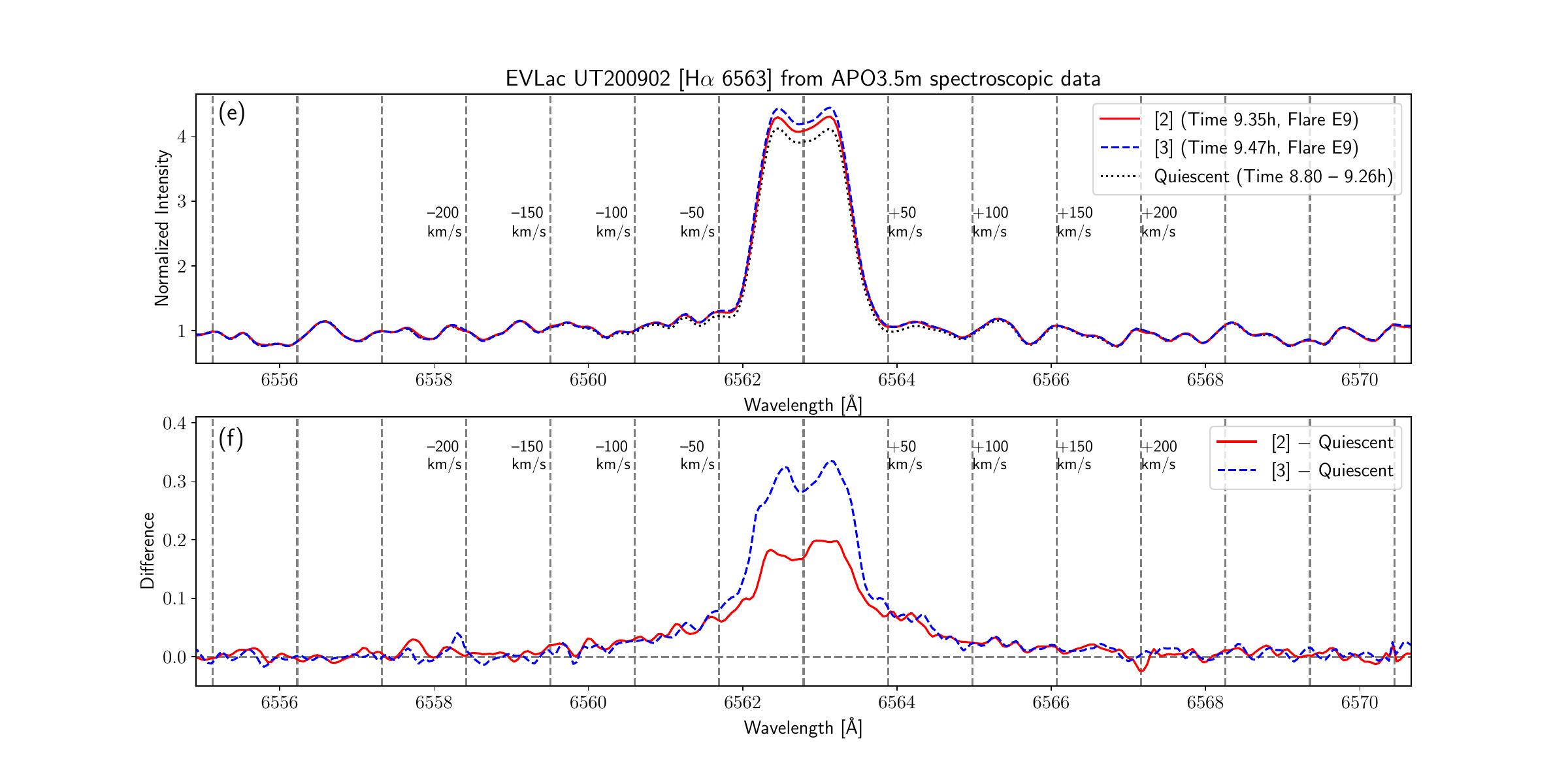}{0.58\textwidth}{\vspace{0mm}}
     \hspace{-0.06\textwidth}
       \fig{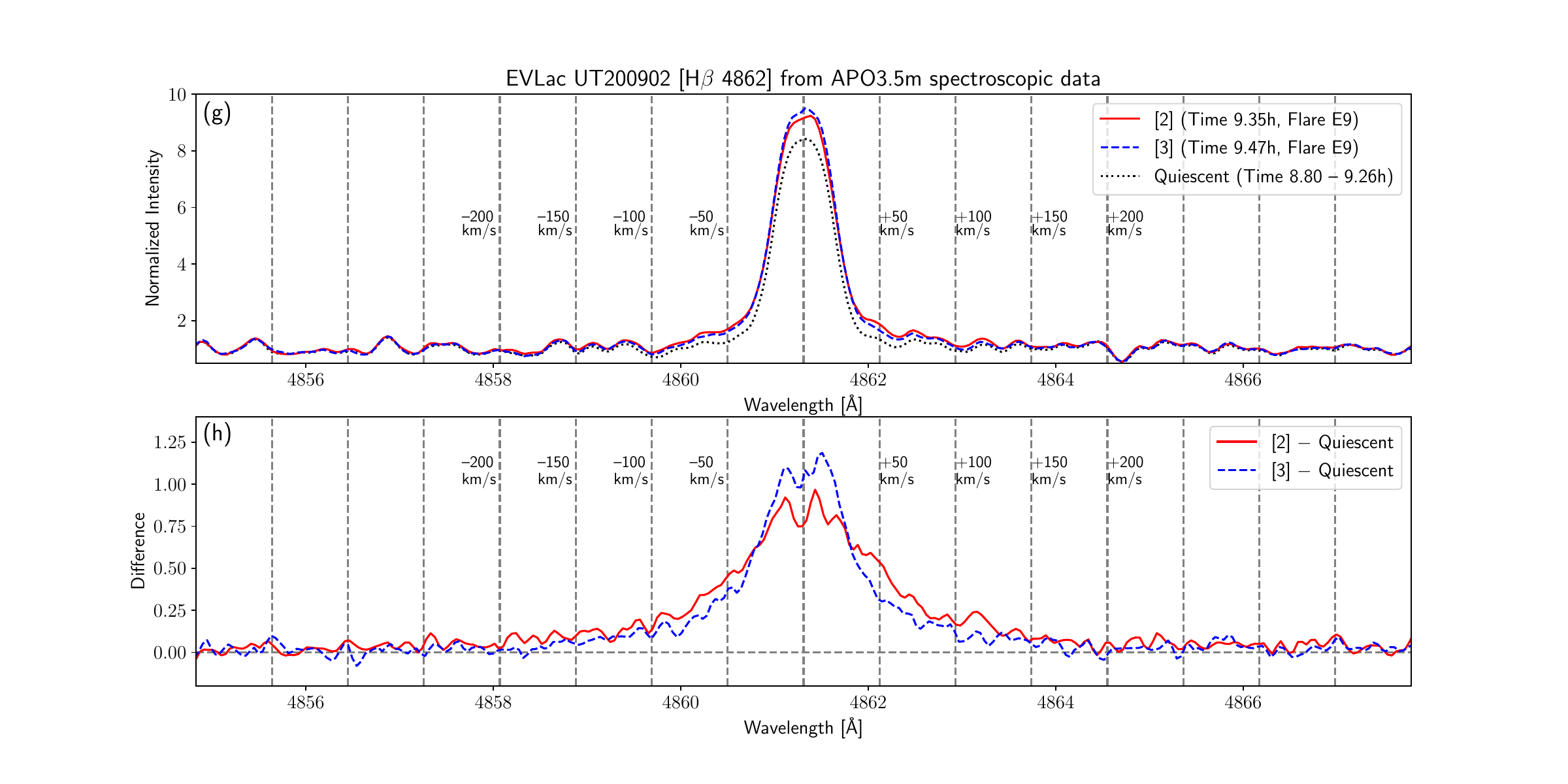}{0.58\textwidth}{\vspace{0mm}}
    }
     \vspace{-0.5cm}
     \caption{
   \color{black}\textrm{  
Line profiles of the H$\alpha$ \& H$\beta$ emission lines  during Flares E8 \& E9 
on 2020 September 2 (at the time [1]--[4]) from APO3.5m spectroscopic data, which are plotted similarly with Figure \ref{fig:spec_HaHb_YZCMi_UT190127}.
The profiles at the time [1]\&[4] are in (a)--(d), while those at at the time [2]\&[3] are in (e)--(h).
 } \color{black}
     }
   \label{fig:spec_HaHb_EVLac_UT200902}
   \end{center}
 \end{figure}

\begin{figure}[ht!]
   \begin{center}
      \gridline{  
     \hspace{-0.07\textwidth}
    \fig{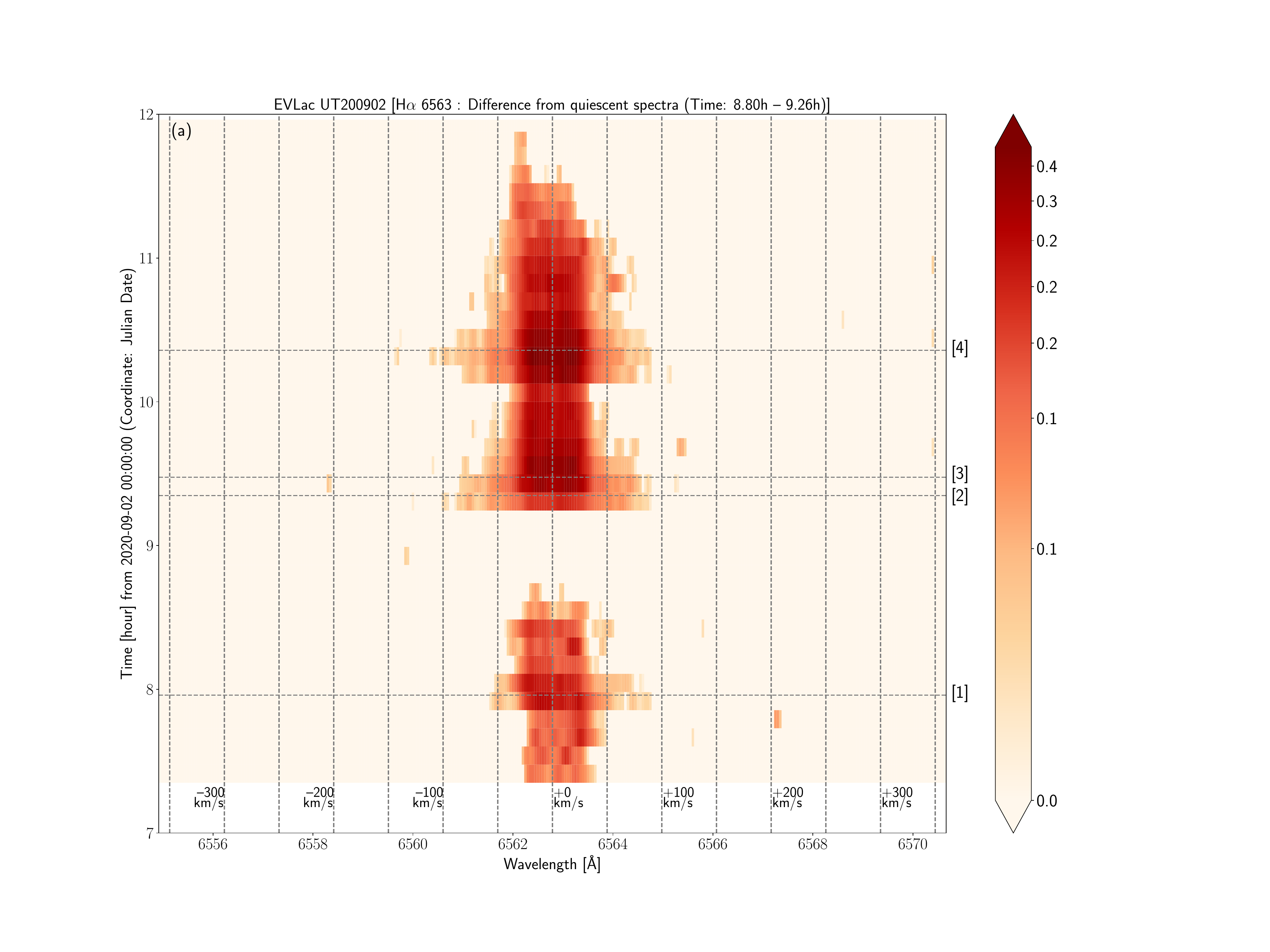}{0.63\textwidth}{\vspace{0mm}}
     \hspace{-0.11\textwidth}
    \fig{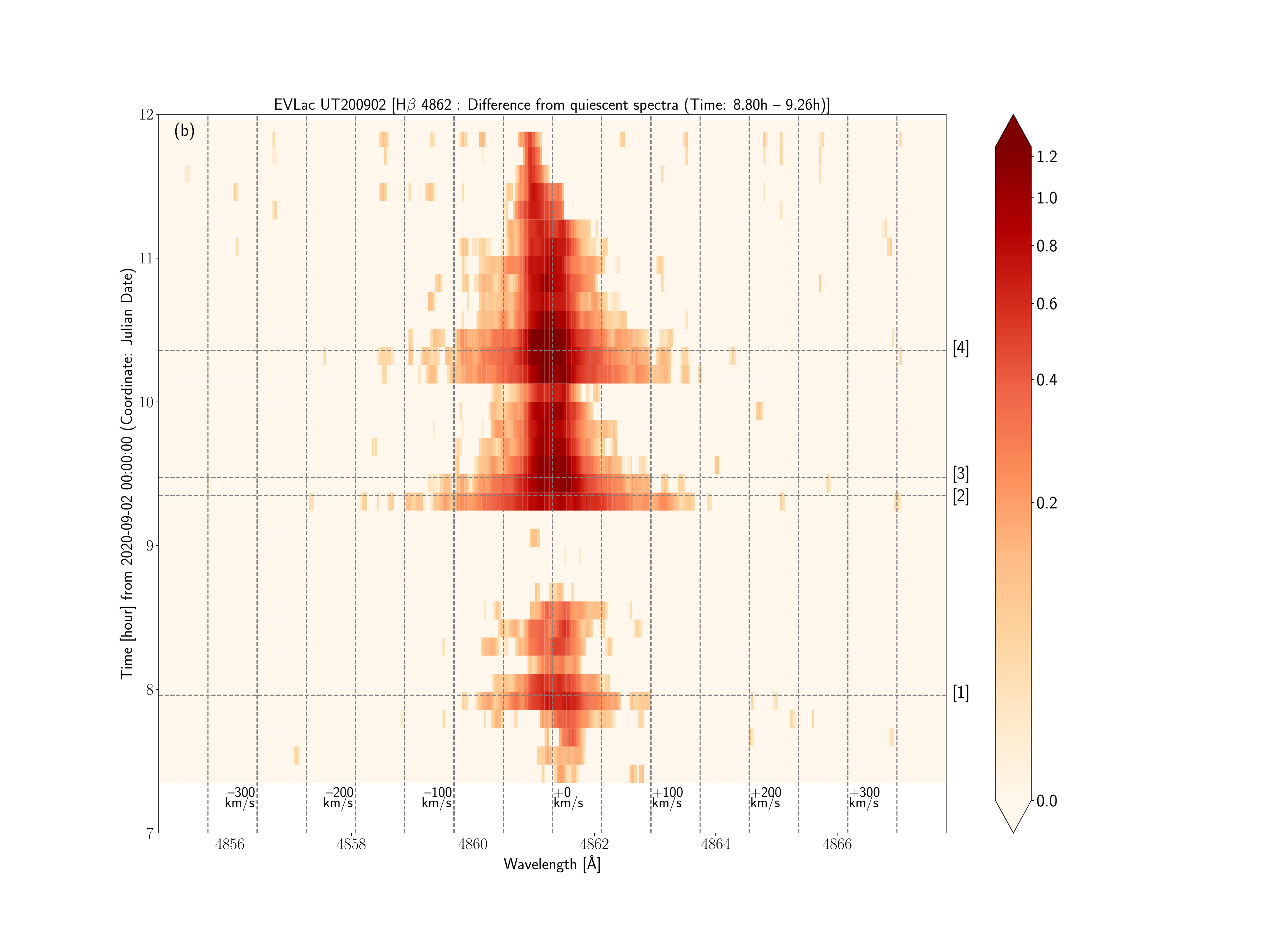}{0.63\textwidth}{\vspace{0mm}}
    }
     \vspace{-0.5cm}
     \caption{
         \color{black}\textrm{  
Time evolution of the H$\alpha$ \& H$\beta$ line profiles covering Flares E8 \& E9
on 2020 September 2, which are plotted similarly with Figure \ref{fig:map_HaHb_YZCMi_UT191212}.
The grey horizontal dashed lines indicate the time [1] -- [4], 
which are shown in Figure \ref{fig:lcEW_HaHb_EVLac_UT200902} (light curves) and Figure \ref{fig:spec_HaHb_EVLac_UT200902} (line profiles).
}\color{black}
     }
   \label{fig:map_HaHb_EVLac_UT200902}
   \end{center}
\end{figure}

The H$\alpha$ \& H$\beta$ line profiles during Flares E8 \& E9 are shown in
Figures \ref{fig:spec_HaHb_EVLac_UT200902} \& \ref{fig:map_HaHb_EVLac_UT200902}. 
There were no clear red or blue wing asymmetries in the H$\alpha$ and H$\beta$ lines during Flares E8 \& E9. 
During Flare E9, 
the H$\alpha$ and H$\beta$ lines show the relatively symmetric line broadenings
with $\pm$150 km s$^{-1}$ and $\pm$200 km s$^{-1}$, respectively 
(time [2]--[4] in Figures \ref{fig:spec_HaHb_EVLac_UT200902} \& \ref{fig:map_HaHb_EVLac_UT200902}).

\clearpage
 
\subsection{Flare A1 observed on 2019 May 17} 
\label{subsec:results:2019-May-17}

\begin{figure}[ht!]
   \begin{center}
   \gridline{
    \fig{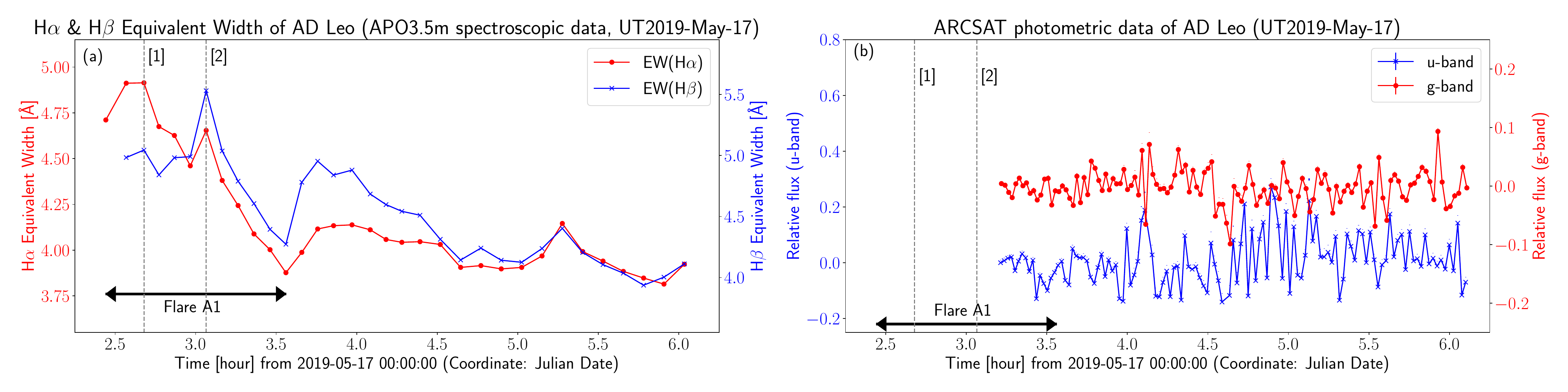}{1.0\textwidth}{\vspace{0mm}}
 }
     \vspace{-5mm}
     \caption{
Light curves of \color{black}\textrm{AD Leo }\color{black} on 2019 May 17 showing Flare A1, which are plotted 
similarly with Figures \ref{fig:lcEW_HaHb_YZCMi_UT191212} (a)\&(b).
The grey dashed lines with numbers ([1] \& [2]) correspond to the time shown 
with the same numbers in Figures \ref{fig:spec_HaHb_ADLeo_UT190517} \& \ref{fig:map_HaHb_ADLeo_UT190517}.
     }
   \label{fig:lcEW_HaHb_ADLeo_UT190517}
   \end{center}
 \end{figure}

On 2019 May 17, one flare (Flares A1) 
was detected on AD Leo in H$\alpha$ \& H$\beta$ lines 
as shown in Figure \ref{fig:lcEW_HaHb_ADLeo_UT190517} (a).  
Flare A1 already started when the observation started.
The H$\alpha$ \& H$\beta$ equivalent widths increased up to 4.9\AA~and 5.5\AA, respectively, and $\Delta t^{\rm{flare}}_{\rm{H}\alpha}$ is $>$1.4 hours (Table \ref{table:list1_flares}).
For most of the time of Flare A1, there was no photometric observation of ARCSAT, so we do not know whether there were the continuum brightness changes during Flare A1  (Figure \ref{fig:lcEW_HaHb_ADLeo_UT190517} (b)).
 \color{black}\textrm{ 
$L_{\rm{H}\alpha}$, $L_{\rm{H}\beta}$, $E_{\rm{H}\alpha}$, and $E_{\rm{H}\beta}$ values are estimated and listed in Table \ref{table:list1_flares}.
The flare peak luminosities and flare energies described here can be lower limit values since Flare A1 already started when the spectroscopic observation started.
} \color{black}

\begin{figure}[ht!]
   \begin{center}
           \gridline{  
     \hspace{-0.06\textwidth}
    \fig{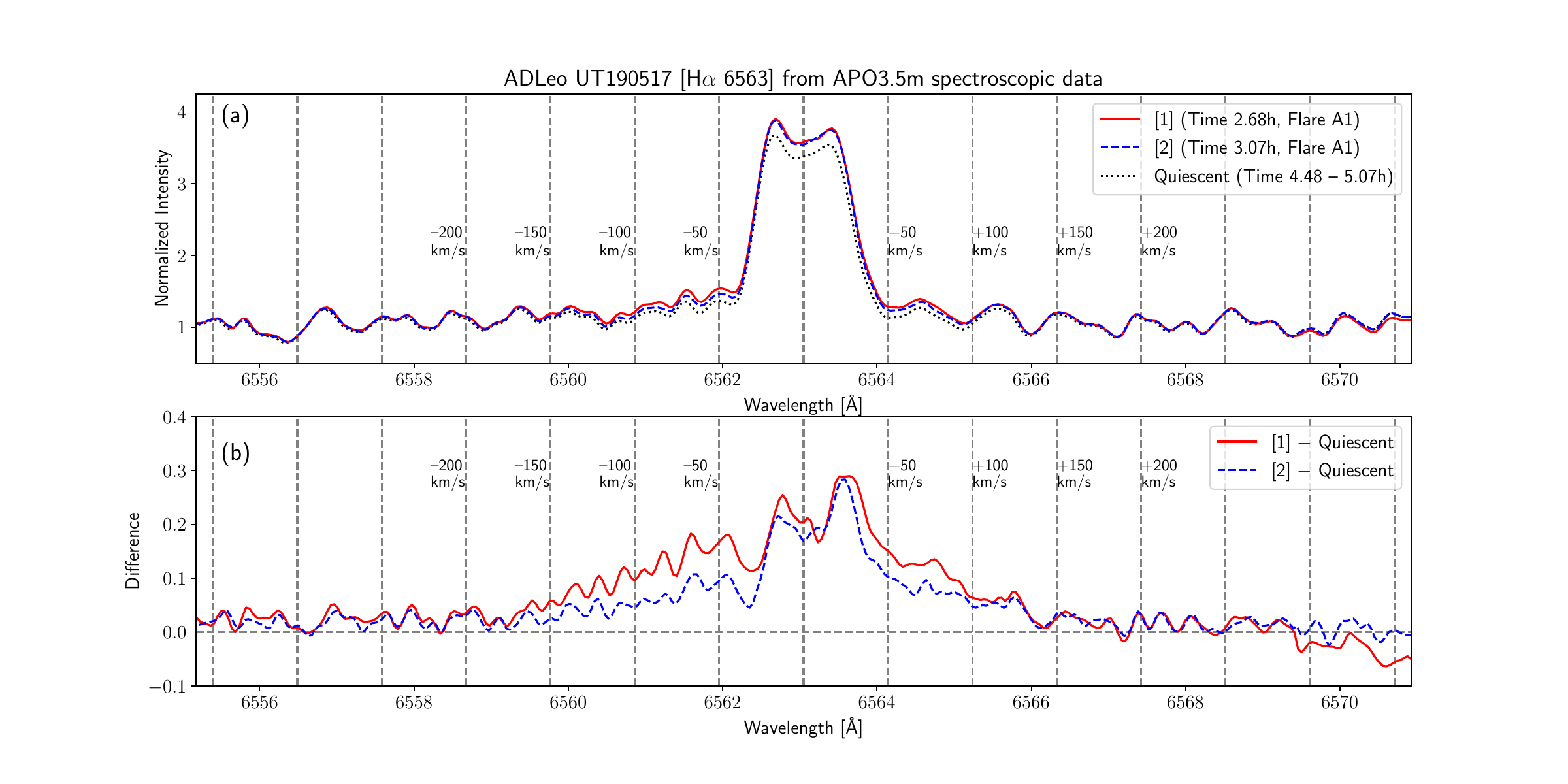}{0.58\textwidth}{\vspace{0mm}}
     \hspace{-0.06\textwidth}
       \fig{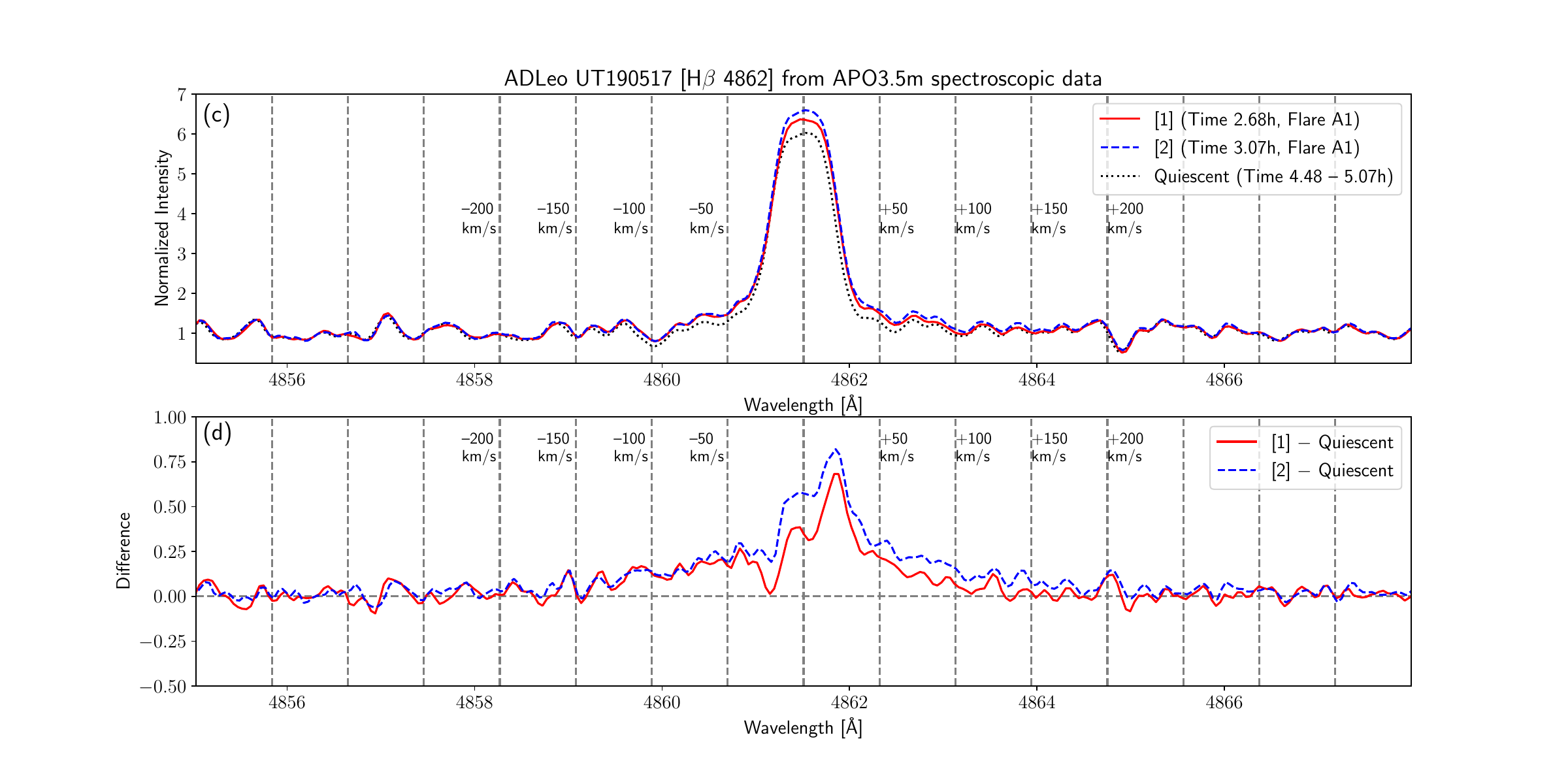}{0.58\textwidth}{\vspace{0mm}}
    }
     \vspace{-0.5cm}
     \caption{
   \color{black}\textrm{  
Line profiles of the H$\alpha$ \& H$\beta$ emission lines during Flare A1 
on 2019 May 17 (at the time [1] and [2]) from APO3.5m spectroscopic data, which are plotted similarly with Figure \ref{fig:spec_HaHb_YZCMi_UT190127}.
 } \color{black}
     }
   \label{fig:spec_HaHb_ADLeo_UT190517}
   \end{center}
 \end{figure}

\begin{figure}[ht!]
   \begin{center}
      \gridline{  
     \hspace{-0.07\textwidth}
   \fig{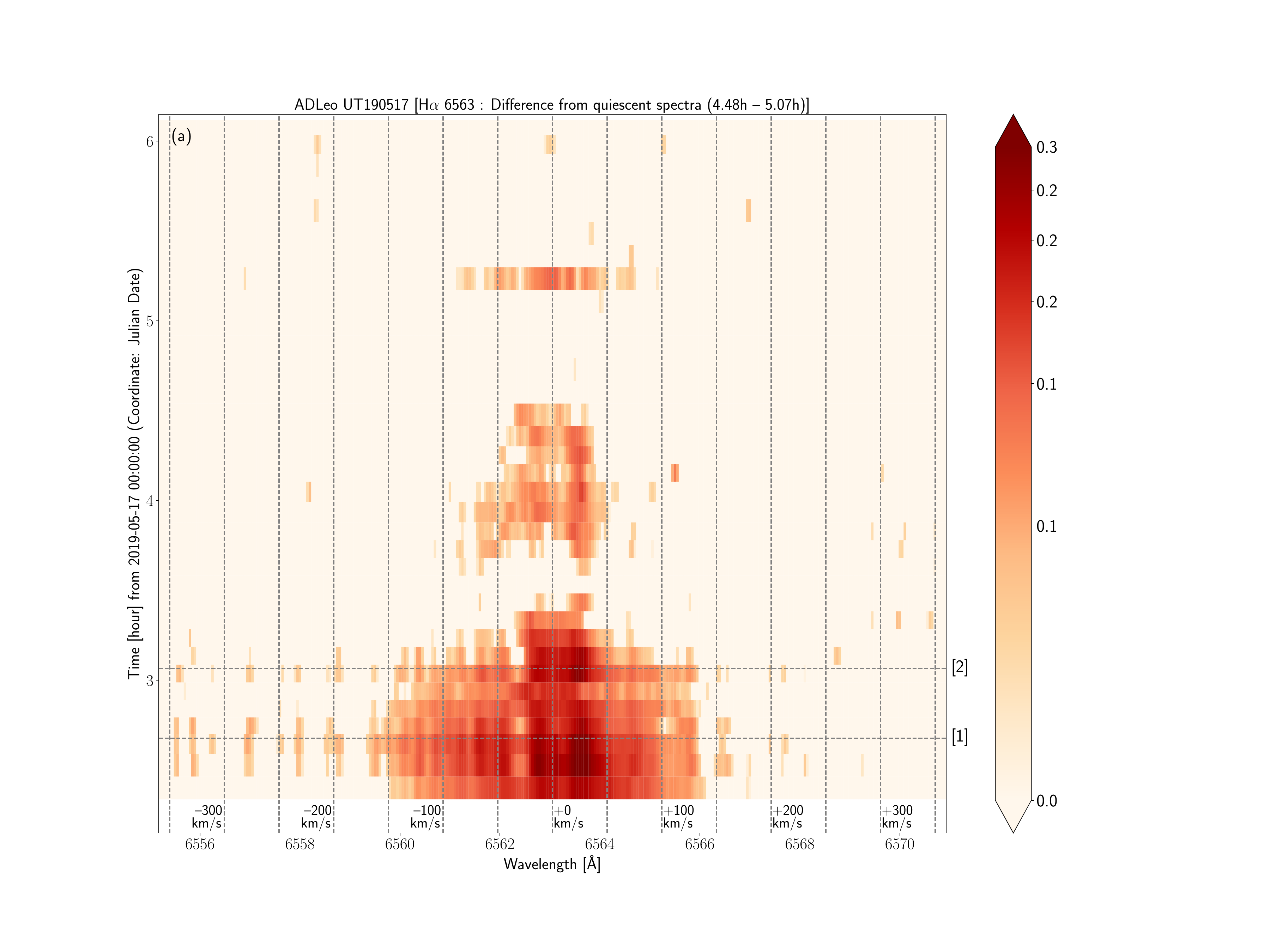}{0.63\textwidth}{\vspace{0mm}}
     \hspace{-0.11\textwidth}
    \fig{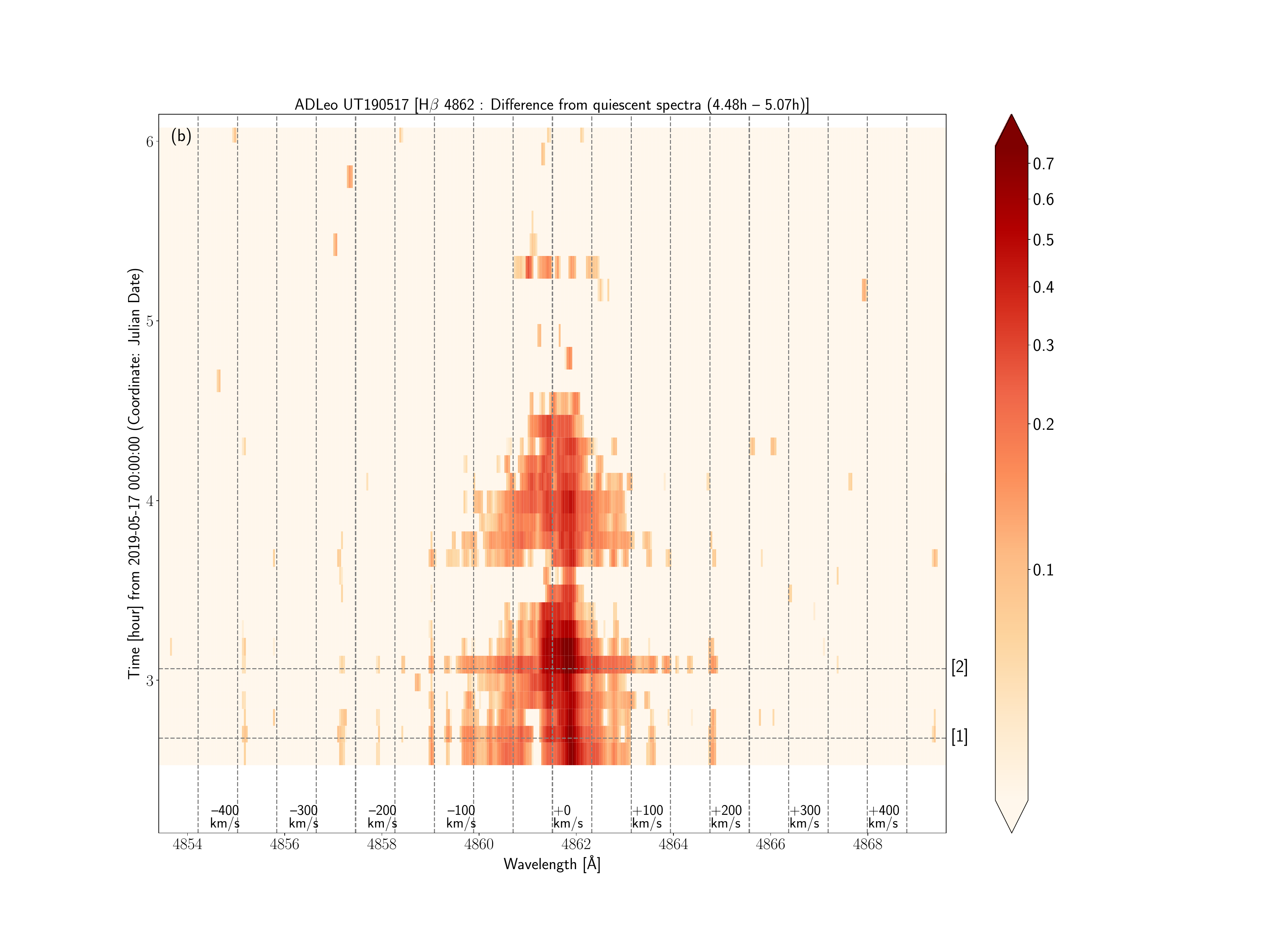}{0.63\textwidth}{\vspace{0mm}}
    }
     \vspace{-0.5cm}
     \caption{
         \color{black}\textrm{  
Time evolution of the H$\alpha$ \& H$\beta$ line profiles covering Flare A1
on 2019 May 17, which are plotted similarly with Figure \ref{fig:map_HaHb_YZCMi_UT191212}.
The grey horizontal dashed lines indicate the time [1] \& [2], 
which are shown in Figure \ref{fig:lcEW_HaHb_ADLeo_UT190517}  (light curves) and Figure \ref{fig:spec_HaHb_ADLeo_UT190517} (line profiles).
}\color{black}
     }
   \label{fig:map_HaHb_ADLeo_UT190517}
   \end{center}
 \end{figure}

 The H$\alpha$ \& H$\beta$ line profiles during Flare A1 is shown in
Figures \ref{fig:spec_HaHb_ADLeo_UT190517} \& \ref{fig:map_HaHb_ADLeo_UT190517}. 
During Flare A1, 
the H$\alpha$ and H$\beta$ lines showed the line broadenings
with $\pm$200 km s$^{-1}$ and $\pm$150 km s$^{-1}$, respectively. 
At around time [1], the blue wing of H$\alpha$ line could be very slightly
enhanced, but it is not so clear and we do not judge this flare showed blue wing asymmetry (Figure \ref{fig:spec_HaHb_ADLeo_UT190517} (b)).

 \subsection{Flare A2 observed on 2019 May 18} 
\label{subsec:results:2019-May-18}

\begin{figure}[ht!]
   \begin{center}
   \gridline{
    \fig{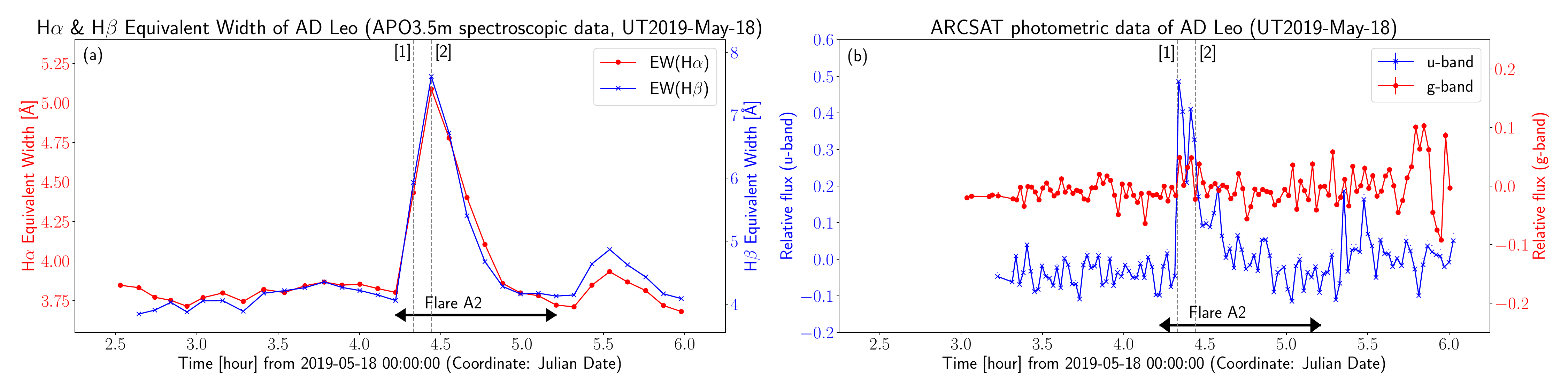}{1.0\textwidth}{\vspace{0mm}}
    }
     \vspace{-5mm}
     \caption{
     \color{black}\textrm{  
Light curves of \color{black}\textrm{AD Leo }\color{black} on 2019 May 18 showing Flare A2, which are plotted 
similarly with Figures \ref{fig:lcEW_HaHb_YZCMi_UT191212} (a)\&(b).
The grey dashed lines with numbers ([1] \& [2]) correspond to the time shown 
with the same numbers in Figures \ref{fig:spec_HaHb_ADLeo_UT190518} \& \ref{fig:map_HaHb_ADLeo_UT190518}.
 } \color{black}
     }
   \label{fig:lcEW_HaHb_ADLeo_UT190518}
   \end{center}
 \end{figure}

On 2019 May 18, one flare (Flares A2) 
was detected on AD Leo in H$\alpha$ \& H$\beta$ lines 
as shown in Figure \ref{fig:lcEW_HaHb_ADLeo_UT190518} (a).  
The H$\alpha$ \& H$\beta$ equivalent widths increased up to 5.1\AA~and 7.6\AA, respectively, and $\Delta t^{\rm{flare}}_{\rm{H}\alpha}$ is 1.0 hour (Table \ref{table:list1_flares}).
In addition to these enhancements in Balmer emission lines, the continuum brightness observed with ARCSAT $u$- \& $g$-bands increased by $\sim$50\% and $\sim$4--5\%, respectively, during Flare A2 (Figure \ref{fig:lcEW_HaHb_ADLeo_UT190518} (b)).
 \color{black}\textrm{ 
$L_{u}$, $L_{g}$, $E_{u}$, $E_{g}$, $L_{\rm{H}\alpha}$, $L_{\rm{H}\beta}$, $E_{\rm{H}\alpha}$, and $E_{\rm{H}\beta}$ values are estimated and listed in Table \ref{table:list1_flares}.
} \color{black}

\begin{figure}[ht!]
   \begin{center}
               \gridline{  
     \hspace{-0.06\textwidth}
    \fig{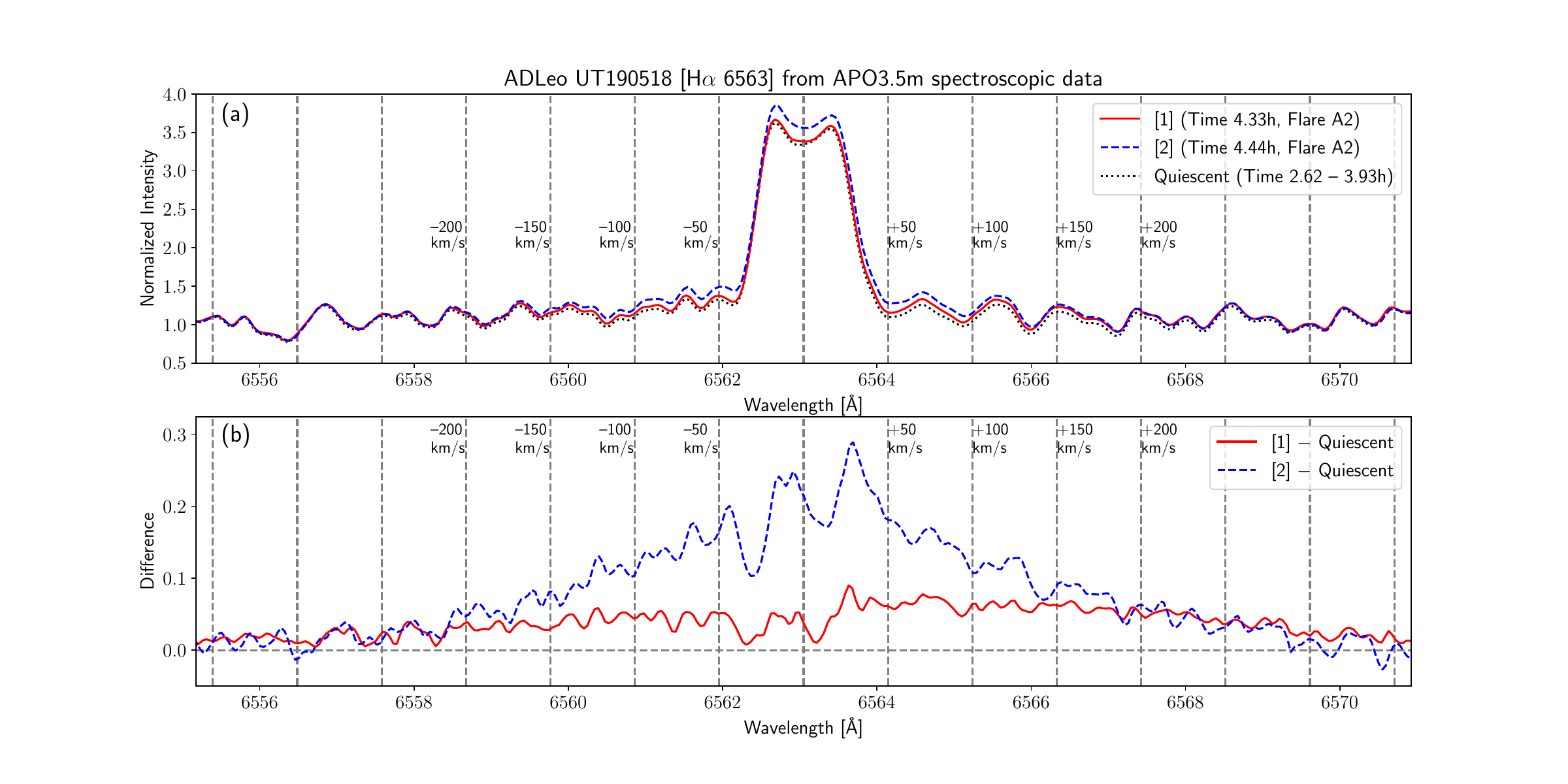}{0.58\textwidth}{\vspace{0mm}}
     \hspace{-0.06\textwidth}
       \fig{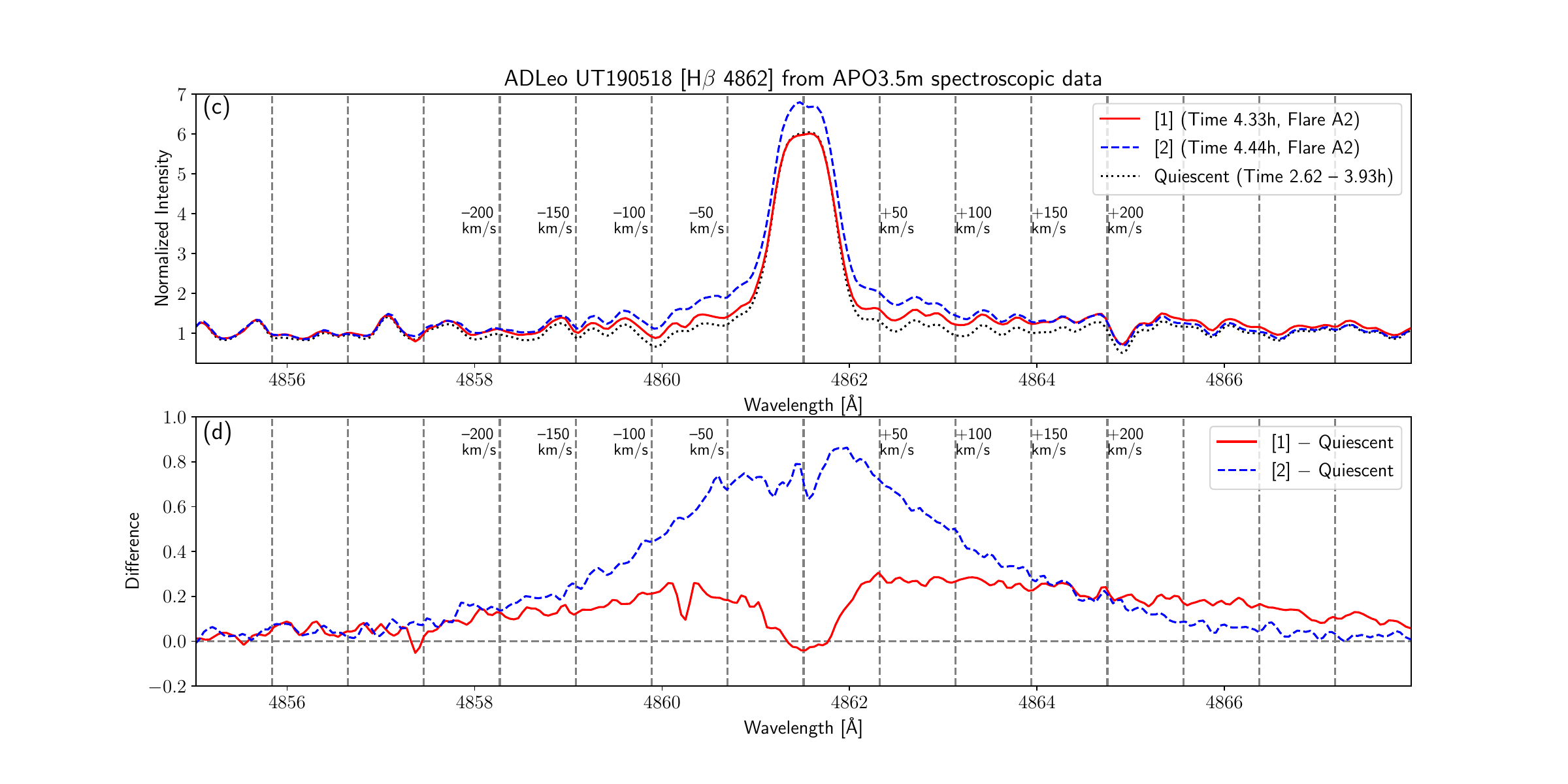}{0.58\textwidth}{\vspace{0mm}}
    }
     \vspace{-0.5cm}
     \caption{
   \color{black}\textrm{  
Line profiles of the H$\alpha$ \& H$\beta$ emission lines during Flare A2 
on 2019 May 18 (at the time [1] and [2]) from APO3.5m spectroscopic data, which are plotted similarly with Figure \ref{fig:spec_HaHb_YZCMi_UT190127}.
 } \color{black}
     }
   \label{fig:spec_HaHb_ADLeo_UT190518}
   \end{center}
 \end{figure}

\begin{figure}[ht!]
   \begin{center}
      \gridline{  
     \hspace{-0.07\textwidth}
   \fig{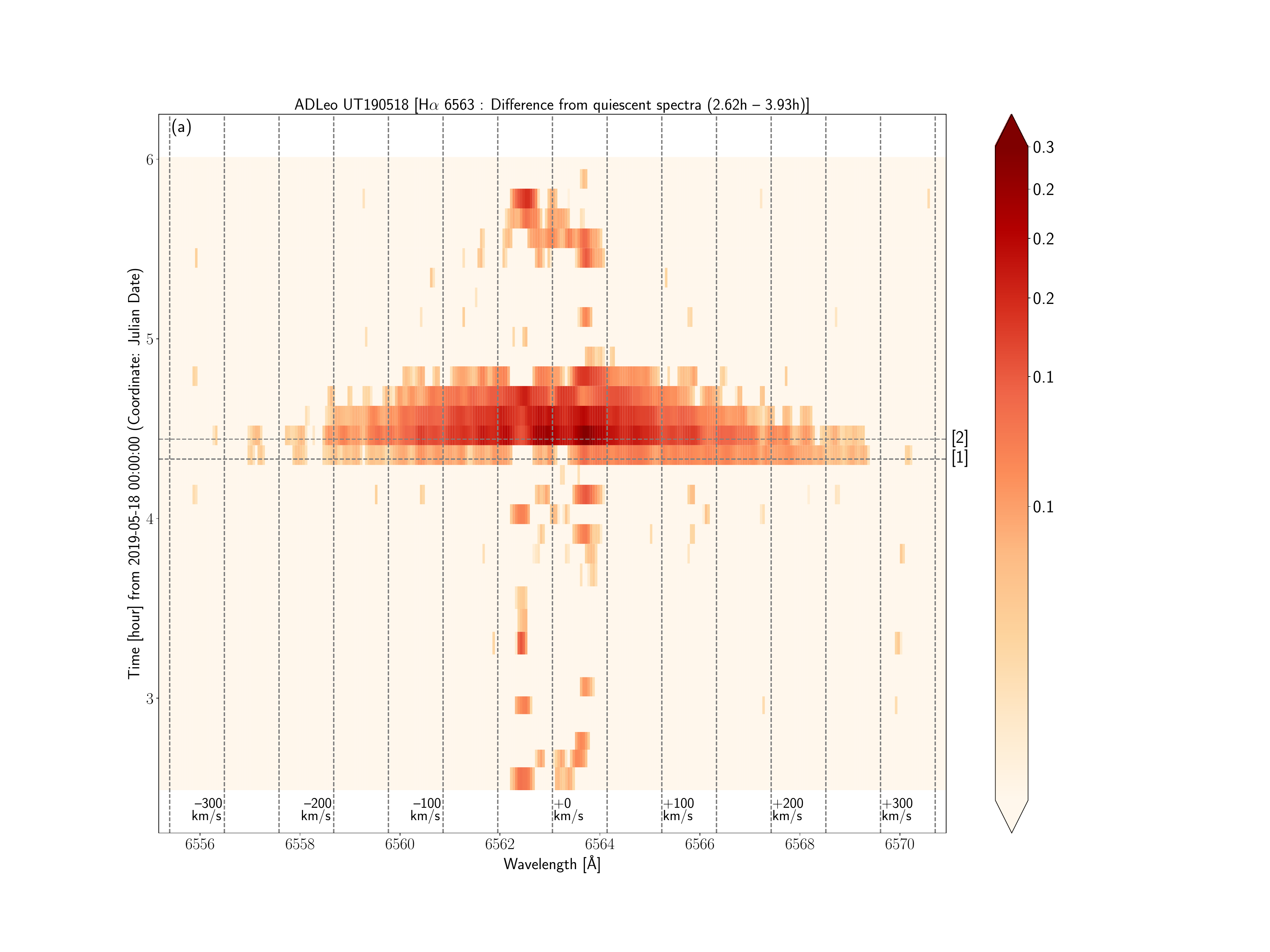}{0.63\textwidth}{\vspace{0mm}}
     \hspace{-0.11\textwidth}
    \fig{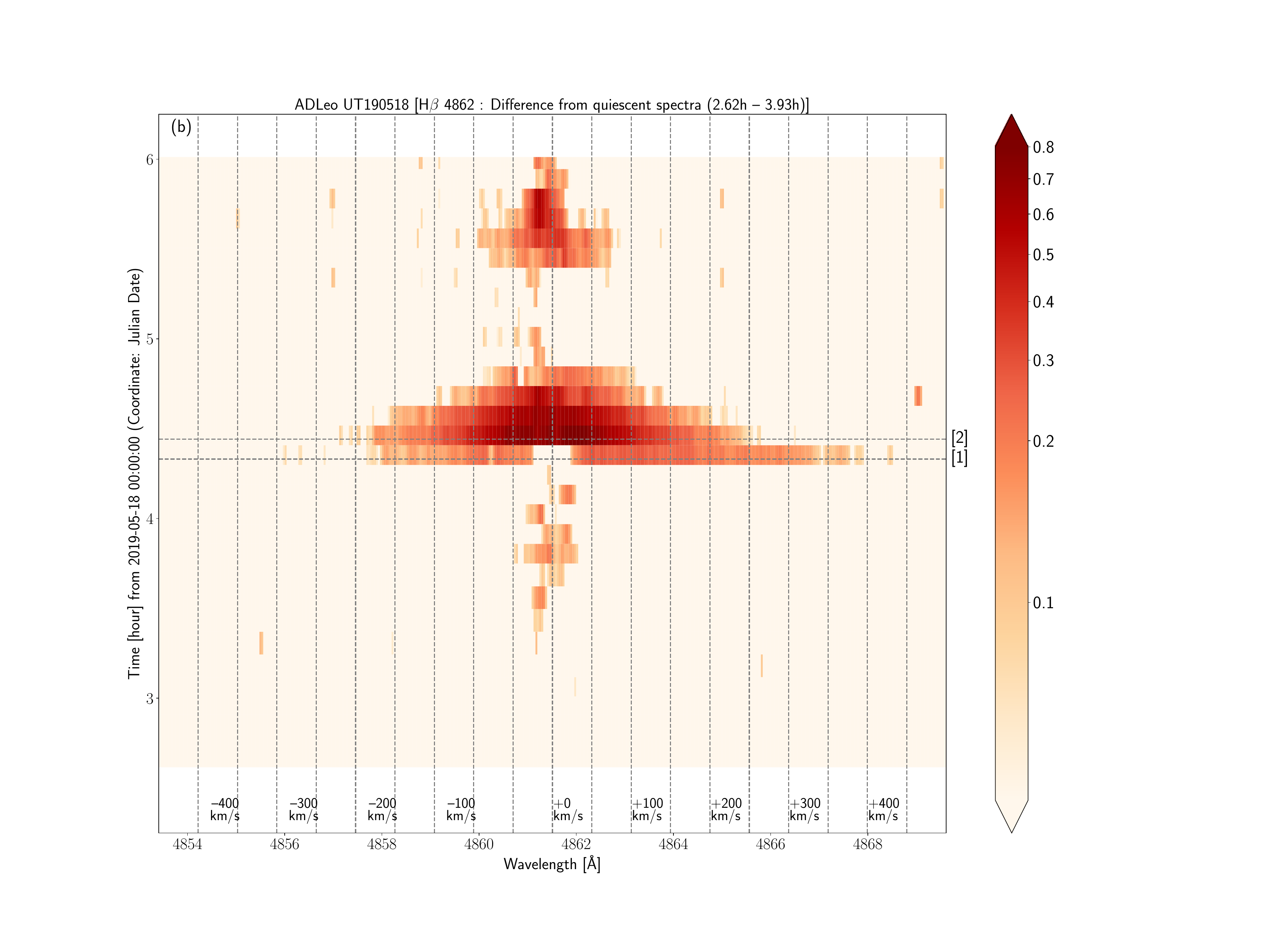}{0.63\textwidth}{\vspace{0mm}}
    }
    
     \vspace{-0.5cm}
     \caption{
         \color{black}\textrm{  
Time evolution of the H$\alpha$ \& H$\beta$ line profiles covering Flare A2
on 2019 May 18, which are plotted similarly with Figure \ref{fig:map_HaHb_YZCMi_UT191212}.
The grey horizontal dashed lines indicate the time [1] \& [2], 
which are shown in Figure \ref{fig:lcEW_HaHb_ADLeo_UT190518}  (light curves) and Figure \ref{fig:spec_HaHb_ADLeo_UT190518} (line profiles).
}\color{black}
     }
   \label{fig:map_HaHb_ADLeo_UT190518}
   \end{center}
 \end{figure}

The H$\alpha$ \& H$\beta$ line profiles during Flare A2 is shown in
Figures \ref{fig:spec_HaHb_ADLeo_UT190518} \& \ref{fig:map_HaHb_ADLeo_UT190518}. 
During Flare A2, 
the H$\alpha$ and H$\beta$ lines show the line broadenings
with -250 -- +300 km s$^{-1}$ and -250 -- +400 km s$^{-1}$, respectively. 
Especially at around time [1], the red wing of H$\alpha$ and H$\beta$ lines were lightly enhanced. (Figures \ref{fig:spec_HaHb_ADLeo_UT190518}(b)). This red wing asymmetry was more clearly seen in H$\beta$ line than in H$\alpha$ line (Figures \ref{fig:spec_HaHb_ADLeo_UT190518}(d)).

\clearpage

\begin{thebibliography}{}
\expandafter\ifx\csname natexlab\endcsname\relax\def\natexlab#1{#1}\fi
\providecommand{\url}[1]{\href{#1}{#1}}
\providecommand{\dodoi}[1]{doi:~\href{http://doi.org/#1}{\nolinkurl{#1}}}
\providecommand{\doeprint}[1]{\href{http://ascl.net/#1}{\nolinkurl{http://ascl.net/#1}}}
\providecommand{\doarXiv}[1]{\href{https://arxiv.org/abs/#1}{\nolinkurl{https://arxiv.org/abs/#1}}}

\bibitem[{{Aarnio} {et~al.}(2012){Aarnio}, {Matt}, \& {Stassun}}]{Aarnio+2012}
{Aarnio}, A.~N., {Matt}, S.~P., \& {Stassun}, K.~G. 2012, \apj, 760, 9,
  \dodoi{10.1088/0004-637X/760/1/9}

\bibitem[{{Airapetian} {et~al.}(2016){Airapetian}, {Glocer}, {Gronoff},
  {H{\'e}brard}, \& {Danchi}}]{Airapetian+2016}
{Airapetian}, V.~S., {Glocer}, A., {Gronoff}, G., {H{\'e}brard}, E., \&
  {Danchi}, W. 2016, Nature Geoscience, 9, 452, \dodoi{10.1038/ngeo2719}

\bibitem[{{Airapetian} {et~al.}(2020){Airapetian}, {Barnes}, {Cohen},
  {Collinson}, {Danchi}, {Dong}, {Del Genio}, {France}, {Garcia-Sage},
  {Glocer}, {Gopalswamy}, {Grenfell}, {Gronoff}, {G{\"u}del}, {Herbst},
  {Henning}, {Jackman}, {Jin}, {Johnstone}, {Kaltenegger}, {Kay}, {Kobayashi},
  {Kuang}, {Li}, {Lynch}, {L{\"u}ftinger}, {Luhmann}, {Maehara}, {Mlynczak},
  {Notsu}, {Osten}, {Ramirez}, {Rugheimer}, {Scheucher}, {Schlieder},
  {Shibata}, {Sousa-Silva}, {Stamenkovi{\'c}}, {Strangeway}, {Usmanov},
  {Vergados}, {Verkhoglyadova}, {Vidotto}, {Voytek}, {Way}, {Zank}, \&
  {Yamashiki}}]{Airapetian+2020}
{Airapetian}, V.~S., {Barnes}, R., {Cohen}, O., {et~al.} 2020, International
  Journal of Astrobiology, 19, 136, \dodoi{10.1017/S1473550419000132}

\bibitem[{{Allred} {et~al.}(2005){Allred}, {Hawley}, {Abbett}, \&
  {Carlsson}}]{Allred+2005}
{Allred}, J.~C., {Hawley}, S.~L., {Abbett}, W.~P., \& {Carlsson}, M. 2005,
  \apj, 630, 573, \dodoi{10.1086/431751}

\bibitem[{{Allred} {et~al.}(2006){Allred}, {Hawley}, {Abbett}, \&
  {Carlsson}}]{Allred+2006}
---. 2006, \apj, 644, 484, \dodoi{10.1086/503314}

\bibitem[{{Allred} {et~al.}(2015){Allred}, {Kowalski}, \&
  {Carlsson}}]{Allred+2015_ApJ}
{Allred}, J.~C., {Kowalski}, A.~F., \& {Carlsson}, M. 2015, \apj, 809, 104,
  \dodoi{10.1088/0004-637X/809/1/104}

\bibitem[{{Alvarado-G{\'o}mez} {et~al.}(2018){Alvarado-G{\'o}mez}, {Drake},
  {Cohen}, {Moschou}, \& {Garraffo}}]{Alvarado-Gomez+2018}
{Alvarado-G{\'o}mez}, J.~D., {Drake}, J.~J., {Cohen}, O., {Moschou}, S.~P., \&
  {Garraffo}, C. 2018, \apj, 862, 93, \dodoi{10.3847/1538-4357/aacb7f}

\bibitem[{{Antolin}(2020)}]{Antorin2020}
{Antolin}, P. 2020, Plasma Physics and Controlled Fusion, 62, 014016,
  \dodoi{10.1088/1361-6587/ab5406}

\bibitem[{{Aulanier} {et~al.}(2013){Aulanier}, {D{\'e}moulin}, {Schrijver},
  {Janvier}, {Pariat}, \& {Schmieder}}]{Aulanier+2013}
{Aulanier}, G., {D{\'e}moulin}, P., {Schrijver}, C.~J., {et~al.} 2013, \aap,
  549, A66, \dodoi{10.1051/0004-6361/201220406}

\bibitem[{{Battersby}(2019)}]{Battersby+2019}
{Battersby}, S. 2019, PNAS, 116, 23368, \dodoi{10.1073/pnas.1917356116}

\bibitem[{{Bellotti} {et~al.}(2023){Bellotti}, {Morin}, {Lehmann}, {Folsom},
  {Hussain}, {Petit}, {Donati}, {Lavail}, {Carmona}, {Martioli}, {Romano
  Zaire}, {Alecian}, {Moutou}, {Fouqu{\'e}}, {Alencar}, {Artigau}, {Boisse},
  {Bouchy}, {Cadieux}, {Cloutier}, {Cook}, {Delfosse}, {Doyon}, {H{\'e}brard},
  {Kochukhov}, \& {Wade}}]{Bellotti+2023_A+A}
{Bellotti}, S., {Morin}, J., {Lehmann}, L.~T., {et~al.} 2023, \aap, 676, A56,
  \dodoi{10.1051/0004-6361/202346845}

\bibitem[{{Benz} \& {G{\"u}del}(2010)}]{Benz+2010}
{Benz}, A.~O., \& {G{\"u}del}, M. 2010, \araa, 48, 241,
  \dodoi{10.1146/annurev-astro-082708-101757}

\bibitem[{{Berlicki}(2007)}]{Berlicki+2007}
{Berlicki}, A. 2007, in Astronomical Society of the Pacific Conference Series,
  Vol. 368, The Physics of Chromospheric Plasmas, ed. P.~{Heinzel},
  I.~{Dorotovi{\v{c}}}, \& R.~J. {Rutten}, 387.
\newblock \doarXiv{0704.2436}

\bibitem[{{Bicz} {et~al.}(2022){Bicz}, {Falewicz}, {Pietras}, {Siarkowski}, \&
  {Pre{\'s}}}]{Bicz+2022_ApJ}
{Bicz}, K., {Falewicz}, R., {Pietras}, M., {Siarkowski}, M., \& {Pre{\'s}}, P.
  2022, \apj, 935, 102, \dodoi{10.3847/1538-4357/ac7ab3}

\bibitem[{{Brasseur} {et~al.}(2023){Brasseur}, {Osten}, {Tristan}, \&
  {Kowalski}}]{Brasseur+2023_ApJ}
{Brasseur}, C.~E., {Osten}, R.~A., {Tristan}, I.~I., \& {Kowalski}, A.~F. 2023,
  \apj, 944, 5, \dodoi{10.3847/1538-4357/acab59}

\bibitem[{{Brown} {et~al.}(2013){Brown}, {Baliber}, {Bianco}, {Bowman},
  {Burleson}, {Conway}, {Crellin}, {Depagne}, {De Vera}, {Dilday}, {Dragomir},
  {Dubberley}, {Eastman}, {Elphick}, {Falarski}, {Foale}, {Ford}, {Fulton},
  {Garza}, {Gomez}, {Graham}, {Greene}, {Haldeman}, {Hawkins}, {Haworth},
  {Haynes}, {Hidas}, {Hjelstrom}, {Howell}, {Hygelund}, {Lister}, {Lobdill},
  {Martinez}, {Mullins}, {Norbury}, {Parrent}, {Paulson}, {Petry}, {Pickles},
  {Posner}, {Rosing}, {Ross}, {Sand}, {Saunders}, {Shobbrook}, {Shporer},
  {Street}, {Thomas}, {Tsapras}, {Tufts}, {Valenti}, {Vander Horst}, {Walker},
  {White}, \& {Willis}}]{Brown+2013}
{Brown}, T.~M., {Baliber}, N., {Bianco}, F.~B., {et~al.} 2013, \pasp, 125,
  1031, \dodoi{10.1086/673168}

\bibitem[{{Buzulukova} \& {Tsurutani}(2022)}]{Buzulukova-Tsurutani_2022}
{Buzulukova}, N., \& {Tsurutani}, B. 2022, Frontiers in Astronomy and Space
  Sciences, 9, 1017103, \dodoi{10.3389/fspas.2022.1017103}

\bibitem[{{Canfield} {et~al.}(1990){Canfield}, {Penn}, {Wulser}, \&
  {Kiplinger}}]{Canfield+1990}
{Canfield}, R.~C., {Penn}, M.~J., {Wulser}, J.-P., \& {Kiplinger}, A.~L. 1990,
  \apj, 363, 318, \dodoi{10.1086/169345}

\bibitem[{{Chen} {et~al.}(2021){Chen}, {Zhan}, {Youngblood}, {Wolf},
  {Feinstein}, \& {Horton}}]{Chen+2021}
{Chen}, H., {Zhan}, Z., {Youngblood}, A., {et~al.} 2021, Nature Astronomy, 5,
  298, \dodoi{10.1038/s41550-020-01264-1}

\bibitem[{{Cliver} {et~al.}(2022){Cliver}, {Schrijver}, {Shibata}, \&
  {Usoskin}}]{Cliver+2022}
{Cliver}, E.~W., {Schrijver}, C.~J., {Shibata}, K., \& {Usoskin}, I.~G. 2022,
  Living Reviews in Solar Physics, 19, 2, \dodoi{10.1007/s41116-022-00033-8}

\bibitem[{{Collins} {et~al.}(2017){Collins}, {Kielkopf}, {Stassun}, \&
  {Hessman}}]{Collins+2017}
{Collins}, K.~A., {Kielkopf}, J.~F., {Stassun}, K.~G., \& {Hessman}, F.~V.
  2017, \aj, 153, 77, \dodoi{10.3847/1538-3881/153/2/77}

\bibitem[{{Cranmer}(2017)}]{Cranmer2017}
{Cranmer}, S.~R. 2017, \apj, 840, 114, \dodoi{10.3847/1538-4357/aa6f0e}

\bibitem[{{Crespo-Chac{\'o}n} {et~al.}(2006){Crespo-Chac{\'o}n}, {Montes},
  {Garc{\'\i}a-Alvarez}, {Fern{\'a}ndez-Figueroa}, {L{\'o}pez-Santiago}, \&
  {Foing}}]{Crespo-Chacon+2006}
{Crespo-Chac{\'o}n}, I., {Montes}, D., {Garc{\'\i}a-Alvarez}, D., {et~al.}
  2006, \aap, 452, 987, \dodoi{10.1051/0004-6361:20053615}

\bibitem[{{Crosley} \& {Osten}(2018)}]{Crosley+2018}
{Crosley}, M.~K., \& {Osten}, R.~A. 2018, \apj, 856, 39,
  \dodoi{10.3847/1538-4357/aaaec2}

\bibitem[{{Davenport}(2016)}]{Davenport+2016}
{Davenport}, J. R.~A. 2016, \apj, 829, 23, \dodoi{10.3847/0004-637X/829/1/23}

\bibitem[{{Davenport} {et~al.}(2012){Davenport}, {Becker}, {Kowalski},
  {Hawley}, {Schmidt}, {Hilton}, {Sesar}, \& {Cutri}}]{Davenport+2012}
{Davenport}, J. R.~A., {Becker}, A.~C., {Kowalski}, A.~F., {et~al.} 2012, \apj,
  748, 58, \dodoi{10.1088/0004-637X/748/1/58}

\bibitem[{{Davenport} {et~al.}(2020){Davenport}, {Mendoza}, \&
  {Hawley}}]{Davenport+2020}
{Davenport}, J. R.~A., {Mendoza}, G.~T., \& {Hawley}, S.~L. 2020, \aj, 160, 36,
  \dodoi{10.3847/1538-3881/ab9536}

\bibitem[{{Davenport} {et~al.}(2014){Davenport}, {Hawley}, {Hebb},
  {Wisniewski}, {Kowalski}, {Johnson}, {Malatesta}, {Peraza}, {Keil},
  {Silverberg}, {Jansen}, {Scheffler}, {Berdis}, {Larsen}, \&
  {Hilton}}]{Davenport+2014}
{Davenport}, J. R.~A., {Hawley}, S.~L., {Hebb}, L., {et~al.} 2014, \apj, 797,
  122, \dodoi{10.1088/0004-637X/797/2/122}

\bibitem[{{Drake} {et~al.}(2016){Drake}, {Cohen}, {Garraffo}, \&
  {Kashyap}}]{Drake+2016}
{Drake}, J.~J., {Cohen}, O., {Garraffo}, C., \& {Kashyap}, V. 2016, in Solar
  and Stellar Flares and their Effects on Planets, ed. A.~G. {Kosovichev},
  S.~L. {Hawley}, \& P.~{Heinzel}, Vol. 320, 196--201,
  \dodoi{10.1017/S1743921316000260}

\bibitem[{{Drake} {et~al.}(2013){Drake}, {Cohen}, {Yashiro}, \&
  {Gopalswamy}}]{Drake+2013}
{Drake}, J.~J., {Cohen}, O., {Yashiro}, S., \& {Gopalswamy}, N. 2013, \apj,
  764, 170, \dodoi{10.1088/0004-637X/764/2/170}

\bibitem[{{Drake} \& {Ulrich}(1980)}]{Drake_Ulrich_1980}
{Drake}, S.~A., \& {Ulrich}, R.~K. 1980, \apjs, 42, 351, \dodoi{10.1086/190654}

\bibitem[{{Eason} {et~al.}(1992){Eason}, {Giampapa}, {Radick}, {Worden}, \&
  {Hege}}]{Eason+1992}
{Eason}, E. L.~E., {Giampapa}, M.~S., {Radick}, R.~R., {Worden}, S.~P., \&
  {Hege}, E.~K. 1992, \aj, 104, 1161, \dodoi{10.1086/116305}

\bibitem[{{Emslie} {et~al.}(2012){Emslie}, {Dennis}, {Shih}, {Chamberlin},
  {Mewaldt}, {Moore}, {Share}, {Vourlidas}, \& {Welsch}}]{Emslie+2012}
{Emslie}, A.~G., {Dennis}, B.~R., {Shih}, A.~Y., {et~al.} 2012, \apj, 759, 71,
  \dodoi{10.1088/0004-637X/759/1/71}

\bibitem[{{Fan}(2018)}]{Yuhong+2018_ApJ}
{Fan}, Y. 2018, \apj, 862, 54, \dodoi{10.3847/1538-4357/aaccee}

\bibitem[{{Fausnaugh} {et~al.}(2019){Fausnaugh}, {Burke}, {Caldwell},
  {Jenkins}, {Smith}, {Twicken}, {Vanderspek}, {Doty}, {Ting}, \&
  {Villasenor}}]{Fausnaugh+2019}
{Fausnaugh}, M.~M., {Burke}, C.~J., {Caldwell}, D.~A., {et~al.} 2019, {TESS
  Data Release Notes: Sector 7, DR9, NASA/TM-2019-220170}, NASA

\bibitem[{{Fausnaugh} {et~al.}(2021){Fausnaugh}, {Burke}, {Caldwell},
  {Jenkins}, {Smith}, {Twicken}, {Vanderspek}, {Doty}, {Ting}, \&
  {Villasenor}}]{Fausnaugh+2021}
---. 2021, {TESS Data Release Notes: Sector 34, DR50, NASA/TM—20210012992},
  NASA

\bibitem[{{Feinstein} {et~al.}(2020){Feinstein}, {Montet}, {Ansdell}, {Nord},
  {Bean}, {G{\"u}nther}, {Gully-Santiago}, \& {Schlieder}}]{Feinstein+2020}
{Feinstein}, A.~D., {Montet}, B.~T., {Ansdell}, M., {et~al.} 2020, \aj, 160,
  219, \dodoi{10.3847/1538-3881/abac0a}

\bibitem[{{Fisher} {et~al.}(1985){Fisher}, {Canfield}, \&
  {McClymont}}]{Fisher+1985}
{Fisher}, G.~H., {Canfield}, R.~C., \& {McClymont}, A.~N. 1985, \apj, 289, 414,
  \dodoi{10.1086/162901}

\bibitem[{{Flores Soriano} \& {Strassmeier}(2017)}]{FloresSoriano+2017}
{Flores Soriano}, M., \& {Strassmeier}, K.~G. 2017, \aap, 597, A101,
  \dodoi{10.1051/0004-6361/201629338}

\bibitem[{{Fuhrmeister} {et~al.}(2011){Fuhrmeister}, {Lalitha}, {Poppenhaeger},
  {Rudolf}, {Liefke}, {Reiners}, {Schmitt}, \& {Ness}}]{Fuhrmeister+2011}
{Fuhrmeister}, B., {Lalitha}, S., {Poppenhaeger}, K., {et~al.} 2011, \aap, 534,
  A133, \dodoi{10.1051/0004-6361/201117447}

\bibitem[{{Fuhrmeister} {et~al.}(2008){Fuhrmeister}, {Liefke}, {Schmitt}, \&
  {Reiners}}]{Fuhrmeister+2008}
{Fuhrmeister}, B., {Liefke}, C., {Schmitt}, J.~H.~M.~M., \& {Reiners}, A. 2008,
  \aap, 487, 293, \dodoi{10.1051/0004-6361:200809379}

\bibitem[{{Fuhrmeister} {et~al.}(2018){Fuhrmeister}, {Czesla}, {Schmitt},
  {Jeffers}, {Caballero}, {Zechmeister}, {Reiners}, {Ribas}, {Amado},
  {Quirrenbach}, {B{\'e}jar}, {Galad{\'\i}-Enr{\'\i}quez}, {Guenther},
  {K{\"u}rster}, {Montes}, \& {Seifert}}]{Fuhrmeister+2018}
{Fuhrmeister}, B., {Czesla}, S., {Schmitt}, J.~H.~M.~M., {et~al.} 2018, \aap,
  615, A14, \dodoi{10.1051/0004-6361/201732204}

\bibitem[{{Gaia Collaboration} {et~al.}(2018){Gaia Collaboration}, {Brown},
  {Vallenari}, {Prusti}, {de Bruijne}, {Babusiaux}, {Bailer-Jones}, {Biermann},
  {Evans}, {Eyer}, {Jansen}, {Jordi}, {Klioner}, {Lammers}, {Lindegren},
  {Luri}, {Mignard}, {Panem}, {Pourbaix}, {Randich}, {Sartoretti}, {Siddiqui},
  {Soubiran}, {van Leeuwen}, {Walton}, {Arenou}, {Bastian}, {Cropper},
  {Drimmel}, {Katz}, {Lattanzi}, {Bakker}, {Cacciari}, {Casta{\~n}eda},
  {Chaoul}, {Cheek}, {De Angeli}, {Fabricius}, {Guerra}, {Holl}, {Masana},
  {Messineo}, {Mowlavi}, {Nienartowicz}, {Panuzzo}, {Portell}, {Riello},
  {Seabroke}, {Tanga}, {Th{\'e}venin}, {Gracia-Abril}, {Comoretto},
  {Garcia-Reinaldos}, {Teyssier}, {Altmann}, {Andrae}, {Audard},
  {Bellas-Velidis}, {Benson}, {Berthier}, {Blomme}, {Burgess}, {Busso},
  {Carry}, {Cellino}, {Clementini}, {Clotet}, {Creevey}, {Davidson}, {De
  Ridder}, {Delchambre}, {Dell'Oro}, {Ducourant},
  {Fern{\'a}ndez-Hern{\'a}ndez}, {Fouesneau}, {Fr{\'e}mat}, {Galluccio},
  {Garc{\'\i}a-Torres}, {Gonz{\'a}lez-N{\'u}{\~n}ez}, {Gonz{\'a}lez-Vidal},
  {Gosset}, {Guy}, {Halbwachs}, {Hambly}, {Harrison}, {Hern{\'a}ndez},
  {Hestroffer}, {Hodgkin}, {Hutton}, {Jasniewicz}, {Jean-Antoine-Piccolo},
  {Jordan}, {Korn}, {Krone-Martins}, {Lanzafame}, {Lebzelter}, {L{\"o}ffler},
  {Manteiga}, {Marrese}, {Mart{\'\i}n-Fleitas}, {Moitinho}, {Mora}, {Muinonen},
  {Osinde}, {Pancino}, {Pauwels}, {Petit}, {Recio-Blanco}, {Richards},
  {Rimoldini}, {Robin}, {Sarro}, {Siopis}, {Smith}, {Sozzetti}, {S{\"u}veges},
  {Torra}, {van Reeven}, {Abbas}, {Abreu Aramburu}, {Accart}, {Aerts},
  {Altavilla}, {{\'A}lvarez}, {Alvarez}, {Alves}, {Anderson}, {Andrei},
  {Anglada Varela}, {Antiche}, {Antoja}, {Arcay}, {Astraatmadja}, {Bach},
  {Baker}, {Balaguer-N{\'u}{\~n}ez}, {Balm}, {Barache}, {Barata}, {Barbato},
  {Barblan}, {Barklem}, {Barrado}, {Barros}, {Barstow}, {Bartholom{\'e}
  Mu{\~n}oz}, {Bassilana}, {Becciani}, {Bellazzini}, {Berihuete}, {Bertone},
  {Bianchi}, {Bienaym{\'e}}, {Blanco-Cuaresma}, {Boch}, {Boeche}, {Bombrun},
  {Borrachero}, {Bossini}, {Bouquillon}, {Bourda}, {Bragaglia}, {Bramante},
  {Breddels}, {Bressan}, {Brouillet}, {Br{\"u}semeister}, {Brugaletta},
  {Bucciarelli}, {Burlacu}, {Busonero}, {Butkevich}, {Buzzi}, {Caffau},
  {Cancelliere}, {Cannizzaro}, {Cantat-Gaudin}, {Carballo}, {Carlucci},
  {Carrasco}, {Casamiquela}, {Castellani}, {Castro-Ginard}, {Charlot},
  {Chemin}, {Chiavassa}, {Cocozza}, {Costigan}, {Cowell}, {Crifo}, {Crosta},
  {Crowley}, {Cuypers}, {Dafonte}, {Damerdji}, {Dapergolas}, {David}, {David},
  {de Laverny}, {De Luise}, {De March}, {de Martino}, {de Souza}, {de Torres},
  {Debosscher}, {del Pozo}, {Delbo}, {Delgado}, {Delgado}, {Di Matteo},
  {Diakite}, {Diener}, {Distefano}, {Dolding}, {Drazinos}, {Dur{\'a}n},
  {Edvardsson}, {Enke}, {Eriksson}, {Esquej}, {Eynard Bontemps}, {Fabre},
  {Fabrizio}, {Faigler}, {Falc{\~a}o}, {Farr{\`a}s Casas}, {Federici},
  {Fedorets}, {Fernique}, {Figueras}, {Filippi}, {Findeisen}, {Fonti},
  {Fraile}, {Fraser}, {Fr{\'e}zouls}, {Gai}, {Galleti}, {Garabato},
  {Garc{\'\i}a-Sedano}, {Garofalo}, {Garralda}, {Gavel}, {Gavras}, {Gerssen},
  {Geyer}, {Giacobbe}, {Gilmore}, {Girona}, {Giuffrida}, {Glass}, {Gomes},
  {Granvik}, {Gueguen}, {Guerrier}, {Guiraud}, {Guti{\'e}rrez-S{\'a}nchez},
  {Haigron}, {Hatzidimitriou}, {Hauser}, {Haywood}, {Heiter}, {Helmi}, {Heu},
  {Hilger}, {Hobbs}, {Hofmann}, {Holland}, {Huckle}, {Hypki}, {Icardi},
  {Jan{\ss}en}, {Jevardat de Fombelle}, {Jonker}, {Juh{\'a}sz}, {Julbe},
  {Karampelas}, {Kewley}, {Klar}, {Kochoska}, {Kohley}, {Kolenberg},
  {Kontizas}, {Kontizas}, {Koposov}, {Kordopatis}, {Kostrzewa-Rutkowska},
  {Koubsky}, {Lambert}, {Lanza}, {Lasne}, {Lavigne}, {Le Fustec}, {Le
  Poncin-Lafitte}, {Lebreton}, {Leccia}, {Leclerc}, {Lecoeur-Taibi},
  {Lenhardt}, {Leroux}, {Liao}, {Licata}, {Lindstr{\o}m}, {Lister}, {Livanou},
  {Lobel}, {L{\'o}pez}, {Managau}, {Mann}, {Mantelet}, {Marchal}, {Marchant},
  {Marconi}, {Marinoni}, {Marschalk{\'o}}, {Marshall}, {Martino}, {Marton},
  {Mary}, {Massari}, {Matijevi{\v{c}}}, {Mazeh}, {McMillan}, {Messina},
  {Michalik}, {Millar}, {Molina}, {Molinaro}, {Moln{\'a}r}, {Montegriffo},
  {Mor}, {Morbidelli}, {Morel}, {Morris}, {Mulone}, {Muraveva}, {Musella},
  {Nelemans}, {Nicastro}, {Noval}, {O'Mullane}, {Ord{\'e}novic},
  {Ord{\'o}{\~n}ez-Blanco}, {Osborne}, {Pagani}, {Pagano}, {Pailler},
  {Palacin}, {Palaversa}, {Panahi}, {Pawlak}, {Piersimoni}, {Pineau}, {Plachy},
  {Plum}, {Poggio}, {Poujoulet}, {Pr{\v{s}}a}, {Pulone}, {Racero}, {Ragaini},
  {Rambaux}, {Ramos-Lerate}, {Regibo}, {Reyl{\'e}}, {Riclet}, {Ripepi}, {Riva},
  {Rivard}, {Rixon}, {Roegiers}, {Roelens}, {Romero-G{\'o}mez}, {Rowell},
  {Royer}, {Ruiz-Dern}, {Sadowski}, {Sagrist{\`a} Sell{\'e}s}, {Sahlmann},
  {Salgado}, {Salguero}, {Sanna}, {Santana-Ros}, {Sarasso}, {Savietto},
  {Schultheis}, {Sciacca}, {Segol}, {Segovia}, {S{\'e}gransan}, {Shih},
  {Siltala}, {Silva}, {Smart}, {Smith}, {Solano}, {Solitro}, {Sordo}, {Soria
  Nieto}, {Souchay}, {Spagna}, {Spoto}, {Stampa}, {Steele},
  {Steidelm{\"u}ller}, {Stephenson}, {Stoev}, {Suess}, {Surdej}, {Szabados},
  {Szegedi-Elek}, {Tapiador}, {Taris}, {Tauran}, {Taylor}, {Teixeira},
  {Terrett}, {Teyssand ier}, {Thuillot}, {Titarenko}, {Torra Clotet}, {Turon},
  {Ulla}, {Utrilla}, {Uzzi}, {Vaillant}, {Valentini}, {Valette}, {van Elteren},
  {Van Hemelryck}, {van Leeuwen}, {Vaschetto}, {Vecchiato}, {Veljanoski},
  {Viala}, {Vicente}, {Vogt}, {von Essen}, {Voss}, {Votruba}, {Voutsinas},
  {Walmsley}, {Weiler}, {Wertz}, {Wevers}, {Wyrzykowski}, {Yoldas},
  {{\v{Z}}erjal}, {Ziaeepour}, {Zorec}, {Zschocke}, {Zucker}, {Zurbach}, \&
  {Zwitter}}]{GaiaCollaboration+2018}
{Gaia Collaboration}, {Brown}, A.~G.~A., {Vallenari}, A., {et~al.} 2018, \aap,
  616, A1, \dodoi{10.1051/0004-6361/201833051}

\bibitem[{{Gendreau} {et~al.}(2016){Gendreau}, {Arzoumanian}, {Adkins},
  {Albert}, {Anders}, {Aylward}, {Baker}, {Balsamo}, {Bamford}, {Benegalrao},
  {Berry}, {Bhalwani}, {Black}, {Blaurock}, {Bronke}, {Brown}, {Budinoff},
  {Cantwell}, {Cazeau}, {Chen}, {Clement}, {Colangelo}, {Coleman},
  {Coopersmith}, {Dehaven}, {Doty}, {Egan}, {Enoto}, {Fan}, {Ferro}, {Foster},
  {Galassi}, {Gallo}, {Green}, {Grosh}, {Ha}, {Hasouneh}, {Heefner}, {Hestnes},
  {Hoge}, {Jacobs}, {J{\o}rgensen}, {Kaiser}, {Kellogg}, {Kenyon}, {Koenecke},
  {Kozon}, {LaMarr}, {Lambertson}, {Larson}, {Lentine}, {Lewis}, {Lilly},
  {Liu}, {Malonis}, {Manthripragada}, {Markwardt}, {Matonak}, {Mcginnis},
  {Miller}, {Mitchell}, {Mitchell}, {Mohammed}, {Monroe}, {Montt de Garcia},
  {Mul{\'e}}, {Nagao}, {Ngo}, {Norris}, {Norwood}, {Novotka}, {Okajima},
  {Olsen}, {Onyeachu}, {Orosco}, {Peterson}, {Pevear}, {Pham}, {Pollard},
  {Pope}, {Powers}, {Powers}, {Price}, {Prigozhin}, {Ramirez}, {Reid},
  {Remillard}, {Rogstad}, {Rosecrans}, {Rowe}, {Sager}, {Sanders}, {Savadkin},
  {Saylor}, {Schaeffer}, {Schweiss}, {Semper}, {Serlemitsos}, {Shackelford},
  {Soong}, {Struebel}, {Vezie}, {Villasenor}, {Winternitz}, {Wofford},
  {Wright}, {Yang}, \& {Yu}}]{Gendreau+2016}
{Gendreau}, K.~C., {Arzoumanian}, Z., {Adkins}, P.~W., {et~al.} 2016, in
  Society of Photo-Optical Instrumentation Engineers (SPIE) Conference Series,
  Vol. 9905, Space Telescopes and Instrumentation 2016: Ultraviolet to Gamma
  Ray, ed. J.-W.~A. {den Herder}, T.~{Takahashi}, \& M.~{Bautz}, 99051H,
  \dodoi{10.1117/12.2231304}

\bibitem[{{Gershberg}(2005)}]{Gershberg+2005}
{Gershberg}, R.~E. 2005, {Solar-Type Activity in Main-Sequence Stars}
  (Springer-Verlag Berlin Heidelberg), \dodoi{10.1007/3-540-28243-2}

\bibitem[{{Gontikakis} {et~al.}(1997){Gontikakis}, {Vial}, \&
  {Gouttebroze}}]{Gontikakis+1997_A&A}
{Gontikakis}, C., {Vial}, J.~C., \& {Gouttebroze}, P. 1997, \aap, 325, 803

\bibitem[{{Gopalswamy} {et~al.}(2003){Gopalswamy}, {Shimojo}, {Lu}, {Yashiro},
  {Shibasaki}, \& {Howard}}]{Gopalswamy+2003}
{Gopalswamy}, N., {Shimojo}, M., {Lu}, W., {et~al.} 2003, \apj, 586, 562,
  \dodoi{10.1086/367614}

\bibitem[{{Graham} \& {Cauzzi}(2015)}]{Graham+2015}
{Graham}, D.~R., \& {Cauzzi}, G. 2015, \apjl, 807, L22,
  \dodoi{10.1088/2041-8205/807/2/L22}

\bibitem[{{Graham} {et~al.}(2020){Graham}, {Cauzzi}, {Zangrilli}, {Kowalski},
  {Sim{\~o}es}, \& {Allred}}]{Graham+2020}
{Graham}, D.~R., {Cauzzi}, G., {Zangrilli}, L., {et~al.} 2020, \apj, 895, 6,
  \dodoi{10.3847/1538-4357/ab88ad}

\bibitem[{{Grayver} {et~al.}(2022){Grayver}, {Bower}, {Saur}, {Dorn}, \&
  {Morris}}]{Grayver+2022}
{Grayver}, A., {Bower}, D.~J., {Saur}, J., {Dorn}, C., \& {Morris}, B.~M. 2022,
  \apjl, 941, L7, \dodoi{10.3847/2041-8213/aca287}

\bibitem[{{Guarcello} {et~al.}(2019){Guarcello}, {Micela}, {Sciortino},
  {L{\'o}pez-Santiago}, {Argiroffi}, {Reale}, {Flaccomio},
  {Alvarado-G{\'o}mez}, {Antoniou}, {Drake}, {Pillitteri}, {Rebull}, \&
  {Stauffer}}]{Guarcello+2019}
{Guarcello}, M.~G., {Micela}, G., {Sciortino}, S., {et~al.} 2019, \aap, 622,
  A210, \dodoi{10.1051/0004-6361/201834370}

\bibitem[{{G{\"u}del}(2004)}]{Guedel2004_A&ARv}
{G{\"u}del}, M. 2004, \aapr, 12, 71, \dodoi{10.1007/s00159-004-0023-2}

\bibitem[{{G{\"u}del} {et~al.}(2004){G{\"u}del}, {Audard}, {Reale}, {Skinner},
  \& {Linsky}}]{Guedel+2004}
{G{\"u}del}, M., {Audard}, M., {Reale}, F., {Skinner}, S.~L., \& {Linsky},
  J.~L. 2004, \aap, 416, 713, \dodoi{10.1051/0004-6361:20031471}

\bibitem[{{G{\"u}del} {et~al.}(1996){G{\"u}del}, {Benz}, {Schmitt}, \&
  {Skinner}}]{Guedel+1996}
{G{\"u}del}, M., {Benz}, A.~O., {Schmitt}, J. H.~M.~M., \& {Skinner}, S.~L.
  1996, \apj, 471, 1002, \dodoi{10.1086/178027}

\bibitem[{{Gunn} {et~al.}(1994){Gunn}, {Doyle}, {Mathioudakis}, {Houdebine}, \&
  {Avgoloupis}}]{Gunn+1994}
{Gunn}, A.~G., {Doyle}, J.~G., {Mathioudakis}, M., {Houdebine}, E.~R., \&
  {Avgoloupis}, S. 1994, \aap, 285, 489

\bibitem[{{Haisch}(1989)}]{Haisch_1989_A+A}
{Haisch}, B.~M. 1989, \aap, 219, 317

\bibitem[{{Hamaguchi} {et~al.}(2023){Hamaguchi}, {Reep}, {Airapetian},
  {Toriumi}, {Gendreau}, \& {Arzoumanian}}]{Hamaguchi+2023_ApJ}
{Hamaguchi}, K., {Reep}, J.~W., {Airapetian}, V., {et~al.} 2023, \apj, 944,
  163, \dodoi{10.3847/1538-4357/acae8b}

\bibitem[{{Hawley} {et~al.}(2014){Hawley}, {Davenport}, {Kowalski},
  {Wisniewski}, {Hebb}, {Deitrick}, \& {Hilton}}]{Hawley+2014}
{Hawley}, S.~L., {Davenport}, J. R.~A., {Kowalski}, A.~F., {et~al.} 2014, \apj,
  797, 121, \dodoi{10.1088/0004-637X/797/2/121}

\bibitem[{{Hawley} \& {Fisher}(1992)}]{Hawley+1992}
{Hawley}, S.~L., \& {Fisher}, G.~H. 1992, \apjs, 78, 565,
  \dodoi{10.1086/191640}

\bibitem[{{Hawley} \& {Pettersen}(1991)}]{Hawley+1991}
{Hawley}, S.~L., \& {Pettersen}, B.~R. 1991, \apj, 378, 725,
  \dodoi{10.1086/170474}

\bibitem[{{Hawley} {et~al.}(2007){Hawley}, {Walkowicz}, {Allred}, \&
  {Valenti}}]{Hawley+2007}
{Hawley}, S.~L., {Walkowicz}, L.~M., {Allred}, J.~C., \& {Valenti}, J.~A. 2007,
  \pasp, 119, 67

\bibitem[{{Hawley} {et~al.}(1995){Hawley}, {Fisher}, {Simon}, {Cully},
  {Deustua}, {Jablonski}, {Johns-Krull}, {Pettersen}, {Smith}, {Spiesman}, \&
  {Valenti}}]{Hawley+1995}
{Hawley}, S.~L., {Fisher}, G.~H., {Simon}, T., {et~al.} 1995, \apj, 453, 464,
  \dodoi{10.1086/176408}

\bibitem[{{Heinzel}(2019)}]{Heinzel+2019}
{Heinzel}, P. 2019, in The Sun as a Guide to Stellar Physics, ed. O.~{Engvold},
  J.-C. {Vial}, \& A.~{Skumanich} (Elsevier), 157--183,
  \dodoi{10.1016/B978-0-12-814334-6.00006-6}

\bibitem[{{Heinzel} {et~al.}(1994{\natexlab{a}}){Heinzel}, {Gouttebroze}, \&
  {Vial}}]{Heinzel+1994_A&A}
{Heinzel}, P., {Gouttebroze}, P., \& {Vial}, J.~C. 1994{\natexlab{a}}, \aap,
  292, 656

\bibitem[{{Heinzel} {et~al.}(1994{\natexlab{b}}){Heinzel}, {Karlicky}, {Kotrc},
  \& {Svestka}}]{Heinzel+1994_SoPh}
{Heinzel}, P., {Karlicky}, M., {Kotrc}, P., \& {Svestka}, Z.
  1994{\natexlab{b}}, \solphys, 152, 393, \dodoi{10.1007/BF00680446}

\bibitem[{{Heinzel} \& {Rompolt}(1987)}]{Heinzel+1987_SolPhys}
{Heinzel}, P., \& {Rompolt}, B. 1987, \solphys, 110, 171,
  \dodoi{10.1007/BF00148210}

\bibitem[{{Hilton}(2011)}]{Hilton2011}
{Hilton}, E.~J. 2011, PhD thesis, University of Washington

\bibitem[{{Hirayama}(1986)}]{Hirayama+1986}
{Hirayama}, T. 1986, in NASA Conference Publication, Vol. 2442, NASA Conference
  Publication, 149--153

\bibitem[{{Holman}(2012)}]{Holman_2012_ApJ}
{Holman}, G.~D. 2012, \apj, 745, 52, \dodoi{10.1088/0004-637X/745/1/52}

\bibitem[{{Holman} {et~al.}(2011){Holman}, {Aschwanden}, {Aurass}, {Battaglia},
  {Grigis}, {Kontar}, {Liu}, {Saint-Hilaire}, \& {Zharkova}}]{Holman+2011_SSRv}
{Holman}, G.~D., {Aschwanden}, M.~J., {Aurass}, H., {et~al.} 2011, \ssr, 159,
  107, \dodoi{10.1007/s11214-010-9680-9}

\bibitem[{{Honda} {et~al.}(2018){Honda}, {Notsu}, {Namekata}, {Notsu},
  {Maehara}, {Ikuta}, {Nogami}, \& {Shibata}}]{Honda+2018}
{Honda}, S., {Notsu}, Y., {Namekata}, K., {et~al.} 2018, \pasj, 70, 62,
  \dodoi{10.1093/pasj/psy055}

\bibitem[{{Hong} {et~al.}(2020){Hong}, {Li}, {Ding}, \& {Zhou}}]{Hong+2020}
{Hong}, J., {Li}, Y., {Ding}, M.~D., \& {Zhou}, Y.-H. 2020, \apj, 890, 115,
  \dodoi{10.3847/1538-4357/ab6d05}

\bibitem[{{Hori} {et~al.}(1997){Hori}, {Yokoyama}, {Kosugi}, \&
  {Shibata}}]{Hori+1997}
{Hori}, K., {Yokoyama}, T., {Kosugi}, T., \& {Shibata}, K. 1997, \apj, 489,
  426, \dodoi{10.1086/304754}

\bibitem[{{Houdebine} {et~al.}(1993){Houdebine}, {Foing}, {Doyle}, \&
  {Rodono}}]{Houdebine+1993}
{Houdebine}, E.~R., {Foing}, B.~H., {Doyle}, J.~G., \& {Rodono}, M. 1993, \aap,
  274, 245

\bibitem[{{Houdebine} {et~al.}(1990){Houdebine}, {Foing}, \&
  {Rodono}}]{Houdebine+1990}
{Houdebine}, E.~R., {Foing}, B.~H., \& {Rodono}, M. 1990, \aap, 238, 249

\bibitem[{{Howard} {et~al.}(2019){Howard}, {Corbett}, {Law}, {Ratzloff},
  {Glazier}, {Fors}, {del Ser}, \& {Haislip}}]{Howard+2019}
{Howard}, W.~S., {Corbett}, H., {Law}, N.~M., {et~al.} 2019, \apj, 881, 9,
  \dodoi{10.3847/1538-4357/ab2767}

\bibitem[{{Howard} {et~al.}(2020){Howard}, {Corbett}, {Law}, {Ratzloff},
  {Galliher}, {Glazier}, {Gonzalez}, {Vasquez Soto}, {Fors}, {del Ser}, \&
  {Haislip}}]{Howard+2020_ApJ}
---. 2020, \apj, 902, 115, \dodoi{10.3847/1538-4357/abb5b4}

\bibitem[{{Huang} {et~al.}(2019){Huang}, {Xu}, {Sadykov}, {Jing}, \&
  {Wang}}]{Huang+2019}
{Huang}, N., {Xu}, Y., {Sadykov}, V.~M., {Jing}, J., \& {Wang}, H. 2019, \apjl,
  878, L15, \dodoi{10.3847/2041-8213/ab2330}

\bibitem[{{Hunt-Walker} {et~al.}(2012){Hunt-Walker}, {Hilton}, {Kowalski},
  {Hawley}, \& {Matthews}}]{Hunt-Walker+2012}
{Hunt-Walker}, N.~M., {Hilton}, E.~J., {Kowalski}, A.~F., {Hawley}, S.~L., \&
  {Matthews}, J.~M. 2012, \pasp, 124, 545, \dodoi{10.1086/666495}

\bibitem[{{Ichimoto} \& {Kurokawa}(1984)}]{Ichimoto+1984}
{Ichimoto}, K., \& {Kurokawa}, H. 1984, \solphys, 93, 105,
  \dodoi{10.1007/BF00156656}

\bibitem[{{Ikuta} {et~al.}(2023){Ikuta}, {Namekata}, {Notsu}, {Maehara},
  {Okamoto}, {Honda}, {Nogami}, \& {Shibata}}]{Ikuta+2023_ApJ}
{Ikuta}, K., {Namekata}, K., {Notsu}, Y., {et~al.} 2023, \apj, 948, 64,
  \dodoi{10.3847/1538-4357/acbd36}

\bibitem[{{Inoue} {et~al.}(2023){Inoue}, {Maehara}, {Notsu}, {Namekata},
  {Honda}, {Namizaki}, {Nogami}, \& {Shibata}}]{Inoue+2023_ApJ}
{Inoue}, S., {Maehara}, H., {Notsu}, Y., {et~al.} 2023, \apj, 948, 9,
  \dodoi{10.3847/1538-4357/acb7e8}

\bibitem[{{Jackman} {et~al.}(2021){Jackman}, {Wheatley}, {Acton}, {Anderson},
  {Bayliss}, {Briegal}, {Burleigh}, {Casewell}, {G{\"a}nsicke}, {Gill},
  {Gillen}, {Goad}, {G{\"u}nther}, {Henderson}, {Hodgkin}, {Jenkins}, {Pugh},
  {Queloz}, {Raynard}, {Tilbrook}, {Watson}, \& {West}}]{Jackman+2021}
{Jackman}, J. A.~G., {Wheatley}, P.~J., {Acton}, J.~S., {et~al.} 2021, \mnras,
  504, 3246, \dodoi{10.1093/mnras/stab979}

\bibitem[{{Jardine} {et~al.}(2020){Jardine}, {Collier Cameron}, {Donati}, \&
  {Hussain}}]{Jardine+2020}
{Jardine}, M., {Collier Cameron}, A., {Donati}, J.~F., \& {Hussain}, G.~A.~J.
  2020, \mnras, 491, 4076, \dodoi{10.1093/mnras/stz3173}

\bibitem[{{Johns-Krull} \& {Valenti}(2000)}]{Johns-Krull+2000}
{Johns-Krull}, C.~M., \& {Valenti}, J.~A. 2000, in Astronomical Society of the
  Pacific Conference Series, Vol. 198, Stellar Clusters and Associations:
  Convection, Rotation, and Dynamos, ed. R.~{Pallavicini}, G.~{Micela}, \&
  S.~{Sciortino}, 371

\bibitem[{{Johnson} {et~al.}(2021){Johnson}, {Czesla}, {Fuhrmeister},
  {Sch{\"o}fer}, {Shan}, {Cardona Guill{\'e}n}, {Reiners}, {Jeffers},
  {Lalitha}, {Luque}, {Rodr{\'\i}guez}, {B{\'e}jar}, {Caballero}, {Tal-Or},
  {Zechmeister}, {Ribas}, {Amado}, {Quirrenbach}, {Cort{\'e}s-Contreras},
  {Dreizler}, {Fukui}, {L{\'o}pez-Gonz{\'a}lez}, {Hatzes}, {Henning},
  {Kaminski}, {K{\"u}rster}, {Lafarga}, {Montes}, {Morales}, {Murgas},
  {Narita}, {Pall{\'e}}, {Parviainen}, {Pedraz}, {Pollacco}, \&
  {Sota}}]{Johnson+2021}
{Johnson}, E.~N., {Czesla}, S., {Fuhrmeister}, B., {et~al.} 2021, \aap, 651,
  A105, \dodoi{10.1051/0004-6361/202040159}

\bibitem[{{Kochukhov}(2021)}]{Kochukhov+2021}
{Kochukhov}, O. 2021, \aapr, 29, 1, \dodoi{10.1007/s00159-020-00130-3}

\bibitem[{{Kotani} {et~al.}(2023){Kotani}, {Shibata}, {Ishii}, {Yamasaki},
  {Otsuji}, {Ichimoto}, \& {Asai}}]{Kotani+2023_ApJ}
{Kotani}, Y., {Shibata}, K., {Ishii}, T.~T., {et~al.} 2023, \apj, 943, 143,
  \dodoi{10.3847/1538-4357/acac76}

\bibitem[{{Kowalski}(2016)}]{Kowalski2016}
{Kowalski}, A.~F. 2016, in Solar and Stellar Flares and their Effects on
  Planets, ed. A.~G. {Kosovichev}, S.~L. {Hawley}, \& P.~{Heinzel}, Vol. 320,
  259--267, \dodoi{10.1017/S1743921316000028}

\bibitem[{{Kowalski} {et~al.}(2022){Kowalski}, {Allred}, {Carlsson}, {Kerr},
  {Tremblay}, {Namekata}, {Kuridze}, \& {Uitenbroek}}]{Kowalski+2022}
{Kowalski}, A.~F., {Allred}, J.~C., {Carlsson}, M., {et~al.} 2022, \apj, 928,
  190, \dodoi{10.3847/1538-4357/ac5174}

\bibitem[{{Kowalski} {et~al.}(2017){Kowalski}, {Allred}, {Daw}, {Cauzzi}, \&
  {Carlsson}}]{Kowalski+2017}
{Kowalski}, A.~F., {Allred}, J.~C., {Daw}, A., {Cauzzi}, G., \& {Carlsson}, M.
  2017, \apj, 836, 12, \dodoi{10.3847/1538-4357/836/1/12}

\bibitem[{{Kowalski} {et~al.}(2019){Kowalski}, {Butler}, {Daw}, {Fletcher},
  {Allred}, {De Pontieu}, {Kerr}, \& {Cauzzi}}]{Kowalski+2019}
{Kowalski}, A.~F., {Butler}, E., {Daw}, A.~N., {et~al.} 2019, \apj, 878, 135,
  \dodoi{10.3847/1538-4357/ab1f8b}

\bibitem[{{Kowalski} {et~al.}(2010){Kowalski}, {Hawley}, {Holtzman},
  {Wisniewski}, \& {Hilton}}]{Kowalski+2010}
{Kowalski}, A.~F., {Hawley}, S.~L., {Holtzman}, J.~A., {Wisniewski}, J.~P., \&
  {Hilton}, E.~J. 2010, \apjl, 714, L98, \dodoi{10.1088/2041-8205/714/1/L98}

\bibitem[{{Kowalski} {et~al.}(2013){Kowalski}, {Hawley}, {Wisniewski}, {Osten},
  {Hilton}, {Holtzman}, {Schmidt}, \& {Davenport}}]{Kowalski+2013}
{Kowalski}, A.~F., {Hawley}, S.~L., {Wisniewski}, J.~P., {et~al.} 2013, \apjs,
  207, 15, \dodoi{10.1088/0067-0049/207/1/15}

\bibitem[{{Kuridze} {et~al.}(2015){Kuridze}, {Mathioudakis}, {Sim{\~o}es},
  {Rouppe van der Voort}, {Carlsson}, {Jafarzadeh}, {Allred}, {Kowalski},
  {Kennedy}, {Fletcher}, {Graham}, \& {Keenan}}]{Kuridze+2015}
{Kuridze}, D., {Mathioudakis}, M., {Sim{\~o}es}, P.~J.~A., {et~al.} 2015, \apj,
  813, 125, \dodoi{10.1088/0004-637X/813/2/125}

\bibitem[{{Kuridze} {et~al.}(2016){Kuridze}, {Mathioudakis}, {Christian},
  {Kowalski}, {Jess}, {Grant}, {Kawate}, {Sim{\~o}es}, {Allred}, \&
  {Keenan}}]{Kuridze+2016}
{Kuridze}, D., {Mathioudakis}, M., {Christian}, D.~J., {et~al.} 2016, \apj,
  832, 147, \dodoi{10.3847/0004-637X/832/2/147}

\bibitem[{{Labrosse} {et~al.}(2010){Labrosse}, {Heinzel}, {Vial}, {Kucera},
  {Parenti}, {Gun{\'a}r}, {Schmieder}, \& {Kilper}}]{Labrosse+2010}
{Labrosse}, N., {Heinzel}, P., {Vial}, J.~C., {et~al.} 2010, \ssr, 151, 243,
  \dodoi{10.1007/s11214-010-9630-6}

\bibitem[{{Lacy} {et~al.}(1976){Lacy}, {Moffett}, \& {Evans}}]{Lacy+1976}
{Lacy}, C.~H., {Moffett}, T.~J., \& {Evans}, D.~S. 1976, \apjs, 30, 85,
  \dodoi{10.1086/190358}

\bibitem[{{Lalitha} {et~al.}(2013){Lalitha}, {Fuhrmeister}, {Wolter},
  {Schmitt}, {Engels}, \& {Wieringa}}]{Lalitha+2013_A&A}
{Lalitha}, S., {Fuhrmeister}, B., {Wolter}, U., {et~al.} 2013, \aap, 560, A69,
  \dodoi{10.1051/0004-6361/201321419}

\bibitem[{{Lammer} {et~al.}(2007){Lammer}, {Lichtenegger}, {Kulikov},
  {Grie{\ss}meier}, {Terada}, {Erkaev}, {Biernat}, {Khodachenko}, {Ribas},
  {Penz}, \& {Selsis}}]{Lammer+2007}
{Lammer}, H., {Lichtenegger}, H. I.~M., {Kulikov}, Y.~N., {et~al.} 2007,
  Astrobiology, 7, 185, \dodoi{10.1089/ast.2006.0128}

\bibitem[{{Leitzinger} \& {Odert}(2022)}]{Leitzinger_Odert_2022_Review}
{Leitzinger}, M., \& {Odert}, P. 2022, Serbian Astronomical Journal, 205, 1,
  \dodoi{10.2298/SAJ2205001L}

\bibitem[{{Leitzinger} {et~al.}(2022){Leitzinger}, {Odert}, \&
  {Heinzel}}]{Leitzinger+2022}
{Leitzinger}, M., {Odert}, P., \& {Heinzel}, P. 2022, \mnras, 513, 6058,
  \dodoi{10.1093/mnras/stac1284}

\bibitem[{{Li} {et~al.}(2019){Li}, {Ding}, {Hong}, {Li}, \& {Gan}}]{Li+2019}
{Li}, Y., {Ding}, M.~D., {Hong}, J., {Li}, H., \& {Gan}, W.~Q. 2019, \apj, 879,
  30, \dodoi{10.3847/1538-4357/ab245a}

\bibitem[{{Libbrecht} {et~al.}(2019){Libbrecht}, {de la Cruz Rodr{\'\i}guez},
  {Danilovic}, {Leenaarts}, \& {Pazira}}]{Libbrecht+2019}
{Libbrecht}, T., {de la Cruz Rodr{\'\i}guez}, J., {Danilovic}, S., {Leenaarts},
  J., \& {Pazira}, H. 2019, \aap, 621, A35, \dodoi{10.1051/0004-6361/201833610}

\bibitem[{{Liefke} {et~al.}(2010){Liefke}, {Fuhrmeister}, \&
  {Schmitt}}]{Liefke+2010_A&A}
{Liefke}, C., {Fuhrmeister}, B., \& {Schmitt}, J.~H.~M.~M. 2010, \aap, 514,
  A94, \dodoi{10.1051/0004-6361/201014012}

\bibitem[{{Linsky}(2019)}]{Linsky2019}
{Linsky}, J. 2019, {Host Stars and their Effects on Exoplanet Atmospheres},
  Vol. 955 (Springer International Publishing),
  \dodoi{10.1007/978-3-030-11452-7}

\bibitem[{{Longcope}(2014)}]{Longcope_2014}
{Longcope}, D.~W. 2014, \apj, 795, 10, \dodoi{10.1088/0004-637X/795/1/10}

\bibitem[{{Loyd} {et~al.}(2022){Loyd}, {Mason}, {Jin}, {Shkolnik}, {France},
  {Youngblood}, {Villadsen}, {Schneider}, {Schneider}, {Llama},
  {Ramiaramanantsoa}, \& {Richey-Yowell}}]{Loyd+2022}
{Loyd}, R.~O.~P., {Mason}, J.~P., {Jin}, M., {et~al.} 2022, \apj, 936, 170,
  \dodoi{10.3847/1538-4357/ac80c1}

\bibitem[{{Lynch} \& {Edmondson}(2013)}]{Lynch+2013_ApJ}
{Lynch}, B.~J., \& {Edmondson}, J.~K. 2013, \apj, 764, 87,
  \dodoi{10.1088/0004-637X/764/1/87}

\bibitem[{{Lynch} {et~al.}(2016){Lynch}, {Edmondson}, {Kazachenko}, \&
  {Guidoni}}]{Lynch+2016_ApJ}
{Lynch}, B.~J., {Edmondson}, J.~K., {Kazachenko}, M.~D., \& {Guidoni}, S.~E.
  2016, \apj, 826, 43, \dodoi{10.3847/0004-637X/826/1/43}

\bibitem[{{Lynch} {et~al.}(2023){Lynch}, {Wood}, {Jin}, {T{\"o}r{\"o}k}, {Sun},
  {Palmerio}, {Osten}, {Vidotto}, {Cohen}, {Alvarado-G{\'o}mez}, {Drake},
  {Airapetian}, {Notsu}, {Veronig}, {Namekata}, {Winslow}, {Jian}, {Vourlidas},
  {Lugaz}, {Al-Haddad}, {Manchester}, {Scolini}, {Farrugia}, {Davies},
  {Nieves-Chinchilla}, {Carcaboso}, {Lee}, \& {Salman}}]{Lynch+2022_WP}
{Lynch}, B.~J., {Wood}, B.~E., {Jin}, M., {et~al.} 2023, in Bulletin of the
  American Astronomical Society, Vol.~55, 254,
  \dodoi{10.3847/25c2cfeb.2dd884d5}

\bibitem[{{Maehara} {et~al.}(2015){Maehara}, {Shibayama}, {Notsu}, {Notsu},
  {Honda}, {Nogami}, \& {Shibata}}]{Maehara+2015}
{Maehara}, H., {Shibayama}, T., {Notsu}, Y., {et~al.} 2015, Earth, Planets, and
  Space, 67, 59, \dodoi{10.1186/s40623-015-0217-z}

\bibitem[{{Maehara} {et~al.}(2012){Maehara}, {Shibayama}, {Notsu}, {Notsu},
  {Nagao}, {Kusaba}, {Honda}, {Nogami}, \& {Shibata}}]{Maehara+2012}
{Maehara}, H., {Shibayama}, T., {Notsu}, S., {et~al.} 2012, \nat, 485, 478,
  \dodoi{10.1038/nature11063}

\bibitem[{{Maehara} {et~al.}(2021){Maehara}, {Notsu}, {Namekata}, {Honda},
  {Kowalski}, {Katoh}, {Ohshima}, {Iida}, {Oeda}, {Murata}, {Yamanaka},
  {Takagi}, {Sasada}, {Akitaya}, {Ikuta}, {Okamoto}, {Nogami}, \&
  {Shibata}}]{Maehara+2021}
{Maehara}, H., {Notsu}, Y., {Namekata}, K., {et~al.} 2021, \pasj, 73, 44,
  \dodoi{10.1093/pasj/psaa098}

\bibitem[{{Medina} {et~al.}(2020){Medina}, {Winters}, {Irwin}, \&
  {Charbonneau}}]{Medina+2020}
{Medina}, A.~A., {Winters}, J.~G., {Irwin}, J.~M., \& {Charbonneau}, D. 2020,
  \apj, 905, 107, \dodoi{10.3847/1538-4357/abc686}

\bibitem[{{Milligan} {et~al.}(2014){Milligan}, {Kerr}, {Dennis}, {Hudson},
  {Fletcher}, {Allred}, {Chamberlin}, {Ireland}, {Mathioudakis}, \&
  {Keenan}}]{Milligan+2014}
{Milligan}, R.~O., {Kerr}, G.~S., {Dennis}, B.~R., {et~al.} 2014, \apj, 793,
  70, \dodoi{10.1088/0004-637X/793/2/70}

\bibitem[{{Mitra-Kraev} {et~al.}(2005){Mitra-Kraev}, {Harra}, {G{\"u}del},
  {Audard}, {Branduardi-Raymont}, {Kay}, {Mewe}, {Raassen}, \& {van
  Driel-Gesztelyi}}]{Mitra-Kraev+2005a}
{Mitra-Kraev}, U., {Harra}, L.~K., {G{\"u}del}, M., {et~al.} 2005, \aap, 431,
  679, \dodoi{10.1051/0004-6361:20041201}

\bibitem[{{Miyake} {et~al.}(2019){Miyake}, {Usoskin}, \&
  {Poluianov}}]{Miyake+2019}
{Miyake}, F., {Usoskin}, I., \& {Poluianov}, S. 2019, {Extreme Solar Particle
  Storms; The hostile Sun} (IOP Publishing), \dodoi{10.1088/2514-3433/ab404a}

\bibitem[{{Morin} {et~al.}(2008){Morin}, {Donati}, {Petit}, {Delfosse},
  {Forveille}, {Albert}, {Auri{\`e}re}, {Cabanac}, {Dintrans}, {Fares},
  {Gastine}, {Jardine}, {Ligni{\`e}res}, {Paletou}, {Ramirez Velez}, \&
  {Th{\'e}ado}}]{Morin+2008}
{Morin}, J., {Donati}, J.~F., {Petit}, P., {et~al.} 2008, \mnras, 390, 567,
  \dodoi{10.1111/j.1365-2966.2008.13809.x}

\bibitem[{{Moschou} {et~al.}(2019){Moschou}, {Drake}, {Cohen},
  {Alvarado-G{\'o}mez}, {Garraffo}, \& {Fraschetti}}]{Moschou+2019}
{Moschou}, S.-P., {Drake}, J.~J., {Cohen}, O., {et~al.} 2019, \apj, 877, 105,
  \dodoi{10.3847/1538-4357/ab1b37}

\bibitem[{{Muheki} {et~al.}(2020{\natexlab{a}}){Muheki}, {Guenther},
  {Mutabazi}, \& {Jurua}}]{Muheki+2020}
{Muheki}, P., {Guenther}, E.~W., {Mutabazi}, T., \& {Jurua}, E.
  2020{\natexlab{a}}, \aap, 637, A13, \dodoi{10.1051/0004-6361/201936904}

\bibitem[{{Muheki} {et~al.}(2020{\natexlab{b}}){Muheki}, {Guenther},
  {Mutabazi}, \& {Jurua}}]{Muheki+2020_EVLac}
---. 2020{\natexlab{b}}, \mnras, 499, 5047, \dodoi{10.1093/mnras/staa3152}

\bibitem[{{Mullan} {et~al.}(2006){Mullan}, {Mathioudakis}, {Bloomfield}, \&
  {Christian}}]{Mullan+2006_ApJS}
{Mullan}, D.~J., {Mathioudakis}, M., {Bloomfield}, D.~S., \& {Christian}, D.~J.
  2006, \apjs, 164, 173, \dodoi{10.1086/502629}

\bibitem[{{Namekata} {et~al.}(2022{\natexlab{a}}){Namekata}, {Ichimoto},
  {Ishii}, \& {Shibata}}]{Namekata+2022_ApJ}
{Namekata}, K., {Ichimoto}, K., {Ishii}, T.~T., \& {Shibata}, K.
  2022{\natexlab{a}}, \apj, 933, 209, \dodoi{10.3847/1538-4357/ac75cd}

\bibitem[{{Namekata} {et~al.}(2022{\natexlab{b}}){Namekata}, {Maehara},
  {Honda}, {Notsu}, {Nogami}, \& {Shibata}}]{Namekata+2022_IAUS}
{Namekata}, K., {Maehara}, H., {Honda}, S., {et~al.} 2022{\natexlab{b}}, arXiv
  e-prints, arXiv:2211.05506.
\newblock \doarXiv{2211.05506}

\bibitem[{{Namekata} {et~al.}(2017{\natexlab{a}}){Namekata}, {Sakaue},
  {Watanabe}, {Asai}, \& {Shibata}}]{Namekata+2017_PASJ}
{Namekata}, K., {Sakaue}, T., {Watanabe}, K., {Asai}, A., \& {Shibata}, K.
  2017{\natexlab{a}}, \pasj, 69, 7, \dodoi{10.1093/pasj/psw111}

\bibitem[{{Namekata} {et~al.}(2017{\natexlab{b}}){Namekata}, {Sakaue},
  {Watanabe}, {Asai}, {Maehara}, {Notsu}, {Notsu}, {Honda}, {Ishii}, {Ikuta},
  {Nogami}, \& {Shibata}}]{Namekata+2017_ApJ}
{Namekata}, K., {Sakaue}, T., {Watanabe}, K., {et~al.} 2017{\natexlab{b}},
  \apj, 851, 91, \dodoi{10.3847/1538-4357/aa9b34}

\bibitem[{{Namekata} {et~al.}(2020){Namekata}, {Maehara}, {Sasaki}, {Kawai},
  {Notsu}, {Kowalski}, {Allred}, {Iwakiri}, {Tsuboi}, {Murata}, {Niwano},
  {Shiraishi}, {Adachi}, {Iida}, {Oeda}, {Honda}, {Tozuka}, {Katoh}, {Onozato},
  {Okamoto}, {Isogai}, {Kimura}, {Kojiguchi}, {Wakamatsu}, {Tampo}, {Nogami},
  \& {Shibata}}]{Namekata+2020_PASJ}
{Namekata}, K., {Maehara}, H., {Sasaki}, R., {et~al.} 2020, \pasj, 72, 68,
  \dodoi{10.1093/pasj/psaa051}

\bibitem[{{Namekata} {et~al.}(2022{\natexlab{c}}){Namekata}, {Maehara},
  {Honda}, {Notsu}, {Okamoto}, {Takahashi}, {Takayama}, {Ohshima}, {Saito},
  {Katoh}, {Tozuka}, {Murata}, {Ogawa}, {Niwano}, {Adachi}, {Oeda},
  {Shiraishi}, {Isogai}, {Seki}, {Ishii}, {Ichimoto}, {Nogami}, \&
  {Shibata}}]{Namekata+2022_NatAst}
{Namekata}, K., {Maehara}, H., {Honda}, S., {et~al.} 2022{\natexlab{c}}, Nature
  Astronomy, 6, 241, \dodoi{10.1038/s41550-021-01532-8}

\bibitem[{{Namekata} {et~al.}(2022{\natexlab{d}}){Namekata}, {Maehara},
  {Honda}, {Notsu}, {Okamoto}, {Takahashi}, {Takayama}, {Ohshima}, {Saito},
  {Katoh}, {Tozuka}, {Murata}, {Ogawa}, {Niwano}, {Adachi}, {Oeda},
  {Shiraishi}, {Isogai}, {Nogami}, \& {Shibata}}]{Namekata+2022_ApJL}
---. 2022{\natexlab{d}}, \apjl, 926, L5, \dodoi{10.3847/2041-8213/ac4df0}

\bibitem[{{Namizaki} {et~al.}(2023){Namizaki}, {Namekata}, {Maehara}, {Notsu},
  {Honda}, {Nogami}, \& {Shibata}}]{Namizaki+2023_ApJ}
{Namizaki}, K., {Namekata}, K., {Maehara}, H., {et~al.} 2023, \apj, 945, 61,
  \dodoi{10.3847/1538-4357/acb928}

\bibitem[{{Neupert}(1968)}]{Neupert+1968}
{Neupert}, W.~M. 1968, \apjl, 153, L59, \dodoi{10.1086/180220}

\bibitem[{{Notsu} {et~al.}(2013b){Notsu}, {Shibayama}, {Maehara}, {Notsu},
  {Nagao}, {Honda}, {Ishii}, {Nogami}, \& {Shibata}}]{Notsu+2013}
{Notsu}, Y., {Shibayama}, T., {Maehara}, H., {et~al.} 2013b, \apj, 771, 127,
  \dodoi{10.1088/0004-637X/771/2/127}

\bibitem[{{Notsu} {et~al.}(2019){Notsu}, {Maehara}, {Honda}, {Hawley},
  {Davenport}, {Namekata}, {Notsu}, {Ikuta}, {Nogami}, \&
  {Shibata}}]{Notsu+2019}
{Notsu}, Y., {Maehara}, H., {Honda}, S., {et~al.} 2019, \apj, 876, 58,
  \dodoi{10.3847/1538-4357/ab14e6}

\bibitem[{{Odert} {et~al.}(2020){Odert}, {Leitzinger}, {Guenther}, \&
  {Heinzel}}]{Odert+2020}
{Odert}, P., {Leitzinger}, M., {Guenther}, E.~W., \& {Heinzel}, P. 2020,
  \mnras, 494, 3766, \dodoi{10.1093/mnras/staa1021}

\bibitem[{{Odert} {et~al.}(2017){Odert}, {Leitzinger}, {Hanslmeier}, \&
  {Lammer}}]{Odert+2017}
{Odert}, P., {Leitzinger}, M., {Hanslmeier}, A., \& {Lammer}, H. 2017, \mnras,
  472, 876, \dodoi{10.1093/mnras/stx1969}

\bibitem[{{Okada} {et~al.}(2020){Okada}, {Ichimoto}, {Machida}, {Tokuda},
  {Huang}, \& {UeNo}}]{Okada+2020}
{Okada}, S., {Ichimoto}, K., {Machida}, A., {et~al.} 2020, \pasj, 72, 71,
  \dodoi{10.1093/pasj/psaa014}

\bibitem[{{Okajima} {et~al.}(2016){Okajima}, {Soong}, {Balsamo}, {Enoto},
  {Olsen}, {Koenecke}, {Lozipone}, {Kearney}, {Fitzsimmons}, {Numata},
  {Kenyon}, {Arzoumanian}, \& {Gendreau}}]{Okajima2016a}
{Okajima}, T., {Soong}, Y., {Balsamo}, E.~R., {et~al.} 2016, in Society of
  Photo-Optical Instrumentation Engineers (SPIE) Conference Series, Vol. 9905,
  Space Telescopes and Instrumentation 2016: Ultraviolet to Gamma Ray, ed.
  J.-W.~A. {den Herder}, T.~{Takahashi}, \& M.~{Bautz}, 99054X,
  \dodoi{10.1117/12.2234436}

\bibitem[{{Okamoto} {et~al.}(2021){Okamoto}, {Notsu}, {Maehara}, {Namekata},
  {Honda}, {Ikuta}, {Nogami}, \& {Shibata}}]{Okamoto+2021}
{Okamoto}, S., {Notsu}, Y., {Maehara}, H., {et~al.} 2021, \apj, 906, 72,
  \dodoi{10.3847/1538-4357/abc8f5}

\bibitem[{{Oks} \& {Gershberg}(2016)}]{Oks+2016}
{Oks}, E., \& {Gershberg}, R.~E. 2016, \apj, 819, 16,
  \dodoi{10.3847/0004-637X/819/1/16}

\bibitem[{{Osten} {et~al.}(2006){Osten}, {Hawley}, {Allred}, {Johns-Krull},
  {Brown}, \& {Harper}}]{Osten+2006}
{Osten}, R.~A., {Hawley}, S.~L., {Allred}, J., {et~al.} 2006, \apj, 647, 1349,
  \dodoi{10.1086/504889}

\bibitem[{{Osten} \& {Wolk}(2015)}]{Osten+2015}
{Osten}, R.~A., \& {Wolk}, S.~J. 2015, \apj, 809, 79,
  \dodoi{10.1088/0004-637X/809/1/79}

\bibitem[{{Otsu} {et~al.}(2022){Otsu}, {Asai}, {Ichimoto}, {Ishii}, \&
  {Namekata}}]{Otsu+2022}
{Otsu}, T., {Asai}, A., {Ichimoto}, K., {Ishii}, T.~T., \& {Namekata}, K. 2022,
  \apj, 939, 98, \dodoi{10.3847/1538-4357/ac9730}

\bibitem[{{Panos} {et~al.}(2018){Panos}, {Kleint}, {Huwyler}, {Krucker},
  {Melchior}, {Ullmann}, \& {Voloshynovskiy}}]{Panos+2018}
{Panos}, B., {Kleint}, L., {Huwyler}, C., {et~al.} 2018, \apj, 861, 62,
  \dodoi{10.3847/1538-4357/aac779}

\bibitem[{{Parenti}(2014)}]{Parenti_2014_LRSP}
{Parenti}, S. 2014, Living Reviews in Solar Physics, 11, 1,
  \dodoi{10.12942/lrsp-2014-1}

\bibitem[{{Paudel} {et~al.}(2019){Paudel}, {Gizis}, {Mullan}, {Schmidt},
  {Burgasser}, {Williams}, {Youngblood}, \& {Stassun}}]{Paudel+2019}
{Paudel}, R.~R., {Gizis}, J.~E., {Mullan}, D.~J., {et~al.} 2019, \mnras, 486,
  1438, \dodoi{10.1093/mnras/stz886}

\bibitem[{{Paudel} {et~al.}(2021){Paudel}, {Barclay}, {Schlieder}, {Quintana},
  {Gilbert}, {Vega}, {Youngblood}, {Silverstein}, {Osten}, {Tucker}, {Huber},
  {Do}, {Hamaguchi}, {Mullan}, {Gizis}, {Monsue}, {Col{\'o}n}, {Boyd},
  {Davenport}, \& {Walkowicz}}]{Paudel+2021_ApJ}
{Paudel}, R.~R., {Barclay}, T., {Schlieder}, J.~E., {et~al.} 2021, \apj, 922,
  31, \dodoi{10.3847/1538-4357/ac1946}

\bibitem[{{Pillitteri} {et~al.}(2022){Pillitteri}, {Argiroffi}, {Maggio},
  {Micela}, {Benatti}, {Reale}, {Colombo}, \& {Wolk}}]{Pillitteri+2022}
{Pillitteri}, I., {Argiroffi}, C., {Maggio}, A., {et~al.} 2022, \aap, 666,
  A198, \dodoi{10.1051/0004-6361/202244268}

\bibitem[{{Prigozhin} {et~al.}(2016){Prigozhin}, {Gendreau}, {Doty}, {Foster},
  {Remillard}, {Malonis}, {LaMarr}, {Vezie}, {Egan}, {Villasenor},
  {Arzoumanian}, {Baumgartner}, {Scholze}, {Laubis}, {Krumrey}, \&
  {Huber}}]{Prigozhin2016a}
{Prigozhin}, G., {Gendreau}, K., {Doty}, J.~P., {et~al.} 2016, in Society of
  Photo-Optical Instrumentation Engineers (SPIE) Conference Series, Vol. 9905,
  Space Telescopes and Instrumentation 2016: Ultraviolet to Gamma Ray, ed.
  J.-W.~A. {den Herder}, T.~{Takahashi}, \& M.~{Bautz}, 99051I,
  \dodoi{10.1117/12.2231718}

\bibitem[{{Raassen} {et~al.}(2007){Raassen}, {Mitra-Kraev}, \&
  {G{\"u}del}}]{Raassen+2007}
{Raassen}, A.~J.~J., {Mitra-Kraev}, U., \& {G{\"u}del}, M. 2007, \mnras, 379,
  1075, \dodoi{10.1111/j.1365-2966.2007.11983.x}

\bibitem[{{Reep} \& {Airapetian}(2023)}]{Reep+2023_arXiv}
{Reep}, J.~W., \& {Airapetian}, V.~S. 2023, arXiv e-prints, arXiv:2306.03765,
  \dodoi{10.48550/arXiv.2306.03765}

\bibitem[{{Reid} \& {Hawley}(2005)}]{Reid+2005}
{Reid}, I.~N., \& {Hawley}, S.~L. 2005, {New light on dark stars : red dwarfs,
  low-mass stars, brown dwarfs} (Springer-Verlag Berlin Heidelberg),
  \dodoi{10.1007/3-540-27610-6}

\bibitem[{{Reiners} \& {Basri}(2007)}]{Reiners+2007}
{Reiners}, A., \& {Basri}, G. 2007, \apj, 656, 1121, \dodoi{10.1086/510304}

\bibitem[{{Remillard} {et~al.}(2022){Remillard}, {Loewenstein}, {Steiner},
  {Prigozhin}, {LaMarr}, {Enoto}, {Gendreau}, {Arzoumanian}, {Markwardt},
  {Basak}, {Stevens}, {Ray}, {Altamirano}, \& {Buisson}}]{Remillard+2022}
{Remillard}, R.~A., {Loewenstein}, M., {Steiner}, J.~F., {et~al.} 2022, \aj,
  163, 130, \dodoi{10.3847/1538-3881/ac4ae6}

\bibitem[{{Ricker} {et~al.}(2015){Ricker}, {Winn}, {Vanderspek}, {Latham},
  {Bakos}, {Bean}, {Berta-Thompson}, {Brown}, {Buchhave}, {Butler}, {Butler},
  {Chaplin}, {Charbonneau}, {Christensen-Dalsgaard}, {Clampin}, {Deming},
  {Doty}, {De Lee}, {Dressing}, {Dunham}, {Endl}, {Fressin}, {Ge}, {Henning},
  {Holman}, {Howard}, {Ida}, {Jenkins}, {Jernigan}, {Johnson}, {Kaltenegger},
  {Kawai}, {Kjeldsen}, {Laughlin}, {Levine}, {Lin}, {Lissauer}, {MacQueen},
  {Marcy}, {McCullough}, {Morton}, {Narita}, {Paegert}, {Palle}, {Pepe},
  {Pepper}, {Quirrenbach}, {Rinehart}, {Sasselov}, {Sato}, {Seager},
  {Sozzetti}, {Stassun}, {Sullivan}, {Szentgyorgyi}, {Torres}, {Udry}, \&
  {Villasenor}}]{Ricker+2015}
{Ricker}, G.~R., {Winn}, J.~N., {Vanderspek}, R., {et~al.} 2015, Journal of
  Astronomical Telescopes, Instruments, and Systems, 1, 014003,
  \dodoi{10.1117/1.JATIS.1.1.014003}

\bibitem[{{Rodrigo} \& {Solano}(2020)}]{Rodrigo+2020}
{Rodrigo}, C., \& {Solano}, E. 2020, in Contributions to the XIV.0 Scientific
  Meeting (virtual) of the Spanish Astronomical Society, 182

\bibitem[{{Rodrigo} {et~al.}(2012){Rodrigo}, {Solano}, \&
  {Bayo}}]{Rodrigo+2012}
{Rodrigo}, C., {Solano}, E., \& {Bayo}, A. 2012, {SVO Filter Profile Service
  Version 1.0}, IVOA Working Draft 15 October 2012,
  \dodoi{10.5479/ADS/bib/2012ivoa.rept.1015R}

\bibitem[{{Saar} \& {Linsky}(1985)}]{Saar+1985}
{Saar}, S.~H., \& {Linsky}, J.~L. 1985, \apjl, 299, L47, \dodoi{10.1086/184578}

\bibitem[{{Sakaue} \& {Shibata}(2021)}]{Sakaue-Shibata_2021}
{Sakaue}, T., \& {Shibata}, K. 2021, \apj, 919, 29,
  \dodoi{10.3847/1538-4357/ac0e34}

\bibitem[{{Scargle} {et~al.}(2013){Scargle}, {Norris}, {Jackson}, \&
  {Chiang}}]{Scargle+2013}
{Scargle}, J.~D., {Norris}, J.~P., {Jackson}, B., \& {Chiang}, J. 2013, \apj,
  764, 167, \dodoi{10.1088/0004-637X/764/2/167}

\bibitem[{{Scheucher} {et~al.}(2018){Scheucher}, {Grenfell}, {Wunderlich},
  {Godolt}, {Schreier}, \& {Rauer}}]{Scheucher+2018}
{Scheucher}, M., {Grenfell}, J.~L., {Wunderlich}, F., {et~al.} 2018, \apj, 863,
  6, \dodoi{10.3847/1538-4357/aacf03}

\bibitem[{{Sch{\"o}fer} {et~al.}(2022){Sch{\"o}fer}, {Jeffers}, {Reiners},
  {Zechmeister}, {Fuhrmeister}, {Lafarga}, {Ribas}, {Quirrenbach}, {Amado},
  {Caballero}, {Anglada-Escud{\'e}}, {Bauer}, {B{\'e}jar},
  {Cort{\'e}s-Contreras}, {Alonso}, {Dreizler}, {Guenther}, {Herbort},
  {Johnson}, {Kaminski}, {K{\"u}rster}, {Montes}, {Morales}, {Pedraz}, \&
  {Tal-Or}}]{Schofer+2022}
{Sch{\"o}fer}, P., {Jeffers}, S.~V., {Reiners}, A., {et~al.} 2022, \aap, 663,
  A68, \dodoi{10.1051/0004-6361/201936102}

\bibitem[{{Segura} {et~al.}(2010){Segura}, {Walkowicz}, {Meadows}, {Kasting},
  \& {Hawley}}]{Segura+2010}
{Segura}, A., {Walkowicz}, L.~M., {Meadows}, V., {Kasting}, J., \& {Hawley}, S.
  2010, Astrobiology, 10, 751, \dodoi{10.1089/ast.2009.0376}

\bibitem[{{Seki} {et~al.}(2021){Seki}, {Otsuji}, {Ishii}, {Asai}, \&
  {Ichimoto}}]{Seki+2021}
{Seki}, D., {Otsuji}, K., {Ishii}, T.~T., {Asai}, A., \& {Ichimoto}, K. 2021,
  Earth, Planets, and Space, 73, 58, \dodoi{10.1186/s40623-021-01378-4}

\bibitem[{{Seki} {et~al.}(2019){Seki}, {Otsuji}, {Ishii}, {Hirose}, {Iju},
  {UeNo}, {Cabezas}, {Asai}, {Isobe}, {Ichimoto}, \& {Shibata}}]{Seki+2019}
{Seki}, D., {Otsuji}, K., {Ishii}, T., {et~al.} 2019, Sun and Geosphere, 14,
  95, \dodoi{10.31401/SunGeo.2019.02.01}

\bibitem[{{Shibata} \& {Magara}(2011)}]{Shibata+2011}
{Shibata}, K., \& {Magara}, T. 2011, Living Reviews in Solar Physics, 8, 6,
  \dodoi{10.12942/lrsp-2011-6}

\bibitem[{{Shibata} \& {Yokoyama}(1999)}]{Shibata_Yokoyama_1999}
{Shibata}, K., \& {Yokoyama}, T. 1999, \apjl, 526, L49, \dodoi{10.1086/312354}

\bibitem[{{Shibata} \& {Yokoyama}(2002)}]{Shibata_Yokoyama+2002}
---. 2002, \apj, 577, 422, \dodoi{10.1086/342141}

\bibitem[{{Shibata} {et~al.}(2013){Shibata}, {Isobe}, {Hillier}, {Choudhuri},
  {Maehara}, {Ishii}, {Shibayama}, {Notsu}, {Notsu}, {Nagao}, {Honda}, \&
  {Nogami}}]{Shibata+2013}
{Shibata}, K., {Isobe}, H., {Hillier}, A., {et~al.} 2013, \pasj, 65, 49,
  \dodoi{10.1093/pasj/65.3.49}

\bibitem[{{Shibayama} {et~al.}(2013){Shibayama}, {Maehara}, {Notsu}, {Notsu},
  {Nagao}, {Honda}, {Ishii}, {Nogami}, \& {Shibata}}]{Shibayama+2013}
{Shibayama}, T., {Maehara}, H., {Notsu}, S., {et~al.} 2013, \apjs, 209, 5,
  \dodoi{10.1088/0067-0049/209/1/5}

\bibitem[{{Shoji} \& {Kurokawa}(1995)}]{Shoji+1995}
{Shoji}, M., \& {Kurokawa}, H. 1995, \pasj, 47, 239

\bibitem[{{Sinha} {et~al.}(2019){Sinha}, {Srivastava}, \&
  {Nandy}}]{Sinha+2019_ApJ}
{Sinha}, S., {Srivastava}, N., \& {Nandy}, D. 2019, \apj, 880, 84,
  \dodoi{10.3847/1538-4357/ab2239}

\bibitem[{{Smith} {et~al.}(2001){Smith}, {Brickhouse}, {Liedahl}, \&
  {Raymond}}]{Smith+2001}
{Smith}, R.~K., {Brickhouse}, N.~S., {Liedahl}, D.~A., \& {Raymond}, J.~C.
  2001, \apjl, 556, L91, \dodoi{10.1086/322992}

\bibitem[{{Stelzer} {et~al.}(2022){Stelzer}, {Caramazza}, {Raetz}, {Argiroffi},
  \& {Coffaro}}]{Stelzer+2022}
{Stelzer}, B., {Caramazza}, M., {Raetz}, S., {Argiroffi}, C., \& {Coffaro}, M.
  2022, \aap, 667, L9, \dodoi{10.1051/0004-6361/202244642}

\bibitem[{{Sun} {et~al.}(2022){Sun}, {T{\"o}r{\"o}k}, \& {DeRosa}}]{Sun+2022}
{Sun}, X., {T{\"o}r{\"o}k}, T., \& {DeRosa}, M.~L. 2022, \mnras, 509, 5075,
  \dodoi{10.1093/mnras/stab3249}

\bibitem[{{Takahashi} {et~al.}(2016){Takahashi}, {Mizuno}, \&
  {Shibata}}]{Takahashi+2016}
{Takahashi}, T., {Mizuno}, Y., \& {Shibata}, K. 2016, \apjl, 833, L8,
  \dodoi{10.3847/2041-8205/833/1/L8}

\bibitem[{{Takasao} {et~al.}(2020){Takasao}, {Mitsuishi}, {Shimura}, {Yoshida},
  {Kunitomo}, {Tanaka}, \& {Ishihara}}]{Takasao+2020}
{Takasao}, S., {Mitsuishi}, I., {Shimura}, T., {et~al.} 2020, \apj, 901, 70,
  \dodoi{10.3847/1538-4357/abad34}

\bibitem[{{Tang}(1983)}]{Tang+1983}
{Tang}, F. 1983, \solphys, 83, 15, \dodoi{10.1007/BF00148240}

\bibitem[{{Tei} {et~al.}(2018){Tei}, {Sakaue}, {Okamoto}, {Kawate}, {Heinzel},
  {UeNo}, {Asai}, {Ichimoto}, \& {Shibata}}]{Tei+2018}
{Tei}, A., {Sakaue}, T., {Okamoto}, T.~J., {et~al.} 2018, \pasj, 70, 100,
  \dodoi{10.1093/pasj/psy047}

\bibitem[{{Thalmann} {et~al.}(2015){Thalmann}, {Su}, {Temmer}, \&
  {Veronig}}]{Thalmann+2015}
{Thalmann}, J.~K., {Su}, Y., {Temmer}, M., \& {Veronig}, A.~M. 2015, \apjl,
  801, L23, \dodoi{10.1088/2041-8205/801/2/L23}

\bibitem[{{Tilley} {et~al.}(2019){Tilley}, {Segura}, {Meadows}, {Hawley}, \&
  {Davenport}}]{Tilley+2019}
{Tilley}, M.~A., {Segura}, A., {Meadows}, V., {Hawley}, S., \& {Davenport}, J.
  2019, Astrobiology, 19, 64, \dodoi{10.1089/ast.2017.1794}

\bibitem[{{Tokovinin} {et~al.}(2013){Tokovinin}, {Fischer}, {Bonati},
  {Giguere}, {Moore}, {Schwab}, {Spronck}, \& {Szymkowiak}}]{Tokovinin+2013}
{Tokovinin}, A., {Fischer}, D.~A., {Bonati}, M., {et~al.} 2013, \pasp, 125,
  1336, \dodoi{10.1086/674012}

\bibitem[{{Toriumi} {et~al.}(2020){Toriumi}, {Airapetian}, {Hudson},
  {Schrijver}, {Cheung}, \& {DeRosa}}]{Toriumi+2020}
{Toriumi}, S., {Airapetian}, V.~S., {Hudson}, H.~S., {et~al.} 2020, \apj, 902,
  36, \dodoi{10.3847/1538-4357/abadf9}

\bibitem[{{T{\"o}r{\"o}k} {et~al.}(2011){T{\"o}r{\"o}k}, {Panasenco}, {Titov},
  {Miki{\'c}}, {Reeves}, {Velli}, {Linker}, \& {De Toma}}]{Torok+2011_ApJ}
{T{\"o}r{\"o}k}, T., {Panasenco}, O., {Titov}, V.~S., {et~al.} 2011, \apjl,
  739, L63, \dodoi{10.1088/2041-8205/739/2/L63}

\bibitem[{{Tristan} {et~al.}(2023){Tristan}, {Notsu}, {Kowalski}, {Brown},
  {Wisniewski}, {Osten}, {Vrijmoet}, {White}, {Carter}, {Grady}, {Henry},
  {Hinojosa}, {Lomax}, {Neff}, {Paredes}, \& {Soutter}}]{Tristan+2023_ApJ}
{Tristan}, I.~I., {Notsu}, Y., {Kowalski}, A.~F., {et~al.} 2023, \apj, 951, 33,
  \dodoi{10.3847/1538-4357/acc94f}

\bibitem[{{Usoskin} \& {Kovaltsov}(2021)}]{Usoskin+2021}
{Usoskin}, I.~G., \& {Kovaltsov}, G.~A. 2021, \grl, 48, e94848,
  \dodoi{10.1029/2021GL094848}

\bibitem[{{Vanderspek} {et~al.}(2018){Vanderspek}, {Doty}, {Fausnaugh},
  {Villasenor}, {Jenkins}, {Berta-Thompson}, {Burke}, \&
  {Ricker}}]{Vanderspek+2018}
{Vanderspek}, R., {Doty}, J.~P., {Fausnaugh}, M., {et~al.} 2018, {TESS
  Instrument Handbook}, Space Telescope Science Institute, Baltimore

\bibitem[{{Vernazza} {et~al.}(1981){Vernazza}, {Avrett}, \&
  {Loeser}}]{Vernazza+1981}
{Vernazza}, J.~E., {Avrett}, E.~H., \& {Loeser}, R. 1981, \apjs, 45, 635,
  \dodoi{10.1086/190731}

\bibitem[{{Veronig} {et~al.}(2021){Veronig}, {Odert}, {Leitzinger}, {Dissauer},
  {Fleck}, \& {Hudson}}]{Veronig+2021}
{Veronig}, A.~M., {Odert}, P., {Leitzinger}, M., {et~al.} 2021, Nature
  Astronomy, 5, 697, \dodoi{10.1038/s41550-021-01345-9}

\bibitem[{{Vial} \& {Engvold}(2015)}]{Vial+2015_ASSL}
{Vial}, J.-C., \& {Engvold}, O. 2015, Astrophysics and Space Science Library,
  Vol. 415, {Solar Prominences} (Springer International Publishing
  Switzerland), \dodoi{10.1007/978-3-319-10416-4}

\bibitem[{{Vida} {et~al.}(2019){Vida}, {Leitzinger}, {Kriskovics}, {Seli},
  {Odert}, {Kov{\'a}cs}, {Korhonen}, \& {van Driel-Gesztelyi}}]{Vida+2019}
{Vida}, K., {Leitzinger}, M., {Kriskovics}, L., {et~al.} 2019, \aap, 623, A49,
  \dodoi{10.1051/0004-6361/201834264}

\bibitem[{{Vida} {et~al.}(2016){Vida}, {Kriskovics}, {Ol{\'a}h}, {Leitzinger},
  {Odert}, {K{\H{o}}v{\'a}ri}, {Korhonen}, {Greimel}, {Robb}, {Cs{\'a}k}, \&
  {Kov{\'a}cs}}]{Vida+2016}
{Vida}, K., {Kriskovics}, L., {Ol{\'a}h}, K., {et~al.} 2016, \aap, 590, A11,
  \dodoi{10.1051/0004-6361/201527925}

\bibitem[{{Vidotto}(2021)}]{Vidotto_2021}
{Vidotto}, A.~A. 2021, Living Reviews in Solar Physics, 18, 3,
  \dodoi{10.1007/s41116-021-00029-w}

\bibitem[{{Villadsen} \& {Hallinan}(2019)}]{Villadsen+2019}
{Villadsen}, J., \& {Hallinan}, G. 2019, \apj, 871, 214,
  \dodoi{10.3847/1538-4357/aaf88e}

\bibitem[{{{\v{S}}vestka} {et~al.}(1962){{\v{S}}vestka}, {Kopeck{\'y}}, \&
  {Blaha}}]{Svestka+1962}
{{\v{S}}vestka}, Z., {Kopeck{\'y}}, M., \& {Blaha}, M. 1962, Bulletin of the
  Astronomical Institutes of Czechoslovakia, 13, 37

\bibitem[{{Wang} {et~al.}(2003){Wang}, {Hildebrand}, {Hobbs}, {Heimsath},
  {Kelderhouse}, {Loewenstein}, {Lucero}, {Rockosi}, {Sandford}, {Sundwall},
  {Thorburn}, \& {York}}]{Wang+2003}
{Wang}, S.-i., {Hildebrand}, R.~H., {Hobbs}, L.~M., {et~al.} 2003, in Society
  of Photo-Optical Instrumentation Engineers (SPIE) Conference Series, Vol.
  4841, Instrument Design and Performance for Optical/Infrared Ground-based
  Telescopes, ed. M.~{Iye} \& A.~F.~M. {Moorwood}, 1145--1156,
  \dodoi{10.1117/12.461447}

\bibitem[{{Warmuth} \& {Mann}(2016)}]{Warmuth+2016}
{Warmuth}, A., \& {Mann}, G. 2016, \aap, 588, A115,
  \dodoi{10.1051/0004-6361/201527474}

\bibitem[{{Watanabe} {et~al.}(2017){Watanabe}, {Kitagawa}, \&
  {Masuda}}]{Watanabe+2017}
{Watanabe}, K., {Kitagawa}, J., \& {Masuda}, S. 2017, \apj, 850, 204,
  \dodoi{10.3847/1538-4357/aa9659}

\bibitem[{{Wollmann} {et~al.}(2023){Wollmann}, {Heinzel}, \&
  {Kab{\'a}th}}]{Wollmann+2023_A&A}
{Wollmann}, J., {Heinzel}, P., \& {Kab{\'a}th}, P. 2023, \aap, 669, A118,
  \dodoi{10.1051/0004-6361/202244544}

\bibitem[{{Wood} {et~al.}(2016){Wood}, {Howard}, \& {Linton}}]{Wood+2016}
{Wood}, B.~E., {Howard}, R.~A., \& {Linton}, M.~G. 2016, \apj, 816, 67,
  \dodoi{10.3847/0004-637X/816/2/67}

\bibitem[{{Wood} {et~al.}(2021){Wood}, {M{\"u}ller}, {Redfield}, {Konow},
  {Vannier}, {Linsky}, {Youngblood}, {Vidotto}, {Jardine},
  {Alvarado-G{\'o}mez}, \& {Drake}}]{Wood+2021}
{Wood}, B.~E., {M{\"u}ller}, H.-R., {Redfield}, S., {et~al.} 2021, \apj, 915,
  37, \dodoi{10.3847/1538-4357/abfda5}

\bibitem[{{Woods} {et~al.}(2006){Woods}, {Kopp}, \& {Chamberlin}}]{Woods+2006}
{Woods}, T.~N., {Kopp}, G., \& {Chamberlin}, P.~C. 2006, Journal of Geophysical
  Research (Space Physics), 111, A10S14, \dodoi{10.1029/2005JA011507}

\bibitem[{{Wu} {et~al.}(2022){Wu}, {Chen}, {Tian}, {Zhang}, {Shi}, {He}, {Lu},
  {Xu}, \& {Wang}}]{Wu+2022}
{Wu}, Y., {Chen}, H., {Tian}, H., {et~al.} 2022, \apj, 928, 180,
  \dodoi{10.3847/1538-4357/ac5897}

\bibitem[{{Yamashiki} {et~al.}(2019){Yamashiki}, {Maehara}, {Airapetian},
  {Notsu}, {Sato}, {Notsu}, {Kuroki}, {Murashima}, {Sato}, {Namekata},
  {Sasaki}, {Scott}, {Bando}, {Nashimoto}, {Takagi}, {Ling}, {Nogami}, \&
  {Shibata}}]{Yamashiki+2019}
{Yamashiki}, Y.~A., {Maehara}, H., {Airapetian}, V., {et~al.} 2019, \apj, 881,
  114, \dodoi{10.3847/1538-4357/ab2a71}

\bibitem[{{Yashiro} \& {Gopalswamy}(2009)}]{Yashiro+2009}
{Yashiro}, S., \& {Gopalswamy}, N. 2009, in Universal Heliophysical Processes,
  ed. N.~{Gopalswamy} \& D.~F. {Webb}, Vol. 257, 233--243,
  \dodoi{10.1017/S1743921309029342}

\bibitem[{{Zacharias} {et~al.}(2013){Zacharias}, {Finch}, {Girard}, {Henden},
  {Bartlett}, {Monet}, \& {Zacharias}}]{Zacharias+2013}
{Zacharias}, N., {Finch}, C.~T., {Girard}, T.~M., {et~al.} 2013, \aj, 145, 44,
  \dodoi{10.1088/0004-6256/145/2/44}

\bibitem[{{Zhu} {et~al.}(2019){Zhu}, {Kowalski}, {Tian}, {Uitenbroek},
  {Carlsson}, \& {Allred}}]{Zhu+2019}
{Zhu}, Y., {Kowalski}, A.~F., {Tian}, H., {et~al.} 2019, \apj, 879, 19,
  \dodoi{10.3847/1538-4357/ab2238}

\bibitem[{{Zic} {et~al.}(2020){Zic}, {Murphy}, {Lynch}, {Heald}, {Lenc},
  {Kaplan}, {Cairns}, {Coward}, {Gendre}, {Johnston}, {MacGregor}, {Price}, \&
  {Wheatland}}]{Zic+2020}
{Zic}, A., {Murphy}, T., {Lynch}, C., {et~al.} 2020, \apj, 905, 23,
  \dodoi{10.3847/1538-4357/abca90}

\end{thebibliography}



\end{document}